\documentclass[10pt,a4paper,twoside,openright,titlepage,fleqn,%
               headinclude,,footinclude,BCOR5mm,%
               numbers=noenddot,cleardoublepage=empty,%
               tablecaptionabove]{scrreprt}
\usepackage[italian,american]{babel}
\usepackage[utf8]{inputenc}
\DeclareUnicodeCharacter{2212}{-}
\usepackage[T1]{fontenc}
\usepackage{amsmath,amssymb,amsthm}
\usepackage{varioref}
\usepackage[hyperref,natbib]{biblatex}
\usepackage{chngpage}
\usepackage{calc}
\usepackage{listings}
\usepackage{graphicx}
\usepackage[dvipsnames]{color}
\usepackage{subfig}
\usepackage{bm}
\usepackage{multicol}
\usepackage{makeidx}
\usepackage{fixltx2e}
\usepackage{relsize}
\usepackage{lipsum}
\usepackage[eulerchapternumbers,subfig,beramono,pdfspacing,listings]{classicthesis}
\usepackage{arsclassica}
\usepackage{slashed}
\usepackage{mathtools}
\usepackage{amsfonts}
\usepackage{cancel}
\usepackage[normalem]{ulem}
\usepackage{cleveref}
\usepackage{tikz}
\usetikzlibrary{patterns}
\usetikzlibrary{arrows.meta}
\usepackage{environ}
\usepackage{float}
\usepackage{multirow}
\usepackage{enumitem}
\usepackage{booktabs}

\newcommand*{\diffdchar}{\mathrm{d}}
\newcommand*{\dd}{\mathop{\diffdchar\!}}
\newcommand{\me}{\mathrm{e}}
\newcommand{\mev}{\text{ MeV}}
\newcommand{\gev}{\text{ GeV}}
\newcommand{\tev}{\text{ TeV}}

\newcommand{\red}{\textcolor{BrickRed}}

\DeclarePairedDelimiter\abs{\lvert}{\rvert}%

\makeatletter
\newsavebox{\measure@tikzpicture}
\NewEnviron{scaletikzpicturetowidth}[1]{%
  \def\tikz@width{#1}%
  \def\tikzscale{1}\begin{lrbox}{\measure@tikzpicture}%
  \BODY
  \end{lrbox}
  \pgfmathparse{#1/\wd\measure@tikzpicture}%
  \edef\tikzscale{\pgfmathresult}%
  \BODY
}
\makeatother
\usepackage{tcolorbox}

\newcommand\sbullet[1][.5]{\mathbin{\vcenter{\hbox{\scalebox{#1}{$\bullet$}}}}}

\newcommand{\myName}{Daniel T. Murnane}
\newcommand{\myTitle}{The Landscape of Composite Higgs Models}
\newcommand{\mySubTitle}{Thesis submitted for \hspace{3em} Doctorate of Philosophy}

\newcommand{\myGroup}{Centre of Excellence in Particle Physics, Faculty of Physical Sciences, University of Adelaide}
\newcommand{\myUrl}{\url{http://coepp.org.au}}
\newcommand{\myTime}{2019, November}

\newcommand{\mail}[1]{\href{mailto:#1}{\texttt{#1}}}

\let\orgtheindex\theindex
\let\orgendtheindex\endtheindex
\def\theindex{%
	\def\twocolumn{\begin{multicols}{2}}%
	\def\onecolumn{}%
	\clearpage
	\orgtheindex
}
\def\endtheindex{%
	\end{multicols}%
	\orgendtheindex
}

\makeindex

\definecolor{lightergray}{gray}{0.99}

\lstnewenvironment{code}%
{\setkeys{lst}{columns=fullflexible,keepspaces=true}%
\lstset{basicstyle=\small\ttfamily}%
}{}

\lstnewenvironment{sidebyside}{%
    \global\let\lst@intname\@empty 
    \setbox\z@=\hbox\bgroup 
    \setkeys{lst}{columns=fullflexible,%
    linewidth=0.45\linewidth,keepspaces=true,%
    breaklines=true,%
    breakindent=0pt,%
    boxpos=t,%
    basicstyle=\small\ttfamily
}%
    \lst@BeginAlsoWriteFile{\jobname.tmp}%
}{%
    \lst@EndWriteFile\egroup 
        \begin{center}%
            \begin{minipage}{0.45\linewidth}%
                \hbox to\linewidth{\box\z@\hss} 
            \end{minipage}%
            \qquad 
            \begin{minipage}{0.45\linewidth}%
            \setkeys{lst}{frame=none}%
                \leavevmode \catcode`\^^M=5\relax 
                \small\input{\jobname.tmp}%
            \end{minipage}%
        \end{center}%
}

\graphicspath{{Graphics/}}

\bibliography{Bibliography,BibliographyExperimental}

\defbibheading{bibliography}{%
\cleardoublepage
\manualmark
\phantomsection
\addcontentsline{toc}{chapter}{\tocEntry{\bibname}}
\chapter*{\bibname\markboth{\spacedlowsmallcaps{\bibname}}
{\spacedlowsmallcaps{\bibname}}}}     

  \DeclareCiteCommand{\citeyearpar}[\mkbibparens] 
  {\boolfalse{citetracker}%
   \boolfalse{pagetracker}%
   \usebibmacro{prenote}} 
  {\printtext[bibhyperref]{\printfield{year}}} 
  {\multicitedelim} 
  {\usebibmacro{postnote}} 

\makeatletter 
  \DeclareCiteCommand{\citetalias} 
  {\usebibmacro{prenote}} 
  {\usebibmacro{citeindex}%
   \bibhyperref{\@citealias{\thefield{entrykey}}}} 
  {\multicitedelim} 
  {\usebibmacro{postnote}} 
\makeatother 

\usepackage{chngcntr}
\counterwithout{equation}{chapter}

\begin{document}
\pagenumbering{roman}
\pagestyle{plain}
\begin{titlepage}
\pdfbookmark{Titlepage}{Titlepage}
\changetext{}{}{}{((\paperwidth  - \textwidth) / 2) - \oddsidemargin - \hoffset - 1in}{}
\null\vfill
\begin{center}
\large
\sffamily

\bigskip

{\Large\spacedlowsmallcaps{\myName}} \\

\bigskip

{\huge\spacedlowsmallcaps{\myTitle} \\
}

\bigskip
    
\vspace{9cm}

\begin{tabular} {cc}
\parbox{0.4\textwidth}{\includegraphics[width=5cm]{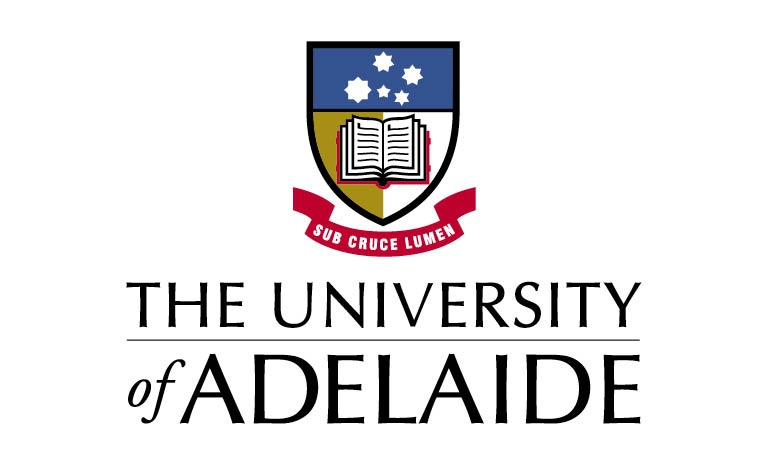}}
&
\parbox{0.48\textwidth}{{\Large\spacedlowsmallcaps{\mySubTitle}} \\ 

					{\normalsize
					
					\myGroup \\
					\myUrl \\
					\myTime}}
			\end{tabular}
\end{center}
\vfill
\end{titlepage}

\thispagestyle{empty}

\hfill

\vfill

\noindent\myName:
\textit{\myTitle,} \mySubTitle,
\textcopyright\ \myTime.

\medskip
\noindent{\spacedlowsmallcaps{Website}}: \\
\url{http://adelaide.academia.edu/DanielMurnane}

\medskip
\noindent{\spacedlowsmallcaps{E-mail}}: \\
\mail{daniel.murnane@adelaide.edu.au}

\vspace{1cm}
\hrule
\bigskip

\clearpage

\pdfbookmark{Abstract}{Abstract}
\begingroup
\let\clearpage\relax
\let\cleardoublepage\relax
\let\cleardoublepage\relax

\chapter*{Abstract}
While the Standard Model (SM) of particle physics contains the most precise set of predictions ever devised by humanity, that precision comes at a cost. The strange nature of the Higgs particle requires its parameters to be tuned so precisely that if the SM is indeed the true description of reality, one is forced to wonder how such a miracle as galactic structure and life could occur. Instead, we search in this work for a \textit{natural} explanation. The concept of naturalness is comprehensively explored, and a new tuning measure proposed, with an aim to place it on well-defined Bayesian footing. We then turn this measure on to the analysis of a class of intriguing new physics - Composite Higgs models. These effective models are the result of a plethora of underlying theories, and they allow the production of a naturally light Higgs particle, appearing as the SM Higgs at low energy. We establish the background required to appreciate the N-site 4D Composite Higgs model, and subsequently focus on the simplest incarnations of this class. A global fit is performed on the Minimal 4D Composite Higgs model (M4DCHM), with strong exclusion bounds placed on collider search channels. We analyse any improvement in tuning that could be gained from several extensions to this model. The Leptonic M4DCHM is explored, with a composite tau lepton embedded in various representations. The possibility of a dark matter candidate existing in the Next-to-Minimal 4DCHM is considered. Ultimately, we are able to define what, if any, benefit to naturalness can come to the Composite Higgs sector by introducing these extensions.

\selectlanguage{american}

\endgroup			

\vfill

\clearpage
\pdfbookmark{Acknowledgements}{Acknowledgements}

\begingroup
\let\clearpage\relax
\let\cleardoublepage\relax
\let\cleardoublepage\relax

\chapter*{Acknowledgements}

Drawing the blood from this particular stone was hugely aided by the companionship I had throughout my PhD years. My various officemates Zach Koumi, Andre Scaffidi, Ankit Beniwal, Filip Rajec, Skye Platten, Nicolas Ivancevic, have bemoaned the process with me, and we have suffered sympathy pains together. Ethan Carragher was a delight to teach and learn from, and helped me rediscover many joys in the material. He also helped to produce much of the recent published material, along with Peter Stangl, and earlier James Barnard. I thank Peter for reading this thesis and sharing deep insights, and James for grounding me in this topic.  All of the Adelaide physics faculty were endlessly helpful, and I thank Fabien Voisin for very late nights of cluster debugging.

When in Denmark, the faculty of CP3 at Southern Denmark University were so welcoming. I thank Francesco Sannino for hosting me, Anders Eller Thomsen for also reading this thesis, and the rest of the institute for giving me a home away from home.

I deeply thank my two supervisors Anthony Williams and Martin White. Martin was a hurricane of activity that seemed to have endless energy and resources to help me, and Tony was the eye of that hurricane - peaceful and thoughtful. 

I thank the George Glass gang for giving me the perfect excuses to procrastinate, travel, and party. In particular, I thank Nic and Trephina for giving me a roof during some of the rougher PhD years. I also thank Natascha's family for providing a roof and bed. And of course, thanks Natascha, for keeping me alive and happy, and giving meaning to the whole damn thing.

Finally, thanks to my family. Dad, Molly and Annie. It's a shame you just missed out on it, Mum.

\endgroup

\pdfbookmark{Declaration}{Declaration}

\begingroup
\let\clearpage\relax
\let\cleardoublepage\relax
\let\cleardoublepage\relax

\chapter*{Declaration}

I certify that this work contains no material which has been accepted for the award of any other degree or diploma in my name, in any university or other tertiary institution and, to the best of my knowledge and belief, contains no material previously published or written by another person, except where due reference has been made in the text. In addition, I certify that no part of this work will, in the future, be used in a submission in my name, for any other degree or diploma in any university or other tertiary institution without the prior approval of the University of Adelaide and where applicable, any partner institution responsible for the joint-award of this degree.

I acknowledge that copyright of published works contained within this thesis resides with the copyright holder(s) of those works.
I also give permission for the digital version of my thesis to be made available on the web, via the University's digital research repository, the Library Search and also through web search engines, unless permission has been granted by the University to restrict access for a period of time. I acknowledge the support I have received for my research through the provision of an Australian Government Research Training Program Scholarship.
\begin{flushright}
\includegraphics[scale=0.05]{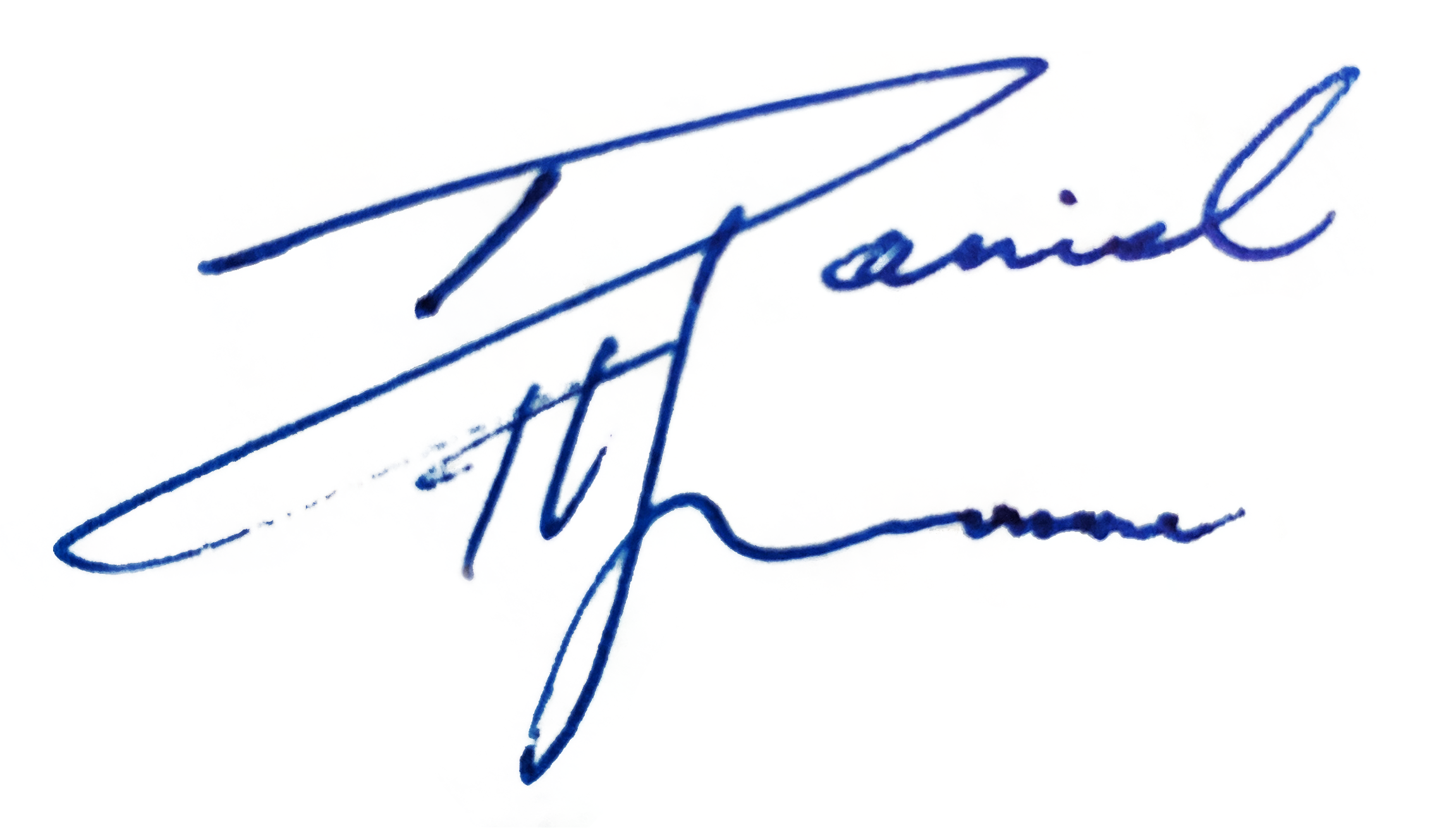}
\hspace*{0.5cm} - Daniel Thomas Murnane
\end{flushright}
\endgroup

\pagestyle{scrheadings} 
\clearpage
\phantomsection
\pdfbookmark{\contentsname}{tableofcontents}
\setcounter{tocdepth}{1}
\begingroup 
    \let\clearpage\relax
    \let\cleardoublepage\relax

    \tableofcontents
\endgroup
\markboth{\spacedlowsmallcaps{\contentsname}}{\spacedlowsmallcaps{\contentsname}} 

\begingroup 
    \let\clearpage\relax
    \let\cleardoublepage\relax
\endgroup

\cleardoublepage

\listoffigures
\listoftables
\pagenumbering{arabic}
\chapter{Introduction}
\label{chp:introduction}

\begin{figure}
\centering
\includegraphics[scale=0.2]{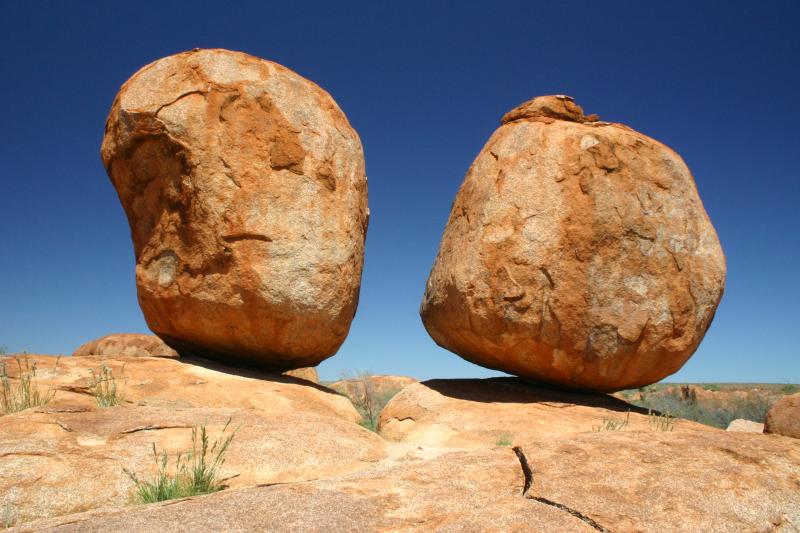}
\caption{A pair of balancing rocks (also called precarious boulders) in the \textit{Karlu Karlu} (Devil's Marbles) Conservation Park}\label{fig:devils_marbles}
\end{figure}

In the centre of Australia's Northern Territory, in the \textit{Karlu Karlu} Conservation Reserve, there are two huge boulders resting on a crest of a slope (see \cref{fig:devils_marbles}). These rocks are balancing on a raised point, and it seems like a small push would topple them. The theory of how they came to be is captured by a Dreamtime story of the local  \textit{Alyawarre} people. The story tells of the \textit{Arrange} - the Devil Man - who walked through the reserve, wearing a belt made of hair. As he walked, he would spit, and spin the belt to pass the time. The hairs and spit that fell turned to stone as they landed on the earth. These Devil's Marbles constitute the many precariously balanced rocks across the reserve.

The Standard Model (SM) has delivered overwhelmingly accurate predictions of some of the most precise experiments ever conducted. And with the discovery of the Higgs boson, it has \textit{explained} the outcome of those experiments with an elegant mechanism - the spontaneous breaking of the Electroweak (EW) Symmetry. But introducing the elementary scalar Higgs boson comes at a price - it must receive quantum corrections to its mass from all scales of the SM. It is reasonable to assume that the range of validity of the SM is cut off at some scale, as we know there are high energy phenomena it cannot explain - gravity, for example. At such a scale, there must be some ultraviolet completion to the SM. We can then state the behaviour of the Higgs relative to this scale. For example, a one-loop contribution to the Higgs mass from the top quark appears as
\begin{align}
\delta \mu^2 = -\frac{3y_t^2}{8\pi^2}\Lambda^2 + \mathcal{O}(\log(\Lambda)) + ...
\end{align}
with $y_t$ a function of the top-Higgs Yukawa coupling, and $\Lambda$ the scale where the current Standard Model is no longer applicable. We would hope that, for the SM to be a broad and useful theory, $\Lambda \gg \Lambda_\text{EW}$, where the EW scale is of order $100$ GeV. In the Planck scale limit, this leads to corrections of 43 orders of magnitude greater than the Higgs mass. This obviously requires a coupling function of order $10^{-43}$ to suppress the correction. If this is the true description of reality, this unnatural situation may have come into being at the moment of the universe's creation, or evolved dynamically. 

Either way, like the Devil's Marbles, the Higgs boulder has settled precariously on a slope that one would not expect, had it simply been dropped. The dynamics of the fall must have been chosen so precisely, and tuned to perfection, to lead to this scenario. Any lower, it would topple and we would live in a world without mass. Any higher, we would have supermassive particles and short neutron decay times, preventing structure in the universe. This is not to say that either the Dreamtime theory or the SM Higgs mechanism are naive. Indeed, they are both elegant and intuitive. But one must wonder \textit{why} these boulders, if they were simply dropped from the sky, came to rest so perfectly. 

This balancing act is called the Hierarchy Problem, and it presents a glaring philosophical shortcoming of the otherwise-elegant SM. Other problems certainly exist in the SM - the hierarchy of heavy quarks to light quarks, for example. But this hierarchy is not a requirement for life. It seems that we \textit{need} the miraculous $\mathcal{O}(\Lambda_{m_\text{electron}}) < \mathcal{O}(\Lambda_\text{EW}) \ll \mathcal{O}(\Lambda_\text{Planck})$ ratio to survive, if the SM is to be believed. 

%
%

Of course, one can cite the almost universal belief that the SM is an effective theory. That there is likely a larger, more complicated model with the SM as its low-energy limit. But in the future one could imagine the creation of a consistent theory including the Standard Model and General Relativity that requires astronomical fine-tuning to account for the gravitational hierarchy problem. In this case, we would have no excuse that the theory is ``undoubtedly effective''. Instead, we might continue searching for another theory that describes the universe equally well, motivated by nothing more than naturalness. This may include new forces, new fields and new parameters. Each additional feature may lower the overall tuning, but Occam would take issue at these additions to the economical SM. Is introducing epicycles to reduce fine-tuning a satisfying approach, and how does one quantify this trade-off?

In this thesis, I will attempt to walk the line between increasing the content of some extension to the SM Higgs model and increasing its naturalness. The extension of interest here is in describing the Higgs as a composite object, effectively described as the naturally light bound state from some high-energy spontaneous symmetry breaking. This paradigm stems from two eras of research. The first was inspired by results of the QCD-Chiral Symmetry connection, when new sets of (techni)quarks and (techni)colors were introduced in the 1980s to include a Higgs boson as a new, composite bound state, analogous to the pion. This research programme stalled due to the endless introduction of hierarchies and tunings, which seemed to defeat the purpose of the exercise. The second era emerged after the discovery of a duality between weakly-interacting 5D theories, and strongly-interacting 4D theories - the \textit{AdS-CFT correspondence}. Motivated by the ability to unify the EW fields and the Higgs field in 5D, models were created that appeared effectively as composite Higgs models in 4D. Modern techniques (such as the ability to calculate in the 5D paradigm, understanding of partial compositeness, and lattice techniques) mean that 4D Composite Higgs Models (4DCHM) are much better understood than their ancestors. This understanding brings the possibility of lower tuning in the Higgs sector.


Exactly how much lower is one concern of this thesis. Accordingly, we focus on two simple symmetry breaking processes: the minimal, and the next-to-minimal 4D Composite Higgs Model (M4DCHM, NM4DCHM). The minimal case prescribes two composite sectors with $SO(5) \rightarrow SO(4) \rightarrow SU(2)_L \times U(1)_Y \rightarrow U(1)_\text{EM}$ symmetry breaking. $SO(5) \rightarrow SO(4)$ can be considered the smallest group breaking that preserves the observed behaviour of the Standard Model. From only symmetry arguments, and with a specific choice of matter embedding, we can determine the generic contributions to Standard Model masses and Higgs potential. Within the minimal model, we also explore varying representations of fermions, as well as including leptons as composite (LM4DCHM). The NM4DCHM follows the same 2-site prescription, but with a larger symmetry of $SO(6) \rightarrow SO(5) \rightarrow SU(2)_L \times U(1)_Y \rightarrow U(1)_\text{EM}$. This process generates one extra degree of freedom, which can be used to describe a potential dark matter candidate - an EW singlet scalar particle with a mass naturally below the $1\tev$ scale. 

CHMs are a powerful way to describe the low mass of the SM Higgs - a pseudo-Nambu-Goldstone Boson that parameterises an approximate non-linear global symmetry. Unfortunately, the M4DCHM suffers from one shortcoming that renders its tuning somewhat larger. This comes from a fine cancellation required to reproduce the correct EW vacuum expectation value, called the "double tuning" problem. Naively including composite matter that gives the composite Higgs its potential tends to break EW symmetry at the composite scale - well above the relatively low $246\gev$ that we measure in the LHC. In this work, we explore several potential solutions to this problem. The first is the inclusion of partially composite leptons that carry an accidental cancellation of the sort we require. The second is the NM4DCHM discussed above. 

To quantify precisely the degree to which the double tuning is mitigated, we introduce a higher order tuning measure. This is an improvement on the usual Barbieri-Giudice tuning, which simply finds the maximum amount of deviation of observables as the model's parameters are changed. The higher order tuning described herein automatically accounts for more complex tuning of the sort found in CHMs.

This work is broadly divided into two parts: first, a pedagogical introduction in chapters \ref{chp:standard_model} and \ref{sec:composite_higgs} which, along with appendices \ref{sec:group_theory} - \ref{sec:tuning_appendix}, form an overview of 4D Composite Higgs models assuming basic understanding of QFT; and second, a description in chapters \ref{sec:fine_tuning} - \ref{sec:NMCHM}, along with appendices \ref{chp:particle_content} and \ref{sec:form_factors} of the research done during this thesis on tuning in the M4DCHM, LM4DCHM and NM4DCHM.

We begin in \cref{chp:standard_model} by describing standard results and methods that will be necessary to understand the Composite Higgs paradigm: the Standard Model, linear and non-linear sigma models, the Coleman-Weinberg effective potential and the CCWZ description of non-linear realisation of symmetries. We review the current state of the SM, and current attempts to address the Hierarchy Problem. As a warm-up, Dynamical Electroweak Symmetry Breaking is reviewed, and shown to be insufficient.

In \cref{sec:composite_higgs}, we construct more and more realistic CHMs - beginning with a \textit{very} minimal non-linear sigma model, and culminating in a two-site 4DCHM with calculable potential and explicit representations of the matter sector. At this point, the double tuning will be naively calculated, and we will explore in \cref{sec:fine_tuning} descriptions of tuning that will lead to a consistent measure of all orders and sources of tuning. Calculating the Higgs potential of 4DCHMs is computationally intensive, and we review sophisticated methods of exploring the parameter space in \cref{sec:scanning_techniques}. 

In \cref{sec:M4DCHM}, we give a reference of the particular model structure used in this work, and the experienced reader could jump to this section. The tuning and scanning methods are used to explore the M4DCHM, in order to perform the first convergent global fit of this model in the literature. This model is extended to include leptons in non-fundamental representations, and \cref{sec:LCHM} with \cref{chp:particle_content} serve as a reference for these models. We report possible improvements in tuning theorised by accidental cancellations between representations. Finally, in \cref{sec:NMCHM}, we examine what tuning reduction could come from extending to the next-to-minimal model. This contains the possibility of the Higgs existing as an admixture of scalar states, and this freedom may lead to a lower tuning. We discuss the possibility of the admixture containing a dark matter candidate, and what properties it may possess. The models considered in this work range from minimal in description, to hugely parameterised, and to understand how the tuning behaves under each addition of complexity is to understand how the field may progress. An increase in tuning with less minimal models is to defeat one main purpose of considering these models in the first place. A decrease or stability of tuning with less minimal models may strongly justify the continued development of this class of Composite Higgs model.

\chapter{Fundamentals}
\label{chp:standard_model}
\section{Visible Symmetries}

\paragraph{Our goal}
\parbox{0.8\textwidth}{To establish the symmetries and fields that we know effectively exist at low energies. }
\vspace*{1em}

The Standard Model is the answer that particle physics gives to the question of a hundred years of extremely precise  experimental observation. The elegantly compact Lagrangian

\begin{align}
\mathcal{L} &= \bar{\psi}^f(\slashed{\partial}_\mu - i g^f_1 \slashed{W}^a T^a - i g^f_2 \slashed{B} -i g^f_3 \slashed{G}^c T^c) \psi^f \nonumber\\
& - \frac{1}{4} G_c^{\mu\nu} G^c_{\mu\nu} - \frac{1}{4} W^{\mu\nu}_aW^a_{\mu\nu}  - \frac{1}{4} B^{\mu\nu}B_{\mu\nu} \\
&+ Y^{ff'} \psi^f_L \Phi \psi^{f'}_R + \left[(\partial_\mu - i g^f_1 W^a T^a - i g^f_2 B)\Phi\right]^2 + m \Phi \Phi^\dagger - \frac{\lambda}{4}(\Phi \Phi^\dagger)^2 \nonumber
\end{align}

can be made to produce observables that match all (barring a few notable exceptions) collider and precision experiments to within experimental and theoretical uncertainty. Each term is described in \cref{tab:matter_fields} and \cref{tab:gauge_fields}. As a disclaimer, this thesis is not concerned with solving the problem of experimental mismatch, but rather with tackling the incompleteness inherent in the Lagrangian above.

The SM Lagrangian achieves this monumental feat by Gauge Symmetry Unification, and subsequently, Symmetry Hiding. The first is by requiring the 15 matter fields contained within $\psi^f$ to obey a unified local symmetry group\footnote{See \cref{sec:group_theory} for a review of group and representation theory}
\begin{align}
\mathcal{G}_\text{SM} = SU(3)_\textnormal{colour} \times SU(2)_\textnormal{weak} \times U(1)_\textnormal{hypercharge}
\end{align}
The particular representations these matter fields appear in have been experimentally determined to be the fundamental representations, as given in \cref{tab:matter_fields}, \footnote{Note the particular convention used here, $Y = Q - T_3$} where each field $\psi$ is a Dirac fermion field composed of independent Weyl fields $\psi_L$ and the charge conjugate of $\psi_R$, $\psi_R^c$. We might also add the instanton and sphaleron as quantised solutions (if not quite "particles") to this list of physical matter arising from the SM.

\begin{table}
\begin{center}
\begin{tabular}{ @{} r l l l @{} }
\toprule
$\psi^f$ & $SU(3)_c$ irrep & $SU(2)_W$ irrep &  $U(1)_Y$ charge\\ \midrule
$l^\mu_1 = (e^\mu_L, \nu^\mu_e)$ & $\bm{1}$ & $\textbf{2}$ & $-1/2$\\
$l^\mu_2 = (\mu^\mu_L, \nu^\mu_\mu)$ & $\bm{1}$ & $\textbf{2}$ & $-1/2$\\
$l^\mu_3 = (\tau^\mu_L, \nu^\mu_\tau)$ & $\bm{1}$ & $\textbf{2}$ & $-1/2$\\ \midrule
$e^\mu_R$ & $\bm{1}$ & $\textbf{1}$ & $-1$\\
$\mu^\mu_R$ & $\bm{1}$ & $\textbf{1}$ & $-1$\\
$\tau^\mu_R$ & $\bm{1}$ & $\textbf{1}$ & $-1$\\ \midrule
$q^\mu_1 = (u^\mu_L, d^\mu_L)$ & $\bm{3}$ & $\bm{2}$ & $1/6$\\
$q^\mu_2 = (c^\mu_L, s^\mu_L)$ & $\bm{3}$ & $\bm{2}$ & $1/6$\\
$q^\mu_3 = (t^\mu_L, b^\mu_L)$ & $\bm{3}$ & $\bm{2}$ & $1/6$\\ \midrule
$u^\mu_R = (u^\mu_r, u^\mu_g, u^\mu_b)$ & $\bm{3}$ & $\bm{1}$ & $2/3$\\
$d^\mu_R = (d^\mu_r, d^\mu_g, d^\mu_b)$ & $\bm{3}$ & $\bm{1}$ & $-1/3$\\
$c^\mu_R = (c^\mu_r, c^\mu_g, c^\mu_b)$ & $\bm{3}$ & $\bm{1}$ & $2/3$\\
$s^\mu_R = (s^\mu_r, s^\mu_g, s^\mu_b)$ & $\bm{3}$ & $\bm{1}$ & $-1/3$\\
$t^\mu_R = (t^\mu_r, t^\mu_g, t^\mu_b)$ & $\bm{3}$ & $\bm{1}$ & $2/3$\\
$b^\mu_R = (b^\mu_r, b^\mu_g, b^\mu_b)$ & $\bm{3}$ & $\bm{1}$ & $-1/3$\\
\bottomrule
\end{tabular}\caption{Matter fields in the SM Lagrangian. There are $3$ colours $\times$ $2$ chiralities $\times$ $6$ species $\times$ $4$ Lorentz directions $= 144$ components in the quark sector, and $9$ $\times$ $4$ Lorentz $=36$ components in the lepton sector. 
}
\label{tab:matter_fields}
\end{center}
\end{table}

There are many ways to slice these fields, but the usual convention is divide the $15$ fermion ``species" into three ``generations" of five ``flavours". The requirement of local invariance gives rise to gauge fields, enabling the matter field dynamics to remain invariant under $\mathcal{G}_\text{SM}$ gauge transformations. In order to transform correctly, these are embedded in adjoint and trivial representations, given in table \ref{tab:gauge_fields}.

\begin{table}
\begin{center}
\begin{tabular}{@{} r l l l @{}}
\toprule
Gauge field & $SU(3)_c$ irrep & $SU(2)_W$ irrep &  $U(1)_Y$ charge\\ \midrule
$W_{\mu,ij} = W_\mu^a \sigma^a_{ij}$ & $\bm{1}$ & $\bm{3}$ & $0$\\
$B_{\mu,ij} = B_\mu Y_{ij}$ & $\bm{1}$ & $\bm{1}$ & $1$\\
$G_{\mu,ij} = G_\mu^c \lambda^c_{ij}$ & $\bm{8}$ & $\bm{1}$ & $0$ \\ \midrule
$\Phi = \frac{1}{\sqrt{2}}(h^+ + ih^-, H + ih^{0})$ & $\bm{1}$ & $\bm{2}$ & $1/2$\\
\bottomrule
\end{tabular}\caption{Bosonic fields (gauge fields and the Higgs field) in the SM Lagrangian. 
}
\label{tab:gauge_fields}
\end{center}
\end{table}

Finally, note each gauge field $F^a_\mu$ has a field strength tensor $F^a_{\mu\nu} = \partial_\mu F_\nu^a - \partial_\nu F_\mu^a + g f^{abc} F^b_\mu F^c_\nu$, where $f^{abc}$ are the structure constants for the gauge group. Given this set of matter fields, gauge fields, and local symmetries, we notice there are several accidental and approximate \textit{global} symmetries that appear. These should be reproduced in any physics beyond the Standard Model. 

A subtle accidental symmetry is an $SO(4)$ symmetry present in the Higgs sector, called "custodial symmetry", about which much more will be said in \cref{sec:composite_higgs}. 

A much more obvious accidental symmetry is that of quark and lepton number conservation. Lepton number conservation arises from each family of lepton being constrained to interact via terms of the form
\begin{align}
\mathcal{L} \supset -ig\bar{l}^i\slashed{W}^\mu l + l^i_L \phi e_R^i
\end{align}
that is, the kinetic term, and the Yukawa term (where off-diagonal $Y^{ff'}$ are zero). These lead to interactions of the sort
\begin{align}
e^- \rightarrow \nu_L + W^-, \qquad \mu^+ + \mu^- \rightarrow Z
\end{align}
where we can see that the lepton number for each family (the total of leptons minus antileptons) is conserved. These are three separate symmetries - one for each lepton family. There is no mechanism to exchange one family for another in the SM Lagrangian. On the other hand, the quark Lagrangian contains off-diagonal Yukawa matrices, allowing for processes such as
\begin{align}
u \rightarrow s + W^+
\end{align}
thus, a more relaxed invariance exists for the total number of quarks plus antiquarks, across all three families\footnote{Historically, this is called baryon number}. These two sets of symmetries - leptons and baryons - may be broken by sphaleron processes, although the difference of the two will still be conserved \cite{PhysRevD.28.2019}. 

A broader set of important symmetries exist in the limit that the Standard Model is Yukawaless\cite{Willenbrock:2004hu,Grossman:2010gw}. These are called flavour symmetries, since each flavour of field can be rotated by a unitary transformation about its generations as
\begin{align}
q^i_L &\rightarrow U^{ij}_{q_L} q_L^j \nonumber\\
u^i_R &\rightarrow U^{ij}_{u_R} u_R^j \nonumber\\
d^i_R &\rightarrow U^{ij}_{d_R} d_R^j\\
l^i_L &\rightarrow U^{ij}_{l_L} l_L^j \nonumber\\
e^i_R &\rightarrow U^{ij}_{e_R} e_R^j \nonumber
\end{align}
In the Yukawaless limit, this is an exact $\left[U(3)\right]^5$ symmetry\footnote{There is a tendency to say the "massless" limit of the SM. This would be the case for either Yukawa matrices of zeros, or if the Higgs mass is real, thus not breaking the EW group. However, even for a Higgs without a vev (i.e. giving "massless" fermions), the terms $q_L^i Y^{ij} \Phi u_R^j + ...$ still generically break $[U(3)]^5$. We will speak of a Yukawaless limit for precision.}. While the Yukawa couplings break this symmetry down to just $B$ and $L_{e,\mu,\tau}$ number conservation, there is an important approximate symmetry remaining from the extremely light $u$, $d$ and $s$ quarks. This hidden symmetry is the historical motivation for the chiral symmetry model, non-linear sigma model, Technicolor, and many of the results in this work. 


%

If we consider only the fields symmetric under the strong force (neglecting the electroweak symmetry), we see that the Standard Model is classically conformal. That is, it has no explicit scale. Indeed, even the renormalisation group equations relating the coupling strength at any scale $\alpha_S(Q)$ to that of some reference scale $\alpha_S(\mu_0)$ only requires the ratio of the two scales \cite{deur2016qcd}
\begin{align}
\frac{Q^2}{\alpha_S^2}\frac{\partial \alpha_S}{\partial Q^2}= -\frac{1}{4\pi} \beta_0 && \implies && \frac{4\pi}{\alpha_S(\mu_0^2)} - \frac{4\pi}{\alpha_S(Q^2)} = \beta_0 \ln\left(\frac{\mu_0^2}{Q^2}\right)\label{eq:coupling_equation}
\end{align}
Where $\beta_0 = 11 - \frac{2}{3}n_F$. This expression is obtained using a first-order perturbative expansion of QCD, and as such fails once the couplings become strong. However, we can find the scale of its failure $\Lambda$ by taking the limit as the coupling goes to infinity. That is, where
\begin{align}
\frac{4\pi}{\alpha_S(\Lambda^2)} &= \frac{4\pi}{\alpha_S(\mu_0^2)} - \beta_0 \ln\left(\frac{\mu_0^2}{\Lambda^2}\right)=0 \nonumber\\
\implies \Lambda^2 & \sim \mu_0^2 \exp \left( -\frac{4\pi}{\beta_0 \alpha_S(\mu_0^2)}\right)
\end{align}
We can determine the coupling constant at a perturbative scale, say at the mass of the Z, $m_Z \approx 91\; \gev$, using electron-positron annihilation data. This gives a coupling of around $0.12$ \cite{dissertori2016determination}. At this scale, only five of the quark flavours are "active" (that is, lighter than the reference scale), giving $\beta_0 = 23$. Then the scale at which perturbativity has certainly failed is
\begin{align}
\Lambda_{\text{QCD}} &= 91 \exp\left(-\frac{6\pi}{23\times 0.12}\right) \gev \sim \mathcal{O}(10^{-1} \gev)
\end{align}
Finding the boundary between perturbativity and non-perturbativity produced an explicit scale, a phenomenon called "dimensional transmutation"\cite{PhysRevLett.31.851}. Perturbativity is just a mathematical tool, and the scale may a priori have no physical meaning. However, by using tools beyond perturbation theory, the boundary turns out to be a phase transition between hadronic quarks and quark-gluon plasma.

Lattice calculations have shown the scale of QCD confinement (that is, the scale at which quarks, locally symmetric under $SU(3)_c$, will always be found in colourless bound states) to be $160$MeV \cite{PhysRevLett.48.1140}. These same calculations show the scale of Chiral Symmetry condensation (that is,  the scale where an effective field $\varphi = q_R q_L$ globally symmetric under $SU(3)_L \times SU(3)_R$ attains a vacuum expectation value $\langle q_R q_L \rangle = v$ and hides the full symmetry) to be $260$MeV. Although not precisely the same scale, there naively appears to be a connection between these two phenomena, given that they could have independently occurred at any scale up to the Planck scale. The identification of these two phase transitions is the key to the Standard Model, and a playbook for Dimensional Transmutation (how a scale can appear spontaneously), effective field theories (how a strong local symmetry can be studied as a weak global one), and non-linear symmetries (how a symmetry can be hidden, in order to generate mass).

Compare this emergent phenomenon with the case of the weakly gauged $SU(2)_L \times U(1)_Y$. Without confinement, the electroweak group requires an external scalar to play the part of a condensate, and induce a phase transition. This is the famous Higgs mechanism that hides the full $SU(2)_L \times U(1)_Y$ symmetry in the physical vacuum. 

\section{Hidden Symmetries}

\paragraph{Our goal}
\parbox{0.8\textwidth}{To provide the set of tools and intuitions required to understand the interplay between hidden global symmetries and gauged symmetries, known as "collective symmetry breaking".}
\vspace*{1em}

\subsection{A Prototype}

\paragraph{Our goal}
\parbox{0.8\textwidth}{To see geometrically how a symmetry can be hidden by a re-parameterisation.}
\vspace*{1em}

We consider a spacetime-independent (i.e. global) transformation $g$ belonging to group $G$ that acts on a field multiplet $\phi$, where the field has vacuum expectation values (vev) under $G$ with magnitude and direction given by $\{\vec{f}\}_G$. To be clear, if the (classical or quantised) Lagrangian of the field is given by a kinetic energy $T$ and a potential energy $V$
\begin{align}
\mathcal{L} = T - V
\end{align}
then the vev(s) are the solution(s) to
\begin{align}
\frac{\partial V}{\partial \phi}|_{\langle\phi\rangle = \{\vec{f}\}} = 0, \qquad s.t. \qquad \frac{\partial^2 V}{\partial \phi^2}|_{\langle\phi\rangle = \{\vec{f}\}} > 0
\end{align}
If the fields are redefined as having zero vev, i.e. $\phi \rightarrow \phi' = \phi + \vec{f}$, which leaves the redefined Lagrangian invariant under some smaller group $h \in H$, then we say this set $\{\vec{f}\}_H$ is invariant under the subgroup $H < G$. Casually, we say that the vev has spontaneously broken the global symmetry $G \rightarrow H$.

\begin{figure}
\center
\subfloat[A 2-component Mexican hat potential, with $V$ presented as the $z-$axis \label{fig:simple_symmetries_a}]{\includegraphics[scale=0.1]{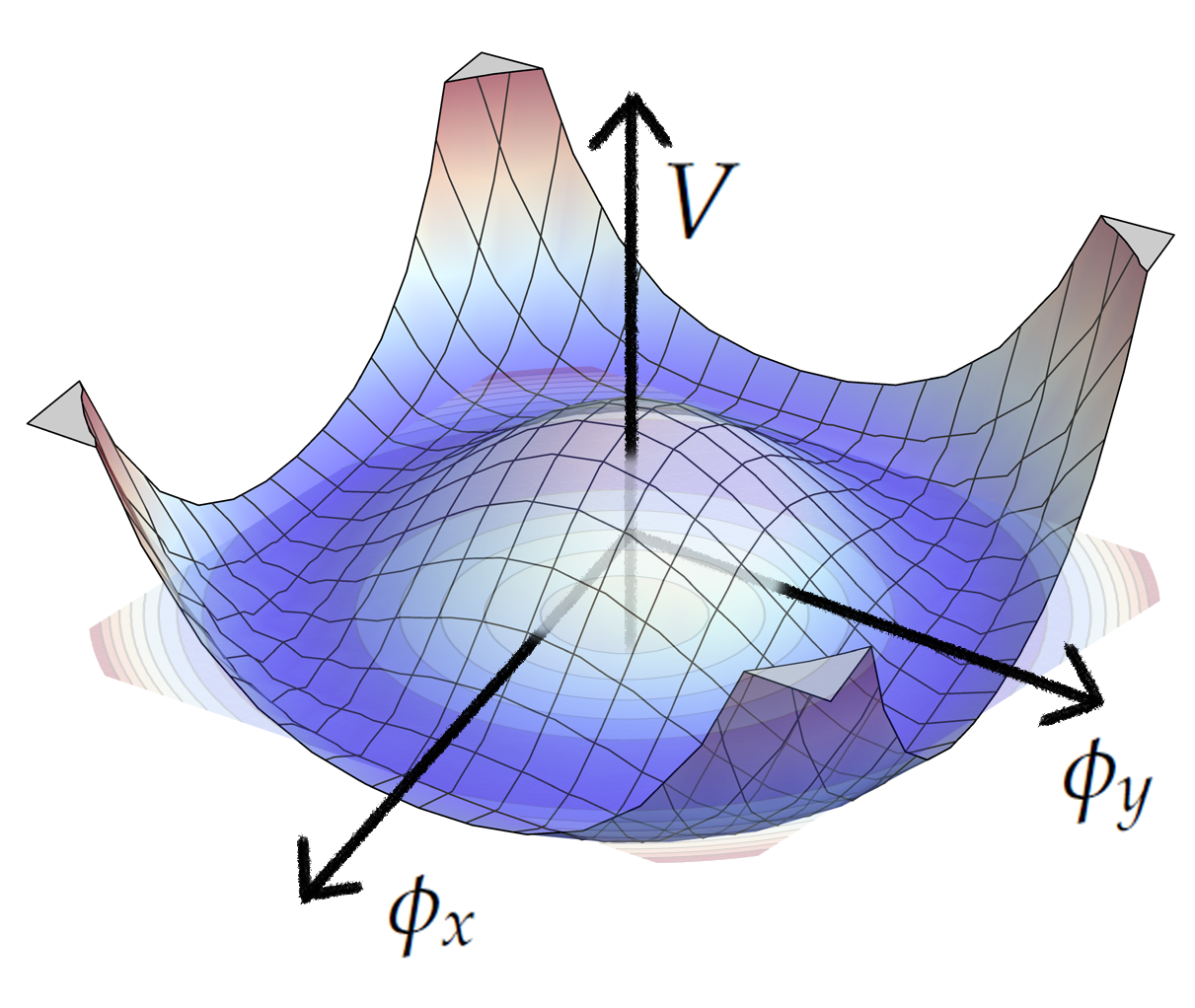}}\hspace{0.2\textwidth}%
\subfloat[The shifted fields vs. the potential, both as a contour and as the $z-$axis \label{fig:simple_symmetries_b}]{\includegraphics[scale=0.1]{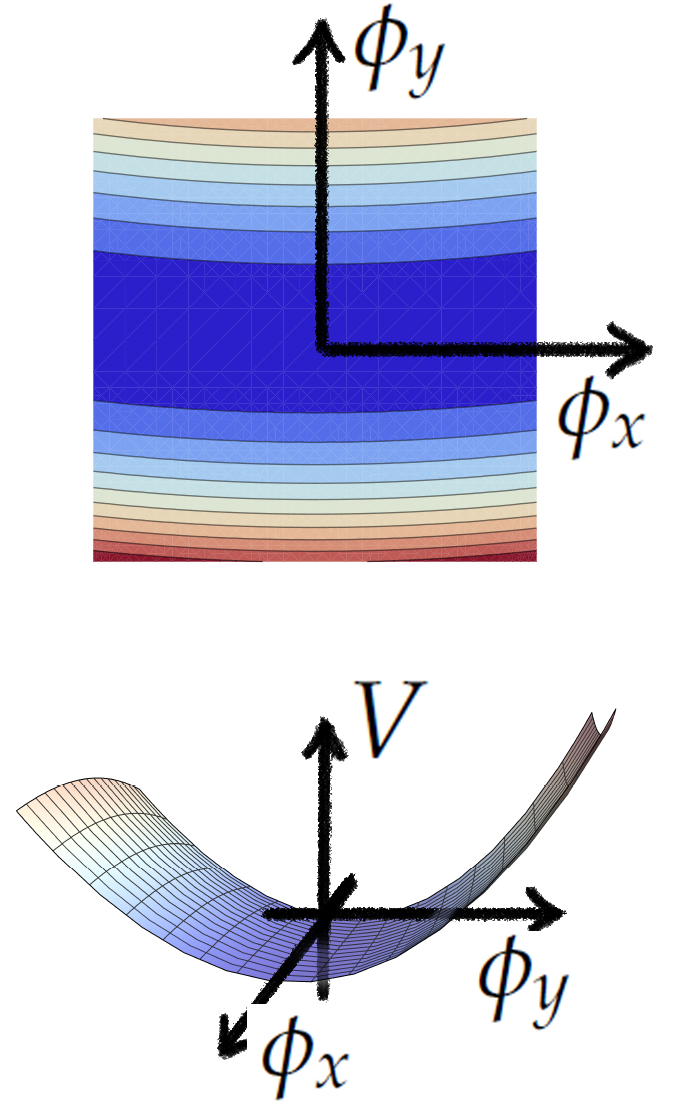}}\\
\subfloat[A 3-component Mexican hat potential, as above, with $V$ as the temperature of each contour \label{fig:simple_symmetries_c}]{\includegraphics[scale=0.1]{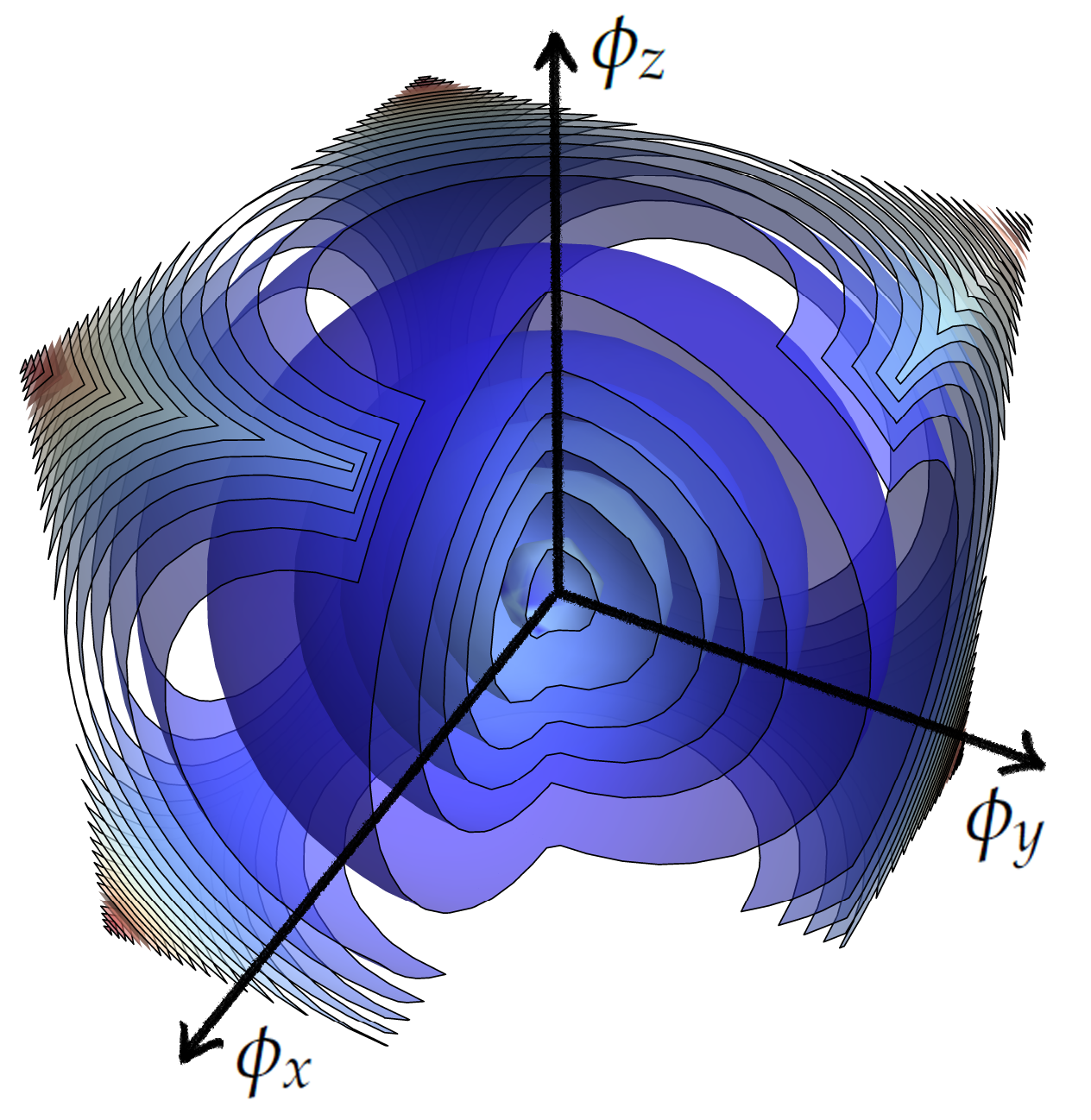}}\hspace{0.2\textwidth}%
\subfloat[The shifted fields vs. the potential, where now there is a remaining $SO(2)$ symmetry about the $z-$axis \label{fig:simple_symmetries_d}]{\includegraphics[scale=0.1]{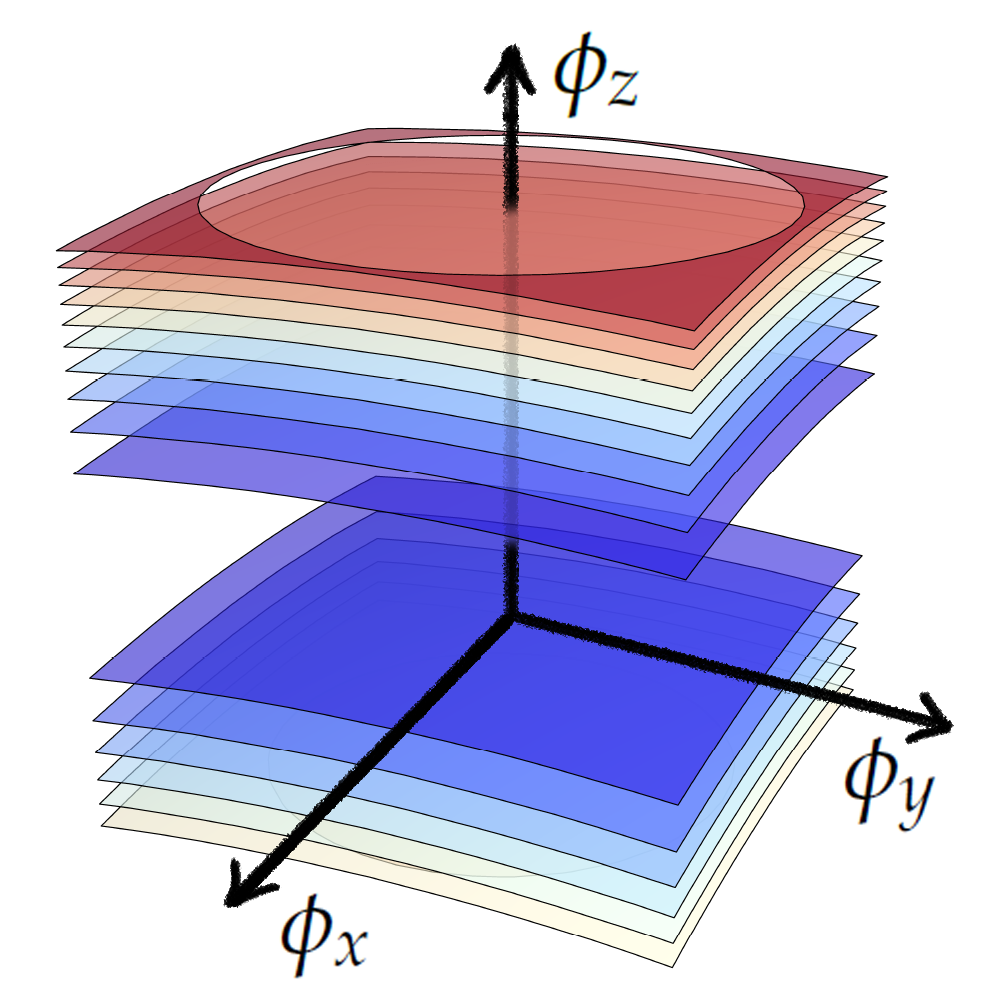}}
\caption{Visualisation of some low-dimensional symmetries in the classical vacuum and then shifted to a physical vacuum.}
\label{fig:simple_symmetries}
\end{figure}

For example, consider two fields $\phi_x, \phi_y$ with the same properties, in the following Lagrangian
\begin{align}
\mathcal{L} &= (\partial \phi_x)^2 + (\partial \phi_y)^2 - \frac{\lambda}{4!} (\phi_x^2 + \phi_y^2 - v^2)^2 \nonumber\\
&= (\partial \phi)^T(\partial \phi) - \frac{\lambda}{4!} (\phi^T \phi- v^2)^2\label{eq:O(N)_lagrangian}
\end{align}
Clearly, the first line was able to be condensed by writing the fields as $\phi = (\phi_x, \phi_y)$. This multiplet can be transformed by $\phi \rightarrow O \phi$ leaving the Lagrangian invariant, where $O^T = O^{-1}$. Then the symmetry of the classical Lagrangian is $G = O(2)$. The potential of this Lagrangian is called a Mexican hat, and is sketched in \cref{fig:simple_symmetries_a}. Applying the definition of the vev, we find
\begin{align}
\langle\phi_x^2\rangle + \langle\phi_y^2\rangle = |\vec{f}|^2 = v^2
\end{align}
Observe that there is no one unique vacuum, instead there is a continuous set of equivalent vacua, one of which is chosen by nature to be the physical ground state. To explore the properties of the "broken" vacuum, we shift to one of the vacua $\vec{f}_0 = (v,0)$ with an arbitrary redefinition of the fields: $\phi_x \rightarrow \phi_x' = \phi_x, \phi_y \rightarrow \phi_y' = \phi_y+ v $. The "broken" Lagrangian is then
\begin{align}
\mathcal{L} = (\partial \phi')^T (\partial \phi') - \frac{\lambda}{4!}\left((\phi^{\prime T} \phi')^2 + 4v^2\phi_y^{\prime 2} - 4v\phi_y^{\prime 3} - 4v\phi'_y \phi_c^{\prime 2}\right)
\end{align}
where we have picked up some extra interactions, $\phi_x$ is now massless and there is only a parity symmetry ($\mathbb{Z}_2 \sim O(1)$) remaining for $\phi_x$. This behaviour is sketched in \cref{fig:simple_symmetries_b}. The case for $\phi = (\phi_x, \phi_y, \phi_z)$ is also visualised. Given this multiplet, then the Lagrangian in \cref{eq:O(N)_lagrangian} has an $O(3)$ invariance. If we shift to the ground state, $\vec{f}_0 = (0,0,v)$, then the potential appears as in \cref{fig:simple_symmetries_c} and \cref{fig:simple_symmetries_d}. The potential is locally linear in the directions of $\phi_x,\phi_y$ - indicating masslessness - and therefore the two fields are still related by an $O(2)$ transformation.    We will return to this model many times, with increasing sophistication, and generalise it from the classical case to the quantum one. For now, it suffices to say that for small oscillations around the ground state of the model, we appear to have lost two degrees of freedom ($SO(3)$ rotations require three parameters, $SO(2)$ rotations only one). This resulted in two massless, degenerate scalars $\phi_x', \phi_y'$.  In doing so, we effectively hid the symmetries still very much obeyed by $\phi$, due to the change of variable in the Lagrangian. This brings us to the most famous example of hidden symmetry: the Higgs mechanism.
%

\subsection{The Higgs Mechanism}\label{sec:higgs_mechanism}

\paragraph{Our goal:}
\parbox{0.8\textwidth}{To give an informal look at a case of higher symmetries being hidden in nature.}
\vspace*{1em}

%

In \cref{sec:CCWZ}, we will explore symmetry hiding in a language that can be generalised easily - that of the Coleman-Callan-Wess-Zumino (CCWZ) construction. For now we will think of the mechanism quite geometrically, inspired by \cite{Alonso:2015fsp}. This is practical due to the low dimensionality of the SM Higgs symmetries. We propose four scalar fields $\psi_i$ arranged into a complex doublet
\begin{align}
H = \frac{1}{\sqrt{2}}\left(\begin{matrix}
\psi_1 + i \psi_2\\
\psi_3 - i \psi_4
\end{matrix}\right)
\end{align}
We can supply the Lagrangian of this field with a potential $V$ that depends only on polynomials of $\frac{1}{2}HH^\dagger$
\begin{align}
V = V(\frac{1}{2}HH^\dagger) = V(\psi_1^2 + \psi_2^2 + \psi_3^2 + \psi_4^2) = V(|\psi|^2)
\end{align}
and our Lagrangian thus has a four-dimensional rotational symmetry $SO(4)$. The local isomorphism $SO(4) \sim SU(2)_L \times SU(2)_R$ \footnote{We also have an extra degree of freedom since the potential is also invariant under parity, $O(4) \sim SO(4) \times O(1)$} becomes clear if we rearrange the fields as 
\begin{align}
\Sigma = (\tilde{H}, H)
\end{align}
where $\tilde{H}=(i\sigma_2)H^*$ is the ``charge conjugate" of $H$. This field transforms as a ``chiral bi-doublet"
\begin{align}
\Sigma \rightarrow U_L \Sigma U_R^\dagger
\end{align}
where $U_{L,R}$ are two-by-two unitary matrices. 

If we impose the condition on the potential
\begin{align}
\frac{\partial V}{\partial \psi}|_{|\psi| = v} = 0 
\end{align}
This equates, to first order\cite{75845}, to a condition on the fields
\begin{align}
\psi_1^2 + \psi_2^2 + \psi_3^2 + \psi_4^2 = v^2
\end{align}
If we were to redefine one or more of the of the fields to oscillate around this minimum, then we would have a higher-dimensional case of the situation in \cref{fig:simple_symmetries_d}, where the full $SO(4)$ symmetry is non-linearly realised by the Higgs fields, and are constrained to linear $SO(3)$ transformations. This remaining symmetry is known as "custodial" symmetry, and protects $M_W = M_Z \cos\theta_W$ at tree level. In the language of chiral symmetry $SU(2)_L \times SU(2)_R$ is non-linearly realised, with a linear $SU(2)_V \sim SO(3)$ diagonal (vector) subgroup, by the condition
\begin{align}
\langle\Sigma \rangle = \frac{v}{\sqrt{2}}\left(\begin{matrix}
1 & 0\\
0 & 1
\end{matrix}\right)
\end{align}
Specifically, we can reparameterise the Dim$(SO(4)) = 6$ linear transformations as Dim$(SO(3))=3$ linear transformations and $3$ ``Goldstone boson" fields. This is always possible when a global symmetry is allowed to be non-linearly realised. The details of this will be discussed in \cref{sec:NLSM} on the non-linear sigma model. 

\begin{figure}
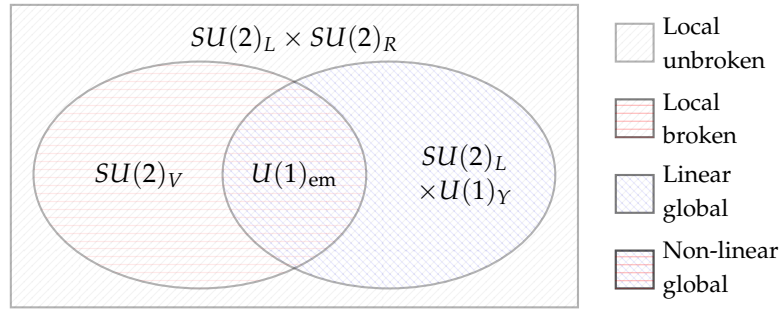

\centering
\tikz[scale=1, every node/.style={transform shape}]{
\draw [pattern=north east lines, pattern color=gray, thick, opacity=0.3] (1,0.25) rectangle (8.5,4.25);
\draw [pattern=north west lines, pattern color=blue, thick, opacity=0.3](6,2) ellipse (2.2 and 1.5);
\draw [pattern=horizontal lines, pattern color=red, thick, opacity=0.3] (3.5,2) ellipse (2.2 and 1.5);
\node [align=center] at (4.75, 3.8) {$SU(2)_L \times SU(2)_R$};
\node [align=center] at (2.7,2) {$SU(2)_V$};
\node [align=center] at (4.75,2) {$U(1)_\text{em}$};
\node [align=center] at (7,2) {$SU(2)_L$\\ $ \times U(1)_Y$};
\draw [pattern=north east lines, pattern color=gray, thick, opacity=0.3] (9,0.5) rectangle (9.5,1);
\draw [pattern=north west lines, pattern color=blue, thick, opacity=0.3] (9,0.5) rectangle (9.5,1);
\draw [pattern=horizontal lines, pattern color=red, thick, opacity=0.3] (9,0.5) rectangle (9.5,1);
\node [align=left, right] at (9.5,0.75) {\small Non-linear \\ \small global};
\draw [pattern=north east lines, pattern color=gray, thick, opacity=0.3] (9,1.5) rectangle (9.5,2);
\draw [pattern=north west lines, pattern color=blue, thick, opacity=0.3] (9,1.5) rectangle (9.5,2);
\node [align=left, right] at (9.5,1.75) {\small Linear \\ \small global};
\draw [pattern=north east lines, pattern color=gray, thick, opacity=0.3] (9,2.5) rectangle (9.5,3);
\draw [pattern=horizontal lines, pattern color=red, thick, opacity=0.3] (9,2.5) rectangle (9.5,3);
\node [align=left, right] at (9.5,2.75) {\small Local \\ \small broken};
\draw [pattern=north east lines, pattern color=gray, thick, opacity=0.3] (9,3.5) rectangle (9.5,4);
\node [align=left, right] at (9.5,3.75) {\small Local \\ \small unbroken};
}\caption{The Higgs mechanism sketched using the convention established by S. Weinberg in \cite{weinberg1975}} \label{fig:higgs_mech}
\end{figure}

Up to this point, we have only hidden the full symmetry. If we were to gauge a part of the global $SU(2)_L \times SU(2)_R$ symmetry that wasn't the linear subgroup $SU(2)_V$, then we would explicitly break the subgroup's symmetry. Choosing to gauge the electroweak subgroup $SU(2)_L \times U(1)_Y$, where $SU(2)_L$ is the same as in the global group, and $U(1)_Y$ is derived from the electromagnetic charge of each field, does exactly this. The intersection of the global and local symmetries is the remaining linear symmetry
\begin{align}
SU(2)^{\textnormal{local}}_L \times U(1)_Y^{\textnormal{local}} \cap SU(2)_V^{\textnormal{global}} = U(1)_{\textnormal{em}}^{\textnormal{local}}
\end{align}
The transformations of the global subgroup should be incorporated into the covariant derivative, along with the gauge terms. In fact, a particular parameterisation of the global tranformations called the "Maurer-Cartan form" behaves \textit{exactly} like the gauge fields. We can, again, reparameterise our $3$ Goldstone bosons as a Maurer-Cartan form, absorbing them into the definition of the $SU(2)_L \times U(1)_Y$ gauge fields, giving them three longitudinal degrees of freedom. To be very clear, it is the Goldstone mechanism of non-linear realisation of global symmetry, combined with a subgroup gauging, that gives this Higgs mechanism\footnote{The Higgs mechanism is, unfortunately, often introduced as a spontaneous symmetry breaking of $SU(2)_L \times U(1)_Y$, although this is blatantly untrue. Indeed, \cite{Elitzur:1975im} (updated recently by \cite{Splittorff:2003ye}) showed that it is generally impossible to spontaneously break a local symmetry. Only collective symmetry breaking can break a local symmetry, and this is an explicit breaking due to the global symmetry pattern. The source of this anguish for students is likely that the $SU(2)_L$ group is common to both the global and local symmetries. However there is no reason in the SM for nature to have chosen this coincidence a priori.}. All the symmetry structures in this work use this basic pattern in various levels of complexity, which is sketched in \cref{fig:higgs_mech}.

\subsection{Goldstone's Theorem}

\paragraph{Our goal:}
\parbox{0.8\textwidth}{To sketch the proof of an extremely important feature common to all hidden symmetries.}
\vspace*{1em}



The Higgs mechanism is a special case of hidden global degrees of freedom realised as masses by a shift of vacuum. However, there is no reason that this must be the fate of all hidden symmetries. Goldstone's Theorem describes the structure of symmetry hiding in the absence of gauge fields \cite{PhysRev.127.965}:
\begin{center}
A continuous global symmetry $G$ hidden by a vacuum expectation value, linearly represented by a subgroup $H$, will always be accompanied by the presence of $\text{dim}(G) - \text{dim}(H)$ massless "Goldstone" bosons.
\end{center}
A proof of this theorem will be outlined here (based on the strategies found in \cite{RevModPhys.46.7,aitchison2003gauge,lancaster2014quantum}), although the full quantum-theoretic argument contains many subtleties. At the semi-classical level, a much more elegant proof is provided by the CCWZ construction, described later in \cref{sec:CCWZ}.

Consider a set of scalar fields $\phi^a$, described by a Lagrangian invariant under spacetime-independent $U(1)$ transformations
\begin{align}
\mathcal{L} &= (\partial_\mu \phi^a)^\dagger (\partial^\mu \phi^a) - \mu^2 \phi^{a\dagger} \phi^a + \frac{\lambda}{4!}\left(\phi^{a\dagger} \phi^a\right)^2, & & \phi^a \rightarrow \me^{i\alpha} \phi^a.
\end{align}
The associated Noether current for this field is
\begin{align}
j^\mu(x) &= \frac{\partial \mathcal{L}}{\partial(\partial_\mu \phi^a)}\delta\phi^a = (\partial^\mu \phi^a)^\dagger i\alpha\phi^a - (\partial^\mu \phi^a)i\alpha\phi^{a\dagger},
\end{align}
since
\begin{align}
\phi^{'a} = \phi^a + \delta\psi^a &= \phi^a + i\alpha\psi^a, \qquad \text{and} & & \phi^{'a\dagger} = \phi^{a\dagger} + \delta\psi^{a\dagger} = \phi^{a\dagger} -i\alpha\psi^{a\dagger}.
\end{align}
By the Euler-Lagrange equations, 
\begin{align}
\partial_\mu j^\mu (x) = 0\label{eq:lorentz_invariance}
\end{align}
as can be quickly verified. Now, let one of these fields have a non-zero vacuum expectation value $\langle 0 | \phi^0 | 0\rangle \neq 0$. But then we have broken the formalism of what a particle is (i.e. excitations of a field that is expanded around a vacuum state). If the vacuum is non-zero, our Taylor expansion of $n-$particle interactions is non-physical. Let's examine how to think of this state.

We can define a Noether charge for the field $\phi$ to be the integral over the charge density $j^0(x)$
\begin{align}
& Q = \int d^3x j^0(x)
\end{align}
although this may or may not be a well-defined value when integrated over all space. The meaning of $Q$ is made more clear when it is rigorously taken as a quantum operator $\hat{Q}$ acting on a vacuum state
\begin{equation}
\hat{Q}|0\rangle = \begin{cases}
0, & \text{if}\ \langle 0 | \hat{Q}\hat{Q} |0\rangle = \int^\infty_\infty d^3 x \langle 0 | \hat{j}^0(0) \hat{Q}| 0\rangle = 0\\
\text{undefined}, & \text{if}\ \langle 0 | \hat{Q}\hat{Q} |0\rangle = \infty
\end{cases}
\end{equation}
which is the Fabri-Picasso theorem, proven in \cite{PhysRevLett.16.408.2}. For our purposes, we will just consider $Q|0\rangle \neq 0$.

We could consider a field that is \textit{not} invariant under $\hat{Q}$ (i.e. it does not commute with $Q$) but does have a vanishing vev - $\langle 0 | \phi' | 0 \rangle = 0$
\begin{align}
0 \neq \langle 0 | [Q,\phi']| 0 \rangle & :=  \langle 0 | \phi | 0 \rangle \nonumber\\
& = \langle 0 | \left[ \int d^3x j^0(x), \phi'\right] | 0 \rangle \nonumber\\	
& = \int d^3 x \langle 0 | \left[ j^0(t,\bm{0}), \phi'\right] | 0 \rangle \label{eq:norm}
\end{align}
Where the last line comes from the Poincare invariance of the vacuum - that is, the vacuum expectation shouldn't depend on the location the commutator is evaluated at. This is an assumption about the nature of quantum vaccua in general and, provided we are in an stable vacuum, is fair. We use \cref{eq:lorentz_invariance} to show that the vev is constant in time
\begin{align}
\frac{\partial}{\partial t}\int d^3 x \langle 0 | \left[ j^0(t,\bm{0}), \phi\right] | 0 \rangle &= - \int d^3 x \langle 0 | \left[\bm{\nabla}\cdot \bm{j}(x) ,\phi \right]| 0 \rangle \nonumber\\
&= - \int d \bm{S}\cdot \langle 0 | \left[\bm{j}(x) ,\phi \right]| 0 \rangle
\end{align}
which we propose vanishes for a surface taken to infinity. That is, the interaction of $\bm{j}(x)$ and $\phi(0)$ from the origin to any point $x$ on a very large surface should tend to zero.
\begin{align}
\implies \frac{\partial}{\partial t}\int d^3 x \langle 0 | \left[ j^0(t,\bm{0}), \phi\right] | 0 \rangle &= 0
\end{align}

The time-independence of the vev leads to massless states. To see this, consider inserting a set of states into \cref{eq:norm}
\begin{align}
0 & \neq \int d^3 x \sum_n \left( \langle 0 | j_0(x) | n\rangle \langle n | \phi(0) | 0\rangle - \langle 0 | \phi(0) n \rangle \langle n | j_0(x) | 0 \rangle \right) \nonumber\\
&= \int d^3 x \sum_n \left( \langle 0 | j_0(x) | n\rangle \langle n | \phi(0) | 0\rangle\me^{-ip_n\cdot x} - \langle 0 | \phi(0) n \rangle \langle n | j_0(x) | 0 \rangle \me^{ip_n \cdot x} \right)
\end{align}
then, integrating over 3-space gives
\begin{align}
0 \neq \sum_n \delta^3(\bm{x}) \left( \langle 0 | j_0(x) | n\rangle \langle n | \phi(0) | 0\rangle\me^{ip_{n,0} x_0} - \langle 0 | \phi(0) n \rangle \langle n | j_0(x) | 0 \rangle \me^{-p_{n,0} \cdot x_0} \right)
\end{align}
which is also $t$-independent. There must be, therefore, at least one $|n\rangle$ state such that
\begin{align}
\langle 0 | j_0(0)| n\rangle \neq 0
\end{align}
and is $t$-independent. $t$-independence implies zero energy in this state, by the Schr\"odinger equation. There is also the condition that the 3-momentum of such a state must be zero, in order for $\delta(\bm{p_n}) \neq 0$. A massive particle can have positive energy even with zero momentum (from its rest energy). For a state's energy $\rightarrow 0$ as $\bm{p_n} \rightarrow 0$, we require \textit{massless} particle state or states. That is the essence of Goldstone's Theorem. In order to calculate the norm of $\phi$, which is invariant under $Q$-charged transformations, but is not a particle in the usual sense, we need to introduce massless states. The number of these states is equal to the dimension of the current, which, we will see soon, is precisely the dimension of the group generated by the charges.

There are three conditions for Goldstone's Theorem to hold. The first is that there is a continuous symmetry of the Lagrangian. One needs to be careful with this point as we will see in \cref{sec:chiral_symmetry}, where global symmetries may not be symmetries of the action - called quantum anomalies. The second condition is that the Lagrangian is "manifestly covariant". This is not true for a gauge symmetry, and it allows the "Higgs loophole" found in the previous section. The third is that there must be a potential that allows the non-linear shift $\phi \rightarrow \phi'$. This can be added by hand, as it is in the SM, or generated radiatively. The latter is known as the Coleman-Weinberg process.


\section{The Coleman-Weinberg Process}
\label{sec:coleman_weinberg}

\paragraph{Our goal:}
\parbox{0.8\textwidth}{To derive a procedure commonly used to give potential terms to fields which do not have them at tree-level.}
\vspace*{1em}

There are several tools available for studying the effective potential of a Lagrangian. Recall that an effective potential of a field $\varphi$ linearly coupled to some "classical source" of current in
\begin{align}
\mathcal{L} = \mathcal{L}_\text{kinetic} + \varphi(x) J(x)
\end{align} 
is the first term $V_\text{eff}$ of the effective action $\Gamma_\text{eff}$ in momentum space
\begin{align}
\Gamma_\text{eff} &= \int d^4 x \left[V_\text{eff}(\varphi) + \frac{1}{2} (\partial\varphi)^2 Z(\varphi) + ...\right] \label{eq:effective_action_expansion}
\end{align}
Because the Coleman-Weinberg (CW) process \cite{coleman1973} is integral to the study of Composite Higgs models, in-depth explanations can be found in \cref{chp:colemanweinberg}, including a dictionary of assumed knowledge. In the appendix, we derive that the functional derivative of the effective action, with respect to the classical field gives the source term
\begin{align}
\frac{\delta \Gamma}{\delta \phi_c(x)} &= J
\end{align}
We define spontaneous symmetry breaking (SSB) as a phenomenon that can occur when the classical field is non-zero, even in the absence of a source term
\begin{align}
\frac{\delta \Gamma[\phi_c]}{\delta \phi_c(x)}|_{\phi_c \neq 0} &= 0 \label{eq:ssb}
\end{align}
Note that the effective action is intrinsically quantum mechanical, despite being defined in terms of what we are calling the ``classical field". $\phi_c$ could be also termed the ``mean field", and includes all quantum corrections to the classical equation of motion
\begin{align}
\frac{\delta S[\phi]}{\delta \phi(x)} = J
\end{align}
\cref{eq:ssb} gives a rigorous requirement for SSB. However, we can connect it to our usual understanding as a minimum of the potential if we impose translational invariance on the vacuum state $\langle \phi \rangle_0 = \phi_c$, using the expansion in \cref{eq:effective_action_expansion}
\begin{align}
\partial_\mu\phi_c = 0 && \implies && \frac{\partial V_\text{eff}}{\partial \phi_c}|_{\phi_c \neq 0} = 0
\end{align}
For this to be stable, we require the solution to be a minimum, and call the solution the vacuum expectation value.

In general, once we have the effective potential, finding a minimum is straightforward. Obtaining the effective potential in the first place is not as obvious. The potential contains infinite quantum loop corrections
\begin{align}
V_\text{eff}(\phi_c) &= \sum\limits_{n=1}^\infty \frac{1}{n!} \Gamma^{(n)}(0,...,0)[\phi_c]^n \label{eq:effective_potential_sum}
\end{align}

\begin{figure}[h]\
\centering
\includegraphics[scale=0.15]{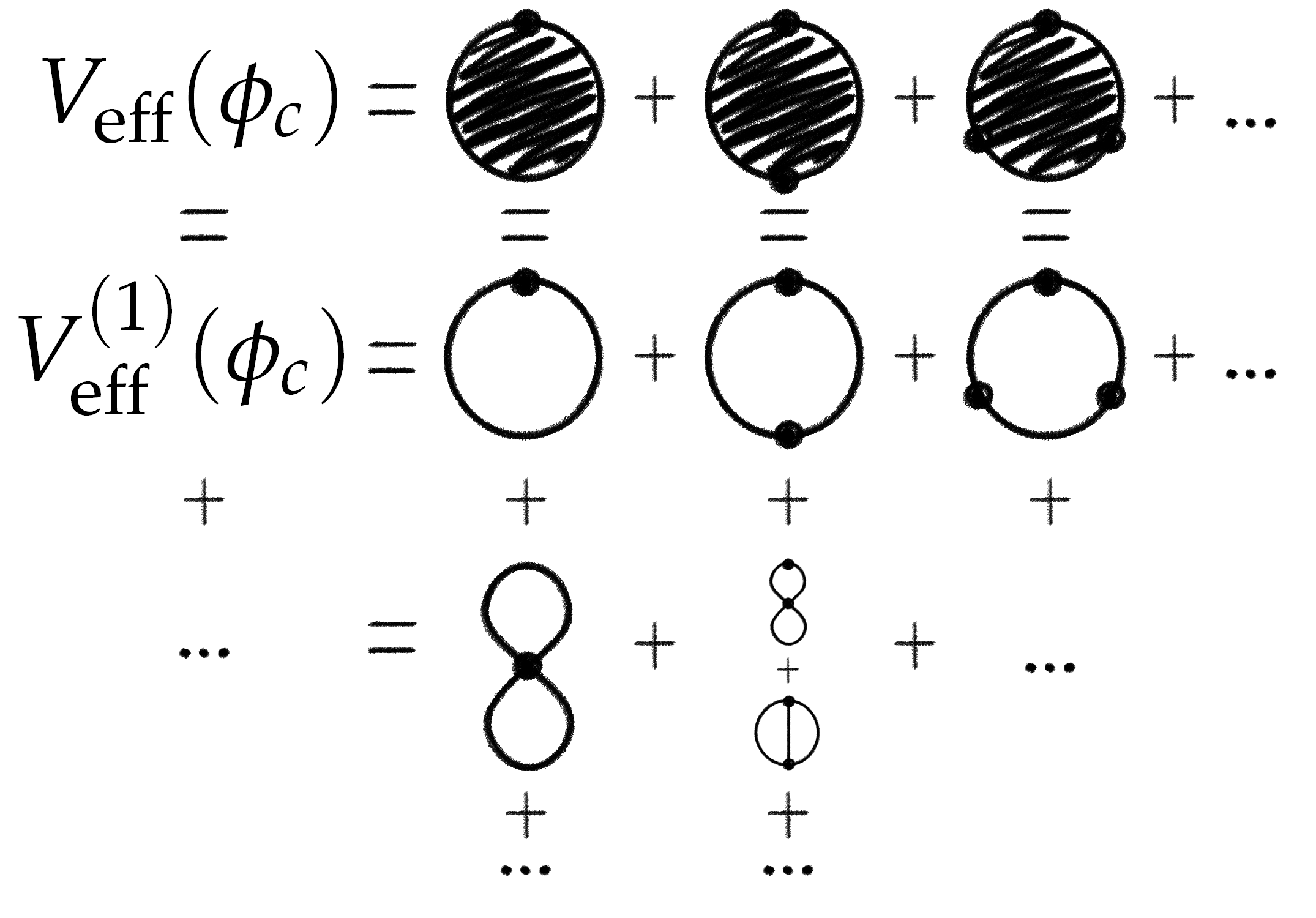}
\caption{Diagrammatic expansion of the quantum Coleman-Weinberg effective potential. Each order of the $n$-leg 1PI summation is itself a series of $m$-loop diagrams. \cite{coleman1973} argues that the first layer of the summation $V_\text{eff}^{(1)}$ is a sufficient approximation.}\label{fig:CW_interactions}
\end{figure}

Each order of $\Gamma^{(n)}$ is itself a sum of all 1PI diagrams with $n$ vanishing external momenta. This is sketched in \cref{fig:CW_interactions}. 

We will assume that the one-loop-order of \cref{eq:effective_potential_sum} - that is $n=1$ - is adequate for calculating the radiative correction to the tree-level potential, as argued in the original paper. In fact, it can be shown that this corresponds to an order $\mathcal{O}(h^0)$ correction to the classical action, with more loops giving $\mathcal{O}(h,h^2,...)$ contributions. Proving the adequacy of this truncation is outside the scope of this section, but is perfectly well-explained in \cite{bhattacharjee2013quantum}. For now, we will be content with seeing clearly the connection between 1PI diagrams, and the effective potential, knowing that summing the 1-loop 1PI diagrams gives an excellent approximation of the quantum corrections. 

Explicit calculations of loop diagrams requires knowledge of a particular model, so we shall put this aside until we have a library of Lagrangians leading to SSB.

\section{Hidden Chiral Symmetry}
\label{sec:chiral_symmetry}

\paragraph{Our goal:}
\parbox{0.8\textwidth}{To give an instructive example of non-linear symmetry hiding that occurs in nature.}
\vspace*{1em}

\subsection{The Sigma Model}
\label{sec:sigma_model}
%

\paragraph{Our goal:}
\parbox{0.8\textwidth}{To briefly show the structure of a global symmetry in a physical model.\footnote{I will broadly follow the (extremely intuitive) approach of \cite{zee2010quantum}.}}
\vspace*{1em}

Consider the following Lagrangian of two massive fermion (that is, Dirac spinor) fields in a multiplet $\Psi = (\psi_1, \psi_2)$. The kinetic term can be conveniently written as
\begin{align}
\mathcal{L} &= \bar{\Psi} i\slashed{\partial} \Psi \label{eq:nonchiral_lagrangian} \\
&= \bar{\Psi}_L i\slashed{\partial} \Psi_L + \bar{\Psi}_R i\slashed{\partial} \Psi_R \label{eq:chiral_lagrangian}
\end{align}
where one can reduce the Dirac spinor representation to the direct sum of two chiral representations (which would be Weyl fermions in the $SU(2)\times SU(2)$ Lorentz group). This reducibility is always true. Proving it is beyond our scope, but is eloquently explained by D. Kaplan \cite{Lellouch:2011zz}. For reference, we give a dictionary of chirality below. 

\vspace{2em}
\begin{tcolorbox}[colback = black!2!white]
\paragraph{Dictionary of chiral symmetry}\label{dic:chiral_symmetry}

We define left- and right-handed Dirac spinors in terms of the $\gamma_5$ matrix, which in $D=4$ is given by 
\begin{align}
\gamma^5 &= \left(\begin{matrix}
0 & \mathbb{1}_{2\times 2}\\
\mathbb{1}_{2\times 2} & 0
\end{matrix}\right)
\end{align}
Defining two projection matrices
\begin{align}
P_{\pm} = \frac{\mathbb{1}_{4\times 4}\pm \gamma^5}{2}\label{eq:projection_definition}
\end{align}
which clearly have the properties (i) $P_+ + P_- = 1$, $P_+ - P_- = \gamma^5$, (ii) $P_{\pm}^2 = P_{\pm}$ and (iii) $P_+ P_- = 0$, then we can project to the "chiral basis"
\begin{align}
\begin{split}
\Psi_L = P_- \Psi, \qquad \bar{\Psi}_L = \bar{\Psi} P_+\\
\Psi_R = P_+ \Psi, \qquad \bar{\Psi}_R = \bar{\Psi} P_-
\end{split}\label{eq:projection_identities}
\end{align}
One sees explicitly the reducibility of the Dirac representation, as $\Psi_L + \Psi_R = (P_- + P_+) \Psi = \Psi$. Inserting this into the \cref{eq:nonchiral_lagrangian} gives \cref{eq:chiral_lagrangian} once we note that $\{\gamma^0, \gamma^5\}=0$ leads to same-handed kinetic coupling. Conversely, property (iii) leads to opposite-handed mass coupling as in \cref{eq:mass_transformation}.
\end{tcolorbox}
\vspace{2em}

The Lagrangian of \cref{eq:chiral_lagrangian} exhibits an $SU(2)_L \times SU(2)_R$ global ``chiral" symmetry\footnote{Beware: The space-time Lorentz group is isomorphic to $SU(2)\times SU(2)$. The internal chiral symmetry is $SU(N_f)_L \times SU(N_f)_R$, with $N_f$ the number of fermion species of equal mass in the model.}. Given that the transformations are parameterised by
\begin{align}
\Psi_L \rightarrow \me^{i \alpha_L^i T^i} \Psi_L && \Psi_R \rightarrow \me^{i \alpha_R^i T^i} \Psi_R
\end{align}
for generators related to the Pauli matrices $\tau^i = 2 T^i$ given in \cref{sec:group_theory}. Then the charges of the fields are given by
\begin{align}
\Psi_L = (\frac{1}{2},0) && \Psi_R = (0,\frac{1}{2})
\end{align}
Consider a case such that we \textit{know} that \cref{eq:chiral_lagrangian} represents the underlying theory\footnote{Clearly, the physical situation being described is low-energy QCD chiral symmetry, but it is instructive to see the process in full generality.}. The chiral symmetry thus has powerful predictive properties, but we cannot directly add a mass term without dashing that power, since
\begin{align}
m \bar{\Psi}_L \Psi_R \xrightarrow{g_L, g_R} m\bar{\Psi}_L \me^{-i\alpha_L^i T^i} \me^{i\alpha_R^j T^j} \Psi_R \neq m \bar{\Psi}_L \Psi_R  \qquad \text{ for, } \alpha_L^i \neq \alpha_R^i\label{eq:mass_transformation}
\end{align}
A simple solution for \textit{effectively} introducing mass is to propose a new multiplet consisting of four real scalar fields, which transforms as a bidoublet under $SU(2)_L \times SU(2)_R$
\begin{align}
\Sigma = \sigma + i\bm{\tau}\cdot \bm{\pi} \sim (\frac{1}{2},\frac{1}{2}), && \Sigma \rightarrow \me^{i\alpha_L^iT^i} \Sigma \me^{-i\alpha_R^jT^j}
\end{align}
By coupling this meson multiplet as a Yukawa-type term to the fermion, we may be able to linearly preserve a subgroup of the overall symmetry, with the full symmetry hidden at low energy - that is, at relatively small perturbations from the vacuum state. Consider the interacting Lagrangian 

\begin{align}
\mathcal{L}_\text{int} &= \frac{1}{4}\text{Tr}\left(\partial_\mu \Sigma \partial^\mu \Sigma^\dagger\right) \bar{\Psi}_L + \bar{\Psi}_L i \slashed{\partial} \Psi_L + \bar{\Psi}_R i\slashed{\partial} \Psi_R + g(\bar{\Psi}_L\Sigma \Psi_R + \bar{\Psi}_R\Sigma^\dagger \Psi_L)
\end{align}

Given the above transformation rules, this is clearly $SU(2)_L\times SU(2)_R$ invariant (e.g. $\bar{\Psi}_L\Sigma\Psi_R \rightarrow \bar{\Psi}_L \me^{-i\alpha_L^iT^i}\me^{i\alpha_L^iT^i}\Sigma\me^{-i\alpha_R^jT^j} \me^{i\alpha_R^jT^j}\Psi_R = \bar{\Psi}_L\Sigma\Psi_R$). We now connect this abstract theory to reality. The fermion fields can represent the nucleon doublet, and the meson multiplet is precisely a linear combination of the three light pions $\{\pi^0, \pi^+, \pi^-\}$ of the hadronic model, and a $\sigma$ particle. The $\sigma$ does not appear in our low-energy experiments, so we would like to integrate it out. Therefore we must find a way to give it a relatively large mass. We know experimentally that there are three fields with the same quantum numbers - a triplet of pions. Thus, the meson multiplet must decompose to some multiplets under a smaller group - this is the linearly realised subgroup $SU(2)_I$ called isospin. This can be more intuitively written when noting $SU(2)_L \times SU(2)_R \sim SO(4)$, then the meson decomposes as
\begin{align}
\underbrace{(\bm{\frac{1}{2}},\bm{0}) \otimes (\bm{0},\bm{\frac{1}{2}})}_{SU(2)_L \times SU(2)_R} \rightarrow \underbrace{\bm{1} \oplus \bm{0} }_{SU(2)_{I}} && \sim && \underbrace{\bm{4}}_{SO(4)} \rightarrow \underbrace{\bm{3} \oplus \bm{1}}_{SO(3)} && 
\implies \Phi = (\pi_1, \pi_2, \pi_3, \sigma)^T
\end{align}
Given this decomposition, we need to ensure we can write our Yukawa term in $SO(4)$ language. To do so, note from the identities in \cref{eq:projection_identities} that
\begin{align}
\bar{\Psi}_L \Sigma \Psi_R + \bar{\Psi}_R \Sigma^\dagger \Psi_L &= \bar{\Psi} P_+ (\sigma + i\bm{\pi}\cdot \bm{\tau}) P_+ \Psi + \bar{\Psi} P_- (\sigma - i \bm{\pi}\cdot \bm{\tau}) P_- \Psi \nonumber\\
&= \bar{\Psi} P_+ (\sigma + i\bm{\pi}\cdot \bm{\tau})\Psi + \bar{\Psi} P_- (\sigma - i \bm{\pi}\cdot \bm{\tau})\Psi \nonumber\\
&= \bar{\Psi} (P_+ + P_-) \sigma \Psi + \bar{\Psi} (P_+ - P_-) i \bm{\pi}\cdot \bm{\tau} \Psi \nonumber\\
&= \bar{\Psi}(\sigma + i\tau^i \pi^i \gamma_5)\Psi
\end{align}
where in the second line we used $P_\pm^2 = P_\pm$. Then we can write the chiral Lagrangian using $SO(4)$ language
\begin{align}
\mathcal{L}_{\text{int}} = \bar{\Psi} \left[ i\slashed{\partial} + g(\sigma + i\tau^i \pi^i \gamma_5) \right] \Psi + \mathcal{L}( \pi^i, \sigma)\label{eq:LSM_interacting}
\end{align}
We have thus introduced a Yukawa-type term. But, just like in the SM with a vev-less Higgs, our fermions are still massless. We generate a mass term by giving the meson fourplet $\Phi = (\bm{\pi},\sigma)$ a vev
\begin{align}
\mathcal{L}(\pi^i, \sigma) &= \frac{1}{2}(\partial \sigma)^2 + \frac{1}{2} (\partial \bm{\pi})^2 \nonumber\\
& +\frac{\mu^2}{2}(\sigma^2 + \bm{\pi}^2) - \frac{\lambda}{4}(\sigma^2 + \bm{\pi}^2)^2 \label{eq:GB_lagrangian}
\end{align}
This should look familiar. It is the precisely the Higgs potential and kinetic Lagrangian. Indeed, just as in that mechanism, we proceed to expand about an arbitrary minimum $\vec{F} = (0,0,0,f)$ to find any physically equivalent vacuum such that
\begin{align}
\langle 0 | \Phi | 0 \rangle = \vec{F}, \qquad \text{where} \qquad f = \sqrt{\frac{\mu^2}{\lambda}}\label{eq:f_def}
\end{align}
Again, the choice of direction of $\vec{F}$ is arbitrary and we require only that $\vec{F}^2 = f^2$. Expanding then about a \textit{shifted} field $\sigma'$, where $\sigma = \sigma' + f$,
\begin{align}
\mathcal{L}'_{\text{int}} = \bar{\Psi} \left[ i\slashed{\partial} + g(\sigma' + i\tau^i \pi^i \gamma_5) \right] \Psi  + gf \bar{\Psi} \Psi + \mathcal{L}( \pi^i, \sigma) \label{eq:LSM_int}
\end{align}
and
\begin{align}
\mathcal{L}'(\Phi) &= \frac{1}{2} (\partial \sigma')^2 + \frac{1}{2} ( \partial \bm{\pi})^2 \nonumber \\
& + \frac{\mu^2}{2}\left(\sigma'^2 + f^2 + 2\sigma' f + \bm{\pi}^2\right) - \frac{\lambda}{4}\left(\sigma'^2 + f^2 + 2\sigma' f + \bm{\pi}^2\right)^2 \nonumber\\
&= ... - \frac{\lambda}{4}\left( \sigma'^4 + f^4 + 4\sigma'^2 f^2 + \bm{\pi}^4 + 2\sigma'^2 f^2 + 4\sigma'^3 f\right. \nonumber\\
&+\left. 2\sigma'^2 \bm{\pi}^2 + 4\sigma' f^3 + 2 f^2\bm{\pi}^2 + 2\sigma' f \bm{\pi}^2\right)
\end{align}
We can use the solution of the minimum in \cref{eq:f_def} to write  
\begin{align}
\mathcal{L}'(\Phi)  &= - \frac{\lambda}{4}(\sigma'^4 + \bm{\pi}^4) - \lambda\sigma'^3 f - \frac{\lambda}{2}\sigma'^2\bm{\pi}^2 + \cancelto{0}{\red{\frac{\mu^2}{2}\bm{\pi}^2 - \frac{\lambda f^2}{2} \bm{\pi}^2}} \qquad (\text{by} \; \cref{eq:f_def}) \nonumber\\
&  + (\frac{\mu^2}{2} - \frac{3\lambda f^2}{2})\sigma'^2
+  \frac{\lambda f}{2}\sigma'\bm{\pi}^2 - \cancelto{0}{\red{\lambda \sigma' f^3 + \mu^2 \sigma' f}} + \text{const.} \nonumber \\
&= - \frac{\lambda}{4}(\sigma'^4 + \bm{\pi}^4) - \lambda\sigma'^3 f - \frac{\lambda}{2}\sigma'^2\bm{\pi}^2  - \mu^2 \sigma'^2\\
&+  \frac{\lambda f}{2}\sigma'\bm{\pi}^2 + \text{const.} \nonumber
\end{align}
We have transferred the overall mass term to only $\sigma$, allowing us to remove only the $\sigma$ as non-interacting, leaving the physical $\pi_1,\pi_2,\pi_3$ as massless and the fermions with mass $m_\psi = gf$. 
Compare this situation to the Higgs mechanism. In that case, we would then rotate to a gauge such that the triplet does not appear in the GB Lagrangian, instead appearing as mass terms for the gauge fields. We do not have gauge fixing freedom here. However, we can at least enforce only the presence of a massless triplet. We know this must be possible by Goldstone's Theorem, as we have broken $SO(4) \rightarrow SO(3)$, giving three massless modes. We will soon learn that the solution to the removal of $\sigma$ requires a non-linear parameterisation of the fields, that can be done elegantly using a "Goldstone Matrix". However, let's see how it can be done intuitively first.

\subsection{The Non-linear Sigma Model}
\label{sec:NLSM}

\paragraph{Our goal:}
\parbox{0.8\textwidth}{To outline the prototypical model of collective symmetry breaking in a physical model.\footnote{There are many differing conventions in this model, and the following references were used to distinguish the rigorous from the not: \cite{pich1998effective,Donoghue:2017pgk,ma2016lecture,santiago2009physics,DELRIOGAZTELURRUTIA1990319}}}
\vspace*{1em}

We chose in \cref{eq:f_def} a particular vacuum for the theory. This didn't affect the physics, as all the vaccua are equivalent. In expanding about that vacuum, we enforced the condition $\langle \sigma^2 \rangle + \langle \bm{\pi}^2 \rangle = f^2$. To remove the now massive $\sigma'$ we can integrate it out. This was our aim from the beginning. The easiest approach to this is to simply solve for the $\bm{\pi}$ fields, given $\sigma^2 = f^2 - \bm{\pi}^2$. 
Redefining $\sigma$ as $\sqrt{f^2 - \bm{\pi}^2}$ gives, from \cref{eq:GB_lagrangian},
\begin{align}
\mathcal{L}(\bm{\pi}) &= \frac{1}{2}(\partial \bm{\pi})^2 + \frac{1}{2}\frac{(\bm{\pi}\cdot \partial\bm{\pi})^2}{f^2 - \bm{\pi}^2} \left( + \frac{\mu^4}{4\lambda} \right)
\end{align}
by the chain rule, and where we will ignore the constant potential term. This definition is non-linear in the pions, and is commonly called the "square root representation". To express the Lagrangian in a form that is simply polynomial, expand around $\bm{\pi}^2 = 0$ to remove the field from the denominator
\begin{align}
\mathcal{L}(\bm{\pi}) = \frac{1}{2}\left[(\partial \bm{\pi})^2 + \frac{(\bm{\pi}\cdot\partial \bm{\pi})^2}{f^2} + \bm{\pi}^2	\frac{(\bm{\pi}\cdot \partial \bm{\pi})^2}{f^4} + \bm{\pi}^4 \frac{(\bm{\pi}\cdot \partial \bm{\pi})^2}{f^6} + ... \right] \label{eq:sqrt_expansion}
\end{align}
If we are sufficiently caffeinated, we would be able to find a more compact form for $\bm{\pi}$ that exploits this exponential-like series. To exponentiate a multiplet of fields requires it to be "dressed" as a matrix. The basis to choose should be clear from the polar re-definition of sigma: the basis of Pauli matrices. Then we might attempt to dress our multiplet as
\begin{align}
U = \exp{\left( \frac{i}{f} \bm{\pi}\cdot \bm{\tau}\right)} .
\end{align}
Let's check this intuition with the most generic Lagrangian we can form
\begin{align}
\mathcal{L} &= \text{Tr}[(\partial_\mu U)^\dagger (\partial^\mu U)]\label{eq:generic_GB_lagrangian}\\
&= \text{Tr}\left[\partial_\mu\left(1 + \frac{i}{f} \bm{\pi\cdot\tau} - \frac{1}{2! f^2} (\bm{\pi\cdot\tau})^2 - \frac{i}{3! f^3} (\bm{\pi\cdot\tau})^3 + ...\right)^\dagger \right.\nonumber\\
&\times \left.\partial^\mu\left(1 + \frac{i}{f} \bm{\pi\cdot\tau} - \frac{1}{2! f^2} (\bm{\pi\cdot\tau})^2 - \frac{i}{3! f^3} (\bm{\pi\cdot\tau})^3 + ...\right)\right] \nonumber
\end{align}
Due to non-commutativity, we must be careful with taking the derivatives, and apply the hermitian conjugate, reversing the order of the terms
\begin{align}
\mathcal{L} &= \text{Tr}\left[ \left( - \frac{i}{f}\partial_\mu \pi - \frac{1}{2f^2} \pi \partial_\mu \pi - \frac{1}{2f^2} \partial_\mu \pi \pi + \frac{i}{6f^3}\pi^2 \partial_\mu \pi + \frac{i}{6f^3}\pi \partial_\mu \pi \pi + \frac{i}{6f^3}\partial_\mu \pi \pi^2\right) \right. \nonumber\\
& \times \left.\left( \frac{i}{f} \partial^\mu \pi - \frac{1}{2f^2} \partial^\mu \pi \pi - \frac{1}{2f^2} \pi \partial^\mu \pi  - \frac{i}{6f^3} \partial^\mu \pi \pi^2 - \frac{i}{6f^3} \pi \partial^\mu \pi \pi - \frac{i}{6f^3} \pi^2 \partial^\mu \pi \right) \right]
\end{align}
using the notation $\bm{\pi} = \pi \cdot \tau$. Expanding the expression gives
\begin{align}
\mathcal{L} &= \text{Tr}\left[\frac{1}{f^2}(\partial_\mu\pi)^2  -
\frac{1}{6 f^3 \times f} \left( 4 (\partial_\mu \pi)^2 \pi^2 + 2 \partial_\mu \pi \pi \partial_\mu \pi \pi \right) \right. \nonumber\\
& \qquad \qquad + \left.\frac{1}{2f^2 \times f^2} \left( (\partial_\mu \pi)^2 \pi^2 + \partial_\mu \pi \pi \partial_\mu \pi \pi \right) \right]\nonumber \\
&= \frac{2}{f^2}(\partial\bm{\pi})^2 + \frac{2}{3f^4}(\bm{\pi}\cdot \partial \bm{\pi})^2 - \frac{2}{3f^4}\bm{\pi}^2 (\partial \bm{\pi})^2 + ... \label{eq:exp_expansion}
\end{align}
where the $\mathcal{O}\left(\frac{1}{f^3}\right)$ terms cancel due to the cyclic property of the trace. Note how we evaluate the dot products of Pauli matrices in the trace
\begin{align}
\text{Tr}[\left((\bm{\pi}\cdot\bm{\tau})(\partial \bm{\pi}\cdot \bm{\tau})\right)^2 - (\bm{\pi}\cdot \bm{\tau})^2(\partial \bm{\pi} \cdot \bm{\tau})^2] &\equiv 4(\bm{\pi}\cdot \partial \bm{\pi})^2 - 4\bm{\pi}^2 (\partial \bm{\pi})^2 \;. 
\end{align}
This comes from a trace identity for Tr$[(\bm{\pi}\cdot \bm{\tau})(\partial \bm{\pi} \cdot \bm{\tau})(\bm{\pi}\cdot \bm{\tau})(\partial \bm{\pi} \cdot \bm{\tau})]$, where
\begin{align}
\text{Tr}[(\pi^a \tau^a)(\partial\pi^b \tau^b)(\pi^c\tau^c)(\partial\pi^d \tau^d)] &= \pi^a \partial\pi^b \pi^c \partial \pi^d \text{Tr}[\tau^a \tau^b \tau^c \tau^d] \nonumber\\
&= 2 \pi^a \partial\pi^b \pi^c \partial \pi^d (\delta^{ab}\delta^{cd} - \delta^{ac}\delta^{bd} + \delta^{ad}\delta^{bc}) \nonumber\\
&= 2\left( 2(\bm{\pi}\cdot \partial\bm{\pi})^2 - \bm{\pi}^2(\partial\bm{\pi})^2\right) \;.
\end{align}
Combining this with the other component Tr$[(\bm{\pi}\cdot \bm{\tau})^2 (\partial\bm{\pi}\cdot \bm{\tau})^2]$
\begin{align}
\text{Tr}[(\pi^a\tau^a)(\pi^b\tau^b)(\partial\pi^c\tau^c)(\partial\pi^d\tau^d)] &= \pi^a\pi^b\partial\pi^c\partial\pi^d \text{Tr}[\tau^a\tau^b\tau^c\tau^d] \nonumber\\
& = 2\pi^a\pi^b\partial\pi^c\partial\pi^d (\delta^{ab}\delta^{cd} - \delta^{ac}\delta^{bd} + \delta^{ad}\delta^{bc}) \nonumber\\
&= 2 \bm{\pi}^2 (\partial\bm{\pi})^2 \, ,
\end{align}\marginpar{Self-derived}
gives the above result, which agrees with the more terse \cite{santiago2009physics}.

The key point is that our square root expansion in \cref{eq:sqrt_expansion} and exponential expansion in \cref{eq:exp_expansion} are not quite equivalent, even once we include a factor of $\frac{f^2}{4}$ in \cref{eq:generic_GB_lagrangian}. The divergence lies in the need for a non-linear field redefinition, and this is where our sigma model becomes non-linear. To match term-for-term, we can redefine in \cref{eq:sqrt_expansion} $\frac{\bm{\tau}\cdot \bm{\pi}}{f} \rightarrow \sin \frac{\bm{\tau}\cdot \bm{\pi}}{f}$, then $\Sigma = \sigma + i \bm{\tau}\cdot \bm{\pi} \rightarrow f\cos\frac{\bm{\tau}\cdot \bm{\pi}}{f} + i f\sin\frac{\bm{\tau}\cdot \bm{\pi}}{f}$. This type of redefinition will be returned to many times in later chapters.

We have stumbled upon a vital ingredient for Composite Higgs models: the Goldstone matrix $U$. This matrix precisely represents the symmetries being non-linearly realised, parameterised as fields.

\section{CCWZ: A General Prescription for Non-linear Symmetry}
\label{sec:CCWZ}

\paragraph{Our goal}
\parbox{0.8\textwidth}{To prove that collective symmetry breaking can be abstracted, and that spontaneous breaking terms can \textit{always} be written in the same form. 
}
\vspace*{1em}

\paragraph{Note}
\parbox{0.85\textwidth}{If the previous intuitive description of non-linear parameterisation of Goldstone Bosons is satisfactory, this section may be skipped. Otherwise, here I give a much deeper description of the connection between linearly and non-linearly realised symmetries, and the perturbative understanding of quantum field theory. }
\vspace*{1em}

We have already seen how a symmetry can be hidden at low energies ("in the IR") by shifting to a different vacuum of a scalar field. This was achieved by introducing at tree level some external potential with a non-zero expectation in the classical vacuum. If the degrees of freedom obey quantum field theory, then this potential can also be generated by the interactions of gauge fields with \textit{massless} scalars. That is, in the tree-level Lagrangian, we not only avoid an artificial quartic term $\frac{\lambda}{4!}\varphi^4$, but we also avoid a dangerous quadratic interaction $\frac{1}{2} \mu \varphi^2$ that would otherwise lead to a hierarchy problem (see next \cref{sec:dissatisfaction}). As we saw, these interactions do end up occuring at low energy, after we shift to the new vaccuum and the symmetry is hidden. But they are still prohibited by the symmetry at higher energies, and we can sleep easy.

Both of these mechanisms can be used to realise symmetries non-linearly, giving mass to particles that would normally be prohibited from being massive, and generating particles much lighter than the scale of the vacuum expectation. The classic example of this is the non-linear realisation of Chiral Symmetry - the approximate symmetry of the up, down and, depending on your taste, strange quarks. By parameterising this approximate global $O(4) \approx U(2)_L \times U(2)_R \approx SU(2)_L \times SU(2)_R \times U(1)_V \times U(1)_A$ symmetry using non-linear fields, and thus hiding the symmetry as $SU(2)_V \times U(1)_B$\footnote{$\{L,R,V,A,B\}\equiv \{$left, right, vectorial, axial, baryon number$\}$ } we can reveal a natural mass splitting between light (pseudo)scalar meson fields and heavy vector meson fields.

This general pattern of non-linearly realised symmetries will be relied upon heavily within this work, and is captured most generally by the Callan-Coleman-Wess-Zumino (CCWZ) construction \cite{CCWZ1,CCWZ2}. This is a way of dealing with collective (i.e. multiple sources of spontaneous global or local) symmetry hiding. Explicit symmetry breaking may also be incorporated with this construction. As we saw in the non-linear $\sigma$ model, the construction requires four ingredients:
\begin{itemize}
\item A set of matter fields $\varphi_i$ that transform in some way under
\item a global symmetry  $G$ of the physical system\footnote{This is required by the CCWZ construction \textit{to be specified}, but this is not a general requirement. That is, CCWZ is a top-down approach, where other bottom-up constructions may constrain NGB interactions directly by observables, e.g. scattering amplitudes. See \cite{low2015adler,low2015adler2} for fascinating derivations of NGB Lagrangians using Adler's zero condition.}, $g\in G: \varphi \rightarrow \varphi'$, but which specifically transform as linear irreducible representations (irreps) under
\item a subgroup\footnote{See \cref{sec:symmetric_space_appendix} for explanation of the conditions on this subgroup, e.g. that it be a stability group of the vacuum} $H \subset G$, $h \in H: \varphi \rightarrow \varphi' = D(h)\varphi$, and may also transform under 
\item a gauge symmetry of the system $\mathcal{H} \subset G$, $\ell(x) \in \mathcal{H}: \varphi(x) \rightarrow \varphi'(x) = D(\ell(x))\varphi(x) $
\end{itemize}
where $D(h,\ell)$ is a realisation of the subgroup $H,\mathcal{H}$, acting linearly on $\varphi_i(x)$, called a representation. Note that a gauge symmetry will be automatically a subgroup of the full global symmetry. Note also the convention of calligraphic lettering for explicitly gauged groups, while global symmetries (even if they are non-linearly realised and therefore spacetime dependent) are denoted by regular lettering.

\subsection{General Form of Non-linear Transformation}

\paragraph{Our goal:}
\parbox{0.8\textwidth}{To identify the operation that parameterises the non-linear group transformation}
\vspace*{1em}

We have our conditions, so we need only find transformations that obey these conditions and we will be well on the way to reproducing the non-linear $\sigma$ model in a general way. To do so, we need to establish a language for the theory of broken groups, which will be a little tedious but is quite necessary. We follow the cue of the original works, as well as \cite{xianyu2013nonlinear}. Provided the group transformations $g \in G$ are equipped with a Lie algebra (as all continuous compact groups must be), then we can write any group member as 
\begin{align}
g = \exp(i w^A S^A), && \text{where, } && A = 1,...,\text{dim}(G)
\end{align}
where $S^A$ are the members of the Lie algebra $\mathfrak{g}$ generating the group $G$, and $w^A$ are the parameters controlling the transformation. We can break the Lie algebra into a set that generates a subgroup $H$, and one that generates the coset $G/H$
\begin{align}
\{S^A\} = \{T^a, X^b\}, && \text{where, } && a = 1,...,\text{dim}(H), && b = 1, ..., \left(\text{dim}(G)-\text{dim}(H)\right)
\end{align}
For the remainder of this work, for simplicity, the range of the indices should be inferred from the generator it denotes. Once a subgroup is chosen, the generators can be ordered as above according to their commutation relations
\begin{align}
[T^a, T^b] = if^{abc} T^c, && [T^a, X^b] = i f^{abc} X^c, && [X^a, X^b] = if^{abc} T^c + i f^{abc} X^c\label{eq:commutation_relations}
\end{align}
for structure constants $f^{abc}$. For $T^a$ to be generators of $H$ is to say
\begin{align}
\me^{iu^a T^a} \in H, && \text{and} && \me^{i\pi^{b} X^b} \in G/H \label{eq:generators}
\end{align}
where $u^a, \pi^a$ are vectors of free parameters. The commutation relations \ref{eq:commutation_relations} should not be a mysterious ansatz. They are simply a rephrasing of the definition of a Lie subalgebra: A Lie subalgebra $\mathfrak{h}$ is a subset of the Lie algebra $\mathfrak{g}$ that is closed under the Lie bracket
\begin{align}
\mathfrak{h} \subset \mathfrak{g} && \iff && \forall \qquad x,y\in \mathfrak{h}, \qquad  [x,y]\in \mathfrak{h}\label{eq:subalgebra_def}
\end{align}
The fact that a subalgebra necessarily generates a subgroup will be clear shortly. Also note that generators not in the subalgebra are not necessarily closed under commutation, and therefore do not necessarily form a subalgebra or generate a subgroup.

Now observe that we can write any $g \in G$ transformation uniquely\footnote{The orthonormality of the generators gives the uniqueness: if $g=\me^{i\pi}\me^{iu}$ and $g=\me^{i\pi'}\me^{iu'}$ then $\pi = \pi', u =u'$. This is vital to the construction.} as
\begin{align}
g =  \me^{i \pi^b X^b}\me^{iu^a T^a} \label{eq:group_structure}
\end{align}
To see why, we can expand this expression with the Baker-Campbell-Hausdorff (BCH) formula, and the partition in \cref{eq:commutation_relations},
\begin{align}
\me^{i\pi^b X^b} \me^{iu^a T^a} &= \me^{i \left\lbrace \pi^b X^b + u^a T^a + \frac{1}{2}\pi^b u^a [X^b,T^a] + \frac{1}{12}\left( [X^b,[X^c,T^a]] + [T^a, [T^b, X^b] \right) + ...\right\rbrace} \label{eq:BCH_start}\\
 &= \me^{i \left\lbrace \pi^b X^b + u^a T^a + \frac{1}{2}\pi^b u^a i f^{abc} X^c + \frac{1}{12}\left( f(\pi^c,u^a,f^{abc}) X^b + f(\pi^b,u^c,f^{b a c})T^a\right) + ...\right\rbrace} \nonumber\\
 &= \me^{if(\pi^a,u^b,f^{abc}X^c + if(\pi^a,u^b, f^{abc})T^c} \nonumber\\
 & \equiv \me^{i \pi'^b X^b + i u'^a T^a} \nonumber\\
 &= \me^{i w^A S^A}\label{eq:BCH_end}
\end{align}
That is, there is a unique expression that allows us to decompose the full group into a series of these two operations, in the order given above. Note that lines \ref{eq:BCH_start} to \ref{eq:BCH_end} can be easily used to show that a subalgebra generates a subgroup, simply by replacing $\pi$ with $u$ and using the definition of the subalgebra (in \cref{eq:subalgebra_def}) to show $\me^{i u_1^a T^a} \me^{iu_2^a T^a} = \me^{i u_3^a T^a} \in H$. Now, the operation of $g$ is not yet defined. It may be \textit{any} transformation, i.e. any function of the $\psi$ fields, \textit{and of other group elements}\footnote{This is a slightly subtle point that we will return to shortly: the group $G$ has an "action" on the "target set" - the vector space of $V \ni \psi$. We can also choose its target set to be the group itself, with the same action, commonly called the "natural action" \citep{arvanitogeorgos2003introduction}[p66]}
\begin{align}
g: \qquad & \psi(x) \rightarrow \psi'(\psi(x), \pi, u)(x), \nonumber\\
& \pi \rightarrow \pi' = \pi' (\psi(x),\pi,u), \; u\rightarrow u' = u' (\psi(x),\pi,u)
\end{align}
where we note that generators may be suppressed for convenience, as
\begin{align}
u := u^a T^a , && \text{and} && \pi := \pi^b X^b.
\end{align}

However, we know that $G$ is a Lie group, which imposes conditions on the transformation. For example, almost every Lie group permits a representation $D$, and the group transformation may be chosen \textit{independently} of the field $\psi$. Additionally, we have already required that $\psi$ be in an irreducible representation (\textit{irrep}) of $H$\footnote{Even if $\psi$ is not in an irrep of $H$, it can then be treated as a vector of irreps $(\psi_1, \psi_2,...)$ due to the linearity of the transformations $(D_1, D_2, ...)$ of $H$, and the results of this section still apply to each irrep.}. That is, $\psi$ transforms by matrix multiplication with a function $D$ of \textit{some} set of $H$-generating parameters $u^{\prime a}$. With these conditions, we can narrow our transformation to 
\begin{align}
g:  \qquad & \psi(x) \rightarrow \psi'(x) =  D(\me^{iu^{\prime a}T^a})\psi(x) =  D(\pi, u) \psi(x), \nonumber\\
& \pi \rightarrow \pi'(\pi, u), \; u \rightarrow u'(\pi, u)
\end{align}
We should check that this realisation satisfies the properties of a group. It clearly has an identity, for $u' = 0$. It has an inverse $g^{-1}: \varphi \rightarrow D(\me^{-iu'^a T^a})\varphi$. And it has closure, given two elements $g_0, g_1 \in G$,
\begin{align}
 g_0\me^\pi &= \me^{\pi'}\me^{u'}, \; \; \text{ and, } \; \;  g_1 \me^{\pi'} = \me^{\pi''}\me^{u''}, \; \; \text{ by eq. } \cref{eq:group_structure} \nonumber\\
 \implies g_1 g_0 \me^\pi &= g_1 \me^{\pi'}\me^{u'} = \me^{\pi''}\me^{u''}\me^{u'} \nonumber\\
&= \me^{\pi''}\me^{u'''} = g_2 \me^{\pi'''}\, .
\end{align}
That is, the group transformations $g_0,g_1\in G$ on some $\pi\in G/H$ compose to another\footnote{Many results that are useful in the CCWZ construction can be found in \cref{sec:representations}.} $g_2\in G$ operating on some other $\pi'''\in G/H$, and the representations compose as they should, by the partition of \cref{eq:commutation_relations},
\begin{align}
D\left(\me^{u'''}\right) = D\left(\me^{u''}\right) D\left(\me^{u'}\right)
\end{align}

To uncover the form of the transformation on other group members (i.e. the "natural action"), consider first that 
\begin{align}
g \me^{i \pi} &\equiv g' \nonumber\\
&= \me^{i \pi'}\me^{i u'}
\end{align}
by the proposition in \cref{eq:group_structure}. Letting $g = h = \me^{i u_h}$ (by the definition in \cref{eq:generators}),
\begin{align}
h \me^{i\pi} = \me^{iu_h} \me^{i\pi} & \left(\equiv g' = \me^{i\pi'} \me^{iu'}\right) \nonumber\\
&= \me^{iu_h} \me^{i \pi} \me^{-i u_h} \me^{i u_h}
\end{align}
simply inserting an identity matrix on the right. We can expand the exponential to examine the behaviour of this adjoint operation
\begin{align}
\me^{iu_h} \me^{i\pi} \me^{-iu_h} &= \me^{iu_h} (\mathbb{1} + i\pi -\frac{1}{2} \pi^2 -...) \me^{-iu_h} \nonumber\\
&= \mathbb{1} + i\me^{iu_h}\pi \me^{-iu_h} - \frac{1}{2}\me^{iu_h}\pi \me^{-iu_h}\me^{iu_h}\pi \me^{-iu_h} -...\label{eq:conjugation_step}
\end{align}
where we simply inserted identities. The key to this derivation, is that we can evaluate these terms using a version of the BCH formula
\begin{align}
\me^{iu_h^a T^a}\pi^{b} X^{b} \me^{-iu_h^c T^c} &= \pi^{a}T^{a} + \frac{1}{2!}[u_h^a T^a, \pi^{b} X^{b}] +  \frac{1}{3!} [u_h^a T^a, [u_h^b T^b, \pi^c X^c]] \nonumber \\
& + ...+ \frac{1}{k!} \underbrace{[u_h^a T^a, [u_h^b T^b, [... ,[u_h^p T^p}_{k \; \text{times}}, \pi^q X^q]...] + ...\\
&= \pi''^a X^a \label{eq:subgroup_conjugation}\\
\implies \me^{iu_h} \me^{i\pi} \me^{-iu_h} &= \me^{i\pi''}
\end{align}
by the partition in \cref{eq:commutation_relations}. We will use this useful identity in later chapters. Therefore, using the uniqueness of proposition in \cref{eq:group_structure}
\begin{align}
h \me^{i\pi} &= \me^{i\pi''} \me^{i u_h} \nonumber\\
&= \me^{i\pi'} \me^{iu'} \nonumber\\
\implies & h = \me^{i u_h} = \me^{iu'} \; \text{and} \;  \pi'' = \pi' = h\pi h^{-1} 
\end{align}
Therefore, we have derived the action of the subgroup on elements in the coset. This turns out to be linear as we desired, since $u' = u_h$ is an independent choice
\begin{align}
h: && \pi \rightarrow h \pi h^{-1}, && \psi \rightarrow D(e^{iu_h})\psi \;.
\end{align}
On the other hand, for a general transformation $g$, we cannot evaluate the conjugation step $g \pi^{\hat{a}} T^{\hat{a}}  g^{-1}$ of \cref{eq:conjugation_step} as a function only of broken generators. The best we can do is
\begin{align}
g: && \me^{i\pi'} \rightarrow g \me^{i\pi} h^{-1}(\pi,g), && \psi \rightarrow D(h(\pi,g)) \psi \;.
\end{align}
That is, $\pi$ transforms in a non-linear way, while $\psi$ is able to still transform linearly (as a representation) but the cost is that it must do so in tandem with the $\pi$ transformation. This should sound familiar, and we will take a quick detour to physics in order to put all of these pieces together.

\vspace{1em}
\begin{tcolorbox}[colback = black!2!white]
\paragraph{Key point} Given \cref{eq:commutation_relations}, we can commute $\pi, e^{iu_h}$ with only a remaining $\pi'$. This leads to the Goldstone bosons transforming linearly under $H$, $\pi \rightarrow h\pi h^{-1}$.
However, we cannot in general commute $\pi, \exp^{i\pi_g}\exp^{i u_g}$. This leads to the Goldstone bosons transforming non-linearly as the Goldstone matrix under $G$, $U \rightarrow g U h^{-1}(\pi,g)$. We \textit{can} perform this commutation if $G/H$ is a symmetric space, which we will come to shortly.
\end{tcolorbox}
\vspace{1em}

\begin{table}
\begin{center}
\small
\hspace*{-1.5cm}\begin{tabular}{ @{} r | lclcl @{}} \toprule
 	& Linear & $\subset$ & Gauge & $\subset$ & Non-linear  \\ \midrule
	\parbox{1.5cm}{Transfor- \\ mation} & \parbox{1cm}{\begin{align*}
		\psi &\rightarrow \psi'(\epsilon, \psi)\\
			&= \me^{i\epsilon^A T^A} \psi \end{align*}} & & \parbox{1cm}{\begin{align*}
		\psi &\rightarrow \psi'(\epsilon(x), \psi)\\
			&= \me^{i\epsilon(x)^A T^A} \psi \end{align*}} & & \parbox{1cm}{\begin{align*}
		\psi &\rightarrow \psi'(\epsilon(x), \psi)\\
			&= D(h(g,\epsilon))\psi \end{align*}}\\
\parbox{1.5cm}{Covariant\\ Derivative} & \parbox{1cm}{\begin{align*}
		D^\mu \psi = \partial^\mu \psi \end{align*}} & &  \parbox{1cm}{\begin{align*}
		& D^\mu \psi = \partial^\mu \psi + iA^\mu_A T^A \psi\\
		& A=1,...,\text{dim}(\mathcal{G}) \end{align*}}
		& & \parbox{1cm}{\begin{align*}
		&D^\mu \psi = \partial^\mu \psi + i w^\mu_a T^a \psi\\
		& a=1,...,\text{dim}(H) \end{align*}}\\
\parbox{1.5cm}{Covariant\\ Fields} & \parbox{1cm}{\begin{align*}
		- \end{align*}}
		& & \parbox{1cm}{\begin{align*}
		& F^{A,\mu\nu} F^A_{\mu\nu}, \text{ or,}\\
		& \text{Tr}\left[ F^{\mu\nu}F_{\mu\nu}\right]\\
		& \text{"Killing Form"} \end{align*}} \footnote{\cite{gibbons2006part} and \cite{arvanitogeorgos2003introduction}, pages 60, 61, and https://physics.stackexchange.com/questions/52452/why-is-the-yang-mills-gauge-group-assumed-compact-and-semi-simple} & & 
\parbox{1cm}{\begin{align*}
&\hat{w}^\mu = \hat{w}^\mu_i T^i = -i U^\dagger \mathcal{D} U\\
&\text{where } U=\exp{(i\pi^i T^i)},\\
&i = 1,...,dim(G/H), \text{ and}\\
& w = -i U^\dagger dU = w^a T^a + \hat{w}^i T^i\\
& \text{"Maurer-Cartan Form"}
\end{align*}}\\ 
\bottomrule
\end{tabular}
\vspace*{1em}
\caption{A summary of the types of transformations in this study. Linear transformations are a subset of gauge transformations, which are a subset of non-linear transformations. The CCWZ construction defines a unique way to construct every non-linear transformation based on a non-linear group $G$ with a linear representation $D(h)$, $h\in H\subset G$. The construction may include a linear group dependent on spacetime (a "gauge group") $\mathcal{G}$}\label{tab:transformations}
\end{center}
\end{table}



For a non-linearly realised symmetry, we would like to represent the transformations themselves as fields. Just as for an infinitesimal gauge transformation
\begin{align}
g: \psi \in \mathcal{G} \sim \delta_{ij} - iq \epsilon_{ij}(x) + \mathcal{O}(\epsilon^2)
\end{align}
we can parameterise the symmetry as a field $\bm{\alpha}(x), \epsilon_{ij} = \alpha^a(x) T^a_{ij}$. Then to preserve the gauge symmetry of some field $\psi_\mu$ we need to introduce gauge fields $A_\mu^a(x)$ that transform as (linear) functions of these "symmetry fields" $\bm{\alpha}(x)$. Similarly, for a more general non-linear symmetry, 
\begin{align}
g: \psi \in G \sim \delta_{ij} - iq \epsilon_{ij}(x,\psi) + \mathcal{O}(\epsilon^2)
\end{align}
then we parameterise this symmetry in fields $\pi^a(x), \epsilon_{ij} = \pi^a T^a_{ij}$. To preserve the global symmetry of some field $\psi_\mu$, we introduce a Goldstone matrix of fields $U_{ij} = \me^{i\pi^a(x) T^a}_{ij}$ that transform non-linearly under the $\pi^a(x)$ field transformations.

We now make connection with the physical understanding that a vacuum expectation leads to non-linear symmetries. Consider an arbitrary multiplet $\Phi(x)$ with a vacuum expectation $\langle \Phi \rangle = f$. At high energies, the field may transform under $g \in G$, leaving the Lagrangian invariant. One can then define the field as a transformation away from its vacuum expectation vector $\vec{f}$, where some transformations will leave the vacuum invariant 
\begin{align}
\Phi &= \me^{i w^A(x) S^A} \vec{f} \nonumber\\
& = \me^{i \pi^b(x) X^b} \me^{i u^a(x) T^a} \vec{f} \nonumber\\
&= \me^{i \pi^b(x)X^b} \vec{f} \nonumber\\
&= \mathcal{U} \vec{f}\label{eq:goldstone_matrix}
\end{align}
In the language we have developed, we would say that the vacuum is invariant under the unbroken subgroup $H$, and therefore the field is entirely parameterised by the broken degrees of freedom $\pi^{\hat{a}}$ and the vacuum direction $\vec{f}$. The matrix $\mathcal{U}$ is the Goldstone matrix - a spacetime-dependent transformation of the vacuum in the direction of the coset $G/H$. 

\begin{figure}
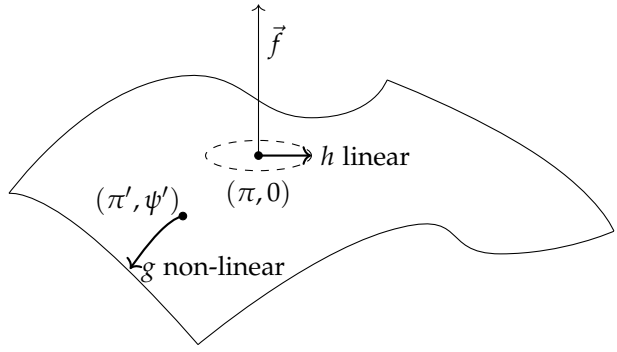

\centering
\tikz{
\draw plot [smooth, tension=1] coordinates {(3.5,0.5) (6,2) (7.5,1.7) (9,2)};
\draw plot [smooth, tension=1] coordinates {(9,2) (8,3) (6,4)};
\draw plot [smooth, tension=1] coordinates {(6,4) (5,3.5) (3,4) (1,2.5)};
\draw plot [smooth, tension=1] coordinates {(1,2.5) (2,2) (3.5,0.5)};
\draw [dashed] (4.3,3) ellipse (0.7 and 0.2);
\draw [fill=black] (4.3,3) circle [radius=0.05];
\draw [->] (4.3,3) -- (4.3,5);
\node [below] at (4.3,2.8) {$(\pi, 0)$};
\node [right] at (4.3,4.5) {$\vec{f}$};
\node [right] at (5,3) {$h$ linear};
\draw [thick, ->] (4.3,3) -- (5,3);
\node at (3.7,1.5) {$g$ non-linear};
\node [left] at (3.4,2.4) {$(\pi', \psi')$};
\draw [fill=black] (3.3,2.2) circle [radius=0.05];
\draw [thick, ->] plot [smooth, tension=1] coordinates {(3.3,2.2) (3,2) (2.6,1.5)};
}\caption{The manifold of matter $\psi$ and symmetry $\pi$ fields, around the origin, with a linear subgroup leaving the vacuum expectation invariant}\label{fig:manifold_symmetry}
\end{figure}

Now we assemble the physics and group theory: we denote the space of physical fields by a manifold with co-ordinates $\left(\pi(x), \psi(x)\right)$, which transforms under $g \in G$. Let the subgroup of transformations which preserve the vacuum $(\pi(x),0)$  be denoted $h \in H \subset G$, called the stability group of the vacuum. In the neighbourhood of this vacuum, all $h \in H$ have a linear representation (by the linearisation lemma \cite{CCWZ1})
\begin{align}
h = \me^{iu}: (\pi(x), \psi(x)) \rightarrow ( \me^{iu}\pi(x) \me^{-iu},D(\me^{iu})\psi(x))
\end{align}
In the neighbourhood of the vacuum, the full group acts on points of the manifold as
\begin{align}
g: (\pi(x), \psi(x)) \rightarrow (\pi'(\pi, g), D(\me^{iu'(\pi,g)})\psi)
\end{align}
We now see the reality that the matter fields $\psi$ and symmetry fields $\pi$ don't make sense when discussed alone - they are tightly connected on a non-linear manifold, with linearity of the Goldstone bosons only guaranteed at the physical vacuum. This idea is sketched in \cref{fig:manifold_symmetry}.

\subsection{Terms Invariant Under Non-Linear Transformations}

\paragraph{Our goal:}
\parbox{0.8\textwidth}{To determine what is the general form of a non-linearly realised symmetry in a Lagrangian.}
\vspace*{1em}

As illustrated in \cref{tab:transformations}, these non-linear transformations depend on the Goldstone fields, which are spacetime dependent. The transformations therefore inherit this spacetime dependence and become local invariances. It follows that the covariant derivative of the Goldstone fields $D_\mu U$ should be altered to account for the local degrees of freedom. We are particularly interested in derivative terms, because terms that enter in the Lagrangian as $f(U^\dagger U) = f(1)$ will be quite boring. Let's explore their behaviour
\begin{align}
g: \partial_\mu U \rightarrow &\partial_\mu(g U h(g,\pi(x))^\dagger) & & \partial_\mu \varphi \rightarrow \partial_\mu(h(g,\pi(x))\varphi)\\
= & g\left(\partial_\mu(U)h^\dagger + U \partial_\mu (h^\dagger)\right) & & =  h\partial_\mu(\varphi) + \partial_\mu(h)\varphi \label{eq:derivative_transformation}
\end{align}
where the dependencies of $h$ are included to emphasise $\partial_\mu (h) \neq 0$. The partial derivative clearly does not transform covariantly. However, we have a precedent for this type of transformation - a gauged matter field. We must introduce a gauge field $\mathcal{W}_\mu$ to cancel the second terms in \cref{eq:derivative_transformation}, such that
\begin{align}
D_\mu U = (\partial_\mu U - i U \mathcal{W}_\mu) \rightarrow g D_\mu U h^\dagger
\end{align}
It turns out that, just as having a gauge symmetry that depends on spacetime requires a spacetime-dependent gauge field, having a non-linear symmetry that depends on $\pi(x)$ fields requires a $\pi(x)$-dependent "gauge" field. We can show that the term $w_\mu = -i U^\dagger \partial_\mu U$ contains the ingredients for this field. 
\begin{align}
w_\mu \rightarrow & -i h U^\dagger g^\dagger \left(g \partial_\mu (U) h^\dagger + g U \partial_\mu (h^\dagger)\right)\\
&= i hU^\dagger \partial_\mu U h^\dagger - ih \partial_\mu h^\dagger\\
&= h( w_\mu  + i\partial_\mu) h^\dagger
\end{align}
this is the right form to cancel the term in \cref{eq:derivative_transformation}. There is one last consideration. Just as in the gauge field case, we must consider how the generators in $U= \me^{\pi^{\hat{a}}T^{\hat{a}}}$ commute with the generators in 
\begin{align}
w_\mu \rightarrow w'_\mu = E_\mu'^a T^a + d_\mu'^{\hat{a}}T^{\hat{a}}\label{eq:mc_definition}
\end{align}
which has generators from both sets under a general transformation. It is straightforward to see that since $h\partial_\mu h = f(T^a)$, this term will belong with the $E_\mu^a$ component. That is,
\begin{align}
E_\mu \rightarrow h(E_\mu  +  ih\partial_\mu)h^\dagger && d_\mu \rightarrow h d_\mu h^\dagger \label{eq:mc_transformation}
\end{align}
Then $E_\mu$ is the required gauge connection, and $d_\mu$ is a linear field. Now the covariant derivative for a field $\varphi$ transforming linearly under representation $r$ of $H$, and non-linearly under $G$ is
\begin{align}
D_\mu\varphi &= (\partial_\mu - i E_\mu^a T^a_r ) \varphi \label{eq:maurer_cartan_covariant_derivative}
\end{align}
This correctly cancels the extra term in \cref{eq:derivative_transformation}, giving an invariant matter field terms $\varphi\partial_\mu \varphi$, and $a=1,...,\text{dim}(H)$ gauge-like fields. It also prescribes how the Goldstone fields should enter the Lagrangian in order to be invariant - in the form of 
\begin{align}
d_\mu^{\hat{a}}T^{\hat{a}} = w_\mu - E_\mu = iU^\dagger \partial_\mu U - E_\mu = iU^\dagger (\partial_\mu U - iU E_\mu) = iU^\dagger D_\mu U \label{eq:maurer_cartan_invariant_term}
\end{align}
which is thus a covariant quantity, giving invariant kinetic terms $d_\mu^\dagger d^\mu$ in the leading-order Lagrangian
\begin{align}
\mathcal{L}_2 = \text{Tr}[d_\mu [U]^\dagger d^\mu [U]]\, .
\end{align}
Note that the $2$ subscript denotes that we have only included the two-derivative terms (and there's only one of these).

%

\subsection{Combining Global and Local Symmetries}
\label{sec:HLS}

\paragraph{Our goal:}
\parbox{0.8\textwidth}{To show a deep connection between gauge and non-linear global symmetry called Hidden Local Symmetry.}
\vspace*{1em}

\begin{figure}
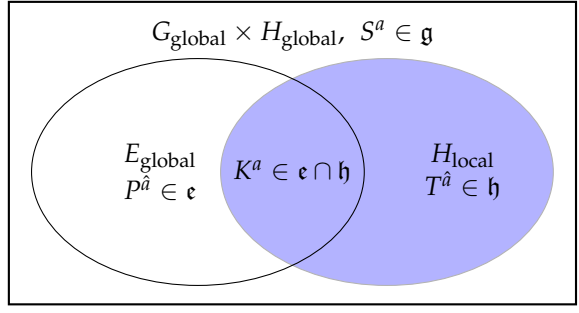

\centering
\tikz{
\draw [thick] (1,0.25) rectangle (8.5,4.25);
\draw [fill=blue, opacity=0.3](6,2) ellipse (2.2 and 1.5);
\draw (3.5,2) ellipse (2.2 and 1.5);
\node [align=center] at (4.75, 3.8) {$G_\text{global}\times H_\text{global},\;$ $S^a \in \mathfrak{g}$};
\node [align=center] at (3,2) {$E_\text{global}$ \\ $P^{\hat{a}} \in \mathfrak{e}$};
\node [align=center] at (4.75,2) {$K^a \in \mathfrak{e} \cap \mathfrak{h}$};
\node [align=center] at (7,2) {$H_\text{local}$ \\ $T^{\hat{a}} \in \mathfrak{h}$};
}
\caption{The notation of "hidden local symmetry"}\label{fig:HLS_notation}
\end{figure}

Consider the line of succession in \cref{tab:transformations}, from linear, to gauge, to non-linear transformations. Let's investigate more closely the case of a system invariant under more than one symmetry. Linear symmetries can be combined ad infinitum without interfering, e.g. successively adding electron, then muon, then tau lepton number conservation to the SM. Naively, gauge symmetries can be included independently as in the SM's seemingly orthogonal $SU(3)_c \times SU(2)_w \times U(1)_Y$. However, we will see in \cref{sec:dynamical_EWSB} that when a gauge coupling becomes strong, loop corrections lead to confinement - effectively mass-like terms. This realises a non-linear symmetry (in the SM, via a bilinear quark condensate) that precisely follows the prescription of the non-linear sigma model. A strong gauge symmetry then implies the presence of a non-linear symmetry. Combined with a weak gauge symmetry, we get situations such as those described in \cref{sec:dynamical_EWSB}: dynamical electroweak symmetry breaking. Combining an arbitrary non-linear symmetry with a gauge symmetry can be described by \cref{fig:HLS_notation}. 

Consider a NGB matrix $\Omega(x)$ transforming linearly as the fundamental of a global group $g \in G_{\text{global}} \equiv G_L$ and the antifundamental of a local group $h \in H_{\text{local}} \equiv H_R \subset G_L$\footnote{For this discussion, $H_R$ is required to be generated by a subset of $G_L$ generators ($H_R \subset G \cong G_L$), though is still a separate symmetry. This is not always true, but we will see why it is a useful requirement in the case of composite models shortly. 
}
\begin{align}
G_L \times H_R: \Omega(x) \rightarrow g \Omega(x) h^\dagger(x)
\end{align}
That is, giving $\Omega(x)$ a vev will spontaneously break $G_L \times H_R$ to the diagonal subgroup $G_{G_L + H_R}$. $\Omega(x)$ can be decomposed into matrices consisting of generators of $H_R$, $T^a \in \mathfrak{h}$ and those in $G_L$ not in $H_R$, $X^{\hat{a}} \in \mathfrak{g} - \mathfrak{h}$
\begin{align}
\Omega(x) = \me^{i\pi_U^{\hat{a}} X^{\hat{a}}} \me^{i\pi_\Xi^{a} T^a} \equiv U(x) \Xi(x)
\end{align}
Note that $\Omega(x)$ is composed of our usual $G/H$ NGB matrix, and an extra term $\Xi(x)$ (called in the literature the "compensator"). By construction, it is in the gauge group $H_R$, so we can choose an appropriate gauge $h(x) = \Xi(x)$ to (wait for it) compensate for it
\begin{align}
\Omega(x) \xrightarrow{h \in H_R} \Omega'(x) =  \Omega(x) h^{-1}(x) = U(x)\label{eq:link_unitary_gauge}
\end{align} 
This is the unitary gauge, since we have removed the unphysical degrees of freedom corresponding to the gauge generators, which have now given mass to the gauge bosons. Unfortunately, applying a $G_L$ transformation does not preserve the unitary gauge
\begin{align}
G_L: \Omega'(x) = U(x) \rightarrow g \Omega'(x) h^{-1}(x) &= \Omega''(x) h^{-1}(x)\\
 &\equiv U'(x) \Xi'(x) \Xi^{-1}(x) \neq U'(x)
\end{align}
in general, since $\Xi'(x)$ could be any element of $H_R$. To again compensate for it, the gauge must be re-fixed to $h(x) = \Xi'(x)$. Thus, to preserve the unitary gauge, each $G_L$ transformation must be non-linear
\begin{align}
G_L: U(x) \rightarrow g U(x) h^{-1}(x,g) = U'(x)
\end{align}
which is the same transformation as for a generic NGB field in the CCWZ construction. The equivalence is not exact: We expanded our symmetry space, and that led to massive vector bosons $(A_H)_\mu$ in the leading-order Lagrangian 
\begin{align}
\mathcal{L} &= f^2 \text{tr} |D_\mu \Omega(x)|^2	\\
\xrightarrow{h \in H_R} \mathcal{L}_\text{unitary} &= f^2 \text{tr} |\partial_\mu U(x)|^2 - \frac{f^2_\Xi g_H^2}{4} (A_H)_\mu 
\end{align}
The vector boson mass terms are therefore proportional to the gauge coupling, and taking $g_H\rightarrow \infty$ allows them to be decoupled. Applying the equations of motions to the vector bosons leads to the solution $i(A_H)_\mu = e_\mu$, as defined in the CCWZ section. In this limit then, the Lagrangian becomes
\begin{align}
\mathcal{L} = f^2 \text{tr} [d^\mu d_\mu^{\dagger}]
\end{align}
and the equivalence to a $G/H$ CCWZ construction is exact. This system of introducing a gauge symmetry that limits to a global symmetry is known as Hidden Local Symmetry (HLS). The HLS method thus allows us to introduce relatively heavy vector resonances without sabotaging the carefully-laid group structure leading to a pNGB Higgs. We will examine some extra components to this HLS discussion in \cref{sec:mooses}.

\section{Dissatisfaction in the Higgs Sector}
\label{sec:dissatisfaction}

\paragraph{Our goal:}
\parbox{0.8\textwidth}{To motivate new sources of collective symmetry breaking beyond the Standard Model.}
\vspace*{1em}

\begin{figure}
\centering
\includegraphics[scale=0.6]{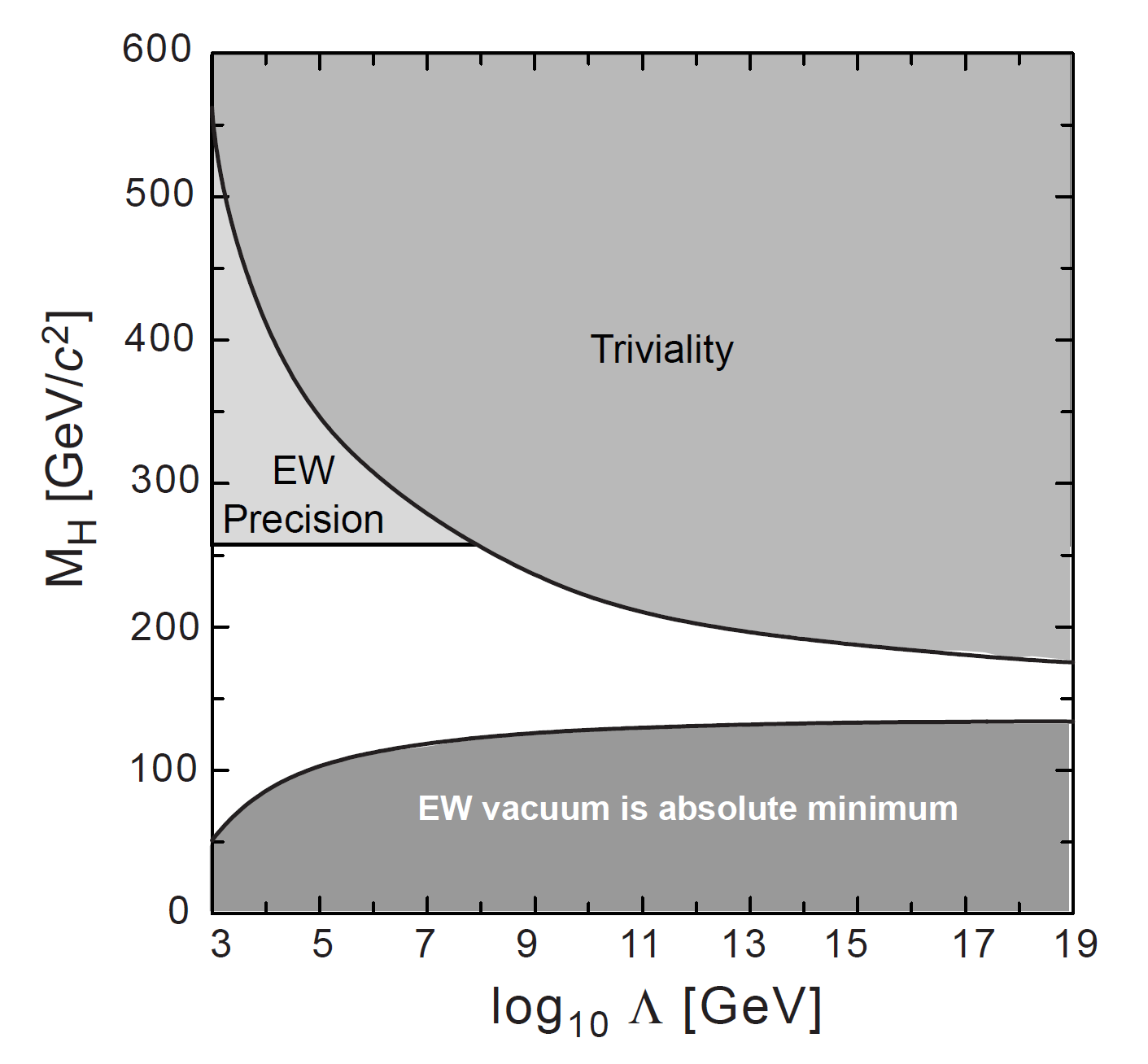}
\caption{An overview on the bounds of the Higgs mass, in order to avoid triviality, vacuum instability, and EW precision exclusion. Reproduced from \cite{Quigg:1999xg}}
\label{fig:higgs_scale}
\end{figure}

While there are no significant experimental holes in the SM Higgs sector, the existence of an elementary scalar field such as the Higgs challenges the elegance of a world described by quantum field theory, in several ways. A general sketch of the Higgs scale structure is given in \cref{fig:higgs_scale}. Although \textit{slightly} out of date, this sketch describes the idea behind bounding the Higgs mass by certain features that would be attractive to have in the Standard Model. We now know the mass of the SM Higgs, so we can use the figure in reverse - to investigate which features appear or are lost at certain scales, assuming an elementary Higgs of mass $m_h = 124.97 \pm 0.28$GeV \cite{ATLAS:2017bxr}. 

\begin{itemize}
\item \textbf{Quantum triviality}

Consider the coupling constant relationship that is obeyed by a renormalisable theory (\cref{eq:coupling_equation}). In the case of the strong force, perturbativity fails at low energy scales and a phase transition to the confined regime requires a different approach. The consistency of QCD at low energies is thus an extremely active field of research. However, for high energies, perturbativity holds and the RGE guarantees the coupling tends to some limit. This is what is meant by "asymptotic freedom". What about the opposite case? The Higgs sector appears to be perturbative at low energies (it couples weakly), and non-perturbative at some high-energy scale.

A quantum field theory is non-trivial if the coupling is bounded at all scales. To see why the name "non-trivial" is used, consider the RGE tending to infinity at some finite scale. This scale must have been experienced by an early universe, and would be a catastrophic moment. To avoid this, we should set the coupling at this scale to zero. But then, by the RGE, it must be non-interacting at all scales. We then have a free field theory that can never be detected: trivial. It may contain gravitational interactions, but it cannot describe our Higgs field, given its recent discovery by weak and Yukawa interactions. There is considerable analysis as to whether an elementary scalar particle can avoid triviality, and the determination involves non-perturbative methods, or at least next-to-leading order calculation \cite{CALLAWAY1988241}. As QED can be tested for triviality and shows a Landau pole, albeit at a scale far beyond the Planck scale, so too can the Higgs sector be searched for poles. This is an unsettled issue, but demands that elementary scalars be handled with caution. 
  
  
\item \textbf{Vacuum stability} 

The Standard Model Higgs has a tree-level potential that causes the field to collapse to an infinite set of identical minima. However, we learned in \cref{sec:coleman_weinberg} that for scalars, potential terms are also generated radiatively. Are these terms negligible, do they reinforce the stability of the potential, or do they introduce new, more energetically favoured minima? This clearly depends on the relationship between the radiative terms introduced by the top quark, and the tree-level vev. Reference \cite{Degrassi2012} gives the most recent, NNLO bounds for this condition. Absolute stability of the EW vacuum up to the Planck scale is guaranteed for a Higgs mass of 
\begin{align}
m_h > 129.4 \pm 1.8 \gev
\end{align}
which, for an upper bound on a Higgs mass at $m_h = 124.97 + 0.28 = 125.25$ excludes absolute vacuum stability by around $2.3\sigma$. It would seem that we inhabit a thin slice of $\{m_t, m_h\}$ parameter space permitting a metastable electroweak vacuum. Metastability here means that, while the vacuum is not the true, global minimum, it provides for an extremely low probability of universal collapse into a non-EW vacuum. The time scale $\tau_{\text{EW}}$ required for the nucleation of a tunnelling bubble large enough to  pervade the universe, is much greater than the age of the universe itself $\tau_{\text{EW}} \gg \tau_{\text{U}}$ \cite{Espinosa:2015kwx}. Instability ($\tau_{\text{EW}} < \tau_U$) is entirely excluded. The extremely small likelihood of living on this slice suggests an appeal to naturalness. Aside from anthropic principles, we would like to suggest that there is a natural reason for this apparent coincidence. There is another major coincidence in the SM besides metastability: tuning of the Higgs mass.

\item \textbf{Tuning correction}

Just as the Higgs vev is radiatively corrected in \textit{just} the right way for a stable universe, so the Higgs mass receives a radiative correction that gives a Higgs mass at the electroweak scale
\begin{align}
m_h^2 = m_{h,\text{bare}}^2 + g_t^2 m_t^2 + \mathcal{O}(g_t^4)
\end{align}
The correction term is of order $\mathcal{O}(100)\gev$, or lower. Unlike metastability, where the vev is measured to be "just right", the Higgs mass correction has a free parameter, $g_t^2$.  Given the top mass squared is of order $\mathcal{O}($10,000$ \gev^2)$, $g_t^2$ must be less than $0.01$. Thus we require an EW-scale correction to be delivered by an extremely small coupling constant. The Higgs mass is then extremely sensitive to changes in this parameter space. Again, we would like a reason for this hierarchy of contributions that is more than an appeal to epicycles. This description of the "Hierarchy Problem" is not meant to be particularly convincing - in section \ref{sec:fine_tuning} we will develop some rigorous techniques for dealing with what is typically an esoteric metaphysical concept.


\item \textbf{Mass spectrum of quarks and leptons - Yukawa structure}

The Higgs sector induces a mass hierarchy below the EW scale that is entirely arbitrary within the SM. The Yukawa matrices contain entries that lead to fermion masses spanning five orders of magnitude, controlled by \textbf{seven?} free parameters (three couplings in the lepton sector, three angles and a phase in the quark sector). While this issue will not be directly addressed in this work, several convincing studies \cite{niehoff2016direct, niehoff2017electroweak} are tackling the flavour hierarchy with solutions similar to those in this work. 
\end{itemize}

\section{Beyond Standard Model Possibilities}

\paragraph{Our goal:}
\parbox{0.8\textwidth}{To give the lay of the land of Higgs sector alternatives. Much deeper detail can be found in \cite{Csaki:2016kln}.}
\vspace*{1em}

None of the previous considerations imply that there \textit{must} be new physics above the EW scale ($\mathcal{O}(100\gev)$). However, the concept of naturalness - the inverse of fine tuning - is highly suggestive of a rock and a hard place. If there is new physics at the $\mathcal{O}(1\tev)$ scale, we will have to answer the Little Hierarchy problem of why the EW scale is suppressed by an order of magnitude, and hopefully the new physics staves off dangerous quadratic corrections. The alternative is far worse - that there is no new physics until the $\mathcal{O}(10\tev)$ or $\mathcal{O}(100\tev)$ scale, or even higher, and our Little Hierarchy Problem is starting to look once more like a Big Hierarchy Problem. These are ultimately philosophical problems - metaphysics rather than physics. But metaphysics can guide physics, casting shadows that we can attempt to shine light on, and maybe find something new in the process. 

\begin{itemize}
\item \textbf{Supersymmetry}

Supersymmetry (SUSY) proposes an extension of the Poincar\'e group of spacetime transformations. By creating a Lie superalgebra, with anticommutation relations with the $SO(3,1)$ generators, spin can be nontrivially incorporated into the spacetime symmetries. Indeed this is the only way that it can be unified with the symmetry paradigm, since the Coleman-Mandula theory prevents it from being combined with internal symmetries. SUSY elegantly provides cancellations that solve the hierarchy problem, and many other attractive features such as unification of gauge couplings. Elegance, however, is no substitute for evidence, and there has been no sign of the predicted SUSY superpartners at the LHC. But there is no cause for SUSY-enthusiasts to despair. Experimental bounds on top-like particles will be discussed in detail in \cref{sec:LCHM}, but in short, the $~1\tev$ region is still perfectly valid ground for heavy partners or resonances. Recent fine tuning studies have carefully analysed precision observables and concluded a tuning of order $10\%$ for top partners in the $2-5\tev$ range \cite{Buckley:2016tbs}. This is comparable to the models described in this work.

\item \textbf{GUT groups}

Instead of enlarging the set of spacetime transformations, one could imagine unifying the gauge groups by enlarging the set of internal transformations. In doing so, one might hope for some dynamics that explains the lightness of the EW scale compared to the GUT scale. The Pati-Salam model \cite{pati1994lepton} and the Georgi-Glashow model \cite{georgi1974unity} enable the quarks and leptons to be embedded into the same representation. These GUT groups ($SU(5)$ and $SO(10)$ are the minimal cases) are gauged, thereby implying many new bosonic particles, and a collective symmetry breaking structure $SU(5) \rightarrow SU(3)_c \times SU(2)_L \times U(1)_Y$, from some Higgs mechanism at the GUT scale $\mathcal{O}(\sim 10^{11}\tev)$. This structure dictates that the EW Higgs sector be embedded as a fiveplet 
\begin{align}
H^a = \left(\begin{matrix}
H^\alpha\\
\phi^+\\
\phi^0
\end{matrix}\right)
\end{align}
which gives the SM Higgs doublet $\phi = (\phi^+, \phi^0)$ and a colour triplet $H^\alpha$. Unfortunately, to balance two scales separated by ten orders of magnitude still requires a great deal of fine tuning. GUT models can somewhat reduce this tuning by incorporating SUSY, but at the cost of introducing vast numbers of parameters. The struggle between Fine Tuning and Occam's Razor is discussed in section \ref{sec:fine_tuning}.

\item \textbf{Higher Dimensions}

Both of the above considerations are simply extensions to the Standard Model's symmetry structure, without affecting its status as a four-dimensional quantum field theory. A classic alternative is to attempt to incorporate gravity into the quantum field theory, but this naively leads to a non-renormalisability. Instead, one could alter the spacetime structure with extra dimensions. Certain constructions (for example 11-dimensional string theory, or 10-dimensional M theory) are compatible with the previous models. These are many decades away from being disprovable. A simpler construction, that could be excluded within the next twenty years is the addition of (an) extra compact dimension(s). These are known as Randall-Sundrum models, and the simplest introduces an extra dimension of finite length. Energy scales are represented as being at one end $L_0$ (EW scale) of the dimension, or the other $L_1$ (Planck scale). By embedding the EW group at $L_0$, and the Higgs at $L_1$, one can describe both with one underlying symmetry. Thus the Higgs and EW gauge bosons are unified as one multiplet of fields - Gauge-Higgs Unification (GHU). This is an extremely attractive theory, which comes with an elegant four-dimensional description in terms of confined fields, thanks to the AdS-CFT correspondence \cite{maldacena1999large}. The weakly-interacting, extra-dimensional field theory then motivates us to consider a strongly-interacting confining field theory: effectively a BSM non-linear sigma model.


\item \textbf{Global symmetry breaking}

Ever since the eight-fold way enabled predictions of complex, strongly-interacting fields, non-linear global symmetries have been employed to describe shortcomings of the EW sector. Prior to the discovery of the Higgs, an alternative existed that could break EW symmetry and deliver masses to the SM particle spectrum: Technicolor.
\end{itemize}

\section{Dynamical Electroweak Symmetry Breaking}%
\label{sec:dynamical_EWSB}

\paragraph{Our goal:}
\parbox{0.8\textwidth}{To describe a historical extension to the Standard Model, and see where it falls short.}
\vspace*{1em}

\begin{figure}
\centering
\includegraphics[scale=0.1]{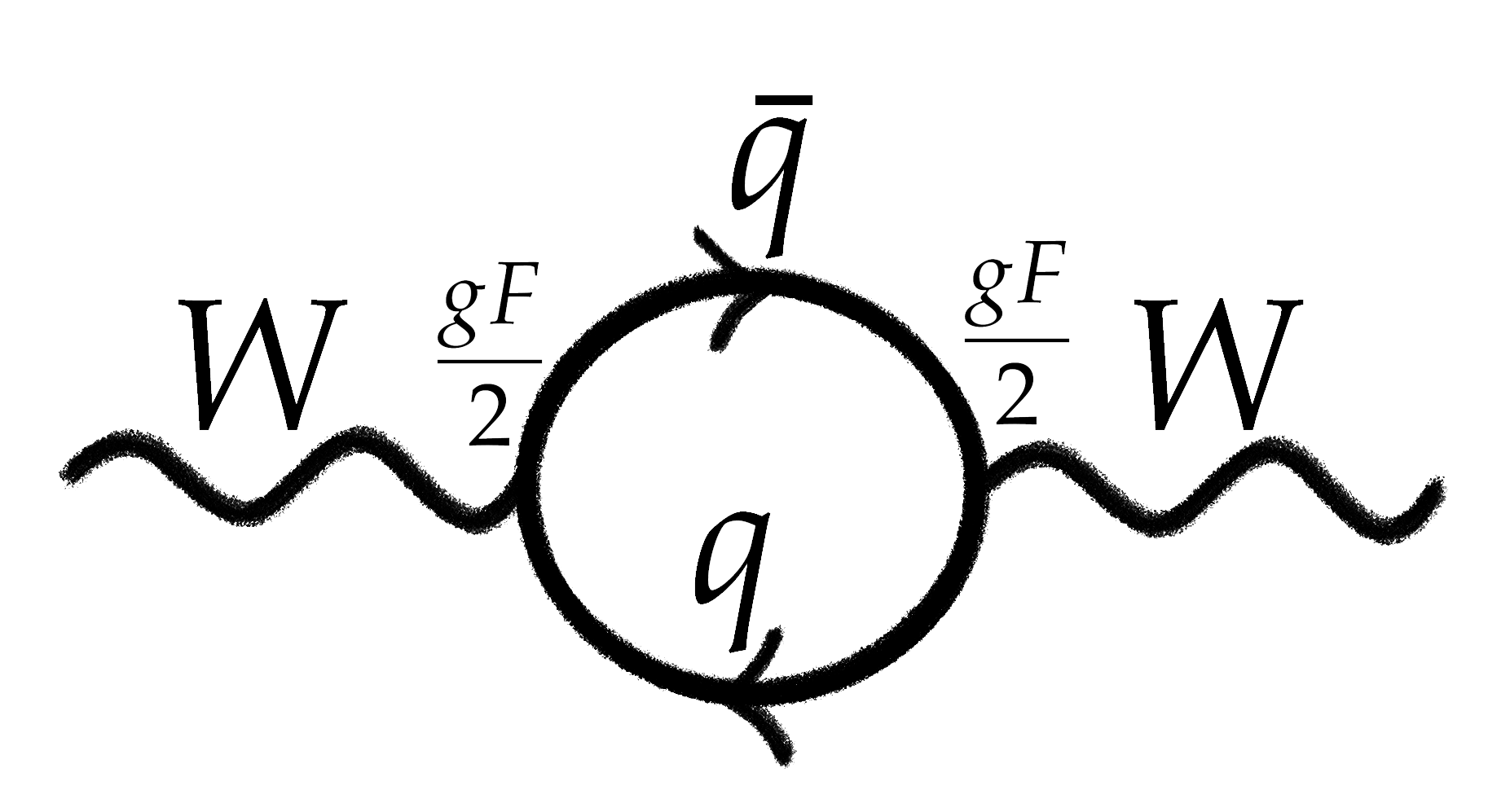}
\caption{A typical contribution to the mass of the W boson}
\label{W_qq_interaction}
\end{figure}

Consider the model\footnote{
In the following, we take a cue from \cite{bardeen1994chiral, hill2003strong}, which in turn makes much use of \cite{lane1994introduction}.} built in \cref{sec:chiral_symmetry}. There, we considered two species of Dirac fermion (equivalent to four species of Weyl fermion), in a sterile situation where gauge interactions were switched off and the fermions were massless. We can build quantum chromodynamics from this model simply by extending the number of fields to six, and introducing a strong non-chiral gauge symmetry $G_S = SU(3)_\text{color}$. Now we can arrange the terms as we did in \cref{eq:chiral_lagrangian}
\begin{align}
\mathcal{L} = \bar{Q}_L \left(i \slashed{\partial} - i g_s\slashed{G}^c\lambda^c\right) Q_L + \bar{Q}_R \left(i \slashed{\partial} - i g_s\slashed{G}^c\lambda^c\right) Q_R
\end{align}
These multiplets transform under under a chiral $SU(6)_L \times SU(6)_R$ transformation 
\begin{align}
Q_{L,R}^T = (Q^T_{L,R})^\alpha_i = \left(\begin{matrix}
u_{L,R}, d_{L,R}, c_{L,R}, s_{L,R}, t_{L,R}, b_{L,R}
\end{matrix}\right)^\alpha \; \rightarrow \;  Q'_{L,R} = e^{i\alpha^a T^a_{L,R}}Q_{L,R}
\end{align}
Rather than a Mexican hat potential, as we imposed "by hand" in the sigma model, the strong coupling generates a condensate of left-and-right quark pairs 
\begin{align}
\langle \Omega | (Q_L)^\alpha_i (Q_R)^\beta_j | \Omega \rangle  = \delta_{ij} \delta_{\alpha\beta} \Delta_Q \simeq \delta_{ij} \delta_{\alpha\beta} 4\pi f_\pi^3
\end{align}
that breaks this chiral symmetry $SU(6)_L \times SU(6)_R$ to its vector subgroup $SU(6)_V$. $f_\pi \approx 93\,$MeV is the experimentally determined pion decay constant. To complete the analogy with the sigma model, we can reparameterise this condensed (and therefore colorless) meson as a new field 
\begin{align}
\Sigma & \equiv \epsilon^{\alpha\beta} (Q_L)^\alpha (Q_R)^\beta \equiv \exp\left(i\pi^a T^a/f_\pi \right)\\
\text{where} \; \Sigma & \xrightarrow{SU(6)_L \times SU(6)_R} L \Sigma R^\dagger
\end{align}
where this form is established in \cite{peskin1980alignment}. Now, we would like to introduce one further feature. If we switch the electroweak interactions back on, then we need to arrange our sixplets into multiplets of degenerate "weak charge" (i.e. chirality, which we've already done) and hypercharge (i.e. separate up-type from down-type). In this representation, the meson field appears as three bidoublets, and we get terms (to first order in the exponential expansion) such as
\begin{align}
\mathcal{L} &\supset \sum\limits_{g=1}^3 (D^\mu\Sigma_g)^\dagger(D_\mu\Sigma_g)\\
& \supset \frac{g}{2} f_\pi W_\mu^+ \partial^\mu \pi^- + \frac{g}{2}f_\pi W_\mu^- \partial^\mu \pi^+ + f_\pi (\frac{g}{2}W^0_\mu + \frac{g'}{2}B_\mu) \partial^\mu \pi^0 \label{chiral_EW}
\end{align}
Observe that the gauge bosons now have a longitudinal component in the form of three linear combinations $\pi^{+,-,0}$ of the Goldstone bosons. (The interactions are of the sort of \cref{W_qq_interaction}.)
This appears exactly like the SM Higgs mechanism, with a electroweak vev of $v = \sqrt{n_g} f_\pi = \sqrt{3}f_\pi$. Then, we can compute the $W, Z$ masses
\begin{align}
m_W &= \frac{g}{2} v, && m_Z= \frac{\sqrt{g^2 + g'^2}}{2} v\\
&\simeq 52.7\mev,  && m_Z \simeq 59.6\mev
\end{align} 
which are, of course, nonsense. Although it is indeed a contribution to the vector boson masses, it is usually negligible in the mass calculation. There are many other inadequacies here: we have $(2n_g)^2-1-3 = 32$ extra Goldstone bosons in the theory, many of which are charged and massive. There is a residual $(SU(3)_u \times SU(3)_d)_L+R$ flavour symmetry which leaves all up-type quark masses degenerate, and the same for down-type quarks. It does not give leptons mass. One might also add the fact that we've \textit{discovered} a Higgs boson to the list of shortcomings, but actually the sigma field around the broken vacuum $\Sigma = \frac{1}{\sqrt{2}}(\sigma(x) + v)$ is consistent with group theoretic properties of the Higgs. Indeed, it must be, as the symmetry breaking is precisely the same as in figure \ref{fig:higgs_mech}. This also gives us a custodial symmetry $SU(2)_L \times SU(2)_R \subset SU(6)_L \times SU(6)_R$, which delivers the correct Weinberg angle. If such a "pionic" Higgs were produced, we would name it a Composite Higgs. 
%
%

Instead, we could imagine an extra set of quarks called \textit{techni}quarks that followed the same process of condensation and Higgs mechanism, but had a vacuum expectation value of $\mathcal{O}(100)\;\gev$ and could thus contribute meaningfully. These techniquarks - at least the ones light enough to be measured at the LHC - should be singlets under the SM color group, otherwise they would have been detected in decay processes \cite{king1992recent}. In this toy model, they could condense under a new, technicolor group $G_\text{TC}$, which expands the set of symmetries to a Technicolor Standard Model (TCSM) of
\begin{align}
SU(2)_W \times U(1)_Y \times SU(3)_c \times SU(N)_\text{TC}
\end{align}

We would insert the techniquarks as we did with the quarks, into $i=1, 2, ... n_f$ electroweak doublets $Q^i_L, Q^i_R$. This gives us a global $SU(2n_f)_L \times SU(2n_f)_R$ symmetry that can be broken by $\left\langle \bar{Q}Q\right\rangle = \sqrt{F_1^2 + F_2^2 + ... + F_{n_f}^2} = F_\pi = 246$GeV. However, there is a glaring shortcoming of this model. In the TCSM, as in chiral symmetry breaking, we have a pion decay constant because the pion is a pNGB due to the non-zero quark masses. Although we described electroweak mass generation, we have completely ignored how to generate the fermion mass spectrum. This requires adding an \textit{extension} to our technicolor symmetry. This ETC ($\mathcal{O}(100)$TeV) symmetry would break to the much lower TC ($\mathcal{O}(1)$TeV) symmetry, giving a mass to the techniquarks, quarks and leptons, which in turn produces a non-zero decay constant and mass for the vector bosons. We are trading in a hierarchical Higgs problem, for a hierarchy of technicolor scales. This is a vastly more elegant hierarchy. The scalar hierarchy requires cancellation of astronomically large parameters, while QCD-like hierarchies can be exponentially generated with ease by renormalisation group running. These allow parameters $p$ to be chosen that give scale differences of $\mathcal{O}(\me^{p})$.


Experiment tells us that a technicolor Standard Model is not the full story. For one, because we have discovered a particle that behaves very closely to the SM Higgs. Any composite Standard Model must include the Higgs as a prediction, or it fails experiment. If the Higgs is included in the TCSM as an elementary scalar, we have simply ignored the hierarchy problem and introduced unnecessary complexity. It makes sense to include the Higgs as a \textit{result} of some underlying dynamics, making it a composite object. Although in this thesis, we will not look at the specifics of this underlying dynamics, this section shows that dynamic electroweak symmetry breaking is a well-motivated and well-understood phenomenon. However, to calculate the full behaviour of any particular underlying model requires either lattice simulation or higher-dimensional correspondence, which are outside the scope of this study. Without knowing the full underlying dynamics, we can make extraordinary progress using only the effective field theory of Composite Higgs models.

\chapter{The Higgs as a Composite Goldstone Boson}
\label{sec:composite_higgs}

\section{Introduction}

\paragraph{Our goal}
\parbox{0.8\textwidth}{To gain a high-level understanding of, and motivation for, a composite Higgs toy model.}
\vspace*{1em}

From the previous chapter, we know that there are potential theoretical shortcomings in the SM. Some of these are simply awaiting more detailed analysis (e.g. quantum triviality), or tighter measurements of Higgs mass and couplings (e.g. vacuum stability). The hierarchy problem is deeply baked into the SM, and doesn't suggest any particular alternative, although Susy solves it elegantly. Technicolor has QCD as a strong blueprint, but it doesn't require the Higgs sector to be implicated. So aside from the elegance of a naturally light pNGB, why should we propose specifically that the Higgs is composite? Because effective composite dynamics appear as the low-energy behaviour of many, seemingly disparate BSM models. This is regardless of whether the underlying theory is some new strongly-interacting force, or a higher-dimensional interaction such as GHU, which will be outlined in \cref{sec:Landscape_4dMCHM}. If significant deviations from couplings are detected at a future International Linear Collider (ILC), or a new resonance is detected at the LHC or a Future Circular Collider (FCC), Composite Higgs models can give structure to the proposed explanation. Whole classes of models can be reduced to models of composite fields obeying some unitary or orthogonal global symmetry. A full description of these groups is given in \cref{sec:group_theory}. From here, we assume a working understanding of group theory, representation theory, and the connection between coset spaces and Goldstone bosons (endearingly called "Cosettology" \cite{sannino2016fundamental}).

\begin{figure}
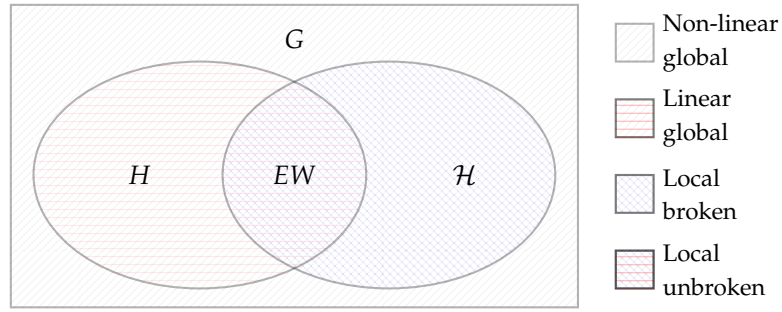

\centering
\tikz[scale=1, every node/.style={transform shape}]{
\draw [pattern=north east lines, pattern color=gray, thick, opacity=0.3] (1,0.25) rectangle (8.5,4.25);
\draw [pattern=north west lines, pattern color=blue, thick, opacity=0.3](6,2) ellipse (2.2 and 1.5);
\draw [pattern=horizontal lines, pattern color=red, thick, opacity=0.3] (3.5,2) ellipse (2.2 and 1.5);
\node [align=center] at (4.75, 3.8) {$G$};
\node [align=center] at (2.7,2) {$H$};
\node [align=center] at (4.75,2) {$EW$};
\node [align=center] at (7,2) {$\mathcal{H}$};
\draw [pattern=north east lines, pattern color=gray, thick, opacity=0.3] (9,0.5) rectangle (9.5,1);
\draw [pattern=north west lines, pattern color=blue, thick, opacity=0.3] (9,0.5) rectangle (9.5,1);
\draw [pattern=horizontal lines, pattern color=red, thick, opacity=0.3] (9,0.5) rectangle (9.5,1);
\node [align=left, right] at (9.5,0.75) {\small Local \\ \small unbroken};
\draw [pattern=north east lines, pattern color=gray, thick, opacity=0.3] (9,1.5) rectangle (9.5,2);
\draw [pattern=north west lines, pattern color=blue, thick, opacity=0.3] (9,1.5) rectangle (9.5,2);
\node [align=left, right] at (9.5,1.75) {\small Local \\ \small broken};
\draw [pattern=north east lines, pattern color=gray, thick, opacity=0.3] (9,2.5) rectangle (9.5,3);
\draw [pattern=horizontal lines, pattern color=red, thick, opacity=0.3] (9,2.5) rectangle (9.5,3);
\node [align=left, right] at (9.5,2.75) {\small Linear \\ \small global};
\draw [pattern=north east lines, pattern color=gray, thick, opacity=0.3] (9,3.5) rectangle (9.5,4);
\node [align=left, right] at (9.5,3.75) {\small Non-linear \\ \small global};
}\caption{A general Goldstone boson breaking pattern}
\label{fig:general_breaking}
\end{figure}

The Higgs sector of the Standard Model is the most general renormalisable Lagrangian that is compatible with $SU(2)_L\times U(1)_Y$ gauge invariance, and no more. It is also the one that nature appears to follow to very good approximation. So we will take our cue from this and form the most general Lagrangian that produces a Goldstone Higgs boson without extending the SM Higgs sector\footnote{This distinguishes the Composite Higgs from theories which presume the SM symmetry structure to be a subgroup of some larger global symmetries (such as in Higgs EFT (HEFT), SM EFT (SMEFT) or the Elementary Goldstone Higgs paradigms) or some larger gauge symmetries (such as a GUT). If the distinction between the SM being a subgroup of a larger group, and gauging some subgroup of a new composite sector, which is isomorphic to the SM bothers you, see a more thorough discussion in section \cref{sec:underlying_dynamics}.}. We do this by assuming some new high-energy physics obeys a global symmetry $G$ and we require a linear subgroup $H$ to include the EW group at the classical level. Schematically, this is sketched in fig. \ref{fig:general_breaking}.
%

%
This should strongly remind you of the SM Higgs mechanism from \cref{sec:higgs_mechanism}. Indeed, as pointed out there, higher order symmetry patterns are just combinations of the Higgs mechanism building block. Recall, in the case of the SM Higgs pattern, we had a global, non-linearly realised group $G$, linearly realised subgroup $H$ and explicit gauging $\mathcal{H}$
\begin{align}
G=SU(2)_L\times SU(2)_R, && H=SU(2)_V, && \mathcal{H} = SU(2)_L \times U(1)_Y,
\end{align}
and the surviving group was $U(1)_\textnormal{em}$. We can calculate the dimensions of each group, given by the number of generators, starting with the dimension of the unbroken group, which we denote as $Q = H \cap \mathcal{H}$. In this case, dim$(Q)$=dim$(U(1)_\textnormal{em}) = 1$, thus 
\begin{align}
\text{dim}(H_0 \cup H_1) &= \textnormal{dim}(H_0) + \textnormal{dim}(H_1) - \textnormal{dim}(H_0 \cap H_1)\\
 &= 3 + 4 - 1 =  6 = \textnormal{dim}(G)
\end{align}
We can see that for the SM Higgs pattern, there are no remaining Goldstone bosons. In other words, all the degrees of freedom in the coset $G/H$ are "eaten" by gauging $\mathcal{H}$. For the higher-energy $G \rightarrow H$ leading to the Higgs doublet, we require that at least four Goldstone bosons remain uneaten.

\subsection{A \textit{Very} Minimal Model}

A \textit{very} minimal case of a Goldstone Higgs is $\textnormal{dim}(G/H) = 4 = \textnormal{dim}(\textnormal{Higgs doublet})$, and $\textnormal{dim}(\mathcal{H}) = \textnormal{dim}(EW) = \textnormal{dim}(H) = 4$. Then, analysing our schematic in \cref{fig:general_breaking}, we see that $\mathcal{H}$ must actually \textit{be} the EW group, and be entirely contained within $H$, which is also the EW group. We also see that $G$ must have four more generators than $H = \mathcal{H} = $ EW. This completely constrains our symmetry pattern to $G = SU(3) \rightarrow SU(2)_L \times U(1)$\footnote{The reader is invited to find another (semi-)simple closed group containing $SU(2)\times U(1)$ with dimension 8. The enclosing group should be semi-simple in order to not contain the electroweak group as a normal subgroup, otherwise the global symmetry will not be explicity broken and the Higgs will not become a \textit{pseudo}-Nambu-Goldstone Boson.}. 

\begin{figure}
\centering
\includegraphics[scale=0.16]{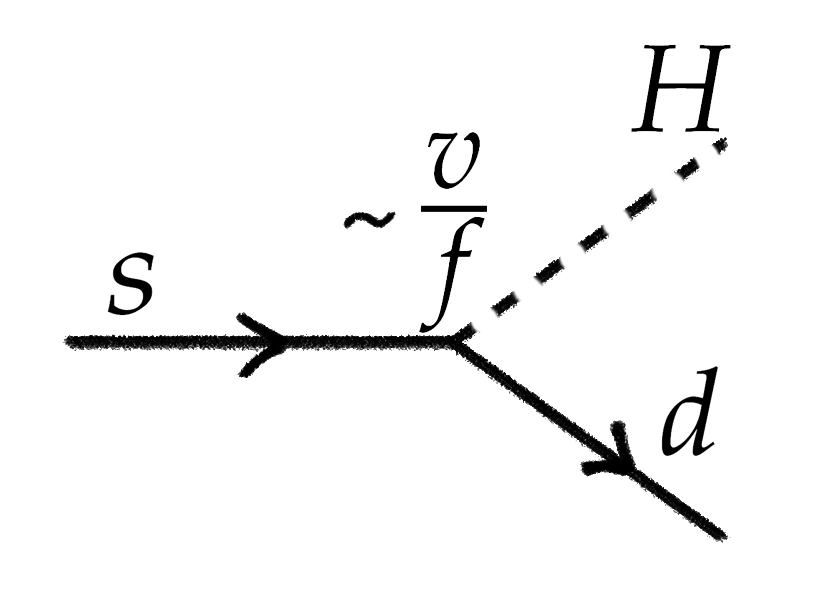}
\caption{A typical FCNC allowed in CHMs without custodial symmetry}
\label{fig:FCNC}
\end{figure}

This \textit{Very} Minimal Composite Higgs Model produces the right number of degrees of freedom to incorporate a Higgs doublet. However, calculating the low-energy behaviour of this model leads to interactions of the sort shown in \cref{fig:FCNC}. These exhibit Flavour Changing Neutral Currents (FCNC)  - a phenomenon extremely suppressed in the SM by the GIM mechanism \cite{PhysRevD.2.1285}. This suppression is supported by experiments of $B^0$-meson decay, where transitions in the off-diagonal of the CKM are extremely rare. 

While the experimental observations show that this \textit{Very} Minimal Composite Higgs Model doesn't work, we can also see why theoretically. We cheated, by suggesting that $H$ need only be dimension four, since we know the global symmetry structure of the Higgs sector is dimension six. It is precisely that $SO(4)$ symmetry that protected the gauge boson masses. This is called Custodial Symmetry, and it is the bare minimum requirement for a Composite Higgs model that reproduces the SM. To see this, consider a common measure that parameterises deviations from the SM EW description
\begin{align}
\rho := \frac{M_W^2}{M_Z^2 \cos^2\theta}
\end{align}
where in the SM, $\theta$ is defined to be the Weinberg angle $\theta_W$ such that $\rho = 1$. This angle, and therefore $\rho$, has been experimentally verified as being the SM one to $1\%$ precision. At first glance, this appears to be only a tree-level relation - derivable from the semi-classical SM Lagrangian. However, it can be shown \cite{diaz2001custodial} that deviations from the tree level SM $\rho$, considering radiative corrections from the first quark generation, are given by
\begin{align}
\Delta \rho &= \frac{3G_F}{8\sqrt{2}\pi^2} \left[m_u^2 + m_d^2  - \frac{2m_u^2 m_d^2}{m_u^2 - m_d^2}\log\frac{m_u^2}{m_d^2}\right]
\end{align}
Given Custodial Symmetry (which implies chiral symmetry  $SU(2)_L \times SU(2)_R$, i.e. $m_u = m_d$), then\footnote{Noting the limit $\frac{1}{1-x}\log\frac{1}{x}\rightarrow 1$} $\Delta\rho \rightarrow 0$ as $\frac{m_d^2}{m_u^2} \rightarrow 1$. So we require Custodial Symmetry to be preserved to the per cent level to reproduce experiment. Then as we step up in the symmetry pattern to include Custodial Symmetry, we can try $G = SO(5), \; H_0=SO(4)$, called the Minimal Composite Higgs Model (MCHM).

\section{Minimal Composite Higgs Model}\label{sec:MCHM}


\begin{tcolorbox}[colback = black!2!white]
\paragraph{Disclaimer of Brevity}\label{par:intuitive_disclaimer}

In \crefrange{sec:MCHM_Group}{sec:MCHM_Gauge}, we will be considering an intuitive approach to the minimal pseudo-Nambu-Goldstone Higgs. This will allow us to quickly reach some important results in \crefrange{sec:MCHM_Potential}{sec:MCHM_Matter_Misalignment}, and then to interpret the phenomenological results of this work in later chapters. From \cref{sec:Landscape_4dMCHM} onwards, we will develop and employ some more rigorous methods, including the full CCWZ construction of \cref{sec:CCWZ}.
\end{tcolorbox}

\subsection{Group Structure}
\label{sec:MCHM_Group}

\paragraph{Our goal}
\parbox{0.8\textwidth}{To find the smallest symmetry structure that reproduces the SM at tree level.}
\vspace*{1em}

Let us now consider 

\begin{align}
H = SO(4)\times U(1)_X, \; \mathcal{H} = SU(2)_L \times U(1)_Y \text{ and, } \implies \; G= SO(5)\times U(1)_X
\end{align}

We were able to derive $G$ as it is the next-largest group able to fit a Higgs doublet into the coset $G/H$. Note that we require the extra degree of freedom provided by $U(1)_X$. To understand why, consider (as always) the EW Higgs mechanism. Why do we not simply break $SU(2)_L \times U(1)_Q$? Because, when writing out the broken Lagrangian, we would get contributions to the diagonal\footnote{Again recalling figure \ref{fig:higgs_mech}, where the unbroken group $U(1)_{em}$ is the intersection of the gauge group and the diagonal global subgroup} from $T_L^3$, giving the unbroken $U(1)$ a charge of $T_L^3 + Q$. But this does not match our matter sector - the electric charge of an electron does not depend on its chirality. So we define a "hypercharge" upstream to be $Y= Q - T_L^3$ to correct for this diagonal contribution. 

In the MCHM, we do precisely the same thing, but even further upstream. Considering only a composite Higgs coupling to the SM requires no additions to the minimal set-up (i.e. no extra $U(1)_X$ group), since the (composite) Higgs is chargeless. However, if composite matter is introduced and couples directly to elementary fields, it must therefore share quantum numbers (e.g. hypercharge). But composite matter will be shown to transform under full representations of $SO(4)$, therefore its charge will receive contributions from $T_L^3$ \textit{and} $T_R^3$. We remove these contributions by defining its (techni)charge to be $X = Y - T_R^3 = Q - T_L^3 - T_R^3$, such that they couple with the SM charges even at the EW-broken scale of the model.

Let us remind ourselves of the goal of the model: We would like a set of four pNGBs non-linearly invariant under $SO(5)$ that replicate the group structure of the SM Higgs sector. That is, linearly invariant under $SO(4)$, in the fundamental representation. The CCWZ construction tells us exactly how to do this: Take a Goldstone matrix $U$ in the fundamental representation and project it in the direction of the vacuum expectation vector $\vec{f}$. This concept of the vev as a vector can be understood geometrically.

\subsection{Geometry of Vacuum Misalignment}\label{sec:VM_geometry}

\paragraph{Our goal}
\parbox{0.8\textwidth}{To provide a geometric interpretation that may help intuition of equivalent vacua.}
\vspace*{1em}

\begin{figure}
\centering
\begingroup
\captionsetup[subfigure]{width=0.4\textwidth}
\subfloat[An arbitrary choice of vacuum alignment $\vec{f}$ in the space of equivalent vacua $S_2$. The blue ring is the intersection of the plane perpendicular to the vev vector, and the classical vacuum symmetry.]{
\tikz[scale=0.8, every node/.style={transform shape}]{
\node[anchor=south west,inner sep=0] (image) at (0,0) {\includegraphics[width=0.5\textwidth]{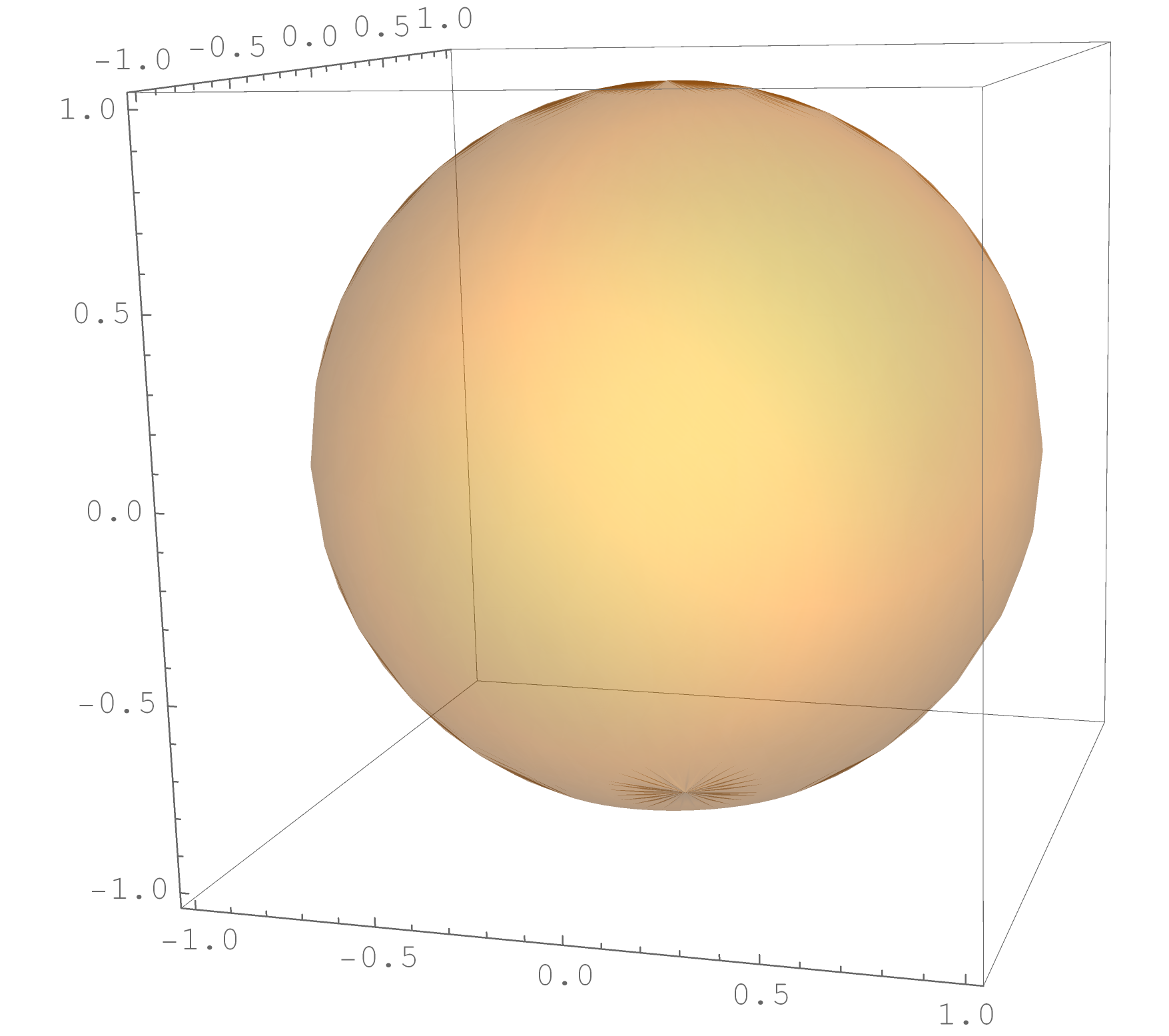}};
    \begin{scope}[x={(image.south east)},y={(image.north west)}]
        \draw [thick, ->] (.58,.55) --(.58,.92);
		\node [left] at (.57,.72) {\large $\vec{f}$};
		\node [above] at (0.7,0.91) {\large $\hat{\pi}_1$};
		\node [left] at (0.5,0.85) {\large $\hat{\pi}_2$};
		\draw [thick, dotted, ->] plot [smooth, tension=0.7] coordinates {(.58,.93) (.7,.91) (.8,.84)};
		\draw [thick, dotted, ->] plot [smooth, tension=0.7] coordinates {(.58,.93) (0.53,0.9) (0.5,0.85)};
		\draw [thick, dotted, ->] plot [smooth, tension=0.7] coordinates {(.64,.6) (0.75,0.57) (0.77, 0.55) (0.75,0.53) (0.64,0.5)};
		\node [right] at (0.78,0.55) {\large $\pi$};
		\draw [fill=blue, opacity=0.3] (0.58,0.55) ellipse (0.31 and 0.08);
    \end{scope}
}}\quad
\subfloat[After gauging, a new vev orientation is manifest, and its invariant space is superimposed in red.]{
\tikz[scale=0.8, every node/.style={transform shape}]{
\node[anchor=south west,inner sep=0] (image) at (0,0) {\includegraphics[width=0.5\textwidth]{group_geom_1}};
    \begin{scope}[x={(image.south east)},y={(image.north west)}]
        \draw [thick, ->] (.58,.55)  -- (.58,.92);
		\node [left] at (.57,.72) {\large $\vec{f}$};
		\node [above] at (0.7, 0.91) {\large $\hat{\pi}_1$};
		\node [left] at (0.5,0.85) {\large $\hat{\pi}_2$};
		\draw [thick, dotted, ->] plot [smooth, tension=0.7] coordinates {(.58,.93) (.7,.91) (.8,.84)};
		\draw [thick, dotted, ->] plot [smooth, tension=0.7] coordinates {(.58,.93) (0.53,0.9) (0.5,0.85)};
		\draw [thick, dotted, ->] plot [smooth, tension=0.7] coordinates {(.58,.8) (.62,.78) (.65,0.76)};
		\node [above] at (0.63,0.8) {$\theta_\text{vm}$};
		\draw [fill=blue, opacity=0.3] (0.58,0.55) ellipse (0.31 and 0.08);
		\draw [fill=red, opacity=0.3, rotate around={-20:(0.575,0.555)}] (0.575,0.555) ellipse (0.31 and 0.1);
		\draw [thick, ->] (.58,.55) --(.7,.89);
		\node [right] at (.64,.7) {\large $\vec{f}'$};
		\draw [green, fill=green, opacity=0.7] (0.53,0.48) circle [radius=0.015];
		\draw [green, fill=green, opacity=0.7] (0.63,0.63) circle [radius=0.015];
    \end{scope}
}}\\
\centering
\subfloat[The vev of the low energy theory, generated by the vacuum misalignment, can be visualised as the mismatch between the two invariant spaces, projected into a plane. The degree of mismatch is precisely $v = f \sin\theta_\text{vm}$ ($= f \sin\frac{h}{f}$ in the minimal case).]{
\begin{scaletikzpicturetowidth}{0.45\textwidth}
\begin{tikzpicture}[scale=\tikzscale]
\draw [ ->] (0,-3) -- (0,3);
\draw [ ->] (-3,0) --(3,0);
\draw [fill=blue, opacity=0.3] (0,0) ellipse (2.5 and 2.5);
\draw [fill=red, opacity=0.5] (0,0) ellipse (1.6 and 2.5);
\draw [thick, <->] (1.6,0) -- (2.5,0);
\node [above] at (2.05,0.1) {\large $|\vec{f}| \sin\theta_\text{vm}$};
\draw [green, fill=green, opacity=0.7] (0,2.5) circle [radius=0.1];
\draw [green, fill=green, opacity=0.7] (0,-2.5) circle [radius=0.1];
\end{tikzpicture}
\end{scaletikzpicturetowidth}
}\quad
\subfloat[Group breaking diagram. Note that this is a null set of transformations due to the \textit{special} nature of the groups. If they included parity transformations, then one could transform between the two green points.]{
\begin{scaletikzpicturetowidth}{0.45\textwidth}
\begin{tikzpicture}[scale=\tikzscale]
\draw [thick] (-1.7,-1.4) rectangle (1.8,1.9);
\draw [thick, fill=blue, opacity=0.3] (-0.5,0) circle [radius=1];
\draw [thick, dashed, red, opacity=0.6] (-0.5,0) circle [radius=1.02];
\draw [thick, fill=red, opacity=0.5] (0.5,0) circle [radius=1];
\node [above] at (0,-1.3) {$SO(3)$};
\node [left,align=right] at (-.5,0) {\small $SO(2)$};
\node [right,align=left] at (0.5,0) {\small $SO(2)$};
\node at (0,0) {$\slashed{0}$};
\draw [dotted, <->] (0.5,1.1) arc (45:135:0.7);
\node [above] at (0,1.3) {$\sin \frac{\langle h\rangle}{f}$};
\end{tikzpicture}
\end{scaletikzpicturetowidth}
}
\endgroup
\caption{The prototypical symmetry-breaking diagram, illustrating the invariant vacuum space $SO(3)/SO(2)\sim S_2$}\label{fig:SO(3)_geometry}
\end{figure}

When we gauge $G$ with $\mathcal{H}$, an important phenomenon occurs: the NGB-matter loop interactions shift the global symmetry such that the remaining gauge symmetry changes from $Q = H \cap \mathcal{H}$ to $Q' = H' \cap \mathcal{H}$. This phenomenon is \textit{vacuum misalignment}, and we can explore it geometrically. The minimal case is that of the Abelian Composite Higgs, well-explained in \cite{Panico:2015jxa}. This case requires an $G = SO(3)$ global symmetry to be broken to an $H = SO(2)$ subgroup. The coset $SO(3)/SO(2)$ is isomorphic to a $2-$sphere, given in \cref{fig:SO(3)_geometry}. This is a general property of such cosets: $SO(N+1)/SO(N) \sim S_N$. The coset represents the space of equivalent vacua. Just as in \cref{fig:simple_symmetries_d}, where transformations in the directions of $(\phi_x, \phi_y)$ left the potential unchanged, motion about the coset space of $(\hat{\pi}_1,\hat{\pi}_2)$ leaves the vacuum unchanged at its physical minimum\footnote{Care should be taken when interpolating between \cref{fig:simple_symmetries} and \cref{fig:SO(3)_geometry}. They indeed represent the same model, however the former is the linear sigma model, that is parameterised by simple Cartesian co-ordinates. The latter is parameterised non-linearly  - specifically the square root parameterisation. Hence its $S_2$ spherical geometry. The isomorphism is complete when the vacua are parameterised in exponential form as generators of $SO(3)/SO(2)$, and we retrieve the CCWZ formalism.}. 

The exponential parameterisation allows us to visualise the equivalent vacuua in the directions of $(\phi_x, \phi_y)$ as angles around the coset
\begin{align}
(x,y,z) \rightarrow (f, \chi, \varphi): && (\phi_x, \phi_y, \phi_z) \rightarrow (\hat{\pi}_1, \hat{\pi}_2, \pi)
\end{align}
Then any choice of $\vec{f}$ will give the same physics, until we explicitly break the vacuum by gauging $H = \mathcal{H} = SO(2)$. We can achieve this by choosing a physical vacuum to gauge at tree level $\vec{f} = (0,\varphi)$. Clearly for this choice, rotations around $\varphi$ will leave $\vec{f}$ unchanged. This is the invariance that we gauge. Prior to any matter interactions, the $\mathcal{H} = H=  SO(2)$ symmetry parameterised by $\pi$ is gauged by a massless gauge boson, leaving two massless NGBs $\hat{\pi}_1, \hat{\pi}_2$. 


Up to this point, the dynamics have been determined entirely by symmetry considerations (inasmuch as NGBs parameterise non-linear global symmetries and gauge fields parameterise local symmetries). At this point, the physical pNGBs may interact with the matter content of the theory, radiatively generating a vacuum expectation $\langle \hat{\pi}\rangle$ that may further break the symmetry. To emphasise: the choice of either the gauged vacuum direction, or the new global symmetry direction coming from the matter embedding, is arbitrary. However, the relative misalignment between them is physical and non-arbitrary. In the figure, this "misaligned" vacuum is given by $\vec{f}'$, with an invariant plane given by the red disc. Inspecting the geometry of the situation, there are no remaining symmetries  - the intersection between the local $S_1$ (red) and global $S_1$ (blue) is only two points unable to be transformed between, due the ``special" nature of the groups.


\begin{figure}
\centering
\subfloat[A projection in the $\{x,y,z\}$ plane of the spaces orthogonal to the chosen vacuum before (blue) and after (red) vacuum misalignment by explicit gauging. The mismatch in two co-ordinates between the spaces is analogous to figures \ref{fig:SO(3)_geometry} (b) and (c).]{
\tikz{
\node [opacity=0.7] (image) at (0,0) {\includegraphics[width=0.45\textwidth]{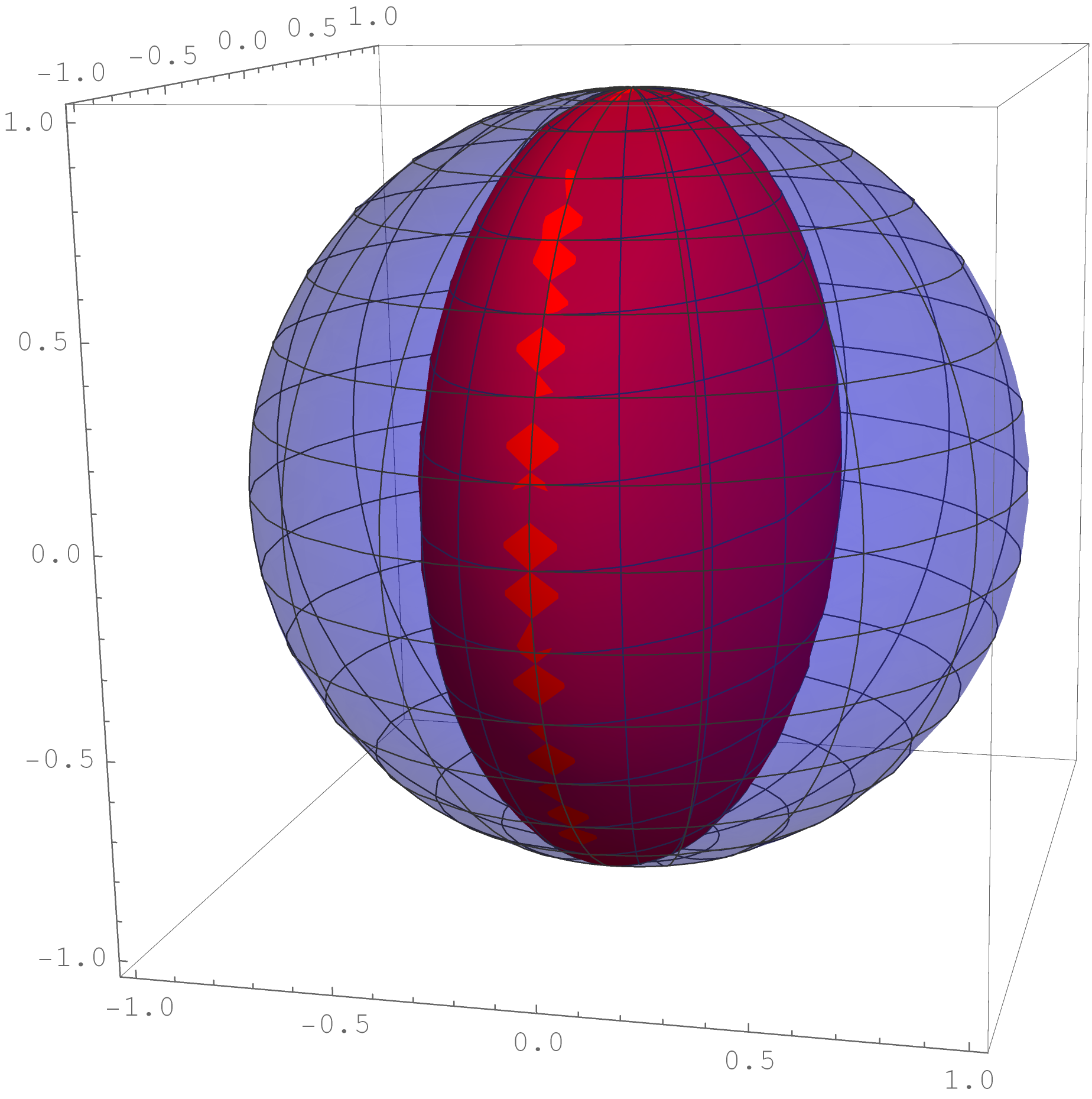}};
\draw [dashed, thick, green] plot [smooth] coordinates {(0.3,-1.7) (0,-0.7) (0, 1.2) (0.5,2.5)};
		\node [above] at (1.2,2.3) {\large $\hat{\pi}_1$};
		\node [above] at (1.5,1.5) {\large $\hat{\pi}'_1$};
		\node [left] at (0,2.3) {\large $\hat{\pi}_2, \hat{\pi}'_2$};
		\draw [thick, dotted, ->] plot [smooth, tension=0.7] coordinates {(0.5,2.6) (1,2.5) (1.7, 2.2)};
		\draw [thick, dotted, ->] plot [smooth, tension=0.7] coordinates {(0.5,2.6) (1,2.2) (1.2, 1.8)};
		\draw [thick, dotted, ->] plot [smooth, tension=0.7] coordinates {(0.5,2.6) (0.2,2.2) (0,1.5)};
}}\quad
\subfloat[The two spaces are identical in the $\hat{\pi}_2$ direction, showing the remaining symmetry. We interpret the misaligned co-ordinates as massive gauge bosons, and the remaining symmetry as a massless gauge boson.]{
\tikz{
\node (image) at (0,0) {\includegraphics[width=0.45\textwidth]{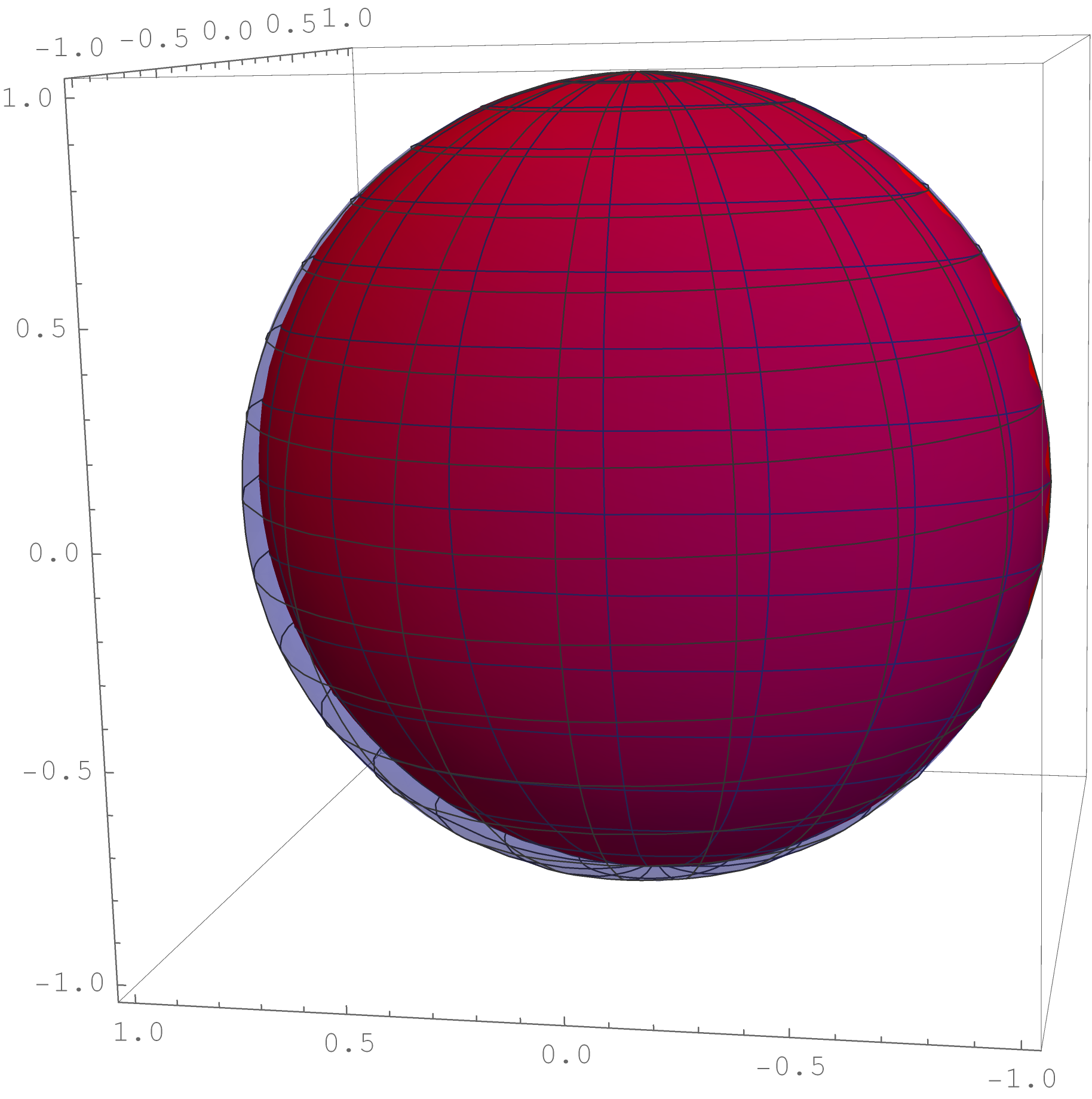}};
\draw [dashed, thick, green] (0.565,0.4) circle [radius=2.15];
\node [right] at (1,2.65) {\large $\hat{\pi}_2, \hat{\pi}'_2$};
\draw [thick, dotted, ->] plot [smooth, tension=0.7] coordinates {(0.5,2.7) (1,2.62) (1.7,2.32)};
}} \; \;
\subfloat[Group breaking diagram]{
\tikz[scale=1.5]{
\draw [thick] (-1.7,-1.4) rectangle (1.8,1.9);
\draw [thick, fill=blue, opacity=0.3] (-0.5,0) circle [radius=1];
\draw [thick, dashed, red, opacity=0.6] (-0.5,0) circle [radius=1.02];
\draw [thick, fill=red, opacity=0.5] (0.5,0) circle [radius=1];
\node [above] at (0,-1.4) {$SO(4)$};
\node [left,align=right] at (-.5,0) {\small $SO(3)$};
\node [right,align=left] at (0.5,0) {\small $SO(3)$};
\node at (0,0) {$SO(2)$};
\draw [dotted, <->] (0.5,1.1) arc (45:135:0.7);
\node [above] at (0,1.3) {$\sin \frac{\langle h\rangle}{f}$};
}}
\caption{Visualisation of $SO(4)/SO(3)$ symmetry breaking, including a misalignment of the vacuum that leaves a $SO(2)$ symmetry remaining.}\label{fig:SO(4)_geometry}
\end{figure}
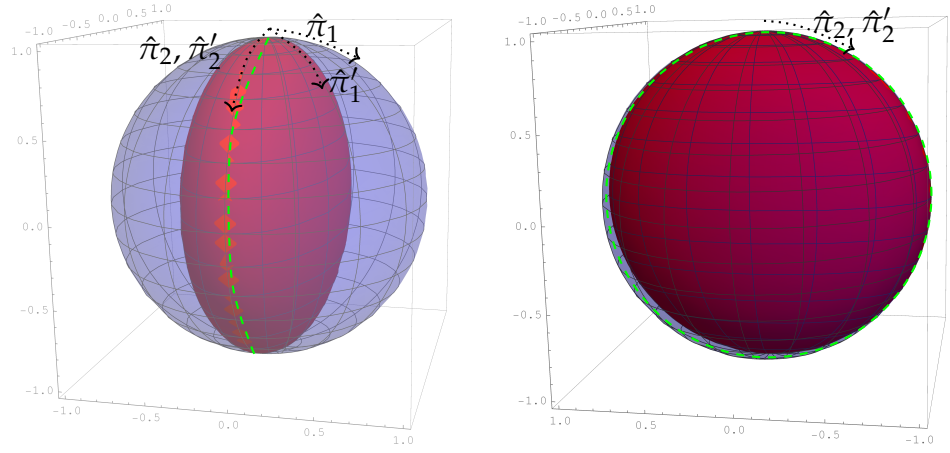
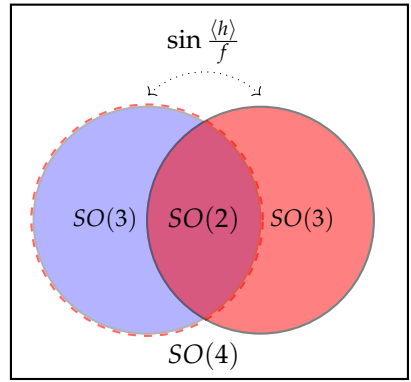

There is thus a phenomenological facet missing from this example, which is that of a remaining symmetry. We can attempt to visualise the more general case by considering a larger global symmetry, since we have a null set as the intersection of the gauged and misaligned groups. But how does vacuum misalignment work for higher-dimensional symmetries? We can quite naturally generalise the $SO(3)$ situation to that of an $SO(4)$ symmetry breaking to $SO(3)$. Just as in the above, we use the fact that $SO(N+1)/SO(N) \sim S_N$ to define the geometric space of equivalent vacua. Any point on the 3-sphere $SO(4)/SO(3) \sim S_3$ can be chosen as a physical vacuum $\vec{f}$. We then explicitly break the $SO(4)$ symmetry as before  - by gauging the $SO(3)$ subgroup. The Lagrangian can be rewritten as invariant under a new vacuum $\vec{f}'$ lying on the new 3-sphere $SO(4)/SO(3)' \sim S_3'$. We can consider the same trick as above, to visualise how the vacuum misalignment appears. The space orthogonal to the vacuum vector is now a 3-volume. A 3-sphere bounds a 3-volume running through its equator into a 2-sphere (which agrees intuitively with the lower-dimensional version). Misaligning the two-spheres now gives  \cref{fig:SO(4)_geometry}. 

A crucial difference to the Abelian Composite Higgs case is that we have a remaining $SO(2)$ symmetry. This depends on the alignment of the physical vacuum with the gauged subgroup. The remaining symmetry may correspond to, for example, the remaining electromagnetic charge invariance in the SM. The geometric argument generalises smoothly to higher dimensions, however it becomes unwieldy after four-dimensional transformations. We have a more efficient machine for examining broken symmetries: the Goldstone matrix.


\subsection{The MCHM Goldstone Matrix}
\label{sec:MCHM_Goldstone}


\paragraph{Our goal}
\parbox{0.8\textwidth}{To describe in generality the physics of NGBs that parameterise an $SO(5)/SO(4)$ vacuum space.}
\vspace*{1em}

We can use our tools of the CCWZ construction to do this. Recall that the Lagrangian containing only Goldstone fields, invariant under the action of the group $G$ is
\begin{align}
\mathcal{L}_2 &= \frac{f^2}{4} \text{Tr}[d_\mu[U] d^\mu [U]],
\end{align}
where the "2" subscript reminds us that this Lagrangian is truncated at two-derivative terms\footnote{The physics of terms with $n$ pairs of derivatives is suppressed, relative to the leading order term, by a magnitude of $\mathcal{O}(\frac{f^{2n}}{\Lambda^{2n}}) \sim \mathcal{O}(100^{n-1})$}. For convenience, we repeat that the Maurer-Cartan form is
\begin{align}
w_\mu  = -iU^\dagger \partial_\mu U \equiv e^a_\mu T^a + d^a_\mu X^a && \implies && d_\mu \equiv d^a_\mu X^a = w_\mu - e_\mu = -i U^\dagger D_\mu U
\end{align}
assuming the CCWZ "standard form" of the Goldstone matrix $U = \exp\left( i\pi^{\hat{a}}X^{\hat{a}}/f\right)$, summing a vector of Goldstone bosons $\pi^{\hat{a}}$ over the broken generators $X^{\hat{a}}$. Now, $d_\mu$ is, in general, an infinite series of commutator terms projected into the broken direction \cite{ballesteros2017exceptional}
\begin{align}
d_\mu &= \sum\limits_{k=0}^\infty \frac{(-i)^k}{f^{k+1}(k+1)!}\text{ad}^k_\pi (\partial_\mu \pi)_X\\
&= \frac{1}{f} \partial_\mu \pi - \frac{i}{2f^2}[\pi,\partial_\mu \pi]_X - \frac{1}{6f^3}[\pi, [\pi, \partial_\mu \pi]]_X + \frac{1}{24f^4}[\pi, [\pi, [\pi, \partial_\mu \pi ]]]_X + ... \label{eq:broken_form_explicit}
\end{align}
Here the $X$ subscript means a "projection" into the broken generators,
\begin{align}
[\pi^a T^a, \pi^b T^b]_X := \pi'^c X^c
\end{align}
where the particular generators for $SO(4)$, $SO(5)$ and $SO(6)$ are given in \cref{sec:representations}. We can continue this line of thought generally and abstractly, and in certain vacua this is convenient. For example, the $d_\mu$ spanning symmetric spaces simplify greatly while remaining general. We discuss the physics of these spaces in \cref{sec:mooses}. However, a more intuitive approach for this first look at $SO(5)/SO(4)$ is to choose a particular vacuum that gives only broken generators in the commutators. Choosing a specific vacuum by parameterising $U \rightarrow U\Phi_0 \equiv \Phi$, as in \cref{eq:goldstone_matrix}, automatically selects only broken generators. For example
\begin{align}
[ \Phi_0^T \pi^a X^a, \partial_\mu \pi^b X^b \Phi_0] &= \Phi_0^T \pi^a \left( if^{abc}T^c + if^{abd}X^d \right)\Phi_0 \partial_\mu \pi^b\\
& = \Phi_0^T \pi^a  if^{abd}X^d \Phi_0 \partial_\mu \pi^b \nonumber\\
\text{since} \qquad \Phi_0 T^c  &= 0
\end{align}
by the definition in \cref{eq:goldstone_matrix}. Indeed, this parameterisation of the Goldstone matrix as $\Phi$ agrees with the physical argument in given prior to \cref{eq:goldstone_matrix}, with the physical Goldstone fields being fluctuations about the vacuum in the direction of the broken generators. In the absence of any gauging or external sources, this is a physically invariant reparameterisation. It is the parameterisation that captures the transition between linear and non-linear sigma model, as the transformation of the fields is from a non-linear realisation to a linear representation:
\begin{align}
g: && U(x) \rightarrow g U(x) h(g,x)^{-1}\;,  && \Phi(x) &= U(x)\Phi_0 \rightarrow  g U(x) h(g,x)^{-1} \Phi_0 \nonumber\\
&&&& & = g U(x) \Phi_0 = g \Phi(x)
\end{align}
simply absorbing the subgroup transformation into the vacuum. The benefit of the explicit vacuum parameterisation is that it allows quick direct calculation of the vector components. Consider a particular choice of the vacuum expectation vector $\vec{f}$ to be
\begin{align}
\vec{f} &\equiv \Phi_0 = (0,0,0,0,1)^T = (\bm{0},1)^T
\end{align}
although of course this is an arbitrary decision, and $\bm{F}$ could be a unit vector in any direction in 5-space. Then parameterising the Goldstone matrix about the explicit, fundamental\footnote{For complete clarity, fundamental in the sense of the regular representation $\textbf{5}$ of $SO(5)$. Furthermore, the vacuum need not be in the fundamental, and other vacuum configurations have been considered in the literature \cite{gripaios2009beyond,PhysRevD.66.072001,Bertuzzo2013,ArkaniHamed:2002qy,Chang:2003zn,Katz:2005au}} vacuum
\begin{align}
\Phi &=\me^{-\frac{i\sqrt{2}}{f}\pi^{\hat{a}}T^{\hat{a}}} \vec{f} \equiv U\Phi_0 \\
&=\exp\left(\begin{array}{cc}
0 & \begin{array}{c}
\frac{\pi _1}{f}\\
\frac{\pi _2}{f}\\
\frac{\pi _3}{f}\\
\frac{\pi _4}{f}
\end{array}\\
\begin{array}{cccc}
-\frac{\pi _1}{f} & -\frac{\pi _2}{f} & -\frac{\pi _3}{f} & -\frac{\pi _4}{f}
\end{array} & 0
\end{array}
\right)
\left(\bm{0},1\right)^T \label{eq:goldstone_definition}
\end{align}
Expanding the exponential, (and noting that $\pi^2 := \pi_1^2 + \pi_2^2 + \pi_3^2 + \pi_4^2$) we get that
\hspace{-1cm}\begin{tabular}{ c }
$\begin{aligned}
\Phi &= \left( \bm{1} + i\pi^{\hat{a}}T^{\hat{a}}/f  + \frac{1}{2}(i\pi^{\hat{a}}T^{\hat{a}}/f)^2 + \frac{1}{6}(i\pi^{\hat{a}}T^{\hat{a}}/f)^3 + ...\right) \Phi_0\\
& = \left(\begin{matrix}
\bm{0}\\
1
\end{matrix}\right) + 1/f\left(\begin{matrix}
\bm{\pi}\\
0
\end{matrix}\right) + \left(\begin{matrix}
\bm{0}\\
\frac{1}{2} \pi^2/f^2
\end{matrix}\right) + \frac{1}{6}\pi^2/f^3\left(\begin{matrix}
\bm{\pi}\\
0
\end{matrix}\right) + \left(\begin{matrix}
\bm{0}\\
\frac{1}{24} \pi^4/f^4
\end{matrix}\right) + ...\\
& = \left(\begin{matrix}
\bm{\pi}\\
0
\end{matrix}\right)\frac{1}{\pi}(\pi/f + \frac{1}{6}\pi^3/f^3 + \frac{1}{120}\pi^5/f^5+...) + \left(\begin{matrix}
\bm{0}\\
1\end{matrix}\right) (1 + \frac{1}{2}\pi^2/f^2 + \frac{1}{24}\pi^4/f^4+...)
\end{aligned}$
\end{tabular}
which we recognise as a trigonometric series
\begin{align}
\Phi & = \left(\begin{matrix}
\bm{\pi}\\
0
\end{matrix}\right) \frac{\sin(\pi/f)}{\pi} + \left(\begin{matrix}
\bm{0}\\
\cos(\pi/f)\end{matrix}\right) \\
& = \frac{\sin(\pi/f)}{\pi}\left(\begin{matrix}
\pi_1 & \pi_2 & \pi_3 & \pi_4 & \pi\cot(\pi/f)
\end{matrix}\right)^T\hspace{3cm}
\end{align}
and for completeness
\begin{align}
U = \left( \begin{matrix}
\mathbb{1}_4  - (1-\cos\frac{\pi}{f}) \frac{\vec{\pi} \vec{\pi}^T}{\pi^2} & \sin\frac{\pi}{f} \frac{\vec{\pi}}{\pi} \\
-\sin\frac{\pi}{f} \frac{\vec{\pi}}{\pi} & \cos\frac{\pi}{f}
\end{matrix}\right)\label{eq:explicit_goldstone_matrix}
\end{align}

We now associate the NGBs with an $SU(2)_L \times SU(2)_R$ invariant Higgs-like doublet.
\begin{align}
H = \left(\begin{matrix}
i\pi_1 + \pi_2\\
i\pi_3 + \pi_4
\end{matrix}\right)
\end{align}
with the usual vacuum $\langle H \rangle = (0,v)^T$.

If you're suspicious of the redundancy of embedding four real fields in a five-by-five matrix just so they transform as a fundamental under $SO(5)$, you're justified. There is a more efficient embedding, which relies on the isomorphism $SO(5) \sim Sp(4)$. A $\textbf{4}$-plet is the fundamental representation of the symplectic group $Sp(4)$, and the spinor representation of $SO(5)$, which is the smallest building block of $SO(5)$
\begin{align}
U_\textbf{4} = \me^{i\frac{\sqrt{2}}{f}\pi_i(x) \hat{T}_{\textbf{4}}^i} = \left(\begin{matrix}
\cos\frac{\pi}{2f} \mathbb{1}_2 & i \sin\frac{\pi}{2f} \frac{\Sigma}{\sqrt{2}\pi}\\
 i \sin\frac{\pi}{2f} \frac{\Sigma^\dagger}{\sqrt{2}\pi} & \cos\frac{\pi}{2f} \mathbb{1}_2
\end{matrix}\right)
\end{align}
where $\Sigma = (H^c, H)$ is a bi-doublet made out of the Higgs-like doublets. In fact, any $SO(2n+1)/SO(2n)$ coset can be parameterised by a $\textbf{2}^n$-plet in the spinor representation of $SO(2n+1)$, which decomposes into two spinorial $\textbf{2}^{n-1}$-plets of $SO(2n)$. In this case, we further decompose under $SU(2)_L\times SU(2)_R$ to see that the two Goldstone doublets $H, H^c$ transform together as a bidoublet. Once we gauge in the SM, the left-right symmetry is lost, and we no longer consider the Higgs as a bidoublet. We denote these representations under each group as
%
\begin{align}
& \begin{matrix}
\small SO(5)/SO(4)  &         & \small SO(5)                 &          & \small Sp(4)/SO(4) &          & \small Sp(4) &\\
U_\textbf{5} & \sim & \Phi_\textbf{5} &  \sim & U_\textbf{4} & \sim & \Phi_\textbf{4} & \sim & \dots \\
\small \text{non-lin.} \textbf{5} & & \textbf{5} & &\small \text{non-lin.} \textbf{4} & & \textbf{4} &
\end{matrix} \nonumber\\[10pt]
 & \begin{matrix}
 \small SO(4)         &                 & \small SU(2)_L \times SU(2)_R & & \small SU(2)_L \times U(1)_Y\\
 H \oplus H^c & \sim & \Sigma & \sim & H_{SM}\\
 \textbf{2} \oplus \bar{\textbf{2}}  & & (\textbf{2}, \bar{\textbf{2}})  & & \textbf{2}_{\frac{1}{2}}
\end{matrix}
\end{align}
For a less terse explanation of spinor and chiral algebra, see Appendix \cref{sec:group_theory}.

As in the SM, we can always choose the unitary gauge of the $SU(2)_L \times U(1)_Y$ subgroup of $SO(4)$ to rotate in the direction of one of the bosons. Then we redefine our $SO(5)$-language and $SU(2)_L\times SU(2)_R$-language Higgs field using $\pi_1,\pi_2,\pi_3 \rightarrow 0$ and $\pi_4 \rightarrow h$ to
\begin{align}
\Sigma \rightarrow (0,0,0,s_h,c_h)^T && H \rightarrow (0,h)^T \label{eq:unitary_gauge}
\end{align}
where $s_h = \sin\frac{h}{f}$, $c_h = \cos\frac{h}{f}$. To be very clear, $H$ (with $H^c$) is a  fundamental of $SO(4)$, hence its linear transformation under the unitary gauge.
 $\Phi$ is a fundamental of $SO(5)$ in the non-linear, square-root parameterisation (from \cref{sec:NLSM}). Thus, we can give the expectation value of $H$ as simply the vev vector
\begin{align}
\langle H \rangle = \langle h \rangle (0,1)^T
\end{align}
but the fiveplet vev is nonlinearly related to the vev
\begin{align}
\langle \Phi \rangle = (0,0,0,s_{\langle h \rangle}, c_{\langle h \rangle})
\end{align}
by spherical rotations in the 4-5 plane. We will determine the value of $\langle h \rangle$ in terms of the EW vev once we gauge the NGBs. Up to this point the expectation value is zero. If they had retained a vev, we could have parameterised them further to remove this vev. In other words, any $SO(4)$ subgroup in $SO(5)$ is equivalent to any other, prior to introducing any interactions.

\subsection{Gauging the Composite Sector}
\label{sec:MCHM_Gauge}
%
%
%

\paragraph{Our goal}
\parbox{0.8\textwidth}{To interact our new Goldstone bosons with the SM electroweak sector.}
\vspace*{1em}

\begin{figure}
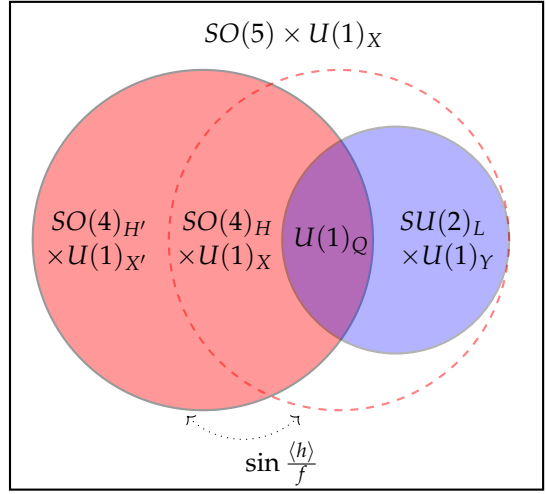

\centering
\tikz[scale=1.5]{
\draw [thick] (-2.9,-2.2) rectangle (1.8,2.1);
\draw [thick, fill=red, opacity=0.4] (-1.2,0) circle [radius=1.5];
\draw [thick, dashed, red, opacity=0.5] (0,0) circle [radius=1.5];
\draw [thick, fill=blue, opacity=0.3] (0.5,0) circle [radius=1];
\node [above] at (-0.4,1.6) {$SO(5)\times U(1)_X$};
\node [left,align=right] at (-1.6,0) {$SO(4)_{H'}$ \\ $\times U(1)_{X'}$};
\node [left,align=right] at (-0.5,0) {$SO(4)_{H}$ \\ $\times U(1)_X$};
\node [right,align=left] at (0.45,0) {$SU(2)_L$ \\ $\times U(1)_Y$};
\node [left] at (0.35,0) {$U(1)_Q$};
\draw [dotted, <->] (-0.35,-1.55) arc (-45:-135:0.7);
\node [above] at (-0.5,-2.25) {$\sin \frac{\langle h\rangle}{f}$};
}
\caption{A sketch of "vacuum misalignment" (VM)  - i.e. the mixing of two groups' generators, which may change due to one-loop corrections, leading to a vev $\langle h \rangle$. Before VM, the vacuum is invariant under $SO(4)_H \sim SU(2)_L \times SU(2)_R$. Afterwards, $SO(4)_{H'}$ mixes $T_L, T_R$ with previously broken generators $X$. }\label{fig:MCHM_group_misalignment_1}
\end{figure}

To use the Higgs doublet in EWSB, it must be connected to the SM, both through gauge couplings and Yukawa couplings. Yukawa coupling the Higgs directly to pairs of SM fermions amounts to an "elementary pNGB Higgs". We will be forbidding this form of interaction for reasons later discussed. However, direct EW gauge interactions will be allowed. We do this by gauging an $SU(2)_L \times U(1)_Y$ subgroup of $SO(4)$ with the same gauge fields used to gauge the SM. Why a subgroup? We have designed the four scalars to assemble as a bi-doublet under $SU(2)_A \times SU(2)_B \sim SO(4)$, which could be the Higgs doublet only if we associate $A,B \rightarrow L,R$, thus obtaining the correct quantum numbers for the SM Higgs. As for why we require it to be a subgroup of $SO(4)$ rather than generically a subgroup of $SO(5)$ (and fiddling with couplings to get the correct quantum numbers), we return to \cref{fig:general_breaking}. Gauging any non-linearly realised generators (i.e. the "broken generators") will enact the Higgs mechanism prematurely. That is, our electroweak gauge bosons will attain mass at tree-level, at the scale of the new physics $\sim 1\tev$. Instead, we will gauge linearly realised generators, preserving the electroweak symmetry at tree-level. Loop corrections will deliver a misalignment of the groups $SU(2)_L \times U(1)_Y$ and $SO(4)_{H'} \times U(1)_{X'}$ of the sort in \cref{fig:MCHM_group_misalignment_1}.

We will gauge the $\mathcal{H} = SU(2)_L \times U(1)_Y$ symmetry of the NGBs in the obvious way - via the covariant derivative\footnote{Again, see Appendix \cref{sec:representations} for dictionary and explanation of generators} 
\begin{align}
\mathcal{L} &= \frac{f^2}{2}[D^\mu \Phi]^\dagger [D_\mu \Phi] \nonumber\\
&= \frac{f^2}{2}\left[\partial^\mu \Phi^\dagger + ig (W^\mu_\alpha T_L^\alpha \Phi)^\dagger + ig' (B_\mu T_R^3 \Phi)^\dagger\right] \left[\partial_\mu \Phi - ig W_\mu^\alpha T_L^\alpha \Phi - ig' B^\mu T_R^3 \Phi \right] \nonumber \\
&= \frac{f^2}{2}\left[\frac{1}{f}\partial^\mu h \left(\begin{matrix}
0\\
0\\
0\\
c_h\\
-s_h
\end{matrix}\right)^T + \frac{g}{2}\sin\frac{h}{f} \left(\begin{matrix}
W_1^\mu\\
W_2^\mu\\
W_3^\mu\\
0\\
0
\end{matrix}\right)^T - \frac{g'}{2} \sin\frac{h}{f} \left( \begin{matrix}
0\\
0\\
B^\mu\\
0\\
0
\end{matrix}\right)^T\right] \nonumber\\
& \times \left[ \frac{1}{f}\partial^\mu h \left(\begin{matrix}
0\\
0\\
0\\
c_h\\
-s_h
\end{matrix}\right) + \frac{g}{2}\sin\frac{h}{f} \left(\begin{matrix}
W^1_\mu\\
W^2_\mu\\
W^3_\mu\\
0\\
0
\end{matrix}\right) - \frac{g'}{2} \sin\frac{h}{f} \left( \begin{matrix}
0\\
0\\
B_\mu\\
0\\
0
\end{matrix}\right) \right] \nonumber\\
&= \frac{1}{2}\partial^\mu h \partial_\mu h + \frac{g^2}{4}f^2\sin^2\frac{h}{f}W^\mu_\alpha W_\mu^\alpha + \frac{g'^2}{4} f^2 \sin^2\frac{h}{f}B^\mu B_\mu - \frac{gg'}{2} f^2 \sin^2\frac{h}{f}W^\mu_3 B_\mu \label{eq:sigma_model_effective_gauge}
\end{align}
We remind that we are in the unitary gauge. Redefining, as in the SM, $W_\mu^\pm = \frac{1}{\sqrt{2}} (W^1_\mu \mp W^2_\mu)$ and $Z_\mu = c_W W^3_\mu - s_W B_\mu$, with $\tan\theta_W= \frac{\sin\theta_W}{\cos\theta_W} = \frac{s_W}{c_W} \equiv \frac{g'}{g}$ we get
\begin{align}
\mathcal{L} = \frac{1}{2}(\partial	_\mu h)^2 + \frac{g^2}{4}f^2 \sin^2 \frac{h}{f} (W^\mu_+ W_\mu^+ + W^\mu_- W_\mu^- + \frac{1}{2c_W^2} Z_\mu Z^\mu)
\end{align}
We can read off the SM EW vev in terms of the Composite Higgs vev
\begin{align}
m_W = \frac{1}{2} g f \sin\frac{\langle h \rangle}{f} = \frac{g}{2}v && \implies && v = f\sin \frac{\langle h \rangle}{f}
\end{align}

This may be considered the bottom-up approach - for the physical pNGB to interact with the SM at low energy, it must obey gauge symmetries enforced by the gauge bosons $W^a, B$. On the other hand, the CCWZ approach is top-down, and will allow us to work in a more general way once extra symmetries are introduced in the following chapters. To observe how the embedding of $SU(2)_L \times U(1)_Y$ in the non-linear $SO(5)$ gives explicit breaking terms, we can take the whole global group as gauged and subsequently "turn off" non-physical gauge fields. In group terms, this means temporarily enlarging
\begin{align}
\mathcal{H} \rightarrow G = SO(5) \times U(1)_X
\end{align}
The Maurer-Cartan form $w_\mu$ (which decomposed the NGBs into linearly and non-linearly transforming components) is then appended with gauge terms that cancel the now-local transformations. Specifically, define fields
\begin{align}
A_\mu := A^A_\mu T^A \rightarrow g(x) (A_\mu + i \partial_\mu ) g(x)^{-1} \label{eq:gauge_definition}
\end{align}
for $g(x) \in G$ as a book-keeping device. $T^A$ is the set of $SO(5)$ generators, as in previous sections. As usual, we segregate this set into $\{T_L^i,T_R^i,\hat{T}^a\}$, which are the generators of $SU(2)_L, SU(2)_R$ and the broken generators of $SO(5)/SO(4)$ respectively. This leads to corresponding terms $A_\mu = \{A^i_{L,\mu}, A^i_{R,\mu}, \hat{A}^a_\mu\}$. That is, we fill out our adjoint multiplets of gauge bosons with spurious terms, such that it would serve to gauge the full $SO(5) \times U(1)_X$.  One could then set unphysical terms to zero after deriving the low-energy couplings
\begin{align}
A_{L,\mu}^i \rightarrow W_\mu^i, & & A_{R,\mu}^3 + X_\mu \rightarrow c_X B_\mu + s_X B_\mu, && A_{R,\mu}^1, A_{R,\mu}^2, \hat{A}^\alpha \rightarrow 0
\end{align} 
with the $X$-boson following a similar prescription as in the SM, $\frac{s_X}{c_X} = \frac{g'}{g_X}$. Then the local Maurer-Cartan form  is 
\begin{align}
\mathcal{W}_\mu = U^\dagger (A_\mu - i\partial) U := d_\mu(\pi, A) + e_\mu(\pi, A)
\end{align}
This expression can be used to give the locally covariant derivative for matter fields and the $G$-invariant NGB terms that can enter the Lagrangian, precisely as in eq.s (\ref{eq:maurer_cartan_covariant_derivative}) and (\ref{eq:maurer_cartan_invariant_term}). We will derive these terms in detail in later sections. 

For now, pursuant to the \nameref{par:intuitive_disclaimer}, let's consider a quick way to connect the effective Lagrangian in \cref{eq:sigma_model_effective_gauge} to the high energy, unbroken physics. First, note that we can write the interactions in \cref{eq:sigma_model_effective_gauge} effectively as
\begin{align}
\mathcal{L}_\text{eff} &= \frac{1}{2}\sum\limits_{i=1}^3 \Pi_{W} W_{L\mu}^i W_L^{i\mu} + \Pi_{WB} W_{L\mu}^3 B^\mu + \frac{1}{2} \Pi_B B_\mu B^\mu \label{eq:effective_gauge_low_energy}
\end{align}
where we have introduced the notation $\Pi_{V} = \Pi_{V}(p^2, h)$, called a form factor\footnote{Although notation may get lazy, the form factors $\Pi$ will in general depend on momentum $p^2$, and the composite Higgs field $h$.}, which contains all dynamical terms and parameters from the high-energy theory that contribute to the vector boson coupling $VV$.  The high-energy bosonic Lagrangian should contain the most general $SO(5) \times U(1)_X$-invariant set of interactions. Given the defintion in  \cref{eq:gauge_definition} and split into left, right and broken fields, we can write our gauge multiplet as
\begin{align}
A_\mu &= \left(
\begin{matrix}
0 &  -\frac{i}{2}A_L^3 - \frac{i}{2}A_R^3 &  \frac{i}{2}A_L^2 + \frac{i}{2}A_R^2 & \frac{i}{2}A_R^1 + \frac{i}{2}A_L^1 & -\frac{i}{\sqrt{2}} \hat{A}^1\\
 \frac{i}{2}A_L^3 + \frac{i}{2}A_R^3 & 0 & -\frac{i}{2}A_L^1 - \frac{i}{2}A_R^1 & \frac{i}{2}A_R^2 - \frac{i}{2}A_L^2 & -\frac{i}{\sqrt{2}} \hat{A}^2\\
 -\frac{i}{2}A_L^2 - \frac{i}{2}A_R^2 & \frac{i}{2}A_L^1 + \frac{i}{2}A_R^1 & 0 & \frac{i}{2} A_R^3 - \frac{i}{2}A_L^3 & -\frac{i}{\sqrt{2}} \hat{A}^3 \\
 \frac{i}{2}A_L^1 - \frac{i}{2}A_R^1 & \frac{i}{2}A_L^2 - \frac{i}{2}A_R^2 & \frac{i}{2}A_L^3 -\frac{i}{2}A_R^3 & 0 &  - \frac{i}{\sqrt{2}} \hat{A}^4\\
 \frac{i}{\sqrt{2}}\hat{A}^1 & \frac{i}{\sqrt{2}}\hat{A}^2 & \frac{i}{\sqrt{2}}\hat{A}^3 & \frac{i}{\sqrt{2}}\hat{A}^4 & 0
\end{matrix}\right)_\mu
\end{align}
Given this $\textbf{10}$ of $SO(5)$, as well as the Goldstone matrix in the $\textbf{5}_0$ of $SO(5)\times U(1)_X$ and a new gauge singlet $X_\mu$ of $U(1)_X$, the most general interaction Lagrangian we can write at quadratic order in the gauge fields is
\begin{align}
\mathcal{L}_\text{eff}^\text{spurious} &= \frac{(P_T)^{\mu\nu}}{2} \left[ \Pi^{(0)} \text{Tr}[A_\mu A^\mu]+ \Pi^{(0)}_X X_\mu X^\mu + \Pi^{(2)} \Phi^T A_\mu A^\mu \Phi  \right. \nonumber \\
& \left. + \Pi^{(4)}\left(\text{Tr}\left[(U^\dagger A_\mu U)_L (U^\dagger A_\nu U)_L - (U^\dagger A_\mu U)_R (U^\dagger A_\nu U)_R\right]\right)\right]
\label{eq:gauge_lagrangian}
\end{align}
where $\Pi^{(n)}$ are form factors containing dynamical terms and parameters in the mass basis, coupling gauge fields to $n$ Goldstone matrices. The notation of \cref{eq:broken_form_explicit} is used again here, where the $L,R$ subscript is a projection in the $T_L^i, T_R^i$ generators. Finally, note that $(P_T)^{\mu\nu}$ is the transverse projection operator that conveniently includes both interaction and kinetic terms:
\begin{align}
(P_T)_{\mu\nu} A^\mu A^\nu := (\eta_{\mu\nu} - \frac{p_\mu p_\nu}{p^2} ) A^\mu A^\nu
\end{align}
Recall that in this spuriously $SO(5)$-invariant gauge Lagrangian, there could not yet be a Higgs vev or masses for $A,X$. Thus we only require the transverse component of the gauge fields. The longitudinal projection operator, which we will use later to fix the gauge, is similar
\begin{align}
(P_L)_{\mu\nu} A^\mu A^\nu := \frac{p_\mu p_\nu}{p^2} A^\mu A^\nu = (\partial_\mu A^\mu) (\partial_\nu A^\nu).
\end{align}
We can use the above matrix to calculate the terms in the $SO(4)$-preserving vacuum, i.e. the limit of no EWSB $ h \rightarrow \langle h \rangle \rightarrow 0$
\begin{align}
\text{Tr}[A_\mu A^\mu] &= A_{L\mu}^i A_L^{i\mu} + A_{R\mu}^i A_R^{i\mu} + \hat{A}_\mu^b \hat{A}^{b\mu}
\end{align}
and, less trivially, 
\begin{align}
\Phi A_\mu A^\mu \Phi^T \rightarrow \Phi_0 A_\mu A^\mu \Phi_0^T &= \frac{1}{2} \hat{A}^b_\mu \hat{A}^{b\mu}
\end{align}
Then we can rewrite \cref{eq:gauge_lagrangian} as
\begin{align}
\mathcal{L}_\text{eff}^\text{spurious}|_{\langle h\rangle=0} &= \frac{1}{2}\Pi^{(0)} (A_{L\mu}^i A_L^{i,\mu} +A_{R\mu}^i A_R^{i,\mu})+ \frac{1}{2} \left( \Pi^{(0)} + \frac{1}{2}\Pi^{(2)} \right) \hat{A}_\mu^b \hat{A}^{b,\mu} + \frac{1}{2}\Pi^{(0)}_X X_\mu X^\mu \nonumber \\
&\equiv  \frac{1}{2}\hat{\Pi}_V (A_{L\mu}^i A_L^{i,\mu} +A_{R\mu}^i A_R^{i,\mu})+ \frac{1}{2} \hat{\Pi}_A  \hat{A}_\mu^b \hat{A}^{b,\mu} + \frac{1}{2} \hat{\Pi}_X X_\mu X^\mu
\end{align}
where we have defined the decomposed generators as vector $V$ and axial $A$ subsets of the generators. This notation will become more meaningful once we add genuine chiral groups to contain these composite gauge fields. For now, it allows a convenient matching between unbroken and broken form factors
\begin{align}
\Pi^{(0)} = \hat{\Pi}_V   && \Pi^{(2)} = 2( \hat{\Pi}_V - \hat{\Pi}_A) \label{eq:broken_vs_unbroken}
\end{align}

Returning to \cref{eq:gauge_lagrangian}, we have that, in the unitary gauge,
%
\begin{align}
\mathcal{L}_\text{eff} &= \frac{P_T^{\mu\nu}}{2} \left(\Pi^{(0)} W_\mu^i W_\nu^i + \Pi^{(2)} \frac{s_h^2}{4} (W^1_\mu W^1_\nu + W^2_\mu W^2_\nu) + (s_X^2 \Pi_X^{(0)} + c_X^2 \Pi^{(0)}) B_\mu B_\nu \right. \nonumber \\
& \left. + \Pi^{(2)} \frac{s_h^2}{4} \left(c_X B_\mu - W^3_\mu\right)^2 + c_h \Pi^{(4)} \left( W_\mu^i W_\nu^i - c_X^2 B_\mu B_\nu\right)\right)\label{eq:effective_gauge_spurious}
\end{align}

Comparing \cref{eq:effective_gauge_low_energy} and \cref{eq:effective_gauge_spurious}, we can derive the form factors
\begin{align}
& \Pi_W = \Pi^{(0)} + \frac{s_h^2}{4}\Pi^{(2)} + c_h \Pi^{(4)} && \Pi_{WB} = -c_X \frac{s_h^2}{4}\Pi^{(2)}\\
& \Pi_B = s_X^2 \Pi^{(0)}_X + c_X^2 (\Pi^{(0)} + \frac{s_h^2}{4}\Pi^{(2)}  - c_h \Pi^{(4)})
\end{align}

The form factors are a way of connecting the high energy model, which contains the full particle content, to the low energy effective model. This is necessary to derive the Higgs potential one would observe at the electroweak scale, in terms of high-energy parameters, as we do in the next section. They can also parametrise deviations from SM values. Expanding \cref{eq:effective_gauge_spurious} around the vacuum (i.e. $p^2 = 0$, $h = \langle h \rangle$), in powers of $p^2$, one gets that
\begin{align}
\mathcal{L}_\text{eff, vacuum} &= P_T^{\mu\nu} \left( \mathcal{O}(q^0)  +  \frac{q^2}{2} (\Pi^{(0)\prime}(0) W_\mu^{a_L} W_\nu^{a_L} \right. \\ 
& \left. + (\Pi^{(0)\prime}(0) + \Pi^{(0)\prime}_X(0))B_\mu B_\nu + \dots ) + \mathcal{O}(q^4, q^6, \dots) \right)
\end{align}
The coefficient of the $W_\mu^{a_L} W_\nu^{a_L}$ term gives the SM isospin coupling strength, and the coefficient of the $B_\mu B_\nu$ gives the hypercharge coupling strength
\begin{align}
g^2 = -\frac{g_0^2}{\Pi^{(0)\prime}(0)} & & g'^2 = \frac{g_0^2}{s_X^2\Pi^{(0)}_X + c_X^2 \Pi^{(0)}}
\end{align}

Note that in \cref{eq:gauge_lagrangian}, if the composite sector is invariant under $L \leftrightarrow R$, the second line cancels, and we can take $\Pi^{(4)} \rightarrow 0$.  For the remainder of this thesis, we will assume that we have this composite chiral invariance (that is, the composite matter is vector-like). 

%

\subsection{Calculating the Higgs Potential and Vacuum Misalignment}
\label{sec:MCHM_Potential}

\paragraph{Our goal} 
\parbox{0.8\textwidth}{To use the Coleman-Weinberg 1-loop procedure to find the Higgs potential in terms of the gauge form factors.}
\vspace*{1em}

From our experience with the Coleman-Weinberg formalism of \cref{sec:coleman_weinberg}, we know that we can approximate the interaction of a weakly-interacting quantised field with a classical background field using a one-loop expansion. We do precisely that here, with the (still massless) EW gauge bosons, and the to-be Higgs field. The expansion is precisely in analogy to \cref{eq:effective_potential_sum}, where we will obtain the propagator and vertex expressions from the gauged Lagrangian \cref{eq:gauge_lagrangian}. 

\begin{figure}
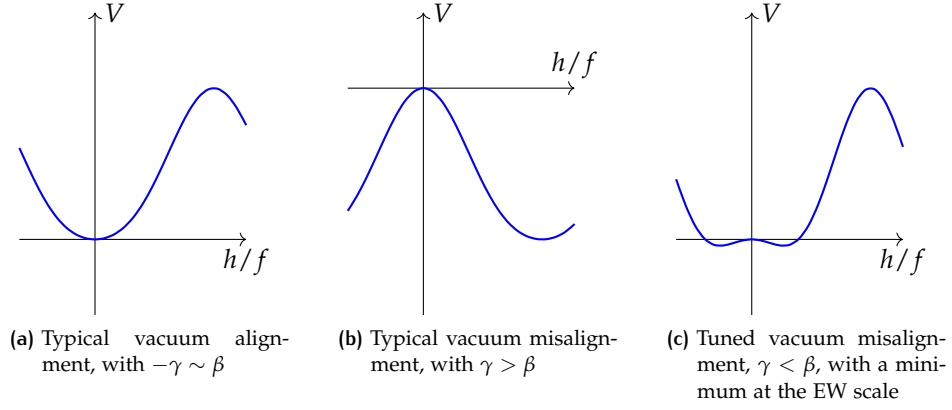

\centering
\subfloat[Typical vacuum alignment, with $- \gamma \sim  \beta$]{
\tikz{
\node [right] at (0,5) {$V$};
\draw [->] (-1,2) --(2,2);
\draw [->] (0,1) --(0,5);
\node [below] at (2,2) {$h/f$};
\draw [domain=-1:2, smooth, variable=\x, blue, thick] plot ({\x}, {sin(deg(\x))^2 + sin(deg(\x))^4 + 2});
}
}\qquad
\subfloat[Typical vacuum misalignment, with $\gamma > \beta$]{
\tikz{
\node [right] at (0,3) {$V$};
\draw [->] (-1,2) --(2,2);
\draw [->] (0,-1) --(0,3);
\node [above] at (2,2) {$h/f$};
\draw [domain=-1:2, smooth, variable=\x, blue, thick] plot ({\x}, {-3*sin(deg(\x))^2  + sin(deg(\x))^4 + 2});
}
}\qquad 
\subfloat[Tuned vacuum misalignment, $\gamma < \beta$, with a minimum at the EW scale\label{fig:minimising_the_potential_c}]{
\tikz{
\node [right] at (0,5) {$V$};
\draw [->] (-1,2) --(2,2);
\draw [->] (0,1) --(0,5);
\node [below] at (2,2) {$h/f$};
\draw [domain=-1:2, smooth, variable=\x, blue, thick] plot ({\x}, {-sin(deg(\x))^2  + 3*sin(deg(\x))^4 + 2});
}
}\caption{The types of vacuum (mis)alignment in Composite Higgs models}\label{fig:minimising_the_potential}
\end{figure}

As a reference and for motivation, we state here the one-loop Higgs potential 
\begin{align}
V(h) =& \int\limits_0^\infty \frac{\dd p^2}{16\pi^2} p^2\left( \frac{6}{2} \log\Pi_{W^+W^-} + \frac{3}{2}\log \left[ \Pi_{BB} \Pi_{W^3 W^3} - \Pi_{W^3 B}^2\right] \right) \label{eq:potential} \\
&- 2 \sum\limits_{\psi=t,b,...} \textnormal{N}^\psi_c \int\limits_0^\infty \frac{\dd p^2}{16\pi^2}p^2 \log \left[ p^2 (1+\Pi_\psi)(1+ \Pi_{\psi^c}) - |M_\psi|^2\right]  \nonumber \\
& \equiv -\gamma s_h^2 + \beta s_h^4 + \mathcal{O}(s_h^6) \, . &
\end{align}
where we expand the potential in the leading and next-to-leading order Higgs field terms. A central goal of this thesis, is to determine areas of parameter space that produces a suitable Higgs potential, that is, in the manner of \cref{fig:minimising_the_potential_c}. We will derive this expression for both the boson and fermion contributions. 

\begin{figure}
\centering
\tikz{
\node at (0,0) {\includegraphics[scale=0.17]{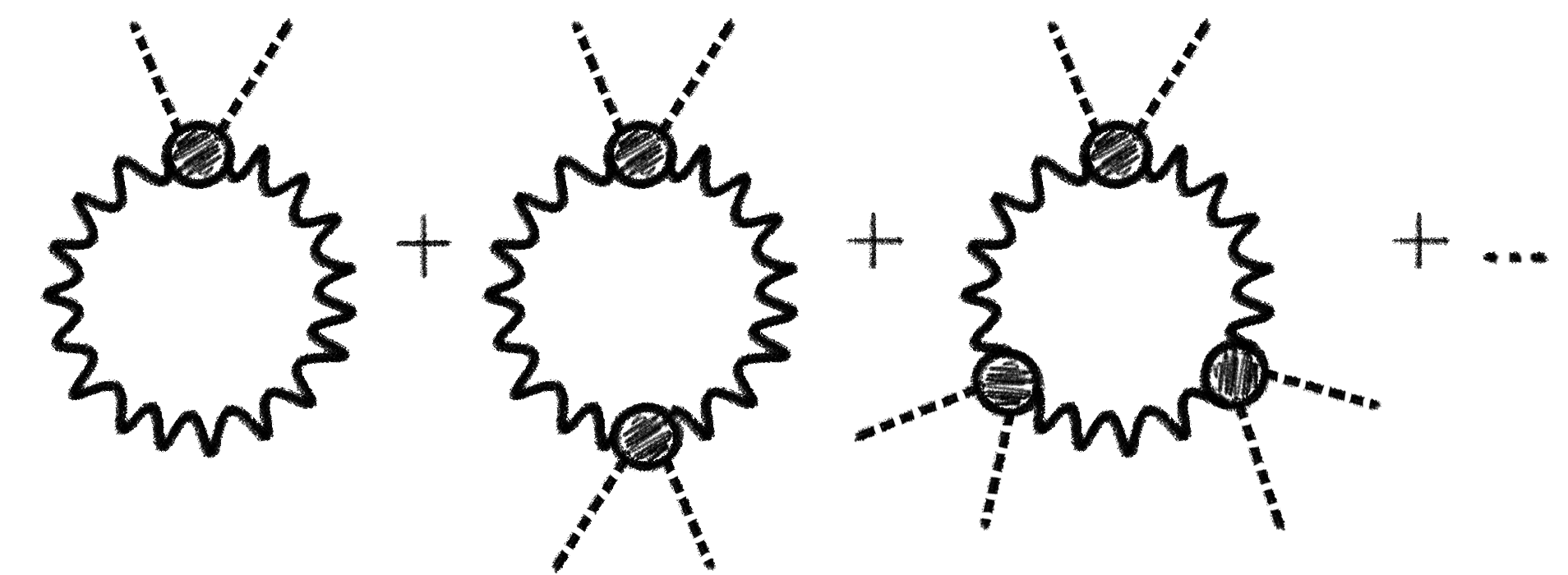}};
\node at (-6.3,1.2) {$\frac{i\Pi_1(q^2)}{4}\sin^2(h/f)$};
\node at (-5.5,-1) {$\frac{i}{\Pi_0(q^2)}$};
}
\caption{Series of gauge boson interactions with the classical background Higgs field. These are contributions to the 1-loop effective potential}\label{fig:gauge_potential}
\end{figure}

To begin, recall the argument made by Coleman and Weinberg, where a series of one-loop diagrams is a good first-approximation to the effective potential. Specifically, we are treating the Higgs field as a classical background field, and ask how the $A^{a_L}_\mu$ fields interact with it, at one-loop order. We get the series of diagrams in \cref{fig:gauge_potential}. The form factors will depend on how the elementary gauge sector interacts with the composite sector. For now, we can get the interactions (vertices and propagators) from the Lagrangian in \cref{eq:gauge_lagrangian}.

We start with the one-loop effective potential %
given in \cref{sec:CW_potential_appendix} (and discussed thoroughly in \cite{kleinert2016particles}) , in terms of the propagators $G_{\mu\nu}$ and vertex functions $\Gamma_{\mu\nu}$ 
\begin{align}
V(\Phi) &= \sum\limits_i^n C_i \int \frac{d^4 p}{(2\pi)^4}\log\left( 1 + i G^i_{\mu\nu} \Gamma^{\mu\nu,i}\right)
\end{align}
This is a general formula, which sums over potential contributions from all particles in the theory, weighted by a symmetry factor $C_i$, to be determined. For now, we focus on the contribution only from the $W^\pm$ bosons, but the contributions of $W^3$ and $B$ are derived in exactly the same way.

The propagator is the inverse of the interaction of $W^\pm_\mu$ with only itself, in \cref{eq:gauge_lagrangian}\footnote{We should first fix the gauge by adding the term $\frac{1}{2g^2\zeta}(\partial^\mu A^{a_L}_\mu)^2$, which introduces a second propagator term.}. We propose that it is
\begin{align}
i G_{\mu\nu} = \frac{i}{\Pi^{(0)}(p^2)}(P_T)_{\mu\nu} - \zeta\frac{ig^2}{p^2}(P_L)_{\mu\nu}
\end{align}
We can see that this is the propagator, by the definitions of the projection operators 
\begin{align}
(P_T)_{\mu\nu} = \eta_{\mu\nu} - \frac{p_\mu p_\nu}{p^2} && (P_L)_{\mu\nu} = \frac{p_\mu p_\nu}{p^2}
\end{align}
\begin{align}
\implies (P_T)_{\mu\nu} (P_T)^{\mu\nu} = d -1 = 3, & & (P_L)_{\mu\nu} (P_L)^{\mu\nu} = 1, &  & (P_L)_{\mu\nu} (P_T)^{\mu\nu} = 0 
\end{align}
then contract the propagator with the self-interaction terms in \cref{eq:gauge_lagrangian}
\begin{align}
&\left(\frac{1}{\Pi^{(0)}(p^2)}(P_T)_{\mu\nu} - \zeta\frac{g^2}{p^2}(P_L)_{\mu\nu}\right)\left( \frac{1}{2} \Pi^{(0)}(p^2) (P_T)^{\mu\nu}  + \frac{1}{2} \frac{p^2}{g^2 \zeta}(P_L)^{\mu\nu}\right)\\
&= \frac{1}{2}(P_T)_{\mu\nu}(P_T)^{\mu\nu} - \frac{1}{2}(P_L)_{\mu\nu}(P_L)^{\mu\nu} = 1
\end{align}
The grey blobs in \cref{fig:gauge_potential} are the interaction of two $W^\pm_\mu$s with the Higgs field
\begin{align}
i\Gamma_{\mu\nu} = \frac{i\Pi^{(2)}(p^2)}{4}\sin^2(h/f)(P_T)_{\mu\nu}
\end{align}

We derive this from the generic Lagrangian composed of a free field Green's function, and terms involving interactions between some gauge field $V_\mu$ and the composite form factors:
\begin{align}
\mathcal{L} & \supset (\partial^2 - m^2)V^2 + \Gamma_{\mu\nu}(h^2)V^\mu V^\nu \nonumber\\
& := \frac{1}{G^{\mu\nu}} V^\mu V^\nu + \Gamma_{\mu\nu}(h^2) V^\mu V^\nu \\
& = \frac{1}{2}(P_T)^{\mu\nu}\Pi^{(0)} W^i_\mu W^i_\nu -\frac{1}{2} \frac{p^2}{2g^2 \zeta}(P_L)^{\mu\nu} W^i_\mu W^i_\nu + \frac{1}{4} (P_T)^{\mu\nu} \Pi^{(2)}(p^2)\sin^2(h/f) W^i_\mu W^i_\nu \nonumber
\end{align}
using our explicit model from \cref{eq:effective_gauge_spurious}. We arrive at 
\begin{align}
V(h) &= C \int \frac{d^4 p}{(2\pi)^4}\log\left( 1 - (\frac{i}{\Pi^{(0)}(p^2)}(P_T)_{\mu\nu} - \zeta\frac{ig^2}{p^2}(P_L)_{\mu\nu})\frac{i\Pi^{(2)}(p^2)}{4}\sin^2(h/f)(P_T)_{\mu\nu}\right) \nonumber\\
& = C \int \frac{d^4 p}{(2\pi)^4}\log \left(1  + \frac{\Pi^{(2)}(p^2)}{4\Pi^{(0)}(p^2)}\sin^2(h/f)\right)\label{eq:cw_formula}
\end{align}
Where in the last line, we threw away terms that did not depend on the Higgs field. These terms only raise or lower the potential, leading to no physically observable effects. 

We will briefly illustrate that the term in \cref{eq:cw_formula} is the sum of one-loop diagrams. This process is derived more thoroughly in \cref{chp:colemanweinberg}. This is an exercise in monitoring factors and negative signs, so let's be careful. By construction, applying the Feynman rules and integrating over momenta gives an extra factor of $-i$, so we correct with $V = i\times$ an integral over the propagator and vertex terms.

Applying the Feynman rules then, we get that a sum of one-loop diagrams is
\begin{align}
V = i\times N_p N_f \int \frac{d^4 p}{(2\pi)^4}\left(\frac{1}{2}iG_{\mu\nu}\Gamma^{\mu\nu} + \frac{1}{4}(iG_{\mu\nu}\Gamma^{\mu\nu})^2 + \frac{1}{6} (iG_{\mu\nu}\Gamma^{\mu\nu})^3 + ...\right)\label{eq:c_w_propagator_vertex}
\end{align}

The $\frac{1}{2n}$ fractions come from the number of transformations that give the same diagram. In the first diagram, the diagram can be flipped without producing distinguishable physics  - this suppresses the amplitude by $\frac{1}{2}$. In the second, the diagram can be flipped or rotated, suppressing by $\frac{1}{2\times 2}$. In the third, the diagram can be rotated in three configurations, or flipped and then rotated to one of three configurations, suppressing by $\frac{1}{2\times 3}=\frac{1}{6}$. This is the dihedral group, with the diagrams treated as a point, a line, an equilateral triangle, a square, etc., with group size $\frac{1}{2n}$.

$N_p$ and $N_f$ are factors based on the number of combinations that can be made from the polarizations and flavours of electroweak gauge boson available. That is, each polarisation of the now-vector-like (as we have gone to the unitary gauge) EW bosons gives a new graph from each loop configuration, as does each flavour. This is $N_p N_f = 3\times 3 = 9$.

Evaluating the propagator and vertex function gives
\begin{align}
V = 9i\int\frac{d^4 p}{(2\pi)^4}\sum\limits_{n=1}^{\infty}\frac{1}{2n}\left(\frac{-\Pi^{(2)}(p^2)}{4\Pi^{(0)}(p^2)}\sin^2(h/f)\right)^n
\end{align}

Pulling the half out the front, this looks like the power series 
\begin{align}
\sum\limits_{n=1}^\infty \frac{z^n}{n} = - \ln(1-z)
\end{align}
with $z = -\frac{\Pi_1}{4\Pi_0}$. Then, summing we get
\begin{align}
V = -\frac{9i}{2}\int\frac{d^4 p}{(2\pi)^4}\ln\left(1+\frac{\Pi^{(2)}(p^2)}{4\Pi^{(0)}(p^2)}\sin^2(h/f)\right)
\end{align}

Now Wick rotate to Euclidean space. This takes $p^2 \rightarrow -p_E^2$ and $d^4 p \rightarrow id^4p_E$ and thus
\begin{align}
V = \frac{9}{2}\int\frac{d^4p_E}{(2\pi)^2}\ln\left(1 + \frac{\Pi^{(2)}(-p_E^2)}{4\Pi^{(0)}(-p_E^2)}\sin^2(h/f)\right)\label{eq:gauge_potential}
\end{align}

This is a general result for the electroweak gauge boson contribution to the Higgs potential, and will be used in later sections. The most important aspect of it, is that it always gives a minimum at the classical vacuum expectation $\langle h \rangle = 0$. This is in contrary to what we require of our minimal model  - that there is a small negative quadratic contribution to the potential, and a large positive quartic contribution. Although to comprehensively show that this expression in \cref{eq:gauge_potential} always preserves the electroweak vacuum requires some machinery we will develop in the proceeding sections, we can argue it at leading order using a large-N effective model. Here N is the number of colours of some underlying strong symmetry. The details of the large-N approximation, and its form factors, can be found in \cref{sec:large_N}. Suffice it to say that the large-N formalism is a good first-order approximation if we are to treat our pNGB Higgs as an effectively composite object, resulting from a confined, QCD-like symmetry\footnote{More will be said about the validity of this assumption, as well as other ways of attaining this global symmetry, in \cref{sec:underlying_dynamics}. Miraculously, we will see that other underlying dynamics manifested as the same global symmetry typically lead to the same functional form of the correlators.}. The functional forms of the form factors are derived in the \cref{sec:form_factors}, and we repeat the results here
\begin{align}
\hat{\Pi}_{V}(p^2) \sim p^2 \sum\limits_n^{N_\rho} \frac{f^2_{\rho_n}}{p^2 + m_{\rho_n}^2}, && \hat{\Pi}_{a}(p^2) \sim p^2 \left(\sum\limits_n^{N_A} \frac{f^2_{a_n}}{p^2 + m_{a_n}^2} + \frac{f^2}{p^2}\right), \label{eq:vector_resonance_sums}
\end{align}
and recalling the matching definition \ref{eq:broken_vs_unbroken}
\begin{align}
\Pi^{(0)}(p^2) = \hat{\Pi}_V(p^2), && \Pi^{(2)}(p^2)= 2(\hat{\Pi}_{V}(p^2) - \hat{\Pi}_{A}(p^2)). 
\end{align}
The sums in \cref{eq:vector_resonance_sums} run over the $N_{\rho,a}$ composite meson resonances $\rho,a$ that are exchanged between electroweak gauge bosons. Note that $f_{\rho_n, a_n} = \langle 0 | J | n \rangle$ is the amplitude of the meson being created from the vacuum by its current $J$, while $f$ is the Goldstone decay constant, of the definition in \cref{eq:goldstone_definition}.

We require the integral in \cref{eq:gauge_potential} to converge at quadratic and quartic orders in $\sin^2(h/f)$. This will be discussed in more detail in \cref{sec:WSR}. For now, simply given the two (Weinberg Sum Rule) conditions
\begin{align}
\sum\limits_n \left( f_{\rho_n}^2 - f_{a_n}^2 \right) = f && \sum\limits_n \left(f^2_{\rho_n} m_{\rho_n}^2 - f_{a_n}^2 m_{a_n}^2 \right) = 0
\end{align}
one can inspect (by expanding in a series for $s_h^2$ and $p^2$) that this leads to convergence of the integral at quadratic and quartic order, respectively
\begin{align}
V &= -\gamma_g s_h^2  + \beta_g s_h^4 + \mathcal{O}(s_h^6) \label{eq:V_gauge_gamma_beta}
\end{align}
where the quadratic and quartic gauge contributions are
\begin{align}
\gamma_g = \frac{9}{8}\int \frac{d^4 p}{(2\pi)^4} \frac{\Pi^{(2)}}{\Pi^{(0)}} && \beta_g = \frac{9}{64}\int \frac{d^4 p}{(2\pi)^4} \left(\frac{\Pi^{(2)}}{\Pi^{(0)}}\right)^2.
\end{align}
For concreteness, we can solve the two conditions for $N_\rho = N_a =1$, giving
\begin{align}
f^2_\rho = f^2\frac{m_{a_1}^2}{m_a^2 -m_\rho^2}, && f_a^2 = f  \frac{m_\rho^2}{m_a^2 - m_\rho^2}\label{eq:large_N_solutions}
\end{align}
We can expand the terms in \cref{eq:vector_resonance_sums}, giving
\begin{align}
\frac{1}{2}\Pi^{(2)} &= \hat{\Pi}_V - \hat{\Pi}_A \nonumber \\
&= p^2 \frac{f_\rho^2}{p^2 + m_\rho^2} - p^2 \left(\frac{f^2_a}{p^2 + m_a^2} + \frac{f^2}{p^2}\right) \nonumber\\
&=p^2 f_\rho^2 \frac{p^2 + m_a^2}{(p^2 + m_a^2)(p^2 + m_\rho^2)}  - p^2 f_a^2 \frac{p^2+m_\rho^2}{(p^2 + m_a^2)(p^2 + m_\rho^2)} - \frac{f^2 (p^2 + m_a^2)(p^2 + m_\rho^2)}{(p^2 + m_a^2)(p^2 + m_\rho^2)}\nonumber
\end{align}
Using the solutions in \cref{eq:large_N_solutions} in the first two terms allows us to factor out $f$
\begin{align}
\frac{1}{2}\Pi^{(2)} &= \frac{f^2 p^2}{(p^2 + m_a^2)(p^2 + m_\rho^2)}\left(\frac{m_a^2(p^2 + m_a^2)}{m_a^2 - m_\rho^2} - \frac{m_\rho^2(p^2 + m_\rho^2)}{m_a^2 - m_\rho^2}  \right) - \frac{f^2 (p^2 + m_a^2)(p^2 + m_\rho^2)}{(p^2 + m_a^2)(p^2 + m_\rho^2)} \nonumber\\
&= \frac{f^2 p^2}{(p^2 + m_a^2)(p^2 + m_\rho^2)}\left(p^2 - \frac{m_\rho^4 -m_a^4}{m_a^2 - m_\rho^2} - \frac{1}{p^2} (p^2 + m_a^2)(p^2 + m_\rho^2) \right) \nonumber\\
&= \frac{f^2}{(p^2 + m_a^2)(p^2 + m_\rho^2)}\left(p^4  + p^2(m_a^2 + m_\rho^2) - p^4 - p^2(m_a^2 + m_\rho^2) - m_a^2 m_\rho^2 \right) \nonumber\\
&= - f \frac{m_a^2 m_\rho^2}{(p^2 + m_\rho^2)(p^2 + m_a^2)}
\end{align}
which is strictly negative. Inserting this behaviour into the potential equation \cref{eq:V_gauge_gamma_beta} leads to a local minimum of  the potential at $\langle h \rangle = 0$. There are also solutions at $\langle h \rangle = f\frac{n\pi}{2}$. The energy scale between the two sets of solution is too great to tunnel in finite time. Regardless, assuming $f> 1\tev$, these solutions do not correctly reproduce EWSB either. This leads to a vev-less electroweak sector, failing to reproduce the SM Higgs mechanism. Suffice it to say that the potential needs matter fields to misalign the vacuum.

\begin{figure}
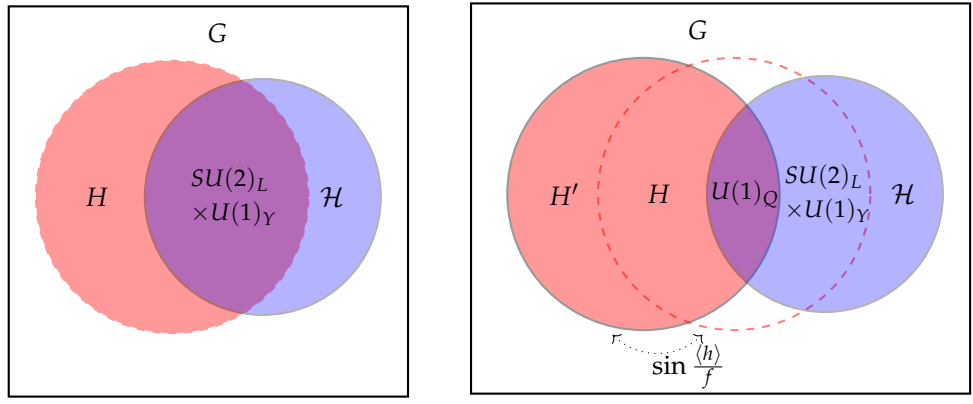

\centering
\subfloat[The tree-level symmetry structure of a generic Composite Higgs model. We must have a vacuum invariant under $H$, with some gauged sector $\mathcal{H}$. The tree-level vacuum must be gauged by at least the electroweak sector $SU(2)_L \times U(1)_Y$ to suppress EWSB.] {
\tikz[scale=1.2]{
\draw [thick] (-1.8,-2.2) rectangle (2.6,2.1);
\draw [thick, dashed, red, fill=red, opacity=0.4] (0,0) circle [radius=1.5];
\draw [thick, fill=blue, opacity=0.3] (1,0) circle [radius=1.3];
\node [above] at (0.5,1.6) {$G$};
\node [left,align=right] at (2,0) {$\mathcal{H}$};
\node [left,align=right] at (-0.6,0) {$H$};
\node [right,align=left] at (0.1,0) {\small $SU(2)_L$ \\ \small $\times U(1)_Y$};
}%
}\qquad
\subfloat[Once composite matter is introduced, the invariant vacuum will misalign. We allow couplings that ensure the electric charge $Q = T^3_L + T^3_R + X$ remains conserved.]{
\tikz[scale=1.2]{
\draw [thick] (-2.9,-2.2) rectangle (2.6,2.1);
\draw [thick, fill=red, opacity=0.4] (-1,0) circle [radius=1.5];
\draw [thick, dashed, red, opacity=0.5] (0,0) circle [radius=1.5];
\draw [thick, fill=blue, opacity=0.3] (1,0) circle [radius=1.3];
\node [above] at (-0.4,1.6) {$G$};
\node [left,align=right] at (-1.6,0) {$H'$};
\node [left,align=right] at (-0.6,0) {$H$};
\node [left,align=right] at (2.1,0) {$\mathcal{H}$};
\node [right,align=left] at (0.45,0) {\small $SU(2)_L$ \\ \small $\times U(1)_Y$};
\node [left] at (0.6,0) {\small $U(1)_Q$};
\draw [dotted, <->] (-0.35,-1.55) arc (-45:-135:0.7);
\node [above] at (-0.5,-2.25) {$\sin \frac{\langle h\rangle}{f}$};
}}
\caption{The Composite Higgs paradigm of vacuum misalignment}\label{fig:MCHM_group_misalignment}
\end{figure}

\subsection{Vacuum Misalignment from Composite Matter}
\label{sec:MCHM_Matter_Misalignment}

\paragraph{Our goal}
\parbox{0.8\textwidth}{To radiatively generate a potential from interactions with composite fermions.}
\vspace*{1em}

Up to this point, we have only considered the $SO(5)/SO(4)$ NGBs, which parameterise the non-linear global symmetries, and gauge fields, which explicitly promote some of those symmetries to local symmetries. These interactions, while indeed explicitly breaking $SO(5)$ to $SU(2)_L \times U(1)_X$ do not induce a non-zero vev in the Higgs field, and thus do not break EW symmetry. More generally, all vector boson fields follow this "vacuum aligning" property \cite{witten1983some}. This is an important feature of BSM models that include pseudo-Nambu Goldstone Bosons as the Higgs or other massive scalars. Therefore, we must attempt to induce a potential with matter fields. To include matter, we could make the naive assumption that the NGBs $\Phi$ interact with SM matter fields $\psi^i$ in the usual SM way - via Yukawa couplings with some function of the $\Phi$ field
\begin{align}
\mathcal{L}_\text{naive} &= y_{ab} \bar{\psi}^a_L f(\Phi) \psi^b_R
\end{align}
This has historically been a valid way to have the SM interact with, say, a Higgs-like technicolor condensate. However, there are phenomenologically unsatisfactory elements with this. For example, the fermion masses are of the form \cite{contino2011higgs}
\begin{align}
m_\psi \sim v \frac{4\pi}{\sqrt{N}}\left(\frac{\Lambda}{\Lambda_{UV}}\right)^{[f(\Phi)]}
\end{align}
It has been shown that the dimension of $f(\Phi)$ must be greater than or equal to two. It is then very difficult to generate the hierarchy of fermion masses, without introducing extreme tuning, or a higher symmetry scale. The latter was suggested as a solution to technicolor, called extended technicolor, or indeed a waterfall of extended symmetries "tumbling" to lower symmetries. Although these are elegantly justified with the "most attractive channel" paradigm \cite{raby1980tumbling}, there is still something epicyclic about solving a large hierarchy problem with multiple small hierarchies. If we introduce complexity, we would prefer to do it because some more fundamental theory suggests it. As it happens, models of extra compact dimensions result in a composite Higgs interpretation. This interpretation will not be given here, but they give a guide for a particular form of composite-elementary interaction:
\begin{align}
\mathcal{L}_f = \lambda \bar{\psi}^a f(\Phi, \Psi^a)\label{eq:partial_compositeness_form}
\end{align}
where the function $f(\Phi, \Psi^a)$,  should now have fermionic quantum numbers, to contract with those of $\bar{\psi}$. A similar analysis can be done as above for these terms and we would see that these couplings can much more naturally give a mass hierarchy amongst the SM fermions\footnote{See \cite{contino2011higgs} for a thorough discussion}. It does come at a price: the interaction presupposes some composite, fermion fields $\Psi^a$ - at least one for each SM multiplet that couples to the Higgs. We require a "composite partner" for each elementary fermion in order to enable interaction with the composite Higgs while correctly accounting for the full SM group, as well as the higher energy global group. 

\vspace{1em}
\begin{tcolorbox}[colback = black!2!white]
\paragraph{Key point} This bears repeating. The composite Higgs paradigm can alleviate the hierarchy problem, however gauge interactions do not produce an electroweak vev - we need matter interactions. But (SM matter)-(composite Higgs) Yukawa interactions lead back to a hierarchy problem, the hierarchy of fermion masses. We solve this by introducing composite matter that mediates the interaction, enabling the SM fermions to have degrees of compositeness, leading to a natural hierarchy.
\end{tcolorbox}
\vspace{1em}

We can see that these matter terms give a potential to the Higgs using the same process as for the gauge fields. First, consider an explicit interaction term at the electroweak scale, with the composite fields in some representation of $SO(5)$. If the coupling is linear, then the elementary fields must also be embedded somehow in the same representation of $SO(5)$, leading to coupling
\begin{align}
\mathcal{L}_f \ni \bar{\psi}_L \Delta_l  \Psi_R + \bar{\psi}_{R,u} \Delta_u \tilde{\Psi}_{L,u} + \bar{\psi}_{R,d} \Delta_d \tilde{\Psi}_{L,d}\label{eq:SM_spurions}
\end{align}
where $\Delta_{L,R}$ is treated as a scalar coupling constant. This is called the \textit{partial compositeness} paradigm. Clearly these terms break the $SO(5)$ symmetry explicitly, as elementary fields do not transform as $SO(5)$ multiplets. At low energy, they should decompose into representations of $SU(2)_L \times U(1)_Y$, and couple with the components of the composite multiplets that share those quantum numbers. Note that we pre-empt this with different partners for the left-handed field, and the right-handed up and down-type fields. In general, any elementary field could couple with the composite sector, provided a partner with the correct quantum numbers is introduced. For most cases however, including only the top quark as partially composite will suffice to demonstrate concepts, and account for the leading order contributions to observables.


As in the gauge case, we are interested in the coupling of fermions to the Higgs field, as a classical background. Given an effective Lagrangian
\begin{align}
\mathcal{L} \supset \Pi_q(p,h)\bar{q}_L\slashed{p}q_L + \Pi_t(p,h)\bar{t}_R\slashed{p}t_R  + M(p,h)\bar{t}_L t_R
\end{align}

where the $\Pi_{q,t}(p,h)$ can be split into Higgs-dependent and -independent parts
\begin{align}
\Pi_{q,t}(p,h) =  \Pi^{q,t}_0(p) + \Pi^{q,t}_1(p,h)
\end{align}

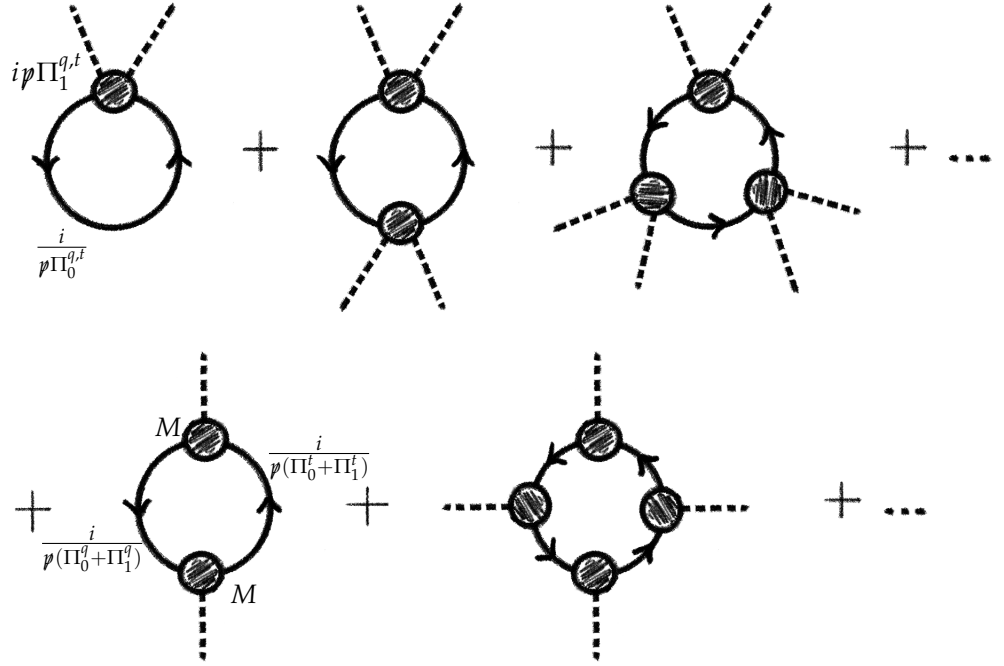
\begin{figure}
\centering
\tikz{
\node at (0,0) {\includegraphics[scale=0.2]{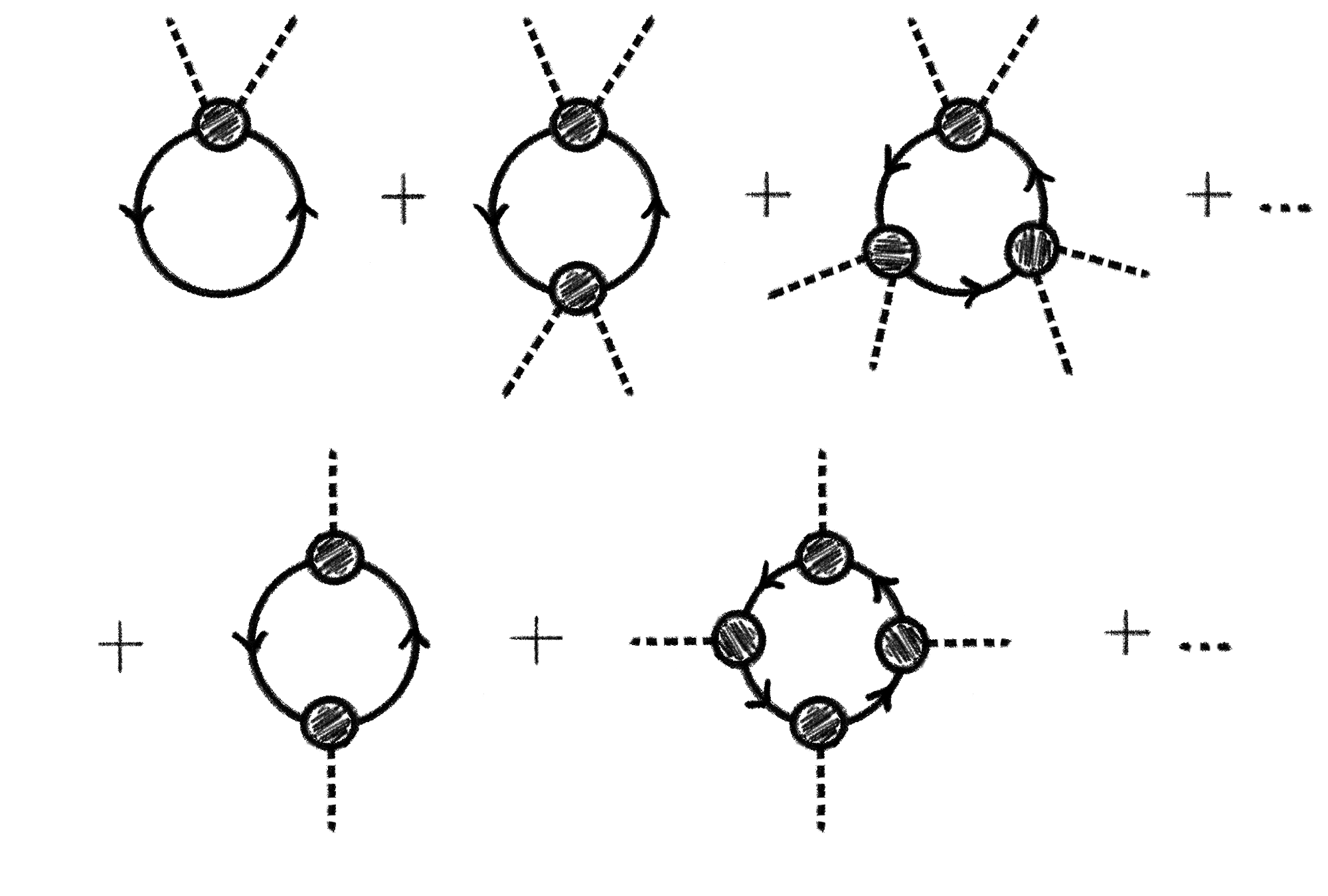}};
\node at (-5.4,1.3) {$\frac{i}{\slashed{p}\Pi_0^{q,t}}$};
\node at (-5.6,3.7) {$i \slashed{p} \Pi_1^{q,t}$};
\node at (-5,-2.6) {$\frac{i}{\slashed{p}(\Pi_0^q + \Pi_1^q)}$};
\node at (-2,-1.4) {$\frac{i}{\slashed{p}(\Pi_0^t + \Pi_1^t)}$};
\node at (-4,-1) {$M$};
\node at (-3,-3.2) {$M$};
} \caption{Series of fermion interactions with the classical background Higgs field. These are contributions to the 1-loop effective potential}\label{fig:fermion_potential}
\end{figure}

as we did in the gauge boson case, we are interested in applying the Feynman rules to diagrams of the sort in \cref{fig:fermion_potential}, where the top diagrams have propagator form factors $\frac{i}{\slashed{p}\Pi^{q,t}_0}$ and vertex form factors $i\slashed{p}\Pi^{q,t}_1$. The bottom diagrams have total propagators $\left(\frac{i}{\slashed{p}(\Pi^q_0 + \Pi^q_1)}\times\frac{i}{\slashed{p}(\Pi^t_0 + \Pi^t_1)}\right)^n$ and total vertices $M^{2n}$. Observe that only powers of $M^2$ will enter the potential, as multiples of two vertices are required to make a loop\footnote{Why not mixed diagrams of both the upper row and lower row in \cref{fig:fermion_potential}? Because the propagator from an $M$-type vertex carries Higgs terms which, if propagated to a $\Pi_1$-type vertex would contract with an outgoing Higgs field and the vertex would become $M$-type. Thus our sums capture those situations, which are indistinguishable.}.

Applying the same reasoning as in the vector boson case, the potential has the form
\begin{align}
V = 2N_ci\int\frac{d^4p}{(2\pi)^4}\left[ \sum\limits_{n=1}^\infty \frac{1}{n}(\frac{-\Pi^q_1}{\Pi^q_0})^n + \sum\limits_{n=1}^\infty \frac{1}{n}(\frac{-\Pi^t_1}{\Pi^t_0})^n + \sum\limits_{n=1}^\infty \frac{1}{n}\left(\frac{-M^2}{p^2\Pi_q\Pi_t}\right)^n\right]
\end{align}
where now only rotations will render the same diagrams, so we only divide by $n$\footnote{A flip will transform a particle into an anti-particle, which is an independent contribution. We have also used that $\slashed{p}^2 = p^2$}. We also have a factor of 2, from the two possible spin configurations. As before, we use the Laurent series identity
to simplify the potential
\hspace{-2cm}\begin{align}
V &= 2N_ci\int\frac{d^4p}{(2 \pi)^4}\left[ -\ln(1 + \frac{\Pi^q_1}{\Pi^q_0}) - \ln(1 + \frac{\Pi^t_1}{\Pi^t_0}) - \ln(1 + \frac{M^2}{p^2\Pi_q \Pi_t})\right] \nonumber\\
& = -2N_ci\int\frac{d^4p}{(2\pi)^4}\left[ \ln\left\lbrace\frac{1}{\Pi^q_0}(\Pi^q_0  + \Pi^q_1)\right\rbrace + \ln\left\lbrace\frac{1}{\Pi^t_0}(\Pi^t_0  + \Pi^t_1)\right\rbrace + \ln\left\lbrace\frac{-1}{p^2}(-p^2 - \frac{M^2}{\Pi_q\Pi_t})\right\rbrace\right] \nonumber\\
\end{align}
We can disregard terms that do not depend on the Higgs field
\begin{align}
& = -2N_ci\int\frac{d^4p}{(2\pi)^4}\left[\ln(\Pi^q) + \ln(\Pi^t) + \ln(-p^2  - \frac{M^2}{\Pi_q\Pi_t}) - \ln(\Pi^q_0) - \ln(\Pi^t_0) - \ln(-p^2)\right] \nonumber\\
& = -2N_ci\int\frac{d^4p}{(2\pi)^4}\ln\left[-\Pi_q\Pi_t(p^2  + \frac{M^2}{\Pi_q\Pi_t})\right] = -2N_ci\int\frac{d^4p}{(2\pi)^4}\ln(-p^2\Pi_q\Pi_t - M^2)
\end{align}
As in section \ref{sec:MCHM_Potential}, we Wick rotate and integrate over the solid angle
\begin{align}
V & = -2N_ci\int\frac{d^4p_E(i)}{(2\pi^2)^4}\ln(-(-p_E^2)\Pi_q\Pi_t  - M^2) \nonumber \\
& = 2N_c\int\frac{d^2p_E}{16\pi^2}p_E^2\ln(p_E^2\Pi_q\Pi_t  - M^2)
\end{align}
Note that one should add a bare term to the form factors for the SM kinetic terms $t_L \slashed{p} t_R$ and so on, giving $\Pi_{q/t} = 1 + \Pi^{q/t}_0(p) + \Pi^{q/t}_1(p,h)$.

To summarise,
\begin{enumerate}
\item Begin the computation with the Coleman-Weinberg formula (in integral form), i.e. \cref{eq:c_w_propagator_vertex}; 
\item Wick rotate the integrands to Euclidean space, by taking $p^2 \rightarrow -p_E^2, d^4 p \rightarrow i d^4 p_E$. Issues with the integral being imaginary are ameliorated by simply moving this outside of the action, and it will thus not affect the potential, which is solely within the Lagrangian density;
\item The Euclidean integral treats all components of $p$ equally, and isotropically. Thus, we can integrate over infinite 4-space by finding the 4-dimensional solid angle. This formula is given in many places as 
\begin{align}
\int\limits_{-\infty}^\infty d^4 p_E = \int\limits_0^\infty d(p_E^2) p_E^2 \Omega_4  = \int\limits_0^\infty (dp_E^\mu p_{E,\mu} ) p_E^2 2 \pi^2
\end{align}
We then have, switching back to Minkowski momenta,
\begin{align}
V(h) =& \int\limits_0^\infty \frac{\dd p^2}{16\pi^2} p^2\left( \frac{6}{2} \log\Pi_{W^+W^-} + \frac{3}{2}\log \left[ \Pi_{BB} \Pi_{W^3 W^3} - \Pi_{W^3 B}^2\right] \right) \label{eq:full_potential} \\
&- 2 \sum\limits_{\psi=t,b} \textnormal{N}_{\psi,c} \int\limits_0^\infty \frac{\dd p^2}{16\pi^2}p^2 \log \left[ p^2 (1+\Pi_\psi)(1+ \Pi_{\psi^c}) - |M_\psi|^2\right]  \nonumber 
\end{align}
The second term includes a factor for the number of colours  $N_{c,\psi}$ of each fermion $\psi$. We then expand the logarithm in powers of the Goldstone fields, to quartic order
\begin{align}
V(h) &= -\gamma s_h^2  + \beta s_h^4 \label{eq:potential_quartic}
\end{align}
\item Take the integral factors of each NGB field (i.e. $\gamma, \beta$) as the coefficients that must be solved for a minimum of the potential, giving an expectation value of the fields in terms of these integrands
\begin{align}
\frac{\partial V}{\partial h} \bigm\lvert_{h=\langle h \rangle} \equiv 0
\end{align}
\end{enumerate}
 The potential is expanded up to quartic order in the Higgs fields, to make connection with the usual SM Higgs potential. The Higgs VEV can then be found at the minimum of \cref{eq:potential_quartic}
\begin{align}
\frac{\partial V}{\partial h} \bigm\lvert_{h=\langle h \rangle} &= -\frac{2\gamma}{f}  s_{\langle h\rangle} c_{\langle h\rangle}  + \frac{4\beta}{f} s^3_{\langle h\rangle} c_{\langle h\rangle} = 0 \nonumber \\
\implies \xi \equiv \sin^2\left( \frac{\langle h \rangle}{f}\right) &= \frac{\gamma}{2\beta} \label{eq:minimum_solution}
\end{align}
We are only interested in the case where the minimum is away from $h = 0$, which corresponds to $0< \frac{\gamma}{2\beta} < 1$. We get the Higgs mass by setting the second derivative of the potential to its expectation value
\begin{align}
\frac{\partial^2 V}{\partial h^2} \bigm\lvert_{h  = \langle h\rangle} = m^2_h &= - \frac{2 \gamma}{f^2} (c^2_{\langle h\rangle}  - s^2_{\langle h\rangle}) - \frac{4\beta}{f^2}  ( 3 s^2_{\langle h\rangle} c^2_{\langle h\rangle} - s^4_{\langle h\rangle}) \nonumber \\
&= - \frac{2\gamma}{f^2}(1- 2s^2_{\langle h\rangle})  + \frac{4\beta}{f^2} (3 s_{\langle h\rangle}^2 - 4 s^4_{\langle h\rangle})
\end{align}
Inserting the minimum solution 
\begin{align}
m^2_h &= -\frac{2\gamma}{f^2}(1- \frac{\gamma}{\beta})  + \frac{4\beta}{f^2} (\frac{3\gamma}{2\beta} - \frac{\gamma^2}{\beta^2}) = - \frac{2\gamma}{f^2}  + \frac{2\gamma^2}{\beta f^2}  + \frac{6\gamma}{f^2} - \frac{4\gamma^2}{\beta f^2}\nonumber\\
&= \frac{4\gamma}{f^2} - \frac{2\gamma^2}{\beta f^2} = \beta (\frac{8\xi}{f^2} - \frac{8\xi^2}{f^2})\nonumber\\
& = \frac{8\beta\xi}{f^2}( 1- \xi)
\end{align}
Note that some works (for example \cite{pomarol2012, de2012}) take $V=- \alpha s_h^2  + \beta' s_h^2 c_h^2$ and then will give a different expression for the minimum in terms of $\alpha = \gamma - \beta, \beta' = -\beta$.
 
We have a functioning model for a Higgs that solves the hierarchy model. We haven't yet defined physically how the composite matter sector couples with the elementary fields, and there are several ways to do this. Let's take a moment to zoom out and see the landscape of options available to us.

\section{Landscape of Composite Higgs Models}

\paragraph{Our goal}
\parbox{0.8\textwidth}{To give a bird's-eye view of pathways leading to or from a composite-like Higgs field.}
\vspace*{2em}

Section \ref{sec:MCHM} sketched a model of global symmetry breaking, from which a SM-like Higgs doublet emerged as the (pseudo-)Goldstone bosons. In general, the results obtained in that section, from the CCWZ Lagrangian terms, to the form of the radiatively-generated potential, and the explicit electroweak gauging of the NGB fields, are applicable to any effective theory of a Composite Higgs. However, there were some missing pieces of the puzzle. We used a large-N approximation to explore the form of the composite vector particle content, when there was no reason that this was a good approximation. We didn't state explicitly the composite fermion \textit{or} vector particle content or how it couples at high energy. We didn't even suggest why it should be called "composite" (vs elementary) or provide a physically meaningful mechanism for including this composite content. We only suggested that it would be required to misalign the Higgs NGBs to pseudo-NGBs and therefore reproduce the SM Higgs mass and vev, and indeed it is. Rather than leap into a concrete realisation of "composite" matter, it will be instructive to see broadly what realisations are available, and why the model is considered "composite".

\begin{figure}[tbp]
\centering
\resizebox{\textwidth}{!}{%
\begin{tikzpicture}[
  font=\footnotesize,
  cat/.style={draw, rounded corners=3pt, align=center, inner sep=5pt, line width=0.6pt},
  ax/.style={-{Latex[length=3mm]}, line width=0.8pt},
]
\draw[ax] (0.2,0) -- (16.4,0);
\node[anchor=north] at (8.3,-0.7)
   {Electroweak vacuum misalignment $\theta$ \quad ($\xi \equiv v^2/f^2 = \sin^2\theta$)};
\foreach \x in {2.0,8.0,13.7} \draw[line width=0.8pt] (\x,-0.12)--(\x,0.12);
\node[anchor=north] at (2.0,-0.12) {$\theta=0,\ \xi=0$};
\node[anchor=north] at (8.0,-0.12) {$0<\xi\lesssim 0.1$};
\node[anchor=north] at (13.7,-0.12) {$\theta=\pi/2,\ \xi=1,\ v=f$};
\foreach \x in {2.0,8.0,13.7} \draw[densely dashed, black!45] (\x,0.12)--(\x,1.05);
\node[cat, text width=3.0cm, fill=black!4, anchor=south] (EH) at (2.0,1.1)
  {\textbf{Elementary Higgs}\\[2pt]
   Standard Model\\(fundamental scalar)\\[2pt]
   $f\to\infty$ (decoupling\\ limit of a pNGB Higgs)\\[2pt]
   $H^2,H^4 \in \mathcal{L}_{\mathrm{tree}}$};
\node[cat, text width=5.2cm, fill=Cerulean!8, anchor=south] (CH) at (8.0,1.1)
  {\textbf{Composite (pseudo-Goldstone) Higgs}\\[2pt]
   $SO(5)/SO(4)$ \ (MCHM)\\
   $SO(6)/SO(5) \cong SU(4)/Sp(4)$\\ (NMCHM $=$ fundamental composite)\\
   generic $SO(N{+}1)/SO(N)$\\[3pt]
   potential radiatively generated:\\
   $H^2,H^4 \notin \mathcal{L}_{\mathrm{tree}}$ \ (Coleman--Weinberg),\\
   light Higgs for $f \gg v$};
\node[cat, text width=3.4cm, fill=black!4, anchor=south] (TC) at (13.7,1.1)
  {\textbf{Technicolor}\\[2pt]
   $SU(2)_L{\times}SU(2)_R \to SU(2)_V$\\
   walking / conformal TC\\
   extended TC (fermion masses)\\[3pt]
   $v=f$: \ no light pNGB ---\\
   would-be Higgs is a heavy\\ $\sigma$-resonance};
\node[cat, text width=5.2cm, fill=BurntOrange!8, anchor=south] (LH) at (8.0,5.7)
  {\textbf{Little Higgs}\ \ {\small(same small $\xi$; different mechanism)}\\[2pt]
   $SU(5)/SO(5)$ \ (Littlest)\qquad
   $[SU(3)/SU(2)]^2$ \ (Simplest)\\[3pt]
   collective symmetry breaking:\\
   $H^2$ protected ($\notin \mathcal{L}$),\quad quartic $H^4$ unsuppressed};
\draw[{Latex[length=2.5mm]}-{Latex[length=2.5mm]}, line width=0.8pt] (CH.north) -- (LH.south)
   node[midway, fill=white, inner sep=2pt, align=center]
   {dimensional deconstruction\\ $+$ custodial $SO(4)$};
\end{tikzpicture}%
}
\caption[An overview of Higgs sectors]{An overview of electroweak-symmetry-breaking scenarios with Higgs dynamics, organised by the vacuum misalignment angle $\theta$ ($\xi = v^2/f^2 = \sin^2\theta$). The composite pseudo-Goldstone Higgs interpolates continuously between the elementary Standard-Model Higgs ($\xi\to0$, $f\to\infty$) and technicolor ($\xi\to1$, $v=f$); little-Higgs models realise the same small-$\xi$ pNGB Higgs but protect its mass through collective symmetry breaking. The categories are not sharp---many models share features. The $H^2/H^4$ labels indicate whether the Higgs mass and quartic terms appear in the tree-level Lagrangian or are radiatively generated.}\label{fig:landscape}
\end{figure}
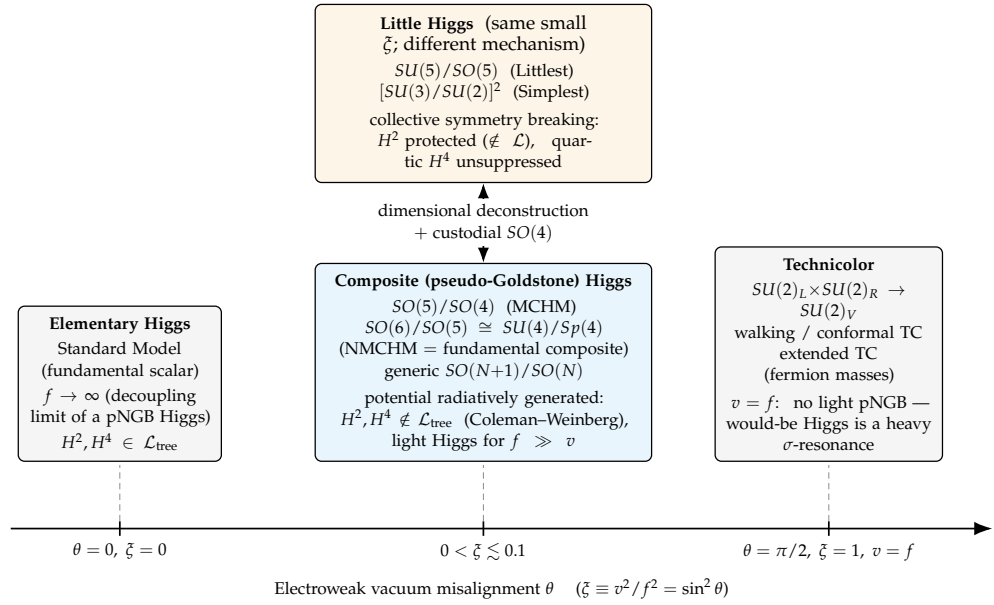

The general types of models that attempt to solve the question of electroweak symmetry breaking with Higgs dynamics are summarised in fig. \ref{fig:landscape}. In particular, note that these are not discrete categories - the Higgs models in this study contain features shared by many models, and are useful tools in the general study of Higgs dynamics. True Composite Higgs models (that is, models where the Higgs is a pNGB of some underlying dynamics) can be considered an interpolation between technicolor and the elementary Higgs, depending on how misaligned the EW group is with any unbroken gauge symmetries in the model.


\subsection{Roads to a pNGB Composite Higgs}
\label{sec:underlying_dynamics}

A Higgs that emerges as a pNGB from hidden global symmetries is a convincing way to ameliorate the Hierarchy Problem. But what does the hidden global symmetry describe - what degrees of freedom does it represent and how are some of these "broken" to a smaller set? And how would composite matter also be described by this group, especially when we are used to matter interacting via gauge fields? Although in this work we are developing an EFT of the pNGB Composite Higgs, we will briefly look at three roads leading to such an EFT. These three modern formulations can be made to produce a global symmetry structure and new matter obeying this symmetry, as required by our EFT. This is not a taxonomy of interesting models - these three are deeply related and each can be reformulated as another. The three roads are technicolor-like models with a pion-like Higgs, Moose models with a little Higgs, and extra-dimensional models with a holographic Higgs. 

We have already seen in \cref{sec:dynamical_EWSB} how the introduction of a new strong sector (e.g. technicolors and techniquarks) can lead both to a radiatively generated EW potential and to a fermion condensation that appears as a light Higgs-like boson. We have a good prototype for this, QCD, which appears below confining scales as a chiral symmetry - a non-linear sigma model. This is a special case of what we shall later describe as a Moose model. The Moose model paradigm provides a template for multiple hidden local symmetries (i.e. QCD) which appear as a chain of connected sites of non-linear sigma models (i.e. chiral symmetry), each with sets of new hadron-like matter. The particular collective breaking of these sites leads to a "little Higgs", as they must each break simultaneously.

\begin{figure}
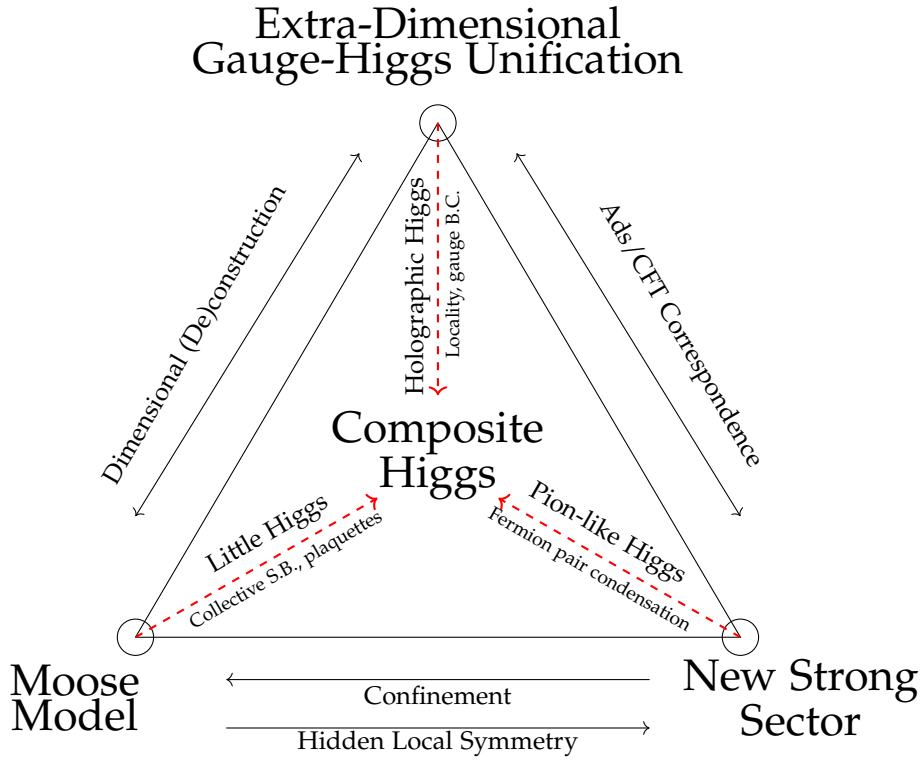

\centering
\tikz[scale=0.8, every node/.style={transform shape}]{
\draw (1,1)  -- (6,9.5)   -- (11,1)  --(1,1);
\draw (1,1) circle [radius=0.3, fill=white];
\draw (6,9.5) circle [radius=0.3, fill=white];
\draw (11,1) circle [radius=0.3, fill=white];
\node [align=center] at (0,0) {\huge Moose \\ \huge Model};
\node [align=center, above] at (6,10) {\huge Extra-Dimensional\\ \huge Gauge-Higgs Unification};
\node [align=center] at (12, 0) {\huge New Strong \\ \huge Sector};
\node [align=center] at (6,4) {\huge Composite \\ \huge Higgs};
\node [align=center, rotate=60] at (2,6) {\large Dimensional (De)construction};
\draw [<->] (1,3) -- (4.7,9);
\node [align=center, rotate=-60] at (10,6) {\large Ads/CFT Correspondence};
\draw [<->] (7.3,9) --(11 ,3);
\node [align=center] at (6,0.05) {\large Confinement};
\draw [<-] (2.5,0.3) -- (9.5,0.3);
\node [align=center] at (6,-0.75) {\large Hidden Local Symmetry};
\draw [->] (2.5,-0.5) -- (9.5,-0.5);
\draw [->, thick, dashed, red] (1,1) -- (5,3.3);
\node [align=center, rotate=30] at (3.2,2.7) {\large Little Higgs};
\node [align=center, rotate=30] at (3.5,2.1) {\small Collective S.B., plaquettes};
\draw [->, thick, dashed, red] (11,1) -- (7,3.3);
\node [align=center, rotate=-30] at (8.8,2.7) {\large Pion-like Higgs};
\node [align=center, rotate=-30] at (8.5,2.1) {\small Fermion pair condensation};
\draw [->, thick, dashed, red] (6,9.5) -- (6,5);
\node [align=center, above, rotate=90] at (6,6.8) {\large Holographic Higgs};
\node [align=center, below, rotate=90] at (6,7) {\small Locality, gauge B.C.};
}\caption{Roads to Composite Higgs}\label{fig:triforce}
\end{figure}

Each site of a Moose model can be treated as a symmetry breaking of successively higher energy scales. For a UV-complete theory, we could take the lowest symmetry breaking scale as the EW scale, and the highest as the Planck scale. As we add more and more sites to the model between these scales, the chiral models become a continuum, and it is more sensible to treat all the linking fields as a single field along an extra finite dimension. We have then built (discretised) a genuine physical dimension using a "dimensional (de)construction" approach. In certain extra-dimensional theories, one can assemble the EW gauge fields, and the Higgs field into one multiplet $A_M$, $M=1,...,5$ leading to Gauge-Higgs Unification (GHU).

Matter is introduced in an extra-dimensional theory quite naturally, with light particles (i.e. elementary) living on the EW brane of the dimension, the Higgs-like multiplet living on the Planck brane, and 5D matter living in the bulk in-between. When projecting to four dimensions, the "holographic" Higgs appears as the massless zero mode of $A_5$  that acquires a mass from radiative corrections - a pNGB. There are several geometries available for the extra dimension, which are compatible with different choices of Moose structures. An $(n,m)$-spacetime of constant curvature is represented by the Lorentz interval
\begin{align}
ds^2 = \sum\limits^n dx_i^2 - \sum\limits^m dt^2_j
\end{align}
with $ds^2$ greater than, equal to, or less than zero corresponding to de Sitter (spherical), Minkowski (flat), or anti-de Sitter (hyperbolic) spacetime geometry. Choosing our five dimensions to be anti-de Sitter spacetime (AdS$_5$) leads to a correspondence between the weakly-interacting GHU model and a conformal field theory (CFT) of strong gauge interactions. The Ads/CFT correspondence allows one to explore models of strong interactions, which would normally be beyond perturbative limits, by formulating them as extra-dimensional models of weak interactions. One can then use holographic methods to translate back to four dimensions. The prototypical strong CFT is QCD, and we are thus back to the beginning. These deep connections are sketched in \cref{fig:triforce}. A full exploration of the connection between these roads is beyond the scope of this work. Suffice it to say that the EFT we develop in this work is well-motivated and supported by several converging models.

\subsection{An Effective 4D pNGB Higgs}
\label{sec:Landscape_4dMCHM}

%

\paragraph{Our goal}
\parbox{0.8\textwidth}{To begin to build an effective Composite Higgs theory from the simplest assumptions, without the simplifications of previous sections.}
\vspace*{2em}

In developing an explicit EFT that gives a pNGB Higgs boson, we require what Marzocca, et al. \cite{marzocca2012} call the "Minimal Higgs Potential (MHP) hypothesis"\footnote{Much of this section will follow the general argument of that work}. This hypothesis assumes an effective model with a pNGB Higgs coming from global symmetry hiding $G \rightarrow H$. It further assumes the pNGB(s) receives radiative corrections to its potential \textit{at one-loop order}, that are both \textit{calculable} and \textit{finite}\footnote{We include both of these related concepts: the first is a condition on the complexity of the model, the second on the convergence of its Coleman Weinberg potential integral terms}. The reason for these assumptions is that we are building an effective model, while any physically realisable UV description would presumably fulfil these minimal assumptions automatically. We will build this effective theory using only the MHP hypothesis, arriving at a class of model called the multi-site Moose model.

Firstly, we must produce four Goldstone bosons to later act as the Higgs doublet. This can be achieved with a non-linear sigma model. In the CCWZ formalism, the non-linear degrees of freedom (i.e. the Goldstone bosons) can be described to quadratic order with a Lagrangian
\begin{align}
\mathcal{L}_\sigma &= \frac{f^2}{4}\text{Tr}[d_\mu d^\mu]
\end{align}
Recalling the description of $d_\mu$ in \cref{eq:mc_definition} and \cref{eq:maurer_cartan_invariant_term}. To recall why this is the correct term, we attempt a transformation of the sort in \cref{eq:mc_transformation}
\begin{align}
\mathcal{L}_\sigma \xrightarrow{G} \mathcal{L}'_\sigma = \frac{f^2}{4}\text{Tr}[d'_\mu d'^\mu] = \frac{f^2}{4}\text{Tr}[(h d_\mu h^\dagger )(h d^\mu h^\dagger)] = \frac{f^2}{4}\text{Tr}[d_\mu d^\mu] = \mathcal{L}_\sigma
\end{align}
using the simple permutation property of the trace. In this form, the Lagrangian is not so useful. We can expand the Maurer-Cartan form in powers of $1/f$, recalling the generic definition for the Goldstone matrix
\begin{align}
\Phi = \exp{\left(i\frac{\sqrt{2}}{f}\pi^{\hat{a}}T^{\hat{a}}\right)}
\end{align}
Then 
\begin{align}
w_\mu &= -iU^\dagger \partial_\mu U \nonumber \\
&= -i\left( 1 + i\frac{\sqrt{2}}{f}\pi^{\hat{a}}T^{\hat{a}} + ...\right)^\dagger \partial_\mu \left(1 + i\frac{\sqrt{2}}{f}\pi^{\hat{a}}T^{\hat{a}} + ...\right) \nonumber \\
&= -i\left( 1 - i\frac{\sqrt{2}}{f}\pi^{\hat{a}}T^{\hat{a}} + ...\right) \left(i\frac{\sqrt{2}}{f}\partial_\mu \pi^{\hat{a}}T^{\hat{a}} + ...\right) \nonumber \\
&= \frac{\sqrt{2}}{f}\partial_\mu \pi^{\hat{a}}T^{\hat{a}} + ...\\
&\equiv d^{\hat{a}}_\mu T^{\hat{a}} + E_\mu^a T^a
\end{align}
Therefore we can match $d^{\hat{a}}_\mu = \frac{\sqrt{2}}{f}\partial_\mu \pi^{\hat{a}}T^{\hat{a}}$ at first order, and find the result 
\begin{align}
\mathcal{L}_\sigma = \frac{1}{2}(\partial_\mu \pi^{\hat{a}})(\partial^\mu \pi^{\hat{a}})
\end{align}
%

At this point, we would also like to gauge the NGBs in the obvious way, with $D_\mu \pi^{\hat{a}} = (\partial_\mu - ig_0 W^a_\mu T_{aL} -ig_0' B_\mu T_{3R}) \pi^{\hat{a}}$, thus
\begin{align}
\mathcal{L}_{\sigma - \text{gauge}} &= \frac{f^2}{4} \left(D_\mu \pi^{\hat{a}}\right)\left(D^\mu \pi^{\hat{a}}\right) - \frac{1}{4}W_{\mu\nu}^{aL}W^{aL\mu\nu} - \frac{1}{4}B_{\mu\nu} B^{\mu\nu}
\end{align}
which is a first order description of the Lagrangian we previously arrived at in \cref{eq:sigma_model_effective_gauge}, albeit in a less rigorous fashion.

The gauging of the $SU(2)_L \times U(1)_Y$ subgroup, explicitly breaks the global $G$, causing the Higgs to become a pNGB.  At this point, we have reproduced the required SM Higgs sector, with the Higgs as a physical particle that we expect to have a mass naturally well below the symmetry breaking scale $f$. To find the mass explicitly, we must match a high energy theory to the low energy theory we have just described. This high energy theory is expected to have resonances, as is our experience of other non-linear sigma models such as nuclear chiral perturbation theory. To deal effectively with the symmetry structure of these resonances we will introduce a language for multiple sites of matter.

\section{The N-Site Model}
\label{sec:mooses}


\paragraph{Our goal}
\parbox{0.8\textwidth}{To develop a toolkit for understanding collective symmetry breaking}
\vspace*{2em}

There are many ways that composite matter could be included as coupling to a pNGB Higgs. In anticipation of our theory having a calculable and finite Higgs potential, we will impose a coupling structure that this matter must obey. This is also well-motivated by the discussion in \cref{sec:underlying_dynamics}, where we are discretising a fifth dimension, with the EW sector located at one boundary, the Higgs at the other, and discretised layers, or \textit{sites}, of composite fermions connecting the two in the bulk. 

\subsection{Moose Diagrams}

\begin{figure}[h!]
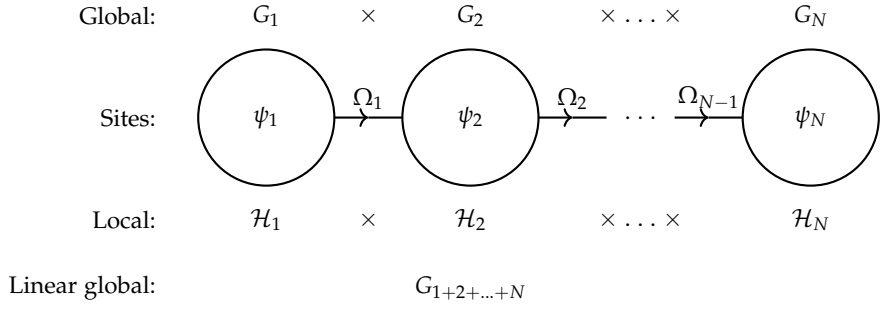

\centering\tikz[scale=0.9, every node/.style={transform shape}]{
\node [left] at (-1.5,1.5) {Global:};
\node [left] at (-1.5,0) {Sites:};
\node [left] at (-1.5,-1.5) {Local:};
\node [left] at (-1.5,-2.5) {Linear global:};
\draw [thick] (0,0) circle [radius=1] node {$\psi_1$};
\node at (0,1.5) {$G_1$};
\node at (0,-1.5) {$\mathcal{H}_1$};
\node at (1.5,1.5) {$\times$};
\node at (1.5,-1.5) {$\times$};
\draw [thick, ->] (1,0) -- (1.5,0);
\draw [thick] (1.5,0) -- (2,0);
\node [above] at (1.5,0) {$\Omega_1$};
\draw [thick] (3,0) circle [radius=1] node {$\psi_2$};
\node at (3,1.5) {$G_2$};
\node at (3,-1.5) {$\mathcal{H}_2$};
\node at (5.5,1.5) {$\times \;  .\; . \; . \; \times$};
\node at (5.5,-1.5) {$\times \; .\; .\; . \; \times$};
\node at (3,-2.5) {$G_{1+2+...+N}$};
\draw [thick,->] (4,0) -- (4.5,0);
\draw [thick] (4.5,0) -- (5,0);
\node [above] at (4.5,0) {$\Omega_2$};
\node at (5.5,0) {$.\; .\; .$};
\draw [thick] (8,0) circle [radius=1] node {$\psi_N$};
\draw [thick,->] (6,0) -- (6.5,0);
\draw [thick] (6.5,0) -- (7,0);
\node [above] at (6.5,0) {$\Omega_{N-1}$};
\node at (8,1.5) {$G_N$};
\node at (8,-1.5) {$\mathcal{H}_N$};
}
\caption{A generic Moose diagram \label{fig:generic_moose}}
\end{figure}

This happens to already be a well-explored model, pioneered in \cite{GEORGI1986274}, and called a \textit{Moose model}. These models are conveniently described by \textit{Moose diagrams}, such as that shown in \cref{fig:generic_moose}. In the original work, these figures captured local symmetries and matter fields. Here, we will be more pedagogical. A \textit{site} in the diagram is denoted by a circle; this is a multiplet of Dirac fermionic matter $\Psi_i$ transforming under a global symmetry $G_i$, and possibly gauged by some vectorial local symmetry. Global symmetries are stated above the sites of the diagram, local symmetries below. A \textit{link} in the diagram is denoted by a line connected to a site or sites; this is a multiplet of scalar fields Yukawa-coupling the fermionic matter. If it attains a vev, we can re-parameterise the scalars as multiplets of Goldstone bosons $\Omega_i$ which non-linearly preserve the global symmetries stated above the diagram. We have seen this paradigm, as a chiral NL$\sigma$M, in \cref{sec:NLSM}. Again, for the sake of pedagogy, to begin we will explicitly state the remaining linear (i.e. unbroken) symmetries below the gauge symmetries in each site, although they can be deduced when the global symmetries and link fields are known. This modern form of the Moose model is well explored in \cite{ArkaniHamed:2001ca,arkani2003effective,arkani2002twisted,
arkani2001electroweak,arkani2002minimal,ArkaniHamed:2002qy}.

Let us quickly examine each of these features by building a Gedanken model using the tools we developed in section \ref{sec:CCWZ}, which we can represent with a Moose diagram. Consider a NL$\sigma$M represented by $\mathcal{L} = \mathcal{L}_\text{matter} + \mathcal{L}_\text{gauge} + \mathcal{L}_\text{NGB}$, which consists of
\begin{enumerate}
\item A set of $k$ sites of fermions multiplets, 
\begin{align}
\mathcal{L}_\text{matter} &= \bar{\psi}^\alpha_1 ( \slashed{\partial}  - m_1)\psi^\alpha_1 + \bar{\psi}^\alpha_2 (\slashed{\partial}  - m_2) \psi^\alpha_2 + ... + \bar{\psi}^\alpha_N ( \slashed{\partial}  - m_N ) \psi^\alpha_N
\end{align}
%
each in a fundamental\footnote{Considering fermions in other representations is a simple extension of the following argument, as is considering non-unitary groups.} of $G_i$, where $G_1 \sim G_2 \sim ... \sim G_k$ for convenience. Denote each fermion $\psi^\alpha_i$, where ($G$ flavour index) $\alpha=1,...,M$ and (site index) $i=1,...,N$. 
\item A set of gauge bosons $A_{i,\mu} = A^a_{i,\mu}S^a$ that promote each $G$ invariance to a local one. This is achieved with "minimal coupling"
\begin{align}
\mathcal{L}_\text{gauge} = ig_1\bar{\psi}^\alpha_1 \slashed{A}^{\alpha\beta}_1 \psi^\beta_1 + ig_2\bar{\psi}^\alpha_2 \slashed{A}^{\alpha\beta}_2 \psi^\beta_2 + ... + ig_N \bar{\psi}^\alpha_N \slashed{A}^{\alpha\beta}_N \psi^\beta_N + \mathcal{L}_\text{gauge-kinetic}
\end{align}
\item A set of $N-1$ Nambu-Goldstone bosons that couple "nearest neighbour" matter fields, 
\begin{align}
\mathcal{L}_\text{NGB} &= \bar{\psi}^\alpha_{1,R} \Omega_{1}^{\alpha\beta} \psi^\beta_{2,L} + \bar{\psi}^\alpha_{2,R} \Omega_2^{\alpha\beta} \psi^\beta_{3,L} + ... \bar{\psi}^\alpha_{N-1,R} \Omega_{N-1}^{\alpha\beta} \psi^\beta_{N,L} + \text{h.c.} \label{eq:moose_gb_lagrangian} \\
& + \mathcal{L}_\text{NGB-kinetic} + \mathcal{L}_\text{NGB-gauge} \nonumber
\end{align}
If the matter fields are locally invariant under each $G_i$, then clearly the NGBs must also be gauged. The $\Omega_i$ fields are a non-linear parameterisation of the Goldstone fields resulting from a $G_i \times G_{i+1} \rightarrow G_{i + (i+1)}$ symmetry hiding. We will derive the particular parameterisation shortly.
The kinetic and gauge terms are
\begin{align}
\mathcal{L}_\text{NGB-kinetic} + \mathcal{L}_\text{NGB-gauge} &=  \text{Tr}\left[  \frac{f^2_1}{4}( D^\mu_1 \Omega_1)^\dagger D_{1,\mu} \Omega_1 +  \frac{f^2_2}{4}( D^\mu_2 \Omega_2)^\dagger D_{2,\mu} \Omega_2 \right. \nonumber\\ 
& \left. + ... +  \frac{f^2_{N-1}}{4}( D^\mu_{N-1} \Omega_{N-1})^\dagger D_{N-1,\mu} \Omega_{N-1}\right] \label{eq:link_general_lagrangian}\\
\text{where} \qquad D^\mu_\alpha \Omega_\alpha &= \partial^\mu \Omega_\alpha  - i g_\alpha A^\mu_\alpha \Omega_\alpha + i g_{\alpha + 1} \Omega_\alpha A^\mu_{\alpha+1}  \nonumber
\end{align}
%
\end{enumerate}


Consider each of the above Lagrangian subsets as a quality that a physical theory may have. Assuming only (1), then the model can be described by a single-site Moose diagram, containing multiple non-interacting massive fermions, transforming under $G_1 \times G_2 \times ... \times G_N$. Already we have imposed structure in the form of mass terms - we will require these fermions to be heavy in our final model, thus they should attain mass at tree level, at the scale of symmetry breaking. We are then dealing with vector-like fermions.


Assuming (1)+(2), then the fermions are simply gauged under $G_1 \times G_2 \times ... \times G_N$, and can thus interact within each site. The fermions already had mass terms, so diagonally gauging the multiplets does not alter the global symmetry structure, which is entirely linear.

Assuming (1)+(3), the symmetry can be hidden by the non-linear transformation properties of $\Omega_i$, which we henceforth will call "link fields". They transform as
\begin{align}
\Omega_i \rightarrow g_{i} \Omega_i g_{i+1}^\dagger \, \qquad g_i \in G_i, \; g_{i+1} \in G_{i+1}\; .
\end{align}
We can see that they non-linearly preserve the global $G_1 \times ... \times G_N$ symmetry. However, just as in the sigma model of \cref{sec:sigma_model}, if we used a linear parameterisation, there would be mass-like couplings between sites $\mathcal{L}_\text{matter} \supset g_i f_i \bar{\psi}_i \psi_{i+1}$. Thus, a transformation of a fermion on the first site $\psi_{1,L} \rightarrow g_{1,L} \psi_{1,L}$ requires a simultaneous transformation in all other fermions, in the linear parameterisation. The linear subgroup is thus the diagonal subgroup of all the sites
\begin{align}
G_1 \times \;.\;.\;.\; \times G_N \rightarrow G_{1+2+...+N} \, ,
\end{align}
which leads to $N-1$ sets of dim$[G]$ NGBs: the $\Omega_i$ multiplets. Note that the presence of tree-level mass terms is a new feature. It destroys our usual chiral symmetry breaking diagram, as in \cref{fig:higgs_mech}, since our theory is not chiral but vector-like. To combine the three Moose features (1)+(2)+(3), we must derive the particular form of $\Omega_i$, which, as usual, requires knowing the form of the broken generators.

\begin{figure}[h!]
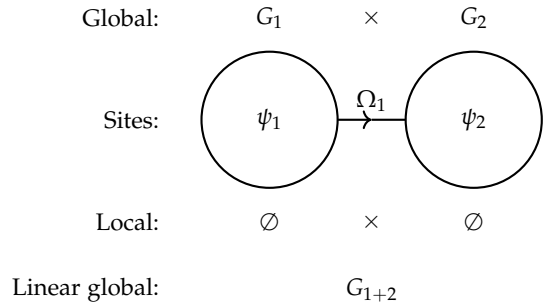

\centering\tikz[scale=0.9, every node/.style={transform shape}]{
\node [left] at (-1.5,1.5) {Global:};
\node [left] at (-1.5,0) {Sites:};
\node [left] at (-1.5,-1.5) {Local:};
\node [left] at (-1.5,-2.5) {Linear global:};
\draw [thick] (0,0) circle [radius=1] node {$\psi_1$};
\node at (0,1.5) {$G_1$};
\node at (0,-1.5) {$\emptyset$};
\node at (1.5,1.5) {$\times$};
\node at (1.5,-1.5) {$\times$};
\node at (1.5,-2.5) {$G_{1+2}$};
\draw [thick, ->] (1,0) -- (1.5,0);
\draw [thick] (1.5,0) -- (2,0);
\node [above] at (1.5,0) {$\Omega_1$};
\draw [thick] (3,0) circle [radius=1] node {$\psi_2$};
\node at (3,1.5) {$G_2$};
\node at (3,-1.5) {$\emptyset$};
}
\caption{A two-site model \label{fig:moose_model_two_site}}
\end{figure}

Consider the Moose diagram given in \cref{fig:moose_model_two_site}. Take the group symmetry of each site $G_i$ to be generated by the same Lie algebra $S_1^a = S_2^a = ...$. Then the unbroken $T^i$ and broken $X^i$ generators of the breaking $G_1 \times G_2 \rightarrow G_{1+2}$ are given by the vectorial and axial generators
\begin{align}
T_1^a = \frac{1}{\sqrt{2}}\left( 
\begin{matrix}
S^a & 0\\
0 & S^a
\end{matrix} \right) & & X_1^a = \frac{1}{\sqrt{2}} \left( \begin{matrix}
S^a & 0\\
0 & -S^a
\end{matrix} \right)\label{eq:T_X_generators}
\end{align}

\vspace{2em}
\begin{tcolorbox}[colback = black!2!white]
\paragraph{Locality of the Moose}\label{sec:locality}

It is interesting to note that the successive links need not couple in this way. We have employed the minimal case: Locality in theory space (i.e. nearest neighbour coupling), and consistent dimension of coupling. The first is strongly motivated by extra-dimensional models, where the sites effectively represent a discretised 5th dimension, and therefore couple according to space-time locality \cite{Kahn:2012as}. Specifically, the Moose diagrams represent a lattice on a warped dimension with AdS$_5$ geometry. The gauge couplings and decay constants in four dimensions correspond to the "warp factor" of the fifth dimension's geometry. Importantly, this locality (and therefore the construction assumed in this work) requires a linear Moose, that is, a directed acyclic graph. A cyclic Moose could more realistically include non-local "hopping" terms. In a general CHM though, there is no reason to enforce such locality a priori. We do so, in order to produce the minimal, calculable model. The second assumption is also motivated by minimalism. Coupling successive links according to the effective form $\Omega_j^\dagger \Omega_{j+1}^q$, with $q$ not necessarily equal to one, leads to a clockwork type mechanism \cite{Ahmed:2016viu}, that quite naturally generates flavour hierarchies, and is worth further investigation.
\end{tcolorbox}
\vspace{2em}

\subsection{Symmetric Spaces}\label{sec:symmetric_spaces}

We have stumbled here on an extremely important special case of broken symmetry. The coset $G_1 \times G_2 / G_{1+2}$ is called a "symmetric space", and it allows the simplification of many concepts used in the CCWZ construction. A symmetric space $K$ is a particular kind of homogeneous space (i.e. a coset $G/H$ that has the properties of a manifold, as all our Lie group cosets have) that is also equipped with an automorphism $\sigma$ with a special property: after applying the automorphism twice to either the coset $G/H$ or the subgroup $H$, we must get back the original set:
\begin{align}
G/H = K && \text{iff} && \exists \; \sigma: \; (G/H,G) \rightarrow (G/H,G) && \text{s.t.} && \sigma^2 = 1
\end{align}
$\left( SU(N) \times SU(N) \right) / SU(N)$ is one example of symmetric space; we can derive from \cref{eq:T_X_generators} the following relations
\begin{align}
[T^a, T^b] = if^{abc} T^c, && [T^a, X^b] = i f^{abc} X^c, && [X^a, X^b] = if^{abc} T^c\label{eq:symmetric_commutation_relations}
\end{align}
Observe the duality between the broken generators and the unbroken generators, which satisfies our definition of a symmetric space automorphism. Specifically, 
\begin{align}
\sigma: \qquad & T^a \rightarrow [T^a, X^b] \propto X^c \nonumber\\
& X^a \rightarrow [X^a, X^b] \propto T^c
\end{align}
In fact, the identity in \cref{eq:symmetric_commutation_relations} holds for any symmetric space. Let's see how this identity simplifies our leading order Lagrangian\footnote{The following is an approach broadly of my own design that gives more exposure to the Maurer-Cartan form. For a more elegant, lateral approach using the automorphism $\sigma$, see for example \cite{thaler2005little}}. Recall the explicit form of the $h$-transforming term $d_\mu$ given in \cref{eq:broken_form_explicit}
\begin{align}
d_\mu &=  \frac{1}{f} \partial_\mu \pi - \frac{i}{2f^2}[\pi,\partial_\mu \pi]_X - \frac{1}{6f^3}[\pi, [\pi, \partial_\mu\pi]]_X + \frac{1}{24f^4}[\pi, [\pi, [\pi, \partial_\mu \pi ]]]_X + ... \label{eq:maurer_cartan_broken_again}
\end{align}
Applying  \cref{eq:symmetric_commutation_relations} to \cref{eq:maurer_cartan_broken_again} we get...

\begin{align}
d_\mu &=  \frac{1}{f} \partial_\mu \pi - \frac{i}{2f^2}\cancelto{0}{\underbrace{[\pi,\partial_\mu \pi]}_{T}}_X - \frac{1}{6f^3}[\pi, [\pi, \partial_\mu\pi]]_X + \frac{1}{24f^4}\cancelto{0}{\underbrace{[\pi, [\pi, [\pi, \partial_\mu \pi ]]]}_{T}}_X + ... \nonumber\\
&=  \partial_\mu(U(x) U(x) ) \label{eq:symmetric_commutation_relation}
\end{align}

Then we can define the term that enters the Lagrangian as
\begin{align}
\mathcal{L} = d_\mu d^\mu \equiv \frac{f^2}{16} \text{tr}[\partial_\mu\Sigma^{-1} \partial^\mu \Sigma], \; \text{where } \;  \Sigma = U(x)U(x) = \me^{i\frac{2\sqrt{2}}{f} \pi^{\hat{a}}X^{\hat{a}}}
\end{align}

The key to the commutation step leading to \cref{eq:symmetric_commutation_relation} is the identity
\begin{align}
\sum\limits_{k=0}^i \left( -1\right)^k {i \choose k} = 0
\end{align}
This can be seen most easily by recalling how each line of Pascal's triangle is built - with pairwise sums of the previous line
\small{\begin{align}
& \sum\limits_{k=0}^i \left( -1\right)^k {i \choose k} = {i \choose 0} - {i \choose 1} + {i \choose 2} - ... + (-1)^{i-1} {i \choose i-1} + (-1)^i {i \choose i} \nonumber\\
&\equiv 1 - \left[{i-1 \choose 0} + {i-1 \choose 1} \right] + \left[{i-1 \choose 1} + {i-1 \choose 2} \right] - ... + (-1)^{i-1} \left[{i-1 \choose i-2} + {i-1 \choose i-1} \right] + (-1)^i \nonumber\\
&={\small \cancelto{A}{1} - \left[\cancelto{-A}{{i-1 \choose 0}} + \cancelto{B}{{i-1 \choose 1}} \right] + \left[\cancelto{-B}{{i-1 \choose 1}} + \cancelto{C}{{i-1 \choose 2}} \right] - ... + (-1)^{i-1} \left[ \cancelto{-H}{{i-1 \choose i-2}} +  \cancelto{I}{{i-1 \choose i-1}} \right] +  \cancelto{-I}{(-1)^i}} \nonumber
\end{align}}
%

For certain symmetric spaces, this may still not be the most economical description. Returning to the case of $G_1 \times G_2 \rightarrow G_{1+2}$, the Goldstone matrix is given by 
\begin{align}
U = \me^{i\frac{\sqrt{2}}{f} \pi^{\hat{a}}X^{\hat{a}}} &=  \left(\begin{matrix}
\me^{\frac{i}{f}\pi^a S^a} & 0\\
0 & \me^{-\frac{i}{f}\pi^a S^a}
\end{matrix}\right) \equiv \left(\begin{matrix}
u(x) & 0\\
0 & u^{-1}(x)
\end{matrix}\right)
\end{align}
Then 
\begin{align}
\Sigma &= \left(\begin{matrix}
u(x)u(x) & 0\\
0 & u^{-1}(x)u^{-1}(x)
\end{matrix}\right) =  \left(\begin{matrix}
\me^{\frac{2i}{f}\pi^a S^a} & 0\\
0 & \me^{-\frac{2i}{f}\pi^a S^a}
\end{matrix}\right) \equiv \left(\begin{matrix}
\Omega(x) & 0\\
0 & \Omega^{-1}(x)
\end{matrix}\right)
\end{align}

One can show that the Goldstone matrix $U$ transforms non-linearly, as we would expect. But in particular, it's non-linearity leads to a symmetric transformation for $u$
\begin{align}
G_1 \times G_2: \qquad u \rightarrow g_1 u k^{-1}(g_1,g_2,x) = k(g_1,g_2,x) u g_2
\end{align}
where $k$ is some non-linear transformation that we recognise from the CCWZ formalism. This is a complicated transformation. Luckily, we don't need $u$, only $uu$, which transforms as
\begin{align}
G_1 \times G_2: \qquad uu = \Omega \rightarrow g_1 \Omega g_2^{-1} \, .
\end{align}
Thus, we recognise $\Omega$ from its transformation property as the link field from the Moose diagrams, \textit{and} the field transforming under a left and right field investigated in \cref{sec:HLS}. Using the cancellations afforded by the symmetric structure, we can write the invariant terms of the Maurer-Cartan form explicitly:
\begin{align}
\mathcal{L}_2 = \frac{f^2}{8}\text{tr}[d_\mu^\dagger d^\mu] = \frac{f^2}{8}\text{tr}[\partial_\mu\Omega^{-1} (x) \partial^\mu \Omega(x)]
\end{align}

We should be appreciative of this closed form - symmetric spaces are one of only a small number of ways to state the closed form of the NGBs in generality. Another was to take a specific vacuum, as we did in \cref{sec:MCHM_Goldstone}.


Now we may explore the combined features (1)+(2)+(3), by gauging the symmetric spaces. We apply the $G^i$ local symmetry.  
Now we may gauge the above theory, leading to (1)+(2)+(3). We apply the $G^i$ local symmetry. At first glance, this would seem to enact the Higgs mechanism on the broken global symmetries $(G \times G)^{N-1}/G^{N-1}$, eating the NGBs, giving mass to $(N-1)\times\text{dim}\left[ G\right]$ of the gauge bosons. In that case, $\text{dim}\left[ G\right]$ of the gauge bosons will remain massless. However, the execution of the Higgs mechanism will depend on how the vacuum (of the link fields) aligns with the gauged generators. We will be using the machinery of Hidden Local Symmetry, developed in \cref{sec:HLS}, where we showed that one can treat a local symmetry as a   non-linear global symmetry of an additional group. In our new Moose notation, we can denote this by 

\begin{figure}[H]
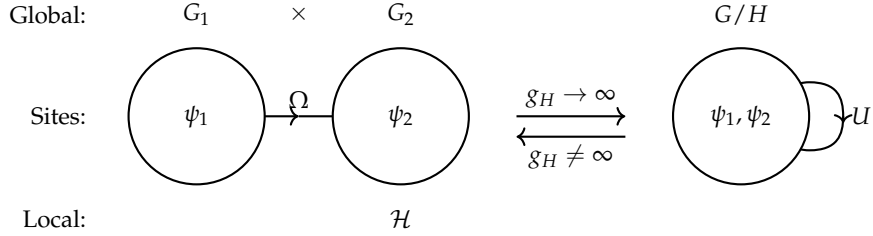

\centering\tikz[scale=0.9, every node/.style={transform shape}]{
\node [left] at (-1.5,1.5) {Global:};
\node [left] at (-1.5,0) {Sites:};
\node [left] at (-1.5,-1.5) {Local:};
\draw [thick] (0,0) circle [radius=1] node {$\psi_1$};
\node at (0,1.5) {$G_1$};
\node at (1.5,1.5) {$\times$};
\node at (3,-1.5) {$\mathcal{H}$};
\draw [thick, ->] (1,0) -- (1.5,0);
\draw [thick] (1.5,0) -- (2,0);
\node [above] at (1.5,0) {$\Omega$};
\draw [thick] (3,0) circle [radius=1] node {$\psi_2$};
\node at (3,1.5) {$G_2$};
\node [above] at (5.5,0) {$g_H \rightarrow \infty$};
\draw [thick, ->] (4.7,0) -- (6.3,0);
\draw [thick, <-] (4.7,-0.3) -- (6.3,-0.3);
\node [below] at (5.5,-0.3) {$g_H \neq \infty$};
\draw [thick,->] plot [smooth, tension=1] coordinates {(8,0) (9,0.5) (9.5,0) (9,-0.5) (8,0)};
\draw [thick, ->] (9.5,0) -- (9.5,-0.1);
\node [right] at (9.5,0) {$U$};
\draw [thick, fill=white] (8,0) circle [radius=1] node {$\psi_1, \psi_2$};

\node at (8,1.5) {$G/H$};
}
\caption{A Moose diagram describing the Hidden Local Symmetry limit of a gauged Moose model \label{fig:HLS_limit}}
\end{figure}

\begin{figure}[H]
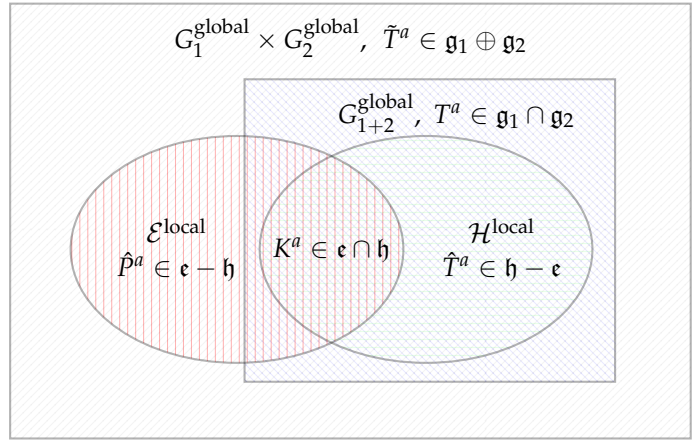

\centering
\tikz{
\draw [pattern=north east lines, pattern color=gray, thick, opacity=0.3] (0.5,-0.5) rectangle (9.5,5.25);
\draw [pattern=north west lines, pattern color=blue, thick, opacity=0.3] (3.6,0.25) rectangle (8.5,4.25);
\draw [pattern=horizontal lines, pattern color=green, thick, opacity=0.3] (6,2) ellipse (2.2 and 1.5);
\draw [pattern=vertical lines, pattern color=red, thick, opacity=0.3] (3.5,2) ellipse (2.2 and 1.5);
\node [align=center] at (6.4, 3.8) {$G_{1+2}^\text{global}, \; T^a \in \mathfrak{g}_1 \cap \mathfrak{g}_2$};
\node [align=center] at (5, 4.8) {$G_1^\text{global}\times G_2^\text{global},\;$ $\tilde{T}^a \in \mathfrak{g}_1 \oplus \mathfrak{g}_2$};
\node [align=center] at (2.7,2) {$\mathcal{E}^\text{local}$ \\ $\hat{P}^a \in \mathfrak{e} - \mathfrak{h}$};
\node [align=center] at (4.75,2) {$K^a \in \mathfrak{e} \cap \mathfrak{h}$};
\node [align=center] at (7,2) {$\mathcal{H}^\text{local}$ \\ $\hat{T}^a \in \mathfrak{h} -\mathfrak{e}$};
}
\caption{The generators of a 2-site Moose model, with $K^a, \hat{T}^a \in T^a$ and $\hat{P}^a \in X^a$ to refer to the sigma model}\label{fig:HLS_Moose_notation}
\end{figure}

\begin{figure}[h!]
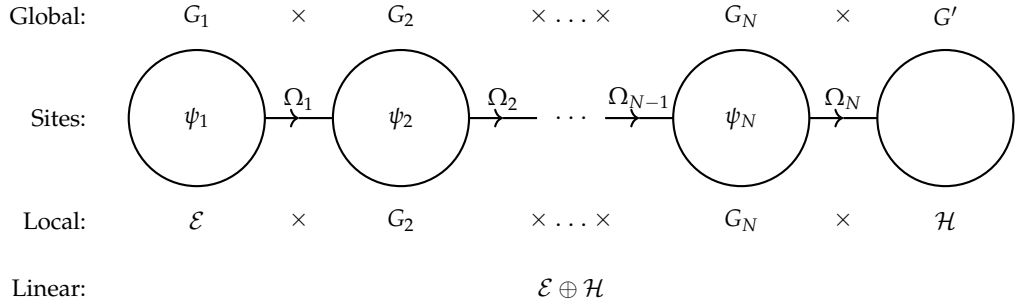

\centering\tikz[scale=0.9, every node/.style={transform shape}]{
\node [left] at (-1.5,1.5) {Global:};
\node [left] at (-1.5,0) {Sites:};
\node [left] at (-1.5,-1.5) {Local:};
\node [left] at (-1.5,-2.5) {Linear:};
\draw [thick] (0,0) circle [radius=1] node {$\psi_1$};
\node at (0,1.5) {$G_1$};
\node at (0,-1.5) {$\mathcal{E}$};
\node at (1.5,1.5) {$\times$};
\node at (1.5,-1.5) {$\times$};
\draw [thick, ->] (1,0) -- (1.5,0);
\draw [thick] (1.5,0) -- (2,0);
\node [above] at (1.5,0) {$\Omega_1$};
\draw [thick] (3,0) circle [radius=1] node {$\psi_2$};
\node at (3,1.5) {$G_2$};
\node at (3,-1.5) {$G_2$};
\node at (5.5,1.5) {$\times \;  .\; . \; . \; \times$};
\node at (5.5,-1.5) {$\times \; .\; .\; . \; \times$};
\node at (5.5,-2.5) {$\mathcal{E} \oplus \mathcal{H}$};
\draw [thick,->] (4,0) -- (4.5,0);
\draw [thick] (4.5,0) -- (5,0);
\node [above] at (4.5,0) {$\Omega_2$};
\node at (5.5,0) {$.\; .\; .$};
\draw [thick] (8,0) circle [radius=1] node {$\psi_N$};
\draw [thick,->] (6,0) -- (6.5,0);
\draw [thick] (6.5,0) -- (7,0);
\node [above] at (6.5,0) {$\Omega_{N-1}$};
\node at (8,1.5) {$G_N$};
\node at (8,-1.5) {$G_N$};
\node at (9.5,1.5) {$\times$};
\node at (9.5,-1.5) {$\times$};
\draw [thick] (11,0) circle [radius=1];
\draw [thick,->] (9,0) -- (9.5,0);
\draw [thick] (9.5,0) -- (10,0);
\node [above] at (9.5,0) {$\Omega_N$};
\node at (11,1.5) {$G'$};
\node at (11,-1.5) {$\mathcal{H}$};
}
\caption{A Moose diagram describing scenario (1), (2) \& (3), $\mathcal{L} = \mathcal{L}_\text{matter} + \mathcal{L}_\text{gauge} + \mathcal{L}_\text{goldstone}$. $\mathcal{E} \oplus \mathcal{H}$ is the diagonal subgroup of those two gauged groups. \label{fig:moose_model_3}}
\end{figure}

We remind that we discriminate between those gauge generators that are shared between sites and those that are not. This is summarised in the diagram from \cref{sec:HLS} with some additional structure we had not included at that stage, in \cref{fig:HLS_Moose_notation}. It can also be written as the Moose diagram in \cref{fig:moose_model_3}.

\begin{figure}[h!]
\centering\tikz[scale=0.9, every node/.style={transform shape}]{
\node [left] at (-1.5,1.5) {Global:};
\node [left] at (-1.5,0) {Sites:};
\node [left] at (-1.5,-1.5) {Local:};
\draw [thick] (0,0) circle [radius=1] node {$\psi_1$};
\node at (0,1.5) {$G_1$};
\node at (0,-1.5) {$\mathcal{E}$};
\node at (1.5,1.5) {$\times$};
\node at (1.5,-1.5) {$\times$};
\draw [thick, ->] (1,0) -- (1.5,0);
\draw [thick] (1.5,0) -- (2,0);
\node [above] at (1.5,0) {$\Omega_1$};
\draw [thick] (3,0) circle [radius=1] node {$\psi_2$};
\node at (3,1.5) {$G_2$};
\node at (3,-1.5) {$\mathcal{H}$};
}
\caption{A two-site model \label{fig:moose_model_two_site_b}}
\end{figure}

The key is that we can produce a hidden local symmetry with the accompanying vector resonances from a sigma model (by taking $g_H \neq \infty$), just as easily as the reverse process. Applying our generator notation, we have the (1) + (2) + (3) case\footnote{Note that the indices are not meant to imply we have the same numbers of each generator. They should simply be understood by context as a sum over all generators.}
\begin{align}
D_\mu \Omega = \partial_\mu \Omega - i g_E \left(\hat{A}_{E,\mu}^a \hat{P}^a + A_{E,\mu}^a K^a\right) \Omega + ig_H \Omega\left( \hat{A}_{H,\mu}^a \hat{T}^a + A_{H,\mu}^a K^a\right)
\end{align}
Applying the HLS definition of $\Omega$ as a product of broken fields and gauged fields 
\begin{align}
\Omega(x) = \me^{\frac{i}{f}\pi_U^{\hat{a}} X^{\hat{a}}} \me^{\frac{i}{f_\Xi}\pi_\Xi^{a} T^a} \equiv U(x) \Xi(x)
\end{align}
to the link field Lagrangian \cref{eq:link_general_lagrangian} we have leading-order gauge field mass-type interactions
\begin{align}
\mathcal{L}_2 \supset & \frac{f_\Xi^2 g_E^2}{4}\left( \frac{f^2}{f_\Xi^2} \hat{A}_{E,\mu}^a \hat{A}_E^{a,\mu} + A_{E,\mu}^a A_E^{a,\mu}\right) +\frac{f_\Xi^2 g_H^2}{4}\left( \hat{A}_{H,\mu}^a \hat{A}_H^{a,\mu} + A_{H,\mu}^a A_H^{a,\mu}  \right) \nonumber \\ 
& - \frac{f_\Xi^2 g_H g_E}{2} A_{H,\mu}^a A_E^{a,\mu}
\end{align}
Terms such as $\text{Tr}[\hat{A}_{E,\mu}^a \hat{P}^a A_{H,\mu}^b K^b]$ go to zero, as we (choose to) use the regular representation
\begin{align}
\text{Tr}[\tilde{T}^a\tilde{T}^b] \propto \delta^{ab}
\end{align}
while only $A_{H,\mu}$ and $A_{E,\mu}$ share a basis. Diagonalising the couplings leads to masses
\begin{align}
m_E^2 = \frac{f^2 g_E^2}{2}, && m_H^2 = \frac{f_\Xi^2 g_H^2}{2}, && m_K^2 = \frac{f_\Xi^2(g_H^2 + g_E^2)}{2}
\end{align}
where the third mass is of the mixed state
\begin{align}
A_{K,\mu}^a = \cos\theta A_{H,\mu}^a - \sin\theta A_{E,\mu}^a, \qquad \sin\theta = \frac{g_E}{\sqrt{g_H^2 + g_E^2}}
\end{align}
and we have an orthogonal, massless state
\begin{align}
A_{0,\mu}^a = \sin\theta A_{H,\mu}^a + \cos\theta A_{E,\mu}^a
\end{align}
The massive gauge bosons at both sites have eaten $N_{\hat{A}_E} + N_{\hat{A}_H} + N_{A_K}$ link field Goldstone bosons, which can then be gauged away in the unitary gauge (\cref{eq:link_unitary_gauge}), while $N_{A_K}$ true Goldstones are still in the spectrum.

\begin{figure}[H]
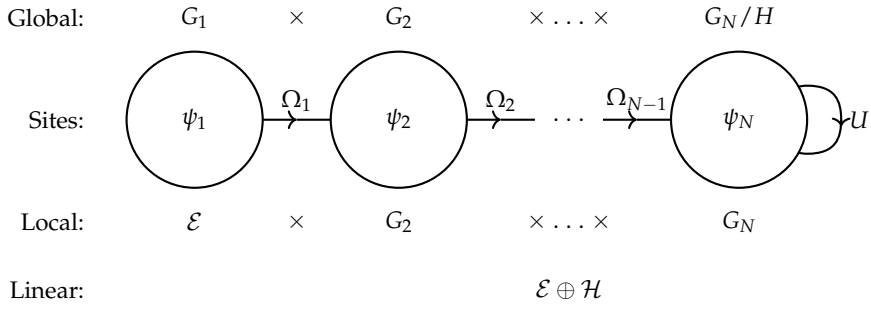

\centering\tikz[scale=0.9, every node/.style={transform shape}]{
\node [left] at (-1.5,1.5) {Global:};
\node [left] at (-1.5,0) {Sites:};
\node [left] at (-1.5,-1.5) {Local:};
\node [left] at (-1.5,-2.5) {Linear:};
\draw [thick] (0,0) circle [radius=1] node {$\psi_1$};
\node at (0,1.5) {$G_1$};
\node at (0,-1.5) {$\mathcal{E}$};
\node at (1.5,1.5) {$\times$};
\node at (1.5,-1.5) {$\times$};
\draw [thick, ->] (1,0) -- (1.5,0);
\draw [thick] (1.5,0) -- (2,0);
\node [above] at (1.5,0) {$\Omega_1$};
\draw [thick] (3,0) circle [radius=1] node {$\psi_2$};
\node at (3,1.5) {$G_2$};
\node at (3,-1.5) {$G_2$};
\node at (5.5,1.5) {$\times \;  .\; . \; . \; \times$};
\node at (5.5,-1.5) {$\times \; .\; .\; . \; \times$};
\node at (5.5,-2.5) {$\mathcal{E} \oplus \mathcal{H}$};
\draw [thick,->] (4,0) -- (4.5,0);
\draw [thick] (4.5,0) -- (5,0);
\node [above] at (4.5,0) {$\Omega_2$};
\node at (5.5,0) {$.\; .\; .$};
\draw [thick,->] plot [smooth, tension=1] coordinates {(8,0) (9,0.5) (9.5,0) (9,-0.5) (8,0)};
\draw [thick, ->] (9.5,0) -- (9.5,-0.1);
\node [right] at (9.5,0) {$U$};
\draw [thick, fill=white] (8,0) circle [radius=1] node {$\psi_N$};
\draw [thick,->] (6,0) -- (6.5,0);
\draw [thick] (6.5,0) -- (7,0);
\node [above] at (6.5,0) {$\Omega_{N-1}$};
\node at (8,1.5) {$G_N/H$};
\node at (8,-1.5) {$G_N$};

}
\caption{A Moose diagram describing scenario (1)+(2)+(3) with $g_H \rightarrow \infty$. \label{fig:moose_model_4}}
\end{figure}

This chain of link fields and gaugings can be extended to an N-site Moose model, as in \cref{fig:moose_model_4}. However, recall that we only want the physical Goldstone bosons corresponding to an overall $G/H$ breaking. Therefore, we would like to gauge away all but those corresponding to a single multiplet of to-be Higgs fields.

The final site can provide the set of physical Goldstones, by taking the limit described in \cref{fig:HLS_limit}. That is, $\Omega_N \xrightarrow{g_N \rightarrow \infty} \phi$. To summarise, our sigma-model fields transform as
\begin{align}
& \Omega_1 \rightarrow g_1 \Omega_1 h_2^{-1}(x) \nonumber\\
& \Omega_i \rightarrow h_i(x) \Omega_i h_{i+1}^{-1}(x) \\
& \Omega_N = \phi \rightarrow h_N(x) \phi h^{-1} \nonumber
\end{align}
We can tie a bow on this section by connecting with our minimal requirement $\Phi$ of \cref{sec:Landscape_4dMCHM} by defining
\begin{align}
\Phi \equiv \Omega_1\Omega_2 ...\Omega_{N-1} \phi
\end{align}
which transforms as we require
\begin{align}
\Phi \rightarrow g_1 \Phi h^{-1}
\end{align}
thus "hiding" our stack of local symmetries as a sigma model. Note that there are $N$ gauged sites, but we have only chosen $N-1$ unitary gauges. Therefore this definition contains an overall gauge freedom that we must fix. There are many to choose from, and I encourage you to see \cite{Stangl:2018kty}. For the remainder of this work, unless otherwise noted, we assume the "site-N holographic gauge"
\begin{align}
&\Omega_k = \mathbb{1} \, , \; 0 < k <  N \,   \\
&\Omega_N = \Phi = \me^{i\frac{\sqrt{2}}{f} \pi^a(x) X^a} \, ,
\end{align}
which does mix the $G/H$ Goldstone bosons with the heavy gauge bosons in $\mathcal{H}$. Again for simplicity, for the remainder of this work, we will take $g_H \rightarrow \infty$ to decouple the last site gauge group $H$, and thus render the holographic NGBs unmixed. However, all other gauge sites are still intact and acquire mass at their respective sigma model breaking scale $f_i$.

Moose models can thus allow composite fermions to interact via massive composite bosons, which we require for a radiatively generated Higgs potential. So far, we have only considered the group theoretic qualities of the model. To see how the composite matter behaves in nature, we need to examine the representations of the matter fields.

\vspace{2em}

\section{General Composite Matter}

\paragraph{Our goal}
\parbox{0.8\textwidth}{To describe an effective Lagrangian for any matter in the Minimal Composite Higgs model.}
\vspace*{2em}

Just as the transformation property of the NGBs determined their interactions and representation, so we consider how bosonic and fermionic resonances can transform under either a non-linear $g \in G$, or a linear $h \in H$. We know we will be coupling our composite fields to the Higgs and SM gauge bosons, so must consider how they are represented in these interactions. To repeat: resonances must be embedded in $H$ and $G$ irreps, and decompose to $SU(2)_L \times SU(2)_R$ irreps. In the minimal case outlined in this section, $H=SO(4)$ and $\mathcal{H}_\text{EW} = SU(2)_L \times SU(2)_R$ are isomorphic, so we will simply consider tranformations under $H = SU(2)_L \times SU(2)_R$ and the gauge transformations will follow. In the next-to-minimal model considered in \cref{sec:NMCHM}, we will discuss the subtlety of the bosonic resonances transforming under $\mathcal{H}_\text{EW}$ as a subgroup of $H$.

Following the discussion in \cite{marzocca2012}, we will deal with bosonic and fermionic resonances in different ways, using the HLS approach for the former and the standard CCWZ approach for the latter. Since bosonic resonances are included by treating them as the massive gauge bosons of some HLS (which is equivalent to other popular meson realisations \cite{ecker1989chiral}), we thus expect massive bosonic resonances to come in adjoint multiplets of $G$. To see how they couple, let's be clear about how they transform under $g \in G$. We will use some machinery that we developed in \cref{sec:CCWZ}.
Our $G_i$-group gauge fields transform under $H$ as
\begin{align}
A_{i,\mu} & \rightarrow h A_{i,\mu} h^{-1}  + \frac{i}{g_i} h \partial_\mu h^{-1} \label{eq:gauge_transformation_rule}
\end{align}
We can decompose each gauge field in a straightforward way
\begin{align}
A_{i,\mu} = A_{i,\mu}^a \tilde{T}^a = A_{i,\mu}^a T^a + A_{i,\mu}^b X^b \equiv \rho_{L,\mu}^a T^a_L + \rho_{R,\mu}^b T^b_R + a_{\mu}^a X^a
\end{align}

Note that we can extract a particular component with the use of the regular representation identity $\text{Tr}[\tilde{T}^a \tilde{T}^b] \propto \delta^{ab}$
\begin{align}
\rho_{L,\mu}^a = \text{Tr}[A_\mu T_L^a] && \rho_{R,\mu}^a = \text{Tr}[A_\mu T_R^a] && a_\mu^a = \text{Tr}[A_\mu X^a]
\end{align}
Applying the decomposition and trace extraction to the transformation rule \cref{eq:gauge_transformation_rule}, we have
\begin{align}
\rho_{L,\mu} = \rho_{L,\mu}^a T^a \rightarrow & \text{Tr}\left[\left(  hA_\mu h^{-1}  + \frac{i}{g_i} h\partial_\mu h^{-1} \right)h T^a_L h^{-1} \right]h T^a h^{-1} \nonumber \\
&= \text{Tr}\left[ A_\mu T^a_L  + \frac{i}{g_i} \partial_\mu h^{-1} h T^a h^{-1} \right]h T^a h^{-1} \nonumber \\
&= h \rho_L h^{-1} + \frac{i}{g_i} \text{Tr}\left[ \partial_\mu h^{-1} h T^a_L \right]h T^a h^{-1} 
\end{align}
Where we have used the fact that generators transform in the adjoint. Now note that $\partial_\mu h^{-1} h \in H$, so we can decompose it as $[\partial_\mu h^{-1} h]^b T^b = [\partial_\mu h^{-1} h]_L^b T_L^b + [\partial_\mu h^{-1} h]_R^b T_R^b $, and treat the terms $[...]_{L}, [...]_{R}$ as vectors that can be projected
\begin{align}
\frac{i}{g_i}\text{Tr}\left[ \partial_\mu h^{-1} h T^a_L \right]h T^a h^{-1}  =\frac{i}{g_i} h (\partial_\mu h^{-1} h)_L^b \delta^{ab} T^a h^{-1} =  \frac{i}{g_i}  (h \partial_\mu h^{-1})_L
\end{align}
since $h = \me^{iu^a T^a} = \me^{i(u^a_L T^a_L + u^a_R T^a_R)}$ and $[T_L, T_R] = 0$ due to $SO(4)$ having $SU(2)_L$ and $SU(2)_R$ as normal subgroups. However, the broken components of the (axial) vector resonances transforms homogeneously since $\text{Tr}[T^b_{L,R} X^a] = 0$. 

To summarise, the N-site $SO(5)/SO(4)$ model has vector resonances $A^A_{i\mu} = \{\rho^a_{i_L, \mu}, \rho^a_{i_R,\mu}, a_i^{\hat{a}}\}$ that transform under $SU(2)_L \times SU(2)_R$ as 
\begin{align}
A_\mu \begin{cases}
\rho_{L,\mu} \rightarrow h \rho_{L,\mu} h^{-1} + \frac{i}{g_{\rho_L}}[h \partial_\mu h^{-1}]_L  \qquad \; \implies   \qquad  \rho_{L,\mu} \sim (\textbf{3},\textbf{1})\\
\rho_{R,\mu} \rightarrow h \rho_{R,\mu} h^{-1}  + \frac{i}{g_{\rho_R}}[h \partial_\mu h^{-1}]_R   \qquad \implies  \qquad  \rho_{R,\mu} \sim (\textbf{1},\textbf{3})\\
a_\mu \rightarrow h a_\mu h^{-1}  \qquad \qquad \qquad  \qquad \qquad \; \; \; \implies \qquad   a_\mu \sim (\textbf{2}, \textbf{2})
\end{cases}
\end{align}

Given the transformation properties of the bosons, we can write the most general two-derivative Lagrangian for the vector resonances
\begin{align}
\mathcal{L}_{\rho} &= \sum\limits_{i=1}^{N_{\rho_L}} \left( -\frac{1}{4} \text{Tr}\left[\rho_{L,\mu\nu}^i \rho^{i,\mu\nu}_L \right] + \frac{f^2_{\rho_L^i}}{2}\text{Tr}\left[\left(g_{\rho_L^i} \rho_{L,\mu}^i - E_{L,\mu} \right)^2\right] \right. \nonumber \\
& \left. + \sum\limits_{j<i} \frac{f^2_{\text{mix}_{ij}}}{2}\text{Tr}\left[\left(g_{\rho_L^i} \rho_{L,\mu}^i - g_{\rho_L^i}\rho_{L,\mu}^j\right)^2\right]\right) + \{L \rightarrow R\}\; ,
\end{align}
and for the axial resonances
\begin{align}
\mathcal{L}_{a}&=  \sum\limits_{i=1}^{N_a} \left(-\frac{1}{4}\text{Tr}\left[a_{\mu\nu}^i a^{i,\mu\nu}\right] + \frac{f_{a^i}}{2\Delta_i^2}\text{Tr}\left[\left( g_{a^i} a^i_\mu - \Delta_i d_\mu\right)^2\right] \right)\; ,
\end{align}
where the field strength tensors are given by
\begin{align}
\rho_{\mu\nu} &= \partial_\mu \rho_\nu - \partial_\nu \rho_\mu -i g_\rho [\rho_\mu, \rho_\nu]\\
a_{\mu\nu} &= (\partial_\mu - iE_\mu)a_\nu - (\partial_\nu - iE_\nu ) a_\mu
\end{align}
recalling the definition of $E_\mu$ in \cref{eq:mc_definition}. The parameters $f_{\rho_L^i}, f_{\rho_R^i}, f_{a^i}, \Delta_i$ can be directly related to the link field decay constants $f_i$, once we decide on the gauging structure. $g_{a^i}$ and $g_{\rho_{L,R}}^i$, the composite gauge couplings, are free parameters. We see from \cref{eq:mc_transformation} that $E_\mu$ is the correct term to include in trace of $\rho$ fields, to cancel the transformed field's $h \partial_\mu h^\dagger$ term.


Dealing with composite fermions is somewhat easier, using the CCWZ ideas developed previously. Again, we ask how the resonances transform under our Moose model symmetries.  Recall from \cref{eq:partial_compositeness_form} that for tuning reasons, we need a linear coupling between each chirality of each elementary field and some function of the composite sector. A direct coupling of elementary fermion $\psi_{L/R}$ and composite fermion $\Psi_{R/L}$ will, however, break the global symmetries we worked hard to craft in \cref{sec:symmetric_spaces}. We can use the Moose formalism to couple fermions between sites via the link fields such that each site is insulated from neighbouring transformations. However, the global product group is hidden in the link fields' transformation properties, and the fermions appear to couple linearly, once the NGBs are eaten in the holographic gauge.

\begin{figure}
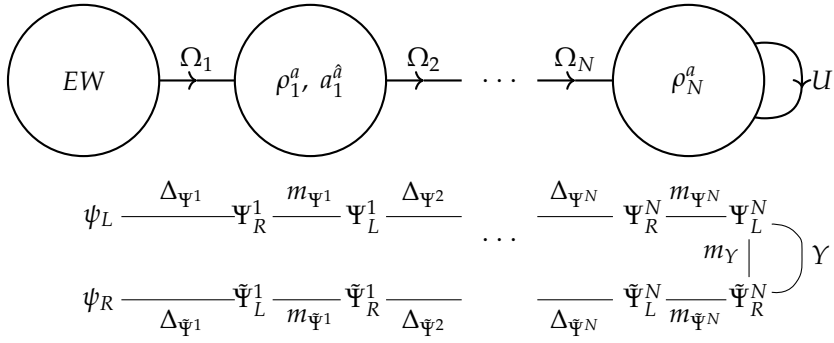

\centering\tikz[scale=1, every node/.style={transform shape}]{
\draw [thick] (0,0.3) circle [radius=1] node {$EW$};
\node at (0.2,-1.5) {$\psi_L$};
\draw (0.5,-1.5) --(2,-1.5);
\node [above] at (1.5,0.3) {$\Omega_1$};
\node [above] at (1.3,-1.5) {$\Delta_{\Psi^1}$};
\node [below] at (1.3,-2.6) {$\Delta_{\tilde{\Psi}^1}$};
\draw (0.5,-2.6) --(2,-2.6);
\node at (0.2,-2.6) {$\psi_R$};
\node at (2.2,-1.5) {$\Psi^1_R$};
\node [above] at (3,-1.5) {$m_{\Psi^1}$};
\node at (3.7,-1.5) {$\Psi^1_L$};
\draw (2.5,-1.5) --(3.4,-1.5);
\draw (2.5,-2.6) --(3.4,-2.6);
\node at (2.2,-2.6) {$\tilde{\Psi}^1_L$};
\node [below] at (3,-2.6) {$m_{\tilde{\Psi}^1}$};
\node at (3.7,-2.6) {$\tilde{\Psi}^1_R$};
\draw (4,-1.5) --(5,-1.5);
\draw (4,-2.6) --(5,-2.6);
\draw [thick, ->] (1,0.3) -- (1.5,0.3);
\draw [thick] (1.5,0.3) -- (2,0.3);
\draw [align=center,thick] (3,0.3) circle [radius=1] node {$\rho_1^a,\; a_1^{\hat{a}}$};
\draw [thick,->] (4,0.3) -- (4.5,0.3);
\draw [thick] (4.5,0.3) -- (5,0.3);
\node [above] at (4.5,0.3) {$\Omega_2$};
\node [above] at (4.5,-1.5) {$\Delta_{\Psi^2}$};
\node [below] at (4.5,-2.6) {$\Delta_{\tilde{\Psi}^2}$};
\node at (5.5,0.3) {$.\; .\; .$};
\node at (5.5,-1.8) {$.\; .\; .$};
\draw [thick,->] plot [smooth, tension=1] coordinates {(8,0.3) (9,0.8) (9.5,0.3) (9,-0.2) (8,0.3)};
\draw [thick, ->] (9.5,0.3) -- (9.5,0.2);
\node [right] at (9.5,0.3) {$U$};
\draw [thick, fill=white] (8,0.3) circle [radius=1] node {$\rho_N^a$};
\draw [thick,->] (6,0.3) -- (6.5,0.3);
\draw [thick] (6.5,0.3) -- (7,0.3);
\node [above] at (6.5,0.3) {$\Omega_{N}$};
\draw (6,-1.5) --(7,-1.5);
\draw (6,-2.6) --(7,-2.6);
\node [above] at (6.5,-1.5) {$\Delta_{\Psi^N}$};
\node [below] at (6.5,-2.6) {$\Delta_{\tilde{\Psi}^N}$};
\node at (7.4,-1.5) {$\Psi^N_R$};
\node at (7.4,-2.6) {$\tilde{\Psi}^N_L$};
\draw (7.7,-1.5) --(8.5,-1.5);
\draw (7.7,-2.6) --(8.5,-2.6);
\node [above] at (8.1,-1.5) {$m_{\Psi^N}$};
\node [below] at (8.1,-2.6) {$m_{\tilde{\Psi}^N}$};
\node at (8.8,-1.5) {$\Psi^N_L$};
\draw (8.8, -1.8) -- (8.8,-2.3);
\node [left] at (8.8, -2) {$m_Y$};
\node at (8.8,-2.6) {$\tilde{\Psi}^N_R$};
\draw plot [smooth, tension=1] coordinates {(9.1,-1.6) (9.4, -1.7) (9.5,-2) (9.4, -2.4) (9.1,-2.5)};
\node [right] at (9.5,-2) {$Y$};
}
\caption{Fermion couplings in the $N$-site Moose (with HLS on site $N$)}\label{fig:fermion_moose_model}
\end{figure}

Prior to electroweak gauging, we would like to couple our composite sector to the elementary (i.e. SM) sector, while non-linearly preserving the $G$-invariance. We also need at least one coupling \textit{between} the two sets of links and the $G/H$ pNGB to induce mass for the SM quarks. This structure can be visualised in \cref{fig:fermion_moose_model} by a chain of Yukawa link-field couplings between elementary fermions $\psi_L, \psi_R$ and composite partners $\Psi^i$, $\tilde{\Psi}^i$, respectively, with a Yukawa connection between the chains on the N$^\text{th}$ site. Coupling constants are given on the line linking each fermion.

\vspace{2em}
\begin{tcolorbox}[colback = black!2!white]
\paragraph{Short-circuit}\label{sec:short_circuit}
I find the analogy of a circuit to be useful in choosing this structure. Consider $\psi_L$ and $\psi_R$ as terminals of a circuit that we wish to close, in order that we can rotate to an effective mass term for the elementary field $\psi$. The mass term should depend on the Higgs potential, which is parameterised by the coupling constants, and it can be thought of as the electric potential (we are lucky with the word here) across the last site. Connecting our sites in this way allows us to use all of the parameters to ensure the potential is finite, a necessity we will show shortly. Including any couplings before the N$^\text{th}$ site, such as $m \bar{\Psi}_R^1 \tilde{\Psi}_L^1$, effectively \textit{short-circuits} this potential calculation, as a mass term for $\psi$ could then be rotated to that did not depend on the Higgs potential. 
\end{tcolorbox}

Following the diagram \cref{fig:fermion_moose_model} then, we see that we must embed the elementary fields in $G$ multiplets, such that they couple with composite sector, as we did in the $G=SO(5)$ case with \cref{eq:SM_spurions}. The Lagrangian describing this situation can be given by
\begin{align}
\mathcal{L}_{e-c} &= \bar{\psi}_L i \slashed{D} \psi_L + \bar{\psi}_R i \slashed{D} \psi_R +  \Delta_{\Psi^1} \bar{\psi}_L\Omega_1 \Psi_R^1 +  \Delta_{\tilde{\Psi}^1} \bar{\psi}_R\Omega_1 \tilde{\Psi}_L^1 \nonumber\\
& + \sum\limits_{k=1}^{N-1} \left( \bar{\Psi}^k \left(i \slashed{D} - m_{\Psi^k} \right)\Psi^k + \bar{\tilde{\Psi}}^k \left( i\slashed{D} - m_{\tilde{\Psi}^k} \right) \tilde{\Psi}^k + \Delta_{\Psi^k} \bar{\Psi}^k \Omega_k \Psi_R^{k+1} + \Delta_{\tilde{\Psi}^k} \bar{\tilde{\Psi}}_R^k \Omega_k \tilde{\Psi}_L^{k+1} \right) \nonumber\\
& - m_{\Psi\tilde{\Psi}} \bar{\Psi}_L^N U P^{(rs)} U^{\dagger} \tilde{\Psi}_R^N + \text{h.c.} \label{eq:general_fermion_lagrangian}
\end{align}
The last term, a mixing with a projection from one representation to the other, can be expanded as a Yukawa-like term plus a mass-like term $Y \bar{\Psi}_L^N \Phi \Phi^\dagger \tilde{\Psi}_R ^N- m_Y \bar{\Psi}_L^N \tilde{\Psi}_R^N$. Recall that we define $\Phi = U \Phi_0$, where $\Phi_0 = (\vec{0},1)^T$ can be used to build projections from one vacuum representation to another. We will investigate the most common $SO(5)$ irreps in \cref{sec:LCHM}. Note that while \cref{eq:general_fermion_lagrangian} contains the most general set of composite fermions, the choice of couplings is far from generic. Specifically, we include only neighbouring site interactions\footnote{To recall why, see the aside \ref{sec:locality} "Locality of the Moose".}, and a single link from $\Psi^N_L$ to $\tilde{\Psi}^N_R$. The latter is motivated by minimality. This is best understood by inspecting the Lagrangian in its mass basis, from which emerges partial compositeness.

\subsection{Partial Compositeness}\label{sec:partial_compositeness}

Consider a two-site model (that is, one elementary site, one composite site), with the fermion interaction terms 
\begin{align}
\mathcal{L} &\supset \Delta_{\Psi} \bar{\psi}_L \Omega \Psi_R + \Delta_{\tilde{\Psi}} \bar{\psi}_R \Omega \tilde{\Psi}_L  - m_\Psi \bar{\Psi}_L \Psi_R -  m_{\tilde{\Psi}} \bar{\tilde{\Psi}}_L \tilde{\Psi}_R  +\mathcal{L}_\text{yuk}  + \text{h.c.}
\end{align}
where $\mathcal{L}_\text{yuk} = - m_{\Psi\tilde{\Psi}} \bar{\Psi}_L U P^{(rs)} U^{\dagger} \tilde{\Psi}_R$.
After choosing the unitary gauge $\Omega \rightarrow \mathbb{1}$, we have an expression
\begin{align}
\mathcal{L} &\supset \Delta_{\Psi} \bar{\psi}_L \Psi_R + \Delta_{\tilde{\Psi}} \bar{\psi}_R \tilde{\Psi}_L  - m_\Psi \bar{\Psi}_L \Psi_R -  m_{\tilde{\Psi}} \bar{\tilde{\Psi}}_L \tilde{\Psi}_R +\mathcal{L}_\text{yuk}  + \text{h.c.}
\end{align}
which we can diagonalise in the linear couplings. Rotating in $(\psi_L, \psi_R, \Psi_L, \Psi_R, \tilde{\Psi}_L, \tilde{\Psi}_R)$-space, we have masses 
\begin{align}
m_{\psi'} = 0 && m_{\Psi'} = \sqrt{m_\Psi^2 + \Delta_\Psi^2} && m_{\tilde{\Psi}'} = \sqrt{m_{\tilde{\Psi}}^2 + \Delta_{\tilde{\Psi}}^2}
\end{align}
for diagonalised fields
\begin{align}
\psi_L' &= \frac{m_\Psi}{m_{\Psi'}} \psi_L + \frac{\Delta_\Psi}{m_{\Psi'}} \Psi_L & \psi_R' &= \frac{m_{\tilde{\Psi}}}{m_{\tilde{\Psi}'}} \psi_R + \frac{\Delta_{\tilde{\Psi}}}{m_{\tilde{\Psi}'}} \tilde{\Psi}_L  \nonumber \\ 
\Psi_L' &= \frac{m_\Psi}{m_{\Psi'}} \Psi_L - \frac{\Delta_\Psi}{m_{\Psi'}} \psi_L & \tilde{\Psi}_R' &= \frac{m_{\tilde{\Psi}}}{m_{\tilde{\Psi}'}} \tilde{\Psi}_R + \frac{\Delta_{\tilde{\Psi}}}{m_{\tilde{\Psi}'}} \psi_R\\
\Psi_R' &= \Psi_R & \tilde{\Psi}'_R &= \tilde{\Psi}_R  \nonumber 
\end{align}
Clearly, this is not the end of the story, as we have not generated a mass for our SM-like elementary fields. Inserting these expressions into the composite Yukawa term, we have the elementary-pNGB interaction
\begin{align}
\mathcal{L} \supset  - \frac{\Delta_\Psi \Delta_{\tilde{\Psi}}}{m_{\Psi'} m_{\tilde{\Psi}'}}\bar{\psi}'_L\left( Y \Phi \Phi^\dagger - m_Y \right)\psi_R' \equiv \sin\theta_\Psi \sin\theta_{\tilde{\Psi}} \left( Y \Phi \Phi^\dagger - m_Y \right)\psi_R' \label{eq:diagonal_yukawa}
\end{align}

The ratio $\Delta_\Psi / m_\Psi = \tan\theta_\Psi$ is called the compositeness of the Standard Model field associated with $\psi_L$. Clearly, taking $\Delta_{\Psi, \tilde{\Psi}} \rightarrow 0$ with $m_{\Psi, \tilde{\Psi}}$ held fixed will give a massless elementary field, as will taking $m_{\Psi, \tilde{\Psi}} \rightarrow \infty$, with  $\Delta_{\Psi, \tilde{\Psi}}$ held fixed. The physical mass terms of the fermion resonances are more complicated than $m_{\Psi'} = \sqrt{m_\Psi^2 + \Delta_\Psi^2}$, as there are contributions from the $\mathcal{L}_\text{yuk}$ term. This will depend explicitly on the choice of representation, which we will derive in \cref{sec:FermionSector}. Note that the massive composite gauge bosons also experience partial compositeness. This is easiest to see in terms of the low energy, decomposed fields.

\subsection{Low Energy Lagrangian}
\label{sec:low_energy_lagrangian}

\paragraph{Our goal}
\parbox{0.8\textwidth}{To connect the high-energy composite model to our low-energy SM description of the Higgs potential.}
\vspace*{2em}


In the Moose model of \cref{sec:mooses}, without explicit breaking of symmetries, the global symmetry $G_{1+2+...+N}$ is non-linearly preserved, as is the linear subgroup $H$, leading to possible Higgs potential terms to be radiatively generated. Thus we break the linear subgroup $SO(4)$ explicitly in two ways
\begin{enumerate}
\item By interacting $SO(5)$-invariant composite fermions with elementary fields \textit{not} in complete $SO(5)$ multiplets, leading to a vev for the NGBs and shifting $SO(4)_\text{tree} \rightarrow SO(4)_\text{1-loop}$, and
\item By gauging an $SU(2)_L \times U(1)_Y$ subgroup of $SO(4)_\text{tree}$, giving the gauge freedom to remove unphysical NGBs associated with the misaligned $SO(4)_\text{tree}$
\end{enumerate}

We have already examined the contribution of the second process to low-energy physics, in \cref{sec:MCHM_Gauge} and \cref{sec:MCHM_Potential}, as it is independent of the fermion sector. Let us examine now the first contribution.  Consider a generic low energy (LE) Lagrangian of a generation of SM quark in momentum space
\begin{align}
\mathcal{L}_{\text{LE}} &= (Z_q + \Pi_{u_L}(p^2, h)) \bar{u}_L\slashed{p} u_L + (Z_u + \Pi_{u_R}(p^2, h))\bar{u}_R \slashed{p} u_R + M_u(p^2, h) \bar{u}_L u_R \nonumber \\
& +  (Z_q + \Pi_{d_L}(p^2, h)) \bar{d}_L\slashed{p} d_L + (Z_d + \Pi_{d_R}(p^2, h))\bar{d}_R \slashed{p} d_R + M_d(p^2, h) \bar{d}_L d_R \label{eq:SM_sources}
\end{align}
where the form factors $\Pi, M$ contain corrections to the bare SM operators $Z_\psi$. Note the inclusion of both components of the doublet $q_L = (u_L, d_L)$. In the following sections, we will consider the lighter component (the bottom quark, in the third generation) only when next-to-leading order effects may be relevant. For example, we examine the contribution of a composite tau lepton to the Higgs potential. With a mass comparable to the tau, the bottom is included in that case. Generally however (for example in the following discussion), only one component will be considered for brevity, except where it is required to fill out the $SU(2)_\text{weak}$-invariant doublet. Moreover, we will specifically consider the third generation, i.e. $u_{L,R} \rightarrow t_{L,R}$.

To connect with the low energy physics, we first re-examine the high energy Lagrangian in \cref{eq:general_fermion_lagrangian}, filling out the elementary electroweak multiplets $q_L, t_R$ as full $SO(5)$ multiplets $q_L^5, q_R^5$ to ensure the preservation of the global group
\begin{align}
\mathcal{L}_{e-c} &= \bar{q}^5_L i \slashed{D} q^5_L + \bar{q}^5_R i \slashed{D} q_R^5 +  \Delta_{\Psi^1} \bar{q}^5_L\Omega_1 \Psi_R^1 +  \Delta_{\tilde{\Psi}^1} \bar{q}^5_R\Omega_1 \tilde{\Psi}_L^1 \nonumber\\
& + \sum\limits_{k=1}^{N-1} \left( \bar{\Psi}^k \left(i \slashed{D} - m_{\Psi^k} \right)\Psi^k + \bar{\tilde{\Psi}}^k \left( i\slashed{D} - m_{\tilde{\Psi}^k} \right) \tilde{\Psi}^k + \Delta_{\Psi^k} \bar{\Psi}^k \Omega_k \Psi_R^{k+1} + \Delta_{\tilde{\Psi}^k} \bar{\tilde{\Psi}}_R^k \Omega_k \tilde{\Psi}_L^{k+1} \right) \nonumber\\
& - m_Y \bar{\Psi}_L^N \tilde{\Psi}_R^N - Y \bar{\Psi}_L^N \Phi^\dagger \Phi \tilde{\Psi}_R^N + \text{h.c.} \label{eq:general_fermion_lagrangian_expanded}
\end{align}
Note that we have expanded the link-field term on the last site to both a linear term and a Yukawa term. The linear term will only be present for certain combinations of representation, and this is explored in \cref{sec:FermionSector}. To derive the form factors of \cref{eq:SM_sources} we should integrate out the heavy resonances $\Psi_i, \tilde{\Psi}_i$. We should be careful, however, since the vacuum of the $SO(5)$-invariant Lagrangian (with unphysical fields supplementing SM fields in the $SO(5)$ multiplets) is not the same as the one with only the SM fields included, namely $\langle h \rangle = 0$ and $\langle h \rangle = v$ respectively. So we will approach the integration in two phases, first by applying the Euler-Lagrange equations in the $SO(4)$-invariant vacuum, and then by throwing away unphysical elementary fields, thereby breaking the $SO(4)$-invariance. 

In the $SO(4)$ vacuum, our $SO(5)$ irreps $r_G$ decompose as $SO(4)$ irreps $r_H$. To see why, let us try ``dressing" our $r_G$-plets with the Goldstone matrix so that they transform linearly under $SO(4)$ We define this as
\begin{align}
\left(\Psi^{r_G}_i\right)_a = U_{ab} \left(\Psi^{r_H}_i\right)_b
\end{align}
That is, the $b^\text{th}$ $H$-component of the $i^\text{th}$ composite fermion can contract with dim$(r_H)$ of the indices of the Goldstone matrix, in order to leave a set of $G$-invariant components $a$. To illustrate this, simply observe:
\begin{align}
\Psi^{r_G}_i = U \Psi^{r_H}_i \rightarrow g U h(g, x)^{-1} h(g,x) \Psi^{r_H}_i = g \Psi^{r_G}_i
\end{align}
which only requires the fermion $\Psi^{r_H}$ to transform correctly under the representation $D(g,x) = h(g,x)$ as established in \cref{sec:CCWZ}. For example, the simplest fermion representation would be a $\textbf{5}$ of $SO(5)$, which decomposes to $\textbf{5} \sim \textbf{4} \oplus \textbf{1} \sim (q^\textbf{4}_1, q^\textbf{4}_2, q^\textbf{4}_3, q^\textbf{4}_4, q^\textbf{1})$:
\begin{align}
\left(q^{\bm{5}}_i\right)_a = U_{ab}\left(q^\textbf{4}_i\right)_b, \qquad \left(t^{\bm{5}}_j\right)_a = U_{a5} t^\textbf{1}_j
\end{align}
where the notation is fiddly but obvious - $q^\textbf{4}_1$ is the first component of the quark fourplet. Note that in embedding the singlet, we give it $SO(5)$ group indices, where before they were redundant, as it transformed trivially.

\vspace{2em}
\begin{tcolorbox}[colback = black!2!white]
\paragraph{Dictionary of fermion dressing}
Consider a fermion $Q^5$ transforming linearly in the fundamental representation of $SO(5)$: $Q^5 \rightarrow gQ^5$. Just as a gauge field can be used to promote this to a local transformation, a Goldstone matrix $U$ is used to promote this to a non-linear transformation. To do this, we "dress" the multiplet with the ansatz
\begin{align}
Q^5 = U^T Q = U^{-1} Q \label{eq:fermion_dressing}
\end{align}
Then the multiplet transforms as
\begin{align}
Q^5 \rightarrow (g U h(g,U)^{-1})^{-1} g Q = h(g,U) U^{-1} Q = h(g,U) Q^5
\end{align}
But we know from our study of the CCWZ mechanism, that $h(g,U)$ is a representation of the subgroup. Therefore, this dressed fermion must decompose into irreps of the subgroup
\begin{align}
U^{T} Q &= \sum\limits_{\oplus}^{r_h} Q^{r_h}
\end{align}
In the case of a fiveplet, we get $U^T Q = Q^5 = Q^4\oplus Q^1 = (Q^4, Q^1)$. This is a general method for decomposing any $G$ irrep into $H$ irreps. Explicitly for the fiveplet,
\begin{align}
Q^4 = U_i^{\dagger J} Q_J, && Q^1 = U_5^{\dagger J} Q_J
\end{align}
\end{tcolorbox}
\vspace{2em}

We are interested only in representations \textbf{4} and \textbf{1}, as they will be coupling with the SM fields. A \textbf{5} gives both a $q^\textbf{4}$ and $t^\textbf{1}$, however a \textbf{10} for example decomposes to \textbf{4} + \textbf{6}, and we thus ignore the sixplet. In the $SO(4)$ vacuum, each field in \cref{eq:general_fermion_lagrangian} decomposes very simply to, for example, $\Psi^5 \rightarrow (\Psi^4, \Psi^1)$. The yukawa-like term behaves non-trivially as 
\begin{align}
\Phi \Psi &\rightarrow \Phi_0 \Psi = \Psi^1\\
\implies & m_Y \bar{\Psi}_L \tilde{\Psi}_R + Y \bar{\Psi}_L \Phi^\dagger \Phi \tilde{\Psi}_R = m_Y \bar{\Psi}^4_L \tilde{\Psi}^4_R + (m_Y + Y) \bar{\Psi}^1_L \tilde{\Psi}^1_R
\end{align}
Applying the EL equations gives the Lagrangian
\begin{align}
\mathcal{L}_{\text{LE}, SO(4)} &\ni \Pi_{q^\textbf{4}}(p^2) \bar{q}_L^\textbf{4} \slashed{p} q^\textbf{4}_L + \Pi_{q^\textbf{1}}(p^2) \bar{q}_L^\textbf{1} \slashed{p} q_L^\textbf{1} + \Pi_{t^\textbf{4}}(p^2) \bar{t}_R^\textbf{4} \slashed{p} t_R^\textbf{4} + \Pi_{t^\textbf{1}}(p^2) \bar{t}_R^\textbf{1} \slashed{p} t_R^\textbf{1} \nonumber\\
&+ M_\textbf{4}(p^2) \bar{q}_L^\textbf{4} t_R^\textbf{4} + M_\textbf{1}(p^2) \bar{q}_L^\textbf{1} t_R^\textbf{1} + h.c. \label{eq:custodial_lagrangian}
\end{align}
where the ``broken" form factors $\hat{\Pi}, \hat{M}$ are given in \cref{sec:form_factors}.

Now we can break the global symmetry, and move away from the $\Phi$ vacuum. To see which fields to remove, consider the elementary fields coupled to $\Phi$ in fiveplets, in the most general way. Imposing only the requirement of linear invariance under $H$, we see that the same irreps interact with each other. That is, we have terms that are functions $f(\slashed{p} , q^\textbf{4}, t^\textbf{4})$ and $f(\slashed{p} , q^\textbf{1}, t^\textbf{1})$. Note that interactions containing only one chirality of fermion must include derivatives, otherwise they go to zero, by identity (iii) of chiral projection given in the dictionary \cref{dic:chiral_symmetry}. However, mass terms such as $\bar{Q}^\textbf{4} T^\textbf{4}$ do not go to zero. There is a subtlety though, where the combination of irreps
\begin{align}
\bar{q}^\textbf{4}t^\textbf{4} + \bar{q}^\textbf{1}t^\textbf{1} &= \bar{q}^I\left[ U^i_I U^{\dagger J}_i + U^5_I U^{\dagger J}_5 \right] t_J\\
&= 0
\end{align}
by orthogonality. Therefore $\bar{Q}^\textbf{1} T^\textbf{1}$ and  $\bar{Q}^\textbf{4} T^\textbf{4}$ are linearly dependent and we can choose just one in the Lagrangian. It is convenient to use the identity
\begin{align}
q^\textbf{1} &= U^{\dagger J}_5 q_J = U^J_I \Phi_0^I q_J = \Phi^J q_J
\end{align}
\begin{align}
\mathcal{L}_{\text{LE},SO(5)} &= \left[\Pi^0_{q^\textbf{5}} \delta_{IJ} + \Pi^2_{q^\textbf{5}} \Phi_I\Phi_J\right]\bar{q}^\textbf{5}_I \slashed{p} q^\textbf{5}_J + \left[\Pi^0_{t^\textbf{5}} \delta_{IJ} + \Pi^2_{t^\textbf{5}} \Phi_I\Phi_J\right]\bar{t}^\textbf{5}_I t^\textbf{5}_J \nonumber\\
& + \left[M^0 \delta_{IJ} + M^2 \Phi_I\Phi_J\right]\bar{q}^\textbf{5}_I \slashed{p} t^\textbf{5}_J  \\
 &= \Pi^0_{q^\textbf{5}}\left(\bar{q}^\textbf{4}_L\slashed{p}q_L^\textbf{4} + \bar{q}_L^\textbf{1}\slashed{p}q_L^\textbf{1}\right) + \Pi^2_{q^\textbf{5}}\left(\bar{q}_L^\textbf{4}\slashed{p}q_L^\textbf{4}\sin^2(h) + \bar{q}_L^\textbf{1}\slashed{p}q_L^\textbf{1}\cos^2(h)\right) + \left\lbrace q_L \rightarrow t_R \right\rbrace \nonumber \\
 &+ M^0\left(\bar{q}^\textbf{4}_L t_R^\textbf{4} + \bar{q}_L^\textbf{1} t_R^\textbf{1}\right) + M^2 \left(\bar{q}_L^\textbf{4} t_R^\textbf{4} + \bar{q}_L^\textbf{1} t_R^\textbf{1}\right) \sin(h)\cos(h) + h.c.\label{eq:effective_g_invariant_lagrangian}
 \end{align}
where we have dressed the fermions with the explicit $\Phi$. Note that it is not necessary to have EW gauged the Lagrangian and chosen the unitary gauge - the explicit decomposition does not depend on gauge. Taking this to the $SO(4)$ vacuum recovers \cref{eq:custodial_lagrangian} and allows us to match the form factors
\begin{align}
\Pi_{q^\textbf{5}}^0 = \Pi_{q^\textbf{4}} && \Pi_{q^\textbf{5}}^2 = \Pi_{q^\textbf{1}} - \Pi_{q^\textbf{4}} && \{\Pi_q \rightarrow \Pi_t, M\} \label{eq:broken_form_factors}
\end{align}
We choose a particular embedding of the elementary fields for the final step
\begin{align}
q_L^5 = \left( b_L \; -ib_L \; t_L \; it_L\; 0 \right) && q_R^5 = \left( 0 \; 0 \; 0 \; 0 \; t_R \right)
\end{align}
Inserting these embeddings into \cref{eq:effective_g_invariant_lagrangian}, given \cref{eq:broken_form_factors} leads to 
\begin{align}
\mathcal{L} &= \Pi^0_{q^\textbf{5}}\bar{q}_L\slashed{p}q_L + \Pi^2_{q^\textbf{5}}\frac{1}{\sqrt{2}}(-i\bar{t}_L)s_h\slashed{p}\frac{1}{\sqrt{2}}(it_L)s_h + \Pi^0_{t^\textbf{5}} \bar{t}_R\slashed{p} t_R + \Pi^2_{t^\textbf{5}}\bar{t}_R c_h \slashed{p} t_R c_h \nonumber\\
& + M^2\frac{1}{\sqrt{2}}(it_L)s_h t_R c_h \nonumber\\
& = \Pi^0_{q^\textbf{5}}\bar{q}\slashed{p}q + \frac{1}{2}\Pi^2_{q^\textbf{5}} s_h^2 \bar{t}_L\slashed{p}t_L + \Pi^0_{t^\textbf{5}}\bar{t}_R\slashed{p}t_R + \Pi^2_{t^\textbf{5}} c_h^2 \bar{t}_R\slashed{p}t_R + \frac{i}{\sqrt{2}}M^2 s_hc_h t_L t_R \nonumber\\
& = \Pi_{q^\textbf{4}}\bar{q}\slashed{p}q + \frac{1}{2}(\Pi_{q_\textbf{1}} - \Pi_{q_\textbf{4}})s_h^2 \bar{t}_L\slashed{p}t_L + \Pi_{t^\textbf{4}}\bar{t}_R\slashed{p}t_R + (\Pi_{t^\textbf{1}} - \Pi_{t^\textbf{4}})c_h^2 \bar{t}_R\slashed{p}t_R \nonumber\\
& + \frac{i}{\sqrt{2}}(M_1 - M_4) s_hc_h \bar{t}_L t_R\\
\end{align}

And so we can match the $SO(5)$-invariant Lagrangian to the broken, Lagrangian that contains only SM degrees of freedom
\begin{align}
\Pi_{u_L} = \frac{1}{2}(2-s_h^2)\Pi_{q^\textbf{4}}  + \frac{1}{2}s_h^2\Pi_{q^\textbf{1}},  & & \Pi_{u_R} = s_h^2\Pi_{t^\textbf{4}} + (1-s_h^2)\Pi_{t^\textbf{1}}\nonumber \\
M_u = \frac{i}{\sqrt{2}}(M_1 - M_4)s_h\sqrt{1 - s_h^2}  \label{eq:decomposed_form_factors}
\end{align}

The final step is to connect the general potential formula to a specific choice of high energy fermion Lagrangian. To be clear, the derivation of this section has been specific to the case of composite fermion resonances where each transforms as a $\textbf{5}$ of $SO(5)$. Other choices will be explored in \cref{sec:LCHM}. We will continue this specific choice for the final step. We can express the form factors as functions of the independent parameters $\Delta^i, m_{Q_r^i}, m_{T_r^i}$, as they are parameterised in \cref{eq:general_fermion_lagrangian}. We apply the Euler-Lagrange equations
\begin{align}
\partial_\mu\frac{\partial \mathcal{L}}{\partial(\partial_\mu \Psi^j)} - \frac{\partial \mathcal{L}}{\partial \psi^j} = 0, \qquad \text{where } \Psi^j = \Psi^i_1, \Psi^i_4, \tilde{\Psi}^i_1, \tilde{\Psi}^i_4
\end{align}
to \cref{eq:general_fermion_lagrangian} (i.e. so that the doublet and singlet representations are explicit). We find that
\begin{align}
\Pi_{q^\textbf{4}} = - \sum\limits_{j=1}^{N_\textbf{4}} \frac{|\Delta_{\Psi^\textbf{4}_j}|^2}{p^2 - m_{\Psi^\textbf{4}_j}^2}, && \Pi_{t^\textbf{4}} = - \sum\limits_{j=1}^{N_\textbf{4}} \frac{|\Delta_{\tilde{\Psi}^\textbf{4}_j}|^2}{p^2 - m_{\tilde{\Psi}^\textbf{4}_j}^2}\\
\Pi_{q^\textbf{1}} = - \sum\limits_{j=1}^{N_\textbf{1}} \frac{|\Delta_{{\Psi}^\textbf{1}_j}|^2}{p^2 - m_{{\Psi}^\textbf{1}_j}^2}, & &  \Pi_{t^\textbf{1}} =  - \sum\limits_{i=1}^{N_\textbf{1}} \frac{|\Delta_{\tilde{\Psi}^\textbf{1}_i}|^2}{p^2 - m_{\tilde{\Psi}^\textbf{1}_i}^2}   \\
M_4 = \sum\limits_{i=1}^{N_\textbf{1}} \frac{\Delta_{\Psi^\textbf{4}_j} \Delta_{\tilde{\Psi}^\textbf{4}_j} m_{\tilde{\Psi}^\textbf{4}_i}}{p^2 - m_{\tilde{\Psi}_i}^2}  & & M_1 = \sum\limits_{j=1}^{N_\textbf{4}} \frac{\Delta_{\Psi^\textbf{1}_j} \Delta_{\tilde{\Psi}^\textbf{1}_j} m_{\tilde{\Psi}^\textbf{1}_i}}{p^2 - m_{\Psi_j}^2} & &\\
\end{align}
where $\Delta_{\Psi_i^\textbf{4}}$ is the coupling between $\Psi^\textbf{4}_i$ and $\Psi^\textbf{4}_{i+1}$ in a decomposed version of \cref{eq:general_fermion_lagrangian}. We can finally insert these explicit expressions into the general formula in \cref{eq:potential} for the CW potential
\begin{align}
V &= -2N \int \frac{dp^2_E}{16\pi^2} p^2_E \log\left[p^2_E \left(1 + \Pi_{q^\textbf{4}} -\frac{1}{2}(\Pi_{q^\textbf{4}} - \Pi_{q^\textbf{1}})s_h^2\right)\right. \nonumber\\
&\left.\times \left(1 + \frac{1}{2}\Pi_{t^\textbf{1}} + \frac{1}{2}(\Pi_{t^\textbf{4}} - \Pi_{t^\textbf{1}})s_h^2\right) - \frac{1}{2}s^2_h(1-s_h^2)(M_4 - M_1)^2\right] \label{eq:potential_again}
\end{align}

We Taylor expand the integrand $f(p^2, s_h^2)$ up to quartic order in $s_h$ 
\begin{align}
f(p^2,s_h^2)&=
p^2 \log \left(\frac{1}{2} p^2 \left(\Pi _{q^\textbf{4}}+1\right) \left(\Pi _{t^\textbf{1}}+2\right)\right)\nonumber\\
&+ s_h^2  \frac{\left(p^2 \left(\left(\Pi _{q^\textbf{1}}-\Pi _{q^\textbf{4}}\right) \left(\frac{\Pi _{t^\textbf{1}}}{2}+1\right)+ \left(\Pi _{q^\textbf{4}}+1\right) \left(\Pi _{t^\textbf{4}}-\Pi _{t^\textbf{1}}\right)\right)- \left(M_4-M_1\right){}^2\right)}{\left(\Pi _{q^\textbf{4}}+1\right) \left(\Pi _{t^\textbf{1}}+2\right)}\nonumber\\
&+ s_h^4 p^2 \left[\frac{\left(2 \left(M_4-M_1\right)^2+ p^2 \left(\Pi _{q^\textbf{1}}-\Pi _{q^\textbf{4}}\right) \left(\Pi _{t^\textbf{4}}-\Pi _{t^\textbf{1}}\right)\right)}{2 p^2 \left(\Pi _{q^\textbf{4}}+1\right) \left(\Pi _{t^\textbf{1}}+2\right)}\right. \nonumber\\
&\left. -\frac{ \left(p^2 \left(\left(\Pi _{q^\textbf{1}}-\Pi _{q^\textbf{4}}\right) \left(\frac{\Pi _{t^\textbf{1}}}{2}+1\right)+ \left(\Pi _{q^\textbf{4}}+1\right) \left(\Pi _{t^\textbf{4}}-\Pi _{t^\textbf{1}}\right)\right)-\left(M_4-M_1\right){}^2\right){}^2}{ 2 p^4 \left(\Pi _{q^\textbf{4}}+1\right){}^2 \left(\Pi _{t^\textbf{1}}+2\right){}^2}\right] \nonumber\\
&+ \mathcal{O}(s_h^6) \label{eq:leading_orders_fermion_potential_contribution}
\end{align}
\normalsize

We now have the explicit, leading-order integrands derived from the high-energy theory. This begs the question of whether they converge, and thus can actually be calculated.

\subsection{Convergence of the Integral}
\label{sec:WSR}

\paragraph{Our goal}
\parbox{0.8\textwidth}{To demonstrate using the explicit expressions for the fermion potential contribution that there are certain generic conditions on the convergence of the Higgs potential. These are analogous to the Weinberg Sum Rules.}
\vspace*{2em}

Given that we have derived the leading order potential $V^{(4)} = -\gamma s_h^2 +\beta s_h^4$ in \cref{eq:leading_orders_fermion_potential_contribution}, then we pick out the $s_h^2$ term for our quadratic integrand $f_{\gamma}$,
\begin{align}
f_{\gamma }=\frac{p^2 \left(\left(\Pi _{q^\textbf{1}}-\Pi _{q^\textbf{4}}\right) \left(\frac{\Pi _{t^\textbf{1}}}{2}+1\right)+\left(\Pi _{q^\textbf{4}}+1\right) \left(\Pi _{t^\textbf{4}}-\Pi _{t^\textbf{1}}\right)\right)-\left(M_4-M_1\right){}^2}{\left(\Pi _{q^\textbf{4}}+1\right) \left(\Pi _{t^\textbf{1}}+2\right)}
\end{align}
Note that each form factor has momentum dependence of the order $1/p^2$, and therefore $f_\gamma$ scales at most as $\mathcal{O}\left(\frac{p^2 \times 1/p^2}{(1/p^4 + 1/p^2 + \textnormal{const.}) }\right) \sim \mathcal{O}(p^0)$. For the integral of the quadratic term to converge, we require
\begin{align}
\lim_{p_E\rightarrow \infty} f_{\gamma_t}(p^2_E) = 0 & &\lim_{p_E\rightarrow \infty} p_E^2 f_{\gamma_t}(p^2_E) = 0 \label{eq:quadratic_WSR}
\end{align}
To see why this is the case, consider expanding $f(p^2)$ as a Laurent series
\begin{align}
f(p^2) &= ... + a_{-4}p^{-4} +a_{-2} p^{-2} + a_0
\end{align}
The integral $\int\limits_0^\infty dp^2 f(p^2)$ diverges quadratically unless $a_n = 0$, $n=0$. It both quadratically and logarithmically diverges unless $a_n = 0$ for $n \leq  -2$. These conditions are equivalent to the expressions in \cref{eq:quadratic_WSR}. We can now use the explicit form of the form factors. Each $\Pi_a$ is of the form
\begin{align}
\Pi_a &= \sum\limits_i\frac{\Delta_i^2}{p^2 - m_i^2} = \delta\sum\limits_i^{N_a} \frac{\Delta^2_i}{1-\delta m_i^2}, \; \text{where } \delta := \frac{1}{p^2}  \nonumber\\
& = \delta\sum\limits_i^{N_a} \Delta^2_i(1+\delta m_i^2 + \mathcal{O}(\delta^2)) \nonumber
\end{align}
We can safely discard $\mathcal{O}(\delta^3)$ terms as they do not contribute to divergence of the integral. Thus, defining 
\begin{align}
c_a &= \sum\limits_i^{N_a} \Delta_i^2, & d_a &= \sum\limits_i^{N_a} \Delta_i^2 m_i^2, \label{eq:sum_definitions}
\end{align}
each form factor appears as
\begin{align}
\Pi_a &\equiv \delta (c_a+ \delta d_a )
\end{align}
Substituting this definition into the quadratic integrand gives
{\small \begin{align}
f_\gamma &= \frac{1}{\left(1-\delta  \left(\delta  d_{q^\textbf{4}}+c_{q^\textbf{4}}\right)\right) \left(2- \delta\left(  d_{t^\textbf{1}}+\delta  c_{t^\textbf{1}}\right)\right)}\left[\frac{1}{\delta }\left[\left(\delta  \left(\delta  d_{q^\textbf{4}}+ c_{q^\textbf{4}}\right)-\delta \left(\delta  d_{q^\textbf{1}}+ c_{q^\textbf{1}}\right)\right)\right.\right. \nonumber\\ 
&\left. (1-\frac{1}{2} \delta   \left(\delta  d_{t^\textbf{1}}+c_{t^\textbf{1}}\right))+\left(1-\delta  \left(\delta  d_{q^\textbf{4}}+c_{q^\textbf{4}}\right)\right) \left(\delta \left(\delta  d_{t^\textbf{1}}+c_{t^\textbf{1}}\right)-\delta  \left(\delta  d_{t^\textbf{4}}+ c_{t^\textbf{4}}\right)\right)]\right. \nonumber\\
&\left.-\left( \delta  \left(\delta  d_{M_4}+c_{M_4}\right)- \delta  \left(\delta  d_{M_1}+c_{M_1}\right)\right){}^2\right]
\end{align}}
To quickly extract the $\delta^0$ and $\delta^1$ terms, simply expand in $\delta$
\begin{align}
f_\gamma &=\frac{1}{2} \left(-c_{q^\textbf{1}}+c_{q^\textbf{4}}+c_{t^\textbf{1}}-c_{t^\textbf{4}}\right) \\
& +\frac{1}{4} \delta \left(2 c_{q^\textbf{4}}(c_{q^\textbf{4}}  - c_{q^\textbf{1}}) + c_{t^\textbf{1}}(c_{t^\textbf{1}}  - c_{t^\textbf{4}}) - 2( d_{q^\textbf{1}} - d_{q^\textbf{4}} + d_{t^\textbf{1}}- d_{t^\textbf{4}} )\right)+ \mathcal{O}(\delta ^2)  \nonumber
\end{align}
Then we can define the two terms $f_\gamma^0, f_\gamma^1$ that correspond to the conditions \ref{eq:quadratic_WSR}. 
\begin{align}
f^0_{\gamma }&=\frac{1}{2} \left(-c_{q^\textbf{1}}+c_{q^\textbf{4}}+c_{t^\textbf{1}}-c_{t^\textbf{4}}\right)\label{eq:WSR_1}\\
f^1_{\gamma}&=\frac{1}{4} \delta \left(2 c_{q^\textbf{4}}(c_{q^\textbf{4}}  - c_{q^\textbf{1}}) + c_{t^\textbf{1}}(c_{t^\textbf{1}}  - c_{t^\textbf{4}}) - 2( d_{q^\textbf{1}} - d_{q^\textbf{4}} + d_{t^\textbf{1}}- d_{t^\textbf{4}} )\right) \label{eq:WSR_2}
\end{align}

Picking out the coefficient of $s_h^4$ gives us $f_\beta$:

\begin{align}
f_{\beta }&=p^2 \left(\frac{\left(2 \left(M_4-M_1\right)^2+ p^2 \left(\Pi _{q^\textbf{1}}-\Pi _{q^\textbf{4}}\right) \left(\Pi _{t^\textbf{4}}-\Pi _{t^\textbf{1}}\right)\right)}{2 p^2 \left(\Pi _{q^\textbf{4}}+1\right) \left(\Pi _{t^\textbf{1}}+2\right)}\right.\nonumber\\
&\left. -\frac{ \left(p^2 \left(\left(\Pi _{q^\textbf{1}}-\Pi _{q^\textbf{4}}\right) \left(\frac{\Pi _{t^\textbf{1}}}{2}+1\right)+ \left(\Pi _{q^\textbf{4}}+1\right) \left(\Pi _{t^\textbf{4}}-\Pi _{t^\textbf{1}}\right)\right)-\left(M_4-M_1\right){}^2\right)^2}{ 2 p^4 \left(\Pi _{q^\textbf{4}}+1\right){}^2 \left(\Pi _{t^\textbf{1}}+2\right)^2}\right)
\end{align}

Using our definitions in \cref{eq:sum_definitions} produces a term up to order $\delta$, 
\begin{align}
f_{\beta} = \frac{1}{8} \delta  \left(-(c_{q^\textbf{4}} - c_{q^\textbf{1}})^2 - (c_{t^\textbf{1}} - c_{t^\textbf{4}})^2 \right) + \mathcal{O}(\delta ^2)
\end{align}
Note there is no constant term. Thus
\begin{align}
f^1_{\beta }&= \frac{1}{8} \delta  \left(-(c_{q^\textbf{4}} - c_{q^\textbf{1}})^2 - (c_{t^\textbf{1}} - c_{t^\textbf{4}})^2 \right)  \\
\implies \lim_{\delta\rightarrow 0} \frac{1}{\delta} f^1_{\beta} &=0 =  -(c_{q^\textbf{4}} - c_{q^\textbf{1}})^2 - (c_{t^\textbf{1}} - c_{t^\textbf{4}})^2 
\end{align}
This gives our first set of sum rules, given that each squared term must go to zero
\vspace*{1em}
\begin{tcolorbox}[colback = black!2!white]
\paragraph{Sum Rules \romannumeral 1\relax}
\begin{align}
\sum\limits_i^{N_S} \Delta _{q^\textbf{1},i}^2-\sum\limits_i^{N_S} \Delta _{q^\textbf{4},i}^2 &= 0\\
\sum\limits_i^{N_S} \Delta _{t^\textbf{1},i}^2-\sum\limits_i^{N_S} \Delta _{t^\textbf{4},i}^2 &= 0 \label{eq:WSR_1b}
\end{align}
\end{tcolorbox}
\vspace*{1em}

This implies the quadratic convergence of $f_\gamma$ in \cref{eq:WSR_1} automatically. Inserting the first SRs into the $f_\gamma$ logarithmic convergence criterion in \cref{eq:WSR_2} gives the second SRs
\vspace*{1em}
\begin{tcolorbox}[colback = black!2!white]
\paragraph{Sum Rules \romannumeral 2\relax}
\begin{align}
\sum\limits_i^{N_S} m_{1,i}^2 \left( \Delta _{t^\textbf{1},i}^2  - \Delta _{q^\textbf{1},i}^2 \right) -  \sum\limits_i^{N_Q} m_{4,i}^2 \left( \Delta _{t^\textbf{4},i}^2 - \Delta _{q^\textbf{4},i}^2\right) = 0
\end{align}
\end{tcolorbox}
\vspace*{1em}

Then using this result with the second sum rule in \cref{eq:WSR_2} gives
\begin{align}
\sum _i(m_{4,i}^2 \Delta _{q^\textbf{4},i}^2- m_{4,i}^2 \Delta _{t^\textbf{4},i}^2) - \sum_i(m_{1,i}^2 \Delta _{q^\textbf{1},i}^2-  m_{1,i}^2 \Delta _{t^\textbf{1},i}^2)=0 \label{eq:WSR_3}
\end{align}

The three rules of \cref{eq:WSR_1}, \cref{eq:WSR_2} and \cref{eq:WSR_3} require $N_Q \geq 1$ and $N_T \geq 1$, and in this case the resonances must either be degenerate in mass,
\begin{align}
m_4^2 = m_1^2 = m && \implies && |\Delta_{q^\textbf{4}}| =|\Delta_{q^\textbf{1}}| = |\Delta_Q| && \text{and} && |\Delta_{t^\textbf{4}}| = |\Delta_{t^\textbf{1}}| = |\Delta_T| 
\end{align}
or in mixing
\begin{align}
|\Delta_{q^\textbf{4}}| =|\Delta_{q^\textbf{1}}| = |\Delta_{t^\textbf{4}}| = |\Delta_{t^\textbf{1}}|
\end{align}
For any general multi-site model, there should be no reason to assume these degeneracies. If the resonances are allowed the generic freedom of the Lagrangian in \cref{eq:general_fermion_lagrangian}, then at least two sets of resonances are required to simultaneously satisfy the Weinberg sum rules. This corresponds to two composite sites. In the language of Panico \& Wulzer, this is the three-site model. However, there is a slightly more minimal model that can be achieved by application of Hidden Local Symmetry. We will detail this model, the Minimal 4D Composite Higgs (M4DCHM) in section \ref{sec:M4DCHM}.

\section{Fine Tuning in  the MCHM}

\paragraph{Our goal}
\parbox{0.8\textwidth}{To understand sources of tuning in the MCHM, and motivate possible extensions to the model.}
\vspace*{2em}

\subsection{Tuning in Any EW Effective Theory}

Consider the general expression for a Higgs that is a pNGB of some higher global symmetry. By naive dimensional analysis, we will have a potential that scales like \cite{Bellazzini:2014yua}
\begin{align}
V(h) &\sim \frac{g_{\text{SM}}^2\Lambda^2}{16\pi^2}\left(-a|h^2| + b \frac{|h|^4}{2f^2}\right)
\end{align}
with the constraints
\begin{align}
v^2 = \frac{a}{b}f^2 = (246\gev)^2 && \text{and} && m_h^2 = 4bv^2 \frac{g_{\text{SM}}^2}{16\pi^2}\frac{\Lambda^2}{f^2} = (125\gev)^2 
\end{align}
Consider that the Barbieri-Giudice measure, given as the maximum derivative of the observables w.r.t the parameters  in log-space
\begin{align}
\Delta_\text{BG} = \text{max}\{ \bigm\lvert \frac{\partial (\ln\mathcal{O}^i)}{\partial (\ln x^j)} \bigm\lvert_{\mathcal{O} = \mathcal{O}_\text{exp}} \} =  \text{max}\{ \bigm\lvert \frac{x^j}{\mathcal{O}^i_\text{exp}}\frac{\partial \mathcal{O}^i}{\partial x^j} \bigm\lvert_{\mathcal{O} = \mathcal{O}_\text{exp}} \} = 
\end{align}
This gives a naive measure of tuning of (for example) the EW vev
\begin{align}
\Delta_{v^2} &= \text{max} \{ \frac{a}{v_\text{exp}}\frac{\partial (v^2)}{\partial a},\frac{b}{v_\text{exp}}\frac{\partial (v^2)}{\partial b}\}| = \frac{1}{(246 \gev)^2}\frac{a}{b}f^2 \propto f^2
\end{align}
So, naively, as the scale increases, the first-order tuning should generally increase. Of course, the particular cancellations between the $a$ and $b$ can give areas of lower tuning even for higher scale.

\subsection{Double Tuning}\label{sec:double_tuning}

For any theory that attempts to capture EW dynamics as an effective description of physics relevant at scale $f$, we see that it will generally need to be tuned proportional to 
\begin{align}
\Delta \propto \frac{f^2}{v^2} = \xi^{-1}
\end{align}
This is generically true for Susy and GUT models. However, after explicit realisations of CHMs are produced, it is realised that there is an extra complication. It was noticed in \cite{panico2011} and explored in \cite{matsedonskyi2012,panico2012} that once the correct scale is obtained, we must further tune to get the correct EW symmetry breaking behaviour. Let us see how this works. As has been derived in section \ref{sec:MCHM_Matter_Misalignment}, the pNGB Higgs doublet receives a potential term from radiative corrections by the gauge bosons and heavy fermions. For simplicity, let's consider only the contribution of the top quark and its composite partner(s). The dominant contribution to the Higgs potential in Composite Higgs models (c.f. Little Higgs) is given in \cref{eq:full_potential} and \cref{eq:decomposed_form_factors}
\begin{align}
V_\text{top}(h) = -2N_c \int \frac{d^2p_E}{16\pi^2} p_E^2\log\left[p^2_E\left(\Pi^{t_L}_0 + \frac{s^2_h}{2}\Pi^{t_L}_1\right)\left(\Pi^{t_R}_0 + c_h^2 \Pi^{t_R}_1\right) - \frac{s_h^2 c_h^2}{2}|M^t_1|^2 \right]\label{eq:top_potential}
\end{align}
If this is to be connected with SM physics, we should find contributions to the SM Higgs' quadratic and quartic couplings. This can be done by expanding \cref{eq:top_potential} in some small parameter. This could be the Higgs field itself $\frac{h}{f}$, as we did in \cref{sec:MCHM_Matter_Misalignment}. For this exploration, let's expand in the compositeness angle $t_{\Psi, \tilde{\Psi}}=\tan\theta_{\Psi, \tilde{\Psi}} = \Delta_{\Psi, \tilde{\Psi}}/m_{\Psi, \tilde{\Psi}}$. Then the leading order contribution to the potential is
\begin{align}
V &= V^{(2)} + \mathcal{O}(t_{\Psi, \tilde{\Psi}}^4)\\
&= \frac{N_c M_t^4}{16\pi^2}f_2(t_\Psi^2,t_{\tilde{\Psi}}^2) \sin^2 (h/f)+ \mathcal{O}(d^4)
\end{align}
since this is the result of interactions of the sort in \cref{fig:fermion_potential}.

Naively this can indeed break EW with a minimum away from the origin, at multiples of $\langle h\rangle=\pi f/2$. But we know that the scale $f$ should be significantly higher than $\langle h\rangle = 246\gev$ otherwise we should have seen evidence for it already. That is, we require $\sin^2\left(\frac{\langle h\rangle}{f}\right)=\sin^2\left(\frac{v}{f}\right)\approx \frac{v^2}{f^2} << 1$. So, we must include contributions to $V$ that give a minimum at $v/f$. 

\begin{align}
V =& V^{(2)} + V^{(4)} + \mathcal{O}(d^6)\\
= & \frac{N_c M_t^4}{16\pi^2}f(t_\Psi^2,t_{\tilde{\Psi}}^2) \sin^2 (h/f)\\
& + \frac{N_c M^4_t}{16\pi^2}\left(f_1(t_\Psi^4,t_{\tilde{\Psi}}^4)\sin^2(h/f) + f_2(t_\Psi^4,t_{\tilde{\Psi}}^4)\sin^4(h/f) \right)+ \mathcal{O}(d^6)\\
\propto & (a^{(2)} t^2 + \tilde{a}^{(2)}\tilde{t}^2)s_h^2\\
&+ (a^{(4)} t^4 + \tilde{a}^{(4)} \tilde{t}^4) s_h^2 + (b^{(4)} t^4 + \tilde{b}^4 \tilde{t}^4) s_h^4
\end{align}

Where $a, b$ are of order one, in the absence of tuning. Thus, we can calculate the potential, making a simplification $t = \tilde{t}$
\begin{align}
\frac{\partial V}{\partial h} \bigm\lvert_{h=v} &= 0 \nonumber \\
& = a^{(2)} + \tilde{a}^{(2)} + a^{(4)} t^2 + \tilde{a}^{(4)} \tilde{t}^2 + 2(b^{(4)} t^2 + \tilde{b}^{(4)} \tilde{t}^2) s_h^2 \nonumber\\
\implies \xi &= \Bigm \lvert \frac{1}{t^2}\frac{ a^{(2)} + \tilde{a}^{(2)}}{2(b^{(4)} + \tilde{b}^{(4)} )}  + \frac{  a^{(4)} + \tilde{a}^{(4)} }{2(b^{(4)} + \tilde{b}^{(4)} } \Bigm \lvert  << 1 \label{eq:tuning_condition}
\end{align}
which requires a tuning (against $a^{(2)}$, for example) of 
\begin{align}
\Delta_\text{BG} = \frac{a^{(2)}}{v^2} \frac{\partial v^2}{\partial a^{(2)}} = \frac{a^{(2)}}{v^2} \frac{f^2}{t^2}\frac{1}{2(b^{(4)} + \tilde{b}^{(4)})} \propto \xi^{-1} t^{-2}
\end{align}
which enhances the naive tuning by $\frac{1}{t^2}$. That is, first there needs to be a tuning of the first term in \cref{eq:tuning_condition} of order $t^2$ to bring it to the same order as the second term, then there needs to be a tuning of the overall expression to be much less than one.  For concreteness, consider the mass of the top in a first-order approximation, as inspected from \cref{eq:diagonal_yukawa}
\begin{align}
m_t \sim \frac{v}{\sqrt{2}} \sin\theta_\Psi \sin \theta_{\tilde{\Psi}} \frac{Y_T}{f}
\end{align}
The most natural configuration is to set the composite sector coupling $\frac{Y_T}{f} \rightarrow 1$ and let the SM fermions be fully composite, $\sin\theta_\Psi, \sin\theta_{\tilde{\Psi}} \rightarrow 1$. Then we exactly recover the SM. If we raise $f$, as we must, we then also tune $\sin\theta$ by the same amount.

I will not go further into the phenomenology of this well-studied ``double tuning", as there is a better way to analyse it. One that does not rely on the choice of parameterisation. We must build the concept of double tuning into the measure of tuning itself. Then, we can't help but include this tuning (and any other that has not yet been noticed). This measure will be described in the next chapter \ref{sec:fine_tuning}, as the phenomenon of double tuning motivates extensions to the top-only MCHM.

\chapter{Naturalness}
\label{sec:fine_tuning}

\section{Gaining Intuition}

Concepts of naturalness can be formally derived from Bayesian arguments - and these will be used later in this section and in appendix \ref{sec:tuning_appendix}. However, the motivation for naturalness has historically been a philosophical one, rather than a statistical one. So let us build an intuition for naturalness and tuning, before deriving them rigorously. We begin with simplicity.

\subsection{Occam's Razor}

Sometime before 300 B.C., Aristotle stated "We may assume the superiority, all things being equal, of the demonstration which derives from fewer postulates or hypotheses". In the 14th century, William of Occam promoted this idea in his work on logic, as the Law of Parsimony. Given a set of data (e.g. hoofbeats), we can choose from a set of models that reproduce the data (e.g. horses approaching, zebras approaching). The models should be able to reproduce \textit{all} of the data, otherwise it should be abandoned (e.g. \sout{octopus approaching}). We should choose the model that requires the least complexity. In this case, we should assume horses are approaching as it only requires the explanation of why there are several horses and why they are running. To suggest zebras requires also explaining why they are not in an African savannah. There is also an element of Occam's Razor that applies to scale: using our ears to measure sound is an everyday scale measurement, and horses are an everyday scale explanation. That is, they are both order $\mathcal{O}(1)$ in our model. We shouldn't consider the contribution of the sun's gravity $\mathcal{O}(10^{28} m_\text{horse})$ or Heisenberg's uncertainty principle $\mathcal{O}(\frac{h}{s^2} \approx 10^{-37}\; \text{horsepower})$. We can summarise this as all our model parameters being of the same order. The further away we move from our data in terms of the scale of our model, the more complex it is, and the more Occam would punish the model-maker. Additionally, not only should we choose the simplest model (the "second level of inference"), we should choose the simplest variation of that model (the "first level of inference"). Does one horse explain the hoofbeats, or are more required? In this case, number of animals is a parameter common to both models.

There is a final intuition to be gained from Occam. If we receive new data (e.g. we're told there was an escape at the zoo), then we should update our belief. We should still choose the simplest idea, but now it's not so obvious. Both models (horses vs. zebras | zoo breakout) still explain all the data, but now the zebra model incorporates the data more simply than it did before. Similarly, the horse model "pays a price" for this new data. We will formalise this idea later, but for now we call this the "Occam Factor" \cite{mackay1992thesis,mackay1992bayesian}. It can be thought of as the sensitivity of a model $\mathcal{H}_i$ to data $d$ across all the reasonable parameters $\bm{x}$. This is quite a counter-intuitive idea, since we usually start with data and then try and fit a scientific hypothesis to that data\footnote{Note that here "data" may be simulated}. However, the Occam Factor is the ratio of one's earlier ("prior") belief about a model compared with one's later ("posterior") belief about a model. This factor is important in determining how "believable" a model is, a measure that we can formalise as the Bayesian Evidence.

\subsection{Bayesian Evidence}

\begin{figure}
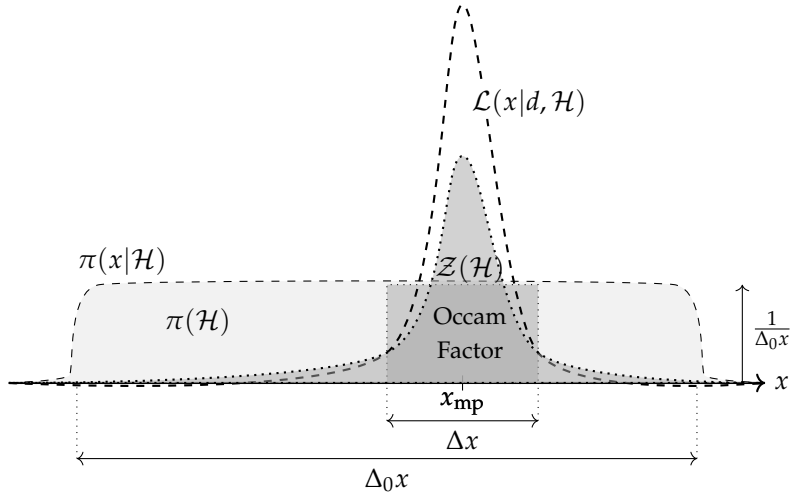

\tikz{
\draw [->, thick] (1,2) -- (11,2);
\draw [<->] (1.9,1) -- (10.1,1);
\node [below] at (6,1) {$\Delta_0 x$};
\draw [<->] (6,1.5) -- (8,1.5);
\node [below] at (7,1.5) {$\Delta x$};
\draw (7,2.1) -- (7,1.9);
\draw [dotted] (6,1.5) -- (6,2);
\draw [dotted] (8,1.5) -- (8,2);
\node [below] at (7,2) {$x_\text{mp}$};
\draw [dashed, fill=gray!20, fill opacity = 0.5] plot [smooth, tension=0.2] coordinates {(1,2) (1.8,2.1) (2.2,3.3) (9.8,3.3) (10.2,2.1) (11,2) } --cycle;
\node at (3.5,2.8) {$\pi(\mathcal{H})$};
\node [above] at (2.5,3.3) {$\pi(x| \mathcal{H})$};
\draw [dotted, fill=gray!70, fill opacity = 0.5] (6,2)  rectangle (8,3.3);
\draw [dotted] (1.9,1) -- (1.9,2);
\draw [dotted] (10.1,1) -- (10.1,2);
\draw [thick, dashed] plot [smooth, tension=0.45] coordinates { (1,2) (6,2.4) (7,7) (8,2.4) (11,2) };
\draw [thick, dotted, fill=gray!70, fill opacity= 0.5] plot [smooth, tension=0.5] coordinates { (1,2) (6,2.4) (7,5) (8,2.4) (11,2) } -- cycle;
\draw [->] (10.7,2) -- (10.7,3.3);
\node [right] at (10.7,2.7) {$\frac{1}{\Delta_0 x}$};
\node [right] at (11,2) {$x$};
\node [below] at (7,2) {$x_\text{mp}$};
\node [above, align=center] at (7.1,2.2) {\small Occam \\ \small Factor};
\node [above] at (7.1,3.2) {$\mathcal{Z}(\mathcal{H})$};
\draw [draw = white, fill = white] (7,5.5) rectangle (8,6);
\node [above, right] at (7,5.7) {$\mathcal{L}(x | d, \mathcal{H})$};
}\caption{Posterior distribution plot, as a uniform prior is ``spent" on the region $\Delta x$, with the purchased area being the Occam Factor. Adapted from \cite{mackay1992thesis}.}\label{fig:occam_factor}
\end{figure}

The Bayesian Evidence of a model (for an introduction to Bayes' Theorem, see \cref{sec:bayes_theorem}) is defined as 
\begin{align}
\mathcal{Z}(\mathcal{H}_i) = p(d|\mathcal{H}_i) = \begin{cases}
\sum\limits_a^N p(d|\mathcal{H}_i,\bm{x}_a) p(\bm{x}_a | \mathcal{H}_i) \text{, if discrete } \bm{x}_a= \bm{x}_1, ... , \bm{x}_N \vspace{10pt}\\ 
\int_{\mathcal{V}} p(d|\mathcal{H}_i, \bm{x}) p(\bm{x} | \mathcal{H}_i) d^n\bm{x} \text{, if continuous } \bm{x} \in \mathcal{V}
\end{cases}
\end{align}
where $\bm{x}$ is a vector of $n$ parameters in a volume of parameter space $\mathcal{V}$, $p(d|\mathcal{H}_i, \bm{x}) \equiv \mathcal{L}_{\mathcal{H}_i}(\bm{x})$ is the likelihood of the model $\mathcal{H}_i$ at $\bm{x}$, and $p(\bm{x}|\mathcal{H}_i)\equiv \pi_{\mathcal{H}_i}(\bm{x})$ is the density of prior belief in the model.
Let's rewrite this with more economical notation
\begin{align}
\mathcal{Z} = \int_\mathcal{V} \mathcal{L}(\bm{x}) \pi(\bm{x}) d^n\bm{x} \label{eq:continuous_evidence}
\end{align}
We see that the prior density at each point in parameter space weights the likelihood at that point. For a concrete example, consider a set of observables $\mathcal{O}_a$ that we assume are distributed normally with standard deviation $\sigma_a$. Let's assume a uniform prior, which we must normalise to one. Therefore the prior density $1/\Delta_0 \bm{x}$ is constant across $\bm{x}$. This prior can then be "spent" once data is found on a region of high likelihood $\Delta \bm{x}$, and the likelihood at the most probable $\mathcal{L}(\bm{x}_{mp})$ determined. The amount "spent" is the Occam Factor. 
\begin{align}
\underbrace{\mathcal{Z} = p(d|\mathcal{H}_i)}_\text{evidence} & \approx \underbrace{p(d|\bm{x}_\text{mp},\mathcal{H}_i)}_\text{max. likelihood} \underbrace{p(\bm{x}_\text{mp} | \mathcal{H}_i) \Delta \bm{x}}_\text{Occam Factor}\\
&= \mathcal{L}(\bm{x}_\text{mp}) \pi(\bm{x})\Delta \bm{x} = \mathcal{L}(\bm{x}_\text{mp}) \frac{\Delta \bm{x}}{\Delta_0 \bm{x}}\label{eq:evidence}
\end{align}
This toy distribution is sketched in \cref{fig:occam_factor}. If the likelihood is also normalised to one, this can be compared between models. We can see this by comparing the posterior belief (see appendix \ref{sec:tuning_appendix}) in hypothesis 1 to hypothesis 2
\begin{align}
K = \frac{p(\mathcal{H}_1|d)}{p(\mathcal{H}_2|d)} &= \frac{p(d|\mathcal{H}_1)}{p(d|\mathcal{H}_2)} \frac{\pi(\mathcal{H}_1)}{\pi(\mathcal{H}_2)}
\end{align}
This ratio $K$ is the Bayes' factor, for normalised priors. The interpretation of this is classically given by Harold Jeffreys, according to the scale given in \cref{tab:jeffreys_scale}. Occam's factor lets us choose the parameter set with the most predictive power for the least complexity, and Bayes' factor incorporates Occam's factor, enabling us also to choose between models. We will thus generalise our understanding of naturalness to Bayes' factor, as the more powerful of the two.

\begin{table}
\centering
\begin{tabular}{@{} rr @{}}
\toprule
$K$ & Strength\\
\midrule
$< 1$ & Dismiss $\mathcal{H}_1$ \\
$1 - 3$ & Inconclusive \\
$3 - 10$ & Substantial \\
$10 - 30$ & Strong \\
$30 - 100$ & Very strong \\
$> 100$ & Decisive\\
\bottomrule
\end{tabular}
\vspace*{1em}
\caption{The Jeffreys scale of a given Bayes' factor.}\label{tab:jeffreys_scale}
\end{table}

Let us make our intuitions rigorous. For a distribution of observables, we assume the Laplace approximation of the likelihood function. This is a multivariate Taylor series to first order around a peak in the likelihood around parameter point $x_\text{max}$
\begin{align}
\log \mathcal{L}_\mathcal{O} (x) \simeq \log \mathcal{L}_\text{max} + \frac{\partial^2 \log \mathcal{L}_\mathcal{O}}{\partial x^i \partial x^j}  \Bigm\lvert_{x_\text{max}} \frac{(x ^i- x^i_\text{max})(x ^j- x^j_\text{max})}{2}
\end{align}
We define the second derivative term to be the inverse of the covariance matrix $A^{-1}$ of the parameters\footnote{$A^{-1}$ is traditionally called the "precision", which exists if the covariance matrix $A$ is invertible. This notation is standard, though a little counter-intuitive.}, which attain their maximum likelihood at the experimental value of the observables
\begin{align}
A^{-1}|_{\mathcal{O}^i = \mathcal{O}^i_\text{exp}} := \frac{\partial^2\mathcal{L}}{\partial x^i \partial^j}\bigm\lvert_{\mathcal{O}^i = \mathcal{O}^i_\text{exp}} =\frac{\partial^2\mathcal{L}}{\partial x^i \partial^j}\bigm\lvert_{x_\text{max}} 
\end{align}
As we gather more data, the central limit theorem tells us that the distribution of an observable (including simulations), will tend to a Gaussian around a local maximum, with some variance $\Delta \mathcal{O}$. For several observables, they will tend to a multidimensional Gaussian with mean $\bm{x}_\text{mp}$ giving $\mathcal{O}(\bm{x}_\text{mp}) = \mathcal{O}_\text{exp}$. Then we can use a change of variables to get
\begin{align}
A^{-1}|_{\mathcal{O}^i = \mathcal{O}^i_\text{exp}} &= \left( \frac{\partial^2 \log \mathcal{L}}{\partial \mathcal{O}^k \partial \mathcal{O}^l} \frac{\partial \mathcal{O}^k}{\partial x^i} \frac{\partial \mathcal{O}^l}{\partial x^j}\right) \Bigm\lvert_{\mathcal{O}^i = \mathcal{O}^i_\text{exp}}\\
& \equiv \left(\Sigma^{-1}_{kl} J_{\mathcal{O}}^{ki} J_{\mathcal{O}}^{lj} \right) \Bigm\lvert_{\mathcal{O}^i = \mathcal{O}^i_\text{exp}}
\end{align}
where $J_\mathcal{O}$ is the Jacobian matrix, and $\Sigma$ is the covariance matrix of the \textit{observables} since
\begin{align}
\log \mathcal{L} &= \log \me^{\frac{1}{2}\left(\mathcal{O}(\bm{x}) - \mathcal{O}_\text{exp}\right)^T \Sigma^{-1}\left( \mathcal{O}(\bm{x}) - \mathcal{O}_\text{exp}\right)}\\
& = \frac{1}{2}\left(\mathcal{O}(\bm{x}) - \mathcal{O}_\text{exp}\right)^T \Sigma^{-1}\left( \mathcal{O}(\bm{x}) - \mathcal{O}_\text{exp}\right)
\end{align}
We take the continuous limit \cref{eq:continuous_evidence} of \cref{eq:evidence} in the neighbourhood of a Gaussian posterior. Note that to get a continuous volume element equivalent to the discrete variance $\Delta x$, we take the square root determinant of the parameter covariance $A$ 
\begin{align}
|A_{ij}|^{1/2} = |(-\nabla_i \nabla_j \log \mathcal{L}(\bm{x}))^{-1}|^{1/2} = |(-\nabla_i \nabla_j \log \mathcal{L}(\bm{x}))|^{-1/2}
\end{align}
For a careful explanation of volumes from vectors, see \cref{sec:tuning_appendix} or reference \cite{sivia1993introduction}. Then, taking \cref{eq:evidence} as continuous,
\begin{align}
\mathcal{Z} = \int_\mathcal{V} \mathcal{L}(\bm{x}_\text{mp}) |A|^{1/2} \pi(\bm{x}|\bm{x}_\text{mp}).
\end{align}

We can analyse the Occam Factor $|A|^{1/2}\pi(\bm{x}|\bm{x}_\text{mp})$ to get a better understanding of this value
\begin{align}
|A|^{1/2} = |\Sigma|^{1/2} |J_\mathcal{O} J_\mathcal{O}^T|^{-1/2}. 
\end{align}
If we take the covariance $\Sigma$ of the \textit{log} of each observable, then we get it in the form of a common measure called the Fisher information
\begin{align}
\Sigma^{-1} = \frac{\partial^2 \mathcal{L}}{\partial \mathcal{O}^i \partial \mathcal{O}^j} \rightarrow \Sigma^{-1}_{\log \mathcal{O}} &= \frac{\partial^2 \mathcal{L}}{\partial (\log\mathcal{O}^i) \partial (\log\mathcal{O}^j)} \nonumber \\ 
\implies J = \frac{\partial \mathcal{O}^i}{\partial x^k} \rightarrow J_{\log\mathcal{O}} &= \frac{\partial \log\mathcal{O}^i}{\partial x^k}
\end{align}
and we call this rescaled value $C = |J_{\log\mathcal{O}} J_{\log\mathcal{O}}^T|^{1/2}$ the sensitivity \cite{fichet2012quantified}. We will explore how it relates to measures of fine tuning in the next sections. For simplicity, we also take the elements of the prior density $\frac{1}{\Delta x_0}$ as flat in log space. Then the infinitesimal elements of the prior density are given analogously to the above as $\frac{1}{|V|^{1/2}} dx^n$. We can use a change of variables to get
\begin{align}
\mathcal{Z} = \mathcal{L}_\text{max} \frac{|\Sigma|^{1/2}}{|V|^{1/2}}\int_\mathcal{V} \frac{1}{C} d \mathcal{O}^n(\bm{x})\label{eq:formal_sensitivity}
\end{align} 
which agrees with \cite{fichet2012quantified}. $|V|^{1/2}$ is called the prior volume. $d\mathcal{O}^n(x)$ is the observable measure that we now integrate over, induced by the change of variable of the Jacobians.

\section{Existing Measures}\label{sec:existing_tuning_measures}

Eq. (\ref{eq:formal_sensitivity}) formalises the concepts of Occam's Razor, and the sensitivity. The latter is often called "fine tuning", as the inverse of "naturalness". We see that the fine tuning is a contribution to the Bayesian evidence, though it's not the only one. We must also take into account the prior volume $|V|^{1/2}$, the uncertainty of our observables $\Sigma$, and the maximum likelihood $\mathcal{L}_\text{max}$. If the first two are held fixed, and we assume the maximum likehood can be arbitrarily close to $1$, then the tuning is the sole contributor to the Bayes factor, as the parameter space is explored. This allows the "first level of inference" - that is, choosing the most likely areas of parameter space for a particular model. If we are comparing two models with similar priors and observables, then the ratio of tuning in neighbourhoods of likely parameter space is a good approximation to the Bayes' factor \textit{between} models, called the "second level of inference"\cite{mackay1992thesis}. Given this justification, let's now look at some special cases of \cref{eq:formal_sensitivity} that have been discovered by intuition.

\subsection{Barbieri-Giudice}

Barbieri and Giudice in \cite{barbieri1988upper} established the classic measure of tuning 
\begin{align}
\Delta_\text{BG} &= \max\limits_{\mathcal{O},x}\frac{\partial \log\mathcal{O}}{\partial \log x}|_{\mathcal{O} = \mathcal{O}_\text{exp}}\label{eq:BG}
\end{align}
which is a first-order form of the Bayesian sensitivity
\begin{align}
\Delta_\text{BG} = C|_{x \rightarrow \log x}
\end{align}
considered observable by observable, parameter by parameter, and differentiated with respect to $\log x$. Importantly, \cite{fichet2012quantified} points out that the tuning measure should be a derivative with respect to $\log x$ \textit{only if} the parameters have logarithmic priors, otherwise they are not consistent with \cref{eq:formal_sensitivity}. This measure is sometimes extended to create vectors of fine tuning, i.e. single-observable Jacobians
\begin{align}
\nabla^{\mathcal{O}_a} &= \left(\begin{matrix}
\Delta_{BG,1}, \Delta_{BG,2}, \cdots,  \Delta_{BG,n_p}
\end{matrix}\right)
\end{align}
and then take the magnitude of this vector over $n_p$ parameters
\begin{align}
\Delta^\mathcal{O}(\vec{x}) = \sqrt{\sum\limits_i^{n_p} (\Delta_{BG,i})^2} = |\nabla^{\mathcal{O}_a}| \label{eq:firstorder}
\end{align}

We can gain some intuition with this measure by applying it to the Standard Model. The most finely-tuned sector of the Standard Model is the Higgs sector (indeed, this is the Hierarchy Problem that motivated this thesis). To quantify this, we use the Z boson mass at tree level of
\begin{align}
m_Z^2 = g\frac{-\mu}{2\lambda}
\end{align}
but we know that the $\mu$ parameter in the SM is, to one-loop order,
\begin{align}
\mu^2 = \mu^2_0 + \Delta\mu^2
\end{align}
where $\mu_0$ is the bare mass of the Higgs, and $\Delta\mu^2$ the one-loop quantum corrections. If the SM is the complete description of reality up to the Planck scale, $m_{pl}$, then the tuning at high energy is
\begin{align}
\Delta_\text{BG} &= \Bigm\lvert\frac{\Delta\mu}{m_Z^2}\frac{\partial m_Z^2}{\partial \Delta\mu}\Bigm\lvert_{m_Z \simeq 100\gev, \Delta\mu \simeq m_{pl}}\nonumber\\
& = \Bigm\lvert\frac{\Delta\mu}{m_Z^2} \frac{g}{2\lambda}\Bigm\lvert_{m_Z \simeq 100\gev, \Delta\mu \simeq m_{pl}}\label{eq:SM_tuning}\\
& \approx \frac{10^{19}}{100^2}\frac{10^{-6}}{10^{-1}} \approx 10^{10} \nonumber
\end{align}
using order of magnitude values. The interpretation of this value is usually given as $\Delta_\text{BG}^{-1} = $ percent precision of the model. When first introduced, to study the Minimal Supersymmetric Standard Model, a $\Delta_\text{BG} = 10 \implies 10\%$ precision of parameters was posited as an acceptable value. The B.G. measure is ultimately a measure of cancellation tuning, so this limit is asking for a universe described by a model that doesn't cancel parameters to observables smaller than the order of magnitude of the parameters themselves. To see this, consider an observable $\mathcal{O} = ax^n - C$, which is a function of parameters $\{a,x,C\}$. The tuning in this model due to the $x$ parameter is 
\begin{align}
\Delta_{\text{BG},x} &= \frac{x}{\mathcal{O}}anx^{n-1} = \frac{x}{\mathcal{O}}n\frac{\mathcal{O}+C}{x} = n(1 + \frac{C}{\mathcal{O}})\label{eq:power_law_immunity}
\end{align}
The BG measure is not very sensitive to power laws - as we would hope, otherwise we could fool it by scaling. It is highly sensitive to tuned cancellations, which the Higgs sector is plagued by. For the SM to have a low BG measure $\Delta_\text{BG} < 10$, by \cref{eq:SM_tuning} the mass of the Z should be of the order of the Planck scale. 

There is a glaring piece of information missing from the tuning that is present in the full Occam factor - the prior. Although we have assumed a log scale (an assumption vital to the BG measure, as proven in the previous section) we haven't used that information in the measure. A more recent development incorporates the prior volume into the tuning.

\subsection{Tuning volume}\label{sec:tuning_volume}

In an effort to address issues with the BG measure, particularly the lack of prior consideration and correlations between observables, \cite{athron2007} proposes a tuning measure of volume ratios. Like \cite{Anderson:1994dz,Anderson:1994tr,Anderson:1995cp,Anderson:1996ew} before them, the authors noticed that the immunity to power laws as in \cref{eq:power_law_immunity} is also a shortcoming. The tuning we arrived at in \cref{eq:power_law_immunity} is independent of the point in parameter space. It is, in a sense, a ``global sensitivity", rather than a local sensitivity. However, one would like to think that the point at $\mathcal{O} = a (1000)^n - C$ is relatively more sensitive than the point at $\mathcal{O} = a 1^n - C$. The solution in \cite{Anderson:1994dz,Anderson:1994tr,Anderson:1995cp,Anderson:1996ew} was to consider both local and global  - the BG measure, normalised by the ``average" global sensitivity
\begin{align}
\bar{\Delta}_\text{BG}^{-1}(x_i) = \frac{\int dx_i p(x_i) x_i \Delta_\text{BG}^{-1}(x_i) }{\int dx_i p(x_i) x_i} \label{eq:anderson_normalisation}
\end{align}
We recognise the numerator as the evidence in each parameter, weighted by the BG tuning.

This concept of normalisation is extended by \cite{athron2007} to handle finite volume elements. That work defines a volume in parameter space $F = V_\text{param} =  [a x_i, b x_i]$. $a$ and $b$ are arbitrary values to be chosen based on the model. Also define a volume in observable space $G = V_\text{obs} = [a \mathcal{O}(\bm{x}), b \mathcal{O}(\bm{x})]$. Then the measure of tuning is defined as the ratio
\begin{align}
\Delta_V = \frac{F}{G}\label{eq:athron_measure}
\end{align}
which can be normalised by
\begin{align}
\bar{\Delta}_V = \frac{\int dx_1 ... dx_n \Delta_V(\textbf{x}) }{\int dx_1... dx_n}
\end{align}
which is clearly a finite analogue to \cref{eq:anderson_normalisation} extended to a volume. To be more precise, the normalisation measure proposed in \cref{eq:anderson_normalisation} is the one-dimensional limit of the term $\int_\mathcal{V} \frac{1}{C} d \mathcal{O}^n (\textbf{x})$ in \cref{eq:formal_sensitivity}, while the volume ratio \cref{eq:athron_measure} is a finite version of the term $\frac{1}{|V|^{1/2}}\int dx^n $ also appearing implicitly (via a change of variable) in \cref{eq:formal_sensitivity}. Evidently, the Bayesian evidence is the rigorous generalisation of many intuitive approaches. I will present one more intuitive approach, which best highlights the limitation of the evidence, as it is written in \cref{eq:formal_sensitivity}.

\section{Higher Order Fine Tuning}

Recall that in \cref{sec:double_tuning}, it was necessary to tune two separate parameters to reach the right order of the EW vev. Later, we will see other representations where it is necessary to tune one parameter to retrieve the vev, while tuning another to retrieve the correct Higgs mass. As pointed out by \cite{Barnard:2015ryq}, it is often the case that the fine-tuning vectors 
\begin{align}
\nabla^{\mathcal{O}_a} &= \left(\begin{matrix}
\Delta_{BG,1}, \Delta_{BG,2}, \cdots,  \Delta_{BG,n_p}
\end{matrix}\right)
\end{align}
are not aligned. That is, the fine-tuning may come from more than one source and the fine-tuning measure should reflect this special double tuning - a higher order tuning. If they are completely orthogonal, then the higher order tuning should be simply the product of each single tuning. If they are completely parallel, the higher tuning should disappear. 

\subsection{An Intuitive Measure}\label{sec:intuitive_measure}

For any two particular tuning vectors $\{\nabla^{\mathcal{O}_a},\nabla^{\mathcal{O}_b}\}$, a quantity displaying these criteria is 
\begin{align}
\Delta_2^{ab} = \begin{vmatrix}
\nabla^{\mathcal{O}_a}\cdot\nabla^{\mathcal{O}_a} & \nabla^{\mathcal{O}_a}\cdot \nabla^{\mathcal{O}_b}\\
\nabla^{\mathcal{O}_a}\cdot\nabla^{\mathcal{O}_b} & \nabla^{\mathcal{O}_b}\cdot \nabla^{\mathcal{O}_b}
\end{vmatrix}^\frac{1}{2}_{\mathcal{O}=\mathcal{O}_\textnormal{exp}\, .}\label{eq:double}
\end{align}
For orthogonal tunings, $\nabla^{\mathcal{O}_a}\cdot \nabla^{\mathcal{O}_b}\rightarrow 0$ and thus $\Delta_2^{ab} \rightarrow \nabla^{\mathcal{O}_a} \nabla^{\mathcal{O}_b}$. For aligned tunings $\nabla^{\mathcal{O}_a} = \lambda \nabla^{\mathcal{O}_b}$, then $\nabla^{\mathcal{O}_a}\cdot \nabla^{\mathcal{O}_b}\rightarrow \lambda \nabla^{\mathcal{O}_a} \nabla^{\mathcal{O}_a}$ and thus $\Delta_2^{ab} \rightarrow 0$. Noting that \cref{eq:double} is the area spanned by any two tuning vectors, this behaviour should be intuitive. 

The total fine tuning $\Delta_2$ should then fulfil the criteria that 
\begin{enumerate}
\item For all observables independent it be a maximum,
\item For only one independent observable it vanish, and
\item \label{item:double_tuning_limit} For the limiting case of two independent observables, it simply be the single double-tuning measure.
\end{enumerate}
For three observables, the measure satisfying these is 
\begin{align}
\Delta_2 = \frac{1}{2}(\Delta_2^{ab} +\Delta_2^{bc}+\Delta_2^{ca})\, .\label{eq:totaldouble}
\end{align}
One can see that for observable $c$ proportional to $b$, $\mathcal{O}_c = \kappa \mathcal{O}_b$, then $\Delta_2^{bc}\rightarrow 0$ and $\Delta_2^{ac} \rightarrow \Delta_2^{ab}$. This comes from both \cref{eq:double} disappearing for aligned tunings, and \cref{eq:BG} being insensitive to a scaling $\kappa$.  Thus, $\Delta_2$ behaves as we would like.

The generalisation of \cref{eq:double} to three observables is also quite straightforward, where we take the volume spanned by three particular tuning vectors:
\begin{align}
\Delta_3^{abc} &= \begin{vmatrix}
\nabla^{\mathcal{O}_a}\cdot\nabla^{\mathcal{O}_a} & \nabla^{\mathcal{O}_a}\cdot \nabla^{\mathcal{O}_b} &\nabla^{\mathcal{O}_a}\cdot \nabla^{\mathcal{O}_c} \\\nabla^{\mathcal{O}_a}\cdot\nabla^{\mathcal{O}_b} & \nabla^{\mathcal{O}_b}\cdot \nabla^{\mathcal{O}_b} &\nabla^{\mathcal{O}_b}\cdot \nabla^{\mathcal{O}_c}\\\nabla^{\mathcal{O}_a}\cdot\nabla^{\mathcal{O}_c} & \nabla^{\mathcal{O}_b}\cdot \nabla^{\mathcal{O}_c} &\nabla^{\mathcal{O}_c}\cdot \nabla^{\mathcal{O}_c}
\end{vmatrix}^\frac{1}{2}_{\mathcal{O}=\mathcal{O}_\textnormal{exp}}\, . \label{eq:triple}
\end{align}
Being a volume, this follows the same behaviour as the double tuning derived above. We sum various triple tunings with the extension
\begin{align}
\Delta_3 &= \frac{1}{n_o-2} \sum\limits_{\{\,a,b,c\,\} | c<b<a}^{\binom{n_o}{3}} \Delta_3^{abc}\, .
\end{align}
In general, the $N$-th order of tuning of a set of $N$ particular observables $\bm{\nabla}_N=(\nabla^{\mathcal{O}_a},\nabla^{\mathcal{O}_b},...)$ is given by 
\begin{align}
\Delta^{ab...}_N = |\bm{\nabla}_N^T \cdot\bm{\nabla}_N|^{\frac{1}{2}}
\end{align}
and the $N$-th higher order tuning over \textit{all} $n_o$ observables is 
\begin{align}
\Delta_N &= \frac{1}{n_o-(N-1)} \sum\limits_{\{\,a,b,...\,\} | ...<b<a}^{\binom{n_o}{N}} \Delta_N^{ab...} \label{eq:HOT}\, .
\end{align}
Finally, we simply sum each order of tuning for a measure of higher order tuning:
\begin{align}
\Delta_\text{HOT} = \sum_{i=1}^{n_o} \Delta_i\, .
\end{align}

\subsection{Bugs and Features}


\subsubsection{Generic Scaling}

It should be clear that our new measure will give strictly greater tuning values compared to the $\Delta_\text{BG}$ measure, due to three factors. The first is obvious: we required it to have $\Delta_\text{BG}$ as its lower limit when all observables tend towards being completely dependent. The other factors are: arbitrary increase of parameters, and arbitrary increase of observables. For random fine-tuning vectors, we would expect the following general dependencies. 

At order one of tuning, the number of observables $N$ will not affect the measure as they are averaged out. In terms of $n_p$ from \cref{eq:firstorder}, $\Delta^\mathcal{O}$ goes as 
\begin{align}
\Delta^\mathcal{O} \sim \sqrt{n_p}\, .
\end{align}
At order two of higher order tuning - that is, double tuning - the measure goes as (\cref{eq:double} and \cref{eq:totaldouble}):
\begin{align}
\Delta_2 &\propto \frac{1}{n_o -1}\left(\begin{matrix}
n_o\\
2
\end{matrix}\right) = \frac{n_o!}{(n_o - 1)2!(n_o-2)!}=\frac{n_o}{2}\sim n_o\, ,\\
\Delta_2 &\propto \begin{vmatrix}
\nabla_a \cdot \nabla_a & \nabla_a \cdot \nabla_b\\
\nabla_a \cdot \nabla_b & \nabla_b \cdot \nabla_b
\end{vmatrix}^{1/2} \sim \sqrt{\nabla_a^2 \nabla_b^2} \sim n_p\, .
\end{align}
assuming mostly orthogonal observables. That is, at second order, the measure scales linearly both with number of parameters and observables. At third order, the measure goes as 

\begin{align}
\label{tripletuning}
\Delta_3 &\propto \frac{1}{n_o -1}\left(\begin{matrix}
n_o\\
3
\end{matrix}\right) \sim n_o^2\\
\Delta_3 &\propto n_p^{3/2}\, .
\end{align}

Higher orders $\Delta_N$ follow this pattern of $\sim n_o^{N-1},n_p^{N/2}$. Of course there is a further scaling of the measure when considering higher numbers of observables. When going from three to four observables, not only do we increase the fine tuning out-of-hand by $(4/3)^2 \approx 1.8$, we also add in the possibility of order-four tuning, which is generically a factor of $\Delta_1$ greater than order-three. In \cref{tab:hot_tuning}, we list some example scaling of each order as a function of number of observables and number of parameters.

\begin{table}
\centering
\hspace*{-8em}\begin{tabular}{ @{} rl | llllllllllllllllll @{} }\toprule
& $n_o$ & \multicolumn{3}{c}{1} & \multicolumn{3}{c}{2} & \multicolumn{3}{c}{3} & \multicolumn{3}{c}{4} & \multicolumn{3}{c}{5} & \multicolumn{3}{c}{6}\\
& $n_p$ & 5 & 10 & 25 & 5 & 10 & 25 & 5 & 10 & 25 & 5 & 10 & 25 & 5 & 10 & 25 & 5 & 10 & 25 \\
\cmidrule{2-20}
 \parbox[t]{2mm}{\multirow{6}{*}{\rotatebox[origin=c]{90}{Order}}} & \small 1 & \small 2.2 &\small 3.2 & \small 5 & \small 2.2 &\small 3.2 & \small 5 & \small 2.2 &\small 3.2 & \small 5 & \small 2.2 &\small 3.2 & \small 5 & \small 2.2 &\small 3.2 & \small 5 & \small 2.2 &\small 3.2 & \small 5 \\ 
& 2 & \small 2.5 &\small 5 &\small  13 &\small 5 &\small 10 &\small  25 &\small  7.5 &\small  15 &\small  38 &\small 10 &\small 20 &\small 50 &\small 13 &\small 25 &\small 63 &\small 15 &\small 30 &\small 75\\
& \small 3 & & & & \small 3.7 & \small 11 & \small42 & \small 11 & \small 32 & \small 125 & \small 22 & \small 63 & \small 250 & \small 37 & \small 105 & \small 417 & \small 56 & \small 158 & \small 625 \\
& \small 4 & & & & & & & \small 6.3 & \small 25 & \small 156 & \small 25 & \small 100 & \small 625 & \small 63 & \small 250 & \small 1.6e3 & \small 125 & \small 500 & \small 3.1e3\\
& \small 5 & & & & & & & & & & \small 11 & \small 63 & \small 625 & \small 56 & \small 316 & \small 3.1e3 & \small 167 & \small 949 & \small 9.3e3\\
& \small 6 & & & & & & & & & & & & & \small 21 & \small 167 & \small 2.6e3 & \small 125 & \small 1e3 & \small 1.6e4\\
\cmidrule{2-20}
 \multicolumn{2}{c}{Sum} & \small 4.7 &\small 8.2 &\small 18 & \small 11 &\small 24 &\small 72 &\small 27 &\small 75 & \small 324 & \small 71 & \small 250 & \small 1.6e3 & \small 191 & \small 867 & \small 7.8e3 & \small 491 & \small 2.6e3 & \small 2.9e4\\
\bottomrule
\end{tabular}
\vspace*{1em}\caption{Example scaling of the higher order tuning, assuming a first-order tuning $\Delta_\text{BG} \sim \mathcal{O}(1)$}\label{tab:hot_tuning}
\end{table}

\begin{figure}
\centering
\tikz{
\node at (5,5) {\includegraphics[scale=0.6]{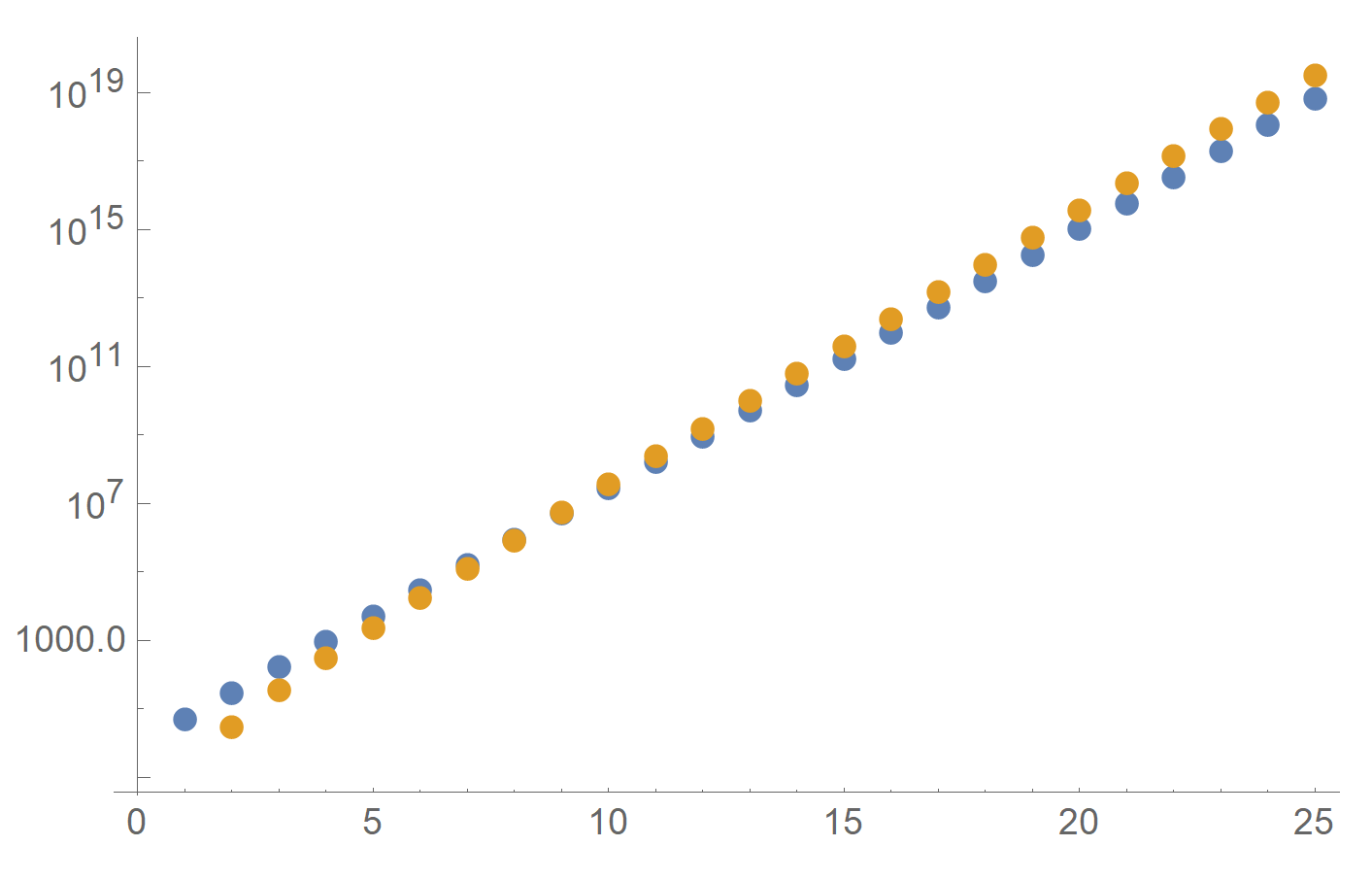}};
\node at (8,4) {\includegraphics[scale=0.6]{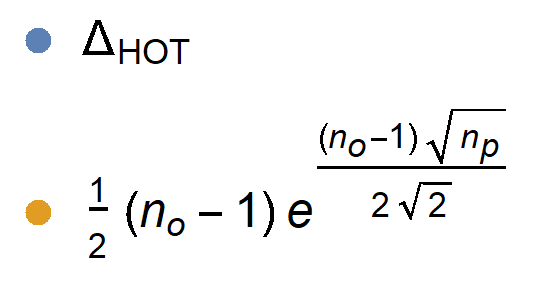}};
\draw [draw = white, fill = white] (6.7,3) rectangle (10,5);
\node [right] at (6.6,4.6) {$ \Delta_\text{HOT}$};
\node [right] at (6.6,3.6) {$ \frac{n_o - 1}{2} \exp\left( \sqrt{\frac{n_p}{2}} \frac{n_o}{2}\right) $};
\node at (5,2) {$n_o$};
\node [rotate=90] at (0.5,5) {Tuning};
}\caption{Exponential scaling of higher order tuning when increasing the number of observables, given a model of $n_p = 25$. }\label{fig:tuning_scaling}
\end{figure}
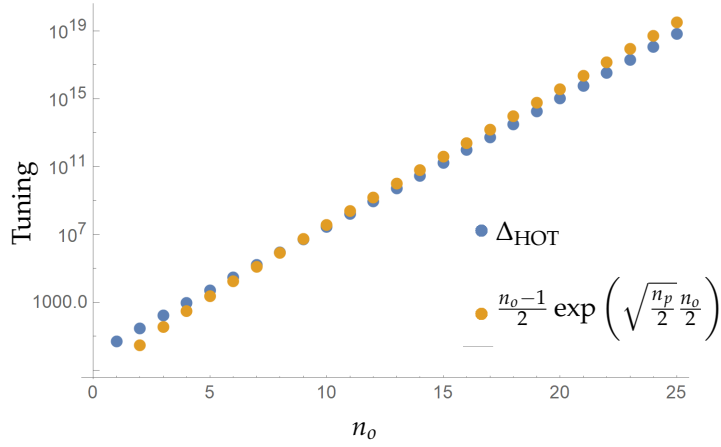

Given $\frac{\partial \mathcal{O}^a}{\partial x^i} \sim \frac{\partial \mathcal{O}}{\partial x}  \sim \mathcal{O}(1)$, the tuning scales as
\begin{align}
\Delta_\text{HOT} \sim n_p^{1/2} \frac{\partial \mathcal{O}}{\partial x} + \frac{1}{2} n_p n_o \left(\frac{\partial \mathcal{O}}{\partial x}\right)^2 + \cdots + \frac{1}{N} n_p^{N/2} {n_o \choose N-1} \left(\frac{\partial \mathcal{O}}{\partial x}\right)^N\label{eq:tuningprediction}
\end{align}
which we recognise has exponential qualities. Indeed, plotting this tuning, as in \cref{fig:tuning_scaling} shows an exponential growth in $n_o$ and $\sqrt{n_p}$. Specifically, we have a growth that can be approximated by 
\begin{align}
\Delta = \frac{n_o - 1}{2} \exp\left( \sqrt{\frac{n_p}{2}} \frac{n_o}{2}\right)
\end{align}
This generic scaling as a function of number of parameters can be considered a punishment for more complicated models, and is an intuitive feature of the model. The scaling with number of observables is, on the other hand, problematic. We should be rewarding a model for attempting to describe more observables, provided the parameter points have not been excluded. The naive preference would be to normalise each order - to average the tunings as we do with the first order. For example, a second order tuning may be normalised to be
\begin{align}
\Delta_\text{norm.}^2 &= {n_o \choose 2}^{-1} \left( \Delta^{ab} + \Delta^{ac} + ... + \Delta^{yz} \right)\label{eq:HOT_normalisation_2}
\end{align}
thereby removing the scaling. However, this does not limit correctly to our criterion \ref{item:double_tuning_limit}, of a single double tuning for dependent observables. In fact, \textit{no} simple normalisation will give this correct limit. Let us see why. 

%

\subsubsection{Dependency Configurations}

\begin{figure}[H]
\centering
\subfloat[We have one set of dependencies $\mathcal{O}_a = \kappa_1\mathcal{O}_b = \kappa_2 \mathcal{O}_d$]{\includegraphics[width=0.25\linewidth]{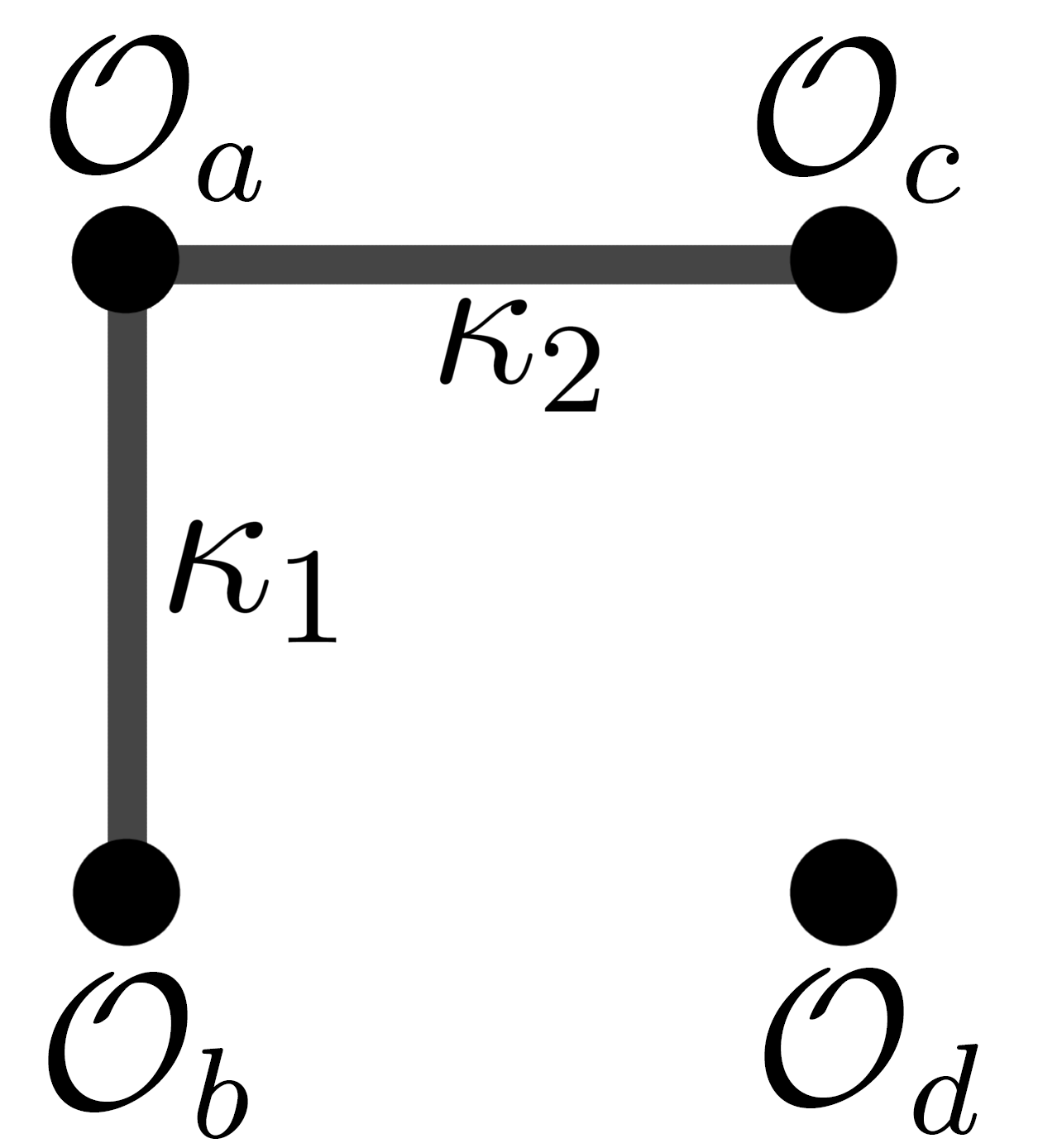}}\qquad\qquad\qquad
\subfloat[We have two dependencies $\mathcal{O}_a = \kappa_1\mathcal{O}_b$ and $\mathcal{O}_c = \kappa_2 \mathcal{O}_d$]{\includegraphics[width=0.25\linewidth]{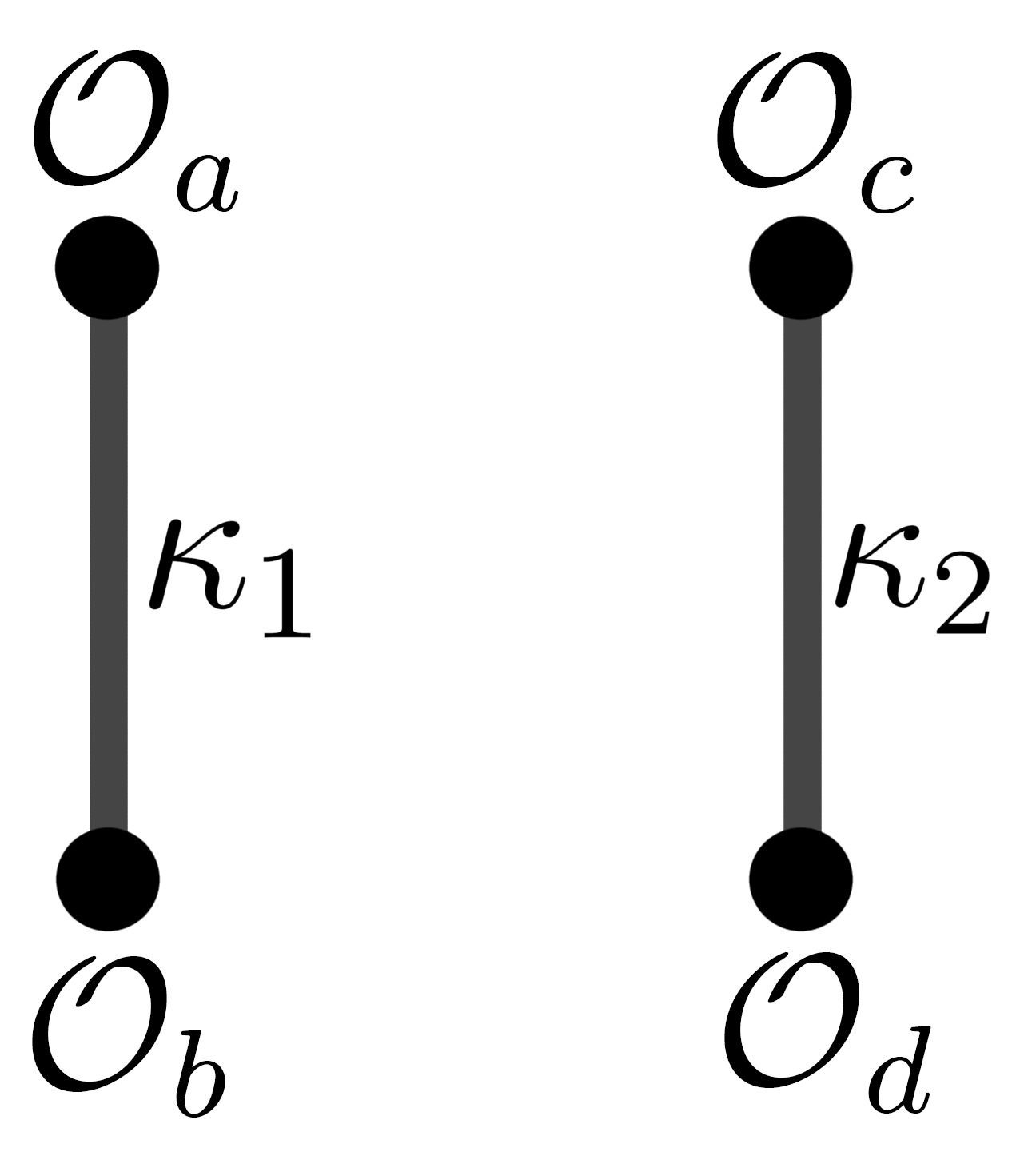}}
\caption{The configurations available for one source of double-tuning amongst four observables}
\label{fig:configs}
\end{figure}

Notice that for more than three observables, the criterion \ref{item:double_tuning_limit} following \cref{eq:totaldouble} is not unique. For example, given four observables $\{a,b,c,d\}$, there are two configurations for observables $a$ and $b$ to be independent. Configuration 1 has all dependency on one observable, configuration 2 has the dependency shared across variables, shown in figure \ref{fig:configs}.

Configuration 1 algebraically satisfies criterion \ref{item:double_tuning_limit} in a simple extension of \cref{eq:totaldouble} to unordered pairs over $n_o$ observables
\begin{align}
\Delta_2 &= \frac{1}{n_o-1} \sum\limits_{\{\,a,b\,\} | b<a}^{\binom{n_o}{2}} \Delta_2^{ab} \xrightarrow{b \rightarrow c \rightarrow d} \Delta_2^{ab} \label{eq:gendoubletotal}\, .
\end{align}
However, calculating \cref{eq:gendoubletotal} for configuration 2 gives more unordered pairs, and thus a factor of $4/3$ above configuration 1. This incorrect limit highlights that we have lost structure by reducing the four-dimensional volume to a collection of two-dimensional areas, and hence no simple normalisation can re-instate that geometric information. We are now at the heart of the matter: Our intuitive Higher Order Tuning is an attempt to capture the behaviour of the Bayesian sensitivity in \cref{eq:formal_sensitivity}, while tackling its major shortcoming. This shortcoming was commented on in \cite{fichet2012quantified} by S. Fichet, where the sensitivity of two observables was calculated as the norm of the wedge product
\begin{align}
C  = \| \nabla \log \mathcal{O}_1 \wedge \nabla \log \mathcal{O}_2 \|,
\end{align}
which is an infinitesimal two-volume in $n_p$-space, as we intuited in sections \ref{sec:tuning_volume} and \ref{sec:intuitive_measure}. Fichet recasts this as a function of the one-dimensional (BG) sensitivities and a correlation $C = C_1 C_2 \sqrt{1 - \rho^2}$. For two sensitivity vectors $\vec{C}_1, \vec{C}_2$ this is clearly the cross product, with $\rho = \cos\theta$ capturing the dependency of the two. If we consider, as above, the case of the vectors tending to total correlation, we get $C \rightarrow 0$, and thus $\int_\mathcal{V} \frac{1}{C} d\mathcal{O}^n(\textbf{x}) \rightarrow \infty$ in \cref{eq:formal_sensitivity}. In that work, Fichet points out that we arrived at \cref{eq:formal_sensitivity} on the \textit{assumption} that each observable is ``'informative", which is quantified by the requirement $C |V|^{1/2} \gg |\Sigma|^{1/2}$. The author suggested to me that this captures the experimental uncertainty in the data. Provided the experimental data is Gaussian, the point at which the measure begins to lose validity is at the limit $\rho_\text{exp} \approx \rho_\text{th}$. This is one possible avenue to solve the problem of vanishing sensitivity.

The other avenue is to take the Bayesian sensitivity as attempting to capture the volume change in observable space given an infinitesimal volume change in parameter space, and define this as the consistent measure of tuning. We thus take the formal measure as \textit{motivation} for the tuning concept explored between eq.s (\ref{eq:double}) and (\ref{eq:HOT}), but normalised in some way as to account for non-informative observables. 

Consider the space of two observable tuning vectors, which are linearly independent. This space is represented by $\mathbb{R}^2$. Now introduce a third tuning vector. The non-informative limit of this observable's tuning is that it is completely redundant - that its tuning vector is identical to an existing observable's tuning vector. In that case we have not added to the dimensionality of the observable space. In fact, any case where the new vector is linearly dependent on the previous two (which is to say that it lies in the same $\mathbb{R}^2$ plane) is non-informative. Clearly, this is a case where orthogonalising the tuning vectors will lead to a set of $k$ linearly independent vectors, and full informativity. This is why we needed to normalise each order of the tuning: to handle the case where an observable tuning vector was linearly dependent.

As elegantly explored in \cite{knill2014cauchy}, we can handle this case by using the pseudo-determinant $\text{Det}(J^T J)$, rather than the usual determinant $\text{det}(J^TJ)$. This is defined as the product of $k$ \textit{non-zero} eigenvalues. Indeed, this is the value we calculate if we were to orthogonalise our observable tuning space and then calculate the (average) of the $k$-volumes. The procedure is thus: i) Project each tuning vector onto some orthogonal basis; ii) Average the vectors along each basis direction; iii) Calculate the volumes at each order $N$ as the rectangular $N$-volume bordered by each combination of $N$ orthogonalised tuning vectors; iv) Sum all orders of tuning, as in \cref{eq:HOT}. This procedure should not surprise you - it is almost precisely that of the Higher Order Tuning definition, but beginning with an orthogonal set of vectors, and thus a simpler normalisation. This is, I believe, a rigorous and informative (in the sense of the Fisher information) extension of the Bayesian evidence. 

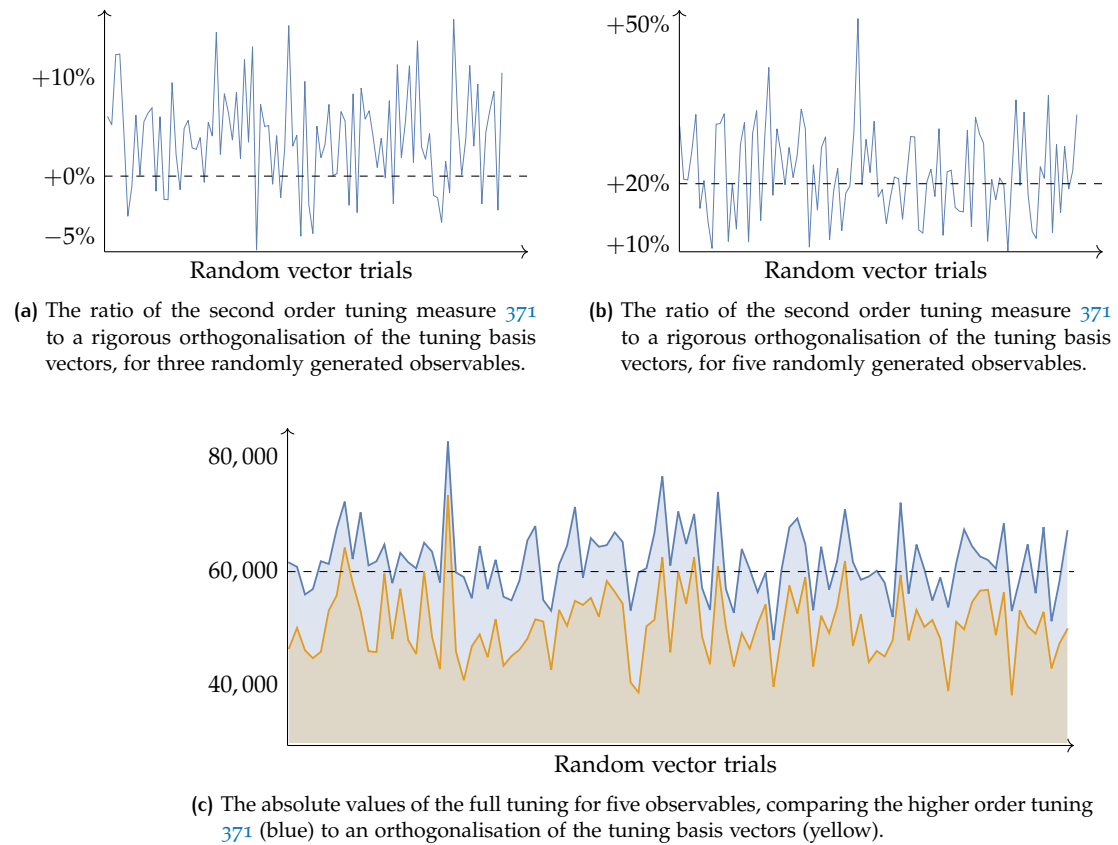
\begin{figure}
\centering
\hspace*{-6em}\subfloat[The ratio of the second order tuning measure \ref{eq:HOT} to a rigorous orthogonalisation of the tuning basis vectors, for three randomly generated observables.]{
\tikz{
\node at (3,1.5) {\includegraphics[scale=0.05]{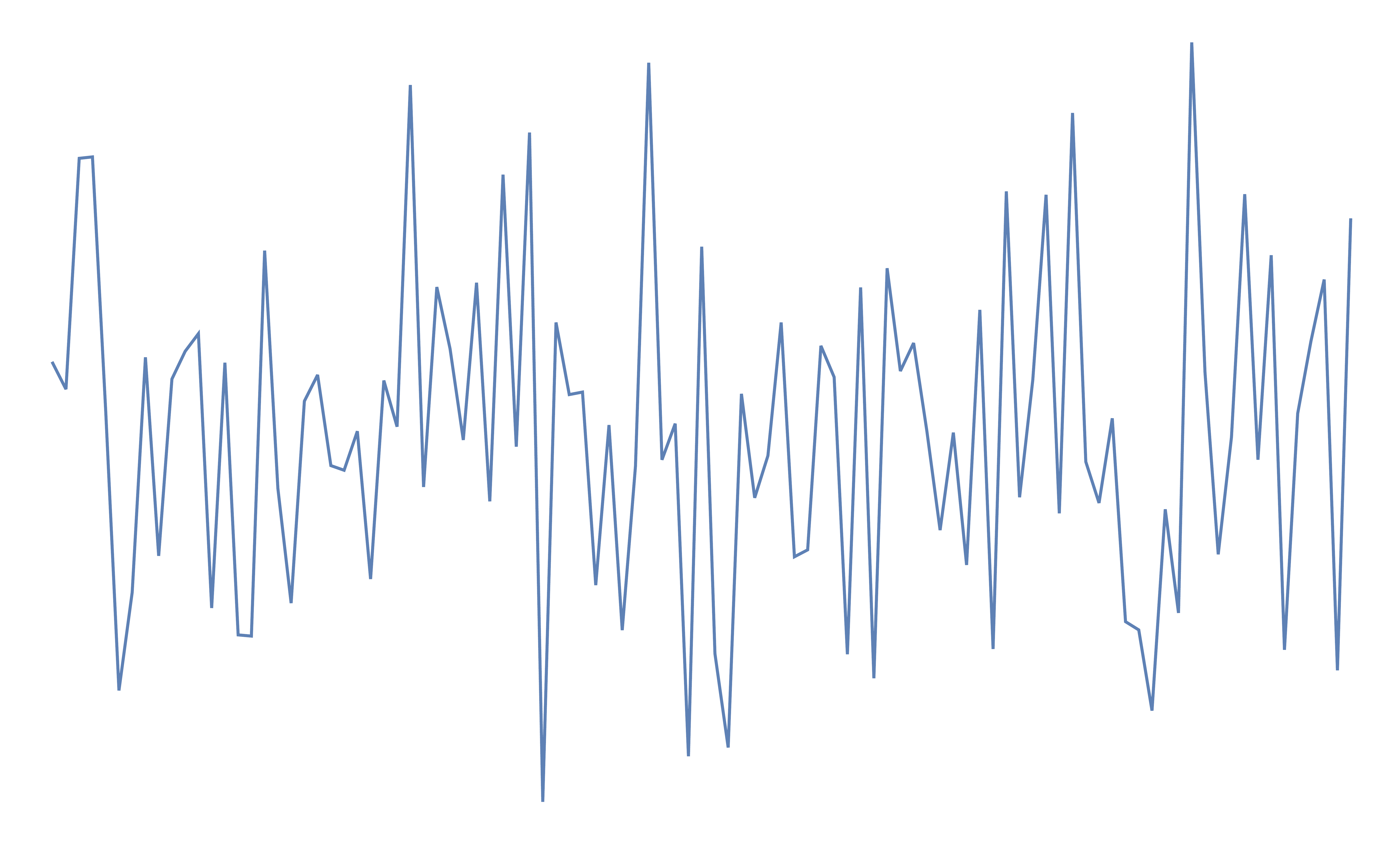}};
\draw [->] (0.4,0)  -- (6,0);
\draw [->] (0.4,0)  -- (0.4,3.2);
\draw [dashed] (0.4,1) -- (6,1);
\node [left] at (0.4,1) {\small $+ 0\%$};
\node [left] at (0.4,0.2) {\small $- 5\%$};
\node [left] at (0.4,2.3) {\small $+ 10\%$};
\node [below] at (3,0) {\small Random vector trials};
}
}\qquad \subfloat[The ratio of the second order tuning measure \ref{eq:HOT} to a rigorous orthogonalisation of the tuning basis vectors, for five randomly generated observables.]{
\tikz{
\node at (3,1.5) {\includegraphics[scale=0.05]{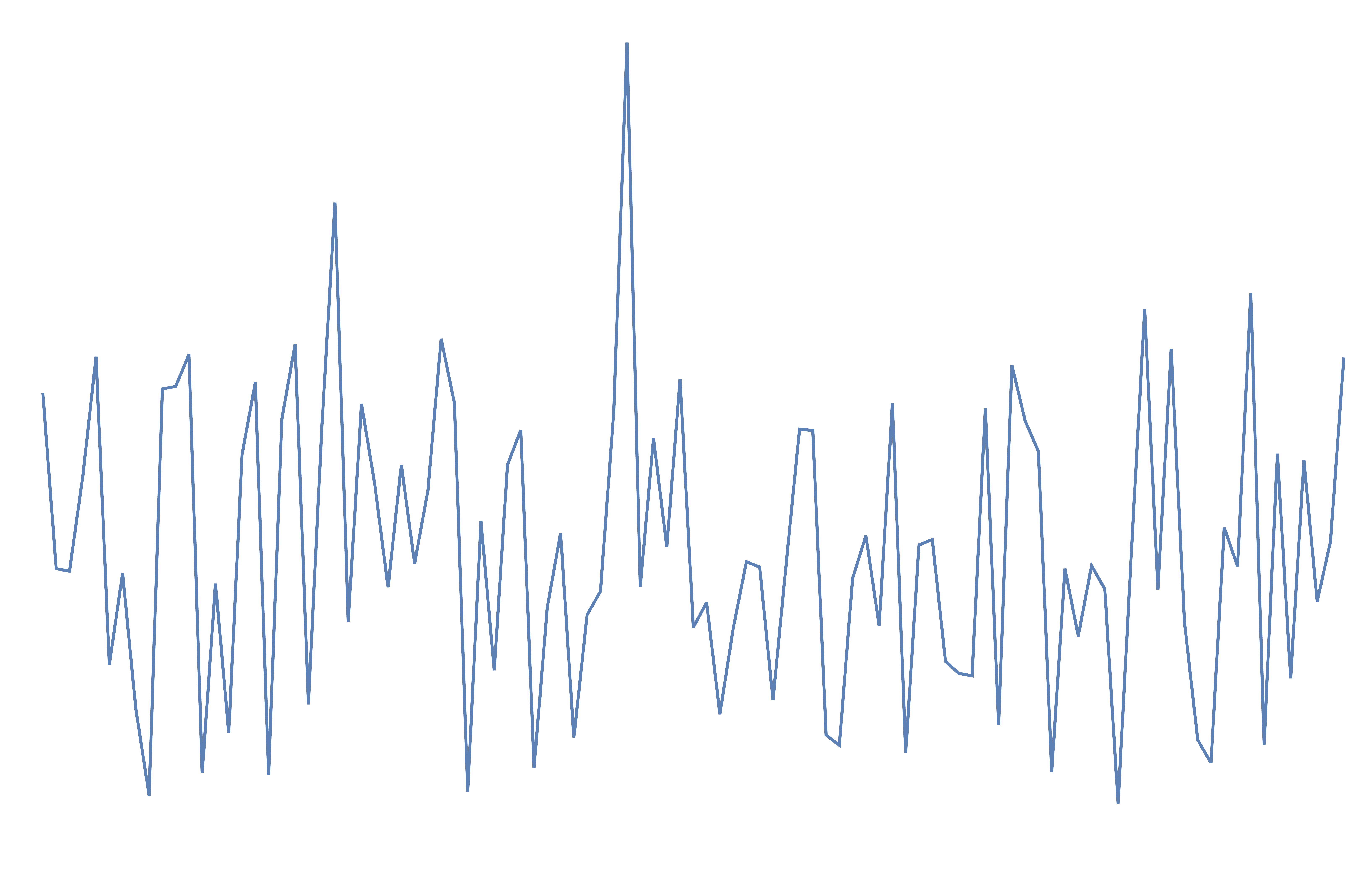}};
\draw [->] (0.4,0)  -- (6,0);
\draw [->] (0.4,0)  -- (0.4,3.2);
\draw [dashed] (0.4,0.9) -- (6,0.9);
\node [left] at (0.4,0.9) {\small $+ 20\%$};
\node [left] at (0.4,0.1) {\small $+ 10\%$};
\node [left] at (0.4,3) {\small $+ 50\%$};
\node [below] at (3,0) {\small Random vector trials};
}
}\\
\subfloat[The absolute values of the full tuning for five observables, comparing the higher order tuning \ref{eq:HOT} (blue) to an orthogonalisation of the tuning basis vectors (yellow).]{
\tikz{
\node at (5.4,2.2) {\includegraphics[scale=0.1]{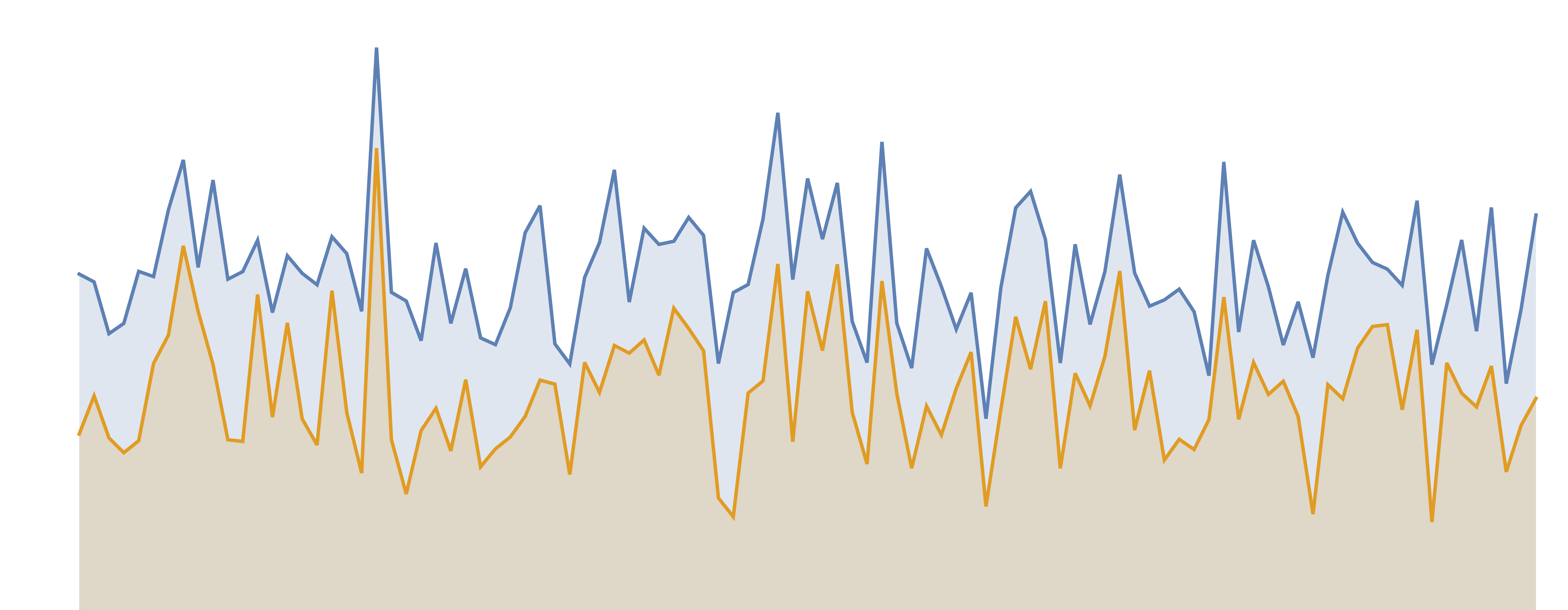}};
\draw [->] (0.4,0)  -- (10.8,0);
\draw [->] (0.4,0)  -- (0.4,4.2);
\draw [dashed] (0.4, 2.3) -- (10.8, 2.3);
\node [left] at (0.4, 3.8) {\small $80,000$};
\node [left] at (0.4, 2.3) {\small $60,000$};
\node [left] at (0.4, 0.8) {\small $40,000$};
\node [below] at (5.4,0) {\small Random vector trials};
}}\caption{Comparisons of the higher order tuning approximation to the orthogonalisation procedure. The approximation generally overestimates tuning, but not significantly.}\label{fig:HOT_comparison}
\end{figure}

It is, however, computationally expensive to orthogonalise the observable tuning vectors at every explored parameter point, compared to the Higher Order Tuning measure. We plot in \cref{fig:HOT_comparison} some example random distributions, comparing the deviation of the HOT measure from the rigorous orthogonalisation procedure. We see that inaccuracy of the HOT measure (due to the multiple possible configurations) is typically below $10 \%$ for three observables. For the purposes of the tuning in this thesis, we accept this as a good approximation. Particularly as for most of this work, we will only be considering up to $5$ tuned observables, where the dependency configuration scaling is not significant. In the full global fit of \cref{sec:M4DCHM}, we do consider considerably more observables, but the computational intensity of diagonalising coupling matrices in that parameter space also requires that we take the simpler HOT measure as a good approximation of tuning.

\chapter{Numerical Exploration}
\label{sec:scanning_techniques}

\section{The 4D Composite Higgs Parameter Space}\label{sec:general_parameters}

In \cref{sec:low_energy_lagrangian} we described a full parameterisation of the composite Higgs potential, required in order to derive the deviation from SM observables. While we laid out the conditions for convergence, the actual calculation of the integrals in \cref{eq:potential_again} cannot normally be done analytically. The exception to this is the (quadratic) contribution of the gauge boson, given in \cref{sec:form_factors}, which is a closed function 
\begin{align}
V_\text{gauge}(g_{\rho^i}, g_{X^i}, g_{G^i},m_{\rho^i}, m_{a^i})
\end{align}
Observe that we have reparameterised here, using convenient mass parameters
\begin{align}
m_{\rho^i}^2 = \frac{g_{\rho^i}^2 f_i^2}{2} && m_{a^i}^2 = \frac{g_{\rho^i}^2(f_i^2 + f_{i+1}^2)}{2}
\end{align}
Also note that in most of the following work, $g_{X^i} \sim g_{G^i} \sim g_{\rho^i}$. Generally then, for $N$ layers of resonances, this analytic contribution is a function of $2N$ mass-type parameters, each of which we take to be of the order of the relevant symmetry breaking scale $f_i$. However, we allow that the lightest layer could be unnaturally below the canonically normalised breaking scale $f = \left( \sum\limits_{k=1}^N \frac{1}{f^2_k} \right)^{-1}$. The upper bound should be the cutoff scale of the composite Higgs model $\Lambda \sim 4\pi f$. The $N$ composite gauge couplings $g_{\rho^i}$ are assumed to be significantly larger than the weak coupling, but perturbative $1 < g_{\rho^i} < 4\pi$.

For the case of the two-site 4D CHM, where the second layer is taken as infinitely heavy, as in this work, there are only two mass parameters $m_\rho, m_a$ and one coupling $g_\rho$ at leading order.

Let us consider the typical parameters of the fermionic sector. To remind, the terms in our (up \textit{and} down-type) fermion Lagrangian which have independent parameters are
\begin{align}
\mathcal{L}_{e-c} &= \bar{q}^5_L i \slashed{D} q^5_L  + \sum\limits_{k=1}^N \left( \bar{\Psi}^k \left(i \slashed{D} - m_{\Psi^k} \right)\Psi^k + \bar{\tilde{\Psi}}^k \left( i\slashed{D} - m_{\tilde{\Psi}^k} \right) \tilde{\Psi}^k\right) \nonumber\\
&+ \bar{q}^5_R i \slashed{D} q_R^5 +  \Delta_{\Psi^1} \bar{q}^5_L\Omega_1 \Psi_R^1 +  \Delta_{\tilde{\Psi}^1} \bar{q}^5_R\Omega_1 \tilde{\Psi}_L^1 + \sum\limits_{k=1}^{N-1}\left(\Delta_{\Psi^k} \bar{\Psi}^k \Omega_k \Psi_R^{k+1} + \Delta_{\tilde{\Psi}^k} \bar{\tilde{\Psi}}_R^k \Omega_k \tilde{\Psi}_L^{k+1} \right) \nonumber\\
& - m_Y \bar{\Psi}_L^N \tilde{\Psi}_R^N - \mathcal{L}_\text{yuk}(Y^a) + \text{h.c.} \label{eq:general_fermion_lagrangian_ch4}
\end{align}
where the Yukawa term is dependent on choice of representation. If both up and down components are included, and the symmetric representation is chosen for both, the Yukawa term depends on four parameters. If only one component is included in the fundamental, the Yukawa term only depends on one parameter. We note that the Yukawa terms have dimension of mass, as do the couplings $m_k = m_{\Psi^k}, m_{\tilde{\Psi}^k}, \Delta_{\Psi^k}, \Delta_{\tilde{\Psi}^k}$ and $m_Y$. For the case of all fermions in the fundamental representation, and considering up and down types, there are $8N+4$ fermionic parameters. For the left doublet not in the fundamental, there are $6N+3$ fermionic parameters, as we can couple the left SM doublet to a composite resonance containing both up and down quantum numbers. We take each of these dimensionful parameters to range over $0 < m_i < 4\pi f_k$.

\section{Sampling Algorithms}

\begin{table}
\centering
\begin{tabular}{ @{} lll@{} }\toprule
Model & $n_p$ & $n_o$\\
\midrule
\small M4DCHM$^{5-5}$ (top only) & 9 & 3 \\
\small NM4DCHM$^{5-5}$ (top only) & 12 & 3 \\
\small M4DCHM$^{5-5-5}$ (simplified\footnote{Where we take the $X$ group terms to be equal to the $SO(5)$ group terms} top and bottom) & 15 & 3 \\
\small M4DCHM$^{5-5-5}$ (full top and bottom) & 22  & 18 \\
\small LM4DCHM$^{5-5-5}_{14-1-10}$ (3rd generation leptonic) & 23 & 5 \\
\small LM4DCHM$^{5-5-5}_{14-14-10}$ (3rd generation leptonic) & 25 & 5 \\
\small LM4DCHM$^{5-5-5}_{5-5-5}$ (3rd generation leptonic) & 27 & 5 \\
\bottomrule
\end{tabular}\caption{An example set of models considered in this work, and the dimension of each parameter space}\label{tab:example_parameter_spaces}
\end{table}

For an $N$-site, M4DCHM model, we can have up to $5N$ gauge parameters, and $8N+4$ parameters for each generation of partially composite SM field. We give some example parameter spaces, as studied in this work, in \cref{tab:example_parameter_spaces}. Random scanning of this parameter set would be an arduous task. To make exploring this theory possible requires either making simplifications to the parameter space, or a more educated method of sampling. In the following chapters, we will present simplified models that attempt to still capture the dominant contributions to composite Higgs physics. In this chapter, we will describe two approaches to more sophisticated sampling.

\subsection{{\tt MultiNest}}
\label{sec:multinest}


\subsubsection{Nested Sampling}

\begin{figure}
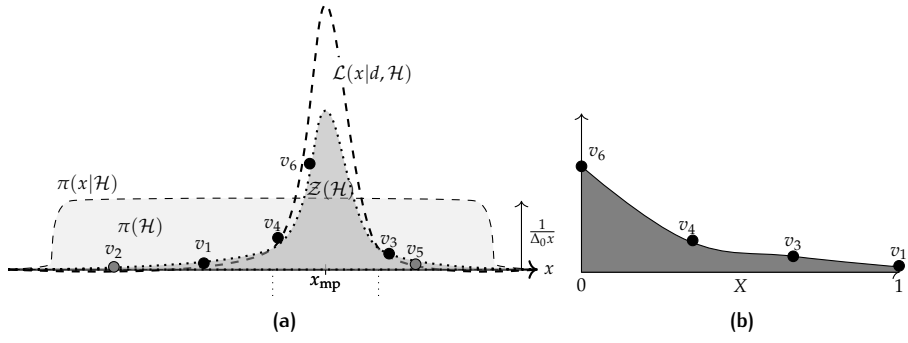

\centering
\subfloat[]{\tikz[scale=0.7,  every node/.style={transform shape}]{
\draw [->, thick] (1,2) -- (11,2);
\draw (7,2.1) -- (7,1.9);
\draw [dotted] (6,1.5) -- (6,2);
\draw [dotted] (8,1.5) -- (8,2);
\node [below] at (7,2) {$x_\text{mp}$};
\draw [dashed, fill=gray!20, fill opacity = 0.5] plot [smooth, tension=0.2] coordinates {(1,2) (1.8,2.1) (2.2,3.3) (9.8,3.3) (10.2,2.1) (11,2) } --cycle;
\node at (3.5,2.8) {$\pi(\mathcal{H})$};
\node [above] at (2.5,3.3) {$\pi(x| \mathcal{H})$};
\draw [thick, dashed] plot [smooth, tension=0.45] coordinates { (1,2) (6,2.4) (7,7) (8,2.4) (11,2) };
\draw [thick, dotted, fill=gray!70, fill opacity= 0.5] plot [smooth, tension=0.5] coordinates { (1,2) (6,2.4) (7,5) (8,2.4) (11,2) } -- cycle;
\draw [->] (10.7,2) -- (10.7,3.3);
\node [right] at (10.7,2.7) {$\frac{1}{\Delta_0 x}$};
\node [right] at (11,2) {$x$};
\node [below] at (7,2) {$x_\text{mp}$};
\node [above] at (7.1,3.2) {$\mathcal{Z}(\mathcal{H})$};
\draw [draw = white, fill = white] (7,5.5) rectangle (8,6);
\draw [fill=black] (4.7,2.12) circle [radius=0.1];
\node [above] at (4.7,2.2) {$v_1$};
\draw [fill=gray] (3,2.05) circle [radius=0.1];
\node [above] at (3,2.1) {$v_2$};
\draw [fill=black] (8.2,2.3) circle [radius=0.1];
\node [above] at (8.2,2.3) {$v_3$};
\draw [fill=black] (6.1,2.6) circle [radius=0.1];
\node [above] at (6,2.6) {$v_4$};
\draw [fill=gray] (8.7,2.1) circle [radius=0.1];
\node [above] at (8.7,2.1) {$v_5$};
\draw [fill=black] (6.7,4) circle [radius=0.1];
\node [left] at (6.6,4) {$v_6$};
\node [above, right] at (7,5.7) {$\mathcal{L}(x | d, \mathcal{H})$};
}}
\subfloat[]{
\tikz[scale=0.7,  every node/.style={transform shape}]{
\draw [->] (2,2) -- (8,2);
\draw [->] (2,2)--(2,5);
\node [below] at (5,2) {$X$};
\node [below] at (2,2) {$0$};
\node [below] at (8,2) {$1$};
\draw [fill=gray] plot [smooth, tension=0.7] coordinates {(8,2.1) (6,2.3) (4,2.6) (2,4)}-- (2,2) -- (8,2) --cycle;
\draw [fill=black] (8,2.12) circle [radius=0.1];
\node [above] at (8,2.12) {$v_1$};
\draw [fill=black] (6,2.3) circle [radius=0.1];
\node [above] at (6,2.3) {$v_3$};
\draw [fill=black] (4.1,2.6) circle [radius=0.1];
\node [above] at (4,2.6) {$v_4$};
\draw [fill=black] (2,4) circle [radius=0.1];
\node [above] at (2.3,4) {$v_6$};
}
}\caption{Posterior distribution plot. Numerical integration of $\mathcal{L}dx =\mathcal{L} dX$ can be achieved with  ordered sampling of (a), assuming a monotonically increasing likelihood function, as in (b). Points $v_2$ and $v_5$ are thrown away as they are not more likely than $v_1$ and $v_4$ respectively.}\label{fig:numerical_evidence}
\end{figure}

The first and second levels of inference discussed in \cref{sec:existing_tuning_measures} can be described as the problem of exploring likely areas of a parameter space, and the problem of obtaining a Bayesian evidence for that parameter space. It is not obvious that the two goals are simultaneously compatible. However, nested sampling \cite{skilling2006nested} exploits a computational convenience to achieve both in the same run-time. Let us re-examine the posterior of a Gaussian function with a uniform prior, in \cref{fig:numerical_evidence}. We are evaluating the integral 
\begin{align}
\mathcal{Z}  = \int_\mathcal{V} \mathcal{L}(x) \pi(x) dx
\end{align}
where we can define a prior volume $\Delta_0 x$, prior density $\pi(x)$, equal to a uniform $\frac{1}{\Delta_0 x}$ in our toy, and therefore a prior mass  defined as
\begin{align}
 X := \int_\lambda \pi(x) dx \xrightarrow{\lambda \rightarrow \mathcal{V}} 1.
\end{align} 
The essence of nested sampling is that the volume defined by $\lambda$ is shrunk from $\mathcal{V}$ to zero according to the requirement that the points it contains are all above the likelihood of the previous $\lambda$. We can use this definition to perform a change of variable on the integral
\begin{align}
\mathcal{Z} = \int\limits_0^1 \mathcal{L}(X) dX
\end{align}
which is now one-dimensional. We have transformed the problem of a potentially complex multidimensional integral, to that of finding progressively more likely contours, and numerically integrating this one-dimensional function. This may not be computationally easier, but it now matches our usual goal of exploring likely parameter space. Thus, we receive a posterior distribution as a by-product. An additional convenience of this procedure is that it is iterative. Rather than dividing up our prior mass and sorting it from lowest to highest, we can sample with $N$ points, throw away the least likely point and replace it (an $\mathcal{O}(N)$ operation). This removes the requirement of constant sorting (an $\mathcal{O}(N\log N)$ operation), hence the moniker "nested". It still remains to decide how to choose successive parameter points. The {\tt MultiNest} package \cite{feroz2008,feroz2013} provides a well-motivated selection procedure.

\subsubsection{{\tt MultiNest} Implementation}

It can be shown \cite{skilling2006nested} that, assuming a posterior of multiple Gaussian peaks, the vast majority of prior volume will be useless in calculating the evidence. We can see this in \cref{fig:numerical_evidence}. The problem scales with the dimension of parameter space. Thus, instead of uniformly decreasing the prior mass $X_{i+1} = X_i - \delta$, which would converge very slowly, by using the sorted list technique we approximately follow a logarithmic progression $X_{i+1} = t_{i+1} X_i$, $t < 1$. This is sketched in  \cref{fig:numerical_evidence}. For a large number of points, these likelihood contours will appear as nested shells. {\tt MultiNest} goes further, and \textit{assumes} the shells as ellipsoidal, with new points chosen from within each successive shell. If this assumption is good, for a particular posterior, the acceptance rate for new points will be greatly increased over random scanning. In this way, a scan is a sequence of shrinking ellipsoids, converging on a point of local maximum likelihood. 

\begin{figure}
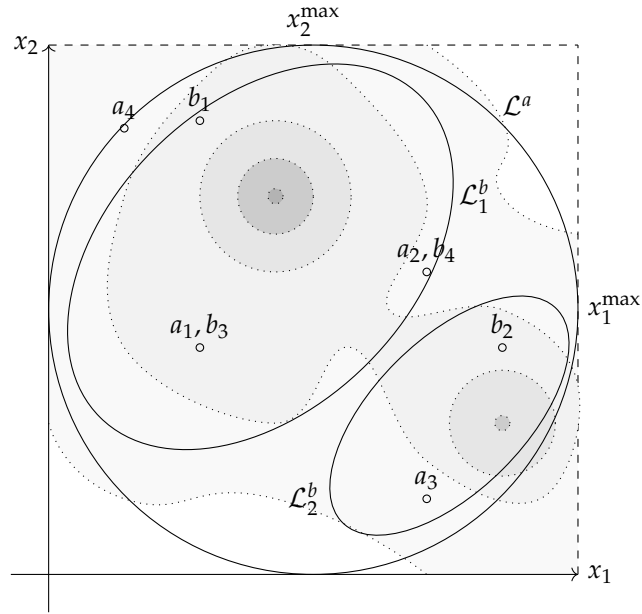

\centering
\tikz[scale=1]{
\fill [gray!5] plot [smooth, tension=1] coordinates {(7,9) (8,8) (8,7) (9,6.5)} -- (9,2) -- plot [smooth, tension=1] coordinates {(7,2)  (5,3) (3,3) (2,4) } -- (2,9) -- cycle;
\draw [->] (1.5,2) -- (9,2);
\draw [->] (2,1.5) -- (2,9);
\draw [dashed] (2,9) -- (9,9);
\draw [dashed] (9,2) --(9,9);
\node [right] at (9,5.5) {$x_1^\text{max}$};
\node [above] at (5.5,9) {$x_2^\text{max}$};
\node [right] at (9,2) {$x_1$};
\node [left] at (2,9) {$x_2$};
\draw [fill=gray!10,dotted] plot [smooth cycle, tension=0.7] coordinates {(3,5) (3,7) (4,8.5) (5,9) (6,8.5) (7,7) (6.5,5.5) (8,5.5) (9,4.5) (8.5,3) (7,3.5) (6,5) (5,4)};
\draw [fill=gray!20, dotted] (5,7) circle [radius=1];
\draw [fill=gray!40,dotted] (5,7) circle [radius=0.5];
\draw [dotted, fill=gray!60] (5,7) circle [radius=0.1];
\draw [dotted] plot [smooth, tension=1] coordinates {(7,9) (8,8) (8,7) (9,6.5)};
\draw [dotted] plot [smooth, tension=1] coordinates {(2,4) (3,3) (5,3) (7,2)};
\draw [dotted, fill=gray!20] (8,4) circle [radius=0.7];
\draw [dotted,fill=gray!40] (8,4) circle [radius=0.1];
\draw (3,7.9) circle [radius=0.05] node [above] {$a_4$};
\draw (7,3) circle [radius=0.05] node [above] {$a_3$};
\draw (4,5) circle [radius=0.05] node [above] {$a_1, b_3$};
\draw (7,6) circle [radius=0.05] node [above] {$a_2, b_4$};
\draw  (5.5, 5.5) circle [radius=3.5];
\node at (8.2,8.2) {$\mathcal{L}^a$};
\draw (8,5) circle [radius=0.05] node [above] {$b_2$};
\draw (4,8) circle [radius=0.05] node [above] {$b_1$};
\draw [ rotate around={135:(4.8,6.2)}] (4.8,6.2) ellipse (2 and 3);
\node [right] at (7.3,7) {$\mathcal{L}^b_1$};
\draw [ rotate around={135:(7.3,4.1)}] (7.3,4.1) ellipse (1 and 2);
\node [left] at (5.7,3) {$\mathcal{L}^b_2$};
}\caption{Ellipsoidal nesting in MultiNest for a population of 4, replacing 2 points per generation. For each generation $a,b,...$, each vector is ordered from highest likelihood to lowest: $\{a_1, a_2, a_3, a_4\}$. Thus $a_3$ and $a_4$ are replaced by $b$ vectors, which are then re-ordered. Elliptical contours are drawn with average likelihoods $\mathcal{L}^a, \mathcal{L}^b,...$. These are integrated over, as in \cref{fig:numerical_evidence} (b).}\label{fig:multinest_clustering}
\end{figure}

There is the question of multiple local maxima, which ellipsoidal nesting has typically struggled with. {\tt MultiNest} uses a technique of clustering to determine whether a single shrinking ellipsoid should be used in the next iteration, or should be split into two separate (and possibly overlapping) ellipsoids. This is sketched in \cref{fig:multinest_clustering}. This package allows parallelisation, which we have used in this study.

{\tt MultiNest} can deliver the valuable Bayesian evidence, to compare between models. For this purpose, {\tt MultiNest} is a gold standard. This goal comes at the cost of relatively slow convergence on local maxima, as well as a plummeting acceptance rate, as it is guided only by the heuristic of ellipsoidal nesting. For much of this work, we will be interested in quickly exploring large parameter spaces with appropriately huge populations of candidate vectors, quickly converging on local maxima. Evolutionary algorithms are a powerful tool for this.

\subsection{{\tt Diver}}

\subsubsection{Evolutionary Algorithms}

Just as in the {\tt MultiNest} algorithm, an evolutionary algorithm (EA) requires a cost function\footnote{Clearly, we use the notation of $\chi^2$ for the cost function as we will be implementing a $\chi^2$ paradigm, but any cost function is valid.} $\chi^2$. Unlike {\tt MultiNest}, we will not be randomly sampling from the parameter space, but instead build new sample vectors from the sample vectors of previous iterations (a.k.a generations). 

The prototypical EA approach is broken into the following steps \cite{Mendes:2004}
\begin{enumerate}
\item \textbf{Seeding} (Once)
Initialise a population of sample vectors (a.k.a individuals)
\item \textbf{Breeding} (Looped):
For each individual (parent) in the generation, we have
\begin{enumerate}
\item \textbf{Crossover}
A new individual (child) is produced, usually as a probabilistic function of an individual or individuals in the population 
\item \textbf{Mutation}
The child vector is perturbed, with some subset of its components operated on by a (usually probabilistic) function
\item \textbf{Selection}
The child is selected to be in the next generation according to some (usually probabilistic) function of its fitness
\end{enumerate}
\item \textbf{Convergence} or \textbf{Time-out} (Once)
Breeding continues until a convergence criterion, or some maximum number of generations, is reached.
\end{enumerate}

Note that the above does not rule crossover and mutation as mutually independent. We will see below that they may begin to resemble each other given certain choices of functions of the population. Generally, however, crossover should capture successful features of the current generation to encourage fast convergence, while mutation should introduce novel features, to discourage suboptimal solutions. As a heuristic, the probability of crossover should be high, while the probability of mutation should be low \cite{Mendes:2004}.

\begin{figure}
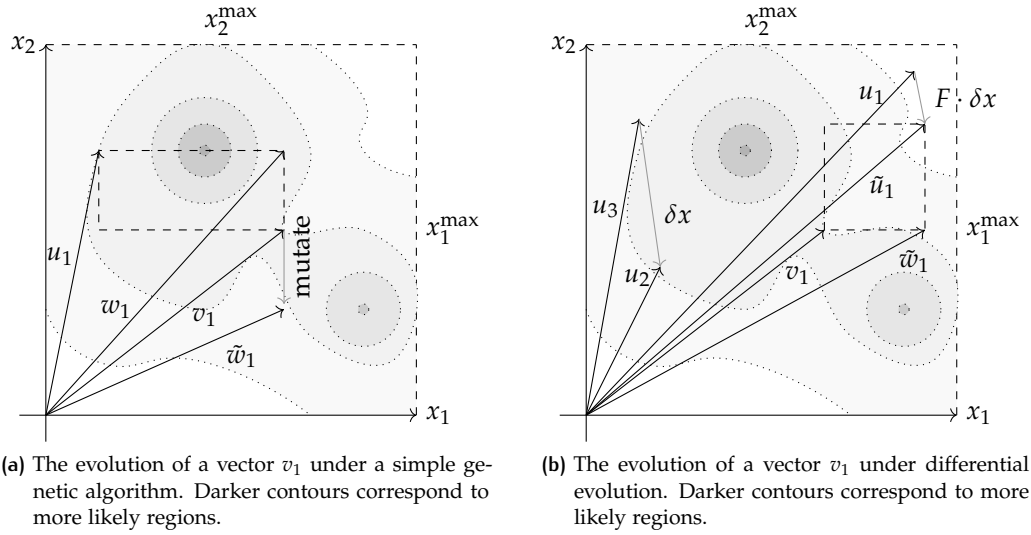

\subfloat[The evolution of a vector $v_1$ under a simple genetic algorithm. Darker contours correspond to more likely regions.\label{fig:general_GA}]{\tikz[scale=0.7]{
\fill [gray!5] plot [smooth, tension=1] coordinates {(7,9) (8,8) (8,7) (9,6.5)} -- (9,2) -- plot [smooth, tension=1] coordinates {(7,2)  (5,3) (3,3) (2,4) } -- (2,9) -- cycle;
\draw [->] (1.5,2) -- (9,2);
\draw [->] (2,1.5) -- (2,9);
\draw [dashed] (2,9) -- (9,9);
\draw [dashed] (9,2) --(9,9);
\node [right] at (9,5.5) {$x_1^\text{max}$};
\node [above] at (5.5,9) {$x_2^\text{max}$};
\node [right] at (9,2) {$x_1$};
\node [left] at (2,9) {$x_2$};
\draw [fill=gray!10,dotted] plot [smooth cycle, tension=0.7] coordinates {(3,5) (3,7) (4,8.5) (5,9) (6,8.5) (7,7) (6.5,5.5) (8,5.5) (9,4.5) (8.5,3) (7,3.5) (6,5) (5,4)};
\draw [fill=gray!20, dotted] (5,7) circle [radius=1];
\draw [fill=gray!40,dotted] (5,7) circle [radius=0.5];
\draw [dotted, fill=gray!60] (5,7) circle [radius=0.1];
\draw [dotted] plot [smooth, tension=1] coordinates {(7,9) (8,8) (8,7) (9,6.5)};
\draw [dotted] plot [smooth, tension=1] coordinates {(2,4) (3,3) (5,3) (7,2)};
\draw [dotted, fill=gray!20] (8,4) circle [radius=0.7];
\draw [dotted,fill=gray!40] (8,4) circle [radius=0.1];
\draw [->] (2,2) -- (3,7);
\node [left] at (2.7,5) {$u_1$};
\draw [->] (2,2) -- (6.5,5.5);
\node [left] at (3.8,4) {$w_1$};
\node [below] at (5,4.2) {$v_1$};
\draw [->] (2,2) -- (6.5,7);
\draw [->] (2,2) -- (6.5,4);
\node [below] at (5.7,3.5) {$\tilde{w}_1$};
\draw [gray!80, ->] (6.5,5.5) -- (6.5,4.1);
\node [below, rotate=90] at (6.5,5) {mutate};
\draw [dashed] (3,7) rectangle (6.5,5.5);
}}\qquad
\subfloat[The evolution of a vector $v_1$ under differential evolution. Darker contours correspond to more likely regions.\label{fig:DE_graph}]{\tikz[scale=0.7]{
\fill [gray!5] plot [smooth, tension=1] coordinates {(7,9) (8,8) (8,7) (9,6.5)} -- (9,2) -- plot [smooth, tension=1] coordinates {(7,2)  (5,3) (3,3) (2,4) } -- (2,9) -- cycle;
\draw [->] (1.5,2) -- (9,2);
\draw [->] (2,1.5) -- (2,9);
\draw [dashed] (2,9) -- (9,9);
\draw [dashed] (9,2) --(9,9);
\node [right] at (9,5.5) {$x_1^\text{max}$};
\node [above] at (5.5,9) {$x_2^\text{max}$};
\node [right] at (9,2) {$x_1$};
\node [left] at (2,9) {$x_2$};
\draw [fill=gray!10,dotted] plot [smooth cycle, tension=0.7] coordinates {(3,5) (3,7) (4,8.5) (5,9) (6,8.5) (7,7) (6.5,5.5) (8,5.5) (9,4.5) (8.5,3) (7,3.5) (6,5) (5,4)};
\draw [fill=gray!20, dotted] (5,7) circle [radius=1];
\draw [fill=gray!40,dotted] (5,7) circle [radius=0.5];
\draw [dotted, fill=gray!60] (5,7) circle [radius=0.1];
\draw [dotted] plot [smooth, tension=1] coordinates {(7,9) (8,8) (8,7) (9,6.5)};
\draw [dotted] plot [smooth, tension=1] coordinates {(2,4) (3,3) (5,3) (7,2)};
\draw [dotted, fill=gray!20] (8,4) circle [radius=0.7];
\draw [dotted,fill=gray!40] (8,4) circle [radius=0.1];
\draw [->] (2,2) -- (3,7.6);
\node [left] at (2.8,5.9) {$u_3$};
\node [left] at (3.4,4.6) {$u_2$};
\draw [->] (2,2) -- (3.4,4.8);
\draw [gray!80, ->] (3,7.6) -- (3.4,4.8);
\node [right] at (3.3,5.8) {$\delta x$};
\node [below] at (7.6,6.7) {$\tilde{u}_1$};
\node [below] at (6,5) {$v_1$};
\draw [->] (2,2) -- (8.2,8.5);
\draw [->] (2,2) -- (6.5,5.5);
\draw [gray!80, ->] (8.2,8.5) -- (8.4, 7.5);
\node [right] at (8.4,8) {$F \cdot \delta x$};
\draw [->] (2,2) -- (8.4,7.5);
\draw [dashed] (6.5,5.5) rectangle (8.4,7.5);
\draw [->] (2,2) -- (8.4,5.5);
\node [above] at (7.4,7.7) {$u_1$};
\node [below] at (8.2,5.3) {$\tilde{w}_1$};
}}\caption{Prototypical genetic algorithm search vs. differential evolution. Compare with \cref{fig:breeding_procedure}}\label{fig:algorithm_graphs}
\end{figure}

A very simple example of this procedure is given in \cref{fig:general_GA}, for two generations. It will be useful to compare this choice of simple genetic operations with the more complex one used in this work: differential evolution.

\subsubsection{{\tt Diver} Implementation}\label{sec:Diver}

In order to produce a well-sampled analysis of the model's fine tuning, we use the {\tt Diver} implementation of the differential evolution algorithm to find physical regions of the model's parameter space \cite{Workgroup:2017htr,storn1997differential}. This has proved particularly useful in finding optimum regions in difficult likelihood functions, such as those encountered in Higgs portal dark matter and supersymmetric examples~\cite{Athron:2017qdc,Athron:2017yua,Cornell:2016gho}.

The algorithm first randomly seeds the parameter space with a \emph{population} of $NP$ vectors $\{v_{i,G}\}$, where $i$ indexes the members of the population, and $G$ indexes the generation. Subsequent generations of the population are then obtained by performing mutation, crossover and selection steps, and these are repeated at each future generation.

\begin{figure}
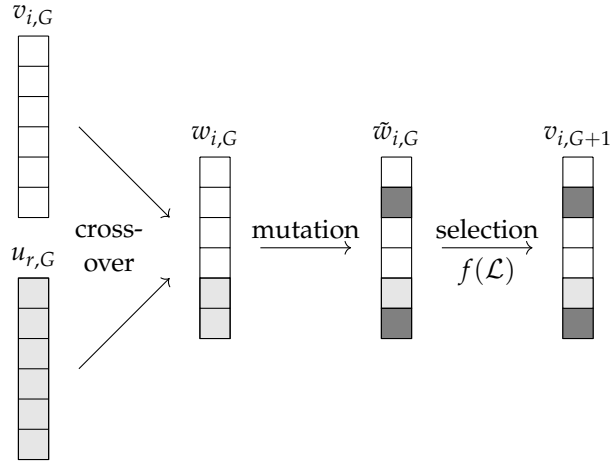
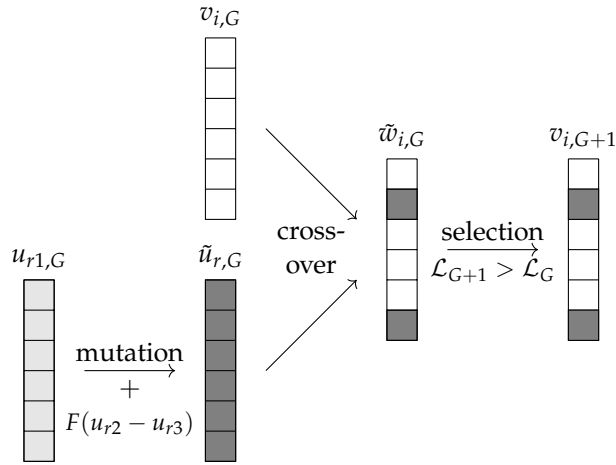

\centering
\subfloat[The prototype for a genetic algorithm, inspired by the breeding of two parents by cutting the gene-vector at one point, and the mutation of their offspring by random choice of genes.]{\tikz[scale=0.4]{
\draw [fill=gray!20] (2,2) rectangle (3,8);
\draw (2,2) grid (3,8);
\node [above] at (2.5, 8) {$u_{r,G}$};
\node [above] at (2.5, 16) {$v_{i,G}$};
\draw (2,10) grid (3,16);
\draw [->] (4,5) -- (7,8);
\draw [->] (4,13) -- (7,10);
\node [align=center] at (5,9) {cross- \\over};
\draw [fill=gray!20] (8,6) rectangle (9,8);
\draw (8,6) grid (9,12);
\node [above] at (8.5,12) {$w_{i,G}$};
\draw [->] (10,9) -- (13,9);
\node [above] at (11.5,9) {mutation};
\draw [fill=gray] (14,6) rectangle (15,7);
\draw [fill=gray!20] (14,7) rectangle (15,8);
\draw [fill=gray] (14,10)  rectangle (15,11);
\draw (14,6) grid (15,12);
\node [above] at (14.5,12) {$\tilde{w}_{i,G}$};
\draw [->] (16,9) -- (19,9);
\node [above] at (17.5,9) {selection};
\node [below] at (17.5,9) {$f(\mathcal{L})$};
\draw (20,6) grid (21,12);
\draw [fill=gray] (20,6)  rectangle (21,7);
\draw [fill=gray!20] (20,7) rectangle (21,8);
\draw [fill=gray] (20,10) rectangle (21,11);
\node [above] at (20.5,12) {$v_{i,G+1}$};
}}\qquad
\subfloat[The differential evolution procedure. A random vector is mutated with the genetic material of other members prior to breeding with the target parent. The crossover rule is a standard random choice of genes. Selection occurs deterministically for a more fit offspring.]{\tikz[scale=0.4]{
\draw [fill=gray!20] (2,2) rectangle (3,8);
\draw (2,2) grid (3,8);
\node [above] at (2.5,8) {$u_{r1,G}$};
\draw [->] (4,5) -- (7,5);
\draw [fill=gray] (8,2) rectangle (9,8);
\draw (8,2) grid (9,8);
\node [above] at (8.5, 8) {$\tilde{u}_{r,G}$};
\node [above] at (8.5, 16) {$v_{i,G}$};
\draw (8,10) grid (9,16);
\draw [->] (10,5) -- (13,8);
\draw [->] (10,13) -- (13,10);
\node [above] at (5.5,5) {mutation};
\node [below, align=center] at (5.5,5) {$+$ \\ \small $F(u_{r2} - u_{r3})$};
\node [align=center] at (11.5,9) {cross- \\over};
\draw [fill=gray] (14,6) rectangle (15,7);
\draw [fill=gray] (14,10)  rectangle (15,11);
\draw (14,6) grid (15,12);
\node [above] at (14.5,12) {$\tilde{w}_{i,G}$};
\draw [->] (16,9) -- (19,9);
\node [above] at (17.5,9) {selection};
\node [below] at (17.5,9) {\small $\mathcal{L}_{G+1} > \mathcal{L}_{G}$};
\draw (20,6) grid (21,12);
\draw [fill=gray] (20,6)  rectangle (21,7);
\draw [fill=gray] (20,10)  rectangle (21,11);
\node [above] at (20.5,12) {$v_{i,G+1}$};
}}\caption{Breeding process for a generic GA (a), and differential evolution (b).}\label{fig:breeding_procedure}
\end{figure}

The \emph{mutation} step produces a set of \emph{donor vectors} $\{\tilde{u}_i\}$ from the current generation of vectors $\{v_{i,G}\}$. The production of each donor vector $\tilde{u}_i$ occurs by choosing three random vectors $u_{r1}$, $u_{r2}$ and $u_{r3}$ from the current generation (on the condition that none of these are the same, and that none of them matches $v_i$). $\tilde{u}_i$ is then taken to be:
\begin{equation}
\tilde{u}_i=u_{r1}+F(u_{r2}-u_{r3})
\end{equation}
where $F$ is a parameter that controls the strength of the differential variation.

The \emph{crossover} step is then used to produce a set of trial vectors $\{\tilde{w}_i\}$ that will potentially form the next generation of vectors. For the $k$th component of the trial vector $\tilde{w}_i$, a random number between 0 and 1 is chosen. If this number is less than or equal to a meta-parameter $Cr$ (chosen in advance of the scan), then the component is taken from the corresponding donor vector $\tilde{u}_i$. Otherwise, the component is taken from the corresponding vector in the current generation $v_i$. After all of the components of $\tilde{w}_i$ have been chosen, one component is reassigned, thus ensuring that the trial vectors and their corresponding vectors in the next generation are always different. That is, a component $k_\text{rand}$ is chosen at random, and the trial vector component $\tilde{w}_i^{k_\text{rand}}$ is set to the donor vector value $\tilde{u}^{k_\text{rand}}_i$, irrespective of its previous value.

Finally, a \emph{selection} step is used to choose the vectors for the next generation. The value of the likelihood function for each vector in the current generation $v_i^j$ is compared with the likelihood for the corresponding trial vector $\tilde{w}_i$, and the points with higher likelihood are retained for the next generation. This procedure as described is sketched in parameter space in \cref{fig:DE_graph} and in gene space in \cref{fig:breeding_procedure}. 

DE algorithms have become the de facto basis of single-objective real-parameter optimisers. They have dominated this form of competition for the past fourteen years, for example in the IEEE Congress on Evolutionary Computation \cite{Molina2018}. A standard improvement to the above procedure is to adapt the meta-parameters during the scan. The {\tt Diver} package optimises the differential evolution algorithm by allowing $Cr$ and $F$ to evolve, called Adaptive Differential Evolution. This occurs in the intuitive way - by sampling $Cr$ and $F$ uniformly in the seeding step, and subsequently propagating those values that lead to lower cost function outputs. We enabled this adaptivity, and in doing so found a suitable set of parameter points (i.e., giving valid EWSB, with SM masses within two $\sigma$ of the measured values) significantly faster than Markov Chain Monte Carlo and MultiNest scanning techniques.

\chapter{The Two-Site Minimal 4D Composite Higgs Model}
\label{sec:M4DCHM}
%

\section{Model Description}

\begin{figure}[H]
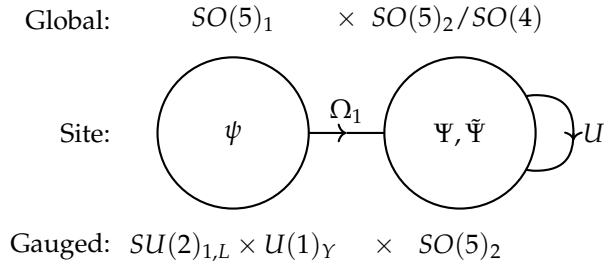

\centering\tikz[scale=1, every node/.style={transform shape}]{
\node [left] at (-1.5,0) {Site:};
\node [left] at (-1.5,1.5) {Global:};
\node [left] at (-1.5,-1.5) {Gauged:};
\draw [thick] (0,0) circle [radius=1] node {$\psi$};
\node at (0,1.5) {$SO(5)_1$};
\node at (1.5,1.5) {$\times$};
\draw [thick, ->] (1,0) -- (1.5,0);
\draw [thick] (1.5,0) -- (2,0);
\node [above] at (1.5,0) {$\Omega_1$};
\draw [thick,->] plot [smooth, tension=1] coordinates {(3,0) (4,0.5) (4.5,0) (4,-0.5) (3,0)};
\draw [thick, ->] (4.5,0) -- (4.5,-0.1);
\draw [thick, fill=white] (3,0) circle [radius=1] node {$\Psi, \tilde{\Psi}$};
\node [right] at (1.7,1.5) {$SO(5)_2 / SO(4)$};
\node [right] at (4.5,0) {$U$};
\node at (0,-1.5) {$SU(2)_{1,L} \times U(1)_Y$};
\node at (2,-1.5) {$\times$};
\node at (3,-1.5) {$SO(5)_{2}$};
}
\caption{The Minimal 4D Composite Higgs Model \label{fig:M4DCHM_moose}}
\end{figure}

We have chiselled out a class of multisite, Moose-type Composite pNGB Higgs models from the large space of composite dynamics, and a great deal of work has been done on this class (\cite{contino2003,agashe2005,contino2007,giudice2007,Anastasiou:2009rv,gripaios2009beyond,Carena:2007ua,de2012,panico2012,carena2014} is a relevant selection). For the remainder of this thesis, we will work with the minimal and next-to-minimal content of this class. We begin with the third generation of quarks as partially composite, called the Minimal 4D Composite Higgs Model (M4DCHM), in \cref{sec:M4DCHM}.  We subsequently include leptons in this model, in \cref{sec:LCHM}, and finally consider the Next-to-Minimal 4D Composite Higgs Model (NM4DCHM) in \cref{sec:NMCHM}. Using the Moose framework, the M4DCHM is sketched in \cref{fig:M4DCHM_moose}. This model delivers a pair of Dirac resonances (e.g. $\Psi_t, \tilde{\Psi}_t$) for each pair of SM Weyl fields (e.g. $t_L, t_R$), which couple via a NL$\Sigma$M link field $\Omega_1$.   They also couple via a second link field $\Omega_2$ to a HLS site\footnote{See \cref{sec:HLS} to refresh the formalism of Hidden Local Symmetry}, which has had its coupling taken to infinity, and thus appears as a simple $SO(5)/SO(4)$ breaking. Finally, there is a gauging of the diagonal group $SO(5)_{2,L} \times SO(5)_{2,R}$ in order to include massive spin-one resonances, and remove ten of the fourteen NGBs from the field content. 

\begin{figure}[H]
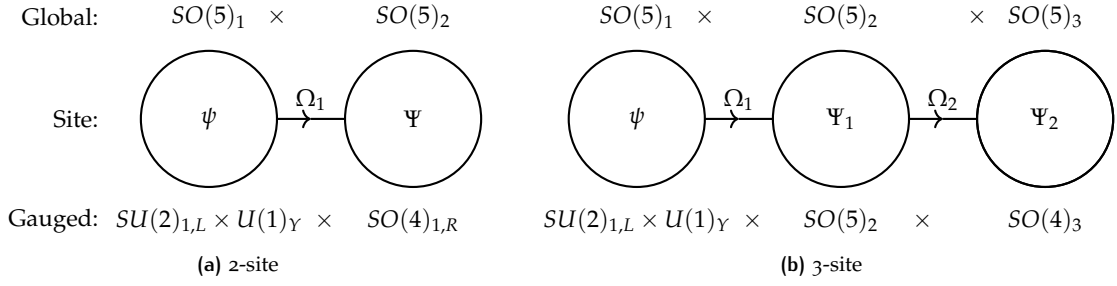

\hspace*{-6em}\subfloat[2-site \label{fig:discrete_2-site_moose}]{\tikz[scale=0.9, every node/.style={transform shape}]{
\node [left] at (-1.5,0) {Site:};
\node [left] at (-1.5,1.5) {Global:};
\node [left] at (-1.5,-1.5) {Gauged:};
\draw [thick] (0,0) circle [radius=1] node {$\psi$};
\node at (0,1.5) {$SO(5)_1$};
\node at (1,1.5) {$\times$};
\draw [thick, ->] (1,0) -- (1.5,0);
\draw [thick] (1.5,0) -- (2,0);
\node [above] at (1.5,0) {$\Omega_1$};
\draw [thick] (3,0) circle [radius=1] node {$\Psi$};
\node at (3,1.5) {$SO(5)_2$};
\node at (0,-1.5) {$SU(2)_{1,L} \times U(1)_Y$};
\node at (1.7,-1.5) {$\times$};
\node at (3,-1.5) {$SO(4)_{1,R}$};
}}\qquad
\subfloat[3-site \label{fig:discrete_3-site_moose}]{\tikz[scale=0.9, every node/.style={transform shape}]{
\draw [thick] (0,0) circle [radius=1] node {$\psi$};
\node at (0,1.5) {$SO(5)_1$};
\node at (1,1.5) {$\times$};
\draw [thick, ->] (1,0) -- (1.5,0);
\draw [thick] (1.5,0) -- (2,0);
\node [above] at (1.5,0) {$\Omega_1$};
\draw [thick] (3,0) circle [radius=1] node {$\Psi_1$};
\node at (3,1.5) {$SO(5)_2$};
\node at (5,1.5) {$\times$};
\draw [thick,->] (4,0) -- (4.5,0);
\draw [thick] (4.5,0) -- (5,0);
\node [above] at (4.5,0) {$\Omega_2$};
\node at (5.5,0) {$.\; .\; .$};
\draw [thick, fill=white] (6,0) circle [radius=1];
\draw [thick] (6,0) circle [radius=1] node {$\Psi_2$};
\node at (3,-1.5) {$SO(5)_2$};
\node at (6,1.5) {$SO(5)_3$};
\node at (0,-1.5) {$SU(2)_{1,L} \times U(1)_Y$};
\node at (1.7,-1.5) {$\times$};
\node at (4.2,-1.5) {$\times$};
\node at (6,-1.5) {$SO(4)_3$};
}}
\caption{The two- and three-site models, to compare with the M4DCHM considered in this work}
\end{figure}

For reference, we note that this is an extension of the 2-site "Discrete CHM (DCHM)", as described by \cite{panico2011} that allows for calculability of the potential. On the other hand, it is a simplification of the 3-site DCHM, where we have enforced certain couplings taken to zero. Indeed, in the M4DCHM, one can interpolate between these two cases by taking $g_2 \rightarrow \infty$, performing the field redefinition $\Psi_{1,L}, \Psi_{2,R} \rightarrow \Psi$ and setting the couplings $\Delta_{1,L}, \Delta_{2,L} = \Delta_1$, and removing the possible coupling of $m_Y \Psi_R \tilde{\Psi}_L$. The two limiting cases are described in figures \ref{fig:discrete_2-site_moose} and \ref{fig:discrete_3-site_moose}. Many of the findings of this thesis apply to good approximation to the 3-site DCHM.

The typical "vanilla" M4DCHM considers only a partially composite top quark in the fundamental representation, as its contribution to the Higgs potential is dominant. In this thesis we will variously consider the following extensions:
\begin{itemize}
\item A partially composite bottom quark (\textit{3rd-generation M4DCHM})
\item Partially composite leptons (\textit{Leptonic M4CHM (LM4DCHM)})
\item Quark and lepton partners in non-fundamental representations (\textit{LM4DCHM$^{q-u-d}_{l-\tau-\nu}$})
\item A promotion of the global symmetry breaking to the Next-to-Minimal $SO(6)/SO(5)$ (\textit{NM4DCHM})
\end{itemize}

This, and the following two chapters will serve as references - all major theoretical results will be re-stated, if required, with little explanation. Prior chapters can be referred back to pedagogically if required. In that spirit, the non-linear realisation of $SO(5)$ in the second site is parameterised by the following field,
\begin{align}
\Phi_1(x) &=  \Phi_0 U & & U= \exp(i\pi^{\hat{a}}T^{\hat{a}}) \label{eq:ngb}\, ,
\end{align}
where the NGB matrix $U$ is given by
\begin{align}
U = \left( \begin{matrix}
\mathbb{1}_5  - (1-\cos\frac{\pi}{f}) \frac{\vec{\pi} \vec{\pi}^T}{\pi^2} & \sin\frac{\pi}{f} \frac{\vec{\pi}}{\pi} \\
-\sin\frac{\pi}{f} \frac{\vec{\pi}}{\pi} & \cos\frac{\pi}{f}
\end{matrix}\right)\label{eq:explicit_goldstone_matrix_m4dchm}
\end{align}

Where a choice of gauge is required, we will specify either taking the holographic site-1 gauge or holographic site-2 gauge, which respectively take
\begin{align}
\Phi := U \phi_0 =  \frac{1}{\hat{h}}\sin\frac{\hat{h}}{f}(h^1,h^2,h^3,h^4,\hat{h}\cot\frac{\hat{h}}{f})^T = \begin{cases}
\Omega_1 \phi_0, \; \Omega_2 = \mathbb{1}, \;  \text{Holo. Site-1}\\
\Omega_2 \phi_0, \; \Omega_1 = \mathbb{1}, \;  \text{Holo. Site-2}
\end{cases}\label{eq:choices_of_gauge}
\end{align}
where $\hat{h} = \sqrt{h^{\hat{a}}h^{\hat{a}}}$. We will also subsequently choose the SM unitary gauge, which allows the complete symmetry breaking chain to be parameterised by the simple fields
\begin{align}
h_4 = \hat{h} &= h \implies&  & & \Phi &= (0,0,0,s_h,c_h)^T & \textnormal{and} & & \langle \Phi \rangle = (0,0,0,\xi,\sqrt{1-\xi^2})^T, \nonumber\\
U &= \left( \begin{matrix}
\mathbb{1} & \vec{0}^T & \vec{0}^T\\
\vec{0} & c_h & s_h \\
\vec{0} & -s_h & c_h
\end{matrix} \right) &&&&&&&\label{eq:phi} 
\end{align}
where $s_h = \sin\frac{h}{f}$, $c_h = \cos\frac{h}{f}$, $\xi=\sin^2\frac{\langle h \rangle}{f}$ and $f$ is the NGB decay constant.

\subsection{Bosonic Sector}

We showed in \cref{sec:WSR} that multiple fermion resonances are required for a finite Higgs potential. If they couple in the partial compositeness paradigm, they will be coupled by link fields that introduce Goldstone degrees of freedom. Given a lack of evidence for massless scalars, these Goldstone degrees must be eaten by gauge bosons, and we thus introduce a site of $SO(5)$ gauging, as included in \cref{fig:M4DCHM_moose}. The NGB and gauge Lagrangian prior to any choice of gauge consists of CCWZ invariant terms of NGBs, the elementary fields $W_\mu, B_\mu$, and the composite fields $\rho_\mu =  \{\rho_{L,\mu}, \rho_{R,\mu}, a_\mu\}$
\begin{align}
\mathcal{L}_\text{bosonic} &= \frac{f_1^2}{2} |D_\mu \Phi_1|^2 + \frac{f_\Omega^2}{4} \text{Tr} |D_\mu \Omega |^2 + \frac{f_{\Omega_X}^2}{4} |D_\mu \Omega_X |^2 \nonumber\\
& - \frac{1}{4} \text{Tr} \left[W_{\mu\nu} W^{\mu\nu}\right] - \frac{1}{4} B_{\mu\nu} B^{\mu\nu}  - \frac{1}{4} \text{Tr} \left[G_{\mu\nu} G^{\mu\nu}\right]\label{eq:MCHM_gauge_lagrangian}\\
& - \frac{1}{4} \text{Tr} \left[\rho_{\mu\nu} \rho^{\mu\nu}\right]  - \frac{1}{4} \rho_{X,\mu\nu} \rho_X^{\mu\nu}
 - \frac{1}{4} \text{Tr} \left[\rho_{G,\mu\nu} \rho_G^{\mu\nu}\right] \nonumber
\end{align} 
where the NGB covariant derivatives are given by
\begin{align}
& D_\mu \Phi_1 = \partial_\mu \Phi_1 - i g_\rho \rho_\mu^A T^A \Phi_1 \nonumber \\
& D_\mu \Omega = \partial_\mu \Omega -i \left( g_0 W^a_\mu T_L^a  \right) \Omega + i \Omega \left( g_\rho\rho_\mu^A T^A \right)\\
& D_\mu \Omega_X = \partial_\mu \Omega_X - i \left( g_0' B_\mu T^3_R \right) \Omega_X + i\left( g_{\rho_X}  \Omega_X \rho_{X,\mu} \right) \nonumber
\end{align}
and the field strength tensors $W_{\mu\nu}, B_{\mu\nu}, \rho_{\mu\nu}$ are given by the usual form
\begin{align}
A_{\mu\nu} = \partial_\mu A_\nu - \partial_\nu A_\mu + ig_A [A_\mu, A_\nu]\;.
\end{align}
Note that we include a composite gluon sector $\rho_G$. This is due to partial compositeness, which implies terms in the low-energy Lagrangian $\Delta \bar{\psi}_L \Psi_R$. The composite fields are thus coloured. At low energy, their symmetries are those of SM QCD. At high energy, the term $\Delta \bar{\psi}_L \Omega_1 \Psi_R$  transforms as $\Delta \bar{\psi}_L G_c^{-1} G_c \Omega G_c'^{-1} G_c' \Psi_R$, where it is not necessary that $G_c = G_c'$. We thus assume two independent colour groups $SU(3) \times SU(3)'$ spontaneously broken to their diagonal $SU(3)_{SM}$ by the vev of $\Omega_1$. The $8$ NGBs are eaten by the axial $SU(3)$ subgroup of gluons, which become massive, leaving the vector SM subgroup gluons massless. I mention this structure in order to ignore it from this point onwards. We will only focus on massive vector bosons coming from the $\rho, \rho_X$ fields, as these are tightly linked to the behaviour of the EW-Higgs sector.

In the holographic gauge, the covariant derivatives of $\Omega$ and $\Omega_X$ in the first line of \cref{eq:MCHM_gauge_lagrangian} goes to 
\begin{align}
\mathcal{L}_{\substack{\text{bosonic,} \\ \text{unitary}}} &= \frac{1}{4} f^2_\Omega \left( W^A_\mu - \rho_\mu^A \right)^2 + \frac{1}{4} f_{\Omega_X}^2 \left( B_\mu - \rho_{X,\mu}\right)^2, \qquad A=1, ..., 10 \nonumber\\
&= \frac{1}{2} m_\rho^2 \left( t_\theta W^i_{L,\mu} - \rho^i_{L,\mu} \right)^2 + \frac{1}{2}m_\rho^2 \left( t_\theta W_{R,\mu}^3 - \rho_{R,\mu}^3 \right)^2 + \frac{1}{2} m_X^2 \left( \frac{g_x}{g_X} B_\mu -  \rho_{X,\mu} \right)^2 \nonumber \\
& + \frac{1}{2} m_\rho^2 \sum\limits_{k=1}^2 \rho_{R,\mu}^k \rho_R^{k,\mu} + \frac{1}{2} m_\rho^2 a^i_\mu a^{i,\mu} \label{eq:MCHM_unitary_gauge}
\end{align}
where we have defined
\begin{align}
m_\rho^2 = \frac{1}{2} g_\rho^2 f_\Omega^2, && m_a^2= \frac{1}{2} g_\rho^2(f_\Omega^2 + f_1^2), && m_X^2 = \frac{1}{2}g_X^2 f_{\Omega_X}^2, && t_\theta = g_0/g_\rho . \label{eq:gauge_parameters}
\end{align}
Thus, we can diagonalise these mixing terms to find the physics fields, which are admixtures of elementary and composite fields
\begin{align}
W_{\text{SM},\mu}^i = c_\theta W_{L,\mu}^i + s_\theta \rho_{L,\mu}^i, && B_{\text{SM},\mu} = \frac{1}{\sqrt{1 + t_{\theta_\rho}^2 + t_{\theta_X}^2}} \left( B_\mu + t_{\theta'_\rho} \rho_{R,\mu}^3 + t_{\theta'_X} X_\mu \right)
\end{align}
where the angles $\theta'_\rho, \theta'_X$ give the ratio of the physical coupling and the bare coupling $t_{\theta'_\rho} = \tan\theta'_\rho = g'_0/g_\rho$ and $t_{\theta'_X} = \tan\theta'_X = g'_0 / g_X$. Therefore the physical couplings are given by
\begin{align}
g = c_\theta g_0 = \left( \frac{1}{g_0^2} + \frac{1}{g_\rho^2}\right)^{-1/2} , && g' = \frac{g_0'}{\sqrt{1 + t_{\theta_\rho}^2 + t_{\theta_X}^2}} = \left( \frac{1}{g_0'^2} + \frac{1}{g_\rho^2} + \frac{1}{g_X^2} \right)^{-1/2}.
\end{align}
We analyse the deviation from SM gauge boson observables, by calculating the low-energy form factors in the elementary Lagrangian
\begin{align}
\mathcal{L}_{\rm eff}={} & {}\frac{1}{2}P^T_{\mu\nu}\left[\Pi_W(p^2,h)W_\mu W_\nu+\Pi_B(p^2,h)B_\mu B_\nu+\Pi_{WB}(p^2,h)W^3_\mu B_\nu\right]\label{eq:Leff}
\end{align}
where $\Pi_i$ and $M$ are the form factors and $P^T$ is the transverse projection operator.  Once a choice has been made for the precise embedding of the elementary fermions, explicit expressions for the form factors can be obtained.

%

The form factor for the $W$ boson is\footnote{The specific derivation of the gauge form factor in the M4DCHM is given in Appendix \cref{sec:form_factors}}
\begin{equation}
\Pi_W=-\frac{p^2(p^2-(1+t_\theta^2)m_\rho^2)}{g_2^2(1+t_\theta^2)(p^2-m_\rho^2)}+\frac{1}{4}s_h^2\left[\frac{2m_\rho^2(m_a^2-m_\rho^2)t_\theta^2}{g_2^2(1+t_\theta^2)(p^2-m_a^2)(p^2-m_\rho^2)}\right]\, ,
\end{equation}
where $g_2$ is the observed $SU(2)_L$ gauge coupling. Plugging into the potential formula \cref{eq:potential} and performing the integral results in a contribution to the $s_h^2$ part of the Higgs potential of
\begin{equation}
\gamma_g=-\frac{9m_\rho^4(m_a^2-m_\rho^2)t_\theta^2}{64\pi^2(m_a^2-(1+t_\theta^2)m_\rho^2)}\ln\left[\frac{m_a^2}{(1+t_\theta^2)m_\rho^2}\right]
\end{equation}
at leading order in $t_\theta$. 

\subsection{Fermion Sector}

The high-energy fermion Lagrangian for each generation of SM quark, with partners in the fundamental representation, prior to any choice of gauge, is described by
\begin{align}
\mathcal{L}_\text{fermionic} &= \bar{\psi}^u_L i \slashed{D} \psi_L + \bar{\psi}^u_R i \slashed{D} \psi^u_R & \text{[elementary]} \nonumber \\
& +  \Delta_{uL} \bar{\psi}^u_L\Omega_1 \Psi^u_R +  \Delta_{uR} \bar{\psi}^u_R\Omega_1 \tilde{\Psi}^u_L & \text{[link]} \nonumber\\
& + \bar{\Psi}^u\left(i \slashed{D} - m_{\Psi^u} \right)\Psi^u + \bar{\tilde{\Psi}}^u \left( i\slashed{D} - m_{\tilde{\Psi}^u} \right) \tilde{\Psi}^u & \text{[composite]} \nonumber\\
& - m_{\Psi^u\tilde{\Psi}^u} \bar{\Psi}^u_L U P^{(rs)} U^{\dagger} \tilde{\Psi}^u_R & \text{[Yukawa]}\nonumber \\
& + \{u \rightarrow d\} + \text{h.c.} \label{eq:fundamental_fermion_lagrangian}
\end{align}
where the covariant derivatives are given by
\begin{align}
D_\mu \psi_L &= \left(\partial_\mu  - i Y_\psi B_\mu - g_3 G^a_\mu \lambda^a \right) \psi_L \nonumber\\
D_\mu \psi_R &= \left(\partial_\mu  - i g_0 W_\mu^a T_L^a i - i Y B_\mu - g_3 G^a_\mu \lambda^a \right) \psi_R \\
D_\mu \Psi &= \left(\partial_\mu  - i g_rho \rho^a_\mu T^a i - i X \rho_{X_\Psi\mu}   - g^c_3 G^a_\mu \lambda^a \right) \Psi \nonumber
\end{align}
Where the Y-charges are given by $\{Y_{q_L}, Y_{u_R}, Y_{d_R}, Y_{l_L}, Y_{e_R}, Y_{\nu_R}\} = \{\frac{1}{6}, \frac{2}{3}, \frac{-1}{3}, \frac{-1}{2}, -1, 0 \}$ and the X-charges by $ \{X_{q_L}, X_{q_R}, X_{l_L}, X_{l_R} \} = \{ \frac{2}{3}, \frac{-1}{3}, -1,0 \}$. In site-1 holographic gauge  - that is, where all Higgs dependence is captured by $\Omega_1$ - the terms involving the link and Yukawa terms go to 
\begin{align}
\mathcal{L}_{\substack{\text{fermionic,} \\ \text{holo. site-1}}} &=  \Delta_{uL} \bar{\psi}^u_L U \Psi^u_R +  \Delta_{uR} \bar{\psi}^u_R U \tilde{\Psi}^u_L  - (m_Y + Y)\bar{\Psi}^u_L\tilde{\Psi}^u_R
\end{align}
While in site-2 holographic gauge (all Higgs dependence captured by $\Omega_2$), they go to 
\begin{align}
\mathcal{L}_{\substack{\text{fermionic,} \\ \text{holo. site-2}}} &=  \Delta_{uL} \bar{\psi}^u_L \Psi^u_R +  \Delta_{uR} \bar{\psi}^u_R \tilde{\Psi}^u_L  - m_Y \bar{\Psi}^u_L \tilde{\Psi}^u_R - Y \bar{\Psi}^u_L (U \phi_0) (U \phi_0)^\dagger \tilde{\Psi}^u_R
\end{align}

In the gauge sector, the choice of gauge is not particularly important. Both are analytically solvable for the quadratic potential gauge contribution. The fermion case is not so straightforward. One route is to choose the site-2 holographic gauge, and we can then find the form factors, analytically expanding in the potential to quartic order in $s_h$. Thus, we get a solution to the Higgs vev in terms of numerical integrals, valid for $h \ll f$ to order $\mathcal{O}(h^4/f^4)$. Another route is to choose the site-1 holographic gauge, and numerically diagonalise the mass mixing matrices. In that case, we must numerically minimise the Higgs potential function to find the Higgs vev. In this chapter, we will use the latter. For the following chapters, the former route is taken, as it can be a faster method, with corrections of order $\mathcal{O}(h^6/f^6) \sim \frac{1}{4000}$. 

As in the gauge case, corrections to SM observables can be described with the effective Lagrangian (for each quark generation)
\begin{align}
\mathcal{L}_{\substack{\text{fermionic,} \\ \text{effective}}} &= \bar{u}_L \slashed{p} (Z_q + \Pi_{u_L} ) u_L + \bar{d}_L \slashed{p} (Z_q + \Pi_{d_L}) d_L + \bar{u}_R \slashed{p} (Z_u + \Pi_{u_R}) u_R \nonumber \\
& + \bar{d}_R \slashed{p} (Z_d + \Pi_{d_R}) d_R + \bar{u}_L M_u u_R + \bar{d}_L M_d d_R + \text{h.c.}\label{eq:effective_SM_fermions}
\end{align}

The process for deriving these form factors is explained in \cref{sec:low_energy_lagrangian}. They are dependent on choice of representation, and will be given in \cref{sec:FermionSector}. They are defined in terms of the broken generators coupling $SO(4)$ irreps coming from $SO(5)$ decompositions
\begin{align}
\hspace*{-2em}\mathcal{L}_\textnormal{eff}|_{h=0} &= \sum\limits_\psi \left[ \sum\limits_{r_4} \bar{\psi}^{(r_4)} \slashed{p} \left(1 + \hat{\Pi}^{(r_4)}_\psi \right) \psi^{(r_4)} + \left( \sum\limits_{r_4} \bar{\psi}_L^{(r_4)} \bar{M}_\psi^{(r_4)} \psi_R^{(r_4)}\right) + \textnormal{h.c.} \right]
\end{align}
It is also necessary to examine how the various representations couple to the Higgs \textit{before} EWSB $\mathcal{L}_\text{eff}^\text{EW}$, but after integrating out the composite fermions. These terms will be given in their respective sections in \cref{sec:FermionSector}, as they are representation-specific.

\section{Model Space}\label{sec:M4DCHM_parameters}

\subsection{Parameters}

A parameterisaton of the N-site MCHM is described in \cref{sec:general_parameters}. Here, we list those relevant to the M4DCHM and its extensions, in \cref{tab:4DCHM_params}. Note that a useful parameterisation in the simplified studies (LM4DCHM, NM4DCHM) is that of 
\begin{align}
d_f =  m_f/\Delta_f \in [0,3]
\end{align}
which is a measure of compositeness, as outlined in \cref{sec:partial_compositeness}. This is implicit in the table of parameters. In \cref{sec:M4DCHM_Results} we will present the results of Diver scans of the model referred to as M4DCHM, including only the 3rd generation of quarks as composite. As mentioned above, we diagonalise the mass mixing matrices of the top and bottom quarks and their partners \cite{niehoff2016direct}. These eigenvalues $m_i^2$ are used to calculate the Coleman-Weinberg potential 
\begin{align}
V_\text{eff}(h) &= \sum \frac{c_i}{64\pi^2} m_i^4(h) \log(m_i^2(h)) && c_i = \{3,6,-12\}
\end{align}
where the sum and constant $c_i$ corresponds to $\{$ neutral gauge, charged gauge, coloured Dirac$\}$ fields, respectively. In this chapter, the potential is numerically minimised to solve for $\langle h \rangle$, enforcing $0 < \langle h \rangle < \frac{f\pi}{2}$ as a valid parameter point. In later chapters, we will use the leading-order analytic solution for the potential, derived in \cref{eq:minimum_solution}
\begin{align}
\sin^2 \left( \frac{\langle h \rangle}{f} \right) = \frac{\gamma}{2\beta}\;  \text{given} \qquad V_\text{eff}(s_h^2, s_h^4) = -\gamma(\Pi_i, M_i) s_h^2 + \beta(\Pi_i, M_i) s_h^4
\end{align}
where the form factors $\Pi_i, M_i$ are also given in Appendix \cref{sec:form_factors}. In that case, we enforce $\gamma / (2\beta) < 1$ as a valid parameter point.

\begin{table}
\hspace*{-3em}\begin{tabular}{@{}lll p{0.3\linewidth}@{}}\toprule
Parameter & Model & Prior Space & Notes\\
\midrule
$f$ & M4DCHM & $[0.5,5]\tev$ & \\
$f_1$ & M4DCHM & $[0.5,5\sqrt{3}]\tev$ & \\
$f_G, f_X$ & M4DCHM & $[0.5,10\sqrt{3}]\tev$ & \\
$g_\rho, g_X, g_G$ & M4DCHM & $[1,4\pi]$ & SM couplings are also varied within theoretical limits\\
$m_\rho = m_{\rho_X}$ & (LM,NM)4DCHM & $[0.5,10]\tev$ & This re-parameterises and simplifies the above row\\
$t_\theta$ & (LM,NM)4DCHM & $[0,1]$ & Described in \cref{eq:gauge_parameters}\\
\multirow{2}{*}{$\Delta^u_R, \Delta^d_R$} & M4DCHM & $\log[0.05,20\pi]\tev$ & \\
 & (LM,NM)4DCHM & $[0.5,m_{\tilde{\Psi}^{u/d}} ]\tev$ & Implemented by re-parameterising with $d_{u,d}$ \\
\multirow{2}{*}{$\Delta_L^u, \Delta_L^d$} & M4DCHM$^\textbf{5}$ & $\log[0.05,20\pi]\tev$ & \\
 & LM4DCHM$^\textbf{5}$, NM4DCHM$^\textbf{6}$ & $[0.5, m_{\Psi^{u/d}} ]\tev$ & \\
$\Delta_L$ & LM4DCHM$^{\textbf{14}-\textbf{X}-\textbf{10}}$ & $[0.5, m_{\Psi}] \tev$ & \\
\multirow{2}{*}{$m_{\tilde{\Psi}^u}, m_{\tilde{\Psi}^d}$} & M4DCHM & $\log[0.05,20\pi]\tev$ & \\
 & (LM,NM)4DCHM & $[0.5,10]\tev$ &  \\
\multirow{2}{*}{$m_{\Psi^u}, m_{\Psi^d}$} & M4DCHM$^\textbf{5}$ & $\log[0.05,20\pi]\tev$  & \\
 &  LM4DCHM$^\textbf{5}$, NM4DCHM$^\textbf{6}$ & $[0.5,10] \tev $ & \\
 $m_{\Psi}$ & LM4DCHM$^{\textbf{14}-\textbf{14}-\textbf{10}}$ & $[0.5,10] \tev$ & \\
 $m_{Y^u}, m_{Y^d}$ & M4DCHM & $\log[0.05,20\pi]\tev$ & \\
  $m_{Y^u} + Y^{u}, m_{Y^d} + Y^{d}$ & M4DCHM & $\log[0.05,40\pi]\tev$ & \\
 $m_Y$ & (LM,NM)4DCHM & $[0.5, 10] \tev$ & \\
  $Y$ & (LM,NM)4DCHM & $[-10, 10] \tev$ & \\
  $\tilde{Y}$ & LM4DCHM$^{\textbf{14}-\textbf{14}-\textbf{10}}$ & $[-10, 10] \tev$ & A second Yukawa interaction is allowed by two $\textbf{14}$ irreps\\
  $\delta, \epsilon$ & NM4DCHM & $[0,\pi/2]$ & Embedding in $\textbf{6}$, see \cref{sec:NMCHM}\\
\bottomrule
\end{tabular}
\vspace*{1em}\caption{List of parameters used in the models of this work. The fermionic parameters apply to each generation of quark or lepton}\label{tab:4DCHM_params}
\end{table}

\subsection{Observables}

Here we describe all physical observables used to constrain the parameter space defined in \cref{tab:4DCHM_params}. Constraints come predominantly from LHC results (both run 1 and run 2), and consist of
\begin{itemize}
\item SM field mass measurements
\item Deviations of the Higgs signal strength from the SM prediction
\item Deviations of the Z-boson signal strength from the SM prediction
\item Precision measurements of EW radiative corrections
\item Direct detection searches of non-SM resonances
\end{itemize}
The results given in this chapter include all these observables, listed in \cref{tab:observables_1}, while in subsequent chapters the smaller subset of SM masses and the electroweak vev are included. These observables are not exhaustively described - for a more thorough description, see \cite{niehoff2016direct} and \cite{Niehoff:2017guu}.

\subsubsection{SM masses}


The mass of each SM fermion ($\psi=t,b,\tau$) can be calculated from the form factors in \cref{eq:effective_SM_fermions} at the electroweak vacuum $(p^2, h) \rightarrow (0,v)$
\begin{equation}
m_\psi=\frac{M_\psi(0,v)}{\sqrt{\Pi_{\psi_L(0,v)}\Pi_{\psi_R}(0,v)}}\, .
\end{equation}

In the following, we will explore three different theories that are distinguished by the choice of embedding for the leptons. For each model, we scan the composite sector parameter space to find points that reproduce measured observables. These observables are the Higgs VEV and mass, and the masses of the top quark, bottom quark and tau lepton. The tau neutrino will be treated as massless, however certain representations of the lepton composite partners can realise a see-saw model \cite{carmona2015}. In the simplified model LM4DCHM, the Higgs VEV only appears in the ratio $v^2/f^2$ and hence we can simply rescale $f$ to give the correct Higgs VEV instead of treating it as an extra input parameter. After performing this rescaling, we take the points that give correct values for the remaining observables and calculate the spectrum of predicted resonances and the expected deviations from the SM Higgs couplings. In the case where we take $f$ as a free parameter, it is constrained to be around $v^2 = f^2 s^2_{\langle h \rangle}$, using the Fermi constant as the observable
\begin{align}
G_\mu^\text{tree} &= \frac{1}{\sqrt{2}v^2}
\end{align}
Comparison of these predictions with current and anticipated collider results will give us limits on the fine tuning of each theory. 

\subsubsection{Higgs signal strength}

The deviations of Higgs couplings from the SM predictions are parameterised as a fraction of the composite Higgs-$\chi$-$\chi$ coupling $c$ with the SM Higgs-$\chi$-$\chi$ coupling $c_\textnormal{SM}$,
\begin{align}
r_\chi &= \frac{c(h\chi\chi)}{c_\textnormal{SM}(h\chi\chi)}\, . \label{eq:coupling_deviation}
\end{align}

The composite sector features several massive vector-boson resonances that are charged under $SU(2)_L\times U(1)_Y$. The quantum numbers and masses are given, to a very good approximation, by {\bf 1}$_{\pm 1}$ with mass $m_{\rho_1}=m_{\rho}$ and {\bf 3}$_{\pm 0}$ with mass $m_{\rho_3}=m_{\rho}/\cos\theta$. The effect is to modify the $hVV$ coupling (where $V$ is a $Z$ or $W$ boson), by
\begin{equation}
r_V=\sqrt{1-\xi}\, .
\end{equation}

There is also a correction to the loop-induced $h\gamma\gamma$ coupling, which is given by \cite{Shifman:1979eb,Carena:2012xa}:
\begin{align}
r_\gamma &= \Bigg\lvert \frac{A_1 r_V + \sum_\psi N^c_\psi Q^2_\psi A_{1/2,\psi} r_\psi}{A_1 + \sum_\psi N^c_\psi Q^2_\psi A_{1/2,\psi}} \Bigg\rvert = \Bigg\lvert \frac{A_1 r_V + \frac{4}{3} A_{1/2,t}r_t + \frac{1}{3} A_{1/2,b}r_b + A_{1/2,\tau}r_\tau}{A_1 + \frac{4}{3} A_{1/2,t} + \frac{1}{3} A_{1/2,b} + A_{1/2,\tau}}\Bigg\rvert\, ,
\end{align}
%
%
where $r_t$, $r_b$ and $r_\tau$ are the modifications to the $htt$, $hbb$ and $h\tau\tau$ couplings that we will describe in the following sections, and $A_{i,\psi}$ is the loop function for particle $\psi$ with spin $i$, charge $Q_\psi$ and number of colours $N^c_\psi$. These are approximately~\cite{Carena:2012xa}:
\begin{align}
\hspace{-3em} A_1 \approx -8.324, & & A_{1/2,t} \approx 1.375, & & A_{1/2,b} \approx -0.072 - 0.095 i, & & A_{1/2,\tau} \approx -0.024 - 0.022 i\, .
\end{align} 

\subsubsection{Z signal strength} 

In the M4DCHM and LM4DCHM, we have a partially composite $b$ quark, which can be constrained by the partial width $\Gamma$ of the $Z \rightarrow b\bar{b}$ process, given as the branching ratio $R_b$ to the $b\bar{b}$ decay
\begin{align}
R_b = \frac{\Gamma(Z \rightarrow b\bar{b})}{\Gamma(Z \rightarrow \text{all hadrons})},
\end{align}
excluding top decays as they decay before hadronisation. This means, in practical terms in this work, only the bottom-contribution to the hadrons is affected by compositeness. The experimental \cite{Agashe:2014kda} and theoretical SM \cite{Freitas:2014hra} value are given by \cite{Angelescu:2015kga}
\begin{align}
R_b^\text{exp} = 0.21629 \pm 0.00066 && \text{and} && R_b^\text{SM} = 0.2158 \pm 0.00015
\end{align}


\subsubsection{Oblique observables}

\begin{figure}[H]
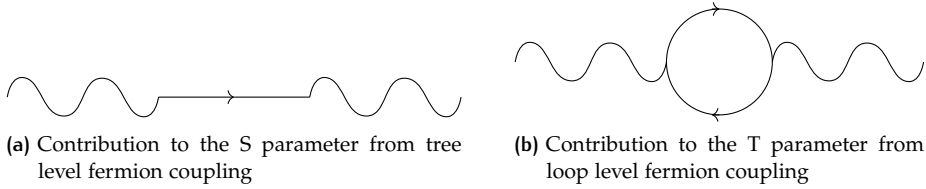

\centering
\subfloat[Contribution to the S parameter from tree level fermion coupling]{\tikz{
\draw plot [smooth, tension=1] coordinates {(0,1) (0.25,1.25) (0.75,0.75) (1.25,1.25) (1.75,0.75) (2,1) };
\draw [->] (2,1) -- (3,1);
\draw (3,1) -- (4,1);
\draw plot [smooth, tension=1] coordinates {(4,1) (4.25,1.25) (4.75,0.75) (5.25,1.25) (5.75,0.75) (6,1) };
}}\qquad
\subfloat[Contribution to the T parameter from loop level fermion coupling]{\tikz{
\draw plot [smooth, tension=1] coordinates {(0,1) (0.25,1.25) (0.75,0.75) (1.25,1.25) (1.75,0.75) (2,1) };
\draw (2.7,1) circle [radius=0.7];
\draw [->] (2.6,1.7) -- (2.7,1.7);
\draw [->] (2.7,0.3) -- (2.6,0.3);
\draw plot [smooth, tension=1] coordinates {(3.4,1) (3.65,1.25) (4.15,0.75) (4.65,1.25) (5.15,0.75) (5.4,1) };
}}\caption{Contributions to $S$ and $T$ deviations from the SM.\label{fig:oblique_corrections}}
\end{figure}

We parameterise BSM contributions to the electroweak radiative corrections by the form factors of the composite-SM gauge coupling $\{\Pi_{ZZ}, \Pi_{\gamma \gamma}, \Pi_{Z\gamma}, \Pi_{WW}\}$. The usual observables given are as a linear combination (also of the Weinberg angle entering $s_w, c_w$)
\begin{align}
\alpha T = \frac{\Pi_{WW}(0)}{m_W^2} - \frac{\Pi_{ZZ}(0)}{m^2_Z}, && \alpha S = 4s_w^2 c_w^2 \left[ \Pi'_{ZZ}(0) - \frac{c_w^2 - s_w^2}{s_w c_w} \Pi'_{Z\gamma}(0) - \Pi'_{\gamma\gamma}(0) \right]
\end{align}
The form factors here are \textit{not} simply those entering the tree-level Lagrangians in previous chapters, but rather include loop corrections (dominated by fermion contributions) of the sort in \cref{fig:oblique_corrections}. We enforce the Standard Model $T$ parameter at tree-level due to the custodial symmetry protecting $m_Z, m_W$. Global fits of the oblique parameters give experimental values of
\begin{align}
T = 0.09 \pm 0.13 && S = 0.05 \pm 0.11
\end{align}

\subsubsection{Resonance searches}
%

We include Drell-Yan search channels from LEP, Tevatron, and run-1 and -2 LHC, for BSM heavy resonances. Bosonic resonances will appear in decays such as $\rho^\pm \rightarrow W^\pm h$, and $\rho^0 \rightarrow t\bar{t}$. These are listed in \cref{tab:observables_2}, \cref{tab:observables_3} and \cref{tab:observables_4}. Masses in the full model can be calculated with the bosonic mass matrices constructed from the couplings in \cref{eq:MCHM_unitary_gauge}, while in the simplified model we assume the lightest masses are simply given by the tree-level relations in \cref{eq:gauge_parameters}.

Fermion resonance searches are included, such as pair production at the Tevatron and run-1 and  -2 LHC, in channels such as $Q \rightarrow tW$. The fermionic resonance masses can also be given by diagonalising the mass matrices constructed. In the simplified models, the masses can be given by the poles and roots of the form factors $\Pi_\Psi$; these are given in \cref{sec:FermionSector}. Note that single production (which is model dependent) and heavy resonance exchanges are not included in the decay channels. In the 3rd generation M4DCHM, there are 8 top-like resonances, 8 bottom-like resonances and 4 exotic resonances, with electromagnetic charges $5/3$ and $-4/3$. The resonances of the LM4DCHM and NM4DCHM are described in subsequent chapters. In general, however, we will focus on the lightest of each type of resonance in order to constrain the model, and do not discuss the full spectrum.

\begin{table}
\centering
\begin{tabular}{ @{}lllll @{} }
        \toprule
        \\
        Channel & Experiment &  $\sqrt{s}$ (TeV)  & Analysis & Ref. \\
        \midrule
        \multirow{2}{*}{$Q \rightarrow qW$} & ATLAS & 8 & EXOT-2014-10 & \cite{Aad:2015tba}\\
         & CMS & 8 & B2G-12-017 & \cite{CMS:2014dka} \\
        \midrule
        \multirow{8}{*}{$Q \rightarrow tW$} & CMS & 7 & B2G-12-004 & \cite{Chatrchyan:2012af}\\
         & ATLAS & 8 & EXOT-2014-17 & \cite{Aad:2015mba}\\
         & CMS & 8 & B2G-12-012 & \cite{Chatrchyan:2013wfa,CMS:vwa}\\
         & CMS & 8 & B2G-13-003 & \cite{CMS:2013una}\\
         & ATLAS & 8 & EXOT-2013-16 & \cite{Aad:2015gdg}\\
         & CMS & 8 & B2G-13-006 & \cite{Khachatryan:2015gza}\\
         & CMS & 13 & PAS-B2G-15-006 & \cite{CMS:2015alb}\\
         & CDF & 1 & 2009 & \cite{Aaltonen:2009nr} \\
        \midrule
        \multirow{3}{*}{$Q \rightarrow bZ$} & CMS & 7 & EXO-11-066 & \cite{CMS:2012jwa}\\
         & CMS & 8 & B2G-13-003 & \cite{CMS:2013una}\\
         & CMS & 8 & B2G-13-006 & \cite{Khachatryan:2015gza} \\
        \midrule
        \multirow{2}{*}{$Q \rightarrow jW$} & CDF & 1 & 10110 & \\
         & ATLAS & 7 & EXOT-2011-28 & \cite{Aad:2012bt} \\
        \midrule
        \multirow{4}{*}{$Q \rightarrow bH$} & ATLAS & 8 & CONF-2015-012 & \cite{ATLAS:2015dka}\\
         & CMS & 8 & B2G-12-019 & \cite{CMS:2012hfa}\\
         & CMS & 8 & B2G-14-001 & \cite{CMS:2014afa}\\
         & CMS & 8 & B2G-13-006 & \cite{Khachatryan:2015gza} \\
        \midrule
        $Q \rightarrow jZ$ & CDF & 1 & 2006 & \cite{Aaltonen:2007je} \\
        \midrule
        \multirow{7}{*}{$Q \rightarrow bW$} & CMS & 7 & EXO-11-050 & \cite{CMS:2012ab,CMS:2011jpa}\\
         & CMS & 7 & EXO-11-099 & \cite{CMS:2012dua,Chatrchyan:2012vu}\\
         & ATLAS & 7 & EXOT-12-07 & \\
         & CMS & 8 & B2G-12-017 & \cite{CMS:2014dka}\\
         & CMS & 8 & B2G-13-005 & \cite{Khachatryan:2015oba}\\
         & ATLAS & 8 & CONF-2015-012 & \cite{ATLAS:2015dka}\\
         & ATLAS & 13 & CONF-2016-102 & \cite{ATLAS:2016cuv} \\
        \midrule
        \multirow{3}{*}{$Q \rightarrow tH$} & CMS & 8 & B2G-13-005 & \cite{Khachatryan:2015oba}\\
         & CMS & 13 & PAS-B2G-16-011 & \cite{CMS:2016dmr}\\
         & ATLAS & 13 & CONF-2016-013 & \cite{TheATLAScollaboration:2016gxs} \\
        \midrule
        \multirow{4}{*}{$Q \rightarrow tZ$} & CMS & 7 & B2G-12-004 & \cite{Chatrchyan:2012af}\\
         & CMS & 7 & EXO-11-005 & \cite{Chatrchyan:2011ay,CMS:2011ara}\\
         & CMS & 8 & B2G-13-005 & \cite{Khachatryan:2015oba}\\
         & ATLAS & 13 & CONF-2016-101 & \cite{ATLAS:2016qlg} \\
        \bottomrule
        \end{tabular}\caption{Decay channels included in the set of constraints in the M4DCHM numerical exploration, with a fermionic initial state.}\label{tab:observables_1}
        \end{table}
        
        \begin{table}
        \centering
	\begin{tabular}{ @{}lllll @{} }
        \toprule
        \\
        Channel & Experiment &  $\sqrt{s}$ (TeV)  & Analysis & Ref. \\
        \midrule
        \multirow{6}{*}{$S \rightarrow HH$} & CMS & 8 & PAS-EXO-15-008 & \cite{CMS:2015zug}\\
         & CMS & 13 & PAS-HIG-16-002 & \cite{CMS:2016tlj}\\
         & ATLAS & 13 & EXOT-2015-11 & \cite{Aaboud:2016xco}\\
         & CMS & 13 & PAS-HIG-16-032 & \cite{CMS:2016vpz}\\
         & CMS & 13 & PAS-B2G-16-008 & \cite{CMS:2016pwo}\\
         & CMS & 13 & PAS-HIG-16-029 & \cite{CMS:2016knm} \\
        \midrule
        \multirow{2}{*}{$S \rightarrow gaga$} & ATLAS & 13 & CONF-2016-059 & \cite{ATLAS:2016eeo}\\
         & CMS & 13 & PAS-EXO-16-027 & \cite{CMS:2016crm} \\
        \midrule
        \multirow{2}{*}{$S \rightarrow WW$} & CMS & 13 & PAS-HIG-16-023 & \cite{CMS:2016jpd}\\
         & ATLAS & 13 & CONF-2016-074 & \cite{ATLAS:2016kjy} \\
        \midrule
        \multirow{4}{*}{$S \rightarrow WZ$} & ATLAS & 8 & HIGG-2014-13 & \cite{Aad:2015nfa}\\
         & CMS & 13 & HIG-16-027 & \cite{Sirunyan:2017sbn,CMS:2016szz}\\
         & ATLAS & 13 & EXOT-2016-11 & \cite{Aaboud:2018ohp}\\
         & CMS & 13 & SMP-18-001 & \cite{Sirunyan:2019ksz,CMS:2018ysc} \\
        \midrule
        \multirow{5}{*}{$S \rightarrow ZZ$} & ATLAS & 13 & CONF-2016-082 & \cite{ATLAS:2016npe}\\
         & ATLAS & 13 & CONF-2016-056 & \cite{ATLAS:2016bza}\\
         & ATLAS & 13 & CONF-2016-079 & \cite{ATLAS:2016oum}\\
         & CMS & 13 & PAS-HIG-16-033 & \cite{CMS:2016ilx}\\
         & CMS & 13 & PAS-B2G-16-010 & \cite{CMS:2016tio} \\
        \midrule
        \multirow{2}{*}{$S \rightarrow \tau{}\tau{}$} & ATLAS & 13 & CONF-2016-085 & \cite{ATLAS:2016fpj}\\
         & CMS & 13 & PAS-HIG-16-006 & \cite{CMS:2016pkt} \\
        \midrule
        \multirow{2}{*}{$S \rightarrow gg$} & CMS & 13 & PAS-EXO-16-032 & \cite{CMS:2016wpz}\\
         & CMS & 13 & PAS-EXO-16-032 & \cite{CMS:2016wpz} \\
        \midrule
        \multirow{5}{*}{$S \rightarrow Zga$} & ATLAS & 13 & EXOT-2016-02 & \cite{Aaboud:2016trl}\\
         & CMS & 13 & PAS-EXO-16-035 & \cite{CMS:2016cbb}\\
         & CMS & 13 & PAS-EXO-16-025 & \cite{CMS:2016mvc}\\
         & ATLAS & 13 & CONF-2016-044 & \cite{ATLAS:2016lri}\\
         & CMS & 13 & PAS-EXO-16-034 & \cite{CMS:2016pax} \\
        \midrule
        $S \rightarrow bb$ & CMS & 13 & PAS-HIG-16-025 & \cite{CMS:2016ncz} \\
        \bottomrule
        \end{tabular}\caption{Decay channels included in the set of constraints in the M4DCHM numerical exploration, with a scalar initial state.}\label{tab:observables_2}
        \end{table}
        
        \begin{table}
        \centering
\begin{tabular}{ @{}lllll @{} }
        \toprule
        \\
        Channel & Experiment &  $\sqrt{s}$ (TeV)  & Analysis & Ref. \\
        \midrule
        \multirow{4}{*}{$V \rightarrow \mu{}\mu{}$} & ATLAS & 8 & EXOT-2012-23 & \cite{Aad:2014cka}\\
         & CMS & 8 & EXO-12-061 & \cite{Khachatryan:2014fba,CMS:2013qca}\\
         & CMS & 13 & PAS-EXO-16-031 & \cite{CMS:2016abv}\\
         & ATLAS & 13 & CONF-2016-045 & \cite{ATLAS:2016cyf} \\
        \midrule
        \multirow{2}{*}{$V \rightarrow qq$} & CMS & 13 & PAS-EXO-16-032 & \cite{CMS:2016wpz}\\
         & CMS & 13 & PAS-EXO-16-032 & \cite{CMS:2016wpz} \\
        \midrule
        \multirow{4}{*}{$V \rightarrow ee$} & CMS & 8 & EXO-12-061 & \cite{Khachatryan:2014fba,CMS:2013qca}\\
         & ATLAS & 8 & EXOT-2012-23 & \cite{Aad:2014cka}\\
         & ATLAS & 13 & CONF-2016-045 & \cite{ATLAS:2016cyf}\\
         & CMS & 13 & PAS-EXO-16-031 & \cite{CMS:2016abv} \\
        \midrule
        \multirow{3}{*}{$V \rightarrow WW$} & CMS & 8 & EXO-13-009 & \cite{Khachatryan:2014gha}\\
         & ATLAS & 8 & EXOT-2013-01 & \cite{Aad:2015ufa}\\
         & ATLAS & 13 & CONF-2016-062 & \cite{ATLAS:2016cwq} \\
        \midrule
        $V \rightarrow l\nu{}$ & ATLAS & 7 & EXOT-2012-02 &  \\
        \midrule
        $V \rightarrow WW+ZH$ & CMS & 13 & PAS-B2G-16-007 & \cite{CMS:2016wev} \\
        \midrule
        \multirow{11}{*}{$V \rightarrow WZ$} & ATLAS & 8 & EXOT-2013-08 & \cite{Aad:2015owa}\\
         & CMS & 8 & EXO-12-024 & \cite{Khachatryan:2014hpa,CMS:2013fea}\\
         & ATLAS & 8 & EXOT-2013-01 & \cite{Aad:2015ufa}\\
         & ATLAS & 8 & EXOT-2013-07 & \cite{Aad:2014pha}\\
         & ATLAS & 13 & EXOT-2016-11 & \cite{Aaboud:2018ohp}\\
         & ATLAS & 13 & CONF-2016-062 & \cite{ATLAS:2016cwq}\\
         & CMS & 13 & PAS-EXO-15-002 & \cite{CMS:2015nmz}\\
         & ATLAS & 13 & EXOT-2016-29 & \cite{Aaboud:2017itg}\\
         & CMS & 13 & PAS-B2G-16-020 & \cite{CMS:2016pfl}\\
         & ATLAS & 13 & CONF-2016-055 & \cite{ATLAS:2016yqq}\\
         & ATLAS & 13 & CONF-2016-082 & \cite{ATLAS:2016npe} \\
        \midrule
        \multirow{4}{*}{$V \rightarrow tb$} & CMS & 8 & B2G-12-010 & \cite{Chatrchyan:2014koa,CMS:pwa}\\
         & CMS & 8 & B2G-12-009 & \cite{Khachatryan:2015edz,CMS:2014kfa}\\
         & CMS & 13 & PAS-B2G-16-009 & \cite{CMS:2016ude}\\
         & CMS & 13 & PAS-B2G-16-017 & \cite{CMS:2016wqa} \\
        \midrule
        \multirow{5}{*}{$V \rightarrow jj$} & CMS & 13 & EXO-15-001 & \cite{Khachatryan:2015dcf,CMS:2015wqf}\\
         & ATLAS & 13 & EXOT-2015-02 & \cite{ATLAS:2015nsi}\\
         & ATLAS & 13 & EXOT-2015-02 & \cite{ATLAS:2015nsi}\\
         & CMS & 13 & EXO-16-056 & \cite{Sirunyan:2018xlo,CMS:2017xrr}\\
         & ATLAS & 13 & EXOT-2018-05 & \cite{Aaboud:2019zxd} \\
        \midrule
        \multirow{6}{*}{$V \rightarrow ZH$} & ATLAS & 8 & EXOT-2013-23 & \cite{Aad:2015yza}\\
         & CMS & 8 & EXO-13-007 & \cite{Khachatryan:2015ywa}\\
         & ATLAS & 13 & CONF-2016-083 & \cite{ATLAS:2016kxc}\\
         & ATLAS & 13 & EXOT-2015-18 & \cite{Aaboud:2016lwx}\\
         & CMS & 13 & PAS-B2G-16-003 & \cite{CMS:2016dzw}\\
         & ATLAS & 13 & CONF-2015-074 & \cite{TheATLAScollaboration:2015ulg} \\
        \bottomrule
        \end{tabular}\caption{Decay channels included in the set of constraints in the M4DCHM numerical exploration, with a vector resonance as an initial state.}\label{tab:observables_3}
        \end{table}
        
        \begin{table}
        \centering
\begin{tabular}{ @{}lllll @{} }
        \toprule
        \\
        Channel & Experiment &  $\sqrt{s}$ (TeV)  & Analysis & Ref. \\
        \midrule
        \multirow{5}{*}{$V \rightarrow WH$} & CMS & 8 & EXO-14-010 & \cite{Khachatryan:2016yji,CMS:2015gla}\\
         & ATLAS & 8 & EXOT-2013-23 & \cite{Aad:2015yza}\\
         & ATLAS & 13 & CONF-2016-083 & \cite{ATLAS:2016kxc}\\
         & CMS & 13 & PAS-B2G-16-003 & \cite{CMS:2016dzw}\\
         & ATLAS & 13 & EXOT-2015-18 & \cite{Aaboud:2016lwx} \\
        \midrule
        \multirow{3}{*}{$V \rightarrow e\nu{}$} & ATLAS & 7 & EXOT-2012-02 & \\
         & ATLAS & 13 & CONF-2016-061 & \cite{ATLAS:2016ecs}\\
         & CMS & 13 & PAS-EXO-15-006 & \cite{CMS:2015kjy} \\
        \midrule
        \multirow{3}{*}{$V \rightarrow \tau{}\nu{}$} & CMS & 8 & EXO-12-011 & \cite{Khachatryan:2015pua,CMS:2015vda}\\
         & CMS & 13 & PAS-EXO-16-006 & \cite{CMS:2016ppa}\\
         & CMS & 13 & PAS-EXO-16-006 & \cite{CMS:2016ppa} \\
        \midrule
        \multirow{3}{*}{$V \rightarrow \tau{}\tau{}$} & ATLAS & 8 & EXOT-2014-05 & \cite{Aad:2015osa}\\
         & CMS & 8 & EXO-12-046 & \cite{CMS:2015ufa}\\
         & CMS & 13 & PAS-EXO-16-008 & \cite{CMS:2016zxk} \\
        \midrule
        \multirow{4}{*}{$V \rightarrow tt$} & ATLAS & 8 & CONF-2015-009 & \cite{ATLAS:2015aka}\\
         & CMS & 8 & B2G-13-008 & \cite{Khachatryan:2015sma,CMS:2015nza}\\
         & CMS & 13 & PAS-B2G-15-003 & \cite{CMS:2016ehh}\\
         & CMS & 13 & PAS-B2G-15-002 & \cite{CMS:2016zte} \\
        \midrule
        \multirow{3}{*}{$V \rightarrow \mu{}\nu{}$} & ATLAS & 7 & EXOT-2012-02 & \\
         & ATLAS & 13 & CONF-2016-061 & \cite{ATLAS:2016ecs}\\
         & CMS & 13 & PAS-EXO-15-006 & \cite{CMS:2015kjy} \\
        \bottomrule
        \end{tabular}\caption{(\textit{cont.}) Decay channels included in the set of constraints in the M4DCHM numerical exploration, with a vector resonance as an initial state.}\label{tab:observables_4}
        \end{table}

\section{Minimal Model Numerical Behaviour}
\label{sec:M4DCHM_Results}


Here we present a brief overview of the first convergent global fit scans of the 3rd-generation M4DCHM. In particular, we are interested in the improvements in tuning and mass constraints gained from recent experimental results, as well as the sophisticated optimisation algorithm. We explore this model space with the Diver evolutionary algorithm detailed in \ref{sec:Diver}, with a convergence criterion of $-\log(\text{likelihood})$ improvement of less than $10^{ -5}$, averaged over $10$ generations. This produced approximately $3000$ sample points with a $-\log(\text{likelihood}) < 200$. This distribution is enough to understand the tuning behaviour of the model, particularly as the higher-order tuning calculation is computationally expensive.

\subsection{Fine Tuning Behaviour}
\begin{figure}
\centering
\subfloat[The BG tuning behaviour of the symmetry breaking scale]{
\centering
\tikz{
\node at (0,0) {\includegraphics[width=0.45\linewidth]{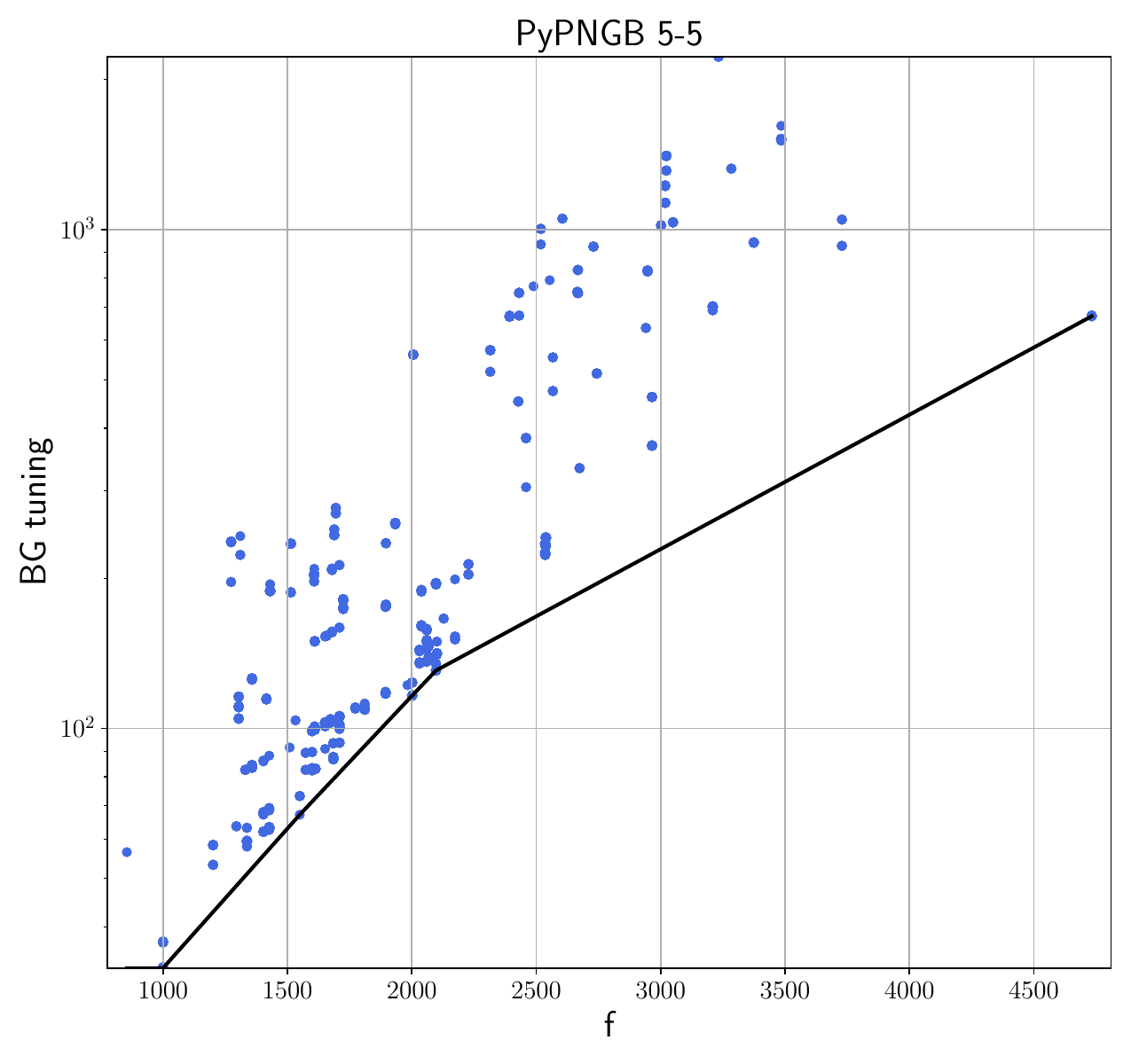}};
\fill [fill=white] (-1,2.5) rectangle (1,2.7);
}\label{fig:MCHM_tuning_f_BG}
}
\; \subfloat[The higher order tuning behaviour of the symmetry breaking scale, which can be used to differentiate points that appear identical in the BG measure]{
\centering
\tikz{
\node at (0,0) {\includegraphics[width=0.45\linewidth]{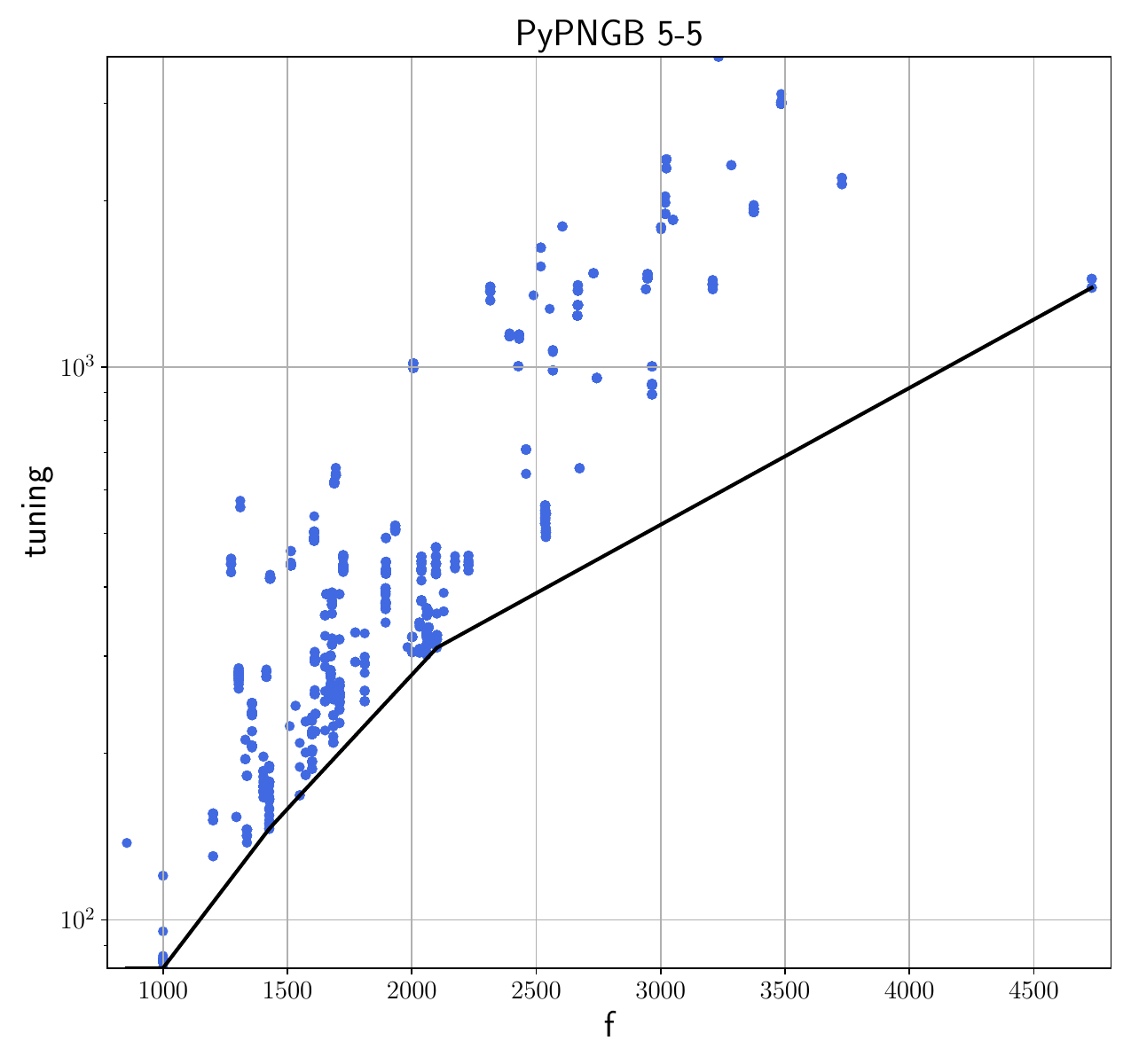}};
\fill [fill=white] (-1,2.5) rectangle (1,2.7);
}
\label{fig:MCHM_tuning_f};
}\\
\subfloat[The tuning behaviour of the lightest fermionic partner in each parameter point]{
\centering
\tikz{
\node at (0,0) {\includegraphics[width=0.8\linewidth]{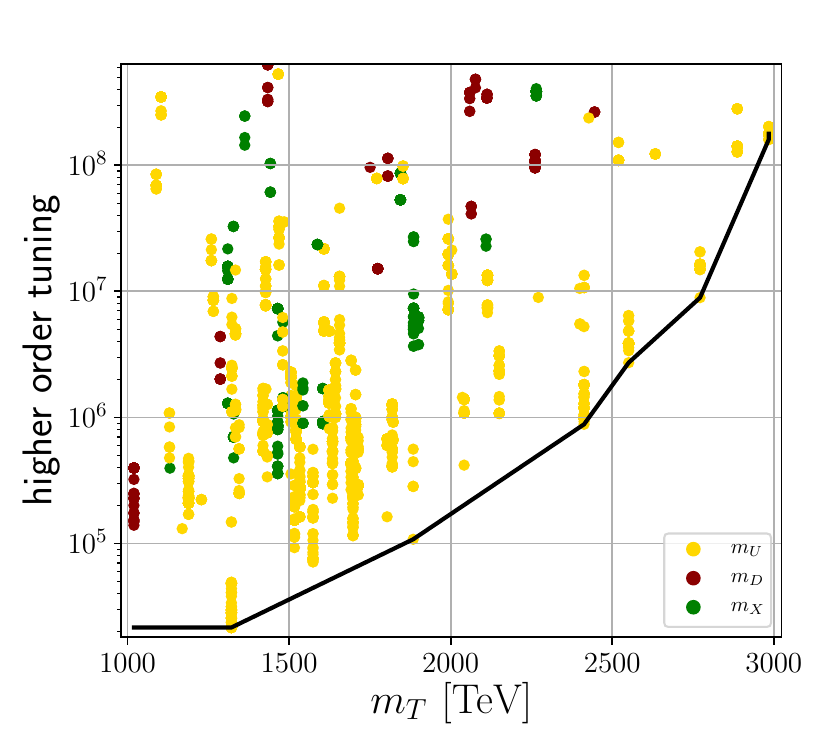}};
}
\label{fig:MCHM_tuning_mT}}
\caption{Tuning of the breaking scale and fermionic composite partner masses}\label{fig:scale_and_partner_masses}
\end{figure}

We strictly enforce experimental constraints by trimming points with \textit{any} observable outside $3\sigma$ of the experimental value. After this cut, we are left with $\sim 2800$ valid points. We present the BG tuning of the mass of the $Z$ boson
\begin{align}
\Delta_{\text{BG}} = \text{max}_{x_i}|\frac{\partial m_Z}{\partial x_i} | 
\end{align}
as a comparison to the results of previous global fit study \cite{niehoff2016direct}. In this work, we find a minimum tuning of $\Delta_\text{BG} = 33$, which agrees with the minimum of the previous work. However, there is a significant difference from that work, where the minimum tuning was found for a breaking scale of $f \sim 600\gev$. In this study, we find the minimum of $\Delta_\text{BG} = 33$ at $f \sim 1000\gev$, as shown in \cref{fig:MCHM_tuning_f_BG}. This reflects that the $3\sigma$ constraint cuts significantly constrain the breaking scale from previous work. The limit placed in \cite{niehoff2016direct} is $f \gtrsim 500 \gev$, and in \cite{Barnard:2015ryq, Falkowski:2013dza} $f \gtrsim 700$. Here, we place $3\sigma$ limits on the following observables
\begin{align}
f & \gtrsim 800\gev \nonumber \\
\{ m_U, m_D, m_X \} & \gtrsim \{1100, 1000, 1200 \} \gev \\
\{m_a, m_\rho \} & \gtrsim \{ 1000, 2500 \} \gev \nonumber
\end{align}
where $m_U, m_D, m_X$ are the masses of the lightest partners with quantum numbers of the top, bottom and an exotic $\textbf{2}_{7/6}$ respectively, and $m_a, m_\rho$ are the masses of the lightest partners of the $Z$ and $W$ respectively. 

While the minimum tuning also agrees with that found in \cite{Barnard:2015ryq}, these constraints significantly update the search bounds on new physics without demanding a larger tuning. The low-tuning bounds placed on the lightest quark partners generally agree with the structure found in \cite{Barnard:2015ryq, Barnard:2017kbb}. There are some interesting deviations. The frontier of least-tuned points is dominated by cases where the top partner is the lightest, with cases of the exotic partner and down partner more tuned by one and two orders of magnitude, respectively. This is in tension with \cite{Barnard:2015ryq} and \cite{de2012}, where the least tuned scenarios typically had a light exotic partner. The tuning also grows much more rapidly than in the bottom-inclusive or top-only previous works \cite{Barnard:2017kbb} and \cite{Barnard:2015ryq}, seen in \cref{fig:MCHM_tuning_mT}. Given a most natural fermion partner mass of $m_U \sim 1300\gev$, we see an order of magnitude increase already for fermion partners with mass $m_U \sim 2100 \gev$. 

To summarise the fermion resonance behaviour, we suggest that the higher order tuning and $3\sigma$ cuts show that the tuning is significantly underestimated as search limits are, in future colliders, placed on masses above $2 \tev$. However, current limits do not exclude masses above $1 \tev$, and indeed these masses can be found for points that are just as natural as sub-$\tev$ fermion resonances.

\begin{figure}
\centering
\hspace*{-3cm}\subfloat[]{
\centering
\includegraphics[width=0.6\linewidth]{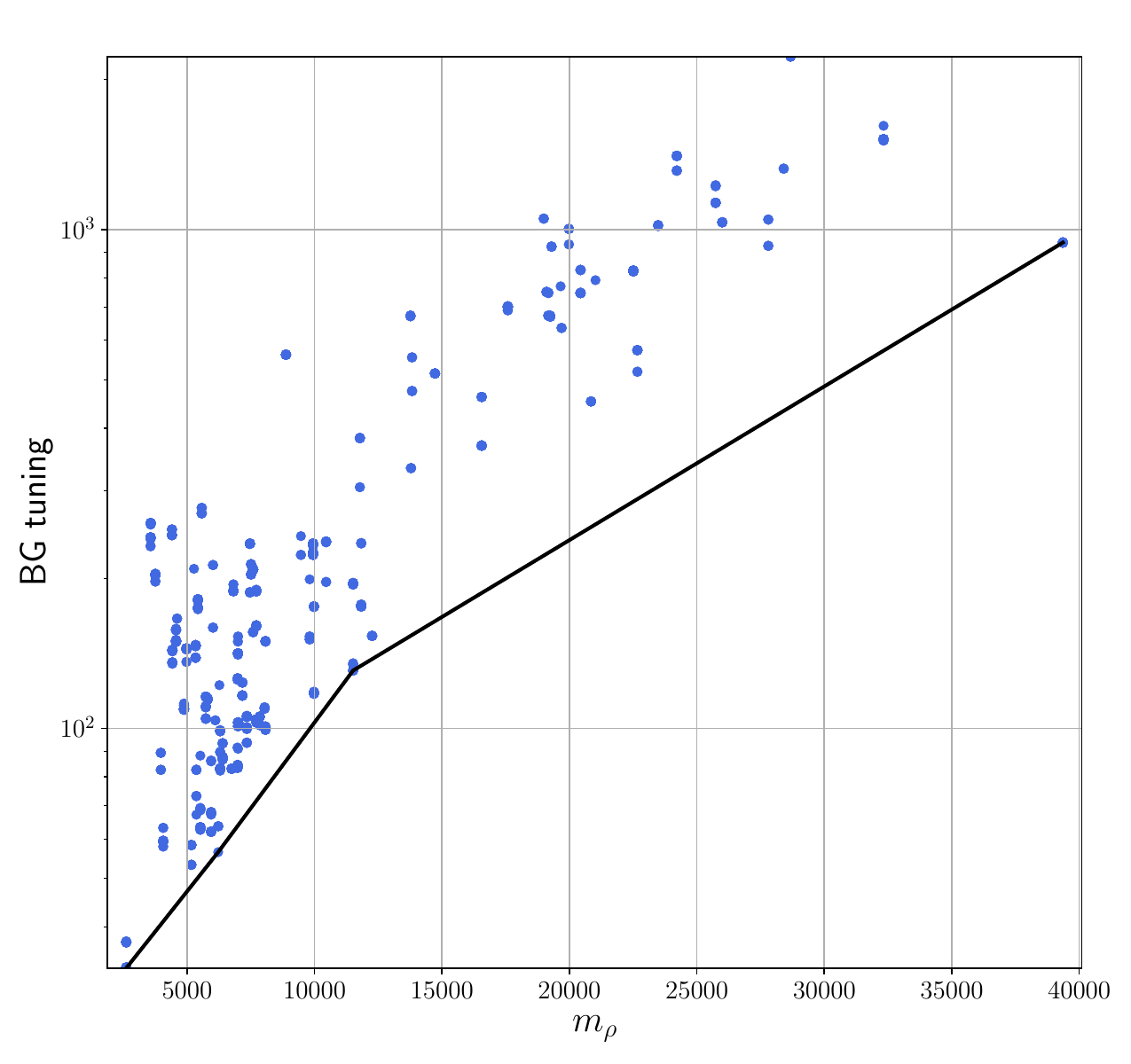}
}
\subfloat[]{
\centering
\includegraphics[width=0.6\linewidth]{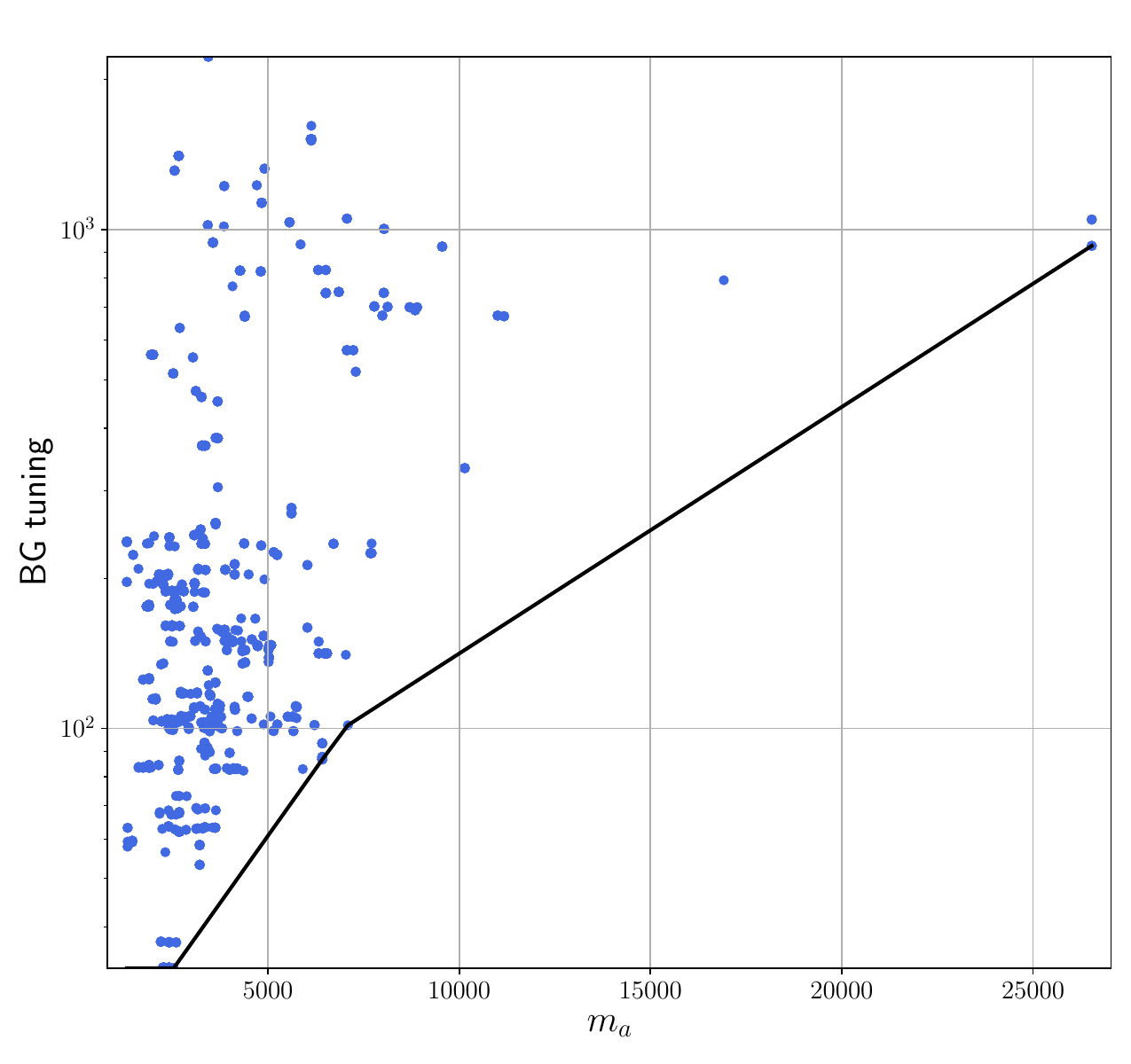}
}\\
\centering
\hspace*{-3cm}\subfloat[]{
\centering
\includegraphics[width=0.6\linewidth]{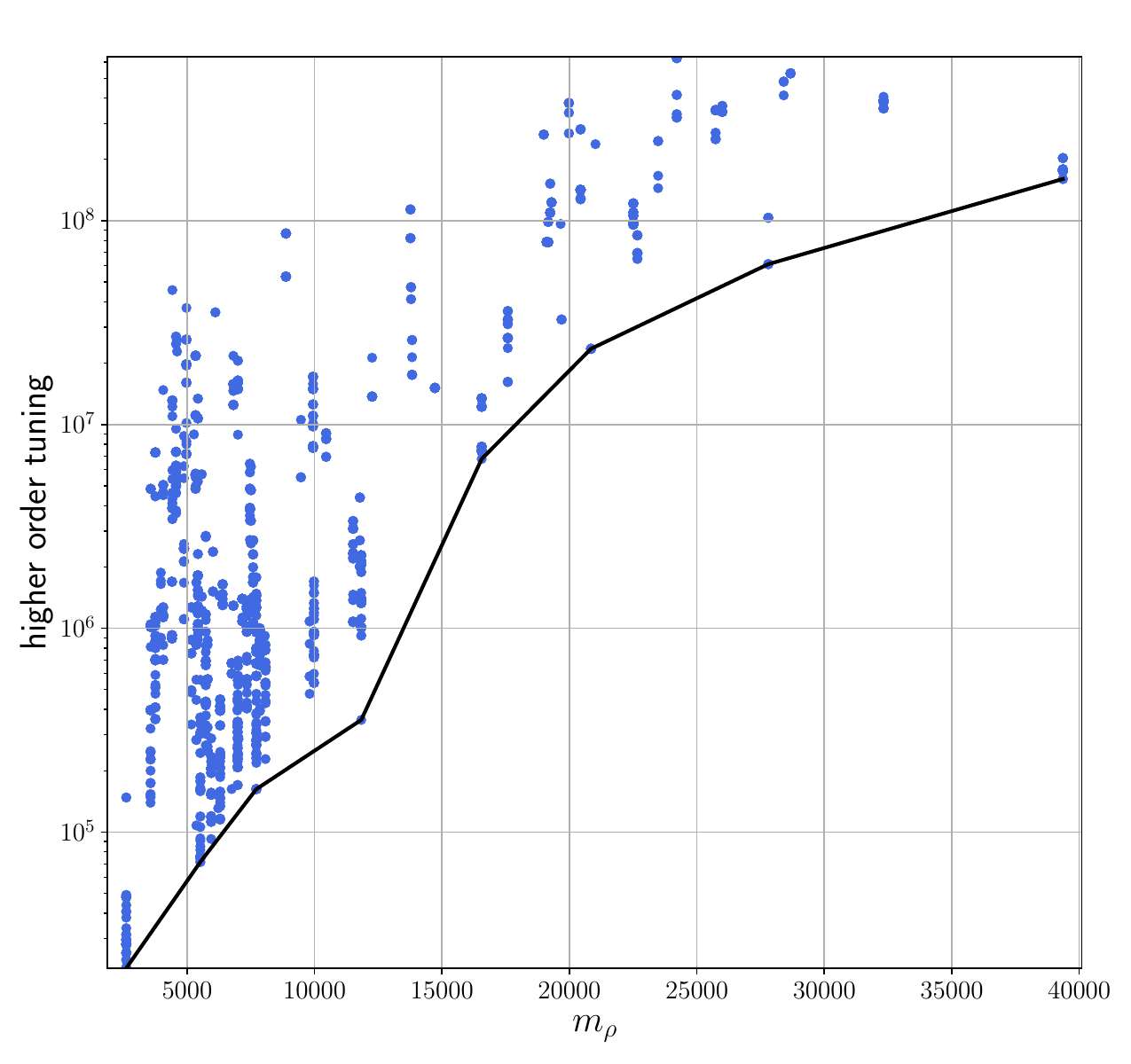}
}
\subfloat[]{
\centering
\includegraphics[width=0.6\linewidth]{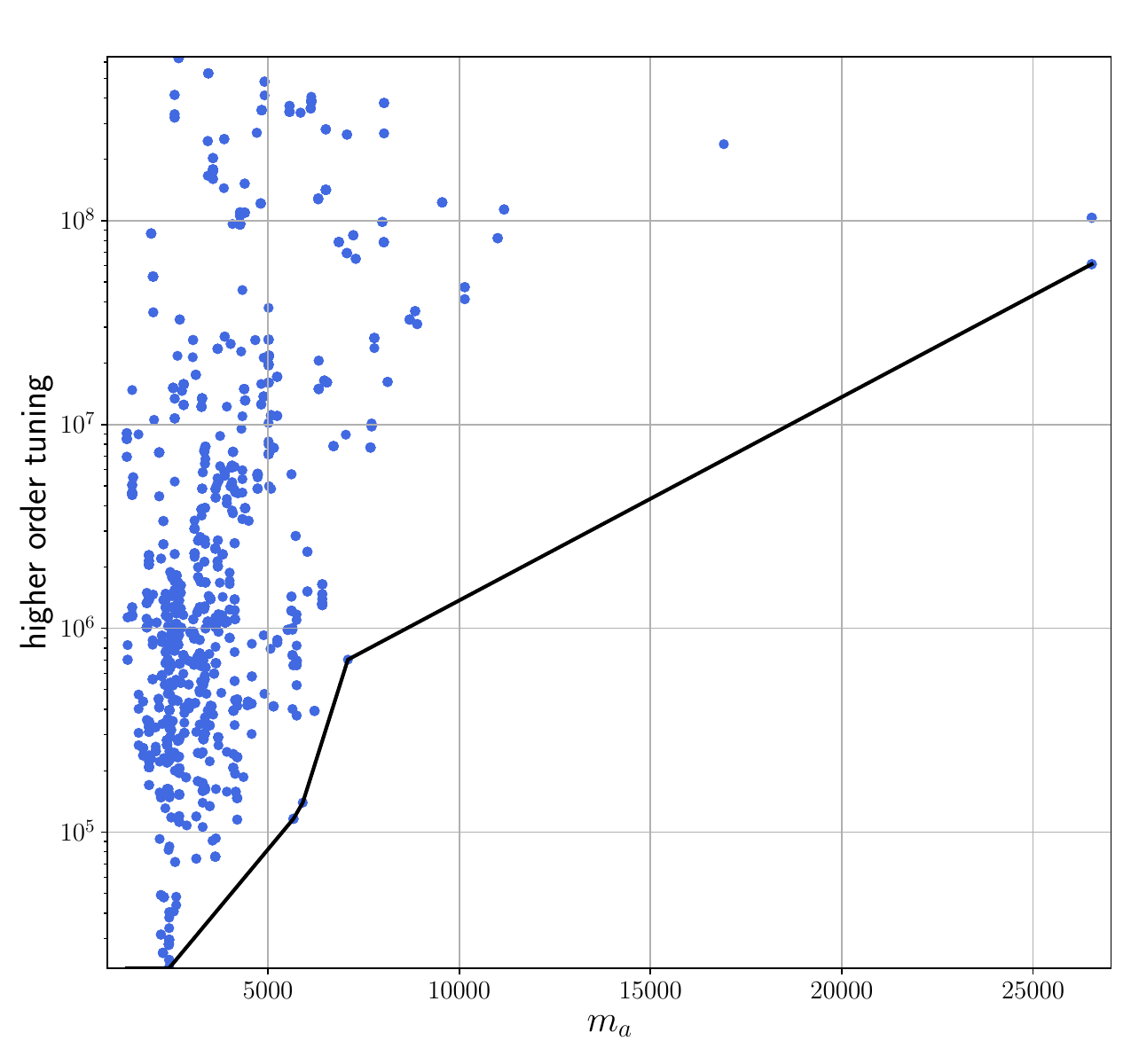}
}
\caption{The tuning behaviour of the lightest composite vector resonances}\label{fig:MCHM_tuning_vectors}
\end{figure}

The experimental constraints on the vector resonances are much more severe, which actually relaxes the tuning pressure on the model - we shouldn't expect to observe vector and axial resonances at LHC energies. Furthermore, the minimal model is still within percent-level tuning (i.e. $\Delta_\text{BG} < 100$) for vector and axial resonances with masses $m_\rho \sim 8\tev$ and $m_a \sim 7 \tev$, as seen in \cref{fig:MCHM_tuning_vectors}. 

\begin{figure}
\centering
\hspace*{0cm}\subfloat[]{
\centering
\includegraphics[width=0.6\linewidth]{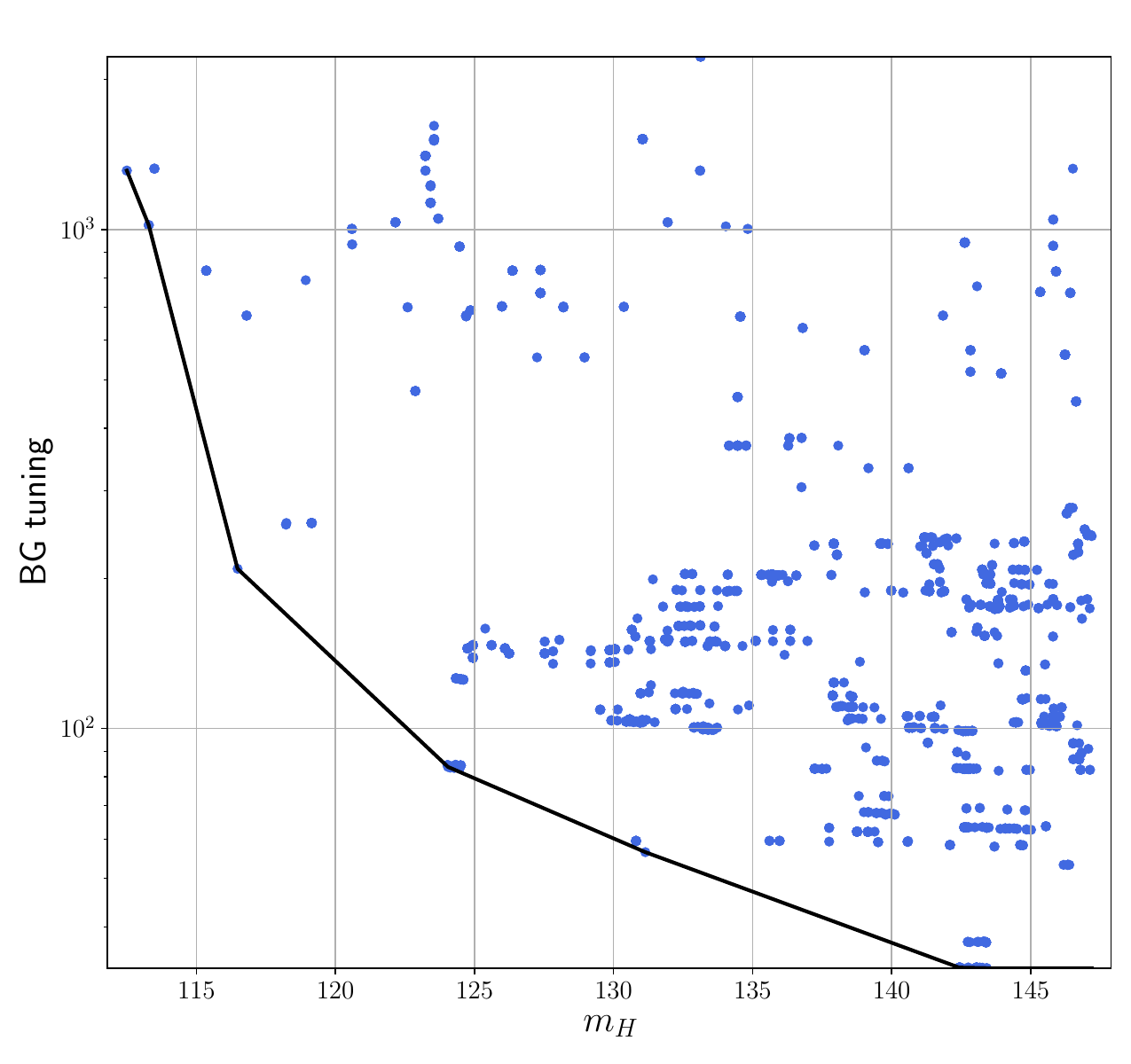}
}
\subfloat[]{
\centering
\includegraphics[width=0.6\linewidth]{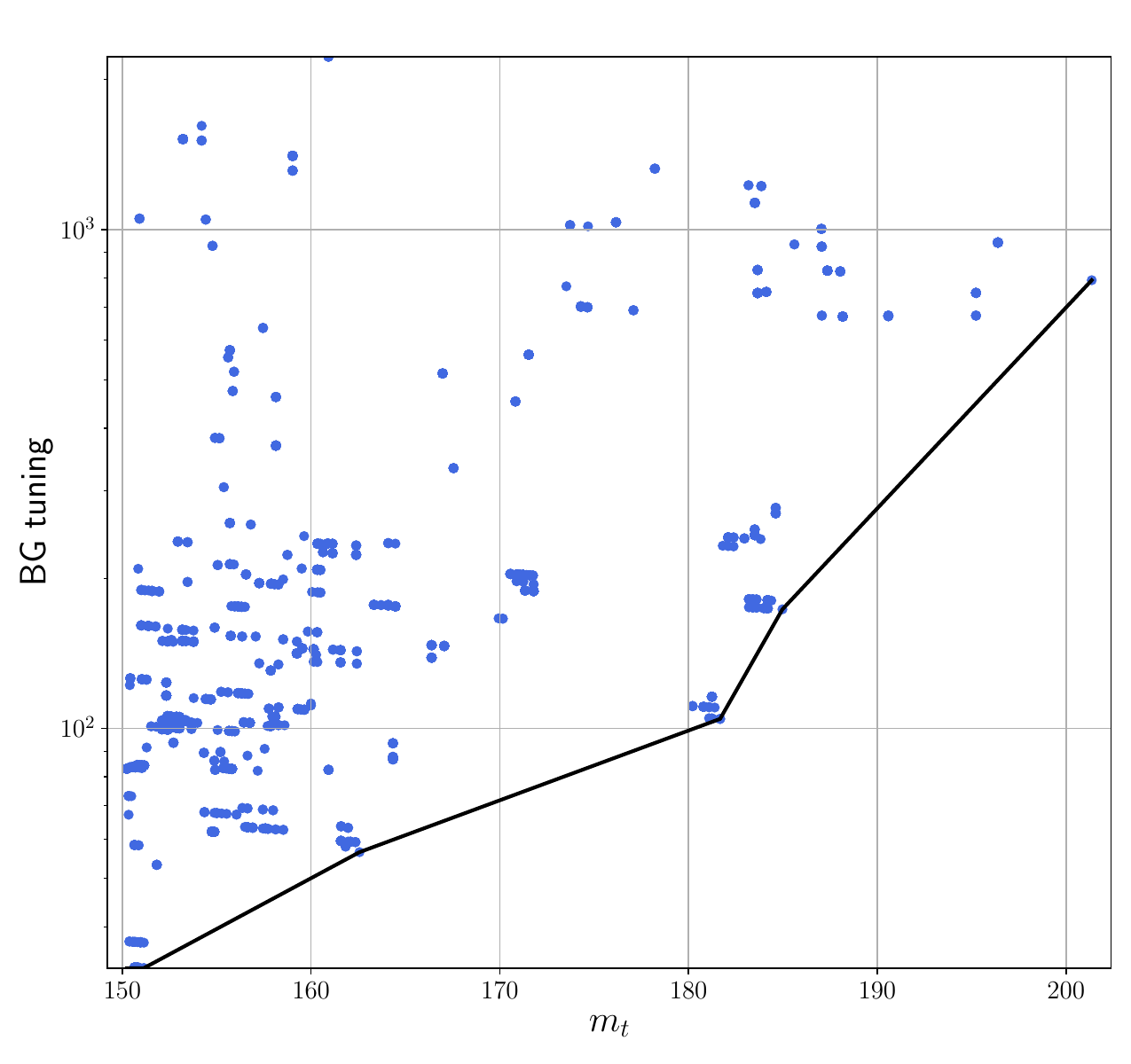}
}
\caption{The tuning behaviour of SM Higgs and top masses}\label{fig:MCHM_tuning_SM}
\end{figure}

We also re-iterate the sentiment in \cite{niehoff2016direct} that important observables, such as the Higgs and top masses, are accompanied with a significant theoretical uncertainty in these parameter points. Their tuning is also systematically lower or higher, as seen in \cref{fig:MCHM_tuning_SM}. Thus, as stronger constraints are placed on these observables, the $3\sigma$ cut on these points may rapidly increase the tuning in the minimal model.

Finally, it should be noted that although the differential evolution scanner produced a convergent global fit, the posterior results are sparsely sampled, and further work should be done to obtain a better understanding of the global behaviour. However, we observed a best likelihood fit of $m_U \sim 2050 \gev$, $m_D \sim 4000 \gev$ in the composite fermion plane, and $m_\rho \sim 7000 \gev$, $m_a \sim 3500\gev$ in the composite vector boson plane. 

\chapter{M4DCHM Extensions: Leptons and Representations}
\label{sec:LCHM}

\section{Motivation}\label{sec:LCHM_motivation}

As we were inspired to consider composite dynamics as a solution to the first-order tuning of the Standard Model, we are inspired to look for extensions to the M4DCHM as a solution to its second-order tuning. In this chapter, we will describe two such extensions by extending the matter sector (see works \cite{Barnard:2015ryq, Barnard:2017kbb, carmona2015, Redi:2012ha, panico2012, matsedonskyi2012} for some history of this development). In the following chapter, we will extend the model's symmetry structure itself.

Recall the potential term of a partially composite fermion (e.g. the top quark), with both elementary chiralities coupling with composite partners in the fundamental representation
\begin{align}
V_\text{top}^{\textbf{5}}(h) & = -2N_c \int \frac{d^2p_E}{16\pi^2} p_E^2\log\left[p^2_E\left(\Pi^{t_L}_0 + \frac{s^2_h}{2}\Pi^{t_L}_1\right)\left(\Pi^{t_R}_0 + c_h^2 \Pi^{t_R}_1\right) - \frac{s_h^2 c_h^2}{2}|M^t_1|^2 \right]\nonumber \\
& = -\gamma s_h^2 + \beta s_h^2
\end{align}
We expanded this in $d_\Psi = \Delta_\Psi/m_\Psi$ and found the leading-order $\gamma$ term to be enhanced by $1/d_\Psi^2$ (where we would expect $d_\Psi^2$ to naturally be less than $1$), relative to $\beta$, thus making the EWSB relation $\alpha/\beta \ll 1$ even more difficult. Note that the Higgs mass is of order
\begin{align}
(m_H^{\textbf{5}})^2 = \frac{8\beta}{f^2} c_{\langle h\rangle}^2 s_{\langle h \rangle}^2  \sim d_\Psi^4 m_\Psi^4 v^2
\end{align}
which is more or less correct, assuming the natural $\Delta_\Psi \approx 1$, so that $d_\Psi^4 m_\Psi^4 = (\Delta_\Psi / m_\Psi)^4 m_\Psi^4 \approx 1$.

Now consider each elementary chirality coupling with multiplets of composite fields in the \textbf{14}, "symmetric" representation. We will discuss the details of this in \cref{sec:FermionSector}. For now, inspect the potential of the somewhat more complicated symmetric representation
\begin{align}
V_\text{top}^{\textbf{14}}(h) &= -2N_c \int \frac{d^2p_E}{16\pi^2} p_E^2\log\left[p^2_E \left( \Pi_0^{t_L} + \frac{1}{2}\Pi_1^{t_L} \left( 1 - \frac{s_h^2}{2} \right) + \Pi_2^{t_L} \left( s_h^2 - s_h^2 \right)\right) \right. \nonumber \\
& \left. \times \left(\Pi_0^{t_R} + \frac{1}{20} \Pi_1^{t_R}\left( 16 - 15 s_h^2 \right) + \frac{1}{20} \Pi_2^{t_R} \left( 4 - 5s_h^2 \right)^2 \right) \right. \nonumber\\
 & \left.   - \frac{1}{80}(s_h^2 - s_h^4) \Bigm\lvert 3 M_1  + 2 M_2 \left( 4 - 5 s_h^2 \right) \Bigm\lvert^2   \right]
\end{align}
Given the explicit form factors in \cref{sec:form_factors}, we can expand this in $d_\Psi$ to quadratic order
\begin{align}
V^{(2)} &= \frac{N_c m_t^4}{16\pi} f(d^2_\Psi, d^2_{\tilde{\Psi}}) ((a + \tilde{a})s_h^2 + (b + \tilde{b})s_h^4)
\end{align}
which still requires relative tuning between $\gamma$ and $\beta$ to achieve $\xi \ll 1$, but now the compositeness factor $d_\Psi$ can be factored out at leading order. This is due to the \textbf{14} having an extra possible coupling term, from the particular transformation properties of the representation.
It turns out that this requires some extra tuning down the line to ensure a Higgs mass of the correct order, since 
\begin{align}
(m_H^{\textbf{14}})^2 = \frac{8\beta}{f^2} c_{\langle h\rangle}^2 s_{\langle h \rangle}^2 \sim d_\Psi^2 m_\Psi^4 v^2
\end{align} 
where now the Higgs mass is carrying an extra $1/d_\Psi$ enhancement. Regardless, we see that different fermion irreducible representation combinations may lead to serendipitous phenomenology. Heavy quarks in the fundamental representation lead to a light Higgs, but tuned EWSB. Heavy quarks in the symmetric representation lead to a heavy Higgs, and a more natural EWSB.

If only we could keep the heavy quarks in fundamental representations, but introduce another field to fix the EWSB tuning. Fortunately, the tau lepton can play this part. Typically, due to their small masses, the leptons contribute negligibly to the Higgs potential. But by including the right-handed tau in the symmetric representation, we have a potential that appears at leading order in $d_\Psi$ and the tau compositeness $d_\mathcal{T}$ as
\begin{align}
V_{t,\tau} &= \left[ \frac{m_\Psi^4}{16\pi^2} \left( N_c (a^{(2)} d_\Psi^2 - \tilde{a}^{(2)}d^2_{\tilde{\Psi}} ) + a_\tau^{(2)} d_{\mathcal{T}}^2 \right) + \alpha_t^{(4)} (d_{\Psi,\tilde{\Psi}}^4) \right] s_h^2 \nonumber\\
& - \left[ \frac{m_\Psi^4}{16 \pi^2} b_\tau^{(2)} d_{\mathcal{T}}^2 + \beta_t(d_{\Psi, \tilde{\Psi}}^4) \right] s_h^4
\end{align}
Now, the $s_h^4$ coefficient includes the normally negligible $b_\tau^{(2)}$ as it is enhanced relative to the quark contribution by $d_{\mathcal{T}}^2$. This extra freedom from lepton couplings $\Delta_\tau, \tilde{\Delta}_\tau$ allows a lighter Higgs, viable EWSB, and potentially heavier fermionic resonances. We will quantitatively investigate the veracity of this claim in this chapter, using the higher order tuning measure developed in \cref{sec:fine_tuning}.


\section{Representing Fermions}\label{sec:FermionSector}

As noted above, the specific fermion form factors that enter \cref{eq:Leff} depend on the way that each composite fermion is embedded in the $SO(5)_1$ group. That is, there is more than one way of representing the fermion multiplet in the Lagrangian such that its terms are invariant under an $SO(5)_1$ rotation. We are interested in all of the lowest dimension representations, the trivial ${\bf 1}$, the fundamental ${\bf 5}$, the antisymmetric ${\bf 10}$ and the symmetric traceless ${\bf 14}$. Note that the spinorial $\textbf{4}$ is ruled out as it leads to FCNC phenomenology. The machinery of embedding and decomposing fields into multiplets is expounded in \cref{chp:particle_content}. We will simply state the relevant embeddings in this section.

\begin{figure}
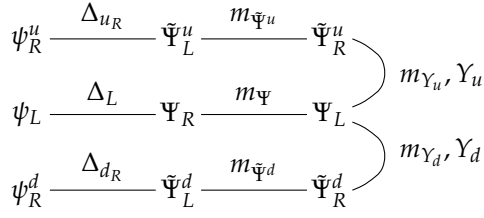

\centering
\tikz{
\node at (1,1) {$\psi^d_R$};
\draw (1.3,1) --(2.7,1);
\node [above] at (2,3) {$ \Delta_{u_R} $};
\node [above] at (2,2) {$ \Delta_L $};
\node [above] at (2,1) {$ \Delta_{d_R} $};
\node at (1,2) {$\psi_L$};
\draw (1.3,2) --(2.7,2);
\node at (1,3) {$\psi^u_R$};
\draw (1.3,3) --(2.7,3);
\node at (3,1) {$\tilde{\Psi}^d_L$};
\draw (3.3,1) --(4.7,1);
\node [above] at (4,1) {$ m_{\tilde{\Psi}^d} $};
\node [above] at (4,2) {$ m_{\Psi} $};
\node [above] at (4,3) {$ m_{\tilde{\Psi}^u} $};
\node at (3,2) {$\Psi_R$};
\draw (3.3,2) --(4.7,2);
\node at (3,3) {$\tilde{\Psi}^u_L$};
\draw (3.3,3) --(4.7,3);
\node at (5,1) {$\tilde{\Psi}^d_R$};
\draw plot [smooth, tension=0.7] coordinates {(5.3,1) (5.5,1.1)  (5.7,1.3) (5.7,1.6) (5.5, 1.8) (5.3,1.9)};
\node [right] at (5.8,1.5) {$ m_{Y_d}, Y_d $};
\node [right] at (5.8,2.5) {$ m_{Y_u}, Y_u $};
\draw plot [smooth, tension=0.7] coordinates {(5.3,2.1) (5.5,2.2)  (5.7,2.4) (5.7,2.7) (5.5, 2.9) (5.3,3)};
\node at (5,2) {$\Psi_L$};
\node at (5,3) {$\tilde{\Psi}^u_R$};
}\caption{Allowed interactions in the minimal model (excluding the \textbf{5}-\textbf{5}-\textbf{5} and \textbf{14}-\textbf{14}-\textbf{14} embedding).}\label{fig:allowed_interactions}
\end{figure}

I remind that our guiding light for forming interactions in the Lagrangian is the Minimal Composite Higgs Hypothesis: include the minimal matter and interaction content that is compatible with the global $SO(5)_0 \times SO(5)_1$ and gauged $SO(5)_1$, while \textit{avoiding divergences in the potential}. This caveat defines the terms we do not include. Let us briefly examine each component of the fermion Lagrangian of \cref{eq:fundamental_fermion_lagrangian}, where all partners were in the fundamental
\begin{align}
\mathcal{L}_\text{fermionic} &= \mathcal{L}_\text{elementary} + \mathcal{L}_\text{composite} + \mathcal{L}_\text{link} + \mathcal{L}_\text{Yukawa}
\end{align}
The elementary Lagrangian contains only kinetic terms, which do not depend on representation. They will remain universal. We also assume that the composite fermions have bare mass, therefore the left and right chiralities must be in the same representation of the gauged $SO(5)_1$. Therefore the composite terms are also universal,
\begin{align}
 \mathcal{L}_\text{elementary} + \mathcal{L}_\text{composite} &= \text{Tr}\left[\bar{\psi}^u_L i \slashed{D} \psi_L\right] + \text{Tr}\left[\bar{\psi}^u_R i \slashed{D} \psi^u_R \right]\\
 & + \text{Tr}\left[\bar{\Psi}^u\left(i \slashed{D} - m_{\Psi^u} \right)\Psi^u \right] + \text{Tr}\left[\bar{\tilde{\Psi}}^u \left( i\slashed{D} - m_{\tilde{\Psi}^u} \right) \tilde{\Psi}^u \right] \nonumber
\end{align}
We include the trace to handle the \textbf{10} and \textbf{14} cases, which are embedded in matrices for convenience. Note that only the form of the covariant derivative will differ between representations, as will be detailed shortly. The link terms would typically depend on representation, but we are \textit{enforcing} that the elementary fields fill full multiplets in matching representations, in order to couple with their respective partner. Tracing over this term handles the case that the terms are in matrix representations. Thus the link terms transform symmetrically in each representation $r$ as
\hspace{-5em}\begin{align}
 \mathcal{L}_\text{link} \supset  \left. \begin{cases}
\Delta_\psi  \text{Tr} \left[\bar{\psi}^{(r)}_L \Omega_1 \Psi^{(r)}_R \right], &  r = \textbf{1} \\
\Delta_\psi  \bar{\psi}^{(r)}_L \Omega_1 \Psi^{(r)}_R, &  r = \textbf{5} \\
\Delta_\psi \text{Tr}\left[ \Omega^{-1}_1 \bar{\psi}^{(r)}_L \Omega_1 \Psi^{(r)}_R\right], & r = \textbf{10},\textbf{14} 
 \end{cases}\right.
\end{align}
Observe that even if $\Psi$ and $\tilde{\Psi}$ are in the same representation, we prevent terms of the sort $\bar{\psi}_L \Omega_1 \tilde{\Psi}_R$. This would reduce the model to the 2-site Discrete CHM, which is logarithmically divergent.
The most significant dependence on choice of representation comes from the Yukawa-type Lagrangian. To begin with, linear couplings are only allowed between fermions of the same representation, for example
\begin{align}
\mathcal{L}_\text{Yukawa} \supset m_Y \text{Tr}[ \bar{\Psi}^\textbf{14}_L \tilde{\Psi}^\textbf{14}_R]
\end{align}
Only interactions involving the NGB matrix can couple different composite representations $r$ and $s$. These terms are functions
\begin{align}
\mathcal{L}_\text{Yukawa}( Y, \Psi^{(r)}_L, \tilde{\Psi}^{(s)}_R, \Omega_2\phi_0 ) 
\end{align}
These terms will be given in their relevant section below, as they depend on combinations of representations. Note that we include \textit{only} $\Psi_L$ and $\tilde{\Psi}_R$ in the $m_Y$ and $Y$ couplings. We do not consider, for example, $m_Y \bar{\tilde{\Psi}}_L \Psi_R$ as this leads to logarithmic divergence. In general, \cref{fig:allowed_interactions} is a guide to the interactions that are allowed. This figure applies to all combinations of representations, except for the case of all fermions in a generation in the fundamental. 

From this point, we will work in the site-2 holographic gauge with
\begin{align}
\Omega_1 \rightarrow \mathbb{1} && \Omega_2 \phi_0 \rightarrow U \phi_0 := \Phi
\end{align}
with our familiar definition of $\Phi$, given explicitly in \cref{eq:choices_of_gauge}. Practically, this means all $s_h$ dependence emergences from the Yukawa sector, rather than the other choice where it emerges from the link sector. They are, of course, equivalent; a matter of taste.


\subsection{Fundamental Representation}\label{sec:fundamental_rep}

For all fermions in the fundamental, we cannot couple the left-handed doublet to a right-handed up-type partner and right-handed down-type partner at the same time. This means introducing a second partner for the left doublet, and having a copy of the elementary doublet couple with that, as we had in \cref{eq:fundamental_fermion_lagrangian},
\begin{align}
\mathcal{L}_\text{link} = \Delta_L^u \bar{\psi}^u_L \Psi_R^u + \Delta_L^d \bar{\psi}^d_L \Psi_R^d + \Delta_R^u \bar{\psi}^u_R \tilde{\Psi}_L^u + \Delta_R^d \bar{\psi}^d_R \tilde{\Psi}_L^d\label{eq:fundamental_link_lagrangian}
\end{align}
where $\Psi^u, \tilde{\Psi}^u \sim \textbf{5}_{X^u}$ and $\Psi^d, \tilde{\Psi}^d \sim \textbf{5}_{X^d}$. In the case of a generation of quarks, $X^u = 2/3$ and $X^d = -1/3$. 
\begin{figure}
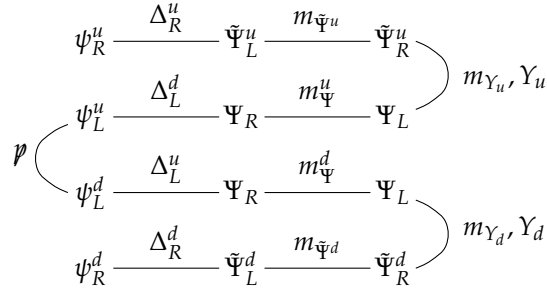

\centering
\tikz{
\node at (1,1) {$\psi^d_R$};
\draw (1.3,1) --(2.7,1);
\node [above] at (2,4) {$ \Delta_R^u $};
\node [above] at (2,2) {$ \Delta_L^u $};
\node [above] at (2,3) {$ \Delta_L^d $};
\node [above] at (2,1) {$ \Delta_R^d $};
\draw plot [smooth, tension=0.7] coordinates {(0.7, 2) (0.5, 2.1)  (0.3,2.3) (0.3,2.6) (0.5, 2.8) (0.7,2.9)};
\node [left] at (0.3,2.5) {$\slashed{p}$};
\node at (1,3) {$\psi^u_L$};
\node at (1,2) {$\psi^d_L$};
\draw (1.3,2) --(2.7,2);
\draw (1.3,3) --(2.7,3);
\node at (1,4) {$\psi^u_R$};
\draw (1.3,4) --(2.7,4);
\node at (3,1) {$\tilde{\Psi}^d_L$};
\draw (3.3,1) --(4.7,1);
\node [above] at (4,1) {$ m_{\tilde{\Psi}^d} $};
\node [above] at (4,2) {$ m_{\Psi}^d $};
\node [above] at (4,3) {$ m_{\Psi}^u $};
\node [above] at (4,4) {$ m_{\tilde{\Psi}^u} $};
\node at (3,2) {$\Psi_R$};
\draw (3.3,2) --(4.7,2);
\node at (3,3) {$\Psi_R$};
\draw (3.3,3) --(4.7,3);
\node at (3,4) {$\tilde{\Psi}^u_L$};
\draw (3.3,4) --(4.7,4);
\node at (5,1) {$\tilde{\Psi}^d_R$};
\draw plot [smooth, tension=0.7] coordinates {(5.3,1) (5.5,1.1)  (5.7,1.3) (5.7,1.6) (5.5, 1.8) (5.3,1.9)};
\node [right] at (5.8,1.5) {$ m_{Y_d}, Y_d $};
\node [right] at (5.8,3.5) {$ m_{Y_u}, Y_u $};
\draw plot [smooth, tension=0.7] coordinates {(5.3,3.1) (5.5,3.2)  (5.7,3.4) (5.7,3.7) (5.5, 3.9) (5.3,4)};
\node at (5,2) {$\Psi_L$};
\node at (5,3) {$\Psi_L$};
\node at (5,4) {$\tilde{\Psi}^u_R$};
}\caption{Allowed interactions in the minimal model, for the \textbf{5}-\textbf{5}-\textbf{5} and \textbf{14}-\textbf{14}-\textbf{14} embedding. The kinetic coupling is included here explicitly to remind that there is indeed coupling between the up-type fermions and down-type.}\label{fig:allowed_interactions_5-5-5}
\end{figure}
The composite Yukawa couplings for a \textbf{5}-\textbf{5}-\textbf{5} generation are
\begin{align}
\mathcal{L}_y &= Y_u(\bar{\Psi}^u_L \Phi)(\Phi^\dagger \tilde{\Psi}_R^u) + Y_d (\bar{\Psi}_L^d \Phi)(\Phi^\dagger \tilde{\Psi}_R^d)
\end{align}
with each fundamental vector constructed thus
\begin{align}
\Psi_\textbf{5} = \left( \begin{array}{ c }
\\
\Psi_\textbf{4} \\ 
\\
\hline
\Psi_\textbf{1}
\end{array} \right), && \text{where,} && \Psi_\textbf{4} = \frac{1}{\sqrt{2}}\left( \begin{matrix}
i \Psi^{--} - i \Psi^{++}\\
\Psi^{--} + \Psi^{++} \\
i\Psi^{-+} - i\Psi^{+-}\\
\Psi^{-+} + \Psi^{+-}
\end{matrix}\right), && \Psi_\textbf{1}= \Psi^{00}\, .
\end{align}
We could of course perform $SO(5)$ rotations of a simple $\Psi_\textbf{5} = (\Psi^1 , \Psi^2 , \Psi^3 , \Psi^4 , \Psi^5)$ representation, but we would like the components to have definite charges under the left and right third generators. This process is detailed in \cref{sec:representations}. Thus, for example, $\Psi^{--}$  has $SU(2)_L\times SU(2)_R$ quantum numbers $(-1/2, -1/2)_X$. By an appropriate choice of $X = 2/3$ charge, this component shares the transformation properties of the left-handed down-type SM quarks. $\Psi^{+-}$ shares quantum numbers with the left-handed up-type quarks. $\Psi^{00}$ shares quantum numbers with the right-handed up, though we allow only the $\tilde{\Psi}^{00}$ to couple with right-handed elementary fields. We apply these embeddings to the elementary fields, as
\begin{align}
\psi_L^u = \frac{1}{\sqrt{2}}\left(\begin{matrix}
i d_L\\
d_L\\
-iu_L\\
u_L\\
0
\end{matrix}
\right), & &
\psi_R^u = \left(\begin{matrix}
0\\
0\\
0\\
0\\
u_R
\end{matrix}
\right), & &
\psi_L^d = \frac{1}{\sqrt{2}}\left(\begin{matrix}
- i u_L\\
u_L\\
i d_L\\
d_L\\
0
\end{matrix}
\right), & &
\psi_R^d = \left(\begin{matrix}
0\\
0\\
0\\
0\\
d_R
\end{matrix}
\right).
\end{align}
One can then see why it is necessary to have two left-handed partners: the down-type $\tilde{\Psi}$ partners must have hypercharge $-1/3$ to couple with our elementary embedding. But then they would not be able to couple with a left-handed partner with hypercharge $2/3$. These fiveplets enter the effective EW Lagrangian
\begin{align}
\mathcal{L}_\text{eff}^\text{EW} &= \sum \left( \bar{\psi} \slashed{p} \Pi^0_\psi \psi + (\bar{\psi} \Phi) \slashed{p} \Pi^1_\psi (\Phi^\dagger \psi) \right) + \sum \left( m_\psi^0 \bar{\psi}_L \psi_R + m_\psi^1 (\bar{\psi}_L \Phi)(\Phi^\dagger \psi_R) \right)
\end{align}
This Lagrangian is used to derive the form factors entering \cref{eq:effective_SM_fermions}
\begin{align}
\Pi_{u_L} = \Pi^0_{\psi_L^u} +  \Pi^0_{\psi_L^d} + \Pi_{\psi_L^u}^1 \frac{s_h^2}{2}, && \Pi_{u_R} = \Pi_{\psi_R^u}^0 + \Pi_{\psi_R^u}^1 c^2_h, && M_u = m_{\psi^u}^1 \frac{s_h c_h}{\sqrt{2}}, \nonumber\\
\Pi_{d_L} = \Pi^0_{\psi_L^u} + \Pi^0_{\psi_L^d} + \Pi_{\psi_L^d}^1 \frac{s_h^2}{2}, && \Pi_{d_R} = \Pi_{\psi_R^d}^0 + \Pi_{\psi_R^d}^1 c^2_h , && M_d = m_{\psi^d}^1 \frac{s_h c_h}{\sqrt{2}}
\end{align}

In $SU(2)_L \times U(1)_Y$ notation (assuming quarks), the first layer of multiplets is given by
\begin{itemize}
\item ${\bf1}_{2/3}=\Psi^{00}_{2/3}$, with mass $m_{{\bf 1}_{2/3}}$ given by a zero of $\Pi_{u_R}(0)$;
\item ${\bf2}_{1/6}=(\Psi^{+-}_{2/3},\, \Psi^{--}_{-1/3})$, with mass $m_{{\bf2}_{1/6}}$ given by a zero of $\Pi_{u_L}(0)$;
\item ${\bf2}_{7/6}=(\Psi^{++}_{5/3},\, \Psi^{-+}_{2/3})$, with mass $m_{{\bf2}_{7/6}}$ given by a pole of $\hat{\Pi}_{q,4}(0)$.
\end{itemize}
Note that the "broken" form factors, derived generally in \cref{sec:low_energy_lagrangian}, are listed in \cref{sec:form_factors}. The modifications to the Higgs-fermion couplings are
\begin{align}
r_\psi = \frac{1-2\xi}{\sqrt{1-\xi}} \;
\end{align}
where $\psi$ are the fermions in the fundamental representation.

\subsection{Antisymmetric Representation}

The antisymmetric, two index tensor \textbf{10} happens to also be a basis of the $SO(5)$ Lie algebra. This representation behaves the same as the fundamental vis \`a vis the tuning contribution. Each multiplet has a larger field content than the fundamental, but note that we only need one left-handed partner now in order to produce the correct hypercharge structure. This is due to the \textbf{6} with hypercharge $2/3$ having components that correspond to both up and down-type quarks. With all fermions in the antisymmetric $\Psi_L, \Psi_R^u, \Psi_R^d \sim \textbf{10}_X$, we have a link Lagrangian for each generation
\begin{align}
\mathcal{L}_\text{link} &= \Delta_L \text{Tr} \left[ \bar{\psi}_L \Psi_R \right]  + \Delta^u_R \text{Tr} \left[ \bar{\psi}^u_R \tilde{\Psi}_L \right]  + \Delta^d_R \text{Tr} \left[ \bar{\psi}^d_R \tilde{\Psi}^d_L \right] 
\end{align}
and the Yukawa terms are given by
\begin{align}
\mathcal{L}_\text{link} &= Y^u \Phi^\dagger \bar{\Psi}_L \tilde{\Psi}_R^u \Phi +  Y^d \Phi^\dagger \bar{\Psi}_L \tilde{\Psi}_R^d \Phi
\end{align}
Note that we throw away the possible terms $\Phi^\dagger \bar{\tilde{\Psi}}^u_L \tilde{\Psi}_R^d \Phi + \Phi^\dagger \bar{\tilde{\Psi}}^d_L \tilde{\Psi}_R^u \Phi$ due to their divergent contributions. There are linear terms of the sort $m_Y \bar{\Psi}_L \tilde{\Psi}_R$. In the case of differing representations, for example $\Psi_L\sim \textbf{10}_X$ and $\Psi_R^u \sim \textbf{5}_X$, that direct coupling would be disallowed. There will still be a Yukawa coupling however
\begin{align}
\mathcal{L}_\text{link} &= Y^u \Phi^\dagger \bar{\Psi}_L \tilde{\Psi}_R^u \label{eq:10-5_interaction}
\end{align}

The antisymmetric representation is the adjoint, and therefore has the generators themselves as a possible choice of basis. However, again we would like each field $\Psi^{T_L^3, T_R^3} $ to have concrete $T_L^3, T_R^3$ quantum numbers, thus we embed the ten fields as
\begin{align}
\Psi_\textbf{10} &= \left( \begin{array}{c | c}
\Psi_\textbf{6} & \Psi_\textbf{4}/\sqrt{2} \\ \hline
- \Psi^T_\textbf{4} / \sqrt{2} & 0
\end{array} \right),  \text{where,} \\
\qquad \Psi_\textbf{6} &= \frac{1}{2}\left( \begin{matrix}
0 & \hat{\Psi}^{0,0}_+  & i (\hat{\Psi}_+^{-1,-1} - \hat{\Psi}_+^{1,1}) & \hat{\Psi}_-^{-1,-1}  + \hat{\Psi}_-^{1,1} \\
 & 0 & \hat{\Psi}_+^{-1,-1}  +  \hat{\Psi}_+^{1,1} & i (- \hat{\Psi}_-^{-1,-1}  + \hat{\Psi}_-^{1,1} ) \\ 
 & & 0 & i \hat{\Psi}^{0,0}_-\\
 & & & 0
\end{matrix}  \right) \nonumber
\end{align}
and $\Psi_\textbf{4}$ is as above. We have defined the convenient combinations 
\begin{align}
\hat{\Psi}^{0,0}_\pm := \Psi^{0,0}_1 \pm \Psi^{0,0}_2, && \hat{\Psi}_\pm^{-1,-1} := \Psi^{-1,0} \pm \Psi^{0,-1}, &&  \hat{\Psi}_\pm^{1,1} := \Psi^{0,1}  \pm \Psi^{1,0}. 
\end{align}
The elementary fields are thus embedded as
\hspace*{-3em}\begin{align}
\psi_L &=  \frac{1}{\sqrt{2}}\left(\begin{matrix}
0& & \cdots& 0 & id_L\\
& & & & d_L\\
\vdots& & \ddots& & -i u_L\\
0 & & & 0 & u_L\\
-id_L & -d_L & i u_L & -u_L & 0
\end{matrix}\right), && \psi_R^u =  \frac{1}{2}\left(\begin{matrix}
0& u_R & & \cdots & 0\\
-u_R & & & & \\
& & \ddots& i u_R & \vdots\\
\vdots & & -iu_R &0 & \\
0 &  & \cdots &  &  0
\end{matrix}\right)\nonumber\\
\psi_R^d &=  \frac{1}{2}\left(\begin{matrix}
0&  & id_R & - d_R & 0\\
 & & d_R & id_R & \\
-id_R& -d_R & &  & \vdots\\
d_R & -id_R &  &\ddots & \\
0 &  & \cdots &  &  0
\end{matrix}\right) 
\end{align}
The unique benefit of the \textbf{10} is clear now: that a $\tilde{\Psi}_6$ with X-hypercharge $2/3$ contains fields of both up and down type, meaning we don't need a new set of $\Psi_L$ partners with $-1/3$ X-hypercharge. We will thus use the \textbf{10} as a convenient representation for the bottom quark in later results. These elementary embeddings enter the effective EW Lagrangian
\begin{align}
\mathcal{L}_\text{eff}^\text{EW} &= \sum \left( \text{Tr}[\bar{\psi} \slashed{p} \Pi^0_\psi \psi] + (\bar{\psi} \Phi) \slashed{p} \Pi^1_\psi (\Phi^\dagger \psi) \right) + \sum \left( \text{Tr}[m_\psi^0 \bar{\psi}_L \psi_R] + m_\psi^1 (\bar{\psi}_L \Phi)(\Phi^\dagger \psi_R) \right)\label{eq:effective_EW_10}
\end{align}
This Lagrangian is used to derive the form factors entering \cref{eq:effective_SM_fermions}
\begin{align}
\Pi_{u_L} &= \Pi^0_{\psi_L} +  \Pi_{\psi_L}^1 \left( \frac{c_h^2}{2} + \frac{s_h^2}{4}\right) , && \Pi_{u_R} = \Pi_{\psi_R^u}^0 + \Pi_{\psi_R^u}^1 \frac{s^2_h}{4}, && M_u =  -m_{\psi^u}^1 \frac{s_h c_h}{4}, \nonumber\\
\Pi_{d_L} &= \Pi^0_{\psi_L} + \Pi_{\psi_L}^1 \frac{c_h^2}{2}, && \Pi_{d_R} = \Pi_{\psi_R^d}^0 + \Pi_{\psi_R^d}^1 \frac{s^2_h}{4}, && M_d = -m_{\psi^d}^1 \frac{s_h c_h}{2\sqrt{2}}
\end{align}

Resonances appearing with SM quantum numbers in the \textbf{10}-\textbf{10}-\textbf{10} case include
\begin{itemize}
\item ${\bf1}_{2/3}=\Psi^{00}_{2/3}$, with mass $m_{{\bf 1}_{2/3}}$ given by a zero of $\Pi_{u_R}(0)$;
\item ${\bf2}_{1/6}=(\Psi^{+-}_{2/3},\, \Psi^{--}_{-1/3})$, with mass $m_{{\bf2}_{1/6}}$ given by a zero of $\Pi_{u_L}(0)$;
\item ${\bf2}_{7/6}=(\Psi^{++}_{5/3},\, \Psi^{-+}_{2/3})$, with mass $m_{{\bf2}_{7/6}}$ given by a pole of $\hat{\Pi}_{q,4}$;
\item ${\bf 3}_{5/3}$, with mass $m_{{\bf 3}_{5/3}}$ given by a pole of $\hat{\Pi}_{q,6}$
\end{itemize}

\subsection{Symmetric Representation}

The symmetric, two index tensor \textbf{14} provides for extra terms in the Lagrangian that lead to a leading-order quartic Higgs contribution to the potential. With all fermions in the symmetric $\Psi_L, \Psi_R^u, \Psi_R^d \sim \textbf{14}_X$, we are actually plagued with the same issue as the \textbf{5}-\textbf{5}-\textbf{5}. That is, given a \textbf{14} with hypercharge of $2/3$, we cannot find a component that has the quantum numbers as the right-handed bottom. Thus, for all fermions in the symmetric, we would include a second left-handed partner, in the manner of \cref{eq:fundamental_link_lagrangian}. For simplicity, here and in the results presented, we will consider down-type fermions coming from a $\textbf{10}_{2/3}$ representation, thus requiring only one left-handed partner
\begin{align}
\mathcal{L}_\text{link} = \Delta_L \text{Tr}\left[  \bar{\psi}_L \Psi_R\right] +  \Delta_R^u \text{Tr}\left[  \bar{\psi}^u_R \tilde{\Psi}^u_L\right] + \Delta_R^d \text{Tr}\left[  \bar{\psi}^d_R  \tilde{\Psi}^d_L\right]  
\end{align}
and the Yukawa terms are given by
\begin{align}
\mathcal{L}_\text{Yukawa} = Y^u_1 \Phi^\dagger & \bar{\Psi}_L \tilde{\Psi}_R \Phi + Y^u_2 (\Phi^\dagger \bar{\Psi}_L \Phi) (\Phi^\dagger \tilde{\Psi}_R \Phi) \nonumber \\
& + Y^d_1 \Phi^\dagger \bar{\Psi}_L \tilde{\Psi}_R \Phi + Y^d_2 (\Phi^\dagger \bar{\Psi}_L \Phi) (\Phi^\dagger \tilde{\Psi}_R \Phi)
\end{align}
where we now have an extra Yukawa parameter $Y_2$ for each generation, from the additional possible term. Note that the Lagrangian interactions between a \textbf{14} irrep and \textbf{10} irrep are identical to those between \textbf{14} irreps. Combining \textbf{14} irreps with \textbf{5} irreps is also done analogously to the \textbf{10}-\textbf{5} interaction in \cref{eq:10-5_interaction}. Though when coupling \textbf{5} with a \textbf{14}, the extra Yukawa term is also allowed. We embed fermions in this representation as
\begin{align}
\Psi_\textbf{14} &= \left( \begin{array}{c | c}
\Psi_\textbf{9} - \Psi_1 \mathbb{1}/(2\sqrt{5}) & \Psi_\textbf{4}/\sqrt{2} \\ \hline
\Psi^T_\textbf{4} / \sqrt{2} & (2/\sqrt{5}) \Psi_1
\end{array} \right), \text{where,} \\
 \qquad \Psi_\textbf{9} &= \frac{1}{2}\left( \begin{matrix}
\hat{\Psi}^{0,0}_{2,+} - \Psi^{0,0}_4 & i\hat{\Psi}^{0,0}_{2,-}  & \hat{\Psi}_+^{1,1} + \hat{\Psi}_+^{-1,-1}  & i(\hat{\Psi}_-^{1,1} - \hat{\Psi}_-^{-1,-1})  \\
 & -\hat{\Psi}^{0,0}_{2,+}  - \Psi^{0,0}_4 & i( \hat{\Psi}_+^{1,1} - \hat{\Psi}_+^{-1,-1} ) &  \hat{\Psi}_-^{1,1}  - \hat{\Psi}_-^{-1,-1} \\ 
 & & \Psi^{0,0}_4 - \hat{\Psi}^{0,0}_{1,-} & i \hat{\Psi}^{0,0}_{1,+} \\
 & & & \Psi^{0,0}_4 + \hat{\Psi}^{0,0}_{1,-} 
\end{matrix}  \right) \nonumber
\end{align}
and the $\Psi_\textbf{4}$ and $\Psi_\textbf{1}$ are as given previously. As in the \textbf{10} case, we have defined some convenient fields: 
\begin{align}
\hat{\Psi}^{0,0}_{1,\pm} := \Psi^{-1,1}_3 \pm \Psi^{1,-1}_4, && \hat{\Psi}^{0,0}_{2,\pm} := \Psi^{1,1} \pm i \Psi^{-1,-1} \nonumber\\
\hat{\Psi}^{-1,-1}_\pm :=  \frac{1}{\sqrt{2}} (\Psi^{-1,0} \pm i \Psi^{0,-1}), && \hat{\Psi}^{1,1}_\pm  = \frac{1}{\sqrt{2}} (\Psi^{0,1} \pm i \Psi^{1,0})
\end{align}
Then the elementary fields can be embedded as
\hspace*{-3em}\begin{align}
\psi_L &=  \frac{1}{\sqrt{2}}\left(\begin{matrix}
0& & \cdots& 0 & id_L\\
& & & & d_L\\
\vdots& & \ddots& & -i u_L\\
0 & & & 0 & u_L\\
id_L & d_L & -i u_L & u_L & 0
\end{matrix}\right), && \begin{matrix}
\psi_R^u = \frac{1}{2\sqrt{5}} u_R \text{diag}(-1,-1,-1,-1,4)\\
\\
\psi_R^d = \frac{1}{2\sqrt{5}} d_R \text{diag}(-1,-1,-1,-1,4)
\end{matrix}
\end{align}

These enter the effective EW Lagrangian
\begin{align}
\mathcal{L}_\text{eff}^\text{EW} &= \sum \left( \text{Tr}[\bar{\psi} \slashed{p} \Pi^0_\psi \psi] + (\bar{\psi} \Phi) \slashed{p} \Pi^1_\psi (\Phi^\dagger \psi) + (\Phi^\dagger \bar{\psi} \Phi) \slashed{p} \Pi^2_\psi (\Phi \psi^\dagger \Phi^\dagger) \right) \nonumber\\
& + \sum \left( \text{Tr}[m_\psi^0 \bar{\psi}_L \psi_R] + m_\psi^1 (\bar{\psi}_L \Phi)(\Phi^\dagger \psi_R) + m_\psi^2 (\Phi^\dagger \bar{\psi} \Phi) (\Phi \psi^\dagger \Phi^\dagger)\right)
\end{align}
As mentioned above, if we take a \textbf{14}-\textbf{14}-\textbf{14} representation, we need two left-handed partners. Instead, for this work, we implement a \textbf{14}-\textbf{14}-\textbf{10} embedding. The form factors entering \cref{eq:effective_SM_fermions} in this representation are
\begin{align}
\hspace*{-1cm}\Pi_{u_L} &= \Pi^0_{\psi_L} +  \Pi_{\psi_L}^1 \left( \frac{c_h^2}{2} + \frac{s_h^2}{4}\right) + \Pi^2_{\psi_L} s_h^2 c_h^2, && \Pi_{u_R} = \Pi_{\psi_R^u}^0 + \Pi_{\psi_R^u}^1 \left( \frac{4}{5}c_h^2 + \frac{s^2_h}{20} \right) + \Pi^2_{\psi_R} \frac{(4 c_h^2 - s_h^2)^2}{20}, \nonumber \\
\hspace*{-1cm}\Pi_{d_L} &= \Pi^0_{\psi_L} + \Pi_{\psi_L}^1 \frac{c_h^2}{2}, && \Pi_{d_R} = \Pi_{\psi_R^d}^0 + \Pi_{\psi_R^d}^1 \frac{s^2_h}{4}, \nonumber\\
\hspace*{-1cm}M_u &=  im_{\psi^u}^1 \frac{3}{4\sqrt{5}} s_h c_h + i m_{\psi^u}^2 \frac{1}{2\sqrt{5}} s_h c_h (4c_h^2 - s_h^2), && M_d =  i m_{\psi^d}^1 \frac{s_h c_h}{2\sqrt{2}} 
\end{align}

Resonances appearing with SM quantum numbers in the \textbf{14}-\textbf{14}-\textbf{10} case include
\begin{itemize}
\item ${\bf1}_{2/3}=\Psi^{00}_{2/3}$, with mass $m_{{\bf 1}_{2/3}}$ given by a zero of $\Pi_{u_R}(0)$;
\item ${\bf2}_{1/6}=(\Psi^{+-}_{2/3},\, \Psi^{--}_{-1/3})$, with mass $m_{{\bf2}_{1/6}}$ given by a zero of $\Pi_{u_L}(0)$;
\item ${\bf2}_{7/6}=(\Psi^{++}_{5/3},\, \Psi^{-+}_{2/3})$, with mass $m_{{\bf2}_{7/6}}$ given by a pole of $\hat{\Pi}_{q,4}$; 
\item ${\bf 3}_{5/3}$ with mass $m_{\textbf{3}_{5/3}}$ given by a pole of $\hat{\Pi}_{q,1}$
\end{itemize}

The remaining modifications to the Higgs couplings are now
\begin{align}
r_\psi &= \frac{(6\xi - 3)Y_1 - 2(20\xi^2 - 23\xi +4)Y_2}{\sqrt{1-\xi}(2(5\xi - 4)Y_2 - 3Y_1)}
\end{align}
where $Y_1, Y_2$ are the Yukawa couplings for the composite partner.

\subsection{Trivial Representation}

The final representation we consider is the simplest, that of a single component multiplet $\mathbb{1}$. This is of phenomenological interest, as a fermion in the singlet representation can be said to be fully composite. This can explain the unusually large mass of, for example, the right-handed top quark. Note that, of course, only right-handed fields can transform trivially, as we require left-handed fields to fill at least $SU(2)$ doublets. The interactions of this representation are straightforward. For example, the link Lagrangian of a \textbf{14}-\textbf{1}-\textbf{10} embedding is given by
\begin{align}
\mathcal{L}_\text{link} = \Delta_L \text{Tr}\left[ U^\dagger \bar{\psi}_L U \Psi_R\right] +  \Delta_R^u \bar{\psi}^u_R \tilde{\Psi}^u_L + \Delta_R^d \text{Tr}\left[ U^\dagger \bar{\psi}^d_R U \tilde{\Psi}^d_L\right]
\end{align}
and the Yukawa terms are given by
\begin{align}
\mathcal{L}_\text{Yukawa} = Y^u \left( \Phi^\dagger \bar{\Psi}_L \Phi \right) \tilde{\Psi}_R \Phi + Y^d \Phi^\dagger \bar{\Psi}_L \tilde{\Psi}_R \Phi
\end{align}
The embedding of this field is trivial, so to speak. The effective EW Lagrangian is essentially identical to \cref{eq:effective_EW_10}. Again, we note that a \textbf{14}-\textbf{1}-\textbf{1} embedding would require two left-handed partners, and so embed the bottom in a \textbf{10}. The form factors of the \textbf{14}-\textbf{1}-\textbf{10} embedding studied in this chapter are given by
\begin{align}
\hspace*{-1cm}\Pi_{u_L} &= \Pi^0_{\psi_L} +  \Pi_{\psi_L}^1 \left( \frac{c_h^2}{2} + \frac{s_h^2}{4}\right) + \Pi^2_{\psi_L} s_h^2 c_h^2, && \Pi_{u_R} = \Pi_{\psi_R^u}^0, && M_u =-m_{\psi^u}^1 \nonumber \\
\hspace*{-1cm}\Pi_{d_L} &= \Pi^0_{\psi_L} + \Pi_{\psi_L}^1 \frac{c_h^2}{2}, && \Pi_{d_R} = \Pi_{\psi_R^d}^0 + \Pi_{\psi_R^d}^1 \frac{s^2_h}{4}, && M_d =  - m_{\psi^d}^1 \frac{s_h c_h}{2\sqrt{2}} 
\end{align}

Resonances appearing with SM quantum numbers in the \textbf{14}-\textbf{1}-\textbf{10} case include
\begin{itemize}
\item ${\bf1}_{2/3}=\Psi^{00}_{2/3}$, with mass $m_{{\bf 1}_{2/3}}$ given by a zero of $\Pi_{u_R}(0)$;
\item ${\bf2}_{1/6}=(\Psi^{+-}_{2/3},\, \Psi^{--}_{-1/3})$, with mass $m_{{\bf2}_{1/6}}$ given by a zero of $\Pi_{u_L}(0)$;
\item ${\bf2}_{7/6}=(\Psi^{++}_{5/3},\, \Psi^{-+}_{2/3})$, with mass $m_{{\bf2}_{7/6}}$ given by a pole of $\hat{\Pi}_{q,4}$; 
\item ${\bf 3}_{5/3}$, with mass $m_{\textbf{3}_{5/3}}$ given by a pole of $\hat{\Pi}_{q,1}$
\end{itemize}
%
%
The modification to the Higgs couplings is now:
\begin{align}
r_\psi = \frac{1-2\xi}{\sqrt{1-\xi}} .
\end{align}

\subsection{Representing Flavor}

To this point, we have not considered an important element that must be reproduced for a viable effective theory - the CKM matrix. Interactions between generations are certainly possible \cite{Barbieri:2012tu}. In any case where more than the third generation is considered, we choose to introduce flavor mixing in the link field terms. That is, given a representation $r$, we will have terms
\begin{align}
\mathcal{L}_\text{link} &= \Delta_L^{ij} \bar{\psi}^i_L \Psi_R + \Delta_R^{ij} \bar{\psi}^i_R \tilde{\Psi}_L
\end{align}
where $\Delta$ is now a $3\times 3$ flavor mixing matrix. It is interesting to note that there can be flavor mixing even between generations in different representations, due to the Yukawa term. We have not investigated the effects on flavor precision tests from different choices of generation representation. However, I note that a choice of, for example, top in a \textbf{5}, and a charm in a \textbf{14} may display similar phenomenology to the introduction of leptons, discussed in the following section.

\section{Composite Leptons}

The M4DCHM can be made to reproduce the SM Higgs sector, even by straightforwardly coupling all matter (excluding the top, in order to use its partners to enforce EWSB) to the low-energy Higgs-like pNGB. For example, each generation of leptons could be included as
\begin{align}
\mathcal{L} \supset & i \bar{l}_L\slashed{D} l_L + i \bar{e}_R \slashed{D} e_R + i\bar{\nu}_{e,R} \slashed{D} \nu_{e,R} \nonumber \\
&- \frac{m_\text{SM}}{v} \bar{l}_L \left( \begin{matrix}
0 \\
h
\end{matrix}\right) e_R -  \frac{m_\text{SM}}{v} \bar{l}_L \left( \begin{matrix}
h \\
0
\end{matrix}\right) \nu_R 
\end{align}
where we allow for the possibility of right-handed neutrinos, even if their mass is set to zero. Considering a non-zero mass, there are methods to implement a see-saw mechanism in the M4DCHM \cite{carmona2015}. For such extensions, however, we must include the leptons under the partial compositeness paradigm. Thus, we carbon copy the quark partners as in \cref{fig:allowed_interactions}, for the lepton sector, introducing three new partners $\Psi^l, \tilde{\Psi}^\tau, \tilde{\Psi}^\nu$. Just as we did for the quarks, we can choose differing representations for the leptons. 

For simplicity, we will consider only three combinations, which characterise three unique phenomenological features. These are the \textbf{5-5-5}, the \textbf{14-14-10}, and the \textbf{14-1-10}, where each irrep corresponds to $l_L-\tau_R-\nu_R$. Consulting the general results of \cref{sec:FermionSector}, we see that with a leptonic \textbf{5-5-5}, the matter content transforms as\footnote{Again, introducing a second partner for the doublet. If we didn't include the right-handed neutrino, we would only have required one partner, of course.}
\begin{align}
\Psi^t, \tilde{\Psi}^t \sim \textbf{5}_{2/3} && \Psi^b, \tilde{\Psi}^b \sim \textbf{5}_{-1/3} && \Psi^\tau, \tilde{\Psi}^\tau \sim \textbf{5}_{-1} && \Psi^\nu, \tilde{\Psi}^\nu \sim \textbf{5}_0
\end{align}
In the leptonic \textbf{14-14-10}, our matter transforms with the quantum numbers
\begin{align}
\Psi^t, \tilde{\Psi}^t \sim \textbf{5}_{2/3} && \Psi^b, \tilde{\Psi}^b \sim \textbf{5}_{-1/3} && \Psi^l, \tilde{\Psi}^\tau \sim \textbf{14}_{-1} && \tilde{\Psi}^\nu \sim \textbf{10}_0
\end{align}
and in the leptonic \textbf{14-1-10}, our matter transforms as
\begin{align}
\Psi^t, \tilde{\Psi}^t \sim \textbf{5}_{2/3} && \Psi^b, \tilde{\Psi}^b \sim \textbf{5}_{-1/3} && \Psi^l \sim \textbf{14}_{-1} && \tilde{\Psi}^\tau \sim \textbf{1}_{-1} && \tilde{\Psi}^\nu \sim \textbf{10}_0
\end{align}
We hold the quark sector fixed, including just a third generation $q_L-t_R-b_R$ in the \textbf{5-5-5} representation. In doing so, we hope to realise the idea that motivated this extension: a reduced tuning by easing the need for light partners, due to the contribution of a lepton partner in the symmetric representation.

In the following section, we present the results of exploring these three scenarios. The parameter space is described in \cref{tab:4DCHM_params}. Common to all three scenarios are the parameters
\begin{align}
\{m_\rho, m_a, t_{g/g'}, \Delta_L^t, \Delta_L^b, & \Delta_R^t, \Delta_R^b, m_{q^t}, m_{q^b}, m_t, m_b, m_{Y_t}, m_{Y_b}, Y_t, Y_b \nonumber\\
& \Delta_R^\tau, \Delta_R^\nu,  m_\tau, m_\nu, Y_\tau, Y_\nu \}
\end{align} 
Leptons embedded in the following irreps imply the additional parameters
\begin{align}
\begin{matrix}
\textbf{5}-\textbf{5}-\textbf{5} & \textbf{14}-\textbf{14}-\textbf{10} & \textbf{14}-\textbf{1}-\textbf{10}\\
\{\Delta_L^\tau, \Delta_L^\nu, m_{l^\tau}, m_{l^\nu}, m_{Y_\tau}, m_{Y_\nu} \} & \{ \Delta_L^l, m_l, m_{Y_l}, \tilde{Y}_\tau \} & \{\Delta_L^l, m_l\}
\end{matrix}
\end{align}

Each parameter point is constrained by four observables - Higgs mass $m_H$, top quark mass $m_t$, bottom quark mass $m_b$, tau lepton mass $m_\tau$. The EW vev is enforced by solving for the overall breaking scale $f = v / s_{\langle h \rangle}$. It \textit{is} however, included in the higher order tuning calculation as a tuned observable. The observable constraints are included as a cost function in \texttt{MultiNest} according to \cref{sec:multinest}. We define a Gaussian likelihood
\begin{align}
\mathcal{L} = \me^{-\chi^2}\; ,
\end{align}
where the $\chi^2$ cost function is given as
\begin{align}
\chi^2 = \frac{(m_H- 125)^2}{2\times 5^2} + \frac{(m_t - 155)^2}{2\times 15^2} + \frac{(m_b - 2.7)^2}{2\times 0.5^2} + \frac{(m_\tau - 1.8)^2}{2\times 0.5^2}
\end{align}

Approximately 80 million points are sampled for each model, with around 40,000 passing initial EWSB conditions. We choose to study the subset that are in the vicinity of the correct SM behaviour by applying mass cuts as follows
\begin{align}
\{120,140,2.2,1.3\} \leq \{m_H, m_t, m_b, m_\tau\} \leq \{130,170,3.2,2.3\}\, .
\end{align}
This gives us a few hundred viable points for each model. We use each of these as the starting point for a Markov Chain Monte Carlo sampling of the same parameter space for each model, giving us a more thorough exploration of each possible preferred region. We use the Metropolis-Hastings algorithm~\cite{hastings}, with step sizes for each parameter given by 0.01 times the current value of the parameter. Our final plots use points from the Metropolis-Hastings output that pass the mass cuts. This manifests in the results as disconnected patches of well-explored parameter space.

\section{Lowered fine tuning?}
\label{sec:LCHM_Results}

\begin{figure}
\centering
\subfloat[]{
\centering
\includegraphics[width=0.45\textwidth]{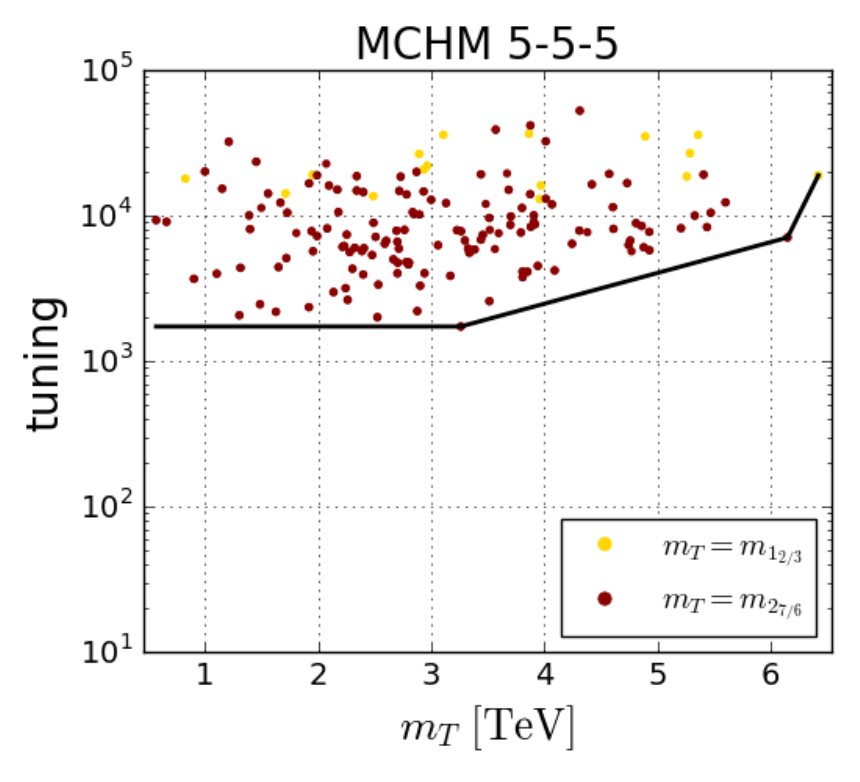}
}
\subfloat[]{
\centering
\includegraphics[width=0.45\textwidth]{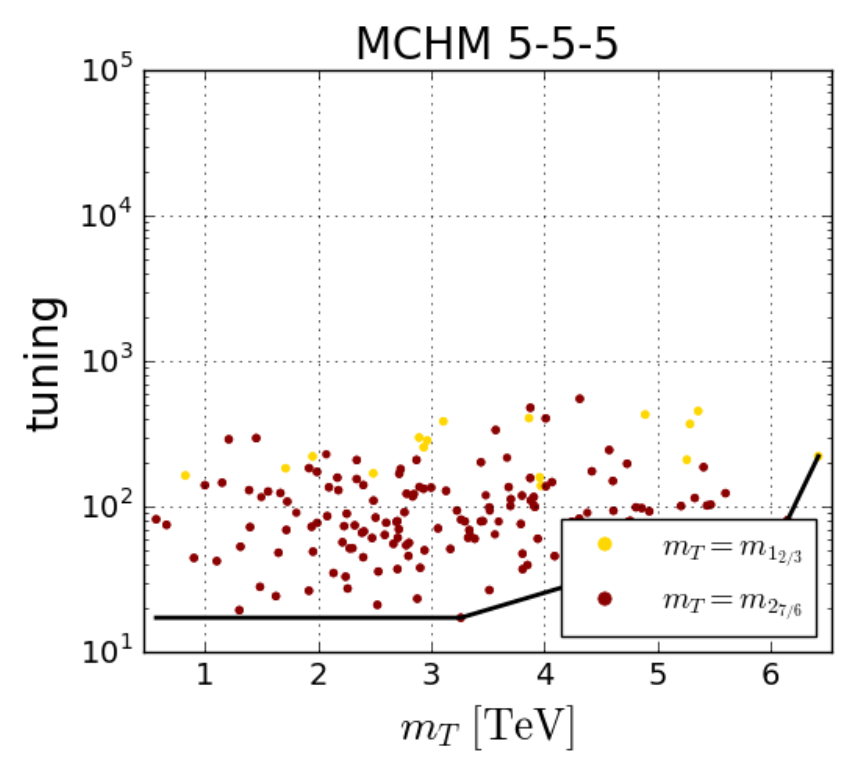}
}
\caption{A comparison of non-normalised (a) and normalised (b) fine tunings in the mass of top partners}
\label{fig:comparison}
\end{figure}

Below, we present the scan results in terms of the fine-tuning found at each viable parameter point. The tuning of each lepton embedding is shown against the lightest vector-boson resonance mass $m_\rho$, the lightest top partner resonance mass $m_T$, the Higgs coupling ratios $r_\chi$ and the vacuum misalignment $\xi \approx v^2/f^2$. A convex hull is provided to understand the general limits of minimal fine tuning (note that given the logarithmic scale, the hull may not always appear to be convex). We observe, in line with the prediction of \cref{eq:tuningprediction}, that the fine-tuning is generally two orders of magnitude higher in this lepton-sensitive case than the top-only case of \cite{Barnard:2015ryq}. If we were interested in comparing with lepton-insensitive models, for example, we could normalise by this factor. Such a normalised plot is given in Figure \ref{fig:comparison}, along with the unnormalised results. For the rest of this section, we stick to using the new measure without additional normalisation, which will permit a relative comparison of our lepton embeddings (since we use the same observables in each case, and the difference between the number of parameters is not significant).

A comparison of our new tuning with less sophisticated tuning measures can be seen in the bottom right panel of \cref{fig:fullplots}, which shows the fine tuning for the LM4DCHM$^{\textbf{5-5-5}}_{\textbf{5-5-5}}$ model as a function of the vacuum misalignment $\xi$. Our measure gives higher values for fine tuning relative to the single tuning $\Delta_1$ or the naive fine-tuning measure $1/\xi$, which is to be expected. In this case, with the leptons and quarks all embedded in fundamental representations of $SO(5)$, the lepton sector is not contributing much at all to the phenomenology of the model, which suffers from the double tuning effect highlighted previously. 

A general note on the results described below is in order. The argument for including composite lepton partners is two-fold \cite{carmona2015} (1) to raise the top partner masses that can be found in the parameter space, irrespective of tuning, and (2) to lower the tuning by making judicious choices of lepton embedding. Regarding the first, we find large parameter volumes that allow for top partner masses $\geq 1$TeV. This is in agreement with \cite{Barnard:2015ryq}, which uses the same sophisticated scanning technique.
Regarding the second, we take the tuning as an order-of-magnitude measure. That is, we consider differences in tuning of less than a factor of ten as being not significant. In this sense, there is already some question of the usefulness in considering leptons as in previous papers, where the most finely tuned was LM4DCHM$^{\textbf{5-5-5}}_{\textbf{5-5-5}}$ with $\Delta \equiv \mathcal{O}(100)$ and the least finely tuned was LM4DCHM$^{\textbf{5-5-5}}_{\textbf{14-1-10}}$ with $\Delta \equiv \mathcal{O}(20)$ \cite{carmona2015}, i.e. a non-significant effect. Our results reflect this, with even smaller differences of up to a factor of two between model tunings. This can be considered a result of the new tuning measure: when tuning dependencies are fully considered, there is no tuning-based preference between lepton embeddings in the lepton-inclusive M4DCHM. However, tunings between models do not equally scale as top partner masses grow, so this may be a point of model distinction as colliders are able to exclude the $\mathcal{O} \equiv 1$TeV partners.

\subsection{LM4DCHM$^{\textbf{5-5-5}}_{\textbf{5-5-5}}$ Fine-tuning}
\begin{figure}
\centering
\subfloat[]{
\centering
\includegraphics[width=0.45\linewidth]{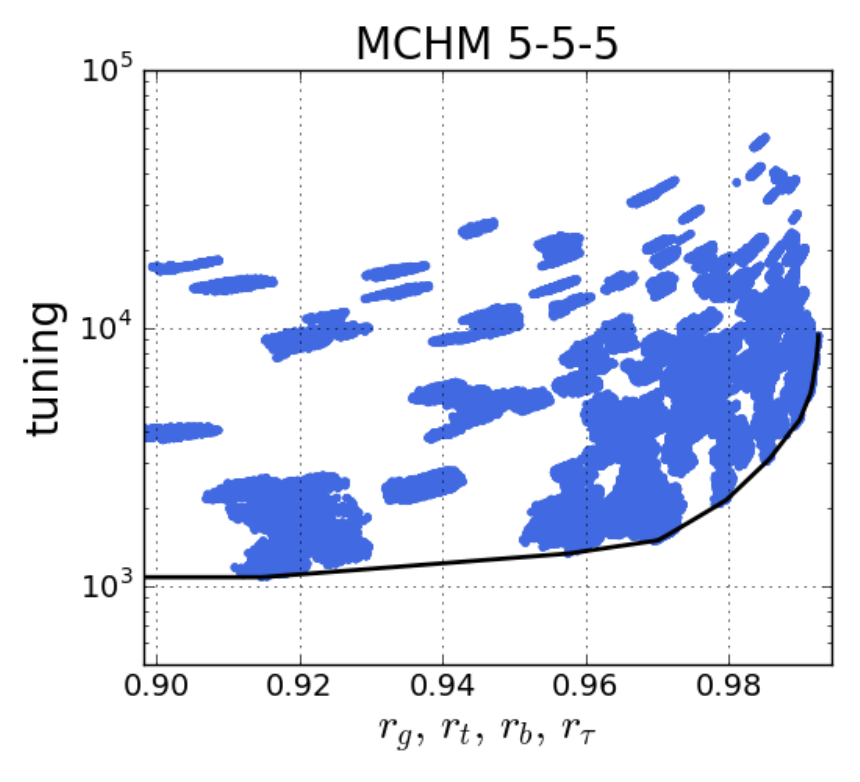} \label{fullrg}
}
\subfloat[]{
\centering
\includegraphics[width=0.45\linewidth]{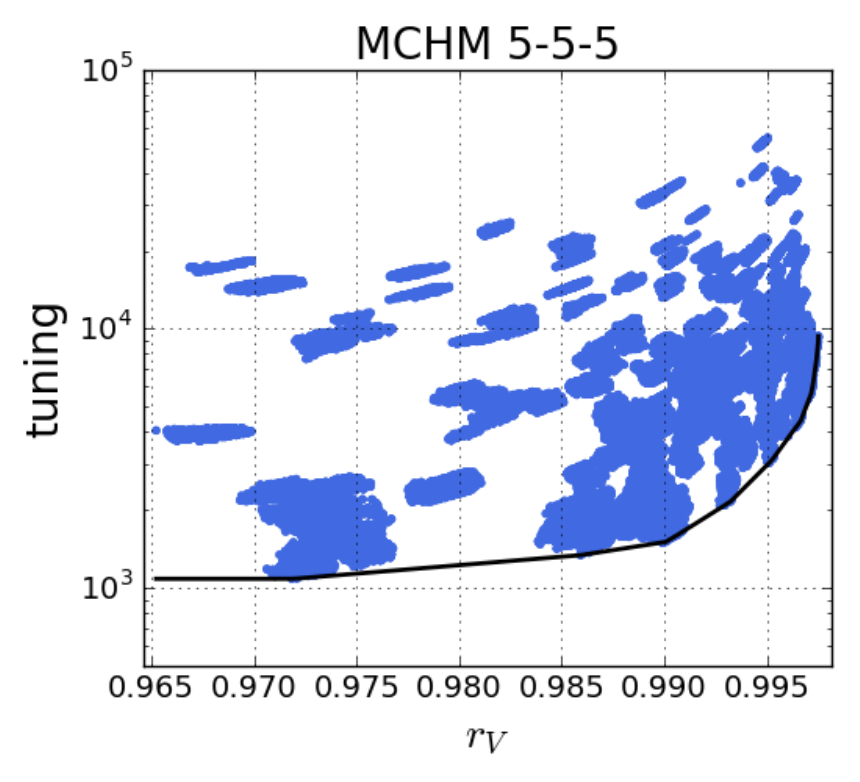} \label{fullrv}
}\\
\subfloat[]{
\centering
\includegraphics[width=0.45\linewidth]{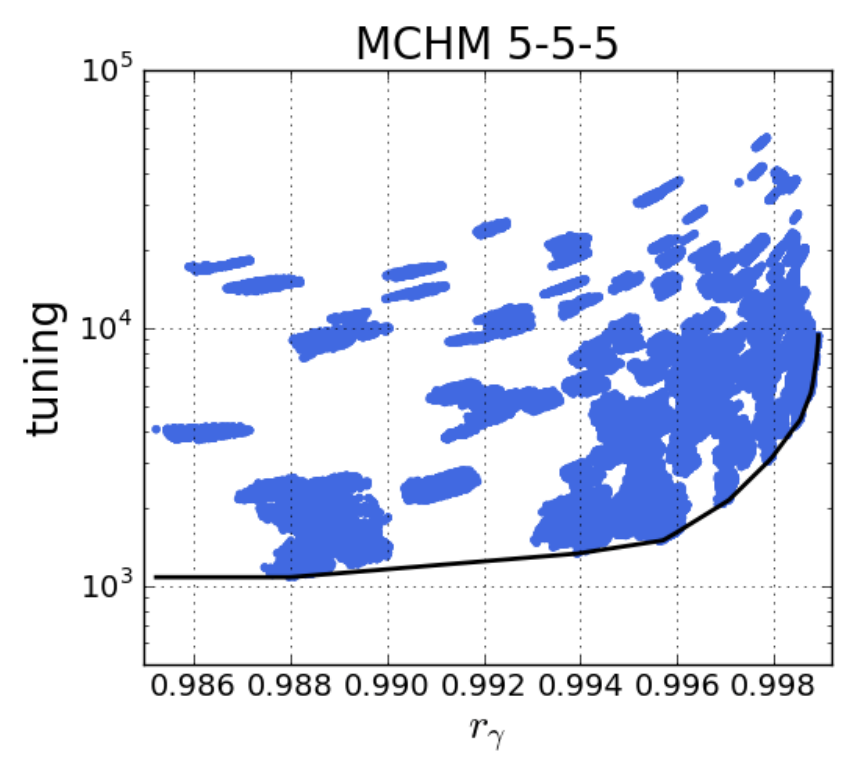} \label{fullry}
}
\subfloat[]{
\centering
\includegraphics[width=0.45\linewidth]{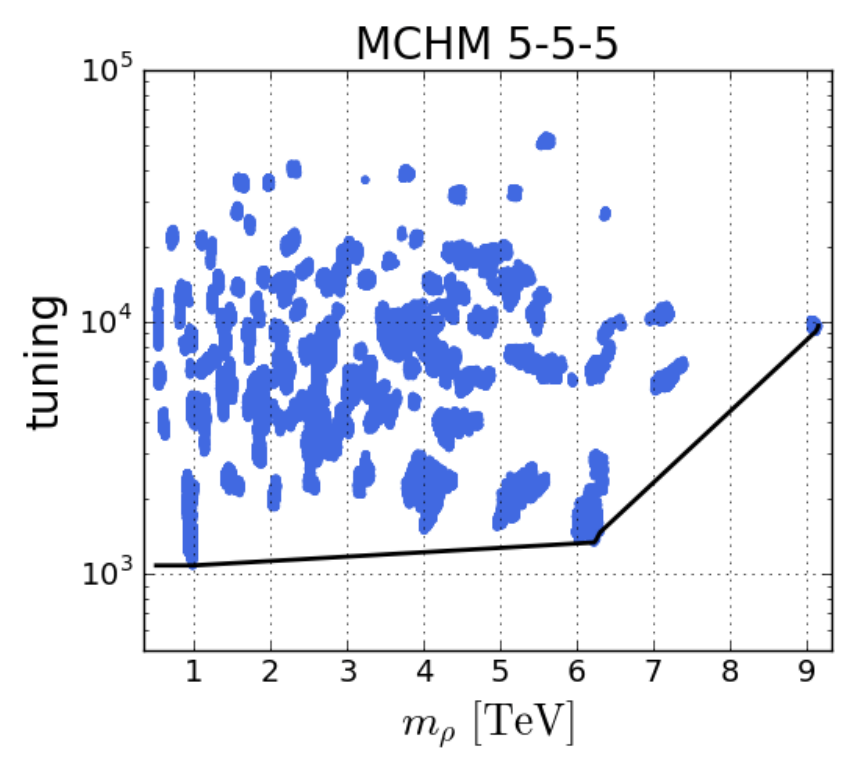} \label{fullmp}
}\\
\subfloat[]{
\centering
\includegraphics[width=0.45\linewidth]{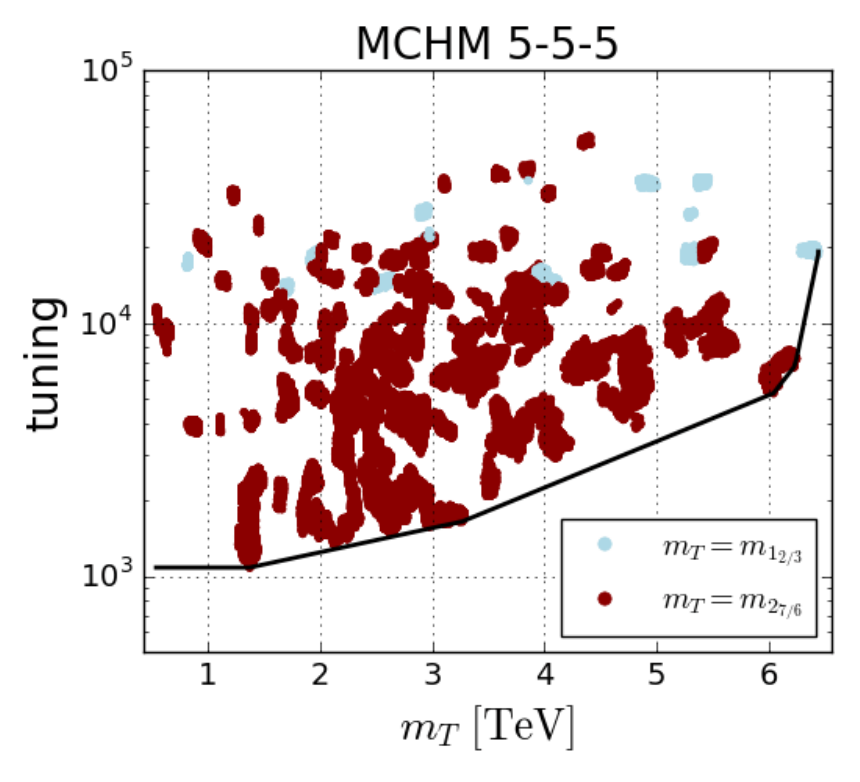} \label{fullmt}
}
\subfloat[]{
\centering
\includegraphics[width=0.45\linewidth]{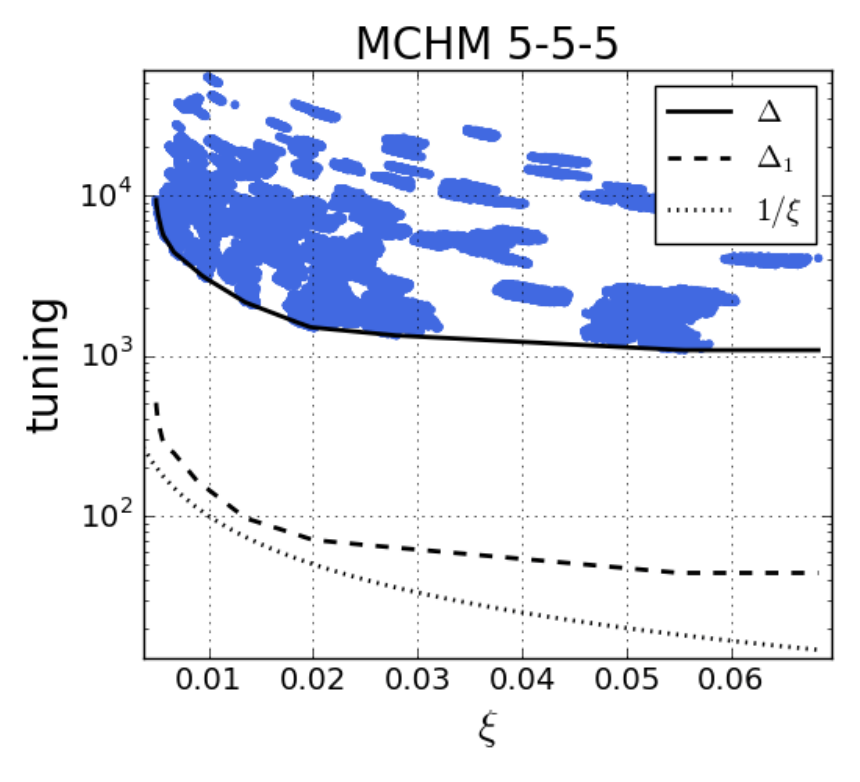} \label{fullxi}
}
\caption{Tuning in the LM4DCHM$^{\textbf{5-5-5}}_{\textbf{5-5-5}}$ model as a function of Higgs coupling ratios, lightest scalar resonance mass, top partner masses, and vacuum misalignment}
\label{fig:fullplots}
\end{figure}

Here we present the results for the fundamental representation, found in \cref{fig:fullplots}. The full tuning is quite severe, partly due to the generic fine tuning reasons explained above, with a minimum tuning of $\Delta = 1082$ at a top partner mass of $m_{2_{7/6}} = 1.37$TeV. However, this model is particularly badly tuned, due to the quark \textit{and} lepton double tuning required to achieve EWSB. Our previous study showed a sharply linear relationship between the lightest top partner mass and the fine tuning of the point for a model that did not include the lepton sector \cite{Barnard:2015ryq}. Our present case, however, is complicated by the fact that the inclusion of the lepton sector introduces both extra parameters and extra potential sources of tuning. These sources include the single tuning associated with reproducing the Higgs VEV and masses of the Higgs and SM fermions, and the new possibilities for multiple tunings across combinations of these observables. It is still true, however, that the fine tuning decreases with lower masses for new particles, a smaller hierarchy between elementary and composite scales, and greater divergence from Standard Model Higgs coupling predictions. There is evidence to suggest that the fine tuning rises more steeply with the lightest partner mass if this mass exceeds $3$ TeV. We also see that points for which the {\bf 2}$_{7/6}$ multiplet is the lightest top partner are significantly less finely tuned than points where it tends to be the {\bf 1}$_{2/3}$. This can be understood from the fact that the {\bf 2}$_{7/6}$ does not mix directly with the elementary top quark, and hence its mass is less constrained and easier to keep light than that of the {\bf 1}$_{2/3}$. A precision of less than 3\% on the Higgs couplings to gluons or fermions would lead to a dramatic increase in the fine tuning of the model. This precision provides the same tuning limits as excluding top partners up to $2.6$ TeV. Currently, however, Run I constraints still allow even the least finely tuned configurations. 

%
\subsection{LM4DCHM$^{\textbf{5-5-5}}_{\textbf{14-14-10}}$ Fine-tuning}
\begin{figure}
\centering
\subfloat[]{
\centering
\includegraphics[width=0.45\linewidth]{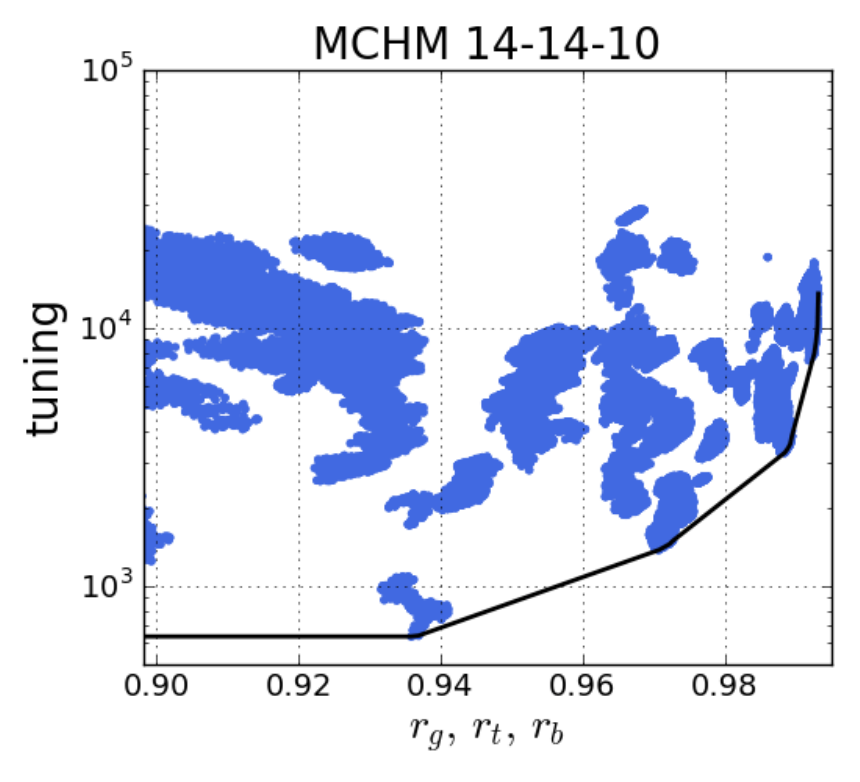}\label{reducedrg}
}
\subfloat[]{
\centering
\includegraphics[width=0.45\linewidth]{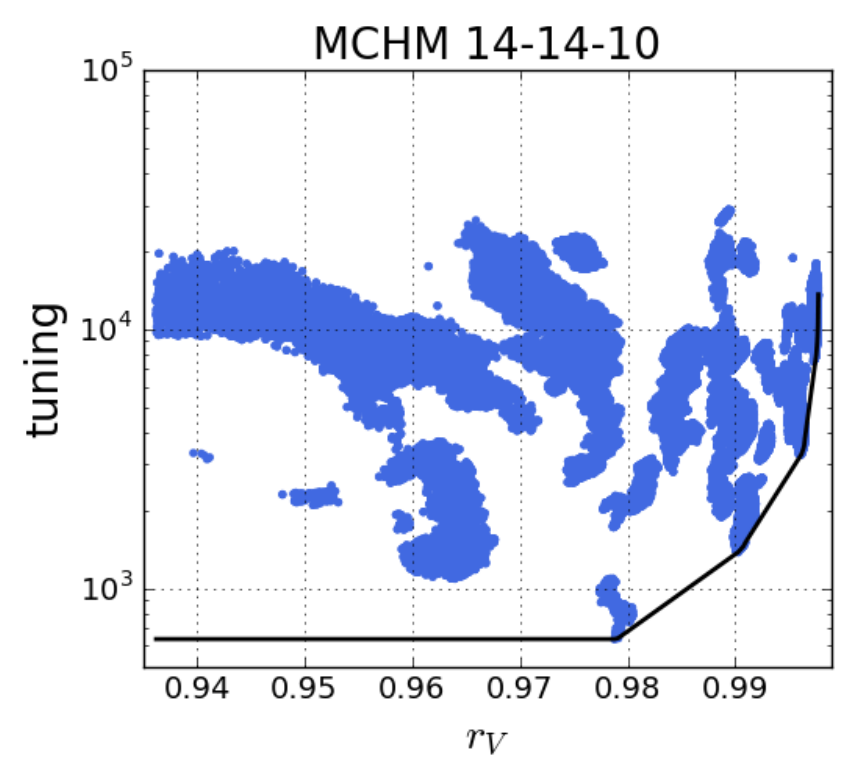}\label{reducedrv}
}\\
\subfloat[]{
\centering
\includegraphics[width=0.45\linewidth]{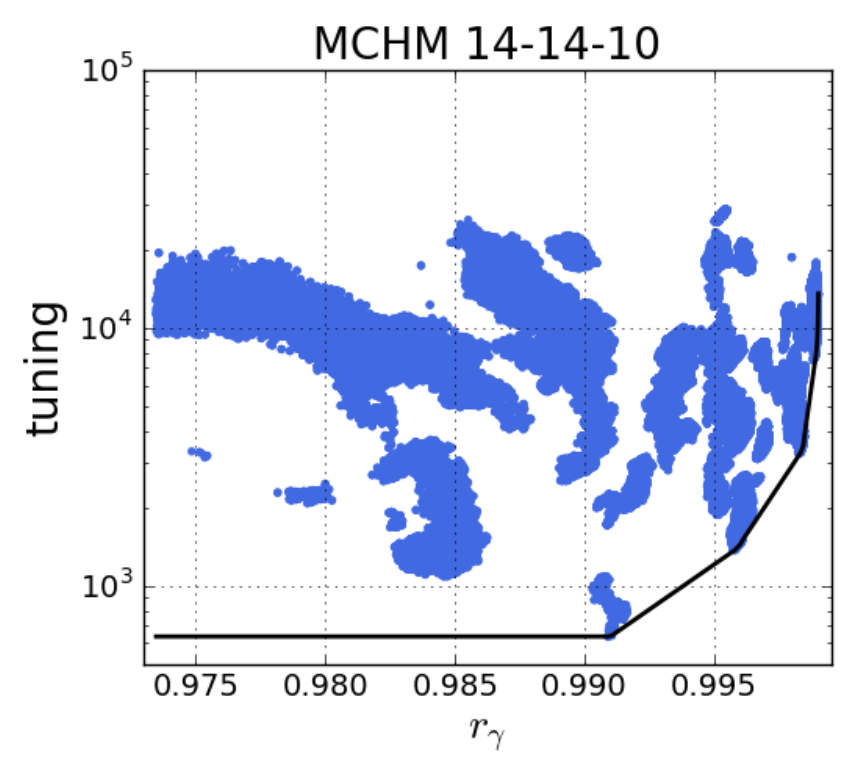}\label{reducedry}
}
\subfloat[]{
\centering
\includegraphics[width=0.45\linewidth]{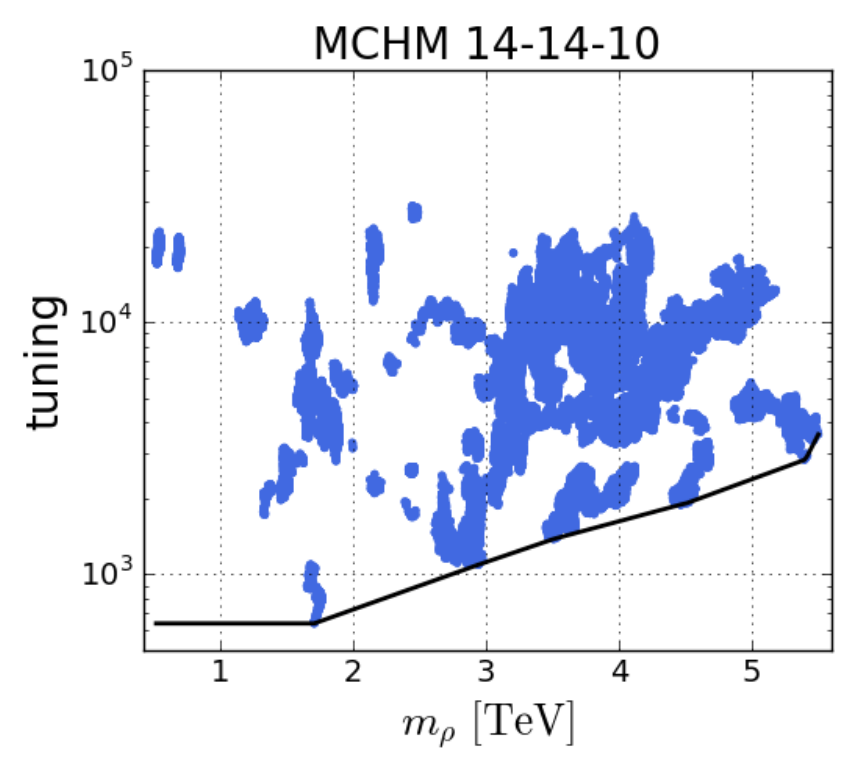}\label{reducedmp}
}\\
\subfloat[]{
\centering
\includegraphics[width=0.45\linewidth]{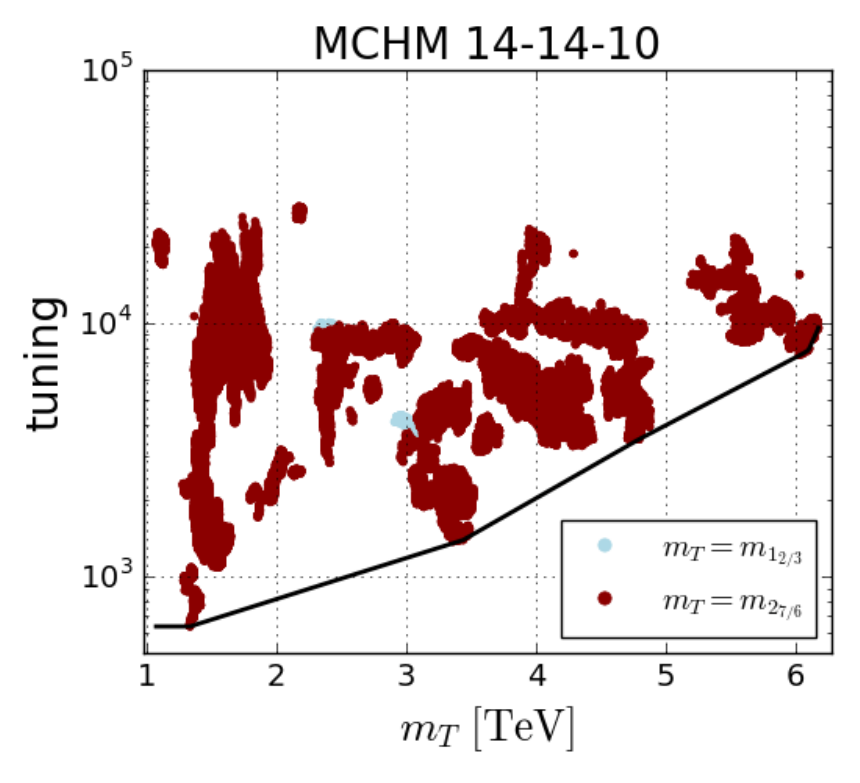}\label{reducedmt}
}
\subfloat[]{
\centering
\includegraphics[width=0.45\linewidth]{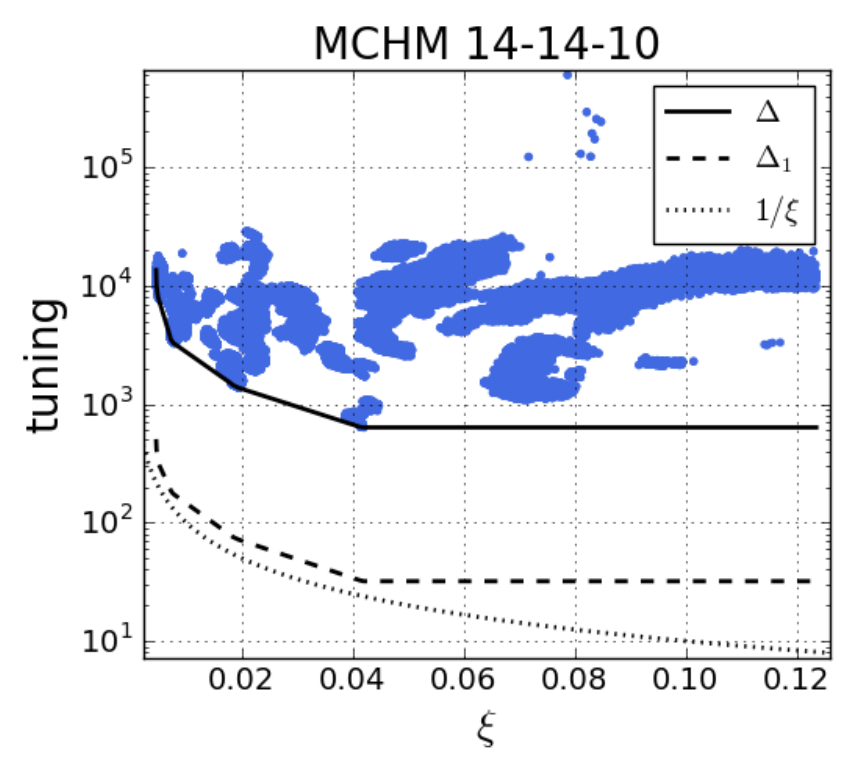}\label{reducedxi}
}\caption{Tuning in the LM4DCHM$^{\textbf{5-5-5}}_{\textbf{14-14-10}}$ model as a function of Higgs coupling ratios, lightest scalar resonance mass, top partner masses, and vacuum misalignment}
\label{fig:reducedplots}
\end{figure}

Here we present the results for the case of symmetric representations for the leptonic doublet and the tau lepton, found in \cref{fig:reducedplots}. We find a lower measure of tuning in this case than for the fundamental, which can be partly attributed to the convenient cancellation of double tuning described in \cref{sec:LCHM_motivation}. A minimum fine tuning was found to be $\Delta = 637$ at a top partner mass of $m_{2_{7/6}} = 1.34$ TeV. The fine tuning again decreases with lower masses for new particles, a smaller hierarchy in scales, and greater divergence from Standard Model Higgs coupling predictions. We see again that in the cases where the $\textbf{2}_{7/6}$ is the lightest top partner, we generally find a lower tuning.

\begin{figure}
\centering
\includegraphics[width=0.4\linewidth]{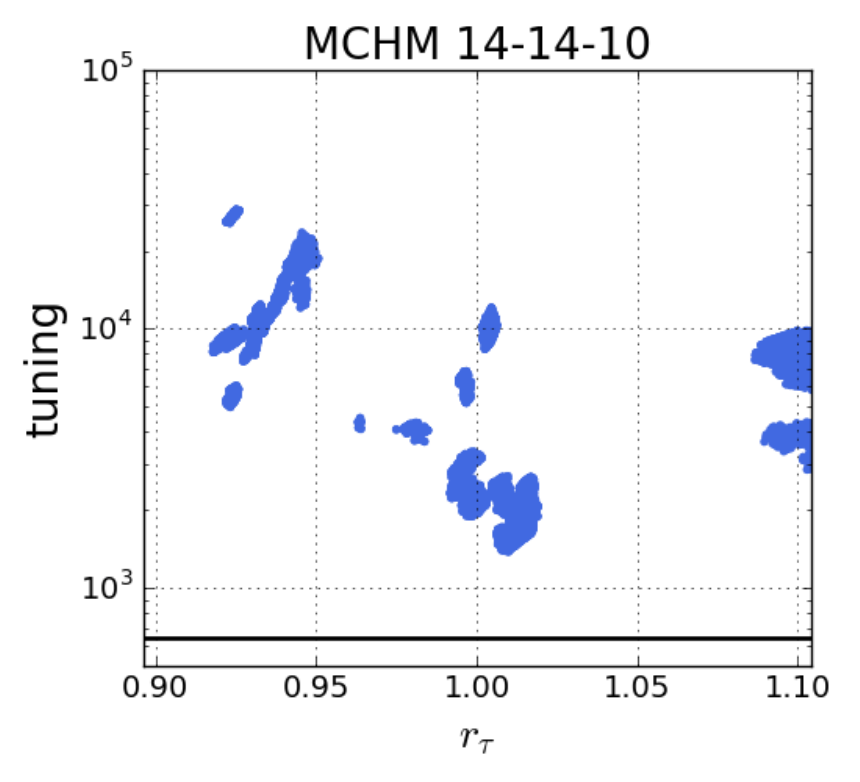}
\caption{The tuning of Higgs-tau coupling modifications}
\label{fig:tau141410plot}
\end{figure}

There is evidence to suggest that, unlike in the LM4DCHM$^{\textbf{5-5-5}}_{\textbf{5-5-5}}$ case, tuning increases more quickly for top partner masses greater than $1$ TeV. We caution, however, that the extreme difficulty of finding viable points in this model leads to a poor sampling density near the convex hull. The tuning is somewhat below that of the LM4DCHM$^{\textbf{5-5-5}}_{\textbf{5-5-5}}$ model for low top partner masses, but may be comparable at higher masses. Again, the reason can be attributed to the tuning measure used. Where previous works consider only the worst tuning in a particular parameter, we consider a cumulative measure that is sensitive to both the cancellation of double tuning, and the LM4DCHM$^{\textbf{5-5-5}}_{\textbf{14-14-10}}$-specific tuning required to achieve low Higgs, top and tau masses that may be more significant at higher top partner masses. Our tuning measure also counts the increase in the number of parameters as a negative feature. Thus, although one can alleviate the double tuning in this model through organising to have a leading order contribution to the quartic Higgs potential term from the leptons, and a sub-leading contribution from the quarks, one has had to introduce additional complexity to do so, thus lessening the attractiveness of the symmetric representation. Due to this, we consider the tuning difference between the previous and current embedding to be not significant.

A measurement of Higgs-top coupling up to 3\% would provide the same tuning constraint as excluding top partners up to $3.4$ TeV. Note that the Higgs-tau coupling modification has a different structure from the other models considered. In this case, the modification is much more forgiving - there exists parameter space with very little modification at low tuning. This is shown in \cref{fig:tau141410plot}.

\subsection{LM4DCHM$^{\textbf{5-5-5}}_{\textbf{14-1-10}}$ Fine-tuning}
\begin{figure}
\centering
\subfloat[]{
\centering
\includegraphics[width=0.45\linewidth]{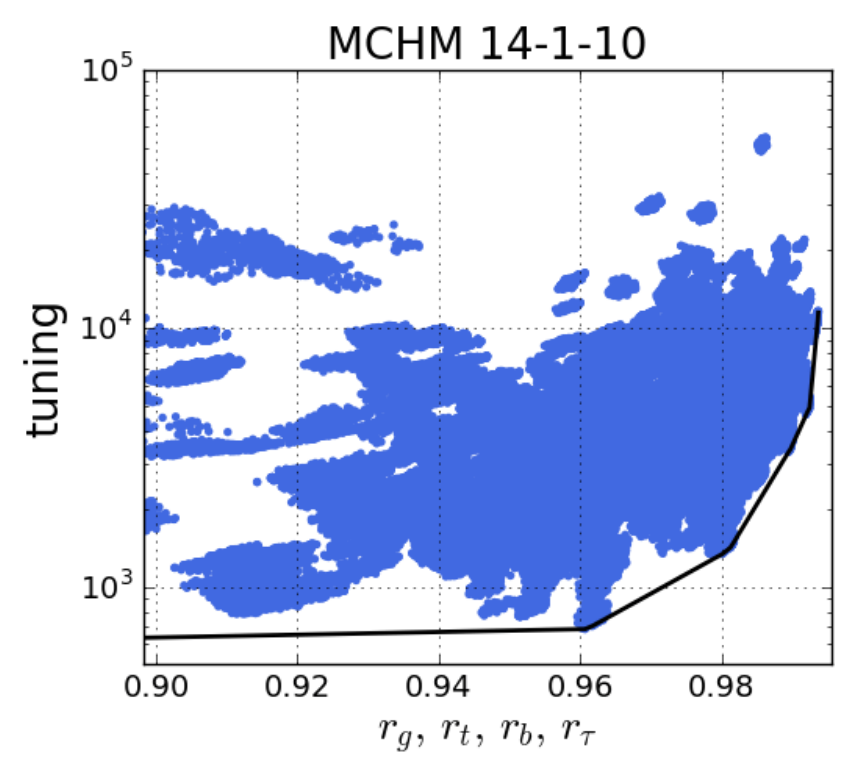}\label{minimalrg}
}
\subfloat[]{
\centering
\includegraphics[width=0.45\linewidth]{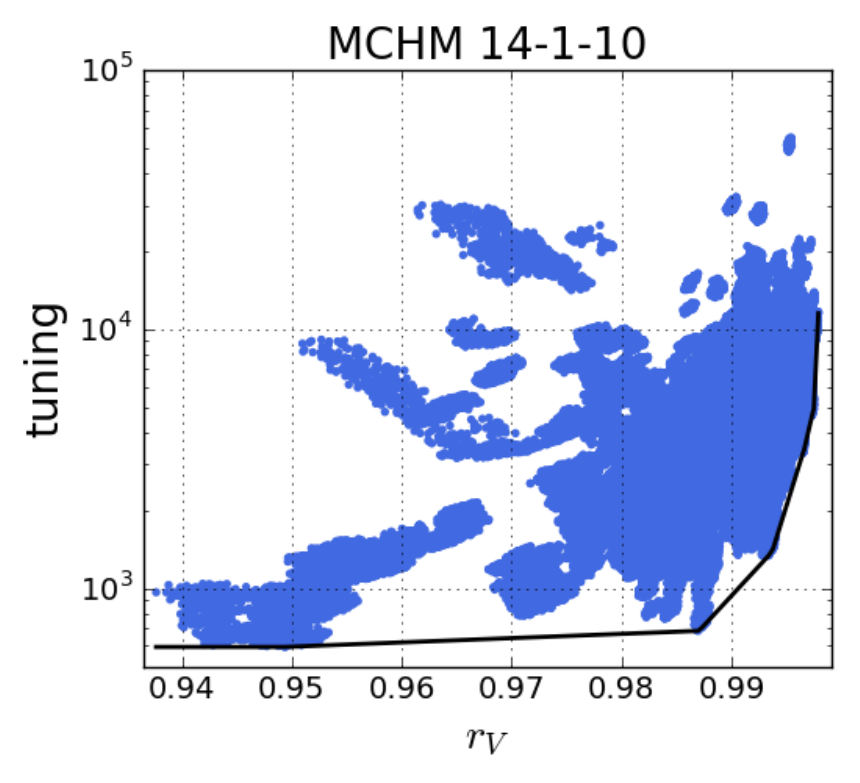}\label{minimalrv}
}\\
\subfloat[]{
\centering
\includegraphics[width=0.45\linewidth]{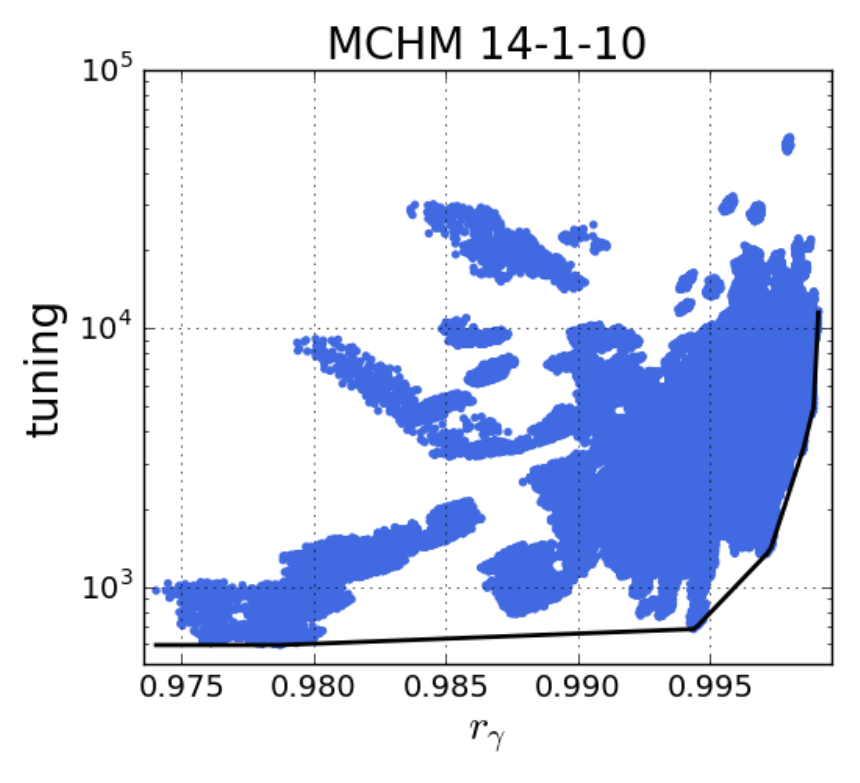}\label{minimalry}
}
\subfloat[]{
\centering
\includegraphics[width=0.45\linewidth]{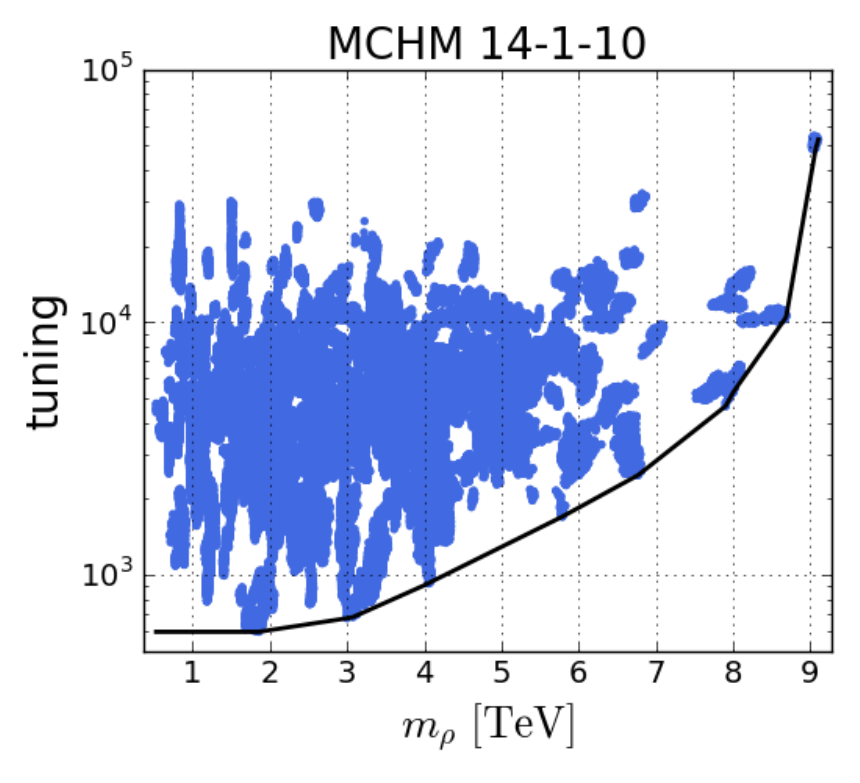}\label{minimalmp}
}\\
\subfloat[]{
\centering
\includegraphics[width=0.45\linewidth]{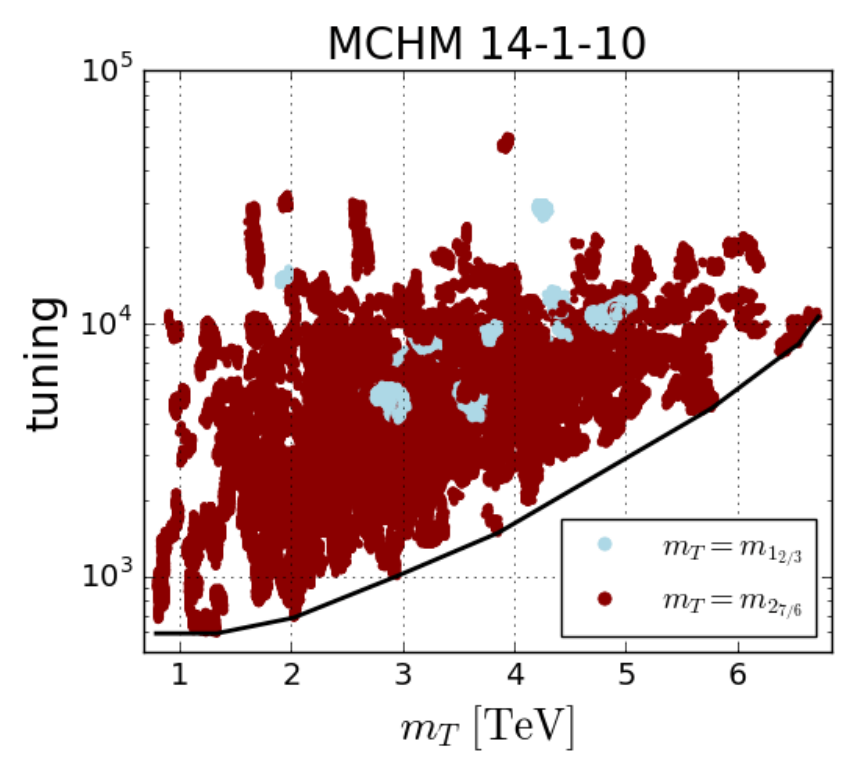}\label{minimalmt}
}
\subfloat[]{
\centering
\includegraphics[width=0.45\linewidth]{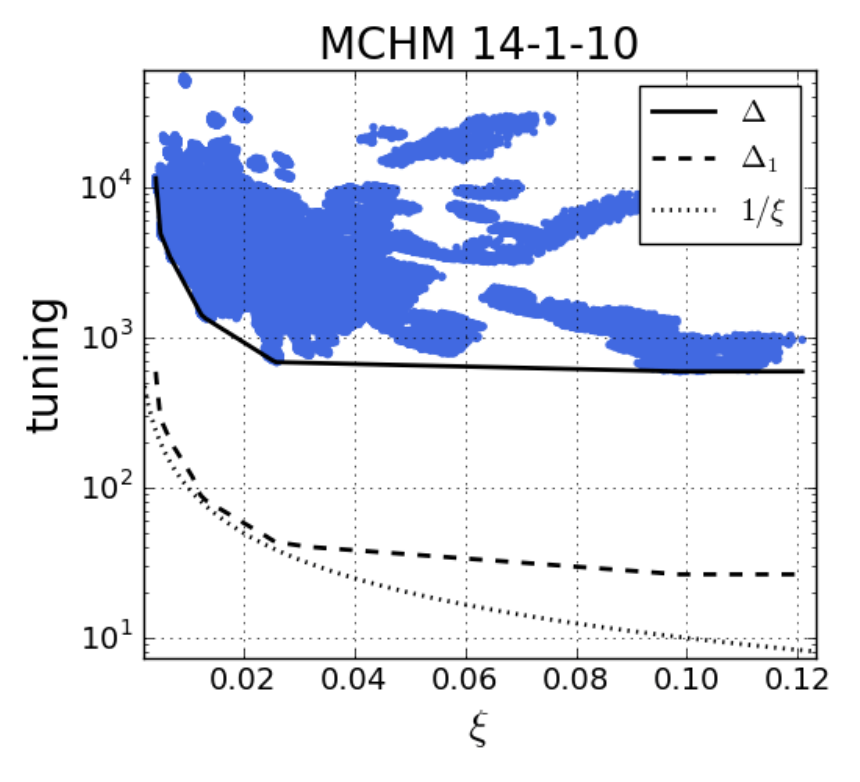}\label{minimalxi}
}
\caption{Tuning in the LM4DCHM$^{\textbf{5-5-5}}_{\textbf{14-1-10}}$ model as a function of Higgs coupling ratios, lightest scalar resonance mass, top partner masses, and vacuum misalignment}
\label{fig:minimalplots}
\end{figure}

Finally, we show the results for the case of a fully composite tau lepton, found in \cref{fig:minimalplots}. The tuning is similar to the previous case, with a minimum tuning of $\Delta = 594$ at a top partner mass of $m_{2_{7/6}} = 1.37$ TeV. As such, by the order-of-magnitude argument, the tuning at these low masses does not prefer this to the previous models. However, where a natural symmetric representation shows a sharp rise in the fine tuning with better top partner mass exclusion limits and more precise Higgs coupling measurements, the present model remains relatively untuned even at top partner masses of $m_{2_{7/6}} = 3.3$ TeV, which corresponds to a coupling ratio precision of $r_\psi \approx 2\%$ (in the LM4DCHM$^{\textbf{5-5-5}}_{\textbf{14-1-10}}$, Higgs coupling modifications have identical fine-tunings regardless of the species of particle being coupled with). This leaves the fully composite tau scenario as the likely most-natural representation once further run-2 data is released. 

\chapter{Next-to-Minimal Composite Higgs Model}
\label{sec:NMCHM}

\section{Motivation}

We continue the crusade to extend the M4DCHM with the minimal content required to lower tuning. The final option we consider is to minimally extend the group structure itself. A next-to-minimal $SO(6)/SO(5)$ coset containing the Higgs doublet has been considered in several instances \cite{gripaios2009beyond,Gripaios:2010mn,Redi:2012ha,Serra:2015xfa,Chala2017,niehoff2017electroweak}. The most significant feature is that now we have five NGBs to play with in the coset. In particular, \cite{Banerjee:2017qod} found that there is a reduced tuning in the NM4DCHM, purportedly from a form of level repulsion - a situation where  the mixing of the Higgs-like scalar $h$ and the new EW singlet $s$ leads to one becoming lighter at the cost of the other becoming heavier. We would like to study the phenomenology of the NM4DCHM without this mixing, in order to focus on the most minimal extension. In this sense, the model discussed here is a simplification of the full NM4DCHM, without reducing completely to the physics of the M4DCHM. We discuss these cases in \cref{sec:GB_Vacuum}. 

There is another intriguing possibility in this extension - that the EW singlet is a dark matter candidate, given the right choice of parameter space. The naturalness of this parameter space can be explored to examine whether a stable DM singlet is a finely tuned feature of the model. This possibility turn out to be highly dependent on the choice of how the right-handed fermion is embedded in the fundamental representation. 

\section{NM4DCHM Construction}

\begin{figure}[H]
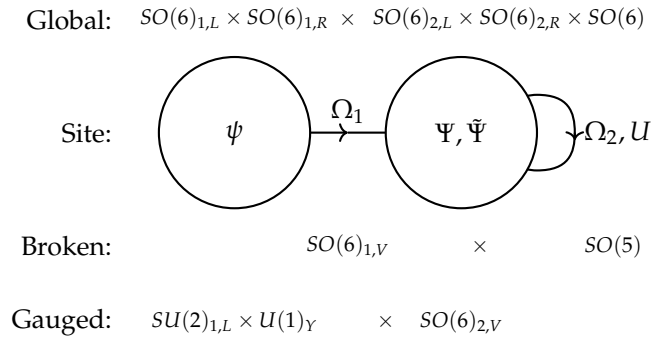

\centering\tikz[scale=1, every node/.style={transform shape}]{
\node [left] at (-1.5,0) {Site:};
\node [left] at (-1.5,1.5) {Global:};
\node [left] at (-1.5,-1.5) {Broken:};
\node [left] at (-1.5,-2.5) {Gauged:};
\draw [thick] (0,0) circle [radius=1] node {$\psi$};
\node at (0,1.5) {\footnotesize $SO(6)_{1,L} \times SO(6)_{1,R}$};
\node at (1.5,1.5) {\footnotesize $\times$};
\node at (1.5,-1.5) {\footnotesize $SO(6)_{1,V}$};
\draw [thick, ->] (1,0) -- (1.5,0);
\draw [thick] (1.5,0) -- (2,0);
\node [above] at (1.5,0) {$\Omega_1$};
\draw [thick,->] plot [smooth, tension=1] coordinates {(3,0) (4,0.5) (4.5,0) (4,-0.5) (3,0)};
\draw [thick, ->] (4.5,0) -- (4.5,-0.1);
\draw [thick, fill=white] (3,0) circle [radius=1] node {$\Psi, \tilde{\Psi}$};
\node [right] at (1.7,1.5) {\footnotesize $SO(6)_{2,L} \times SO(6)_{2,R} \times SO(6)$};
\node at (3.25,-1.5) {\footnotesize $\times$};
\node [right] at (4.5,0) {$\Omega_2, U$};
\node [right] at (4.5,-1.5) {\footnotesize $SO(5)$};
\node at (0,-2.5) {\footnotesize $SU(2)_{1,L} \times U(1)_Y$};
\node at (2,-2.5) {\footnotesize $\times$};
\node at (3,-2.5) {\footnotesize $SO(6)_{2,V}$};
}
\caption{The Next-to-Minimal 4D Composite Higgs Model moose diagram}\label{fig:NM4DCHM_moose}
\end{figure}

We shall focus on what differentiates the NM4DCHM from the M4DCHM. In particular, we still use the two-site construction descibed in the previous two chapters, but now the symmetry structure is given by the moose model in  \cref{fig:NM4DCHM_moose}. The generalisation is quite straightforward - we promote the global $SO(5)_0, SO(5)_1$ to global $SO(6)_0, SO(6)_1$, as well as gauging $SO(6)_{1}$. We again gauge the EW subgroup of $SO(6)_0$, which breaks the global groups to the $SU(2)_L \times U(1)_Y$ subgroup of $SO(6)_{0+1}$. Prior to this gauging, the elementary matter, in incomplete representations of $SO(6)$ still explicitly breaks the $SO(5)$ linear group to $SO(4)$.

The five pNGBs from the spontaneous breaking of the global $SO(6)\rightarrow SO(5)$ symmetry are parameterised as:

\begin{align}
\Phi = e^{\frac{\sqrt{2}}{f}i\pi^{\hat{a}}(x)T^{\hat{a}}}\Phi_0 = \frac{1}{\varphi}\sin\frac{\varphi}{f}\left(h_1, h_2, h_3, h_4, s, \varphi \cot \frac{\varphi}{f}\right)
\end{align}
where $\varphi = \sqrt{h_i h_i + s^2}$, and $\{T^{\hat{a}}\}$ are the broken generators, spanning the coset $SO(6)/SO(5)$. 


After electroweak symmetry breaking, we can simplify the parameterisation by choosing $\pi_1 = \pi_2 = \pi_3 = 0, \pi_4 = h, \pi_5 = s$ in the unitary gauge. We can use the change of basis 
\begin{align}
h = \varphi \cos(\eta/f), && s = \varphi \sin(\eta/f)
\end{align}
to non-linearly recast the two physical fields $h,s$ into the fields $\eta, \varphi$. In the unitary gauge, the Goldstone matrix is
\begin{align}
\hspace*{-2em}{\small U = \left(
\begin{matrix}
\mathbb{1}_{3\times 3} &  & 0 &  \\
 & \cos\left(\frac{\varphi}{f}\right) \cos^2\left(\frac{\eta}{f}\right) + \sin^2\left(\frac{\eta}{f}\right) & -\sin^2\left(\frac{\varphi}{2f}\right) \sin\left(\frac{2\eta}{f}\right) & \cos\left(\frac{\eta}{f}\right)\sin\left(\frac{\varphi}{f}\right)\\
 0 & - \sin^2\left(\frac{\varphi}{2f}\right)  \sin\left(\frac{2\eta}{f}\right)  &  \cos^2\left(\frac{\eta}{f}\right) + \cos\left(\frac{\varphi}{f}\right) \sin^2\left(\frac{\eta}{f}\right) &  \sin\left(\frac{\varphi}{f}\right)  \sin\left(\frac{\eta}{f}\right)      \\
 & -  \cos\left(\frac{\eta}{f}\right) \sin\left(\frac{\varphi}{f}\right) & - \sin\left(\frac{\varphi}{f}\right) \sin\left(\frac{\eta}{f}\right) &  \cos\left(\frac{\varphi}{f}\right)
\end{matrix}
\right)}
\end{align}
which gives a NM4DCHM version of the NGB fundamental multiplet
\begin{align}
\Phi_\text{unitary} = \left( 0,0,0,s_\varphi c_\eta,s_\varphi s_\eta, c_\varphi\right)
\label{eq:unitary_sigma}
\end{align}
noting the shorthand $s_x = \sin\frac{x}{f},$ $c_x = \cos\frac{x}{f}$. 

\subsection{Bosonic Sector}

The gauge sector of the NM4DCHM is a straightforward generalisation of the minimal case. The gauge Lagrangian \cref{eq:MCHM_gauge_lagrangian} can be carbon copied to the moose model described above. The gauged $SO(6)$ is implemented with $15$ gauge bosons
\begin{align}
\rho = \rho_L^i T^i_L + \rho_R^i T^i_R + a_1^a \hat{T}^a + a_2^a X^a
\end{align}
where $\rho_L, \rho_R$ are the vectorial $SO(4)$ gauge bosons of the minimal model. The $a_2$ are the axial gauge bosons of the broken group $SO(6)/SO(5)$. The $a_1$ are axial gauge bosons of the \textit{unbroken} $SO(5)/SO(4)$, which is a phenomenological departure from the minimal model.

The self-interactions in the holographic gauge are given similarly to the minimal case in \cref{eq:MCHM_unitary_gauge}. The NGBs not eaten by the holographic gauge lead to the covariant derivative of SM fields at the EW scale
\begin{align}
\hspace*{-2em}\frac{f^2}{2}(D_\mu \Phi)^T (D^\mu \Phi) =& \frac{f^2}{2}\left[
\left( 0,0,0,\frac{\partial_\mu \varphi}{f} c_\varphi c_\eta - \frac{\partial_\mu \eta}{f} s_\varphi s_\eta,\frac{\partial_\mu \varphi}{f} c_\varphi s_\eta + \frac{\partial_\mu \eta}{f} s_\varphi c_\eta, -\frac{\partial_\mu \varphi}{f} s_\varphi\right)\right.\nonumber \\
& \left. - i g W^{a_L}_\mu T^{a_L} \Phi - i g' B_\mu T^{a_R} \Phi \right]^2 \nonumber \\
= & \frac{1}{2}(\partial \varphi)^2 + \frac{1}{2} (\partial \eta)^2 s_\varphi^2 + \frac{f^2}{8} s^2_\varphi c^2_\eta \left( g^2 W^2 + g'^2 B^2 + 2 g g' B_\mu W^{\mu,(3)}\right) \nonumber \\
&-(\frac{\partial_\mu \varphi}{f} c_\varphi c_\eta-	 \frac{\partial_\mu \eta}{f}s_\varphi s_\eta)\left(\frac{g}{2} W^{\mu,(3)} - \frac{g'}{2} B^\mu\right)s_\varphi c_\eta \label{eq:NMCHM_gauge_lagrangian} 
\end{align}
where $a_L$ runs from $1,2,3$, so $W^{\mu,(3)}$ is the third $W$ field. The third term in \cref{eq:NMCHM_gauge_lagrangian} can be used to match to the SM Higgs-EW Lagrangian
\begin{align}
\mathcal{L}_{\text{Higgs-EW}} &= (D_\mu \Phi_{\text{SM}})^\dagger (D^\mu \Phi_{\text{SM}})\nonumber \\
&= (\partial H)^2 + \frac{1}{4}\left(v + H\right)^2 \left(2g^2 W_\mu^- W^{+\mu} + (g' B_\mu - g A_\mu^3)^2\right)  
\end{align}
We can then identify 
\begin{align}
v &= f\sin\frac{\langle\varphi\rangle}{f}\cos\frac{\langle\eta\rangle}{f}\label{eq:NMCHM_vev}
\end{align}
Note that \cref{eq:NMCHM_gauge_lagrangian} contains kinetic terms that are not canonically normalised. We can rotate within the holographic gauge to redefine
\begin{align}
\sin \eta \rightarrow \sin \left( \frac{\eta}{f\sin(\langle \varphi \rangle )} \right)
\end{align}
such that the kinetic terms do not mix, validated by taking the second derivatives at the GB vevs.

\subsection{Matter content}

Including matter is broadly the same procedure as for the M4DCHM. Considering first the fundamental representation, however, we immediately see a new feature in the embedding of $SO(4)$ fields\footnote{Recall that the custodial $SO(4)$ is still the group that we need our fermions to transform under, before EW gauging.}
\begin{align}
\Psi_\textbf{6} = \left( \begin{array}{ c c c c c | c | c }
 & & \Psi_\textbf{4} & & & \Psi_\textbf{1} & \Psi_\textbf{1}'
\end{array} \right)^T
\end{align}
With two singlets in the fundamental representation, there are two fields that could couple to the right-handed elementary fields. We should embed the elementary field in a way that allows both of these couplings
\begin{align}
\psi_L &= \frac{1}{\sqrt{2}}\left(\begin{matrix}
d_L \\
-i d_L \\
u_L \\
iu_L \\
0\\
0
\end{matrix} \right), &
\psi_R &= \left(\begin{matrix}
0\\
0\\
0\\
0\\
u_R e^{i\delta} \cos\theta\\
u_R \sin\theta
\end{matrix} \right)
\end{align}
Including both right-handed interactions is not only desirable under the maxim of including all possible terms, it is also necessary to prevent the singlet NGB from emerging as massless. $\delta$ appears due to a choice of this top coupling. It is not a physical parameter, and can be removed by a phase transformation under the $SO(2)$ subgroup of $SO(6)$, taking $\me^{i\delta} \rightarrow 1$. $\theta$ \textit{is} however an important artefact of the NM4DCHM, and appears from the choice of composite partner embedding within the $SO(2)$ subgroup. It has two interesting limits. The first is $\theta \rightarrow \pi/2$, in which case the singlet effectively decouples from the observable content, and the model appears in some ways to be the M4DCHM. The subtlety of this limit will be discussed in the next section. The other limit $\theta \rightarrow 0$ is a massless singlet limit, which is prohibited by EW axion constraints.

The two sectors interact via mixing terms in the fermionic Lagrangian, which is the most minimal set of interactions required to generate the SM Yukawas, in the unitary gauge (i.e. using the gauge symmetry to choose, $\langle h_i\rangle = 0, i=\{1,2,3\}$, giving $\Phi$ according to \cref{eq:unitary_sigma})

\begin{align}
\begin{split}
\mathcal{L}_f &= \bar{\psi}_L i \slashed{D} \psi_L + \bar{\psi}_R i \slashed{D} \psi_R + \Delta_{t_L} \bar{\psi}_L \Psi^T_R + \Delta_{t_R} \bar{\psi}_R\Psi^{\tilde{T}}_L \\
& + \bar{\Psi}^T_L(i\slashed{D} - m_T)\Psi^T_R + \bar{\Psi}^{\tilde{T}}_L(i\slashed{D} -m_{\tilde{T}})\Psi^{\tilde{T}}_R -Y_T\bar{\Psi}^T_L \Phi \Phi^\top \Psi^{\tilde{T}}_R - m_{Y_T}\bar{\Psi}^T_L \Psi^{\tilde{T}}_R + \textnormal{h.c.} \label{eq:fundamental_lagrangian}
\end{split}
\end{align}

As before, note the absence of terms $\bar{\Psi}^T_R  \Phi \Phi^\top \Psi^{\tilde{T}}_L$,  $\bar{\Psi}^T_L  \Phi \Phi^\top \Psi^T_R$ and $\bar{\Psi}^{\tilde{T}}_L  \Phi \Phi^\top \Psi^{\tilde{T}}_R$. We impose this absence in order to keep the Higgs potential finite. To compare with the simplified M4DCHM, we include only the top quark and heaviest quark doublet in the analysis. We do not consider partially composite leptons, in order to hold fixed possible sources of tuning.

The elementary terms\footnote{After expanding the 6-plets, one can group the left-handed terms $\{t_L,b_L\}$ into their regular SM doublet $q_L$.} appear in an effective Lagrangian, coming from decomposing the GB and fermion multiplets in the unitary gauge under $SU(2) \times SU(2)$, given in \cite{Redi:2012ha} 
\begin{align}
\begin{split}
\mathcal{L}_\textnormal{fermion} &= \bar{q}_L\Pi_q (p^2, \eta, \varphi)  \slashed{p} q_L + \bar{t}_R\Pi_t (p^2, \eta, \varphi) \slashed{p} t_R +\bar{q}_L M_t(p^2, \eta, \varphi) t_R + \textnormal{h.c} \\
&= \bar{q}_L \left( \frac{\Delta^2}{(\Delta_{t_L})^2} + \Pi_{q_L,0}(p^2) + \frac{1}{2} \frac{s^2_\varphi}{\varphi^2}\Pi_{q_L,1}(p^2)H^c H^c\right) \slashed{p} q_L \\
&+ \bar{t}_R \left( \frac{\Delta^2}{(\Delta_{t_R})^2} + \Pi_{u_R,0}(p^2) + s^2_\theta \Pi_{u_R,1}(p^2) + \left[s_\varphi^2 (c_\theta^2 s_\eta^2 - s_\theta^2)\right] \Pi_{u_R,1}(p^2) \right) \slashed{p} t_R \\
&+\bar{q}_L \frac{M_{u,1}}{\sqrt{2}}\frac{s_\varphi}{\varphi}H^c \left( ic_\theta s_\varphi s_\eta + s_\theta c_\varphi\right) t_R + \textnormal{h.c}
\end{split}\label{eq:fermion_lagrangian}
\end{align}
where the form factors $\Pi_q, \Pi_t, M_t$ have been inserted. These are found via an analogous process to the M4DCHM. A reminder that $q_L$ and $H^c$ are the SM quark doublet and charge conjugate of the normalised SM Higgs doublet, respectively,
\begin{align}
q_L = \left( t_L , b_L \right)^T,  && H^c = \frac{1}{h} i\sigma_2 \left( \begin{matrix}
h^1 - ih^2\\
h^3 - ih^4
\end{matrix}\right)^* = \frac{1}{h} \left( \begin{matrix}
-(h^1 + ih^2)\\
h^3 + ih^4
\end{matrix}\right)
\end{align}
The form factors $\Pi_i$ are given in full in \cref{sec:form_factors}. Here, the elementary bare quark mass terms $\frac{\Delta^2}{(\Delta_{t_L / t_R})^2}$ can be understood as canonically normalised. That is, there is some common scale $\Delta$ that can be factored out once the form factors are found.

%

A final distinction of the NMCHM is the relevance of only one choice of representation (although many composite partners and resonances could be added in this representation) \cite{gripaios2009beyond}. In brief, the three smallest representations under $SO(6) \approx SU(4)$ are the $\textbf{4}$, the $\textbf{6}$, and the $\textbf{10}$. The $\textbf{4}$ does not contain a bidoublet when decomposed under $SU(2)_L \times SU(2)_R$, and thus cannot contain a representation that couples with the SM quark doublet, which must also be incompletely embedded into a $\textbf{4}$. The symmetric traceless $\textbf{10}$ does contain such a bidoublet, however upon embedding the SM quarks in a simple way, we see that there remains a $U(1)_s$ symmetry protecting the scalar singlet. In this case, the singlet will correspond to an electroweak axion, with properties that have been excluded experimentally. Less minimal \textbf{10} embeddings have been shown in reference \cite{Serra:2015xfa} to produce a massive singlet and evade exclusion. Thus, this leaves the $\textbf{6}$ as the simplest representation for the quark partners, and we thus focus on the NM4DCHM$^\textbf{6}$.

\subsection{Goldstone Boson Vacuum Behaviour}\label{sec:GB_Vacuum}

The most significant departure from the minimal model is the inclusion of a second NGB, a singlet under the global $SO(5)$ subgroup. In practical terms, this means that there is now a pNGB mass matrix, a two-dimensional $(\langle h \rangle, \langle s \rangle) \sim (\langle \varphi \rangle, \langle \eta \rangle)$ potential function, and additional SM coupling deviations.  After EWSB, we can write the low energy effective potential for the interactions of the Higgs boson and scalar singlet with gauge fields and fermions \cite{contino2010strong} 
\begin{align}
\begin{split}
\mathcal{L}_\text{eff} &= \frac{1}{2}(\partial_\mu h)^2 + \frac{1}{2}(\partial_\mu s)^2 - V(h,s) + \frac{v^2}{4}\textnormal{Tr} \left[ D_\mu \Sigma^\dagger D^\mu \Sigma \right] \left( 1 + 2a_h \frac{h}{v} + b_h \frac{h^2}{v^2} + b_s \frac{s^2}{v^2} + ... \right) \\
- & m_i \bar{\psi}_{Li} \Sigma\left(1 + c_h \frac{h}{v} + ...\right) \psi_{Ri} - m_i \bar{\psi}_{Li} \left(c_s \frac{s}{v} + ... \right) \psi_{Ri} + h.c. \label{eq:general_lagrangian}
\end{split}
\end{align}

where the GBs eaten by the $W$ and $Z$ bosons are parameterised by $\Sigma=\exp(i\chi^a\sigma^a/v)$. The couplings of $a$, $b$ and $c$ can be obtained as:
\begin{align}
a_h = \sqrt{1-\xi}, && b_h = 1-2\xi, && b_s = 1, && c_h = \frac{1-2\xi}{\sqrt{1-\xi}}, && c_s = i\frac{\xi}{1-\xi}\cot\theta \label{eq:nmchm_couplings}
\end{align}

To compute the vacuum misalignment $\xi$ and therefore the coupling terms, we need to explore the effective potential of the Goldstone bosons. This is generically given by the Coleman-Weinberg formula for the gauge boson and top quark contributions, where the form factors are given in \cref{sec:form_factors}
\begin{align}
V_\textnormal{fermionic} &= \frac{9}{2}\int\frac{d^4p}{(2\pi)^2}\log\Pi_W - 2N_c \int \frac{dp^4}{(2\pi)^4} \ln \left( p^2 \Pi_{t_L} \Pi_{t_R} - \Pi_{t_L t_R}^2\right)\label{eq:fermionic_potential}
\end{align}
As in the MCHM, we require this potential to have a minimum such that it reproduces the electroweak vacuum expectation value (vev). We can attempt to do this at leading order, which would lead to a natural EWSB potential. For example, the potential in \cref{eq:fermionic_potential} can be expanded at leading order in the MCHM Goldstone field as 
\begin{align}
V(h) = \alpha \sin^2\frac{h}{f} + \mathcal{O}(s_h^4)
\end{align}
This has possible minima\footnote{Depending on the sign of $\alpha$} at integer multiples of $\langle h \rangle = \frac{f\pi}{2}$, which is far too high. The case of $\langle h \rangle = 0$ leads to no EWSB. The same obstacle applies to the NMCHM potential, which at leading order in $\varphi, \eta$ is
\begin{align}
V(\varphi, \eta) = \sin^2\frac{\varphi}{f}\left( c_1 + c_2 \sin^2\theta - c_3 \sin^2\theta \right) + \mathcal{O}(s_\varphi^4,s_\eta^4)
\end{align}
where the expressions for the integral terms $c_i$ are given in \ref{sec:form_factors}. This has stationary points at integer multiples of $\langle \varphi \rangle = \frac{f\pi}{2}$. 
Again, this is problematic, as we need the EW vev $v =  f \sin\frac{\langle\varphi\rangle}{f}\cos\frac{\langle\eta\rangle}{f}$ to be at a much lower scale than the typical symmetry breaking scale $f=f \sin\frac{\pi}{2} \cos{0}$.

Therefore, as in the MCHM, we must break EW symmetry by considering higher-order terms that must cancel precisely, requiring the notorious composite Higgs double tuning. We include higher order terms, up to quartic in $\sin\frac{\varphi}{f}\sin\frac{\eta}{f}$ \footnote{We include such seemingly high order terms since $\xi^2 \propto s_{\langle \varphi \rangle}^2 s_{\langle \eta \rangle}^2$, and we must therefore include each field up to consistent order. Note that to obtain \cref{eq:varphi} and \cref{eq:higgs_mass}, it is sufficient to expand to quadratic order $V = s_\varphi^2(c_1 + c_2 s_\theta^2 - c_3 s_\theta^2) - s_\varphi^2 s_\eta^2(c_1 + c_2 c_\theta^2 -c_3 s_\theta^2) + s_\varphi^4 c_3 s_\theta^2$. The singlet mass, on the other hand, requires corrections given by the quartic-order potential.}
\begin{align}
\begin{split}
V(\varphi,\eta) = c_1 \sin^2\frac{\varphi}{f}\cos^2\frac{\eta}{f} + c_2 \sin^2 \frac{\varphi}{f} \left( \sin^2 \theta - \cos^2 \theta \sin^2 \frac{\eta}{f} \right) \\
- c_3 \sin^2 \frac{\varphi}{f} \cos^2 \frac{\eta}{f} \left( \cos^2 \theta \sin^2 \frac{\varphi}{f} \sin^2 \frac{\eta}{f} + \sin^2 \theta \cos^2 \frac{\varphi}{f}\right)\\
 + \mathcal{O}(\sin^{10}_\varphi, \sin^8_\varphi\sin^2_\eta,...,\sin^{10}_\eta)
\end{split}
\label{eq:NMCHM_potential}
\end{align}
To find the classical expectation value, we solve for
\begin{align}
\frac{\partial V}{\partial \eta} \bigm\lvert_{\varphi = \langle \varphi \rangle, \eta = \langle \eta \rangle} & = 0 \nonumber\\
\frac{\partial V}{\partial \varphi} \bigm\lvert_{\varphi = \langle \varphi \rangle, \eta = \langle \eta \rangle} & = 0 \label{eq:vev}
\end{align}
where zeroes will be found both from trivial extrema (i.e. integer multiples of $\varphi, \eta = \frac{f\pi}{2}$) and double tuning extrema (cancellations between terms, requiring tuning of $c_1, c_2, c_3$). 

\subsubsection{The trivial singlet vacuum}

\begin{figure}
\centering
\subfloat[a]{
\tikz{
\node at (2,2) {\includegraphics[scale=0.07]{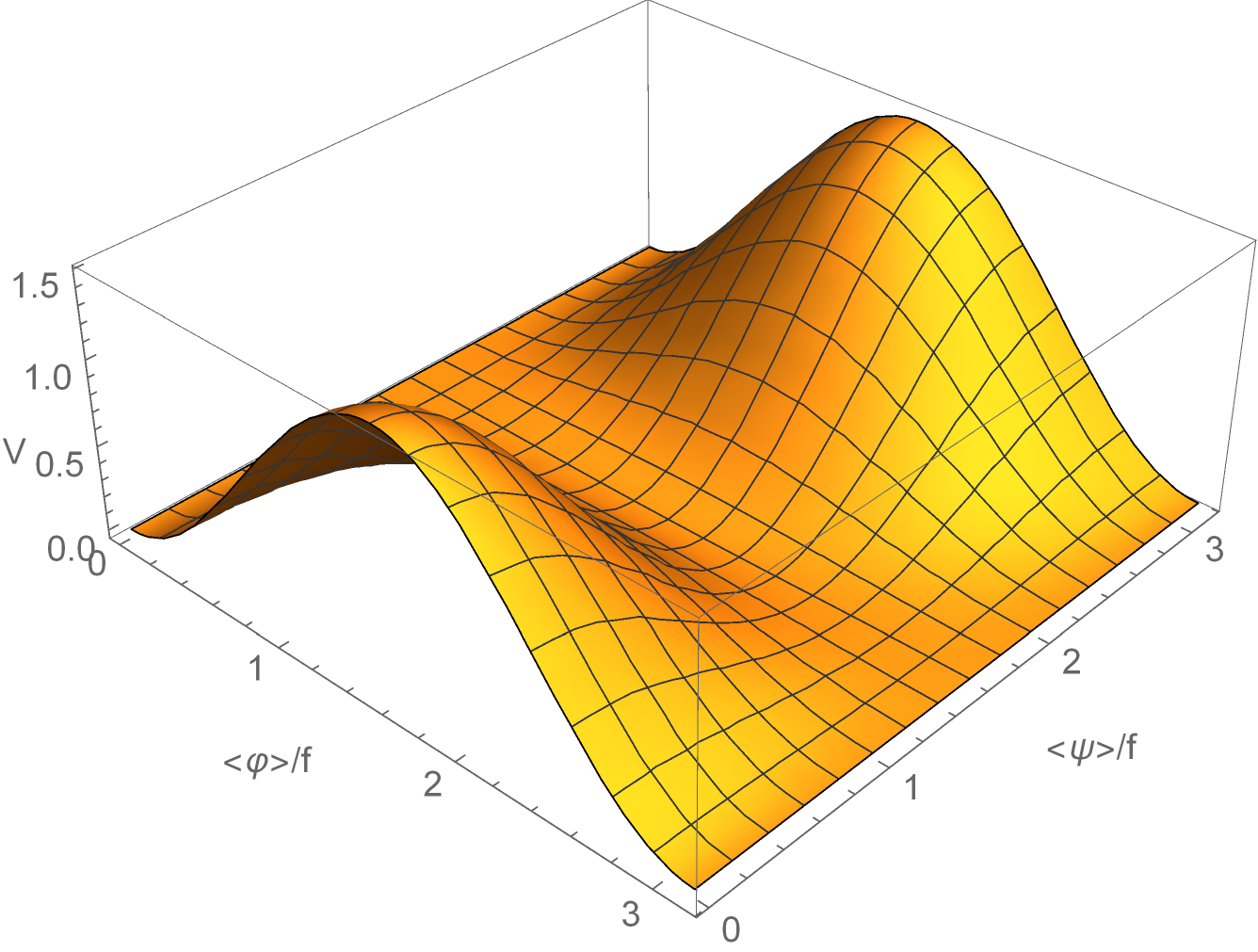}};
\node at (0.2,0.5) {$\varphi/f$};
\node at (4.1,0.5) {$\eta/f$};
}
}
\subfloat[b]{
\tikz{
\node at (2,2) {\includegraphics[scale=0.07]{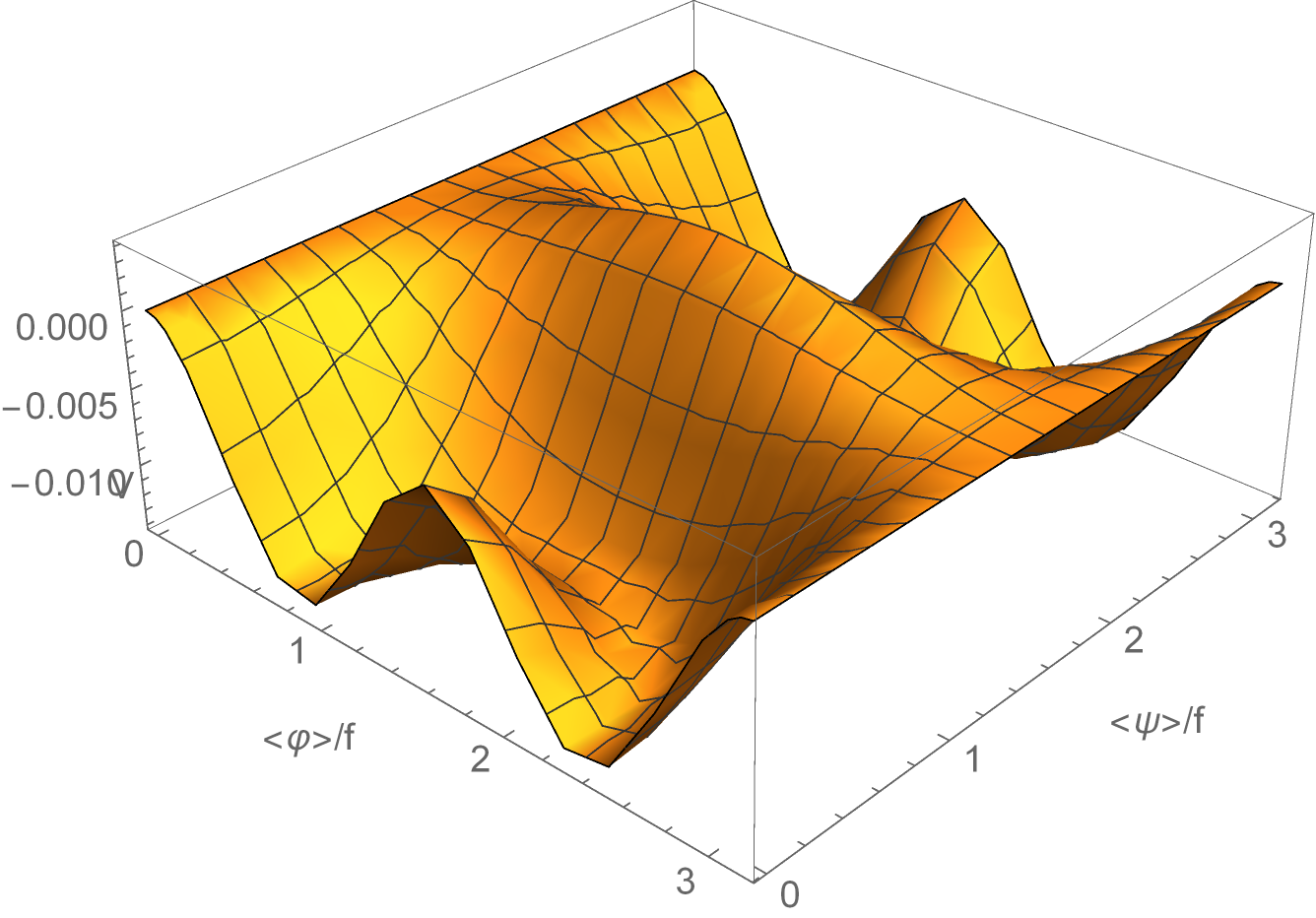}};
\node at (0.2,0.5) {$\varphi/f$};
\node at (4.1,0.5) {$\eta/f$};
}
}\\
\subfloat[The simple cases of minima at $\langle \eta\rangle/f = 0$ and $\langle \varphi \rangle/f = \pi/2$. The general minima $\langle \eta \rangle, \langle \eta \rangle \neq 0$ are given by the dashed line.]{
\tikz{
\node at (5,2) {\includegraphics[scale=0.15,trim={0 0 0 25cm}, clip]{NMCHM/potential_derivative_2_v2}};
\node at (1.2,1) {$\varphi/f$};
\node at (9.3,1) {$\eta/f$};
\node at (1.3,5.5) {$\langle \varphi \rangle \neq 0$};
\node at (4.2,5.5) {$\langle \eta \rangle \neq 0$};
\draw [thick, ->] (1.3,5.2) --(1.3,2.4);
\draw [thick, ->] (4.2,5.2) --(4.2,2.5);
\draw [thick, dashed, white] plot [smooth, tension=1] coordinates {(4.25,0.7) (4.85,1.75) (3.75,2.7)};
}
}\caption{Two examples of the GB potential. On the left, $c_1 = 1, c_2 = 1, c_3 = -0.1, s_\theta = 0.7$ with $\xi = 15.7$, corresponding to no EWSB. On the right, $c_1 = 0.1, c_2 = -0.2, c_3 = 0.1, s_\theta = 0.7$, with $\xi = 0.48$. Satisfying the condition $\xi<1$ allows for the possibility of EWSB.}\label{fig:NMCHM_potential}
\end{figure}

It can be shown that the surface $\langle\eta\rangle =0$ or $\langle \varphi \rangle = \frac{f\pi}{2}$ always contains an extremum of the potential, and either (but not both) can be chosen such that EWSB may still occur realistically. Fig. \ref{fig:NMCHM_potential} illustrates the vacuum configurations, and the validity of this choice is well-discussed in \cite{Redi:2012ha}. Previous work \cite{Niehoff:2017guu,niehoff2017electroweak} on the next-to-minimal model have pointed out that there is a limit of the potential, such that choosing 
\begin{align}
\theta \rightarrow \pi/2 \implies \langle s \rangle \rightarrow 0 \implies
\left( \begin{matrix}
m_{hh} & m_{hs} \\
m_{sh} & m_{ss}
\end{matrix}\right) \rightarrow & \left( \begin{matrix}
m_{hh} & 0 \\
0 & m_{ss}
\end{matrix}\right) \\
&\implies \text{NMCHM} \rightarrow \text{MCHM} \nonumber
\end{align}
While it is true that taking the top partner to only be embedded in the sixth component of the sixplet does lead to this minimal-model-like limit, there is a subtlety. In this work, we assume for simplicity the trivial singlet vacuum $\langle \eta \rangle = 0$, which leads to a diagonal pNGB mass matrix, and a simplified potential. It does not, however, lead to $\theta \rightarrow 0$ and the minimal-model-like limit. We thus retain a massive singlet, and novel phenomenology. One missing feature in this assumption, is that of a $(m_h, m_s)$ "level repulsion" discovered in \cite{Banerjee:2017qod,}. This feature is numerically tested in \cref{sec:NMCHM_results}.

For the choice $\langle \eta \rangle = 0$, a potential extremum is found for
\begin{align}
\sin\frac{\langle\varphi\rangle}{f} &= \sqrt{\frac{c_3 s^2_\theta - c_1 - c_2 s^2_\theta }{2 c_3 s^2_\theta}}\label{eq:varphi}\\
\implies \xi &= \frac{v^2}{f^2} = \frac{c_3 s^2_\theta - c_1 - c_2 s^2_\theta }{2 c_3 s^2_\theta} \nonumber
\end{align}
using the definition in \cref{eq:vev}. This implies that  $0 < \frac{c_3 s_\theta^2 - c_1 - c_2 s_\theta^2}{2c_3 s_\theta^2} < 1$, in order to achieve a non-trivial vev. This can be used as a constraint to rescale $f$ for correct EWSB behaviour\footnote{Note that this constraint is not sufficient for EWSB - it only corresponds to an extremum. The Higgs mass must be found to be positive, to ensure that this solution is a local minimum.}. To better illustrate the possible behaviour of the potential, we show it in \cref{fig:NMCHM_potential} for two different sets of \{$c_1,c_2,c_3,s_\theta$\}. The first plot shows the typical case encountered in much of the parameter space where the extrema are given only by integer multiples of $\varphi, \eta = \frac{f\pi}{2}$, leading to no EWSB. The second plot shows an example of the fine cancellations which occur in a small region of the parameter space, corresponding to the solution in \cref{eq:varphi}. This gives additional minima and maxima, which are a condition of EWSB.

The masses of the two scalars can be found using the second derivatives of \cref{eq:NMCHM_potential}, and the solution for $\langle \varphi \rangle$ in \cref{eq:varphi}
\begin{align}
m_\varphi^2 = m^2_h &= \frac{-4c_1 c_2 -2c_1^2 / s_\theta^2 + 2(c_3^2 - c_2^2)s_\theta^2}{c_3 f^2}\label{eq:higgs_mass}\\
m^2_\eta = m^2_s &= \frac{c_1 - (c_2 + c_3)s_\theta^2}{s_\theta^2 f^2} c_{2\theta}\label{eq:singlet_mass}
\end{align}
Note that we have changed the basis from $\varphi, \eta$ to $h,s$, but that the masses are the same due to $\langle\eta\rangle = 0$ being a stationary point. This can be laboriously shown with liberal application of the chain rule.

We can thus analyse the Higgs mass expression as a function of each of the integral terms. This also gives our top mass term, which can be found by diagonalising the low-energy Lagrangian in \cref{eq:fermion_lagrangian} 

\begin{align}
\abs{m_t}^2 &= \frac{[M_1^u(0)]^2}{\Pi_{t_L}(0) \Pi_{t_R}(0)}s_{\langle\varphi\rangle}^2 c_{\langle\eta\rangle}^2 \left( c_\theta^2 s_{\langle\varphi\rangle}^2 s_{\langle\eta\rangle}^2 +s_\theta^2 c_{\langle\varphi\rangle}^2 \right)\nonumber\\
&= \frac{[M_1^u(0)]^2}{\Pi_{t_L}(0) \Pi_{t_R}(0)} s_\theta^2 \xi (1-\xi)\label{eq:top_mass}
\end{align}

\section{Exploration of Parameter Space}

We numerically explore the simplified next-to-minimal model described above, using the Diver package. This package is presented in \cref{sec:Diver}. Recall the impetus for considering more contributions - for example light quarks in \cref{sec:MCHM} and leptonic contributions in \cref{sec:LCHM} - was differing orders of interactions between representations. As the fundamental representation is our main concern here, we include only the left-handed top and bottom quarks, and right-handed top quark, as composite.  To explore flavor constraints on the NM4DCHM, one would need to include three generations of quark, as in \cite{niehoff2017electroweak}.

The parameter space is spanned by 
\begin{align}
\{ m_\rho, m_a, t_{g/g'}, \Delta_q, \Delta_t, m_q, m_t, m_Y, Y, \theta\}
\end{align}
which is much smaller than the previous study due to far fewer composite fields. The domain of these parameters is presented in \cref{tab:4DCHM_params}. The smaller parameter volume leads to many viable points being found, and so it was not necessary to use a second stage of MCMC scanning as in the composite lepton case. The observable space is also simpler, with a cost function of
\begin{align}
\chi^2 = \frac{(m_H- 125)^2}{2\times 5^2} + \frac{(m_t - 155)^2}{2\times 15^2}
\end{align}
Again, the EW vev is applied as a constraint by solving directly for the breaking scale $f$. In calculating the fine tuning, however, the EW vev is included as a third observable in the higher order tuning.

\section{Lowered fine tuning?}
\subsection{Fine-tuning results}\label{sec:NMCHM_results}

We now present the scan results in terms of the fine-tuning found at each viable parameter point. The tuning of each point is shown against the lightest vector-boson resonance mass $m_\rho$, the lightest top partner resonance mass, the mass of the $SO(6)$ scalar singlet, the Higgs coupling ratios $r_\chi$ and the vacuum misalignment $\xi = v^2/f^2$. A convex hull is provided to understand the general limits of minimal fine-tuning (note that given the logarithmic scale, the hull may not always appear to be convex). In all coupling correction plots, several predicted bounds are included, based on the anticipated precision of the future International Linear Collider (ILC) \cite{tian2016measurement}. Two bounds are included - a pessimistic bound at the $250$GeV baseline ILC, and an optimistic bound from a high-energy, high-luminosity upgrade. These bounds are given in \cref{tab:ILC}.

\begin{figure}
\centering
\subfloat[a]{
\centering
\includegraphics[width=0.45\linewidth]{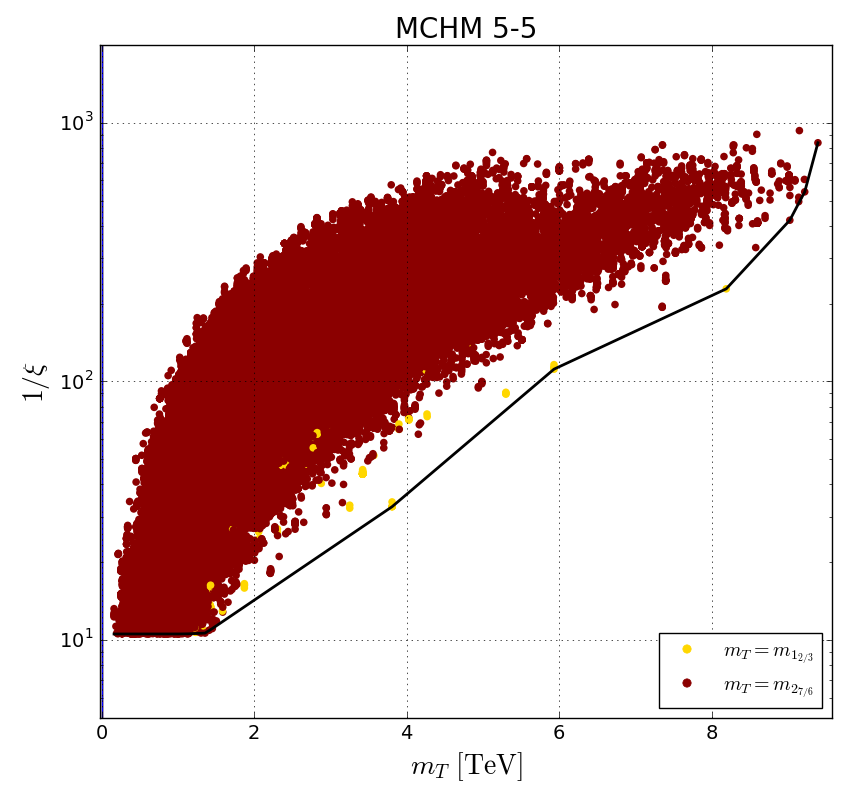}
}
\subfloat[a]{
\centering
\includegraphics[width=0.45\linewidth]{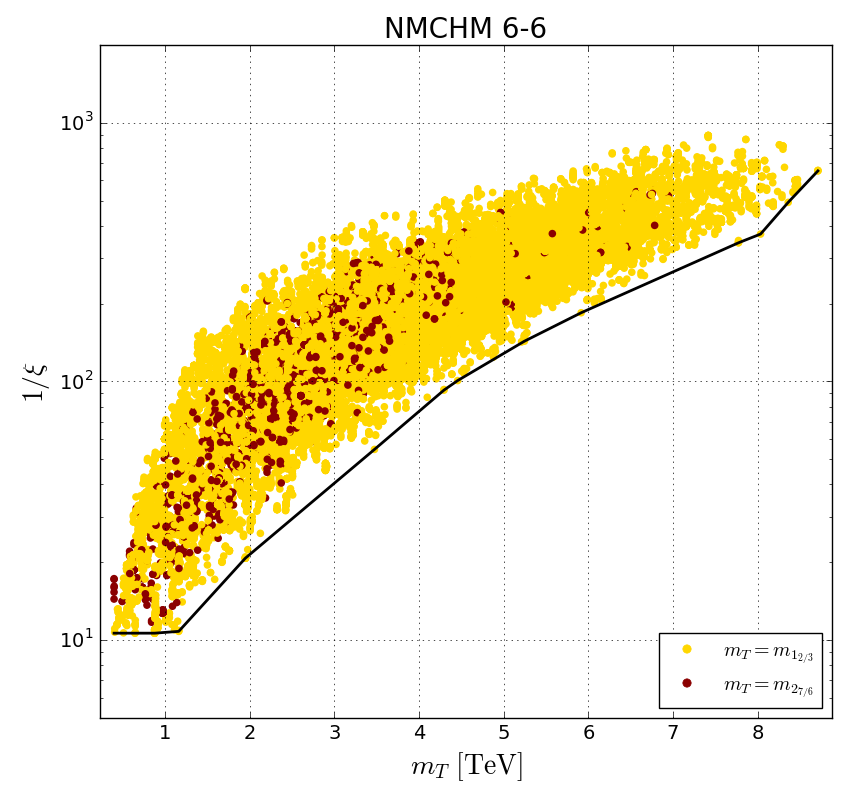}
}
\caption{Comparison of each model's lightest top partner vs. naive tuning}\label{fig:masses_vs_xi}
\end{figure}

Before analysing our results, we note that an earlier study (Reference~\cite{Banerjee:2017qod}) demonstrated that higher top partner masses may be achieved in the NM4DCHM, with no fine-tuning penalty, through a process dubbed ``level repulsion''. If the doublet and singlet in the pNGB sector both get vevs, the model exhibits a tree-level doublet-singlet mixing. If the singlet state is heavier, then the mixing can result in pushing down the dominantly  doublet eigenstate to match the observed Higgs mass at 125 GeV. Before mixing, the masses of both of the states can conceivably be larger, which makes the theory more natural. This earlier result may naively appear to conflict with the results of the previous section, but in fact there is no contradiction once one compares the different scope of the studies and the fine-tuning measure used. The requirement that both the doublet and singlet get a vev corresponds to $\theta$ being close to $\pi/2$, and thus this is a special limit of the more general theory (one that would in fact appear as a fine-tuning contribution in a proper analysis). We assume in our study that the singlet does not acquire a vev, meaning that there is no overlap between our results and the previous study. Indeed, if we examine the naive tuning measure of $1/\xi$ as a function of the lightest top partner (LTP) mass in our study, we find no tuning gain for the NM4DCHM vs the M4DCHM (see \cref{fig:masses_vs_xi}).

\begin{figure}
\centering
\subfloat[]{
\centering
\includegraphics[width=0.45\linewidth]{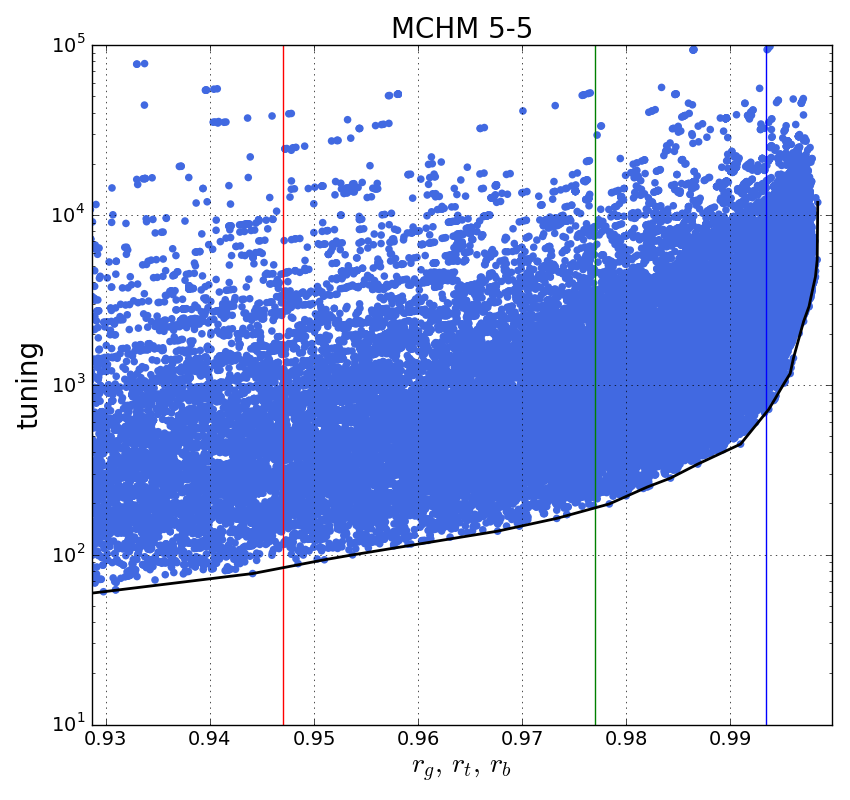}
}
\subfloat[]{
\centering
\includegraphics[width=0.45\linewidth]{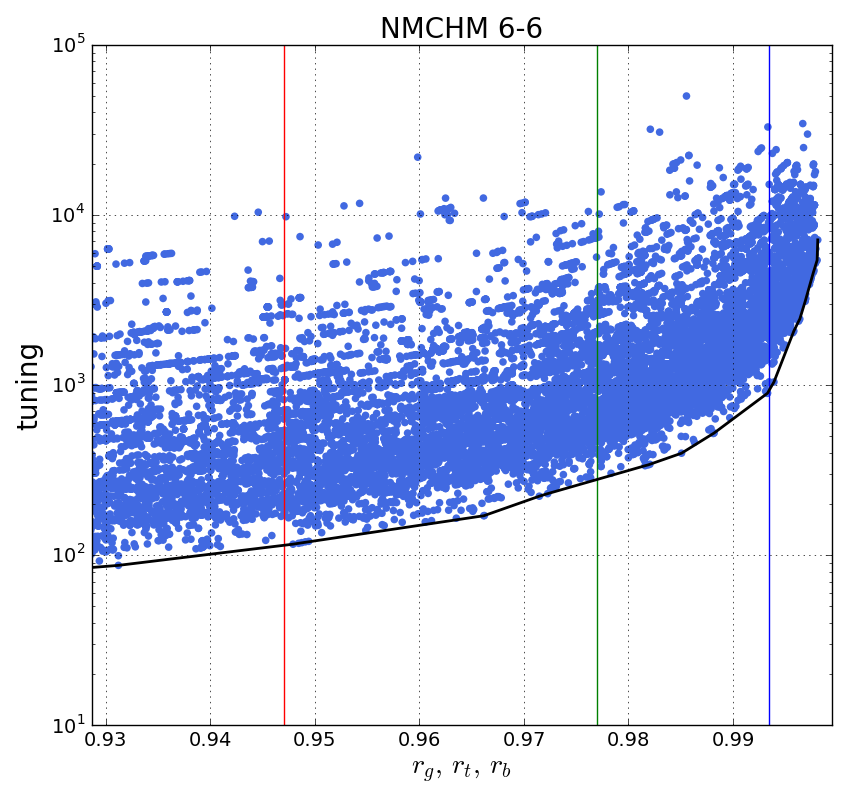}
}
\caption{\label{fig:tuning-vs-couplings} Comparison of higher-order tuning (defined in \cref{eq:firstorder}) in the Higgs-gluon, -top and -bottom coupling deviation (as defined in \cref{eq:coupling_deviation} and \cref{eq:nmchm_couplings}) between the minimal and next-to-minimal models. Precision bounds (denoted by coloured lines) are defined in Table \ref{tab:ILC}. The red line shows the expected precision of a $250$ GeV ILC, green a $500$ GeV ILC, and blue a high-luminosity $1$ TeV ILC.}
\end{figure}

In \cref{fig:tuning-vs-couplings} we show the higher-order tuning as a function of the modification to the Higgs-gluon, Higgs-top and Higgs-bottom couplings, for both the M4DCHM and NM4DCHM models. As one would expect, the minimum fine-tuning available in each model would increase if one were able to measure the Higgs couplings more precisely (assuming that they remain at the SM values). In both models, the impact of a 250 GeV ILC is minimal, but the high-luminosity 1 TeV ILC would increase the minimum fine-tuning by roughly an order of magnitude. We also observe a slightly higher fine-tuning in the NM4DCHM model, relative to the M4DCHM model, regardless of future measurements of the Higgs couplings. This can be attributed to a small punishment for increasing the parameter set from nine to ten. A thorough discussion of parameter set scaling in the higher order tuning measure can be found in reference \cite{Barnard:2017kbb}. This agrees with a first-order expectation, since NM4DCHM observables are generically proportional to M4DCHM observables according to $m_\text{NMCHM} \propto m_\text{MCHM}\sin\theta$, and $\theta$ is a free parameter.

\begin{figure}
\centering
\subfloat[]{
\centering
\includegraphics[width=0.45\linewidth]{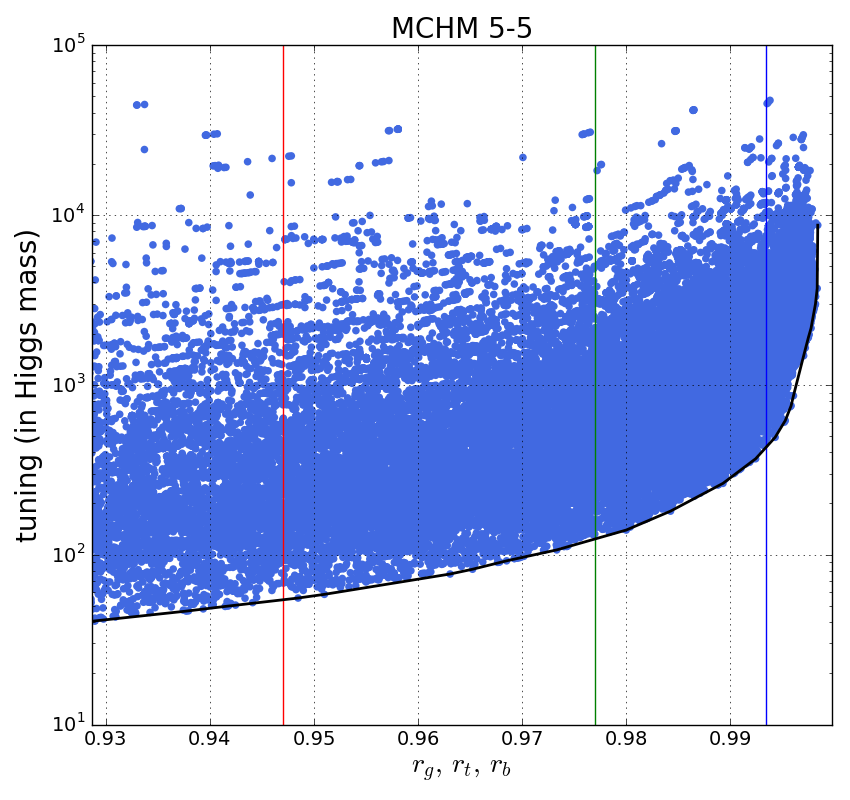}
}
\subfloat[]{
\centering
\includegraphics[width=0.45\linewidth]{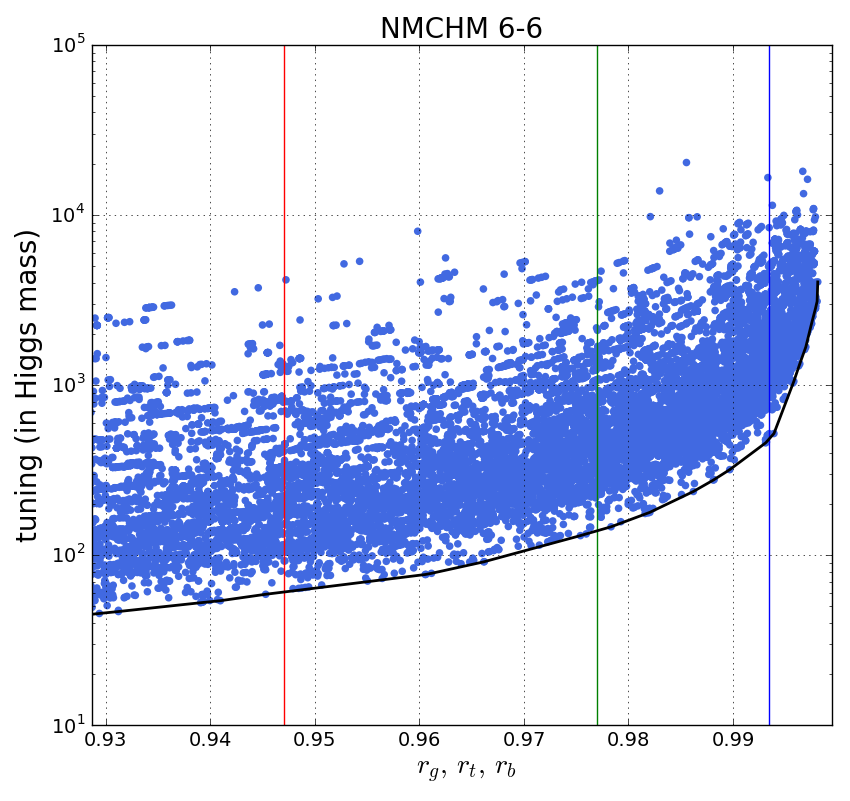}
}
\caption{\label{fig:first-tuning-higgs-vs-couplings} Comparison of the first-order tuning (as defined in \cref{eq:firstorder}) contribution from the Higgs mass, in the Higgs-gluon, -top and -bottom coupling deviation. The red line shows the expected precision of a $250$ GeV ILC, green a $500$ GeV ILC, and blue a high-luminosity $1$ TeV ILC.}
\end{figure}

\begin{figure}
\centering
\subfloat[]{
\centering
\includegraphics[width=0.45\linewidth]{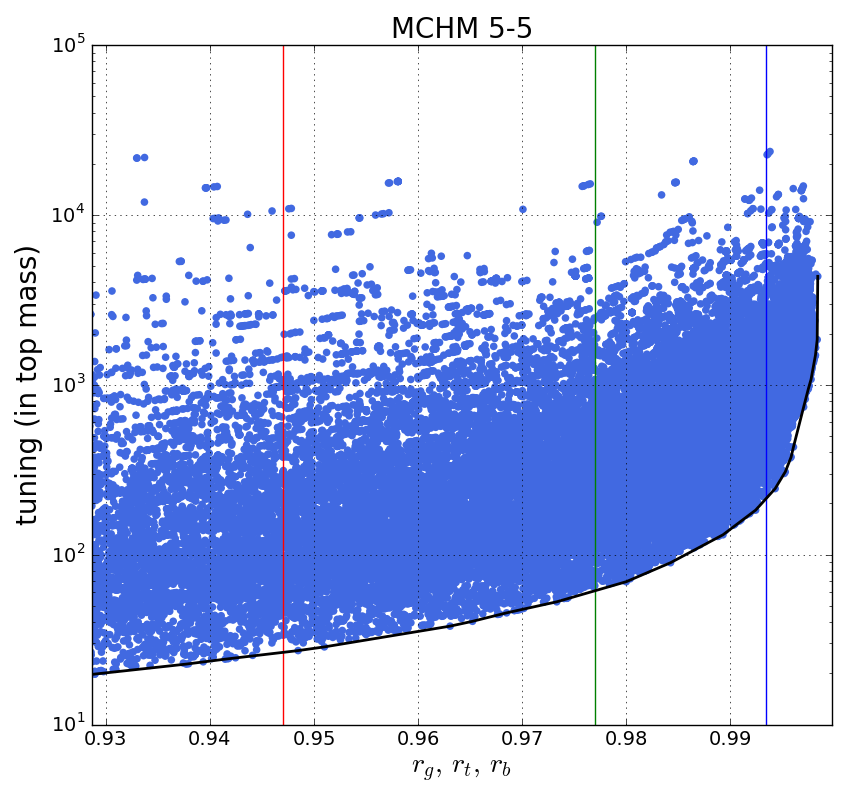}
}
\subfloat[]{
\centering
\includegraphics[width=0.45\linewidth]{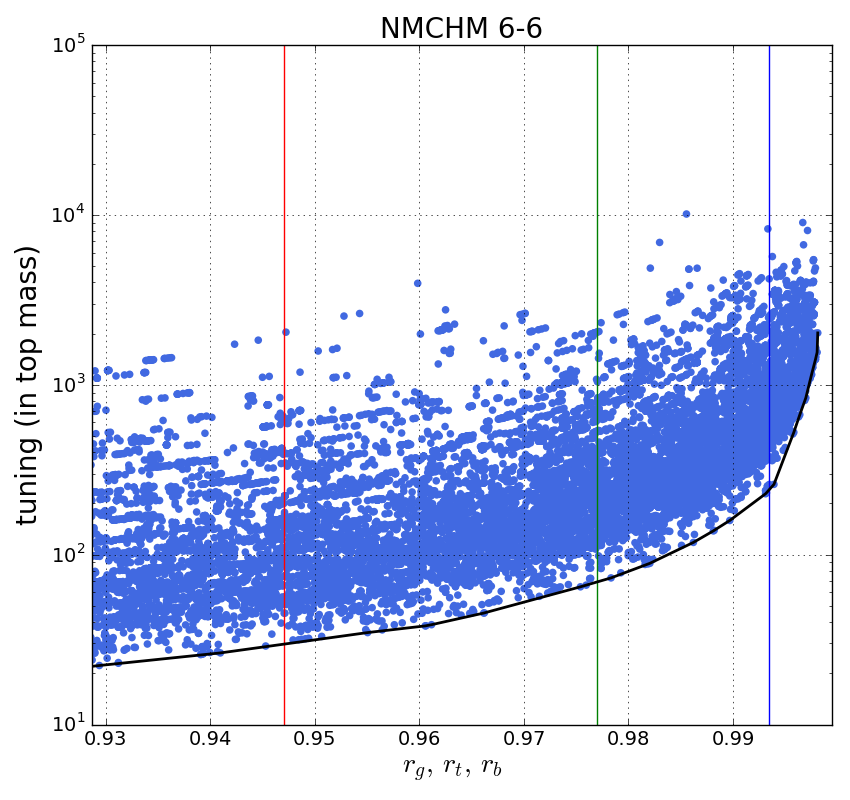}
}
\caption{\label{fig:first-tuning-top-vs-couplings} Comparison of the first-order tuning contribution from the top mass, in the Higgs-gluon, -top and -bottom coupling deviation. The red line shows the expected precision of a $250$ GeV ILC, green a $500$ GeV ILC, and blue a high-luminosity $1$ TeV ILC.}
\end{figure}

\begin{figure}
\centering
\subfloat[]{
\centering
\includegraphics[width=0.45\linewidth]{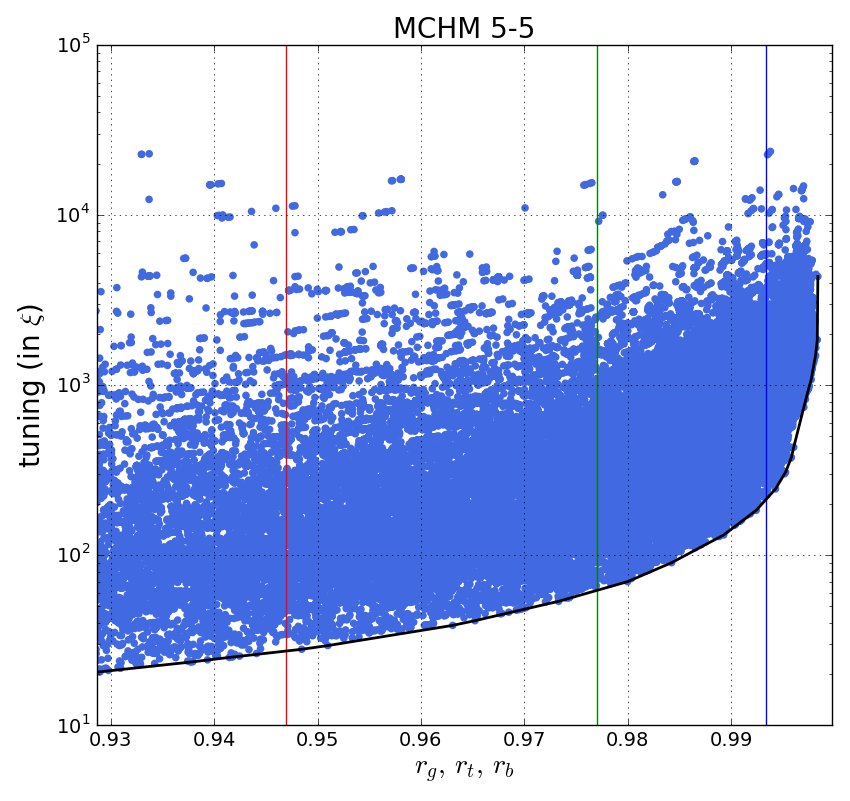}
}
\subfloat[]{
\centering
\includegraphics[width=0.45\linewidth]{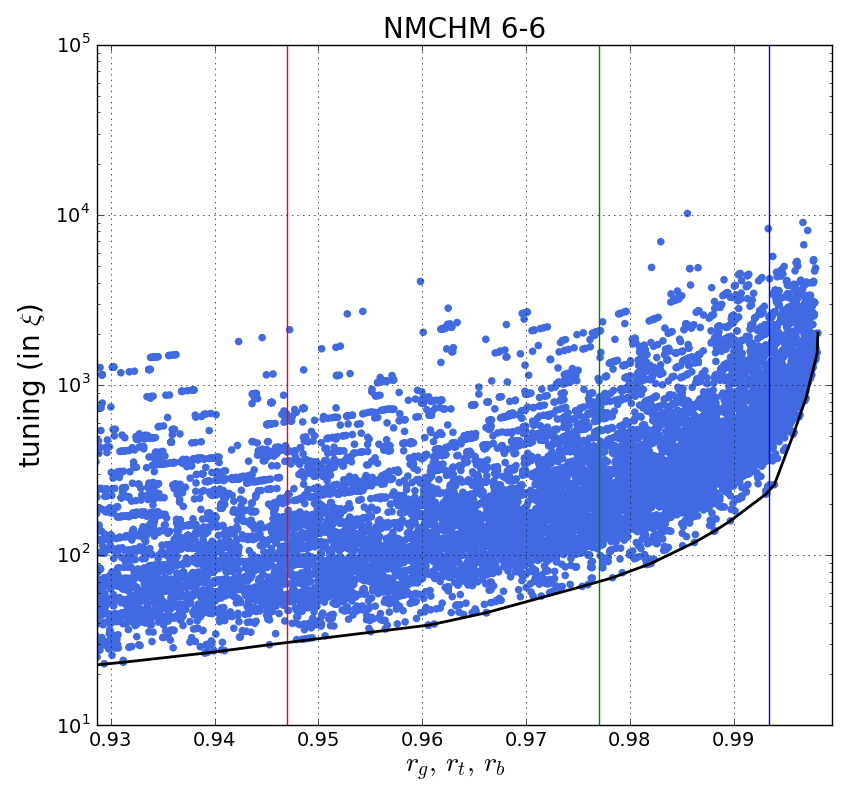}
}
\caption{\label{fig:first-tuning-vac-vs-couplings}Comparison of the first-order tuning contribution from the vacuum misalignment $\xi$, in the Higgs-gluon, -top and -bottom coupling deviation. The red line shows the expected precision of a $250$ GeV ILC, green a $500$ GeV ILC, and blue a high-luminosity $1$ TeV ILC.}
\end{figure}

To understand the different contributions to the higher-order tuning, we show in \cref{fig:first-tuning-higgs-vs-couplings} to \cref{fig:first-tuning-vac-vs-couplings} various first-order tuning contributions (defined in \cref{eq:firstorder}), again plotted as a function of the modification to the Higgs-gluon, Higgs-top and Higgs-bottom couplings. For any given value for the modification of the couplings, we see that the tuning contribution from the Higgs mass is higher than that arising from the top mass and vacuum misalignment contributions. This can be understood from the leading-order relationship between each observable. By \cref{eq:higgs_mass} and \cref{eq:top_mass}, for $\xi << 1$, $m_h \propto \xi^2$, while $m_t \propto \xi$ (recalling that $\frac{1}{f^2} = \frac{\xi}{v^2}$). Thus the first order tuning is expected to be $\nabla^{m_h} \sim 2\nabla^{m_t} \sim 2\nabla^{\xi}$, which agrees with \cref{fig:first-tuning-higgs-vs-couplings} to \cref{fig:first-tuning-vac-vs-couplings}. 

\begin{figure}
\centering
\subfloat[]{
\centering
\includegraphics[width=0.45\linewidth]{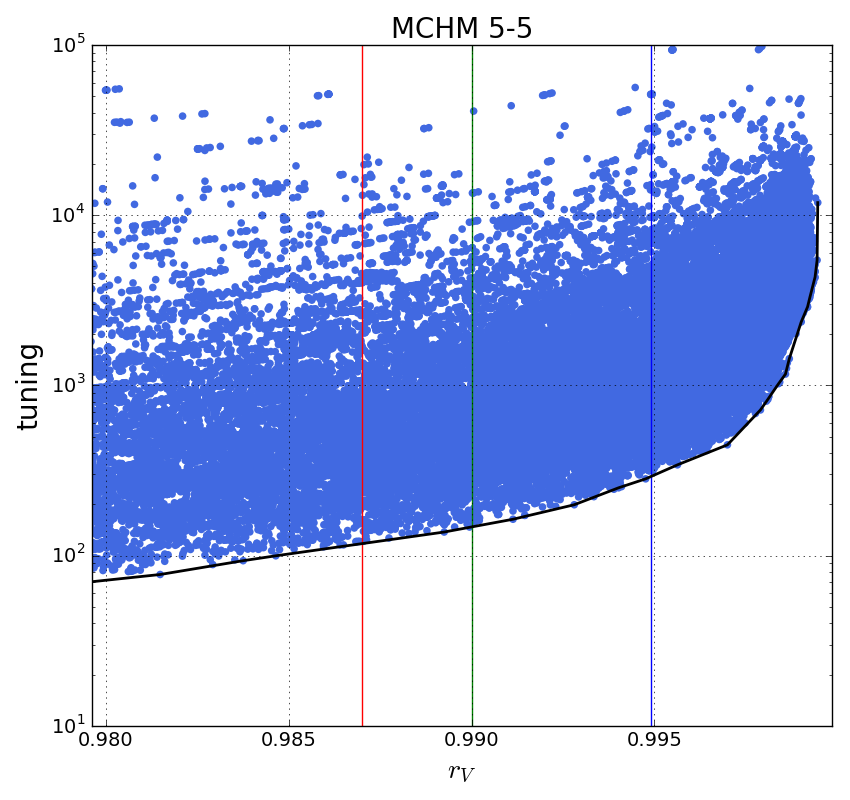}
}
\subfloat[]{
\centering
\includegraphics[width=0.45\linewidth]{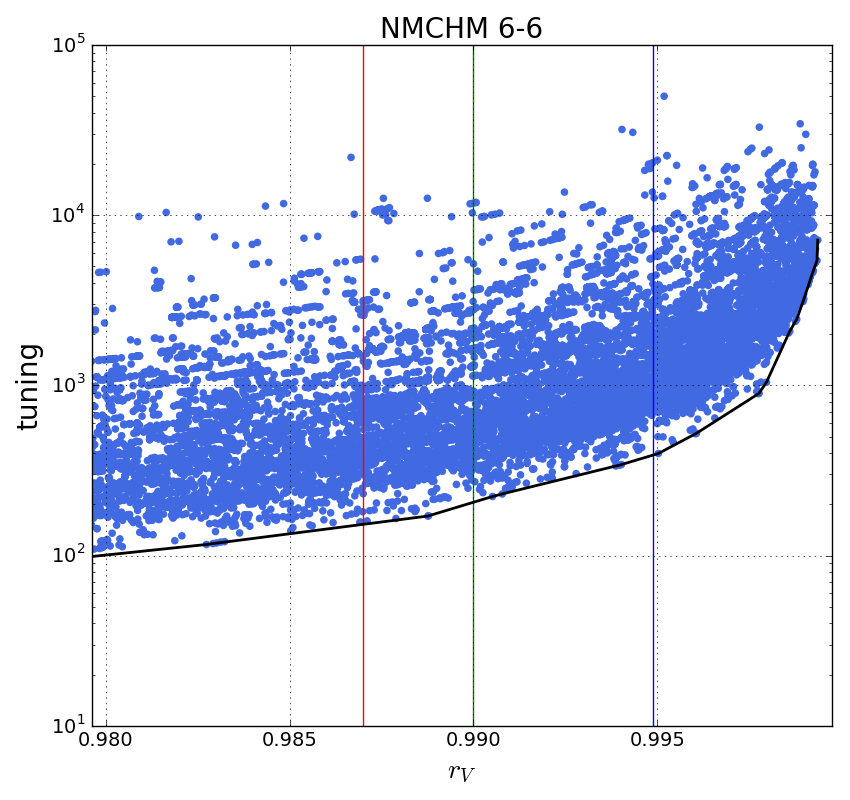}
}
\caption{Comparison of higher-order tuning in the Higgs-vector boson coupling deviation, between the minimal and next-to-minimal models. The red line shows the expected precision of a $250$ GeV ILC, green a $500$ GeV ILC, and blue a high-luminosity $1$ TeV ILC.}\label{fig:tuning-higgs-vs-V} 
\end{figure}

\begin{figure}
\centering
\subfloat[]{
\centering
\includegraphics[width=0.45\linewidth]{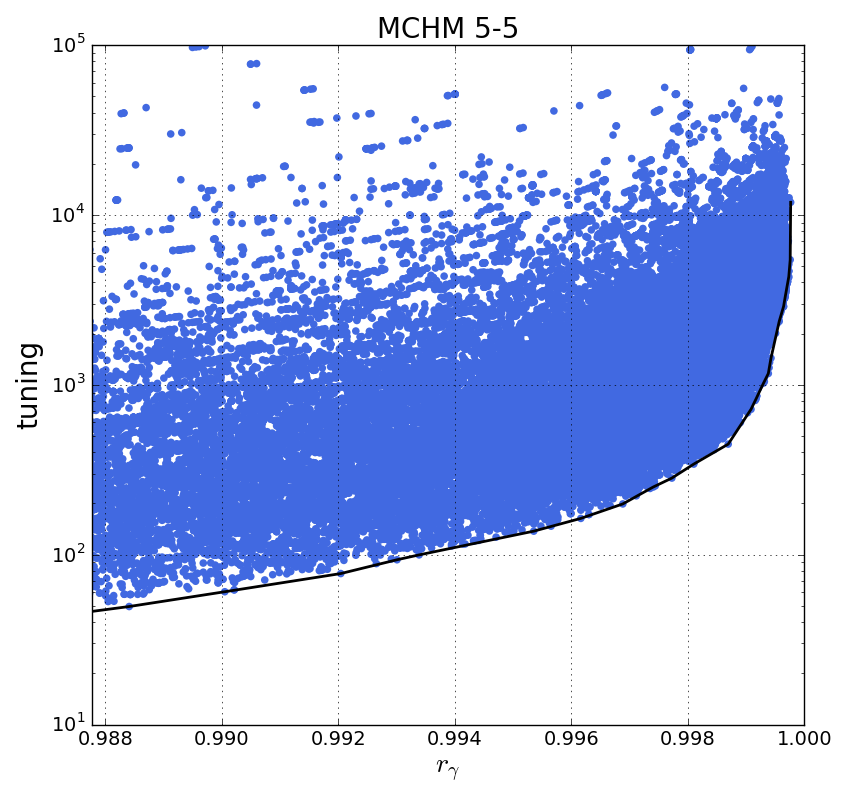}
}
\subfloat[]{
\centering
\includegraphics[width=0.45\linewidth]{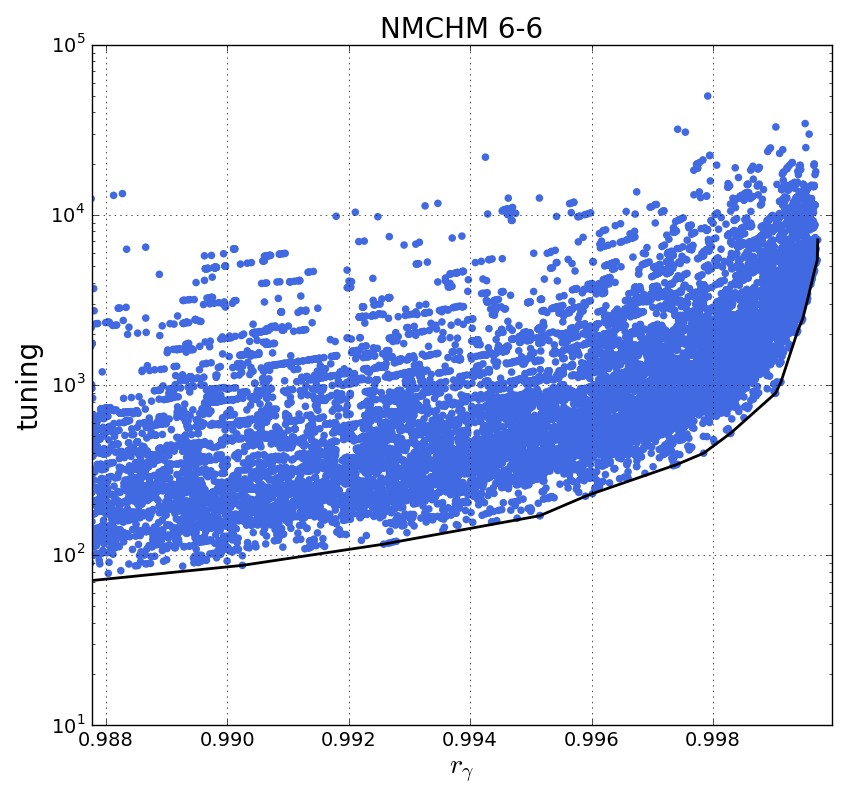}
}
\caption{Comparison of higher-order tuning in the Higgs-photon (loop) coupling deviation, between the minimal and next-to-minimal models. Future ILC bounds are below the cut-off $f>800$ GeV.}\label{fig:tuning-higgs-vs-gamma} 
\end{figure}

In \cref{fig:tuning-higgs-vs-V}, we show the higher-order tuning as a function of the deviation of the Higgs-vector boson couplings. In this case, the impact of the future linear collider is not as pronounced, with a less pronounced increase in the fine-tuning even after the anticipated results of the high-energy, high-luminosity ILC. The situation is even worse for the Higgs-photon coupling (shown in \cref{fig:tuning-higgs-vs-gamma}), where the relative lack of precision of ILC measurements of $r_\gamma$ relative to the other couplings means that there is essentially no impact on the fine-tuning of the model expected from future measurements. This tells us that it is future measurements of the Higgs-gluon, Higgs-top and Higgs-bottom couplings that will be most important in disfavouring composite Higgs scenarios on aesthetic grounds.

\begin{figure}
\centering
\subfloat[]{
\centering
\includegraphics[width=0.45\linewidth]{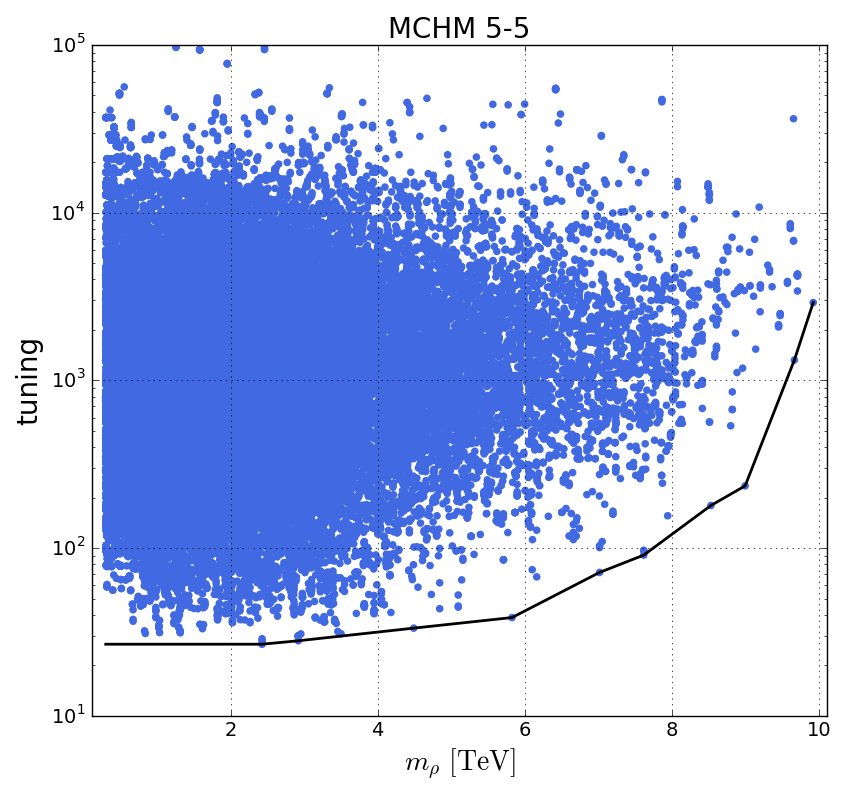}
}
\subfloat[]{
\centering
\includegraphics[width=0.45\linewidth]{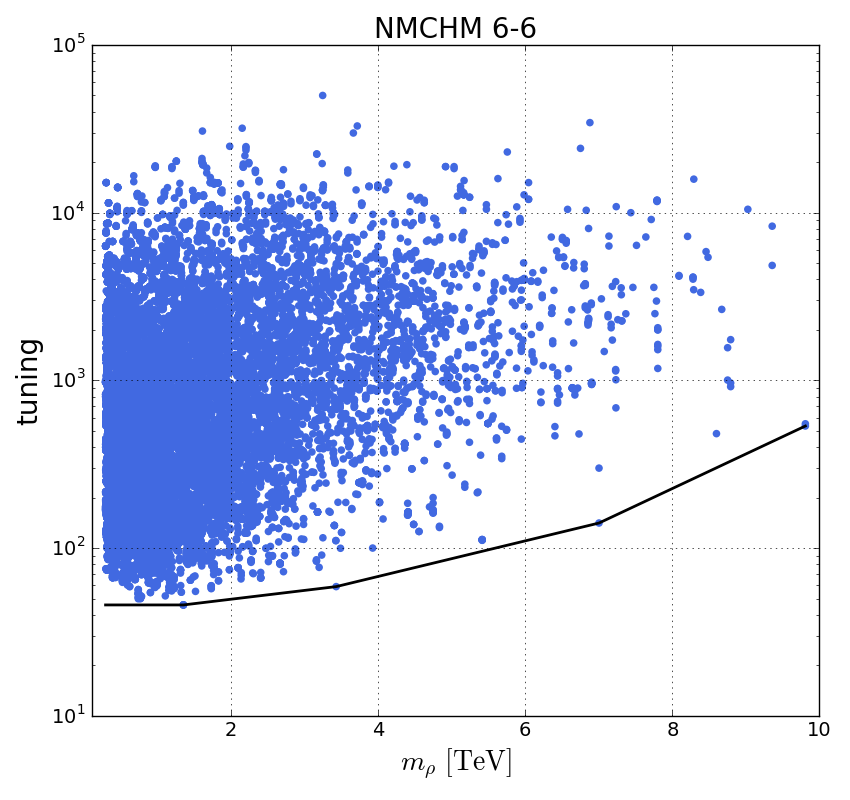}
}
\caption{Comparison of the lightest vector resonance mass vs higher-order tuning, between models.}\label{fig:tuning-vs-mrho}
\end{figure}

\begin{figure}
\centering
\subfloat[]{
\centering
\includegraphics[width=0.45\linewidth]{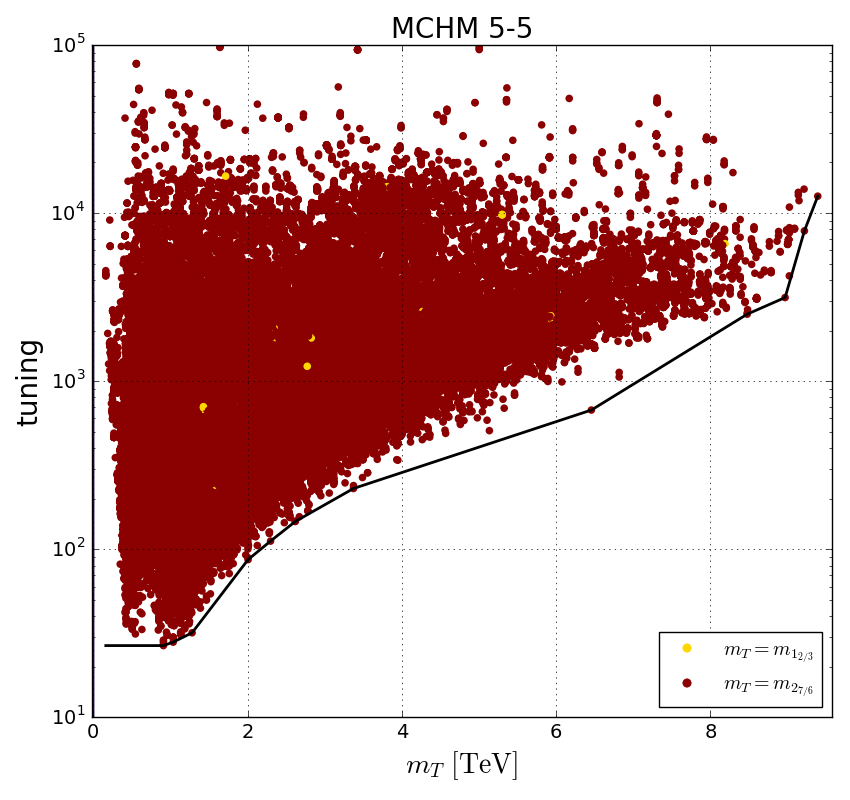}
}
\subfloat[]{
\centering
\includegraphics[width=0.45\linewidth]{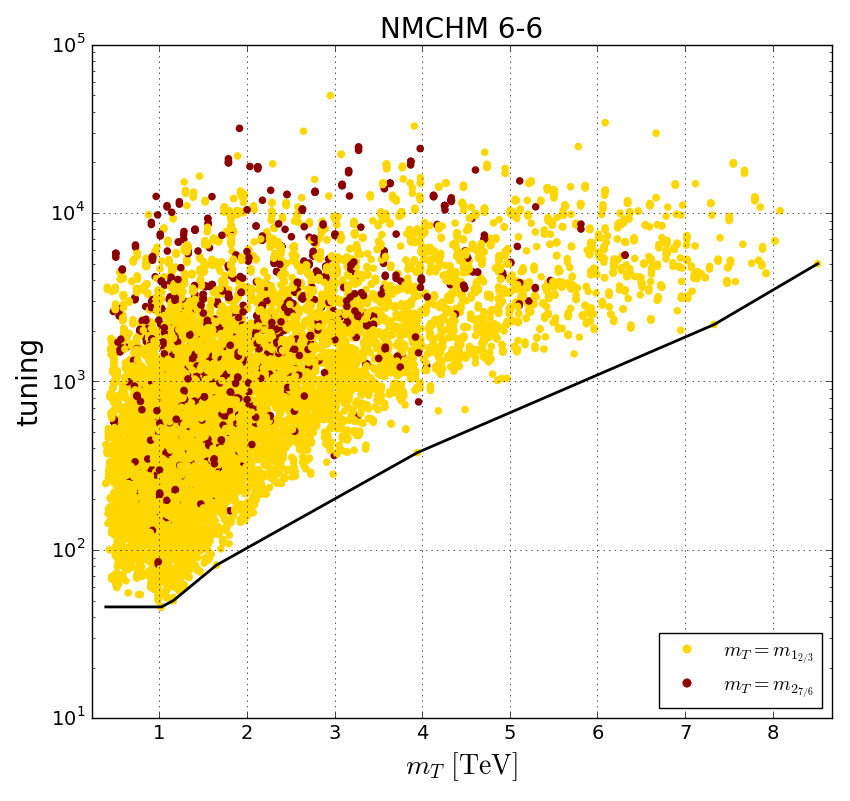}
}
\caption{Comparison of the lightest fermionic resonance mass vs higher-order tuning, between models. Note that the lightest resonance may be either the singlet (yellow) or doublet (maroon).}\label{fig:tuning-vs-mT} 
\end{figure}

Measurements of the Higgs couplings are, of course, only one way to constrain the fine-tuning of the M4DCHM and NM4DCHM. One can also search directly for the fermion and vector resonances. In \cref{fig:tuning-vs-mrho}, we show the higher order tuning as a function of the lightest vector resonance mass, $m_\rho$. A lower bound on this mass would translate directly into a lower bound on the fine-tuning. In this case, the rise in fine-tuning with an increasing lower mass limit is less pronounced for the NM4DCHM, although one would have to have a fairly stringent lower bound to make this difference significant. A steeper rise is apparent in the plots of higher order tuning vs top partner mass $m_T$ shown in \cref{fig:tuning-vs-mT}, although there is not much difference in the behaviour in the NM4DCHM relative to the M4DCHM. Lower limits of around 5 TeV and 9.5 TeV can be expected after 3000 fb$^{-1}$ of 33 TeV and 100 TeV collisions at a future proton--proton collider, respectively \cite{Gershtein:2013iqa,colliderReach,chala2018searches}, which will substantially increase the minimum fine-tuning of both the M4DCHM and NM4DCHM.

\begin{figure}
\centering
\subfloat[]{
\centering
\includegraphics[width=0.45\linewidth]{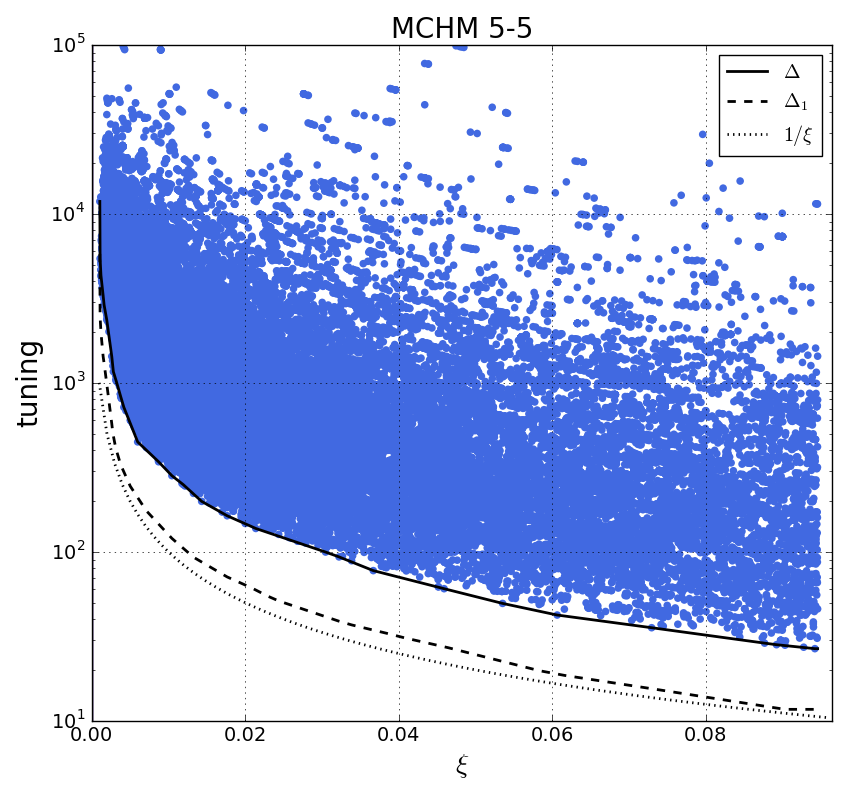}
}
\subfloat[]{
\centering
\includegraphics[width=0.45\linewidth]{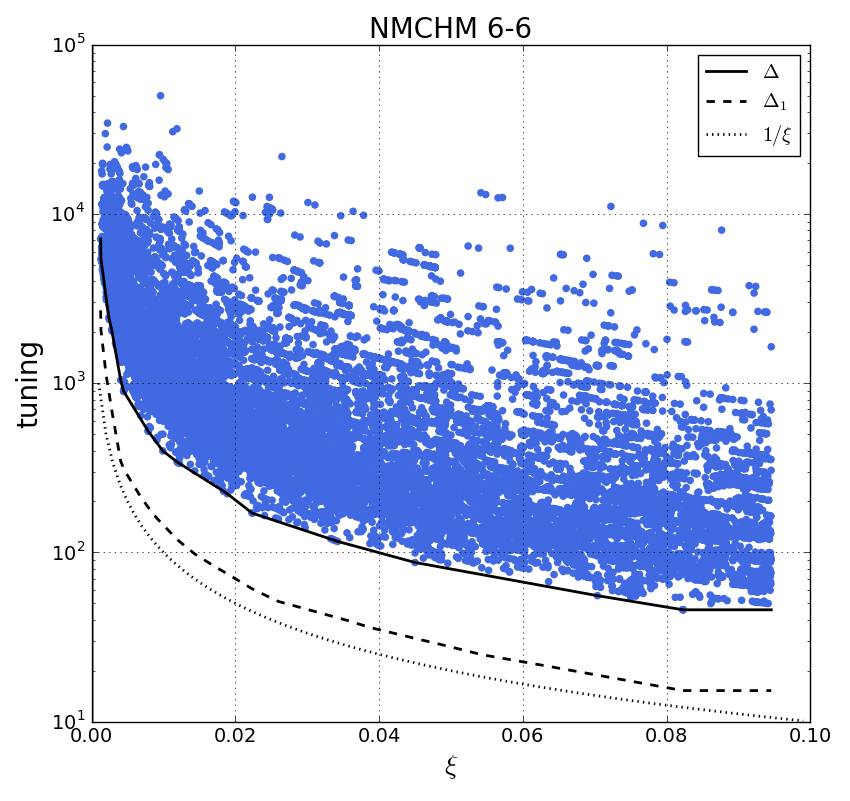}
}
\caption{Comparison of vacuum misalignment vs higher-order tuning, between models.}\label{fig:tuning_vs_xi}
\end{figure}

Finally, we show a comparison of our higher-order tuning measure with less sophisticated tuning measures in \cref{fig:tuning_vs_xi}, which shows the fine-tuning for the NM4DCHM as a function of the breaking scale ratio $\xi$. Our measure gives higher values for fine-tuning relative to the single tuning $\Delta_1$ as defined in \ref{eq:firstorder}, which is to be expected.

\begin{table} 
\begin{center}
\begin{tabular}{c | c | c | c}
Plot & $250\gev$ (red) & $500\gev$ (green) & $1\tev$ HL (blue) \\ \hline
$r_b$ & $5.3\%$ & $2.3\%$ & $0.66\%$\\
$r_Z$ & $1.3\%$ & $1.0\%$ & $0.51\%$\\
$r_\gamma$ & $18\%$ & $8.4\%$ & $2.4\%$
\end{tabular}
\end{center}
\caption{A selection of Higgs coupling deviation exclusion bounds, as predicted in \cite{tian2016measurement}. These are forecasts for ILC precision relative to the SM prediction.}\label{tab:ILC}
\end{table}
%

\section{Singlet Behaviour}

Before concluding,  let us briefly examine the behaviour of the singlet mass in our scan results. Apart from a dependence on the potential integral terms, it depends on the decay constant $f$ and the embedding angle of the right-handed top quark in the $SO(2)$ subgroup of $SO(6)$, $\theta$. We see that there is a critical point determined by the $\cos 2\theta$ factor in \cref{eq:singlet_mass}, with the limits
\begin{align}
m_S^2 \rightarrow \begin{cases}
-\frac{c_1 - c_2 + c_3}{f^2}, & \text{ as } \sin\theta \rightarrow 1\\
0, & \text{ as } \sin\theta \rightarrow \frac{\sqrt{2}}{2} \approx 0.7
\end{cases}
\end{align}
The zero mass case corresponds to the right top embedding being $SO(2)$ symmetrical, leaving this group unbroken and the singlet as a true NGB. It has been shown in reference \cite{gripaios2009beyond} that two-loop contributions from the gauge sector will still give the singlet a small mass, appearing as an electroweak axion. This would be ruled out by experiment.

The $\sin\theta = 1$ limiting case is more interesting. Here, the elementary top quark does not couple with the singlet eigenstate, and \cref{eq:gauge_lagrangian} and \cref{eq:fermion_lagrangian} become (considering only the subset of terms containing the $\eta$ field)
\begin{align}
\mathcal{L}_{\eta} \; \xrightarrow{\theta\rightarrow \pi/2} \; & (\partial \eta)^2 + \frac{g^2 f^2}{4} \sin^2 \frac{\varphi}{f} \cos^2 \frac{\eta}{f} W^2 \nonumber\\
&\stackrel{\eta \rightarrow -\eta}{=} \mathcal{L}_{\eta}
\end{align}

This $Z_2(\eta)$ symmetry is explored in reference \cite{Frigerio:2012uc}, where it is simply assumed. It requires all interactions to preserve $s$-number, which protects the scalar singlet from decay hence making it a suitable candidate for dark matter. In that work, the authors consider four regions of interest: low mass ($m_S < 50\gev$), resonant ($m_S \approx m_H/2$), cancellation ($m_S \sim \sqrt{\frac{\lambda}{2}} f$) and high mass ($m_S >> \sqrt{\frac{\lambda}{2}} f$). Here, $\lambda$ is the four-point coupling of $\eta,\varphi$, appearing in \cref{eq:NMCHM_potential}. In our notation, $\lambda \rightarrow c_1 - c_3$, since
\begin{align}
V(\eta,\varphi) \xrightarrow{s_\varphi^2,s_\eta^2 << 1} (c_1 + c_2 -c_3) \varphi^2 - (c_1 - c_3) \varphi^2 \eta^2 + c_3 \varphi^4 - c_3 \varphi^4 \eta^2
\end{align} 

\begin{figure}
\centering
\subfloat[]{
\centering
\includegraphics[width=0.45\linewidth]{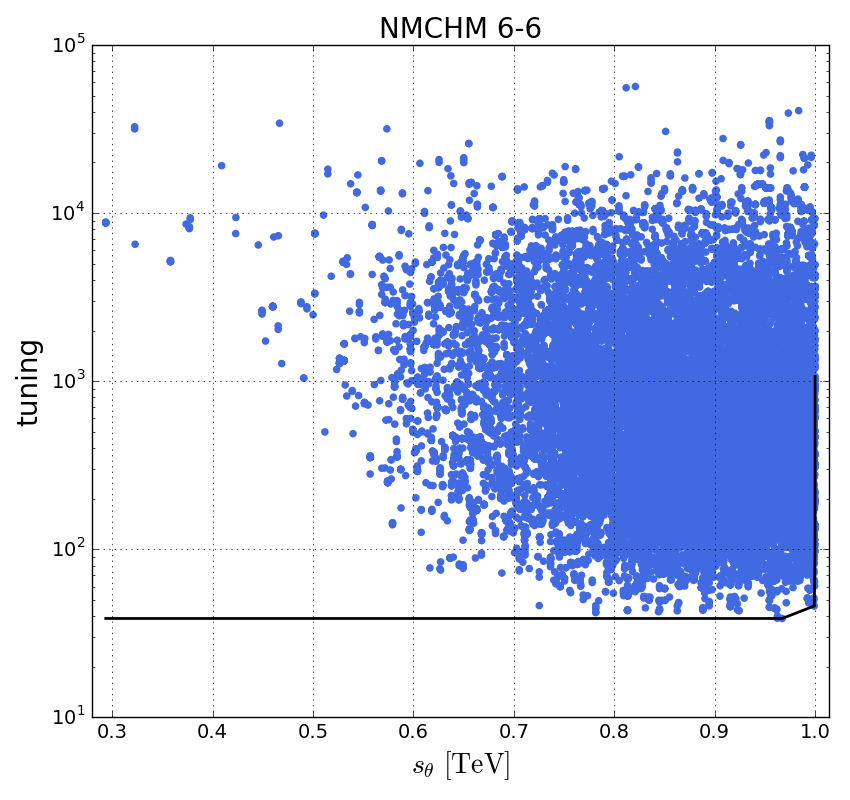}
}
\subfloat[]{
\centering
\includegraphics[width=0.45\linewidth]{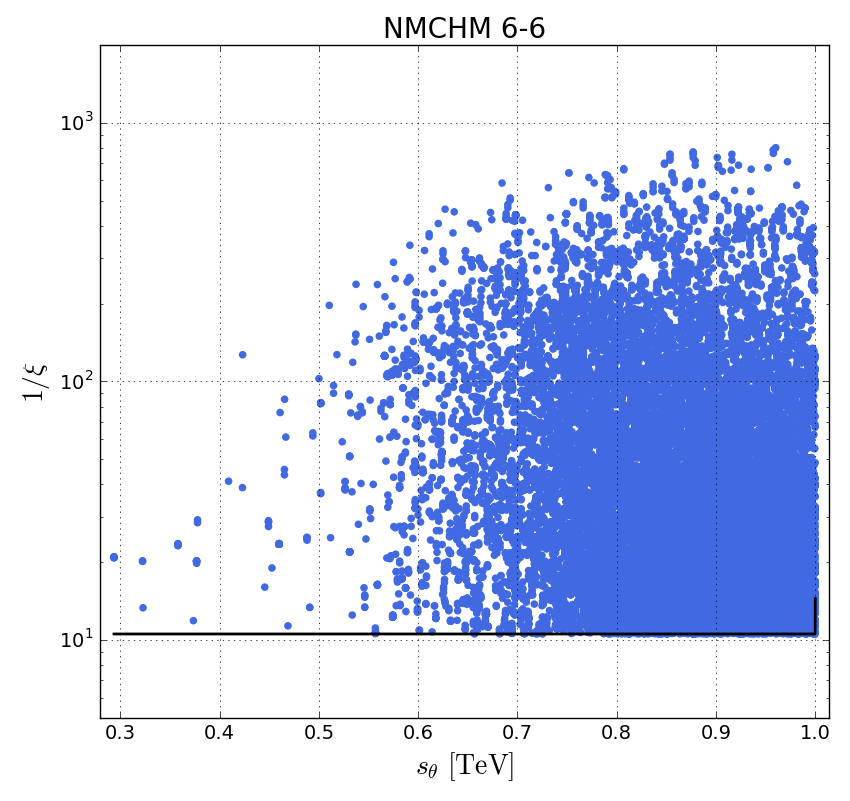}
}
\caption{The top quark mixing parameter $\sin\theta$ vs (top row left) higher order tuning and (top row right) naive tuning.}\label{fig:theta_vs_tuning}
\end{figure}

\begin{figure}
\centering
\subfloat[]{
\centering
\includegraphics[width=0.6\linewidth]{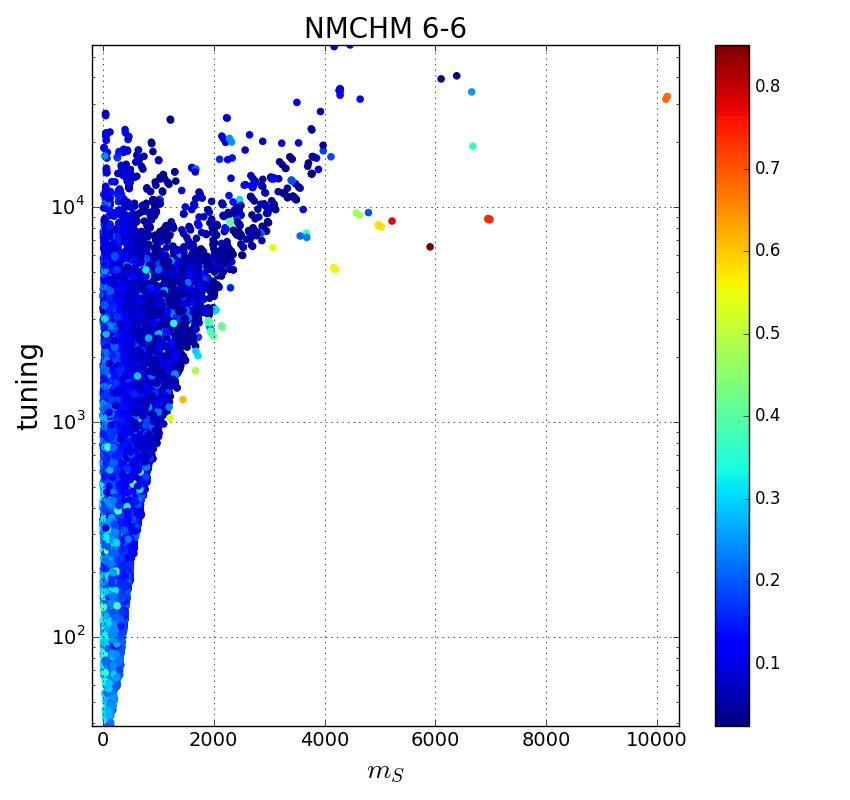}
}
\caption{Mass of the singlet in GeV, with singlet-quark coupling deviation (as defined in \cref{eq:coupling_deviation}) as the third dimension}\label{fig:mSvscoupling}
\end{figure}

In \cref{fig:theta_vs_tuning}, we show both our higher order fine-tuning measure, and the naive measure $1/\xi$, vs $\sin\theta$ for our selected scan points. We see that the NM4DCHM provides points with low fine tuning even as $\sin\theta\rightarrow 1$, and hence a dark matter candidate can easily emerge naturally within this framework. In \cref{fig:mSvscoupling}, we show our higher-order tuning measure vs the singlet mass, with the deviation of the singlet couplings to quarks and gluons from SM Higgs-like couplings shown on the $z$-axis (this deviation is defined in \cref{eq:coupling_deviation}). Higher values on the $z$-axis correspond to a stronger coupling between the singlet and quarks and gluons. We see that obtaining couplings as high as the SM Higgs requires a fine-tuning that is two orders of magnitude greater than the most natural coupling scenario of small coupling.

It is instructive to separate our scan points into the region that has $\theta < \pi/4$,  and that which has $\theta > \pi/4$. Moving from one choice to the other requires a change in the sign of the $c_1 - (c_2 + c_3)s_\theta^2$ term to guarantee a real singlet mass. Specifically, by removing $f$ as a factor in the mass term using the solution for $\xi$, we get the regions in terms of only the integral expressions
\begin{align}
\text{Region 1: } && \theta \in \{0,\pi/4\}, && \implies && \frac{c_2^2 s_\theta^4 - (c_1 - c_3 s_\theta)^2}{2 c_3} > 0\nonumber \\
\text{Region 2: } && \theta \in \{\pi/4, \pi/2\}, && \implies && \frac{c_2^2 s_\theta^4 - (c_1 - c_3 s_\theta)^2}{2 c_3} < 0\label{eq:regions}
\end{align} 

\begin{figure}
\centering
\subfloat[]{
\centering
\includegraphics[width=0.45\linewidth]{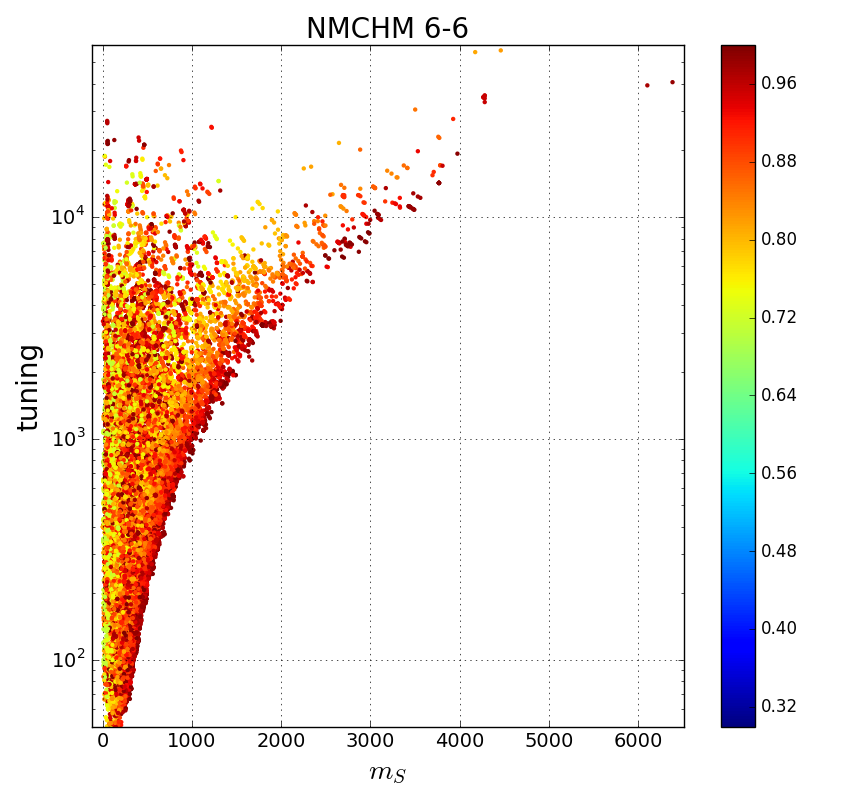}\label{fig:mS_vs_tuning_R2}
}
\subfloat[]{
\centering
\includegraphics[width=0.45\linewidth]{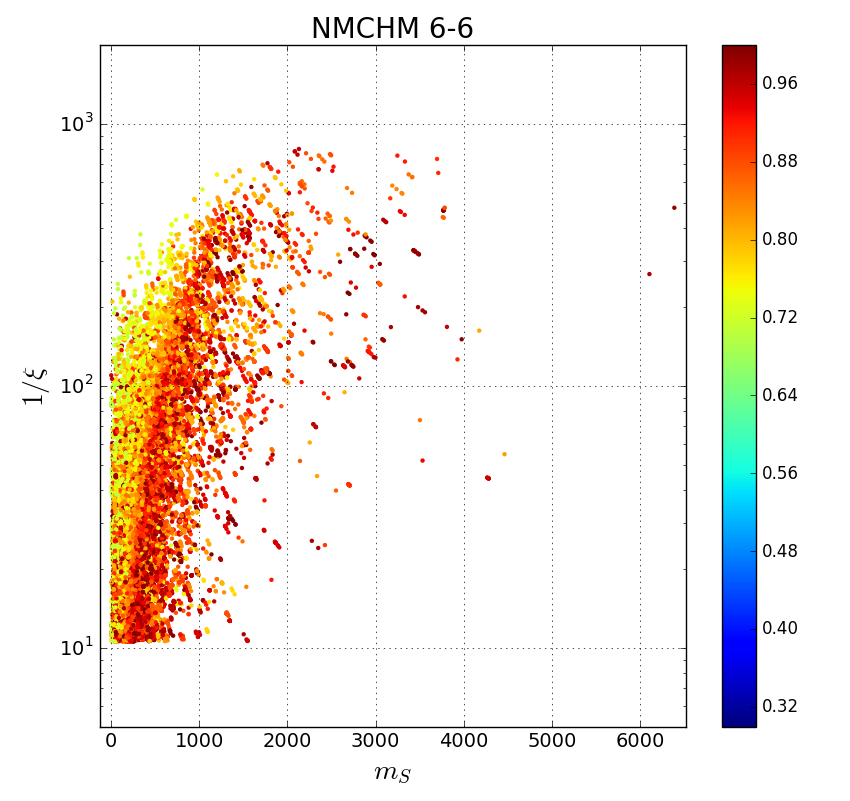}\label{fig:mS_vs_xi_R2}
}
\caption{The singlet mass (in GeV) vs (a) higher order tuning and (b) naive tuning with $\sin\theta$ as the third dimension, for points with $\theta \in \{\pi/4, \pi/2\}$.\label{fig:mS-R2}}
\end{figure}

\begin{figure}
\centering
\subfloat[]{
\centering
\includegraphics[width=0.45\linewidth]{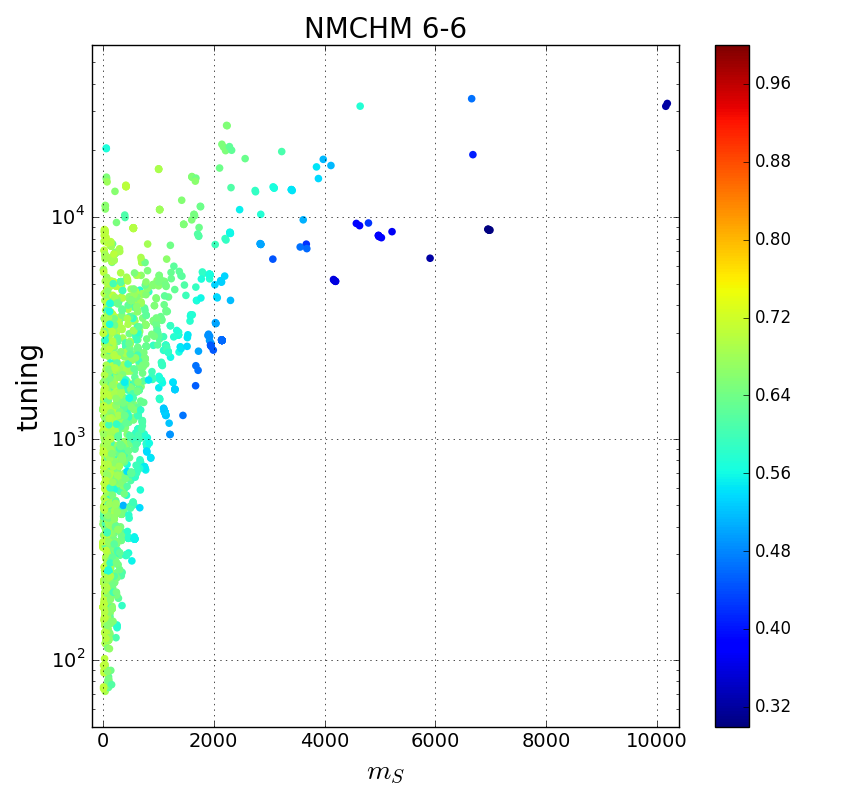}\label{fig:mS_vs_tuning1}
}
\subfloat[]{
\centering
\includegraphics[width=0.45\linewidth]{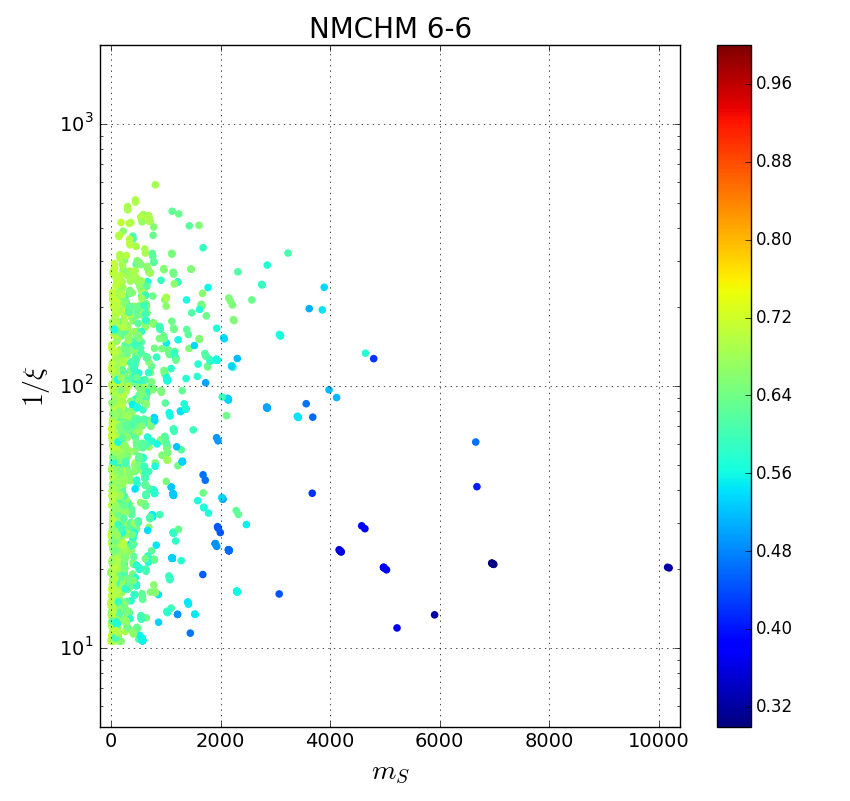}\label{fig:mS_vs_xi1}
}
\caption{The singlet mass (in GeV) vs (a) higher order tuning and (b) naive tuning with $\sin\theta$ as the third dimension, for points with $ \theta \in \{0,\pi/4\}$.\label{fig:mS-R1}}
\end{figure}

In \cref{fig:mS-R2}, we show our higher-order tuning measure, and the naive tuning measure, vs $m_S$ for points with $\theta \in \{\pi/4, \pi/2\}$, indicating that the points of lowest tuning have sin$\theta$ values close to 1. This implies that the $Z_2$ symmetry exists to stabilise a dark matter candidate. Equivalent plots for our $ \theta \in \{0,\pi/4\}$ points are shown in \cref{fig:mS-R1}, in which the contour of lowest fine tuning now exists such as to minimise sin$\theta$. In both cases, the features are pronounced only when considering the higher-order tuning measure which counts multiple contributions to the total fine-tuning properly. We note that the lowest fine-tuning overall is usually encountered for points with $ \theta \in \{0,\pi/4\}$, but that there is not a large difference between the overall fine-tuning vs mass for the two $\theta$ regions. A final note regarding the dark matter candidacy of the singlet; the natural limit of $\sin\theta \approx 1$ suggested by \cref{fig:mS-R2} is based only on considerations of SM mass values. It is not an indication of fine tuning based on cosmological values. Indeed, to achieve the correct relic density of the DM candidate, one may need to be arbitrarily close to the $Z_2$ limit. In this sense, enforcing the limit could be considered a separate source of fine tuning. It is beyond the scope of this paper to provide relic density limits on the singlet-fermion coupling terms. Suffice it to say that given the effective next-to-minimal model, particularly for higher singlet masses, the $\sin\theta \approx 1$ region is preferred by particle mass tuning considerations, and one would be well-motivated to search for UV completions that included this $Z_2$ symmetry explicitly.

\begin{figure}
\centering
\subfloat[]{
\centering
\includegraphics[width=0.45\linewidth]{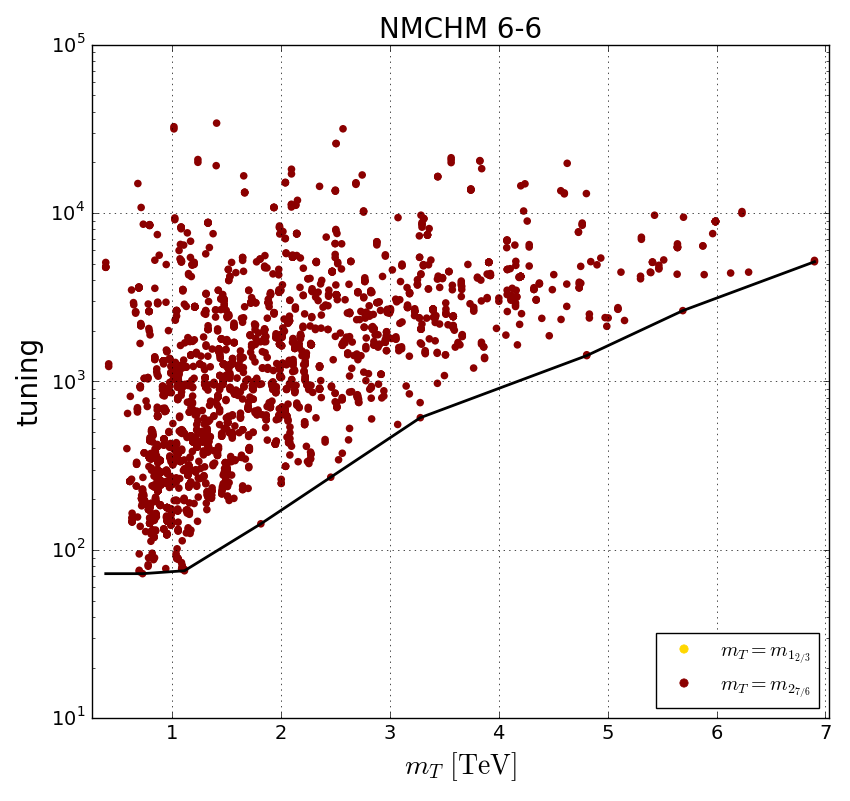}
}
\subfloat[]{
\centering
\includegraphics[width=0.45\linewidth]{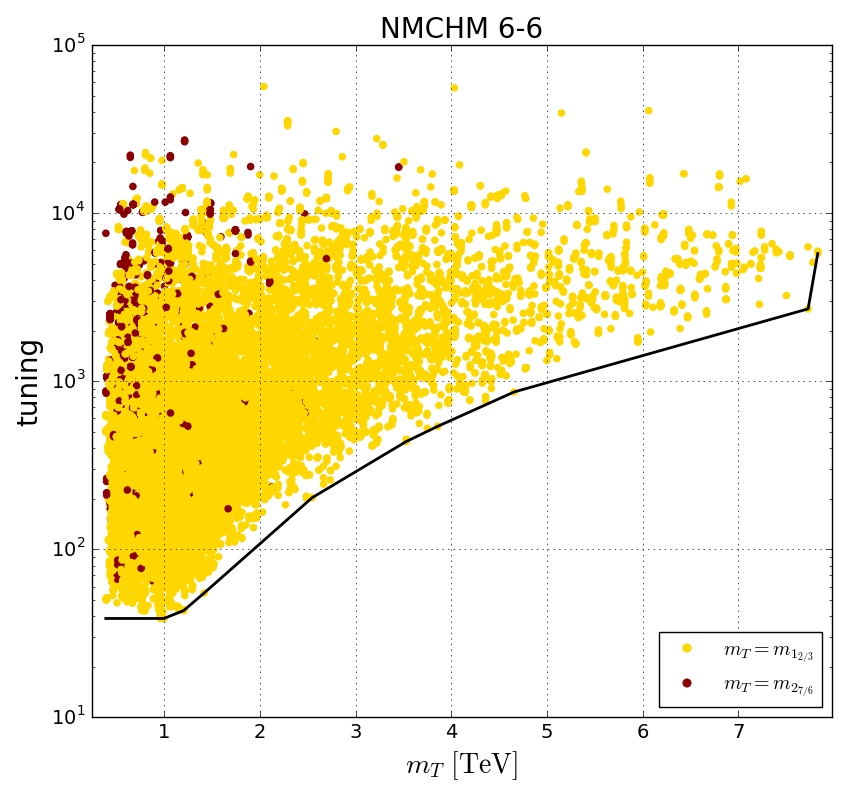}
}
\caption{Top partner masses vs. full tuning, broken into region 1 (left) and region 2 (right), as defined by \cref{eq:regions}.}\label{fig:Tvsth}
\end{figure}

In \cref{fig:Tvsth}, we show the higher-order tuning vs the lightest top partner mass, showing by the colour of each point which of the two top partners is the lightest. The left-hand plot contains only the points with $ \theta \in \{0,\pi/4\}$, whilst the right-hand plot shows the points with $\theta \in \{\pi/4, \pi/2\}$. Our results suggest that a collider observation of a lightest top partner with hypercharge $2/3$ will always allow the identification $\theta \in \{\pi/4, \pi/2\}$ under the assumption that the NM4DCHM is a valid explanation, whereas any observation of the hypercharge will allow the identification of the $\theta$ region for a lightest top partner mass in excess of $3.5\tev$. In turn, this would allow one to infer the singlet's phenomenology, if one were to construct the model with the minimum fine-tuning.

\chapter{Conclusion}
\label{conclusion}

A universe governed by a finely tuned set of rules is not an impossible concept. There are, indeed, encouraging results from string theory that suggest our local universe is but one configuration of quadrillions of other, extant, universes across some multi-dimensional multiverse. Some of these support galactic and chemical structure - and therefore perhaps life, and some do not. This may seem an esoteric possibility, but it is this same consideration of avoiding an easy appeal to the anthropic principle that leads us to question the naturalness of the Standard Model. The fine tuning problem of the SM may be solved in the distant future by some unified theory, either by explaining away the tuning as one possibility of quadrillions, or by proposing a natural mechanism that appears as tuning in our effective SM. A boulder resting on a needle, as our SM Higgs sector appears to be. But to some higher scale theory, it may be explained by the erosion of rock, until a precarious boulder seems like the natural situation. 

The discovery of new physics at a higher scale may assuage the large hierarchy problem, but there remains a small hierarchy problem, between the EW scale and the scale of the new physics. As convincingly explained in \cite{Panico:2015jxa}, a cutoff of the SM much above the TeV scale would lead to an experimental catastrophe, if not a theoretical one. The precision required to probe the microscopic properties of the Higgs quickly becomes impractical beyond a TeV-cutoff. Each run of LHC searches places stronger constraints on this cutoff, such that we are now sandwiched by naturalness - required by the large hierarchy problem to design a BSM theory above the EW scale, and pushed down by the small hierarchy problem to no more than TeV scale if we want to understand the Higgs properties as anything more than input parameters. Both hierarchies are the result of cancellations, and particularly cancellations from multiple sources and combinations of parameters and observables. It behoves us to quantify these finely tuned cancellations given their motivation as the first slide in most BSM seminars.

In this work, we have attempted to address the above concerns. In \cref{chp:standard_model} and \cref{sec:composite_higgs} we begin with an understanding of classical field theory and the SM field content, and develop the notion of non-linearly realised symmetries in a classical field Lagrangian and how it connects to the existence of massless degrees of freedom. We examine the Higgs mechanism, and how misaligning a gauge group and a linear global subgroup at the classical level allows gauge bosons to gain a longitudinal polarisation. Ostensibly, this is as far as we can go in classical field theory. Thus, we introduce the 1-loop Coleman-Weinberg (CW) process as the quantum correction, and we are then able to radiatively generate vacuum misalignment. This allows us to escape the usual Higgs mechanism by explicitly breaking the linear global subgroup at a scale $v$ naturally below the new physics breaking scale $f$. We use this phenomenon as the foundation of all Composite Higgs models in the remainder of the thesis. Indeed, while we use the tools of symmetric spaces and hidden local symmetry to build more complex N-site models, it is the CCWZ formalism combined with the CW process that captures the physics of the Composite Higgs field itself.

We also introduce several phenomena that we expect to follow from a confined strong sector, such as composite matter - a new set of heavy baryons and mesons that should emerge from the composite sites. We show that the natural way to include baryons is with linear coupling to the elementary fermions, leading to partial compositeness. The coupling structure between elementary and composite fermions, inter-site link fields, and the N$^\text{th}$-site NGBs, along with the choice of representations, completely define the matter sector. This allows us to derive the effective field theory's form factors. These appear as corrections to the couplings between SM fields, dependent on the parameters of the high-energy composite Lagrangian. We are then equipped to fit experimental observables by scanning and fine-tuning the input parameters. 

From \cref{sec:fine_tuning} we explored the historical motivation for considerations of naturalness. We were led, both by intuition and Bayesian reasoning, to the measure of Higher Order Tuning (HOT). There were several bugs/features of this measure that ultimately stemmed from how the Bayesian sensitivity limits to zero-informativity in the case of some observable being redundant. We suggested several extensions to HOT that could normalise these features. The first was a generic scaling of the measure with the number of parameters and observables. We determine an analytical function for this scaling, and suggest it could be used as a normalisation for the measure. A more consistent solution is to orthogonalise the observable space. This also solves the other feature - dependency of the measure on the multiplicity of tuning vector configurations. This is not rigorously accounted for in the current HOT. There is a clear connection between the linear dependence of a model's observable tuning vector space and the informativity of the Bayesian sensitivity. Further work should be done to establish this connection, and develop a \textit{more} rigorous HOT that includes some element of observable tuning orthogonalisation. However, it is likely that a completely consistent calculation of an orthogonal observable space (for example using a Gram-Schmidt process) is computationally impractical. Although not done in this work, it may serve as useful to pick a sample of observable points to fully orthogonalise as a benchmark to ensure the HOT approximation is sound. Further results on this topic are yet to be published.

To explore the large space of parameters of the Composite Higgs paradigm, we detailed two powerful computational tools in \cref{sec:scanning_techniques}. The \texttt{Multinest} package uses the convenient parameter point ordering already required for nested sampling to numerically calculate the Bayesian evidence of a model. We did not directly use the evidence in this work, as its purpose is comparison \textit{between} models. The \texttt{Multinest} package is used in earlier studies of this work, while later studies did not result in the Bayesian evidence being calculated. Instead, we use the lowest points of tuning to compare between models, which \textit{is} related to the Bayesian sensitivity and evidence, as mentioned above. Later studies instead use the \texttt{Diver} package, a genetic algorithm that uses the difference of two random vectors, in addition to two parent vectors, to mutate each offspring in the parameter space. The differential evolution technique has been shown in international competitions to outpace other algorithms in optimisation challenges, and we found superior performance in our scans. The \texttt{Diver} package enabled the first convergent global fit study of the Minimal 4D Composite Higgs model.

Chapters \ref{sec:M4DCHM}, \ref{sec:LCHM} and \ref{sec:NMCHM} outline the specific models, methods and results obtained from studies of three extensions to the work originally performed in \cite{Barnard:2015ryq}. These extensions are, respectively: a full global fit of the M4DCHM, inclusion of multiple representations and leptons in the LM4DCHM, and an enlarged symmetry group, called NM4DCHM. The model presented in \cref{sec:M4DCHM} is the simplest Composite Higgs coset that preserves custodial symmetry and therefore suppresses FCNC. We apply the \texttt{Diver} package to the large parameter space in this model, subject to a comprehensive set of direct search and precision constraints. These include reproducing the SM particle masses, minimising deviation from the experimental Higgs and Z-boson couplings, calculation of the oblique EW observables, and constraints on the minimum excluded masses of vector/fermion resonances at LEP, Tevatron and the LHC. 

The constraints were applied with a hard $3 \sigma$ confidence cut, and the remaining points place updated bounds on the search for new physics. In particular, the symmetry breaking scale is excluded below $800\gev$, with the most natural breaking scale around $1000\gev$. We also report that we do not find viable points for composite fermion or vector boson masses below $1000\gev$. The lightest of these should be the partners of the bottom SM quark, and the SM Z boson, although for a given partner mass $m > 1100\gev$ the top partner is naturally the lightest.

The minimal model is extended in the work of \cref{sec:LCHM}. Here we detail how one may be motivated to consider combinations of different fermion embeddings to enact cancellations of sources of tuning. We use the \textbf{5-5-5} embedding (of the lepton doublet, tau lepton, and tau neutrino, respectively) as a benchmark to compare the tuning of the \textbf{14-14-10} and \textbf{14-1-10} embeddings. In particular, we examine the validity of the claim that coupling the elementary quarks to composite partners in the fundamental \textbf{5}, and coupling the elementary leptons to composite partners in the fundamental \textbf{14}, leads to a more natural model. We test this claim with our HOT measure, to properly account for the multiple sources of tuning. This is important, as the tuning that is reduced by the cancellation of the \textbf{5} and \textbf{14}, may be counter to the tuning that is \textit{increased} by adding a larger particle content. Indeed, we find that, although the tuning is lowest in the configuration suggested by \cite{carmona2015}, it remains comfortably within an order of magnitude of the other configurations.

In \cref{sec:NMCHM}, we then test the claim in recent work that the next-to-minimal Composite Higgs model (NM4DCHM) may lead to naturally higher resonance masses. This is an important phenomenological feature, to avoid the strong search constraints that currently exist, and which will likely be improved by future colliders. Some of these future search limits are included in this chapter's work. We apply the same measure as before to the next-to-minimal model and find that it does not have a significantly lower tuning for a fixed lightest top partner (LTP) resonance mass, or a significantly higher LTP mass for a fixed tuning. However, it should be noted that we use a simplification in this work that renders only one non-zero scalar vacuum expectation. It may be that this is the contributing factor to higher LTP masses, and in future work we will include a more comprehensive scalar potential sector. This will, of course, come at an increased computational cost, but the exploration performed in this work should improve further scans by only focussing on parameter volumes of interest. The scalar sector also includes a scalar singlet as a potential dark matter (DM) candidate. We analyse the properties of this scalar singlet, including the tuning of its couplings and mass. We determine that the region of parameter space leading to it being a viable DM candidate is one of low tuning. It is then well-motivated to consider UV completions of the NM4DCHM (for example, strongly coupled sectors of $SU(4) \rightarrow Sp(4)$ breaking) that include a remnant $Z_2$ symmetry protecting the stability of this singlet.

Some cursory Google searches reveal that the Devil's Marbles of the \textit{Karlu Karlu} Conservation Reserve are known as a \textit{degraded nubbin}. Stress cracks in granite form underground cubes which, as the top layer of earth is eroded, peak above the ground. The ground and cubes are subsequently weathered until they are round (weathering is more mechanically stressful for edges and vertices) and appear to be balancing on the layer of granite below. Which they certainly are, but now we have a \textit{natural} explanation for their predicament. The point of this thesis is not that unnatural explanations are necessarily incorrect or inelegant. Often they are the most beautiful at first glance, but are subsequently found to be tuned. The point is that fine tuning has, in the sweeping history of natural science, almost always been a convincing hint that a more fundamental explanation is at play. With a reliable tuning measure in hand, and a landscape of alternative Higgs physics stretching out to the energy horizon, there is a great deal of searching to be done.

\appendix
\chapter{Group Theory}
\label{sec:group_theory}

These appendices reflect the structure of the body of this thesis - earlier appendices are quick, pedagogical reviews of necessary topics that the reader may need to revise. Later appendices are references for results, data and expressions too tedious to present in the body, but necessary for a full understanding of the models in chapters \ref{sec:M4DCHM}, \ref{sec:LCHM}, and \ref{sec:NMCHM}. 

The first review deals with group theory and representation theory. Without a solid understanding of this topic, many of the heuristics used in Composite Higgs "cosetology" are arcane nonsense. In \cref{sec:group_theory_fundamentals}, we will tersely build from basic definitions to an understanding of Lie groups acting on manifolds, and then explore how one represents this action, in \cref{sec:representations}.

\section{Fundamentals}
\label{sec:group_theory_fundamentals}

\begin{quote}
We will have to abandon the philosophy of Democritus and the concept of elementary particles. We should accept instead the concept of elementary symmetries.\\
\hspace*{1em} - \textit{W. Heisenberg}
\end{quote}

The mechanics behind a set of symmetries is developed through group theory. So what is a group?

\subsection{Definition}

A set $G$ is defined to be a \textit{group} if it possesses some operation $\sbullet$ between any two members $x,y \in G$ of the set, and the properties: \textit{closure} ($x\sbullet y\in G$), \textit{associativity} ($(x_1\sbullet x_2) \sbullet y = x_1 \sbullet (x_2 \sbullet y)$), an \textit{identity} ($\exists\; e\in G, \; s.t. \; e\sbullet x = x$), and an \textit{inverse} ($\exists \; x^{-1}\in G, \; s.t. \; x \sbullet x^{-1} = e$).

These can be conveniently summarised and tested with the ansatz
\begin{align}
\forall x,y \in G, \qquad x\sbullet y^{-1} \in G
\end{align}

A group is called \textit{Abelian} if $x\sbullet y = y \sbullet x \; \forall \; x,y \in G$.

A subset $G_\text{gen} = \{T_1, T_2, ..., T_N\}\subset G$ is said to \textit{generate} $G$ if
\begin{align}
\forall \; g_i \in G, \qquad \exists \; T_j, T_k, ... , T_z \in G_\text{gen} \qquad s.t. \qquad T_j^{n_1} \sbullet T_k^{n_2} \sbullet ... \sbullet T_z^{n_z} = g_i
\end{align}
and we denote this generating behaviour as $\langle G_\text{gen} \rangle = G$. From here on, we will drop the operation $\sbullet$ and assume the composition rule $xy := x\sbullet y$.


\subsection{Subgroups}

A subset $H$ of the set of $G$ is called a \textit{subgroup} if $H$ itself satisfies the properties of a group. Every element $g_i$ of $G$ generates an Abelian subgroup $\langle g_i \rangle \subset G$. 

A special subgroup is the \textit{center} $Z$ of $G$ - the set of all elements that commute with all elements of $G$
\begin{align}
Z(G) = \{h \in G \; | \; hg  = gh \; \forall \; g\in G\} 
\end{align}
which is clearly Abelian.

\subsection{Cosets}

Given a subgroup $H$ and some element $g \in G$, we define a subset of $G$ called a left coset
\begin{align}
gH := \{gh_i\} \label{eq:left_coset}
\end{align}
and a right coset as $Hg$. Is a left/right coset a group? Only if $g$ was already a member of $H$. A quick way to see why this is a necessary condition is to request that the left coset contains an inverse. For an element $gh_1 \in gH$, an obvious inverse would be $h_1^{-1} g^{-1}$. However, $h_1^{-1}$ is not a member of the left coset, unless there is some $g h_2 = h_1^{-1} \in H$. But if $g h_2 \in H$, then $g \in H$. So the left coset is either not a group at all, or it is $H$ itself.

We now try to anticipate and clear up some ambiguity: Let the \textit{left} and \textit{right} cosets of $g\in G$ and $H$ be defined as above in \cref{eq:left_coset}. We now define a \textit{set} of cosets to be 
\begin{align}
G/H = \{g_i H \; | \; g_i \in G\}
\end{align}
and call this the \textit{coset space} of $G$ and $H$ (or, simply the \textit{coset}). This terminology is consistent with the body of this thesis, as well as Composite Higgs literature, not to mention all model-building literature. More precisely, this is the \textit{quotient set}. Is the coset space a group? First, define its elements as
\begin{align}
G/H = \{H, aH, bH, ... \}
\end{align}
where $aH$ is a coset where any particular choice of $a \in G$ leads to the same coset. That is, $\exists h$ s.t. $a_i = a_j h_{a_{ij}}$. For example, take $G$ to be the group of integers under addition, and $H$ is the group of even integers under addition. Then adding an odd number to the set of $H$ either by choosing $a_i = 1$ or $a_i = 43$ will lead to the same coset: the set of odd numbers. Then we need a well-defined group operation, which means one that doesn't depend on the choice of representative (e.g. $a_i = 1$ or $43$). One such operation could be
\begin{align}
(aH)\sbullet (bH) := abH
\end{align}
If this was well defined, we would expect any choice of of $a,b$ to be equivalent, $(a_iH)(b_iH) = (a_jH)(b_jH)$. But
\begin{align}
(a_iH)(b_iH)& = a_i b_i H = a_j h_{a_{ij}} b_j h_{b_{ij}}H = a_jh_{a_{ij}}b_jH \nonumber\\
& \stackrel{?}{=} a_jb_j h_{a_{ij}} H \label{eq:commuting_cosets}\\
& = a_j b_j H = (a_jH)(b_jH) \nonumber
\end{align}
The step in \cref{eq:commuting_cosets} only holds true if $h_1 g_i = g_i h_2$. That is, all elements of $G$ commute with the subgroup $H$. If this is the case, $H$ is called a \textit{normal subgroup}, and the coset space is a group called the \textit{quotient group}. No more time will be spent on normal subgroups and quotient groups, because we will rarely have $H$ as normal in $G$, since in that case symmetries are not spontaneously broken.  

\subsection{Continuous Groups}

The generalisation from discrete group elements $g_1, g_2, ... \in G$ to elements $g(w_1, w_2, ...) \in G$ depending on continuous parameters $w_1, w_2, ... \in \mathbb{R}$ is straightforward. The elements $g(w) := g(w_1, w_2, ...)$ are a \textit{continuous group} if they fulfill the properties of a group, and the parameters are continuous. Checking the closure property, we see that 
\begin{align}
g(w) = g(u) g(v)
\end{align}
which implies that $w = f(u,v)$ is a real continuous function of $u$ and $v$. 

\subsection{Lie Groups}

If we constrain $f$ to be an \textit{analytic} function, then we similarly constrain the continuous group $G$ to be a \textit{Lie group} \cite{vvedensky2005group}. This constraint is justified for the vast majority of continuous transformations in the natural world. That is, if we are not dealing with discrete group transformations (reflections, conjugations, etc.), we are dealing with Lie group transformations. More precisely, let's \textit{define} a Lie group to be a group $G$ that has the structure of a differentiable manifold \cite{ludeling2010physics751}. To be formal, there is one further requirement that the maps
\begin{align}
\text{Product} && G \times G \rightarrow G \\
\text{Inverse} && (g_1, g_2) \mapsto g_1 g_2^{-1}
\end{align}
are continuous. The product operation can be thought of as a group member of $G$ acting on the manifold of $G$ to return a position still on the manifold - a geometric formulation of our usual group condition. All the continuous groups that one is likely to find in physics satisfy these conditions. To see why, we need two simple facts. The first is that the general linear group $GL(n, \mathbb{R})$ is a manifold, specifically a \textit{sub}manifold of the set of real $n\times n$ matrices
\begin{align}
M(n, \mathbb{R}) &\sim \mathbb{R}^{n^2}\\
GL(n, \mathbb{R}) &= \{A \in M(n, \mathbb{R}) | \text{det}A \neq 0 \} \subset M(n, \mathbb{R}). \label{eq:GL_matrices}
\end{align}
The second fact is that if a subgroup $H$ of Lie group $G$ is \textit{closed} (that it's a closed subset -that it contains its boundary), then it is a submanifold of $G$, and thus is a Lie subgroup. Combining these two ideas leads us to the submanifolds of the invertible matrices being Lie groups, called the classical Lie groups - special linear $SL(n)$, (special) orthogonal $(S)O(n)$, (special) unitary $(S)U(n)$, and symplectic group $Sp(n)$. These could be over the real or complex space. The ability to describe Lie groups as smooth\footnote{Smoothness from continuity is guaranteed by Hilbert's fifth problem} manifolds gives us many geometric tools to tinker with the classical groups. We will quickly review the ones required to get to the concept of a symmetric group.

\subsection{Tangent Spaces}

%
%

Defining the tangent space can be done without co-ordinates, see \cite{ludeling2010physics751}, section 5.2 for example. It will be much easier if we assume that we have a set of co-ordinates $x^a(t)$ parameterised by $t$ to pass through $g$ in our manifold $G$ at $t=0$. Then a tangent vector $S$ at point $x(0) = g \in G$ has components
\begin{align}
S := \frac{d}{dt}x^a(t) \bigm\lvert_{t=0} \frac{\partial}{\partial x^a} = S^a \partial_a \,. \label{eq:tangent_vectors}
\end{align}
Simply, the tangent vectors $S^a$ are the components of the partial derivatives in each basis co-ordinate, evaluated at some point $g$. These form a tangent vector space $T_g G = \{S\}$. 

We now extend this idea to the tangent spaces of \textit{every} point on the manifold, which forms a \textit{vector field}. The field associates one vector $S(g) \in T_g G$ for every point in $G$
\begin{align}
\mathcal{S} := S^a(x) \partial_a \, .
\end{align}
In a sense, this is a vector space of vector spaces. It is an infinite dimensional space of $d$-dimensional tangent spaces. If we give the space of tangent spaces a bilinear (i.e. composing two members of the space) composition rule, then it is also an algebra. It should be a rule that preserves the product and chain rules, since each vector is a set of derivatives that can act on the manifold. Define the composition rule $[ \cdot , \cdot ]$ of two fields $\mathcal{S}, \mathcal{T}$ to be 
\begin{align}
[\mathcal{S}, \mathcal{T}] = \mathcal{S}(\mathcal{T}) - \mathcal{T}(\mathcal{S})
\end{align}
where the vector fields can act on each other (since we can take derivatives of derivatives)\footnote{Although, note that the $\mathcal{S}(\mathcal{T})$ is not a vector field, since applying it to a point $g$ gives $\mathcal{S}(\mathcal{T})(g) = \mathcal{S}^a (\partial_a (\mathcal{T}^b \partial_b g) = \mathcal{S}^a \partial_a \mathcal{T}^b) \partial_b g + \mathcal{S}^a \mathcal{T}^b \partial_{ab} g$. The second derivative precludes us from expressing this as a linear combination of vector fields.}. Why choose \textit{this} behaviour? It comes down to preserving the product rule. Consider two elements of the manifold $g,h \in G$, which are composed as $gh \in G$. Acting upon them by a field $\mathcal{U} =  [\mathcal{S}, \mathcal{T}]$ gives (using the product rule)
{\small
\begin{align}
\mathcal{U}(gh) &=  [\mathcal{S}, \mathcal{T}](gh) = \mathcal{S}(\mathcal{T}(gh)) - \mathcal{T}(\mathcal{S}(gh)) \nonumber\\
&= \mathcal{S}(g)\mathcal{T}(h) + \mathcal{S}(\mathcal{T}(h))g + \mathcal{S}(\mathcal{T}(g))h + \mathcal{S}(h)\mathcal{T}(g) \nonumber\\
& - \mathcal{T}(h)\mathcal{S}(g) - \mathcal{T}(\mathcal{S}(h))g - \mathcal{T}(\mathcal{S}(g))h - \mathcal{T}(g)\mathcal{S}(h)\\
&= [\mathcal{S}, \mathcal{T}](g)h + [\mathcal{S},\mathcal{T}](h) g \nonumber
\end{align}
}
Which preserves the product rule, and happens to render the composition rule anti-symmetric. This is an important feature. A vector space coupled with an anti-symmetric and bilinear rule, which also satisfies the Jacobi identity (as can be checked), defines a Lie algebra, and the rule is a Lie bracket.

\subsection{Lie Algebra}

Before we define the Lie algebra in terms of generators of a Lie group, we will close the gap between smooth manifolds and Lie groups. There is a subtlety, which is that to use a vector field as a Lie algebra, it shouldn't change "too much" as we move around the manifold. Let us be more precise about what "too much" means. Define a \textit{left translation} $L_g$ by a group element $g\in G$ of a point $h \in \mathcal{M}$ on a manifold as 
\begin{align}
L_g(h) = gh \, ,
\end{align}
which is the obvious definition, if the manifold is the group itself $\mathcal{M} \sim G$. But we will be agnostic about $\mathcal{M}$ for the moment. A vector field is called \textit{left invariant} if 
\begin{align}
(dL_g)(\mathcal{S}(h)) := \mathcal{S}(L_g(h)) = \mathcal{S}(gh) \, .
\end{align}
$dL_g$ reminds us that we are left translating a field of tangent spaces, so it is an infinitesimal translation. What this invariance says is that it shouldn't matter whether we translate a tangent space or take the tangent space of a translated point, our algebra should be the same:

\begin{center}
\tikz[scale=0.4]{
\node at (-3,-3) {$G$};
\draw [->] (-3,-2) --(-3,2);
\node [above] at (0,3) {$dL_g$};
\node [below] at (0,-3) {$L_g$};
\node [left] at (-3,0) {$\mathcal{S}$};
\node [right] at (3,0) {$\mathcal{S}$};
\draw [->] (3,-2) --(3,2);
\draw [->] (-2,-3) --(2,-3);
\draw [->] (-2,3) --(2,3);
\node at (-3,3) {$TG$};
\node at (3,-3) {$G$};
\node at (3,3) {$TG$};
}
\end{center}

Let us now switch to more familiar notation, and show that a Lie group does satisfy this left-invariance, and that this also implies that the tangent space $T_e G$ at the identity of $G$ is a Lie algebra. To do this, begin with a point on a Lie group manifold $h^\mu$, that is operated upon by a small change\footnote{The aim here is intuition. To be more formal, one should act upon a group element $g(\beta)$ parameterised by $\beta$ and a manifold point $h(\alpha)$ parameterised by $\alpha$ with a smooth map $\phi(\beta, \alpha)$, as in \cite{billo}. Then the small change is $\delta g := g(\delta \beta)$.} $\delta g^{\mu\nu}$, with group parameters $\beta$. Then Taylor expanding gives
\begin{align}
h^\mu \rightarrow h^\mu + dh^\mu := (\delta g)^{\mu\nu} h^\nu = h^\mu + \frac{\partial (g^{\mu\nu}h^\nu)}{\partial \beta^i}|_{\beta = 0} \delta \beta^i + \mathcal{O}(\cancelto{0}{\delta \beta^2}) \label{eq:dh}
\end{align}
We throw away the higher order terms. We can rewrite \cref{eq:dh} as
\begin{align}
h^\mu \rightarrow h^\mu + dh^\mu &= (1 + \delta \beta^i \frac{\partial (g^{\rho\nu}h^\nu)}{\partial \beta^i}|_{\beta =0} \frac{\partial}{\partial h^\rho}h^\mu\\
&= (1 + \delta\beta^i S^\rho_i \frac{\partial}{\partial h^\rho} )h^\mu = (1 + \delta \beta^i S_i ) h^\mu
\end{align}
and call $S_i$ the infinitesimal generator of the translation. If it appears like the tangent space in \cref{eq:tangent_vectors}, this is no coincidence. It has components
\begin{align}
S^\rho_i = \frac{\partial (g^{\rho\nu} h^\nu)}{\partial \beta^i} |_{\beta = 0} \, .
\end{align}

\begin{figure}
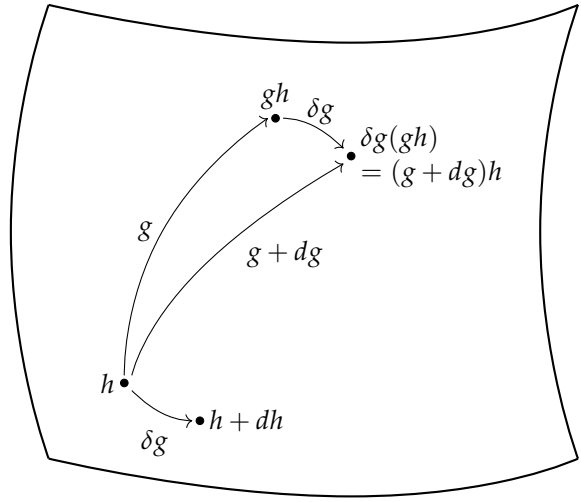

\centering
\tikz{
\draw [thick] plot [smooth, tension=1] coordinates {(0,0) (-0.5,3) (0,6)};
\draw [thick] plot [smooth, tension=1] coordinates {(0,0) (4,-0.5) (7,0)};
\draw [thick] plot [smooth, tension=1] coordinates {(7,0) (6.5,3) (7,6)};
\draw [thick] plot [smooth, tension=1] coordinates {(0,6) (4,5.5) (7,6)};
\draw [fill=black] (1,1) circle [radius = 0.05];
\draw [fill=black] (3,4.5) circle [radius = 0.05];
\draw [fill=black] (4,4) circle [radius = 0.05];
\draw [fill=black] (2,0.5) circle [radius = 0.05];
\draw [->] plot [smooth, tension=1] coordinates {(1,1.1) (1.5,3) (2.9,4.5)};
\draw [->] plot [smooth, tension=1] coordinates {(1.1,1.1) (2,2.5) (3.9,3.9)};
\draw [->] plot [smooth, tension=1] coordinates {(3.1,4.5) (3.5,4.4) (3.9,4.1)};
\draw [->] plot [smooth, tension=1] coordinates {(1.1,0.9) (1.5,0.6) (1.9,0.5)};
\node [left] at (1,1) {$h$};
\node [right] at (2,0.5) {$h + dh$};
\node [below] at (1.4,0.5) {$\delta g$};
\node [left] at (1.5,3) {$g$};
\node [above] at (3,4.5) {$gh$};
\node [above] at (3.6, 4.3) {$\delta g$};
\node [right, align=left] at (4,4) {$\delta g (g h)$ \\ $=(g + dg)h$};
\node [right] at (2.5, 2.7) {$g+dg$};
}\caption{Associativity of infinitesimal and finite transformations}\label{fig:associative_transformation}
\end{figure}

We can see the infinitesimal change in \cref{fig:associative_transformation}. To see how we extend the infinitesimal changes to finite, let's begin with Taylor expanding a differential of the finite transformed point $gh$
\begin{align}
gh + dgh = g + \frac{\partial( g h)}{\partial \beta}|_{\beta = 0} = gh + S \beta \label{eq:finite_transformation_1}
\end{align}
Now, we can use the associativity of the group structure (that is, we can move either way around the arrows in \cref{fig:associative_transformation}) to equate this with
\begin{align}
\delta g(gh) = \beta + d\beta = \beta + \frac{\partial (gh)}{\partial \beta} \delta \beta = S(\beta)\delta \beta \label{eq:finite_transformation_2}
\end{align}
Then we can compare \cref{eq:finite_transformation_1} with \ref{eq:finite_transformation_2} to get the set of partial differential equations
\begin{align}
\frac{d g}{d \beta} = S S(\beta)^{-1}
\end{align}
We also required the invertibility of group elements to get the inverse of the infinitesimal generator. This is called Lie's First Theorem. Differentiating this again leads to Lie's Second Theorem, which states that the infinitesimal generators are constant across the a finite translation. We can use this to get an explicit form for the group elements, with an Abelian group, which we will assume for simplicity. 
Our job is to find out what the coefficients of $\beta$ are in the Taylor expansion
\begin{align}
g(\beta) = g(0) + \frac{dg}{d\beta}|_{\beta = 0} + \frac{1}{2} \frac{d^2g}{d\beta^2}|_{\beta = 0} \beta^2 + ... \label{eq:group_taylor}
\end{align}
We have that 
\begin{align}
g(\beta + \beta') &= g(\beta) g(\beta') \, , \label{eq:abelian_transformation}
\end{align}
then
\begin{align}
\frac{dg(\beta)}{d\beta} &= \frac{dg(\beta + \beta')}{d(\beta + \beta')}\frac{d(\beta + \beta')}{d\beta'} |_{\beta' = 0}\\
&= \frac{dg(\beta_1 + \beta_2)}{d\beta_1}|_{\beta'=0} = \frac{dg(\beta')}{d\beta'}|_{\beta'=0}g(\beta) := S g(\beta)
\end{align}
Therefore we define the infinitesimal generators to be the coefficients in this expansion, and successive differentiations $\frac{d}{d\beta^n}|_{\beta = 0}$ of the group element $g(\beta)$ gives higher orders of $S^n$. We thus define the exponential map the group to be \cref{eq:group_taylor} with coefficients $S$
\begin{align}
g(\beta) &= I + S\beta + \frac{1}{2} S^2 \beta^2 + ...\\
&= \sum\limits_{n=0}^\infty \frac{1}{n!} (S\beta)^n := \me^{S\beta}
\end{align}

%
%

\subsection{Homogeneous Spaces}
%

If $H$ is a Lie subgroup of Lie group $G$, then we call $G/H$ a \textit{homogeneous space}. It is also a manifold, although it's not necessarily a group. This is a familiar heuristic from the body of the thesis, but we will see precisely why it's true. We can equip a manifold $M$ with a \textit{natural action} (like a group) for a group member $a \in G$ and a manifold point $x \in M$ 
\begin{align}
G \times M \rightarrow M: \qquad (a, x) \mapsto ax \, .
\end{align}
We then define a \textit{stabiliser subgroup} $H = \{g \in G | gx_0 = x_0\}$ that corresponds to a particular point $x_0 \in M$. Then, we can formally say that $M$ is diffeomorphic to $G/H$. What does this mean geometrically?

As an example, consider the group of rotations in 3D $SO(3)$, applied to a particular manifold: the 2-sphere $M = S^2$. Given a point on the sphere $x_0$, the isotropy subgroup at that point is the set of transformations which leave the point invariant. This is obviously $H = SO(2) = \{g \in SO(3) | gx_0 = x_0\}$. Thus, $SO(3)/SO(2)$ is diffeomorphic to $S^2$. This holds in any dimension, and is the underlying reason that we stated $SO(N+1)/SO(N) \sim S^N$ in \cref{sec:VM_geometry}.

The reasoning also works the other way. Given a Lie group $G$ and subgroup $H$
\begin{align}
G \times G/H \rightarrow G/H: \qquad (a, gH) \mapsto agH, \; a \in G\, ,
\end{align}
the coset is a manifold that defines the space of points $gH=x \in M$ that leave $G$ transformations invariant under $H$ transformations. This allows us to formally connect the manifold of invariant vacua to the CCWZ coset construction. The physical nature of the vacuum as a space that is invariant in all directions, implies that we should be able to go \textit{further} than requiring our space is homogeneous, but that it is a Riemannian manifold of constant positive curvature - a hypersphere. 

\subsection{Symmetric Spaces}\label{sec:symmetric_space_appendix}


A Riemannian manifold\footnote{A smooth manifold $M$ equipped with a group action $g$} of constant curvature defines a \textit{symmetric space}, which contains more structure than a homogenous space, though is still not necessarily a Lie group manifold \cite{koda2009introduction}. It is defined by a manifold $M$ having geodesics $\gamma(t)$ through every point $p \in M$ that are invariant under \textit{two} applications of some isometry $s_p$
\begin{align}
\forall p \in M \; \; \exists s_p \in I(M,g) \; \; \text{s.t.} \; \; s_p(\gamma(t)) = \gamma(-t) \implies s_p^2 = id.  
\end{align}
We then cast this in very familiar terms by defining the Riemannian manifold $M$ as $G/H$, where $H$ is the isotropy subgroup at $p_0$. Then the isometry becomes an automorphism $\sigma$ defined as
\begin{align}
\sigma : \; \; G \rightarrow G, \; \; a \mapsto s_{p_0} a s_{p_0}\, ,
\end{align}
which defines a space
\begin{align}
G^\sigma = \{a \in G | \sigma(a) = a\} & &  H \subset G^\sigma \subset G
\end{align}
The automorphism that is most convenient to work with is to take
\begin{align}
\sigma: \tilde{T} \in \{T,X\}, && \tilde{T} \mapsto \left[ \left[ \tilde{T},X \right], X\right] = \tilde{T}\, . 
\end{align}
Given that we already have $[T,T]= T$ and $[T,X]=X$ for the Lie generators $T \in \mathfrak{h}$ and $X\in (\mathfrak{g} - \mathfrak{h})$, this leads to the necessary and sufficient condition for a symmetric space
\begin{align}
[X,X] = T \,.
\end{align}
This can be used to derive the explicit Goldstone fields of, for example $G\times G/G$ in \cref{sec:symmetric_spaces}. This is done beautifully in \cite{Stangl:2018kty}.

\section{Representations}
\label{sec:representations}


\subsection{An Example}\label{sec:rep_example}

The rules we assign to the behaviour of quantum fields typically follow the structure of vector spaces. Representation theory is concerned with how to describe the action of an abstract group $G$ on a concrete vector space $V$. However, there's no reason to limit group transformations to vectors themselves, they can also transform functions of those vectors. A simple function $f_1$ of a vector $\vec{v} \in V$ could be\footnote{The following motivation is based on an answer from MathStackExchange \cite{40141}.}
\begin{align}
f_1(\vec{v}) = 7 v_x + 4 v_y  - 2 v_z = (7,4,-2)(v_x, v_y, v_z)^\intercal \,. 
\end{align}
Observe that the function can be written as a vector multiplication. This helps us understand what is happening when we "rotate the function". For example, $SO(3)$ can act upon this function, as a rotation in the $x-y$ plane by angle $\theta$
\begin{align}
f_1(\vec{v}) \rightarrow f_1' &= f_1(\vec{v}') = 7 (v_x c_\theta + v_y s_\theta)  + 4 (v_y c_\theta - v_x s_\theta)   - 2 v_z \nonumber\\
& = (7,4,-2)\left(\begin{matrix}
c_\theta & s_\theta & 0 \\
-s_\theta & c_\theta & 0 \\
0 & 0 & 1
\end{matrix}\right)
\left( \begin{matrix}
v_x \\
v_y \\
v_z
\end{matrix}\right)
 \,.
\end{align}
using the usual convention $c_\theta := \cos\theta, s_\theta := \sin\theta$. This is very intuitive so far, as a linear function transforms in the same way as the vector space itself. Indeed, we are really transforming the 3-dimensional space with basis $(v_x,v_y,v_z)$. Now, consider a quadratic function of vectors
\begin{align}
f_2(\vec{v}) = 2 v_x^2 + 3 v_x v_y + 7v_y^2 + v_z^2 = (2,3,7,0,0,1)(v_x^2,v_x v_y,v_y^2,v_x v_z,v_y v_z,v_z^2)^\intercal
\end{align}
which transforms under an $x-y$ rotation as
\begin{align}
f_2(\vec{v}) \rightarrow f_2' &= f_2(\vec{v}') = 2 (v_x c_\theta + v_y s_\theta)^2 \nonumber\\
&+ 3 (v_x c_\theta + v_y s_\theta)(v_y c_\theta - v_x s_\theta) + 7(v_y c_\theta - v_x s_\theta)^2 + z^2\\
& = (2,3,7,0,0,1)(6 \times 6 \, \text{matrix})(v_x^2,v_x v_y,v_y^2,v_x v_z,v_y v_z,v_z^2)^\intercal\, . \nonumber
\end{align}
We will return to the problem of finding this $6\times 6$ matrix later. This transformation highlights two essential elements of representations. First, notice that $2 v_x^2 + 7v_y^2 \rightarrow 2 v_x^2 + 7v_y^2 + 5 s_\theta( v_x^2 - v_y^2)$, which preserves the Laplacian. In other words, if one could pull out a factor of $(v_x^2 + v_y^2 + v_z^2) := v_\textbf{1}$ from the function, this part would transform trivially. This is possible, by changing basis, with one basis vector being $v_\textbf{1}$. The other five basis vectors $\vec{v}_\textbf{5}$ can be chosen, and these form a subspace that is also closed under $SO(3)$. Specifically
\begin{align}
f_2(\vec{v}) &= 2 v_x^2 + 3 v_x v_y + 7v_y^2 + v_z^2  \nonumber \\
& \equiv ( -\frac{4}{3}, \frac{7}{3} , 3, 0, 0, \frac{10}{3})(v_x^2 - v_y^2, v_y^2 - v_z^2, v_x v_y, v_x v_z, v_y v_z, v_x^2 + v_y^2 + v_z^2)^\intercal \nonumber\\
\rightarrow f_2(\vec{v}')&=  ( -\frac{4}{3}, \frac{7}{3} , 3, 0, 0, \frac{10}{3})
\left( \begin{matrix}
(5\times 5) & \vec{0}^\intercal\\
\vec{0} & 1
\end{matrix}\right) (\vec{v}_\textbf{5}, v_\textbf{1})^\intercal \nonumber
\end{align}
We have thus \textit{decomposed} the $SO(3)$ representation $\textbf{6} = \textbf{5} \oplus \textbf{1}$, which simply stands for a block diagonal matrix. 

The other important element in this example is that we can rewrite the vector
\begin{align}
\hspace*{-2cm} \vec{v} = (v_x^2, & v_x v_y,v_y^2,v_x v_z,v_y v_z,v_z^2) \sim R = \left( \begin{matrix}
v_x^2 & v_x v_y & v_x v_z\\
v_x v_y & v_y^2 & v_y v_z \\
v_x v_z & v_y v_z & v_z^2
\end{matrix} \right) \label{eq:symmetric_matrix_quadratic}\\
&= \left( \begin{matrix}
\frac{1}{3}(2v_x^2 - v_y^2 - v_z^2) & v_x v_y & v_x v_z\\
 & \frac{1}{3}(2v_y^2 - v_x^2 - v_z^2 )& v_y v_z \\
&  & \frac{1}{3}(2v_z^2 - v_x^2 - v_y^2)
\end{matrix} \right) + \frac{1}{3}(v_x^2 + v_y^2 + v_z^2)\mathbb{1}_{3\times 3} \label{eq:decomposed_matrix_quadratic}
\end{align}
There is a lot here. The first step is that \cref{eq:symmetric_matrix_quadratic} says the vector $\vec{v}$ can be expressed as a symmetric matrix $R = v_{\mu\nu}$, which is the tensor product $(v_x, v_y, v_z) \otimes (v_x, v_y, v_z)$. To transform the symmetric matrix requires a different behaviour of the group. In vector form, the transformation was $g_i v_i = g\vec{v}$, with $i = 1,...,6$. In tensor form, the transformation is $g_\mu g_\nu v_{\mu\nu} = g R g^{-1}$. Now it's very clear that we can pull out the trace as a trivial subspace, since $g (\mathbb{1}\text{Tr}[R]) g^{-1} = \mathbb{1} \text{Tr}[R]$, which we do in \cref{eq:decomposed_matrix_quadratic}. This shows that an $N\times N$ symmetric matrix under $SO(N)$ is \textit{not} irreducible - it can be decomposed into two irreducible subspaces: a symmetric traceless, and a trivial representation. We will formally define and describe these concepts in the following sections. For now, just note that a group acts on vector spaces of any dimension, but some may be able to be reduced to smaller spaces, and rewriting them as tensors may make this more convenient.

\subsection{Definition}\label{sec:rep_definition}

A group $G$ can act on sets $X$. Since the quantum world is described by wave functions, we will narrow the acted-upon set to the space of vectors over complex numbers. The action of a Lie group on a manifold had "natural" choice, since a Lie group \textit{is} a manifold! But a Lie group is not a vector space, so we much decide which vector spaces it can act on, and what its action looks like. We require it to be a map from a vector space to itself, therefore the map must follow the same rules of linearity that a vector space contains. We call each acted-upon vector space and its mapping function a \textit{representation}. That is, for some $g \in G$, a representation $D(g)$ of $G$ acting on $n$-dimensional (complex) vector space $V_n$ is uniquely defined by
\begin{align}
& D(g):  && V_n \rightarrow V_n, \nonumber\\
& \text{s.t.} &&  D(g)(v_1 + v_2) = D(g)v_1 + D(g)v_2, \label{eq:rep_definition}\\
& \text{and} && D(g)(k v_1) = k D(g)(v_1) \, . \nonumber
\end{align}
The uniqueness is up to an isomorphism $\phi$, where $D_2 \sim D_1$ if $D_2 = \phi \circ D_1 \circ \phi^{-1}$. 
Clearly a structure that captures this linearity is matrix addition and multiplication. Since we require the representation to follow the rules of a group, the matrices must be invertible, which define the general linear group of \cref{eq:GL_matrices}. In other words, finding a representation of $G$ on $V_n \sim \mathbb{C}^{n}$ is the same as making a group homomorphism from $G$ to $GL(\mathbb{C}, n)$. Assuming a matrix representation $D$ of group elements $g \in G$, then the following can be proven from the definition in \cref{eq:rep_definition} and the $G \sim GL(\mathbb{C}, n)$ homomorphism
\begin{align}
D(g_1 g_2) = D(g_1) D(g_2)\; .
\end{align}
The whole program for representation theory is then classifying all the spaces $V_i^k$ (where the dimension $k$ ranges over all positive numbers, and $i$ ranges over different spaces of the same dimension) that maps $D_i(g)$ act upon as invertible matrices. We can make the program easier, by recognising that some representations can be built out of smaller representations.

\subsection{The Direct Sum}

We define the direct sum of two representations as
\begin{align}
D_1 \oplus D_2 := \left(\begin{matrix}
D_1 & 0 \\
0 & D_2
\end{matrix} \right) \, ,
\end{align}
that is, the block diagonal of each representations matrices. We have already mentioned the isomorphism between representations, which implies a non-diagonal $A$ representation may be \textit{reduced} to a block diagonal direct sum if it can be rotated by a similarity transformation $A \rightarrow B A B^{-1}$. We can define each representation by its trace, called its \textit{character}, which is invariant under these transformations
\begin{align}
\text{Tr}[A] = \text{Tr}[BAB^{-1}]
\end{align}
We can further define a representation by whether it can be reduced to block diagonal form. If it cannot be expressed as a direct sum, we call it \textit{irreducible}. Thus, \textit{almost}\footnote{It should be proven with Schur's First and Second Lemmas, as there are some subtleties.} by definition, all reducible representations can be expressed as direct sums of irreducible representations.

\subsection{The Direct Product}

Formally, a direct product $G\otimes H$ of two groups $G,H$ is the set of ordered pairs $(g,h)$, $g\in G,h \in H$ with the operation of the original groups applied componentwise: $(g_1,h_1)\cdot (g_2,h_2) = (g_1 g_2,h_1 h_2)$. If the groups commute, as is often the case in the physical groups we will work with, then the following representation theory holds:
The direct product of the representations of two groups is a representation of the direct product of those groups. This means we are able to switch between reps and the group itself when taking direct products, without keeping track of the order of switching. 

For example, a direct product $A \otimes B$ can be represented by matrices \cite{vvedensky2005group} 
\begin{align*}
A &= \left(\begin{matrix}
a_{11} & a_{12}\\
a_{21} & a_{22}
\end{matrix}\right) & B & = \left(\begin{matrix}
b_{11} & b_{12} & b_{13}\\
b_{21} & b_{22} & b_{32}\\
b_{31} & b_{32} & b_{33}
\end{matrix}\right)
\end{align*}

as

\begin{align*}
A \otimes B &= \left(\begin{matrix}
a_{11} B & a_{12}B\\
a_{21}B & a_{22}B
\end{matrix}\right)\\
& = \left(\begin{matrix}
a_{11}b_{11} & a_{11}b_{12} & a_{11}b_{13} & a_{12}b_{11} & a_{12}b_{12} & a_{12}b_{13}\\
a_{11}b_{21} & a_{11}b_{22} & a_{11}b_{32} & a_{12}b_{11} & a_{12}b_{12} & a_{12}b_{13}\\
a_{11}b_{31} & a_{11}b_{32} & a_{11}b_{33} & a_{12}b_{11} & a_{12}b_{12} & a_{12}b_{13}\\
a_{21}b_{11} & a_{21}b_{12} & a_{21}b_{13} & a_{22}b_{11} & a_{22}b_{12} & a_{22}b_{13}\\
a_{21}b_{21} & a_{21}b_{22} & a_{21}b_{32} & a_{22}b_{11} & a_{22}b_{12} & a_{22}b_{13}\\
a_{21}b_{31} & a_{21}b_{32} & a_{21}b_{33} & a_{22}b_{11} & a_{22}b_{12} & a_{22}b_{13}
\end{matrix}\right)
\end{align*}

\subsection{The Trivial, the Fundamental, the Antisymmetric and the Symmetric}\label{sec:deconstructing_tensors}

Combining the intuition gained from \cref{sec:rep_example} with the formalism of the direct product, we should start to see that the vector spaces of irreducible representations can be combined to find new vector spaces transforming under irreducible representations. For an N-dimensional fundamental vector $v_i \in V_N$, we have
\begin{align}
v_i \rightarrow v'_i = D_{ij} v_j \, .
\end{align}
Then, there exists an object (called by some the \textit{tensor product}, others the \textit{direct product}, somewhat conflicting) called $v_i v_j$ that transforms as
\begin{align}
v_i v_j := T_{ij} \rightarrow T'_{ij} = D_{ik} D_{jl} T_{kl}
\end{align}
We can formalise the intuition of \cref{sec:rep_example} by defining the tensors
\begin{align}
S_{ij} = &\frac{1}{2}(T_{ij} + T_{ji}) & A_{ij} = & \frac{1}{2}(T_{ij} - T_{ji})\\
\rightarrow & D_{ik} D_{jl} S_{kl}  = S'_{ij} &   \rightarrow & D_{ik} D_{jl} A_{kl} = A'_{ij}\, .
\end{align}
That is, $T$ can be expressed as the direct sum of two objects that do not transform into each, which are then themselves representations. As we saw in \cref{sec:rep_example}, we can also pull out the trace, called the trivial representation, as an invariant object
\begin{align}
T = S \oplus A \oplus \text{Tr}[T] \, ,
\end{align}
and this extends to any order of tensor. 

\subsection{The Adjoint}

A representation that is always available is $m=dim(\mathcal{G})$ (e.g. $m=dim(SU(n))=n^2 - 1$), called the \textit{adjoint representation}. This is also a nice choice because there is a simple algorithm for the generator matrices: $(T^a_\textnormal{adj})_{bc}=f^a_{bc}$. \textit{The adjoint representation transforms just like the fundamental, and like every representation.}\footnote{In this formalism, the proof of $SU(2)\approx SO(3)$ is trivial: $SO(3)$ is (up to a phase) the adjoint representation of $SU(2)$.}

Unfortunately, the phrase ``to transform adjointly" is used for matrices (of any size) which transform as:
\[A_\mu \rightarrow D A_\mu D^{-1}\]

Let's unravel this concept, with the help of \cite{maggiore2004modern}. Consider a vector field in the adjoint representation. As with all representations, it transforms as
\begin{align}
\phi \rightarrow D\phi = (e^{ig\phi^a T^a_\textnormal{adj}})\phi\, .\label{eq:transformvector}
\end{align}

Since the adjoint representation is the same size $m$ as its dimension $N$, we can recast the vector field it acts upon $\phi(x)$ as a matrix $\Phi(x)$. Do this by summing over the generators
\begin{align}
\Phi(x) = \phi^a(x)T^a \, .
\label{vectortotensor}
\end{align}

Importantly, the only thing in the adjoint here is $\phi(x)$. $T^a$ need not be in the adjoint or any other particular representations (i.e. it may be any size matrix), as long as there are $N$ of them, which there always are. Therefore, $\Phi(x)$ can also be a matrix of any size, and still be commonly ``transforming adjointly".

We know how a vector field transforms under a representation: $\phi \rightarrow e^{i\theta^a T^a}\phi$, but how does this new matrix field object transform? Recall the two relations $[T^a,T^b] = f^{abc}T^c$ and $f^{abc}$ is a set of constants, and the second relation $(T^a_\textnormal{adj})^{bc} = -if^{abc}$. Now, \cite{maggiore2004modern} is a little quick to derive the tranformation property, so let's do it carefully
\begin{align}
\Phi &\rightarrow \Phi + \delta\Phi = \Phi+ \delta\phi^a T^a = \Phi+ig\theta^b(T_\textnormal{adj}^b)^{ac}\phi^cT^a\\
& = \Phi+g\theta^b f^{bac}\phi^cT^a  = \Phi+ig[T^b,T^c]\theta^b\phi^c = \Phi+ig\theta^bT^b \Phi -i g\Phi\theta^bT^b\\
& = (1 +ig\theta^aT^a)\Phi(1 -i g\theta^aT^a)\\
&\rightarrow e^{ig\theta^aT^a}\Phi e^{-ig\theta^aT^a} \; \text{as} \; \mathcal{O}(\theta^2)\rightarrow 0\\
& = D\Phi(x)D^{-1}
\label{transformtensor}
\end{align}
This is the transformation law for $\Phi$, where $D$ is actually the \textit{same} matrix as in \cref{eq:transformvector}.

What is the point of creating such an object as $\Phi$? Because in the real world, some quantities are best described by the handy notation of tensors, and we need to know how they transform in representation theory. To find out, we transfer from the mathematician's picture - the "vector picture" where the transformation matrices acting on a vector change in each representation but the transformation rule is always the same - to the physicist's picture - the "tensor picture" where the transformation matrices acting on a tensor are always the same but the transformation rule changes.
%
%
\chapter{The Coleman-Weinberg Process}
\label{chp:colemanweinberg}

\section{A Classical Source}
\label{sec:classicalsource}

%
%
%
%


In the following, I detail the semi-classical argument to derive an effective potential as the sum of 1-particle irreducible (1PI) diagrams with vanishing external momenta. Though rather dry, this conclusion is integral to the concept of vacuum misalignment, and several steps in the argument are frequently ignored by textbooks. In particular, the final connection between the expansion of the effective action in terms of loops, and an expansion in terms of momenta, I have only found in one location.

Consider the Lagrangian, motivated by the notes \cite{lukenotes},
\begin{align}
\mathcal{L}_{\phi - \rho} &= (\partial^\mu \phi^\dagger)(\partial_\mu \phi) - \rho(x)\phi(x)\, .
\end{align}
Then, the Euler-Lagrange equations give
\begin{align}
0 = \partial^\mu \frac{\partial \mathcal{L}}{\partial(\partial^\mu \phi^\dagger)} - \frac{\partial \mathcal{L}}{\partial \phi^\dagger}
& = \partial^\mu(\partial_\mu) + \rho(x) \nonumber \\
\implies \nabla^2 \phi - \frac{\partial^2 \phi}{\partial t^2} &= -\rho \, , \label{eq:EL_charge}
\end{align}
which is the wave equation generated by Maxwell's Equations in the presence of a source $\rho$. That is, $\phi$ describes the waves generated when the stone $\rho$ is dropped. This would generically be called a current $J_\mu(x)$, although as it linearly couples to a scalar field it can be considered the time component $J_0(x) = \rho(x)$ of the current  - a charge distribution. 


Why should one consider a source such as this? The reason is simply that it makes a quantum field theory interesting \cite{cardynotes}. Consider the expectation value of a field $\phi$. To get this, we need the path integral 
\begin{align}
\int D\phi \me^{(i/h)S} & = \int D\phi \me^{(i/h)\int \dd t\int \dd^D x\mathcal{L}}
\end{align}
We should integrate over infinite space, given that the field exists at all points, and relativistically over infinite time:
\begin{align}
\int D\phi \me^{(i/h)S} &= \int D\phi \me^{(i/h)\int^\infty_{-\infty} \dd t\int^\infty_{-\infty} \dd^D x\mathcal{L}} \nonumber\\
&= \int D\phi \me^{(-1/h)\int^\infty_{-\infty} \dd \tau\int^\infty_{-\infty} \dd^D x\mathcal{L}}
\end{align}
after Wick rotating to Euclidean space-time. We would like to know the probability matrix of this field transitioning to one state or another
\begin{align}
\lim_{\tau_f - \tau_i \rightarrow \infty} \left\langle n_i |n_f \right\rangle &= \lim_{\tau_f - \tau_i \rightarrow \infty}\sum_n \me^{-E_n(\tau_f - \tau_i)} \left\langle n|n\right\rangle \nonumber\\
& \approx \me^{-E_0 (\tau_f -\tau_i)}\left\langle 0 | 0 \right\rangle 
\end{align}
but this just says that our system is approximately in a vacuum state at all points in space-time: not very interesting. 

\begin{tcolorbox}[colback = black!2!white]

\paragraph{Dictionary of Path Integral Formalism}
$\;$\\

{\small
\textit{Generating functional} $Z$ \\
A functional integral over the exponent of (the action + a \textit{classical source}). It is differentiated to generate (time-ordered) expectation values and correlation values. 
\\

\textit{Classical source} $J(x)$ \\
If $J(x)$ couples to $\varphi(x)$, then $\varphi(x)$ are the waves, $J(x)$ is the pebble dropped into the pond. Take $\mathcal{L} = (\partial \varphi)^2 - J(x)\varphi$, apply the Euler-Lagrange equations, and see that $\nabla^2 \varphi - \frac{\partial^2 \varphi}{\partial t^2} = - J$.
\\

\textit{Partition function} $Z[J]$ \\
The path integral of the action and interaction of a quantum field 
\begin{align}
Z[J] := \int D\phi \me^{iS + i \int d^d x J(x)\phi(x)}
\end{align}

\textit{Schwinger functional} $W[J]$ \\
$W[J] := \log Z[J]$. This relation generates \textit{connected} correlation functions $\langle \varphi(x_1)... \varphi(x_N)\rangle_c$.
\\

\textit{Green's function} $G^{(n)}$ \\
For a path integral of the sort
\begin{align}
\int \mathcal{D}^N \varphi \me^{i A_{ab}\varphi^a \varphi^b + i J_a \varphi^a}
\end{align}
$G_0^{ab}$ is the inverse function of $A$ (i.e. the free theory),
\begin{align}
G_0^{ab} A_{bc} &= \partial^a_c \; \implies \; Z[J] &= \me^{-i G_0^{ab} J_a J_b/2}
\end{align}
for quadratic theory. Then
\begin{align}
G^{(n)}(x_1,...,x_n) = Z^{-1}[J] \frac{\delta^n}{\delta J(x_1) ... \delta J(x_n)} Z[J]
\end{align}
\\

\textit{Connected Green's function} $G_c$ \\
Obtained from the Schwinger functional
\begin{align}
G_c^{(n)}(x_1,...,x_n) = \frac{\delta}{\delta J(x_1)} ...\frac{\delta}{\delta J(x_n)} W[J]
\end{align}
\\

\textit{Vertex function} \\
A 1PI amputated Green's function, generated by the effective action
\begin{align}
\Gamma[\phi] = W[J] - \int d^4 x \frac{\partial W[J]}{\partial J(x)} J(x)
\end{align}
\\

\textit{Physical propagator} \\
The connected two-point function
\begin{align}
G_c^{(2)} &= \Pi(x-y) =  \langle \Omega | T \phi(x_1) \phi(x_2) S| \Omega\rangle = \left( \frac{\delta^2 \Gamma}{\delta \phi(x_1) \delta \phi(x_2)}\right)^{-1}
\end{align}
}
\end{tcolorbox}

\section{The Generating Functional}

Following \cite{cardynotes}, we introduce a source term $J(x)$ into the Lagrangian and call the path integral of this a "generating functional" $Z[J]$, since it \textit{generates} excitations of the field. Specifically $Z[J]$ is the "partition function",
\begin{align}
Z[J] := \int D\phi \me^{iS + i \int d^d x J(x)\phi(x)}\label{eq:prop_to_expectation_value}
\end{align}
As an aside, note that the functional version of the EL equation of motion in \cref{eq:EL_charge} is given by
\begin{align}
\frac{\delta S[\phi]}{\delta\phi} &= -J \, . \label{eq:EL_functional}
\end{align}
We are able to \textit{generate} physical observables by taking functional derivatives of the functional $Z$. The expectation value at $\phi(x_1)$ is proportional to
\begin{align}
\frac{\delta Z[J]}{\delta J(x_1)}\bigm\lvert_{J=0} &= \int [D\phi] \phi(x_1) \me^{iS[\phi]} \propto \left\langle \phi(x_1) \right\rangle \, .
\end{align}
The correlators can be found with successive functional derivatives
\begin{align}
\frac{\delta ^2 Z[J]}{\delta J(x_1) \delta J(x_2)}\bigm\lvert_{J=0} &= \int [D\phi]\phi(x_1) \phi(x_2) \me^{iS[\phi]} \nonumber\\
& \propto \left\langle \phi(x_1) \phi(x_2) \right\rangle \equiv \langle 0 | T[\hat{\phi}(x_1)\hat{\phi}(x_2)]|0\rangle \, ,
\end{align}
where, for reference, I have included the formal definition of this expectation value as a time ordered expectation of operators on the vacuum. 

We should expect the transition amplitude from the vacuum in the distant past $t_i =-\infty$ to the distant future $t_f = \infty$ to be $1$. So we normalise by the vacuum expectation value of the field $\left\langle \phi(x_i)\right\rangle = Z[0]$.  But we can define a functional $W[J]$ - the \textit{Schwinger} functional - that gives this behaviour automatically
\begin{align}
W[J] &:= \log Z[J]\\
\implies \frac{\delta W[J]}{\delta J(x_1)}\bigm\lvert_{J=0} &= \left(\frac{\delta W[J]}{\delta Z[J]}\frac{\delta Z[J]}{\delta J(x_1)}\right)\bigm\lvert_{J=0}\\
& = \frac{1}{Z[0]}\frac{\delta Z[J]}{\delta J(x_1)}\bigm\lvert_{J=0}
\end{align}
It can be shown that this functional generates only \textit{connected} correlation functions. It can be related to the connected n-point functions and Greens functions by
\begin{align}
G^{(N)}(x_1, ..., x_N)_c &= \left\langle \phi(x_1) ... \phi(x_N)\right\rangle_c\\
&= \frac{\delta W[J]}{\delta Z[J]}\left(\frac{\delta Z^N[J]}{\delta J(x_1)...\delta J(x_N)}\right)\bigm\lvert_{J=0}\\
&= \frac{1}{Z[0]}\left(\frac{\delta Z^N[J]}{\delta J(x_1)...\delta J(x_N)}\right)\bigm\lvert_{J=0}\\
\end{align}
Therefore, given that the Taylor \textit{functional} expansion for functional $f[g]$ around point $x_0$ is defined by 
\begin{align}
f[g] = f[g(x_0)] + \int \dd x \frac{\delta f[g(x)]}{\delta g(x)}\bigm\lvert_{g = g(x_0)} (g(x) - g(x_0)) + ...
\end{align}
then we identify $f \rightarrow W, g\rightarrow J, g(x_0) \rightarrow 0$ and get the expansion around $J = 0$ to be
\begin{align}
W = \sum\limits_{N=1}^\infty \frac{1}{N!} \int d^4 x_1 ... d^4 x^N G^N(x_1,...,x_N)J(x_1)...J(x_N) \label{eq:schwinger_expansion}
\end{align}

\section{Effective Potential}\label{sec:CW_potential_appendix}

We define the \textit{mean field} (or \textit{classical field}) $\phi_c(x)$ as the field expectation as a function of spacetime, where
\begin{align}
\phi_c (x) & := \frac{\delta W}{\delta J(x)} = \frac{1}{Z}\frac{\delta Z[J]}{\delta J(x)} \label{eq:phi_c_def}\\
&= \frac{1}{Z}\int D\phi \me^{i[S[\phi] + J\phi]} \phi(x) = \langle 0 | \hat{\phi}(x) | 0 \rangle \; \text{as} \; J(x) \rightarrow 0 \, ,  \nonumber
\end{align}
using the notation of Zee \cite{zee2010quantum}, where $J\phi \equiv \int d^4 x J(x)\phi(x)$. Note that this gives the correct normalisation for the expectation value, where the relation in \cref{eq:prop_to_expectation_value} was only a proportionality. 

Then, as in the original work \cite{coleman1973}, define the \textit{effective action} as a function of $W$ 
\begin{align}
\Gamma[\phi_c] & := W[J] - \int d^4 x J(x) \phi_c(x) \nonumber \\
\end{align}
The definition gives (with appropriate use of the functional chain rule and product rule, and the definition of \cref{eq:phi_c_def})
\begin{align}
\frac{\delta \Gamma[\phi_c]}{\delta\phi_c(x)}&= \int d^4 x \frac{\delta W}{\delta J}\frac{\delta J}{\delta \phi_c} - \int d^4 x \phi_c \frac{\delta J}{\delta \phi_c} - J \nonumber \\
&= \int d^4 x \phi_c \frac{\delta J}{\delta \phi_c} - \int d^4 x \phi_c \frac{\delta J}{\delta \phi_c} - J =  -J(x) \, .
\end{align}
This shows why we refer to $\Gamma$ as the effective action - this is the quantum analogue to the classical equation of motion in \cref{eq:EL_functional}. 
We can also expand in orders of momentum (in position space)
\begin{align}
\Gamma = \int d^4 x \left\lbrace -V(\phi_c) + \frac{1}{2}(\partial_\mu \phi_c)^2 Z(\phi_c) + ... \right\rbrace \, ,\label{eq:gamma_momentum_expansion}
\end{align}
where $V(\phi_c)$ is called the \textit{effective potential}. Again, this is an intuitive labelling in analogy with the zero-derivative expansion term of the classical action. Spontaneous symmetry breaking occurs if 
\begin{align}
\frac{\delta \Gamma}{\delta \phi_c} &= 0 \; \; \text{and thus} \; \;  \frac{\partial V }{\partial \phi_c} = 0 \; \; \text{for} \; \; \phi_c \neq 0
\end{align}

Typically, textbooks would simply state this the effective potential in \cref{eq:gamma_momentum_expansion} is the sum of 1PI diagrams with vanishing external momenta. But it is important to see how this epiphany is reached. We draw on the brilliant \cite{bhattacharjee2013quantum} to observe the following. Recall that the expansion of $W$ in \cref{eq:schwinger_expansion} gave a sum of connected diagrams, we can do the same for $\Gamma$. The expansion is in terms of 1PI Green's functions\footnote{The derivation of this is given in many places, e.g. \cite{Peskin:1995ev}} $\Gamma^{(N)}$
\begin{align}
\Gamma[\phi_c] &= \sum\limits_{N=1}^\infty \frac{1}{N!} \int d^4 x_1 ... d^4 x_N \Gamma^{(N)}(x_1...x_N)\phi_c(x_1)... \phi_c(x_N) \, .\label{eq:gamma_phi_expansion}
\end{align}
Fourier transforming to momentum space, we have that each 1PI contribution can be expressed as
\begin{align}
\Gamma^{(N)}(x_1...x_N) &= \int \frac{d^4 p_1}{(2\pi)^4} ... \frac{d^4 p_N}{(2\pi)^4} (2\pi)^4 \delta^4 (p_1 + ... + p_N) \me^{i(p_1 \cdot x_1 +... + p_N \cdot  x_N)} \nonumber \\
& \qquad \qquad \times \Gamma^{(N)}(p_1...p_N)
\end{align}
and then we can expand each of \textit{these} $ \Gamma^{(N)}(p_1...p_N)$ in powers of momenta, to find the effective action can be given as
\begin{align}
\hspace*{-1cm}\Gamma[\phi_c] &= \sum\limits_{N=1}^\infty \frac{1}{N!} \int d^4 x_1 ... d^4 x_N \int \frac{d^4 p_1}{(2\pi)^4} ... \frac{d^4 p_N}{(2\pi)^4} \int d^4 x \me^{i(p_1 + ... + p_N) \cdot x} \me^{i(p_1 \cdot x_1 +... + p_N\cdot  x_N)} \nonumber \\
\hspace*{-1cm} & \qquad \qquad \times \left[\Gamma^{(N)}(0,...,0) + ... \right]\phi_c(x_1)... \phi_c(x_N)  \nonumber \\
\hspace*{-1cm} &= \int d^4 x \sum\limits_{N=1}^\infty \frac{1}{N!} \left[ \Gamma^{(N)}(0,...,0)\left(\phi_c(x)\right)^N + ... \right]\\
\hspace*{-1cm} \implies V(\phi_c) &= -\sum\limits_{N=1}^\infty \frac{1}{N!}\Gamma^{(N)}(0,...,0)\left(\phi_c(x)\right)^N \nonumber
\end{align}
where contracting over the many momentum and position variables is menial, and we have Fourier transformed the momentum delta function $(2\pi)^4\delta(p_1 + ... + p_N)$ back to position space. Thus, we see that the effective potential is given by summing each set of 1PI diagrams with $N$ external momenta taken to zero. At this point, the usual Feynman rules can be applied for loop corrections based on the propagators and vertices found in the Lagrangian.

\
\chapter{Large N Approximation}
\label{sec:large_N}

For many purposes, it is useful to know how a QCD-like model behaves for number of colours $N \gg 1$. For example, in this thesis, we naively assume in \cref{sec:MCHM} that the underlying fields constituting the composite particles obey a large-N approximation. The following is based on the work of \cite{Witten:1979kh} and \cite{Hooft1974461}. I will attempt to quickly summarise the logic of these works, using diagrams borrowed from \cite{Witten:1979kh}. The reasoning is based on combinatorics and toying with Feynman diagrams. 

\begin{figure}
\centering
\includegraphics[scale=1]{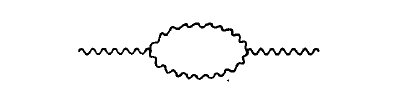}
\caption{Gluon one-loop vacuum polarisation diagram}\label{fig:oneloopsingle}
\end{figure}

First note that, with QCD as a prototype, we would expect the two-point function of a single, tree-level meson current to be
\begin{align}
\frac{f_n^2}{p^2 - m_n^2}
\end{align}
Now, consider the loop diagrams contributing to the current $\langle J_\mu^a J_\nu^a \rangle$ of a composite meson. We will use the QCD gluon as a prototype. For the large-N limit to be smooth, one requires that the loop gluon interaction amplitudes to be independent of the number of colours, otherwise they would blow up to infinity. We propose that this requires a gluon-gluon coupling of $g/\sqrt{N}$, where each vertex in, for example \cref{fig:oneloopsingle}, carries this factor.

\begin{figure}
\centering
\includegraphics[scale=1]{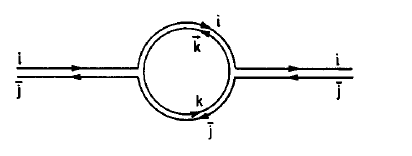}
\caption{The gluon vacuum polarisation diagram of \cref{fig:oneloopsingle} in terms of colour-carrying fermions}\label{fig:oneloopdouble}
\end{figure}

To see why, imagine that a gluon can be represented as a quark-antiquark pair, as in \cref{fig:oneloopdouble}. This is to understand how the colours move through a diagram, and a gluon has the same colour quantum numbers as a meson. Then instead of a squiggly line, we would have two antiparallel arrows.

\begin{figure}
\centering
\includegraphics[scale=1]{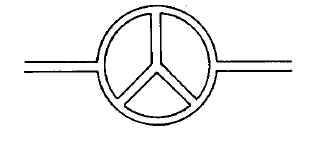}
\caption{A planar diagram of 3-loop contribution to vacuum polarisation}\label{fig:threeloop}
\end{figure}

\begin{figure}
\centering
\includegraphics[scale=1]{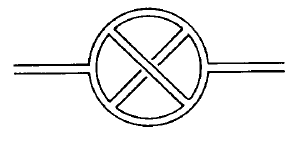}\caption{A non-planar diagram of 2-loop contribution to vacuum polarisation}\label{fig:twoloop}
\end{figure}

Each line carries a colour, so with a hidden colour loop there are N ways of making this diagram. Therefore, with two vertices, the amplitude goes as $(g/\sqrt{N})^2 N = g^2$. Thus, we have a well-behaved theory. Many diagrams survive at large N. For example, the diagram in \cref{fig:threeloop} goes as $(g/\sqrt{N})^6N^3 = g^6$ (three free loops, six vertices). However, the diagram in \cref{fig:twoloop} vanishes at large N. It goes as $(g/\sqrt{N})^6N^2 = g^6/N \rightarrow 0$. This diagram is different because it has overlapping lines - it could not be drawn on a plane. 't Hooft claims that all "planar" diagrams survive in large-N, while non-planar do not. A similar argument (i.e. using the meson lines as gluons) applies to quark loops within a gluon propagator, and to gluon and quark loops in a quark propagator - simply draw a few and count the hidden loops and vertices.

\begin{figure}
\centering
\includegraphics[scale=0.6]{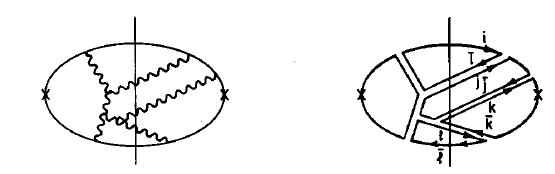}
\caption{A higher-order correction to the vacuum polarisation, which can be cut, leaving only $q\bar{q}$ pairs}\label{fig:quarkcut}
\end{figure}

Looking closely, one can see that cutting a quark-antiquark two-point function (which is what our current represents) generates only $\bar{q}q$ pairs, as in \cref{fig:quarkcut}. This means that a generic current at large N is simply the sum $\sum\limits_n$ of pairs of $\bar{q}q$ meson propagators each with amplitude $\frac{f^2}{p^2-m^2}$ \cite{Witten:1979kh}. Each of these need to be projected in the same direction, in order to be summed. Thus, we have the total amplitude at leading order in $\frac{1}{N}$
\begin{align}
\lim_{N\rightarrow\infty} <J^a_\mu J^a_\nu> = (p^2 \eta_\mu\nu - p_\mu p_\nu) \sum\limits_n\frac{f_n^2}{p^2 - m_n^2}
\end{align}
noted in \ref{sec:MCHM}. Importantly, we could only deduce this by assuming the coupling depended on $1/\sqrt{N}$ and thus sending most other, more complicated, diagrams to zero.

\chapter{Tuning Background: Bayes' Theorem \& Volume Elements}
\label{sec:tuning_appendix}

\section{Bayes' Theorem}\label{sec:bayes_theorem}

\begin{figure}[H]
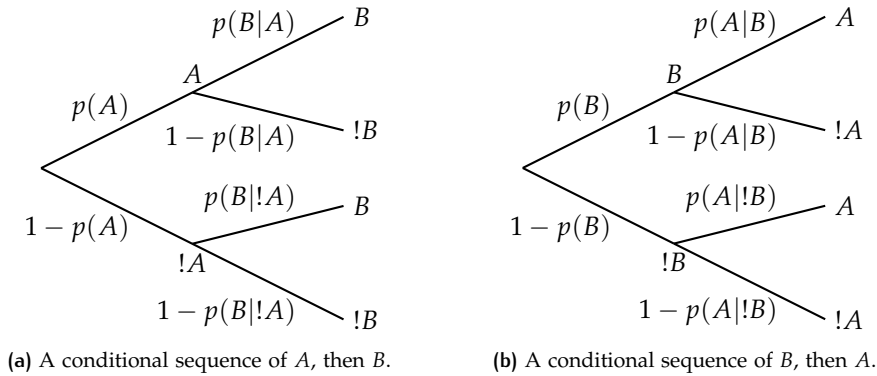

\centering
\subfloat[A conditional sequence of $A$, then $B$. \label{fig:markov_a}]{
\tikz{
\draw [thick] (0,3) -- (2,4);
\draw [thick] (0,3) -- (2,2);
\node [above, left] at (1.3,3.8) {$p(A)$};
\node [below, left] at (1.3,2.2) {$1-p(A)$};
\node [above] at (2,4) {$A$};
\node [below] at (2,2) {$!A$};
\draw [thick] (2,4) -- (4,5);
\draw [thick] (2,4) --(4,3.5);
\draw [thick] (2,2) -- (4,2.5);
\draw [thick] (2,2) -- (4,1);
\node [above, left] at (3.5,4.9) {$p(B|A)$};
\node [below, left] at (3.5,3.4) {$1-p(B|A)$};
\node [above, left] at (3.5,2.6) {$p(B|!A)$};
\node [below, left] at (3.5,1.1) {$1-p(B|!A)$};
\node [right] at (4,5) {$B$};
\node [right] at (4,3.5) {$!B$};
\node [right] at (4,2.5) {$B$};
\node [right] at (4,1) {$!B$};
}}\qquad\qquad
\subfloat[A conditional sequence of $B$, then $A$. \label{fig:markov_b}]{
\tikz{
\draw [thick] (0,3) -- (2,4);
\draw [thick] (0,3) -- (2,2);
\node [above, left] at (1.3,3.8) {$p(B)$};
\node [below, left] at (1.3,2.2) {$1-p(B)$};
\node [above] at (2,4) {$B$};
\node [below] at (2,2) {$!B$};
\draw [thick] (2,4) -- (4,5);
\draw [thick] (2,4) --(4,3.5);
\draw [thick] (2,2) -- (4,2.5);
\draw [thick] (2,2) -- (4,1);
\node [above, left] at (3.5,4.9) {$p(A | B)$};
\node [below, left] at (3.5,3.4) {$1-p(A | B)$};
\node [above, left] at (3.5,2.6) {$p(A | !B)$};
\node [below, left] at (3.5,1.1) {$1-p(A | !B)$};
\node [right] at (4,5) {$A$};
\node [right] at (4,3.5) {$!A$};
\node [right] at (4,2.5) {$A$};
\node [right] at (4,1) {$!A$};
}}\caption{The two possible descriptions of correlated events $A$ and $B$.}\label{fig:markov}
\end{figure}

Bayes' Theorem emerges from the thought experiment of two correlated events, $A$ and $B$. Consider them happening in the order of $A$ then $B$. We can draw this as a Markov chain of events, as in \cref{fig:markov_a}. The probability of $A$ occurring is $p(A)$, and the probability of $B$ occurring given that $A$ has already occurred is $p(B|A)$. The rules of a Markov chain tell us to multiply these two to get the probability that both $A$ and $B$ occur
\begin{align}
p( A \cap B) = p(B|A) p(A)
\end{align}
Note, however, that we can switch the order of $A$ and $B$, as there is nothing inherently causal in the definition. Then we have that
\begin{align}
p( A \cap B) = p(A|B) p(B)
\end{align}
and we can arrive at Bayes' Theorem
\begin{align}
p(B | A) = \frac{p(A | B) p(B)}{p(A)}
\end{align}

\section{Volumes in 3-space}

Let $U$ be a vector space over ${\rm I\!R}^3$, with a basis $\{\hat{e}_1,\hat{e}_2,\hat{e}_3\}$. We thus denote a vector $\textbf{u}$ in $U$ as $\textbf{u} = u^1 \hat{e}_1 + u^2 \hat{e}_2 + u^3 \hat{e}_3 = u^a\hat{e}_a \in U$, or in index notation $u^i = u^a\hat{e}^i_a$, where $\hat{e}^i_a$ is the i-th component of the a-th basis vector.

Since the purpose of this section is to build intuition, let's streamline our notation by defining $\hat{e}_1 = (1,0,0),\; \hat{e}_2 = (0,1,0),\; \hat{e}_3 = (0,0,1)$ then $\textbf{u}=(u^1,u^2,u^3)$, since $\hat{e}^i_a = \delta^i_a$. But when generalising to n-space, it will be useful to return to the more general case.

Now let a parallelogram $P$ in $U$ be defined as the space spanned by the vectors $\textbf{u}_1,\textbf{u}_2$. The area of $P$ is given by the magnitude of the cross product
\begin{align}
\textnormal{Area}(P) &= |\textbf{u}_1\times\textbf{u}_2| =|\epsilon_{ijk} u^j_1 u^k_2| =\sqrt{\epsilon_{ijk}u^j_1 u^k_2 \ \epsilon^{ilm}u_{1l} u_{2m}} \nonumber\\
& = \sqrt{(\delta^l_j\delta^m_k - \delta^m_j\delta^l_k) \ u^j_1 u^k_2 u_{1l} u_{2m}} =\sqrt{\textbf{u}_1^2 \textbf{u}_2 ^2 - (\textbf{u}_1 \cdot \textbf{u}_2)^2 } \nonumber\\
&= \begin{vmatrix}
\textbf{u}_1^2 & \textbf{u}_1 \cdot \textbf{u}_2\\
\textbf{u}_1 \cdot \textbf{u}_2 & \textbf{u}_2 ^2
\end{vmatrix}^{1/2}=\begin{vmatrix}
\left(\begin{matrix}
\textbf{u}_1 \\
\textbf{u}_2
\end{matrix}\right)\left(\begin{matrix}
\textbf{u}_1^T & \textbf{u}_2^T
\end{matrix}\right)
\end{vmatrix}^{1/2}\\
&=\sqrt{\det(AA^T)} \, ,\nonumber
\end{align}
where we have defined the matrix $A=\left(\begin{matrix}
\textbf{u}_1, \cdots , \textbf{u}_m \end{matrix}\right)^\top$, which will be used with the same notation from here on in. Clearly, this matrix is useful for quickly finding the area of a parallelogram. Formally, $AA^T$ is the Gram matrix of the vectors ${\textbf{u}_1,...,\textbf{u}_m}$. The determinant $\det(AA^T)$ is known as the Gramian. It gives the square of the area spanned by vectors. The same procedure works for the volume of a parallelpiped $\mathcal{P}$. That is, where there are three $\textbf{u}$ vectors
\begin{align}
\textnormal{Vol}(\mathcal{P}) &= \textbf{u}_1 \cdot (\textbf{u}_2 \times \textbf{u}_3) = \epsilon_{ijk} \ u_1^i u_2^j u_3^k = \sqrt{\epsilon_{ijk} \ u_1^i u_2^j u_3^k \ \epsilon^{lmn} \ u_{1l} u_{2m} u_{3n}} \nonumber\\
& =\left(\begin{vmatrix}
\delta^l_i & \delta^m_i & \delta^n_i\\
\delta^l_j & \delta^m_j & \delta^n_j\\
\delta^l_k & \delta^m_k & \delta^n_k
\end{vmatrix}u_1^i u_2^j u_3^k u_{1l} u_{2m} u_{3n}\right)^{1/2}
= \begin{vmatrix}
\textbf{u}_1^2 & \textbf{u}_1 \cdot \textbf{u}_2 & \textbf{u}_1 \cdot \textbf{u}_3\\
\textbf{u}_1 \cdot \textbf{u}_2 & \textbf{u}_2^2 & \textbf{u}_2 \cdot \textbf{u}_3\\
\textbf{u}_1 \cdot \textbf{u}_3 & \textbf{u}_2 \cdot \textbf{u}_3 & \textbf{u}_3^2
\end{vmatrix} \nonumber\\
&=\sqrt{\det(AA^T)} \label{eq:volume_three_gramian}
\end{align}
Thus the $\sqrt{\det(AA^T)}$ Gramian formula defines a volume regardless of dimension. To be more precise, a 2-volume (i.e. an area) in 3-space, requires contraction with a 3-index Levi-Civita tensor. In general, the (m-k)-tensor for a k-volume in m-space requires each of the k edges to be contracted with an m-index Levi-Civita tensor.  

\section{Mapping a Volume}

Now we know how the scalar value representing the area or volume is found in 3-space, we will see how this value changes under a mapping to 1, 2 or 3-space. 

Let $\textbf{u}_a = \{\textbf{u}_1,\textbf{u}_2,\textbf{u}_3\}\in U$ be the edges of a parallelepiped in 3-space. Consider a (not necessarily invertible) mapping $f: \; U\rightarrow V, \; f(\textbf{u}_a) \rightarrow \textbf{v}_a \in V$. 

We cannot, in general, say how a volume spanned by $\textbf{u}_a$ maps under $f$. However, we know how an infinitesimal edge transforms under this mapping:

\begin{align}
dv_a^i = \frac{\partial v_a^i}{\partial u_a^j} du_a^j
\end{align}

where $dv^i_a$ is the i-th component of the a-th edge. Clearly we are not summing over the $a$ index on the RHS, only the $j$ index as required by the total derivative formula.

The matrix of coefficients of $du_a^j$ is the a-th Jacobian $(J_a)^i_j$. Now we can see how an infinitesimal volume element $dP$ maps in 3-space
\begin{align}
\textnormal{Vol}(f(dP)) &= \epsilon_{ijk}dv_1^i dv_2^j dv_3^k \nonumber\\
&\left( = \frac{1}{3!}\epsilon^{abc}\epsilon_{ijk}dv_a^i dv_b^j dv_c^k \; \right) \label{eq:three_volume_indices}\\
&= \epsilon_{ijk} (J_1)_l^i du^l_1 (J_2)_m^j du^m_2 (J_3)_n^k du_3^n \nonumber\\
&=\sqrt{\epsilon_{ijk}\epsilon_{opq} (J_1)_l^i du^l_1 (J_2)_m^j du^m_2 (J_3)_n^k du_3^n (J_1)_r^o du^r_1 (J_2)_s^p du^s_2 (J_3)_t^q du_3^t } \nonumber\\
&=\sqrt{\begin{vmatrix}
\delta^o_i & \delta^p_i & \delta^q_i\\
\delta^o_j & \delta^p_j & \delta^q_j\\
\delta^o_k & \delta^p_k & \delta^q_k
\end{vmatrix}(J_1)_l^i (J_2)_m^j (J_3)_n^k (J_1)_r^o  (J_2)_s^p (J_3)_t^q du^l_1 du^m_2 du_3^n du^r_1 du^s_2 du_3^t}\nonumber\\
&=\sqrt{\begin{vmatrix}
(J_1)_l^o & (J_1)_l^p & (J_1)_l^q\\
(J_2)_m^o & (J_2)_m^p & (J_2)_m^q\\
(J_3)_n^o  & (J_3)_n^p  & (J_3)_n^q
\end{vmatrix} (J_1)_r^o  (J_2)_s^p (J_3)_t^q du^l_1 du^m_2 du_3^n du^r_1 du^s_2 du_3^t}\nonumber
\end{align} 

We are almost there, but at this point, the index sums don't allow a simplification of the final line in \cref{eq:three_volume_indices}. To go on, we need our $\textbf{du}_a$ to be orthogonal. Let us simply assume this, but it remains to be shown how the process works if they are not orthogonal. Then, given an orthogonal set $\textbf{du}_a = \{(du_1^1,0,0,...),(0,du_2^2,0,...)\}$ we are able to set $(J_1)^i_l du_1^l = (J_1)^i_1 du^1_1$. Now, $du^1_1$ is just a scalar, and it can be factored out of the square root
\begin{align}
\textnormal{Vol}(f(dP)) &= \sqrt{\begin{vmatrix} 
(J_1)_1^o (J_1)_1^o  & (J_1)_1^p (J_2)_2^p & (J_1)_1^q (J_3)_3^q\\
(J_2)_2^o (J_1)_1^o & (J_2)_2^p (J_2)_2^p & (J_2)_2^q (J_3)_3^q\\
(J_3)_3^o (J_1)_1^o  & (J_3)_3^p (J_2)_2^p  & (J_3)_3^q (J_3)_3^q
\end{vmatrix} du^1_1 du^2_2 du^3_3 du^1_1 du^2_2 du^3_3} \nonumber\\
&= \sqrt{|JJ^T| (\textnormal{Vol}(du))^2}
\end{align} 

This formula is the analogue of the above \cref{eq:volume_three_gramian} for a volume spanned by three vectors. But now, rather than a vector $A$ of vectors $\textbf{u}_a$, we take the Gramian of a vector $J$ of vectors $\textbf{J}_a=\frac{\partial v_a}{\partial \textbf{u}}$
\begin{align}
\textnormal{Gram}(J) &= |JJ^T| \nonumber\\
&= \begin{vmatrix}
\textbf{J}_1 \cdot \textbf{J}_1 & \textbf{J}_1 \cdot \textbf{J}_2 & \textbf{J}_1 \cdot \textbf{J}_3\\
\textbf{J}_2 \cdot \textbf{J}_1 & \textbf{J}_2 \cdot \textbf{J}_2 & \textbf{J}_2 \cdot \textbf{J}_3\\
\textbf{J}_3 \cdot \textbf{J}_1 & \textbf{J}_3 \cdot \textbf{J}_2 & \textbf{J}_3 \cdot \textbf{J}_3
\end{vmatrix}
\end{align}
We now see how to take the ratio of infinitesimal volumes, which is our definition of fine tuning $\mathcal{F}$. Combine the results of sections 2.1 and 2.2 to find the ratio of infinitesimal observable space $dO = f(dP)$ to infinitesimal parameter space $dP$:
\begin{align}
\frac{\textnormal{Vol}(dO)}{\textnormal{Vol}(du)} = |JJ^T| = \Delta
\end{align}
which we define as the maximal tuning, for orthogonal basis vectors.

\chapter{Particle Content}
\label{chp:particle_content}

\section{SO(5) Fermion Embeddings}

\subsection{Generators}

We choose a basis $\tilde{T}^A = \{T_L^a, T_R^a, X^i\}$ for the Lie algebra of $SO(5)$ to be defined by the formulae
\begin{align}
\left[T^a_L\right]_{IJ} &= \frac{i}{2} \left[ \frac{1}{2} \epsilon^{abc} \left( \delta^b_I \delta^c_J - \delta^c_I \delta^b_J \right) + \left( \delta^a_I \delta^4_J - \delta^4_I \delta^a_J \right) \right] \nonumber\\ 
\left[T^a_R\right]_{IJ} &= \frac{i}{2} \left[ \frac{1}{2} \epsilon^{abc} \left( \delta^b_I \delta^c_J - \delta^c_I \delta^b_J \right) - \left( \delta^a_I \delta^4_J - \delta^4_I \delta^a_J \right) \right] \label{eq:SO(5)_generator_formulae}\\
\left[X^a\right]_{IJ} &= \frac{i}{2} \left( \delta^a_I \delta^5_J - \delta^5_I \delta^a_J \right) \nonumber
\end{align}

\subsection{Fundamental Embedding}

In this section, and the following, we choose bases that consists of eigenvectors of $T_L^3$ and $T_R^3$ transformations. This allows us to identify each field with either a partner that can couple to a SM field, or a SM field itself. These fields are constructed as in \cref{sec:rep_example}, where each field is contracted with a basis vector, and thus transforms as a vector itself
\begin{align}
\Psi_\textbf{5} = \Psi^i v^i_\textbf{5}, && \Psi_\textbf{5} \rightarrow \exp\left( i \lambda^A \tilde{T}^A \right) \Psi_\textbf{5}
\end{align} 
One such basis for the fundamental representation is given by
\begin{align}
\hspace{-1cm} v^{--} &= \frac{1}{\sqrt{2}}\left(\begin{matrix}
i \\
1 \\
0 \\
0 \\
0 
\end{matrix}\right), v^{-+} = \frac{1}{\sqrt{2}}\left(\begin{matrix}
0 \\
0 \\
i \\
1 \\
0 
\end{matrix}\right),
v^{+-} = \frac{1}{\sqrt{2}}\left(\begin{matrix}
0 \\
0 \\
-i \\
1 \\
0 
\end{matrix}\right),
v^{++} = \frac{1}{\sqrt{2}}\left(\begin{matrix}
-i \\
1 \\
0 \\
0 \\
0 
\end{matrix}\right),
v^{00} = \frac{1}{\sqrt{2}}\left(\begin{matrix}
0 \\
0 \\
0 \\
0 \\
1 
\end{matrix}\right)
\end{align}
where, as explained in \cref{sec:fundamental_rep}, the superscripts give the $T_L^3, T_R^3$ quantum numbers of each field, either $+\frac{1}{2}$ or $-\frac{1}{2}$. This reflects the decomposition under $SO(4)$ of the fundamental $\textbf{5} \sim \textbf{4} \oplus \textbf{1}$. Then the fundamental embedding is given by 
\begin{align}
\Psi_\textbf{5} = \left( \begin{array}{ c }
\\
\Psi_\textbf{4} \\ 
\\
\hline
\Psi_\textbf{1}
\end{array} \right), && \text{where,} && \Psi_\textbf{4} = \frac{1}{\sqrt{2}}\left( \begin{matrix}
i \Psi^{--} - i \Psi^{++}\\
\Psi^{--} + \Psi^{++} \\
i\Psi^{-+} - i\Psi^{+-}\\
\Psi^{-+} + \Psi^{+-}
\end{matrix}\right), && \Psi_\textbf{1}= \Psi^{00}\, .
\end{align}

\subsection{Antisymmetric Embedding}

We explored in \cref{sec:deconstructing_tensors} how one can always build higher representations by taking the tensor product of fundamental representations, and separating the resulting tensor into an antisymmetric and symmetric tensor, and a trace. The antisymmetric representation is given by
\begin{align}
\Psi_\textbf{10} = \Psi^i v^i_\textbf{10}, && \Psi_\textbf{10} \rightarrow \exp\left( i \lambda^A \tilde{T}^A \right) \Psi_\textbf{10} \exp\left( -i \lambda^A \tilde{T}^A \right)
\end{align}
where the basis vectors are given in terms of the $SO(5)$ generators. The antisymmetric irrep decomposes under $SO(4)\sim SU(2)_L \times SU(2)_R$ as $\textbf{10} = (\textbf{3},\textbf{1}) \oplus (\textbf{1}, \textbf{3}) \oplus (\textbf{2}, \bar{\textbf{2}})$. Therefore we can identify six triplet basis components
\begin{align}
v^{\pm 1,0} = \frac{1}{\sqrt{2}} (T_L^1 \pm i T_L^2) , && v^{0, \pm 1} = \frac{1}{\sqrt{2}} (T_R^1 \pm i T_R^2) , && v^{0,0}_1 = T_L^3, && v^{0,0}_2 = T_R^3
\end{align}
and four bidoublet components
\begin{align}
v^{\pm \frac{1}{2}, \pm \frac{1}{2}} = \frac{1}{\sqrt{2}} (X^1 \pm i X^2) && v^{\pm \frac{1}{2}, \mp \frac{1}{2}} = \frac{1}{\sqrt{2}}(X^3 \pm i X^4)\, .
\end{align}
Then, we repeat the full form of the antisymmetric embedding
\begin{align}
\Psi_\textbf{10} &= \left( \begin{array}{c | c}
\Psi_\textbf{6} & \Psi_\textbf{4}/\sqrt{2} \\ \hline
- \Psi^T_\textbf{4} / \sqrt{2} & 0
\end{array} \right),  \text{where,} \\
\qquad \Psi_\textbf{6} &= \frac{1}{2}\left( \begin{matrix}
0 & \hat{\Psi}^{0,0}_+  & i (\hat{\Psi}_+^{-1,-1} - \hat{\Psi}_+^{1,1}) & \hat{\Psi}_-^{-1,-1}  + \hat{\Psi}_-^{1,1} \\
 & 0 & \hat{\Psi}_+^{-1,-1}  +  \hat{\Psi}_+^{1,1} & i (- \hat{\Psi}_-^{-1,-1}  + \hat{\Psi}_-^{1,1} ) \\ 
 & & 0 & i \hat{\Psi}^{0,0}_-\\
 & & & 0
\end{matrix}  \right) \nonumber
\end{align}
and $\Psi_\textbf{4}$ is as above. We have defined the convenient combinations 
\begin{align}
\hat{\Psi}^{0,0}_\pm := \Psi^{0,0}_1 \pm \Psi^{0,0}_2, && \hat{\Psi}_\pm^{-1,-1} := \Psi^{-1,0} \pm \Psi^{0,-1}, &&  \hat{\Psi}_\pm^{1,1} := \Psi^{0,1}  \pm \Psi^{1,0}. 
\end{align}

\subsection{Symmetric Embedding}

The symmetric embedding is constructed analogously to the antisymmetric
\begin{align}
\Psi_\textbf{14} = \Psi^i v_\textbf{14}^i, && \Psi_\textbf{14} \rightarrow \exp\left( i \lambda^A \tilde{T}^A \right)  \Psi_\textbf{14} \exp\left( -i \lambda^A \tilde{T}^A \right) 
\end{align}
where the basis must now be given by \textit{symmetric} matrices. First, define a symmetric set of matrices $\hat{T} = \{\hat{T}^{ab}, \hat{X}^a, \hat{T}^0\}$
\begin{align}
[\hat{T}^{ab}]_{IJ} &= \frac{1}{\sqrt{2}}(\delta^a_I \delta^b_J + \delta^b_I \delta^a_J), && a< b \; a,b=1,...,4,\\
[\hat{T}^{aa}]_{IJ} &= \frac{1}{\sqrt{2}} (\delta^a_I \delta^a_J - \delta^{a+1}_I \delta^{a+1}_J), && a=1,2,3,\\
[\hat{X}^a]_{IJ} &= \frac{1}{\sqrt{2}} (\delta^a_I \delta^5_J + \delta^5_I \delta^a_J) , && a = 1,...,4,\\
[\hat{T}^0]_{IJ} &= \frac{1}{2\sqrt{5}}\text{diagonal}(1,1,1,1,-4) \, .
\end{align}
This helps us define the 14 basis matrices of the symmetric representation, which decomposes under $SU(2)_L \times SU(2)_R$ as $\textbf{14} \sim (\textbf{3}, \bar{\textbf{3}}) \oplus (\textbf{2}, \bar{\textbf{2}}) \oplus (\textbf{1}, \textbf{1})$. This is spanned by 9 bi-triplet matrices
\begin{align}
\hspace{-2cm} v^{1,1} &= \frac{1}{2\sqrt{2}}(2i\hat{T}^{12} + \hat{T}^{11} - \hat{T}^{22}), \; & v^{1,0} &= \frac{1}{2}(-\hat{T}^{13} - i\hat{T}^{23} - i \hat{T}^{14} + \hat{T}^{24}), \; & v^{1,-1} &= \frac{1}{2\sqrt{2}}(2i\hat{T}^{34} + \hat{T}^{33}) \nonumber\\
\hspace{-2cm} v_1^{0,0} &= \frac{1}{2\sqrt{2}}(-\hat{T}^{11} - \hat{T}^{22} + \hat{T}^{33}), \;  & v^{0,1} & = \frac{1}{2}(-\hat{T}^{13} - i\hat{T}^{23} + i \hat{T}^{14} - \hat{T}^{24}), \;  & v^{0,-1} &= \frac{1}{2}(\hat{T}^{13} - i\hat{T}^{23} + i \hat{T}^{14} + \hat{T}^{24}) \nonumber\\
\hspace{-2cm} v^{-1,-1} &= \frac{1}{2\sqrt{2}}(-2i\hat{T}^{12} + \hat{T}^{11} - \hat{T}^{22}), \; & v^{-1,0} &= \frac{1}{2}(\hat{T}^{13} - i\hat{T}^{23} - i \hat{T}^{14} - \hat{T}^{24}), \; & v^{-1,1} &= \frac{1}{2\sqrt{2}}(-2i\hat{T}^{34} + \hat{T}^{33})\, ,
\end{align} 
four bidoublet components
\begin{align}
v^{\pm \frac{1}{2}, \pm \frac{1}{2}} = \frac{1}{\sqrt{2}} (\mp \hat{X}^1 - i \hat{X}^2) && v^{\pm \frac{1}{2}, \mp \frac{1}{2}} = \frac{1}{\sqrt{2}}(\hat{X}^3 \pm i \hat{X}^4)\, ,
\end{align}
and a singlet
\begin{align}
v_2^{0,0} = \hat{T}^0
\end{align}
We embed fermions in this representation as
\begin{align}
\Psi_\textbf{14} &= \left( \begin{array}{c | c}
\Psi_\textbf{9} - \Psi_1 \mathbb{1}/(2\sqrt{5}) & \Psi_\textbf{4}/\sqrt{2} \\ \hline
\Psi^T_\textbf{4} / \sqrt{2} & (2/\sqrt{5}) \Psi_1
\end{array} \right), \text{where,} \\
 \qquad \Psi_\textbf{9} &= \frac{1}{2}\left( \begin{matrix}
\hat{\Psi}^{0,0}_{2,+} - \Psi^{0,0}_4 & i\hat{\Psi}^{0,0}_{2,-}  & \hat{\Psi}_+^{1,1} + \hat{\Psi}_+^{-1,-1}  & i(\hat{\Psi}_-^{1,1} - \hat{\Psi}_-^{-1,-1})  \\
 & -\hat{\Psi}^{0,0}_{2,+}  - \Psi^{0,0}_4 & i( \hat{\Psi}_+^{1,1} - \hat{\Psi}_+^{-1,-1} ) &  \hat{\Psi}_-^{1,1}  - \hat{\Psi}_-^{-1,-1} \\ 
 & & \Psi^{0,0}_4 - \hat{\Psi}^{0,0}_{1,-} & i \hat{\Psi}^{0,0}_{1,+} \\
 & & & \Psi^{0,0}_4 + \hat{\Psi}^{0,0}_{1,-} 
\end{matrix}  \right) \nonumber
\end{align}
and the $\Psi_\textbf{4}$ and $\Psi_\textbf{1}$ are as given previously. As in the \textbf{10} case, we have defined some convenient fields: 
\begin{align}
\hat{\Psi}^{0,0}_{1,\pm} := \Psi^{-1,1}_3 \pm \Psi^{1,-1}_4, && \hat{\Psi}^{0,0}_{2,\pm} := \Psi^{1,1} \pm i \Psi^{-1,-1} \\
\hat{\Psi}^{-1,-1}_\pm :=  \frac{1}{\sqrt{2}} (\Psi^{-1,0} \pm i \Psi^{0,-1}), && \hat{\Psi}^{1,1}_\pm  = \frac{1}{\sqrt{2}} (\Psi^{0,1} \pm i \Psi^{1,0})
\end{align}

\section{SO(6) Fermion Embeddings}

\subsection{Generators}

We define $T_L$ and $T_R$ as in \cref{eq:SO(5)_generator_formulae}, albeit with an extra empty row and column. Similarly, we redefine $X$ as $T_X$ with an extra empty row and column, as they are now no longer broken generators, but do not have a defined chirality under $SU(2)_L \times SU(2)_R$. The broken generators are now given by a bidoublet $X_B$ and a singlet $X_S$
\begin{align}
\left[X_B^a\right]_{IJ} &= \frac{i}{2}\left( \delta^a_I \delta^6_J - \delta^6_I \delta^a_J \right) \nonumber\\
\left[X_S \right]_{IJ} &= \frac{i}{2} \left( \delta^5_I \delta^6_J - \delta^6_I \delta^5_J \right)
\end{align}

\subsection{Fundamental Embedding}

We define the fundamental representation $\Psi_\textbf{6}$ as above, by its decomposition under $SO(4)$, $\textbf{6} \sim \textbf{4} \oplus \textbf{1} \oplus \textbf{1}$. The basis is identical to the $SO(5)$ construction, with an extra singlet basis vector. This leads to a fundamental embedding
\begin{align}
\Psi_\textbf{6} = \frac{1}{\sqrt{2}}\left( \begin{matrix}
i \Psi^{--} - i \Psi^{++}\\
\Psi^{--} + \Psi^{++} \\
i\Psi^{-+} - i\Psi^{+-}\\
\Psi^{-+} + \Psi^{+-}\\
\sqrt{2}\Psi_1^{00}\\
\sqrt{2}\Psi_2^{00}
\end{matrix}\right)\, .
\end{align}

\chapter{Form Factors and Mixing Matrices}
\label{sec:form_factors}

\section{Composite Lepton MCHM Expressions}

The source term form factors implicitly defined in eq.s (\ref{eq:Leff}) and (\ref{eq:full_potential}) can be written in terms of the decomposed form factor expressions \ref{app:5-5-5}, \ref{app:14-14-10}, \ref{app:14-1-10}. Each representation's form factors generally depend on the four following functions:
\begin{align}
\begin{split}\label{formulas}
A_L(m_1,m_2,m_3,m_4,\Delta) &= \Delta^2\left(m_1^2 m_2^2 + m_1^2 m_4^2 + m_2^2 m_3^2 - p^2(m_1^2 + m_2^2 + m_3^2 +m_4^2) + p^4\right) \\
A_R(m_1,m_2,m_3,m_4,\Delta) &= \Delta^2 \left( m_1^2 m_2^2 + m_2^2 m_3^2 - p^2(m_1^2 + m_2^2 + m_3^2 + m_4^2) + p^4\right)\\
A_M(m_1,m_2,m_3,m_4,\Delta_1,\Delta_2) &= \Delta_1 \Delta_2 m_1 m_2 m_4 (m_3^2 - p^2)\\
B(m_1,m_2,m_3,m_4,m_5) &= m_1^2 m_2^2 m_3^2 - p^2 \left(m_1^2 m_2^2 + m_1^2 m_3^2 + m_2^2 m_3^2 + m_2^2 m_5^2 + m_3^2 m_4^2\right) \\
&+ p^4\left( m_1^2 + m_2^2 + m_3^2 + m_4^2 + m_5^2\right) - p^6
\end{split}
\end{align} 
The precise expressions for the source terms in this study are slightly different from both \cite{Barnard:2015ryq,carena2014}, so we present them in full for each representation. The expressions for the $SO(4)$ decomposed form factors are to be found originally in \cite{carena2014}. They are included here for completeness.

%

\subsection{MCHM$^{\textbf{5-5-5}}_{\textbf{5-5-5}}$}
\label{app:5-5-5}

\textbf{Top quark:}
\begin{align*}
\Pi_{t} &= \frac{\Delta_{q_t}^2}{(m_{Q_t} d_{Q_t})^2} + \hat{\Pi}_{q_t}^{(4)} + \hat{\Pi}_{q_b}^{(4)} + \frac{s_h^2}{2}\left( \hat{\Pi}_{q_t}^{(1)} -\hat{\Pi}_{q_t}^{(4)}\right)\\
\Pi_{t^c} &= \frac{\Delta_t^2}{(m_T d_T)^2} + \hat{\Pi}_t^{(4)} + (1-s_h^2)\left( \hat{\Pi}_t^{(1)} - \hat{\Pi}_t^{(4)}\right)\\
M_t &= \frac{1}{\sqrt{2}}s_h \sqrt{1-s_h^2}\left( \hat{M}_t^{(1)} - \hat{M}_t^{(4)}\right)
\end{align*}
\textbf{Bottom quark:}
\begin{align*}
\Pi_{b} &= \frac{\Delta_{q_b}^2}{(m_{Q_b} d_{Q_b})^2} + \hat{\Pi}_{q_t}^{(4)} + \hat{\Pi}_{q_b}^{(4)} + \frac{s_h^2}{2}\left( \hat{\Pi}_{q_b}^{(1)} -\hat{\Pi}_{q_b}^{(4)}\right)\\
\Pi_{b^c} &= \frac{\Delta_b^2}{(m_B d_B)^2} + \hat{\Pi}_b^{(4)} + (1-s_h^2)\left( \hat{\Pi}_b^{(1)} - \hat{\Pi}_b^{(4)}\right)\\
M_b &= \frac{1}{\sqrt{2}}s_h \sqrt{1-s_h^2}\left( \hat{M}_b^{(1)} - \hat{M}_b^{(4)}\right)
\end{align*}
\textbf{Tau lepton:}
\begin{align*}
\Pi_{\tau} &= \frac{\Delta_{l_\tau}^2}{(m_{L_\tau} d_{L_\tau})^2} + \hat{\Pi}_{l_\tau}^{(4)} + \hat{\Pi}_{l_\nu}^{(4)} + \frac{s_h^2}{2}\left( \hat{\Pi}_{l_\tau}^{(1)} -\hat{\Pi}_{l_\tau}^{(4)}\right)\\
\Pi_{\tau^c} &= \frac{\Delta_\tau^2}{(m_\mathcal{T} d_\mathcal{T})^2} + \hat{\Pi}_\tau^{(4)} + (1-s_h^2)\left( \hat{\Pi}_\tau^{(1)} - \hat{\Pi}_\tau^{(4)}\right)\\
M_\tau &= \frac{1}{\sqrt{2}}s_h \sqrt{1-s_h^2}\left( \hat{M}_\tau^{(1)} - \hat{M}_\tau^{(4)}\right)
\end{align*}
\textbf{Tau neutrino lepton:}
\begin{align*}
\Pi_{\nu} &= \frac{\Delta_{l_\nu}^2}{(m_{L_\nu} d_{L_\nu})^2} + \hat{\Pi}_{l_\nu}^{(4)} + \hat{\Pi}_{l_\tau}^{(4)} + \frac{s_h^2}{2}\left( \hat{\Pi}_{l_\nu}^{(1)} -\hat{\Pi}_{l_\nu}^{(4)}\right)\\
\Pi_{\nu^c} &= \frac{\Delta_\nu^2}{(m_\mathcal{N} d_\mathcal{N})^2} + \hat{\Pi}_\nu^{(4)} + (1-s_h^2)\left( \hat{\Pi}_\nu^{(1)} - \hat{\Pi}_\nu^{(4)}\right)\\
M_\nu &= \frac{1}{\sqrt{2}}s_h \sqrt{1-s_h^2}\left( \hat{M}_\nu^{(1)} - \hat{M}_\nu^{(4)}\right)
\end{align*}
with the $SO(4)$ decomposed form factors given by
\begin{equation}
\begin{aligned}\label{eq:broken5-5-5}
\hat{\Pi}_{q_{t/b}}^{(1)} &= \frac{A_L(m_{T/B}, 0, m_{Y_{T/B}} + Y_{T/B}, 0, \Lambda_{q_{t/b}})}{B(m_{Q_{t/b}}, m_{T/B}, 0, m_{Y_{T/B}} + Y_{T/B}, 0)}, & \hat{\Pi}_{q_{t/b}}^{(4)} &= \frac{A_L(m_{T/B}, 0, m_{Y_{T/B}}, 0, \Lambda_{q_{t/b}})}{B(m_{Q_{t/b}}, m_{T/B}, 0, m_{Y_{T/B}}, 0)}\\
\hat{\Pi}_{t/b}^{(1)} &= \frac{A_R(m_{Q_{t/b}}, 0, m_{Y_{T/B}} + Y_{T/B}, 0, \Lambda_{t/b})}{B(m_{Q_{t/b}}, m_{T/B}, 0, m_{Y_{T/B}} + Y_{T/B}, 0)}, & \hat{\Pi}_{t/b}^{(4)} &= \frac{A_R(m_{Q_{t/b}}, 0, m_{Y_{T/B}}, 0, \Lambda_{t/b})}{B(m_{Q_{t/b}}, m_{T/B}, 0, m_{Y_{T/B}}, 0)}\\
\hat{M}_{t/b}^{(1)} &= \frac{A_M(m_{Q_{t/b}}, m_{T/B}, 0, m_{Y_{T/B}} + Y_{T/B}, \Lambda_{q_{t/b}}, \Lambda_{t/b})}{B(m_{Q_{t/b}}, m_{T/B}, 0, m_{Y_{T/B}} + Y_{T/B}, 0)}, \\
\hat{M}_{t/b}^{(4)} &= \frac{A_M(m_{Q_{t/b}}, m_{T/B}, 0, m_{Y_{T/B}}, \Lambda_{q_{t,b}}, \Lambda_{t/b})}{B(m_{Q_{t/b}}, m_{T/B}, 0, m_{Y_{T/B}}, 0)}
\end{aligned}
\end{equation}

The same expressions apply for the leptonic form factors, with the substitutions $q \rightarrow l, t\rightarrow \tau, b\rightarrow \nu$.

\subsection{MCHM$^{\textbf{5-5-5}}_{\textbf{14-14-10}}$}
\label{app:14-14-10}

The quark expressions are as above.

\textbf{Tau lepton:}
\begin{align*}
\Pi_{\tau} &= \frac{\Delta_l^2}{(m_L d_L)^2} + \hat{\Pi}_l^{(9)} + \left(\hat{\Pi}_l^{(4)} -\hat{\Pi}_l^{(4)}\right)\left(1-\frac{s_h^2}{2}\right) + \frac{1}{4}s_h^2(1-s_h^2)\left(5\hat{\Pi}_l^{(1)} - 8 \hat{\Pi}_l^{(4)} + 4\hat{\Pi}_l^{(9)}\right)\\
\Pi_{\tau^c} &= \frac{\Delta_\tau}{(m_\mathcal{T} d_\mathcal{T})^2} + \hat{\Pi}_\tau^{(9)} + 2\left(\hat{\Pi}_\tau^{(4)} -\hat{\Pi}_\tau^{(9)}\right) \left( \frac{4}{5} - \frac{3}{4}s_h^2 \right) + \frac{1}{5}(4 - 5s_h^2)^2 \left( 5\hat{\Pi}_\tau^{(1)} - 8\hat{\Pi}_\tau^{(4)} + 3\hat{\Pi}_\tau^{(9)}\right) \\
M_\tau &= \frac{3 i}{2\sqrt{5}}\left( \hat{M}_\tau^{(4)} - \hat{M}_\tau^{(9)}\right) s_h \sqrt{1-s_h^2} + \frac{i}{8\sqrt{5}}(4-5 s_h^2)\left(5\hat{M}_\tau^{(1)} - 8\hat{M}_\tau^{(4)} + 3\hat{M}_\tau^{(9)}\right)
\end{align*}
\textbf{Tau neutrino lepton:}
\begin{align*}
\Pi_{\nu} &= \frac{\Delta_l^2}{(m_L d_L)^2} + \hat{\Pi}_l^{(9)} + (1-s_h^2)\left(\hat{\Pi}_l^{(4)} - \hat{\Pi}_l^{(9)}\right)\\
\Pi_{\nu^c} &= \frac{\Delta_\nu^2}{(m_\mathcal{N} d_\mathcal{N})^2} + \hat{\Pi}_\nu^{(6)} + \frac{1}{2}s_h^2 \left( \hat{\Pi}_\nu^{(4)} - \hat{\Pi}_\nu^{(6)}\right)\\
M_\nu &= \frac{-1}{\sqrt{2}}s_h \sqrt{1-s_h^2}\hat{M}_\nu^{(4)}
\end{align*}
with the $SO(4)$ decomposed form factors given by
\begin{equation}
\begin{aligned}\label{eq:broken14-14-10}
\hat{\Pi}_l^{(9)} &= \frac{A_L(m_\mathcal{T},0,m_{Y_\mathcal{T}},0,\Lambda_l)}{B(m_L,m_\mathcal{T},0,m_{Y_\mathcal{T}},0)}, & \hat{\Pi}_l^{(4)} &= \frac{A_L(m_\mathcal{T},m_\mathcal{V},m_{Y_\mathcal{T}} + Y_\mathcal{T}/2, Y_\mathcal{V}/2, \Lambda_l)}{B(m_L, m_\mathcal{T},m_\mathcal{V},m_{Y_\mathcal{T}} + Y_\mathcal{T}/2, Y_\mathcal{V}/2)},\\
\hat{\Pi}_l^{(1)} &= \frac{A_L(m_\mathcal{T}, 0, m_{Y_\mathcal{T}} + (Y_\mathcal{T}+\tilde{Y}_\mathcal{T})4/5,0,\Lambda_l)}{B(m_L,m_\mathcal{T},0,m_{Y_\mathcal{T}} + (Y_\mathcal{T} + \tilde{Y}_\mathcal{T})4/5,0)}\\
\hat{\Pi}_\tau^{(9)} &= \frac{A_R(m_L, 0, m_{Y_\mathcal{T}},0,\Lambda_\tau)}{B(m_L, m_\mathcal{T},0,m_{Y_\mathcal{T}},0)}, & 
\hat{\Pi}_\tau^{(4)} &= \frac{A_R(m_L,m_\mathcal{V},m_{Y_\mathcal{T}} + Y_\mathcal{T}/2, Y_\mathcal{V}/2, \Lambda_\tau)}{B(m_L, m_\mathcal{T},m_\mathcal{V},m_{Y_\mathcal{T}} + Y_\mathcal{T}/2, Y_\mathcal{V}/2)},\\
\hat{\Pi}_\tau^{(1)} &= \frac{A_R(m_L, 0, m_{Y_\mathcal{T}} + (Y_\mathcal{T}+\tilde{Y}_\mathcal{T})4/5,0,\Lambda_\tau)}{B(m_L,m_\mathcal{T},0,m_{Y_\mathcal{T}} + (Y_\mathcal{T} + \tilde{Y}_\mathcal{T})4/5,0)}\\
\hat{\Pi}_\nu^{(4)} &= \frac{A_R(m_L, m_\mathcal{T},Y_\mathcal{V}/2,m_{Y_\mathcal{T}}+Y_\mathcal{T}/2,\Lambda_\nu)}{B(m_L,m_\mathcal{T},m_\mathcal{V},m_{Y_\mathcal{T}} + Y_\mathcal{T}/2,Y_\mathcal{V}/2)}, & \hat{\Pi}_\nu^{(6)} &= \frac{A_R(m_L,0,0,0,\Lambda_\nu)}{B(m_L,m_\mathcal{V},0,0,0)},\\
\hat{M}_\tau^{(9)}&= \frac{A_M(m_L, m_\mathcal{T},0,m_{Y_\mathcal{T}}, \Lambda_l, \Lambda_\tau)}{B(m_L,m_\mathcal{T},0,m_{Y_\mathcal{T}},0)}, & \hat{M}_\tau^{(4)} &= \frac{A_M(m_L,m_\mathcal{T},m_\mathcal{V},m_{Y_\mathcal{T}} + Y_\mathcal{T}/2,\Lambda_l, \Lambda_\tau)}{B(m_L,m_\mathcal{T},m_\mathcal{V},m_{Y_\mathcal{T}} + Y_\mathcal{T}/2, Y_\mathcal{V}/2)},\\
\hat{M}_\tau^{(1)} &= \frac{A_M(m_L, m_\mathcal{T}, 0, m_{Y_\mathcal{T}} + (Y_\mathcal{T} + \tilde{Y}_\mathcal{T})4/5, \Lambda_l, \Lambda_\tau)}{B(m_L, m_\mathcal{T}, 0, m_{Y_\mathcal{T}} + (Y_\mathcal{T} + \tilde{Y}_\mathcal{T})4/5, 0)},\\
\hat{M}_\nu^{(4)} &= -i \frac{A_M(m_L, m_\mathcal{V}, m_\mathcal{T}, Y_\mathcal{V}/2, \Lambda_l, \Lambda_\nu)}{B(m_L, m_\mathcal{T}, m_\mathcal{V}, m_{Y_\mathcal{T}} +Y_\mathcal{T}/2, Y_\mathcal{V}/2)}
\end{aligned}
\end{equation}

\subsection{MCHM$^{\textbf{5-5-5}}_{\textbf{14-1-10}}$}
\label{app:14-1-10}

The quark expressions are as above.

\textbf{Tau lepton:}
\begin{align*}
\Pi_{\tau} &= \frac{\Delta_l^2}{(m_L d_L)^2} + \hat{\Pi}_l^{(9)} + \left(\hat{\Pi}_l^{(4)} -\hat{\Pi}_l^{(4)}\right)\left(1-\frac{s_h^2}{2}\right) + \frac{1}{4}s_h^2(1-s_h^2)\left(5\hat{\Pi}_l^{(1)} - 8 \hat{\Pi}_l^{(4)} + 4\hat{\Pi}_l^{(9)}\right)\\
\Pi_{\tau^c} &= \frac{\Delta_\nu}{(m_\mathcal{T} d_\mathcal{T})^2} + \hat{\Pi}_\tau^{(1)}\\
M_\tau &= \frac{-\sqrt{5}}{4} s_h \hat{M}_\tau^{(1)}
\end{align*}
\textbf{Tau neutrino lepton:}
\begin{align*}
\Pi_{\nu} &= \frac{\Delta_l^2}{(m_L d_L)^2} + \hat{\Pi}_l^{(9)} + (1-s_h^2)\left(\hat{\Pi}_l^{(4)} - \hat{\Pi}_l^{(9)}\right)\\
\Pi_{\nu^c} &= \frac{\Delta_\nu^2}{(m_\mathcal{V} d_\mathcal{V})^2} + \hat{\Pi}_\nu^{(6)} + \frac{1}{2}s_h^2 \left( \hat{\Pi}_\nu^{(4)} - \hat{\Pi}_\nu^{(6)}\right)\\
M_\nu &= \frac{-i}{\sqrt{2}}s_h \sqrt{1-s_h^2}\hat{M}_\nu^{(4)}
\end{align*}
with the $SO(4)$ decomposed form factors given by
\begin{equation}
\begin{aligned}\label{eq:broken14-1-10}
\hat{\Pi}_l^{(9)} &= \frac{A_L(0,0,0,0,\Lambda_l)}{B(m_L,0,0,0,0)}, & \hat{\Pi}_l^{(4)} &= \frac{A_L(0,m_\mathcal{V},0,Y_\mathcal{V}/2,\Lambda_l)}{B(m_L,0,m_\mathcal{V},0,Y_\mathcal{V}/2)},\\
\hat{\Pi}_l^{(1)} &= \frac{A_R(m_\mathcal{T},0,Y_\mathcal{T}\sqrt{4/5},0,\Lambda_l)}{B(m_L,m_\mathcal{T},0,Y_\mathcal{T}\sqrt{4/5},0)},\\
\hat{\Pi}_\tau^{(1)} &= \frac{A_R(m_L,0, Y_\mathcal{T}\sqrt{4/5},0,\Lambda_\tau)}{B(m_L,m_\mathcal{T},0,Y_\mathcal{T}\sqrt{4/5},0)}\\
\hat{\Pi}_\nu^{(4)} &= \frac{A_R(m_L, 0, Y_\mathcal{V}/2,0,\Lambda_\nu)}{B(m_L, 0, m_\mathcal{V}, 0, Y_\mathcal{V}/2)}, &
\hat{\Pi}_\nu^{(6)} &= \frac{A_R(0,0,0,0,\Lambda_\nu)}{B(0,m_\mathcal{V},0,0,0)},\\
\hat{M}_\tau^{(1)} &= -\frac{A_M(m_L, m_\mathcal{T}, 0, Y_\mathcal{T}\sqrt{4/5},\Lambda_l, \Lambda_\tau)}{B(m_L, m_\mathcal{T}, 0, Y_\mathcal{T}\sqrt{4/5},0)},\\
\hat{M}_\nu^{(4)} &= -i\frac{A_M(m_L, m_\mathcal{V}, 0, Y_\mathcal{V}/2, \Lambda_l, \Lambda_\nu)}{B(m_L, 0, m_\mathcal{V}, 0, Y_\mathcal{V}/2)}
\end{aligned}
\end{equation}

\section{NMCHM Expressions}

\textbf{Generic formulae}:
\begin{align}
\hat{\Pi}[m_1,m_2,m_3] &= \frac{(m_2^2 + m_3^2 - p^2)\Delta^2}{p^4 - p^2(m_1^2 + m_2^2 + m_3^2)+m_1^2 m_2^2},\\
\hat{M}[m_1,m_2,m_3] &= \frac{m_1 m_2 m_3 \Delta^2}{p^4 - p^2(m_1^2 + m_2^2 + m_3^2) + m_1^2 m_2^2}
\end{align}
\textbf{Broken correlators}:
\begin{align}
\hat{\Pi}_0^{q_L} &= \hat{\Pi}[m_{T},m_{\tilde{T}},m_{Y_T}], & \hat{\Pi}_1^{q_L} &= \hat{\Pi}[m_{T},m_{\tilde{T}},m_{Y_T} + Y_T] - \hat{\Pi}[m_{T},m_{\tilde{T}},m_{Y_T}], \\
\hat{\Pi}_0^{u_R} &= \hat{\Pi}[m_{\tilde{T}},m_T,m_{Y_T}], & \hat{\Pi}_1^{u_R} &= \hat{\Pi}[m_{\tilde{T}},m_{T},m_{Y_T} + Y_T] - \hat{\Pi}[m_{\tilde{T}},m_{T},m_{Y_T}], \\
\hat{M}_0^{u} &= \hat{M}[m_{T},m_{\tilde{T}},m_{Y_T}], & \hat{M}_1^{u} &= \hat{\Pi}[m_{T},m_{\tilde{T}},m_{Y_T} + Y_T] - \hat{\Pi}[m_{T},m_{\tilde{T}},m_{Y_T}].
\end{align}
\textbf{Full correlators}:
\begin{align}
\Pi_0^q &= \frac{1}{y_{t_L}^2} + \hat{\Pi}_0^{q_L}, &\Pi_1^{q_1} &= \hat{\Pi}_1^{q_L}, \\
\Pi_0^u &= \frac{1}{y_{t_R}^2} + \hat{\Pi}_0^{u_R} + s_{\theta}^2 \hat{\Pi}_1^{u_R}, & \Pi_1^u &= -2\hat{\Pi}_1^{u_R},\\
M_1^u &= \hat{M}_1^u.
\end{align}
\textbf{Potential}:
\begin{align}
V(\varphi,\psi) \approx c_1 s^2_\varphi c^2_\psi + c_2 s^2_\varphi (s^2_\theta - c^2_\theta s^2_\psi) - c_3 s^2_\varphi c^2_\psi (c^2_\theta s^2_\varphi s^2_\psi + s^2_\theta c^2_\varphi)
\end{align}
\textbf{Potential terms}:
\begin{align}
c_1 &= -N_c \int\frac{d^4 p}{(2\pi)^4}\frac{\Pi_1^{q_1}}{\Pi_0^q} + V(h)_\textnormal{gauge}, \qquad  c_2 &= -N_c \int\frac{d^4 p}{(2\pi)^4} \frac{\Pi_1^u}{\Pi_0^u}, \\
c_3 &= -N_c \int\frac{d^4 p}{(2\pi)^4}\frac{(M_1^u)^2}{\left(\Pi_0^q + s^2_\varphi c^2_\psi \Pi_1^{q_1}/2\right)\left(\Pi_0^u + s^2_\varphi c^2_\psi s^2_\theta \Pi_1^u/2\right)},
\end{align}
where 
\begin{align}
V(h)_\textnormal{gauge} & \approx \frac{9}{64\pi^2}\frac{g_0^2}{g_\rho^2}\frac{m_\rho^4 (m_{a_1}^2 -m_\rho^2)}{m_{a_1}^2 - m_\rho^2 (1 + g_0^2/g_\rho^2)}\ln\left[ \frac{m_{a_1}^2}{m_\rho^2(1+ g_0^2/g_\rho^2)}\right]
\end{align}

Now, we can solve the potential $V(\varphi,\psi)$ given a local minimum with a positive second derivative. This leads to $\psi \rightarrow 0$ and $s_\varphi \rightarrow 1$.
%
\clearpage
\nocite{*}
\printbibliography
\clearpage
\manualmark
\markboth{\spacedlowsmallcaps{\indexname}}{\spacedlowsmallcaps{\indexname}}
\phantomsection
\begingroup 
    \let\clearpage\relax
    \let\cleardoublepage\relax
    \let\cleardoublepage\relax
\pagestyle{scrheadings} 
\addcontentsline{toc}{chapter}{\tocEntry{\indexname}}
\printindex
\endgroup 

\end{document}